\documentclass[PhD]{dukethesis2006}

\usepackage{amsmath}
\usepackage{amssymb}
\usepackage{amsthm}
\usepackage{array}
\usepackage{epsfig}
\usepackage{graphicx}
\usepackage{xy}
\usepackage{hyperref}
\usepackage{bm}
\usepackage[toc,page]{appendix}

\newcommand{\trento}{\texttt{T$_R$ENTo} }

\newcommand{\fig}{\begin{figure}[htbp]\centering}

\newcommand{\efig}{\end{figure}} 

\newcommand{\eq}{\begin{equation}}
\newcommand{\eeq}{\end{equation}}
\newcommand{\eqa}{\begin{eqnarray}}
\newcommand{\eeqa}{\end{eqnarray}}

\newcommand{\singlespacing}{%
  \let\CS=\small\renewcommand{\baselinestretch}{1.0}\CS}
\newcommand{\doublespacing}{%
  \let\CS=\small\renewcommand{\baselinestretch}{1.6}\CS}

\author{Wenkai Fan}
\advisor{Steffen Bass}
\member{Ayana Arce}
\member{Berndt M$\ddot{\textrm{u}}$ller}
\member{Phillip Barbeau}
\member{Thomas Mehen}

\department{Physics}
\title{Multi-Stage Heavy Quark Transport in Ultra-relativistic Heavy-ion Collisions}

\begin{document}

\pagenumbering{roman}

\maketitle

\makeabstract

\Copyright

\abstract

The quark gluon plasma (QGP) is one of the most interesting forms of matter providing us with insight on quantum chromodynamics (QCD) and the early universe. It is believed that the heavy-ion collision experiments at the Relativistic Heavy Ion Collider (RHIC) and the Large Hadron Collider (LHC) have created the QGP medium by colliding two heavy nuclei at nearly the speed of light. Since the collision happens really fast, we can not observe the QGP directly. Instead, we look at the hundreds or even thousands of final hadrons coming out of the collision. In particular, jet and heavy flavor observables are excellent probes of the transport properties of such a medium. On the theoretical side, computational models are essential to make the connections between the final observables and the plasma. Previously studies have employed a comprehensive multistage modeling approach of both the probes and the medium. 

In this dissertation, heavy quarks are investigated as probes pf the QGP. First, the framework that describes the evolution of both soft and hard particles during the collision is discussed, which includes initial condition, hydrodynamical expansion, parton transport, hadronization, and hadronic rescattering. It has recently been organized into the Jet Energy-loss Tomography with a Statistically and Computationally Advanced Program Envelope (JETSCAPE) framework, which allows people to study heavy-ion collision in a more systematic manner. 

To study the energy loss of hard partons inside the QGP medium, the linear Boltzmann transport model (LBT) and in medium DGLAP evolution (implemented in the MATTER model) are combined and have achieved a simultaneous description of both charged hadron, D meson, and inclusive jet observables. To further extract the transport coefficients, a Bayesian analysis is conducted which constrains the parameters in the transport models. 

\acknowledgements

I would like to first thank my advisor, Prof. Steffen A. Bass, for his guidance and support during my study at Duke. I am really grateful for all the opportunities he had provided and the discussion we had over the years. I would also like to show gratitude to my committee – both from my preliminary exam and defense, including Prof. Arce, Prof. Barbeau, Prof. Mehen, and Prof. Mueller. 

Next, I would like to thank my collaborators at Goethe University Frankfurt am Main, Dr. Lucia Olivia and Prof. Elena Bratkovskaya. With our project in small systems, I was able to get to know and practice the various models used in our group. I would also like to thank my collaborators at Wayne State University, Dr. Gojko Vujanovic, Dr. Amit Kumar, and Prof. Abhijit Majumder. Without their help, my project on heavy flavor in the JETSCAPE collaboration wouldn't be possible. 

Furthermore, I thank all my former and current colleagues in the Duke QCD group - Dr. Jonnah Bernhard, Dr. Scott Moreland, Dr. Xiao-Jun Yao, Dr. Jean-Francois Paquet, Dr. Ying-Ru Xu, Dr. Weiyao Ke, Dr. Pierre Moreau and Tianyu Dai for their generous help and enlightening discussions. I am really proud to be a contributor to the development and application of the Duke framework for heavy ion collision study. 

In the end, my full gratitude goes to my parents and my wife. It is their love and support, especially during the past three years, helped me go through all the difficult times in my life.

\tableofcontents

\listoffigures

\listoftables

\textspace

\chapter{Introduction} \label{sec:intro}

\pagenumbering{arabic}

\vspace{1in}

Physics is the natural science that studies matter, its fundamental constituents, its motion and behavior through space and time, and the related entities of energy and force. The goal is to understand how the universe behaves. With the development of modern physics since the $20^{th}$ century, the standard model of particle physics is now generally accepted as the fundamental theory which predicts 61 elementary particles categorized into fermions (including quarks and leptons) and bosons (like photons, gluons, and the Higgs boson). Their interactions are divided into three fundamental forces: the electromagnetic force which satisfies $U(1)$ symmetry and is described by quantum electromagnetic dynamics (QED), the weak force which satisfies $SU(2)$ symmetry and can be unified with QED under $U(1)\times SU(2)$ symmetry and the strong force which satisfies $SU(3)$ symmetry and is described by quantum chromodynamics (QCD). The fourth fundamental force, gravity, has yet to be unified with the other three fundamental forces.

In this dissertation, I would like to study the properties of QCD in a special form of matter, namely the quark gluon plasma (QGP). QGP is believed to exist in particle collider experiments by colliding two heavy nuclei at nearly the speed of light. The thermodynamic and transport properties of the QGP can then be inferred from the distribution of the final hadrons produced during the collision. 

\section{Quantum chromodynamics and nuclear matter}

Quantum chromodynamics, which is believed to describe the strong force between elementary particles, has two interesting features called asymptotic freedom and color confinement. Because of confinement, the fundamental degrees of freedom in QCD, namely quarks and gluons, are not observed in the nuclear matter under normal conditions. Instead, composites of quarks and gluons called hadrons are observed. However, under extreme temperature and pressure, hadrons should undergo a phase transition and break into quarks and gluons again due to the asymptotic freedom property of QCD that causes interactions between particles to become weaker as the energy scale increases. 

\subsection{The QCD Lagrangian}

The Lagrangian of QCD can be written concisely as:
\begin{equation}\label{eqn:QCD_lag}
    \mathcal{L}_{QCD}  = \bar{\Psi} (i \gamma_{\mu} D^{\mu} - m) \Psi - \frac{1}{4} F^a_{\mu\nu} F^{\mu\nu}_a.
\end{equation}
$\Psi$ is the spinor of the quark field with $N_c=3$ colors and $N_f$ flavors. $\gamma^{\mu}$ are the Dirac matrices and m is the quark mass matrix. $D_{\mu}=\partial_{\mu}-igT_a A^a_\mu$ is the covariant derivative where $g=\sqrt{4\pi \alpha_s}$ is the coupling strength. The gluon field strength tensor is defined as:
\begin{equation}\label{eqn:gluon_field}
    F^a_{\mu\nu} = (\partial_{\mu} A_{\nu}^a - \partial_{\nu}A^a_{\mu} + g f^a_{bc} A_{\mu}^b A_{\nu}^c),
\end{equation}
where $A_a^\mu$ is the gluon field and $T_a$ is the generator of the local $SU(N_c)$ symmetry. The first term in the equation above is the kinetic term, while the second term is the gluon field self-interaction, which is a unique feature of non-Abelian gauge field theory. 

\begin{figure}[!h]
\centering
\includegraphics[width=0.75\textwidth]{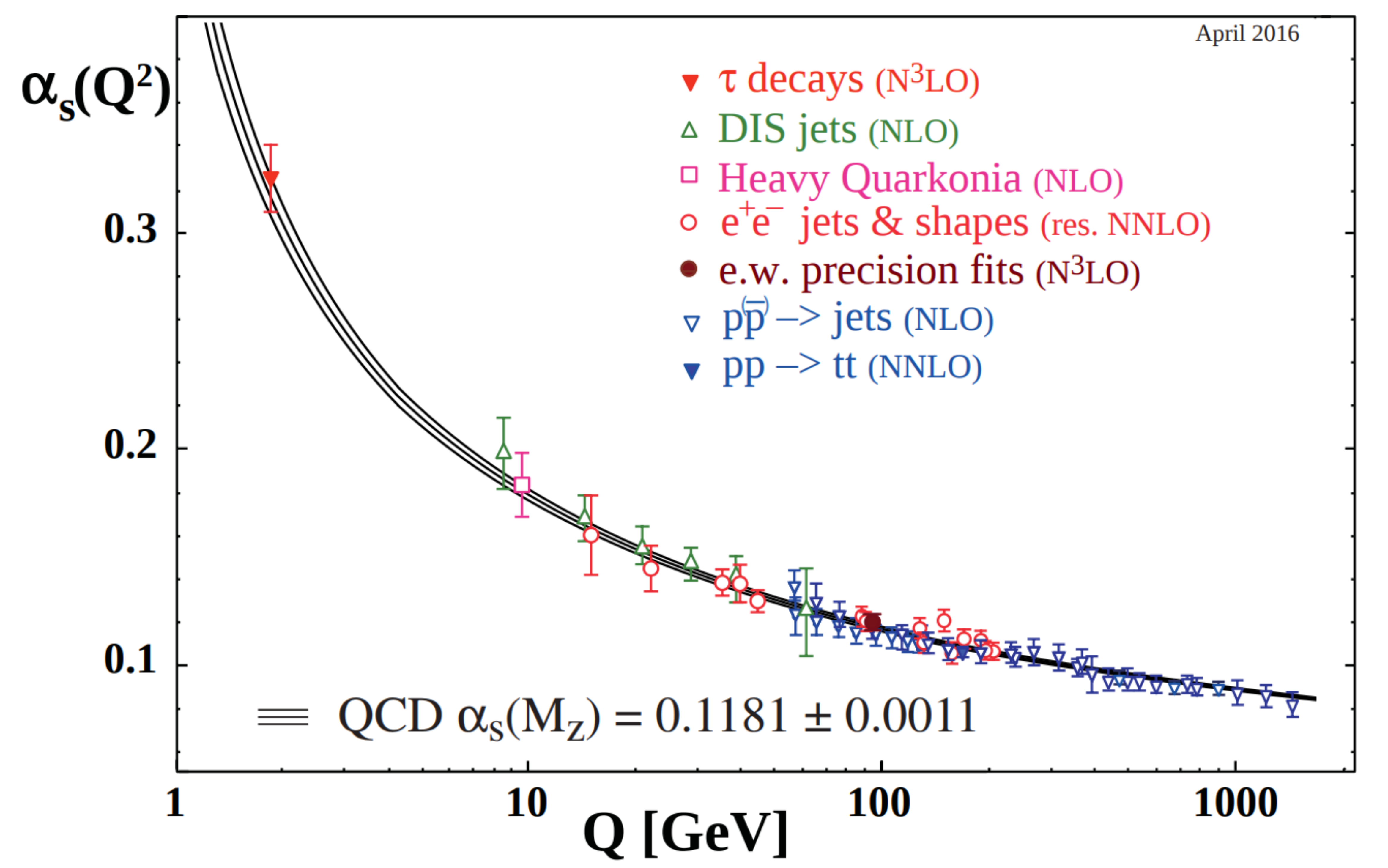}
\caption[Summary of measurements of $\alpha_s$]{Summary of measurements of $\alpha_s$ as a function of energy scale $Q$ from various measurements \cite{particle1998review}.} \label{fig:QCD_alphas}
\end{figure}

One of the most remarkable properties of the QCD is the fact that the strong coupling constant becomes small for processes involving large momentum transfer $Q^2$. Because of the self interacting term in Eq.~\ref{eqn:gluon_field}, the sign of the $\beta$ function is negative. This means the coupling constant becomes small at shorter distance (asymptotic freedom) and large at large distance (color confinement). In Fig.~\ref{fig:QCD_alphas} we can see that experimental measurements do confirm this behavior: 
\begin{equation}
	\alpha_s(Q^2) = \frac{\alpha_s(Q_0^2)}{1 + \frac{\alpha_s(Q_0^2)}{12 \pi} (11 N_c - 2 N_f) \log\left(\frac{Q^2}{Q_0^2}\right)}.
	\label{eqn:alphaS_1}
\end{equation}

If we define a scale parameter $\Lambda_{QCD}$ by $\frac{1}{\alpha_s(Q_0^2)}=\frac{11N_c-2N_f}{12\pi}\log(Q_0^2/\Lambda_{QCD}^2)$, we can further simplify the above expression into:
\begin{equation}
	\alpha_s(Q^2) = \frac{12 \pi}{(11 N_c - 2 N_f) \log\left(\frac{Q^2}{\Lambda_{\rm QCD}^2}\right)}.
	\label{eqn:alphaS_2}
\end{equation}
$\Lambda_{QCD}$ is around $200$~MeV. When the momentum-transfer approaches $\Lambda_{QCD}^2$ from the above, the coupling becomes too large for perturbation theory to be applicable.

\subsection{The QCD phase diagram}

\begin{figure}[!h]
\centering
\includegraphics[width=0.9\textwidth]{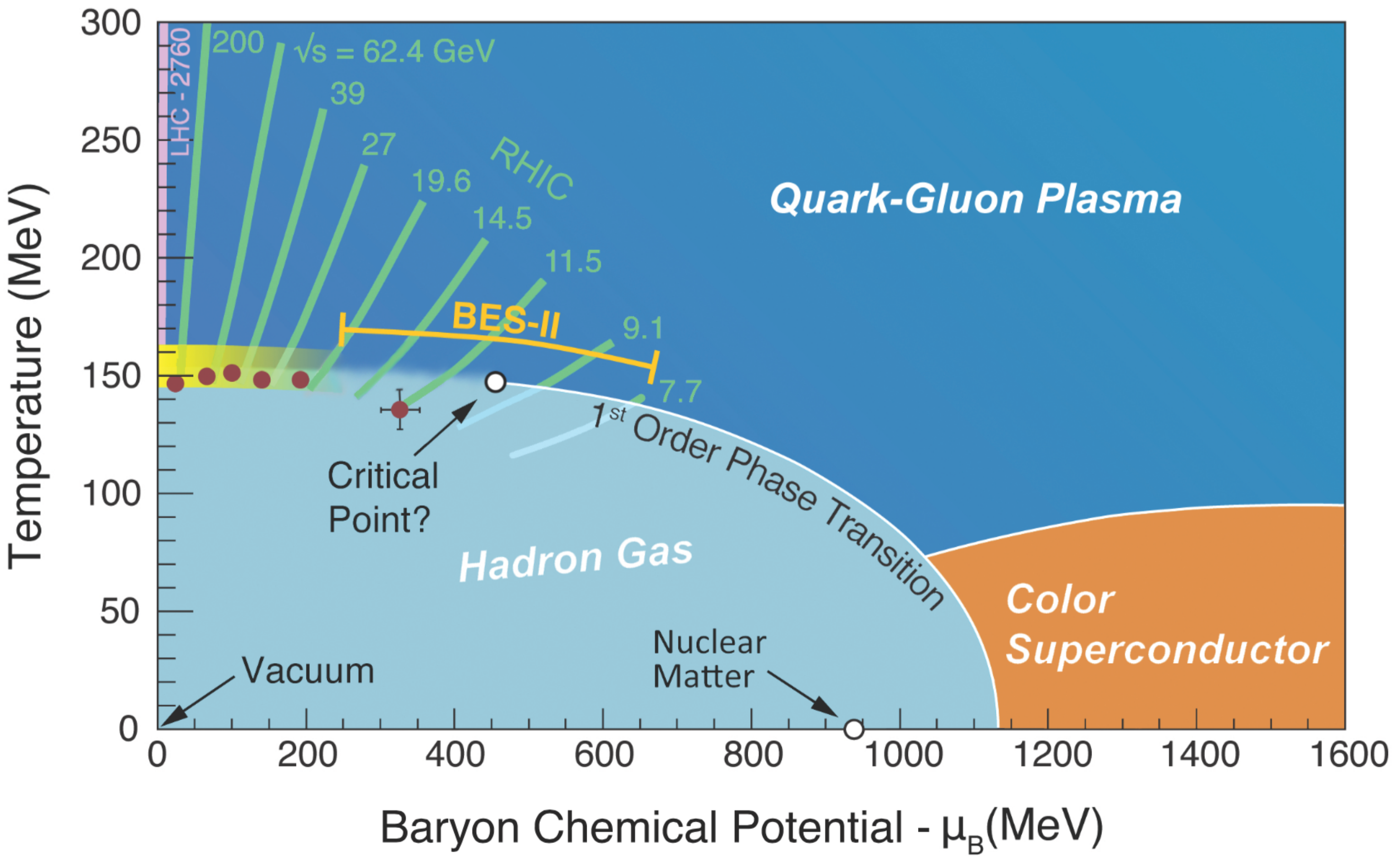}
\caption[Conjectured phase diagram of the nuclear matter]{The conjectured phase diagram of the nuclear matter \cite{20152017reaching}. The red dots are from lattice calculation \cite{bazavov2012freeze}. The pink and green curves indicates the reachable regions by the heavy ion program at the LHC and RHIC. The beam energy scan experiments (denoted by the orange line segment) is trying to determine whether we have a first order phase transition and the critical end point (CEP) on the phase diagram.}\label{fig:QCD_phase_diagram}
\end{figure}

The QCD phase diagram (Fig.~\ref{fig:QCD_phase_diagram}) is the phase diagram that describes the thermodynamics of matter which dominantly interact via strong force. 

\begin{figure}[!h]

\centering
\includegraphics[width=0.75\textwidth]{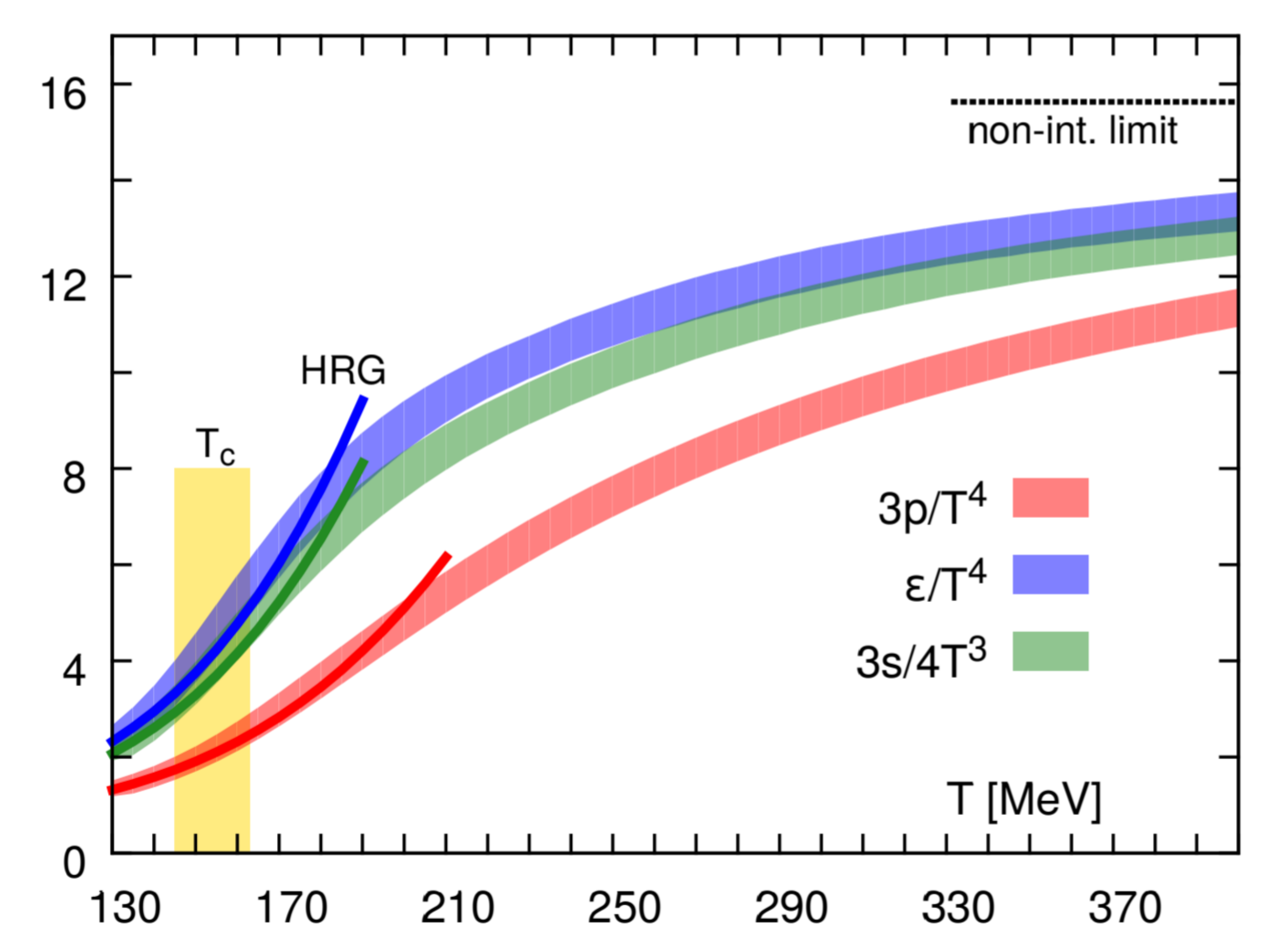}
\caption[Lattice equation of state for (2+1) flavor QCD]{The lattice equation of state for (2+1) flavor QCD taken from \cite{Bazavov:2014pvz}. The pressure, energy density and entropy density as functions of temperature at zero $\mu_B$ are shown as red, blue, and green bands.
The dashed lines denotes non-interacting (Stephan Boltzmann) limit, and the solid
lines show the expected values from a hadron resonance gas.}\label{fig:lattice_phase_transition}
\end{figure}

At asymptotically high temperature, the decrease in the coupling strength should lead to the transition from hadronic matter to a system of deconfined quarks and gluons, called the quark gluon plasma (QGP). First principle lattice QCD calculations \cite{Bazavov:2014pvz} have studied this transition at zero baryon chemical potential with three flavors (up, down, and strange). In Fig.~\ref{fig:lattice_phase_transition}, we can see that this transition is a smooth cross-over and has a pseudo-critical temperature around $150$~MeV. The lattice results agree very well with the hadron gas model at low temperature and approach the non-interacting limit at high temperature. 

At finite baryon densities, the lattice approach is plagued by the well-known sign problem \cite{gattringer2016approaches} and can not produce reliable results. Phenomenological models like the NJL model \cite{BUBALLA2005205} have predicted a first-order phase transition at large baryon chemical potential $\mu_B$ and low temperature $T$. A baryon density this large is not yet achievable in laboratories but is believed to exist at the center of dense celestial bodies like neutron stars. Moreover, suppose this first-order phase transition does exist, there must be a critical endpoint (CEP) on the phase diagram that separates the crossover phase transition at small $\mu_B$ and the first-order phase transition at large $\mu_B$. The beam energy scan program at RHIC \cite{bzdak2020mapping} is searching for such a CEP by colliding different species of heavy nuclei at different collision energies.

\vspace{0.2in} 

\section{Relativistic heavy ion collision}

Relativistic heavy ion collisions are currently the only experimental way to access high energy density QCD medium in a laboratory. Since 2000, the relativistic heavy ion collider (RHIC) has been colliding gold nuclei at $200$~GeV. Shortly after, the large hadron collider (LHC) started colliding lead nuclei at $2.76$~TeV and $5.02$~TeV. Emerging evidence has been pointing to a new state-of-matter: the strongly interacting quark gluon plasma (QGP).

Two heavy nuclei are accelerated to nearly the speed of light and collide head-on. They are highly contracted in the colliding direction which can be thought of two pancakes colliding with each other. The energy density in the overlapping region is so high that nuclear matter should go through the crossover phase transition and dissolve into quarks and gluons. The system would then expand and cool down due to internal pressure and then hadronize into hadrons which various surrounding detectors would detect. 

How do people confirm that this is indeed what happened during the collision? Collective flow and jet quenching were the first supporting observations. In order to explain what they mean, some basic terminologies need to be introduced first.

\subsection{Kinematics in heavy ion collisions}

In ultra relativistic heavy ion collisions, it is common to use a new set of coordinates $(x,y,\eta_s, \tau)$, which are related to the Cartesian coordinates $(x,y,z,t)$ by:
\begin{equation}
    \tau=\sqrt{t^2-z^2},
\end{equation}
\begin{equation}
    \eta_s=\frac{1}{2}\ln\frac{t+z}{t-z}.
\end{equation}

The $z$ direction is where the two nuclei are moving. $\tau$ is called the proper time and $\eta_s$ is called the space-time rapidity. The advantage of using $\tau$ and $\eta_s$ is that their Lorentz transformation is much simpler:
\begin{equation}
    \tau^{'}=\tau,
\end{equation}
\begin{equation}
    \eta_s^{'}=\eta_s+\frac{1}{2}\ln\frac{1+\beta_z}{1-\beta_z},
\end{equation}
where $\beta_z$ is the velocity of a Lorentz boost in the $z$ direction.

The four momentum is parametrized as:
\begin{equation}
    p_x=p_T\cos\phi,
\end{equation}
\begin{equation}
    p_y=p_T\sin\phi,
\end{equation}
\begin{equation}
    m_T=\sqrt{m^2+p_T^2},
\end{equation}
\begin{equation}
    y=\frac{1}{2}\ln\frac{E+p_z}{E-p_z},
\end{equation}
where $p_T$ is the momentum transverse to the $z$ direction. $\phi$ is the azimuth angle. $m_T$ is called the transverse mass and $y$ is called the rapidity. There is also the pseudorapidity defined as 
\begin{equation}
    \eta=\frac{1}{2}\ln\frac{|p|+p_z}{|p|-p_z}=\frac{1}{2}\ln\frac{1+\cos\theta}{1-\cos\theta},
\end{equation}
where $\eta$ is directly related to the polar angle $\theta$ and is close to $y$ when $m_T \ll p_z$.

\subsection{Impact parameter and centrality selection}

Nuclei are extended objects. The radius of heavy nuclei scales approximately to the $1/3$ power of the atomic number. In the center-of-mass (COM) frame of the collision, the nuclei Lorentz contract in the $z$ direction by a factor of about $100$ for gold nuclei at RHIC and larger than $2500$ for lead nuclei at LHC. 

\begin{figure}[!h]
\centering
\includegraphics[width=0.66\textwidth]{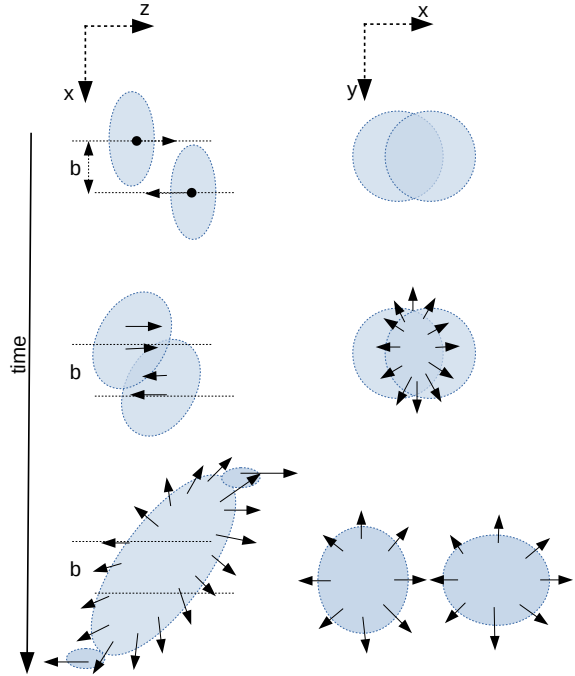}
\caption[Evolution of collision geometry]{\textbf{Left}: A top view of the geometry of the collision system at different times (from top to bottom). \textbf{Right}: The time evolution of the geometry of the collision system in the transverse plane.}\label{fig:reaction_plane}
\end{figure}

As seen from Fig.~\ref{fig:reaction_plane}, since the collision is not always head-on, the overlapping region is like an almond shape in the transverse plane (if we collide two identical nuclei). The transverse distance between the center-of-mass of the two nuclei is defined as the impact parameter $b$. The collision geometry and energy deposition depend largely on $b$. However, in experiments, it is impossible to control or measure the impact parameter. What is used is a proxy called centrality. The idea is that since the collision geometry correlates strongly with the particle production, it is reasonable to assume that the impact parameter $b$ has a negative correlation with the number of final charged particles $N_{ch}$ produced via the collision (multiplicity). Experimentalists make histograms of the multiplicity and binned them into different percentiles. The $0-5\%$ percentile events with the highest multiplicity are associated with the $0-5\%$ centrality and are usually called the most central collisions. Events in the lowest multiplicity percentile are usually called the most peripheral collisions. The map from multiplicity to collision geometry (e.g., the impact parameter) is usually done by some sort of Glauber model \cite{Miller:2007ri}.

\begin{figure}[!h]
\centering
\includegraphics[width=0.66\textwidth]{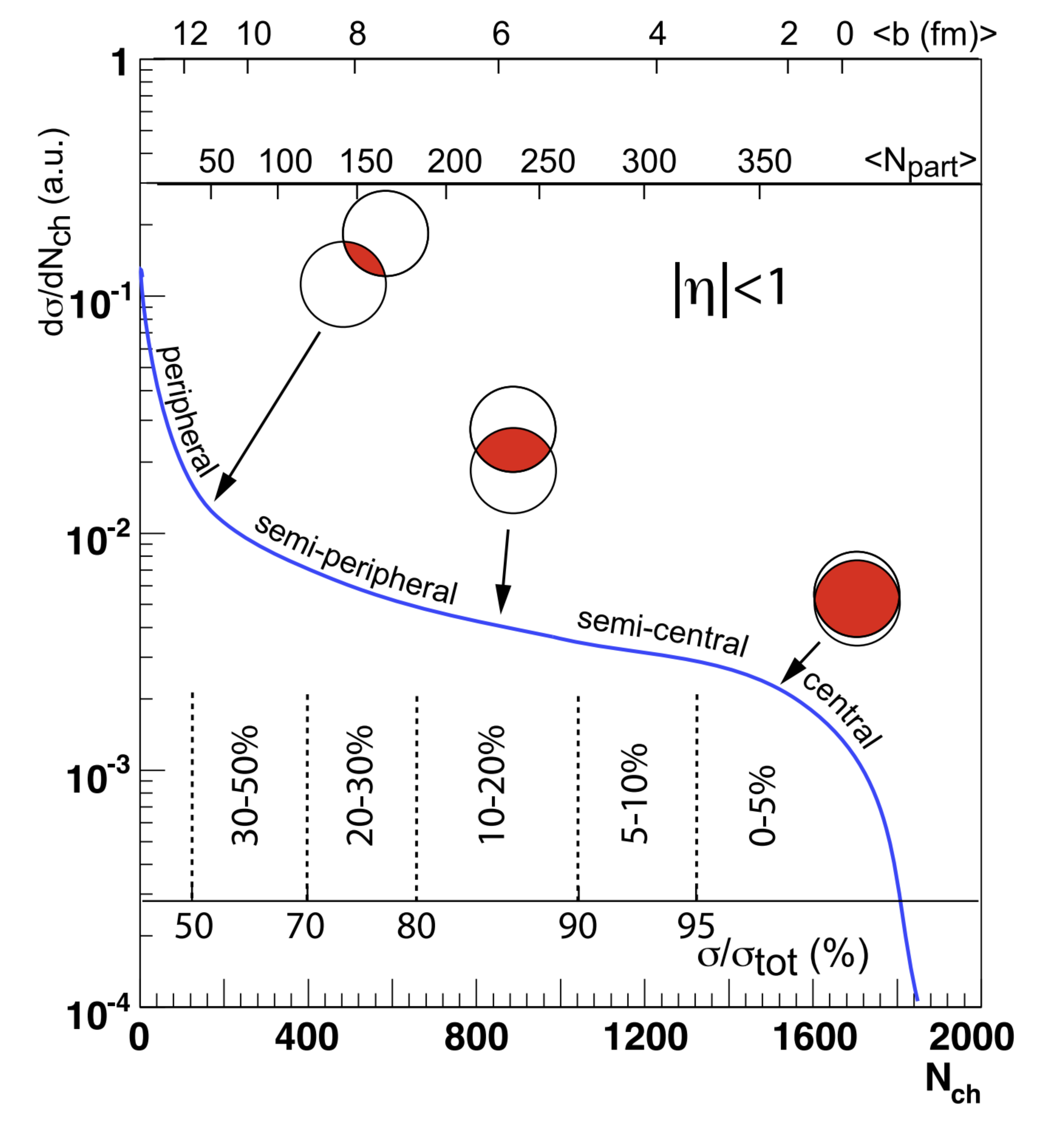}
\caption[Correlation of the final state observable $N_{ch}$ with Glauber calculated quantities]{A cartoon example of the correlation of the final state observable $N_{ch}$ with Glauber calculated quantities ($b$, $N_{part}$).
}\label{fig:glauber_Nch_histogram}
\end{figure}

\subsection{Collective flow}

\begin{figure}[!h]
\centering
\includegraphics[width=0.95\textwidth]{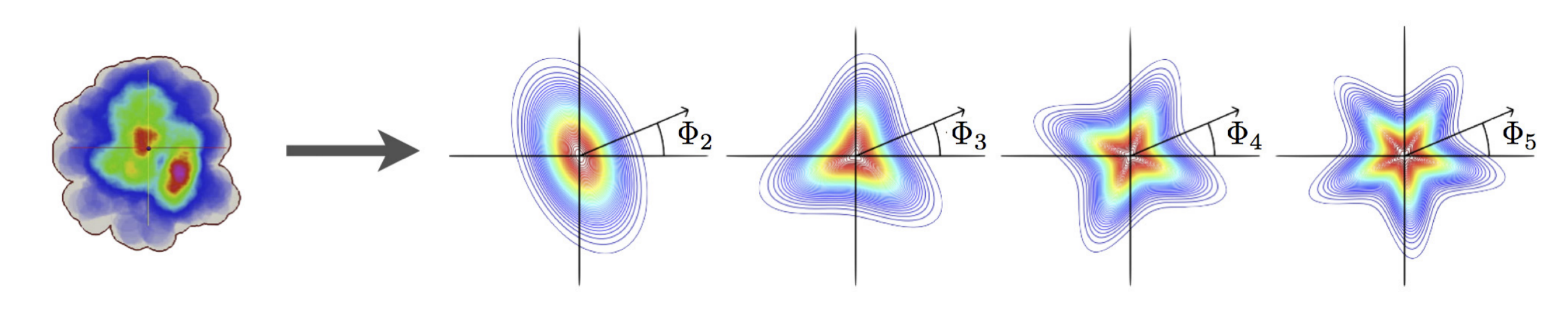}
\caption[Decomposition of one initial condition into its first 4 harmonic deformations]{Decomposition of one initial condition into its first 4 harmonic deformations \cite{qiu2013event}.} \label{fig:harmonic_decomposition}
\end{figure}

\begin{figure}[!h]
\centering
\includegraphics[width=0.7\textwidth]{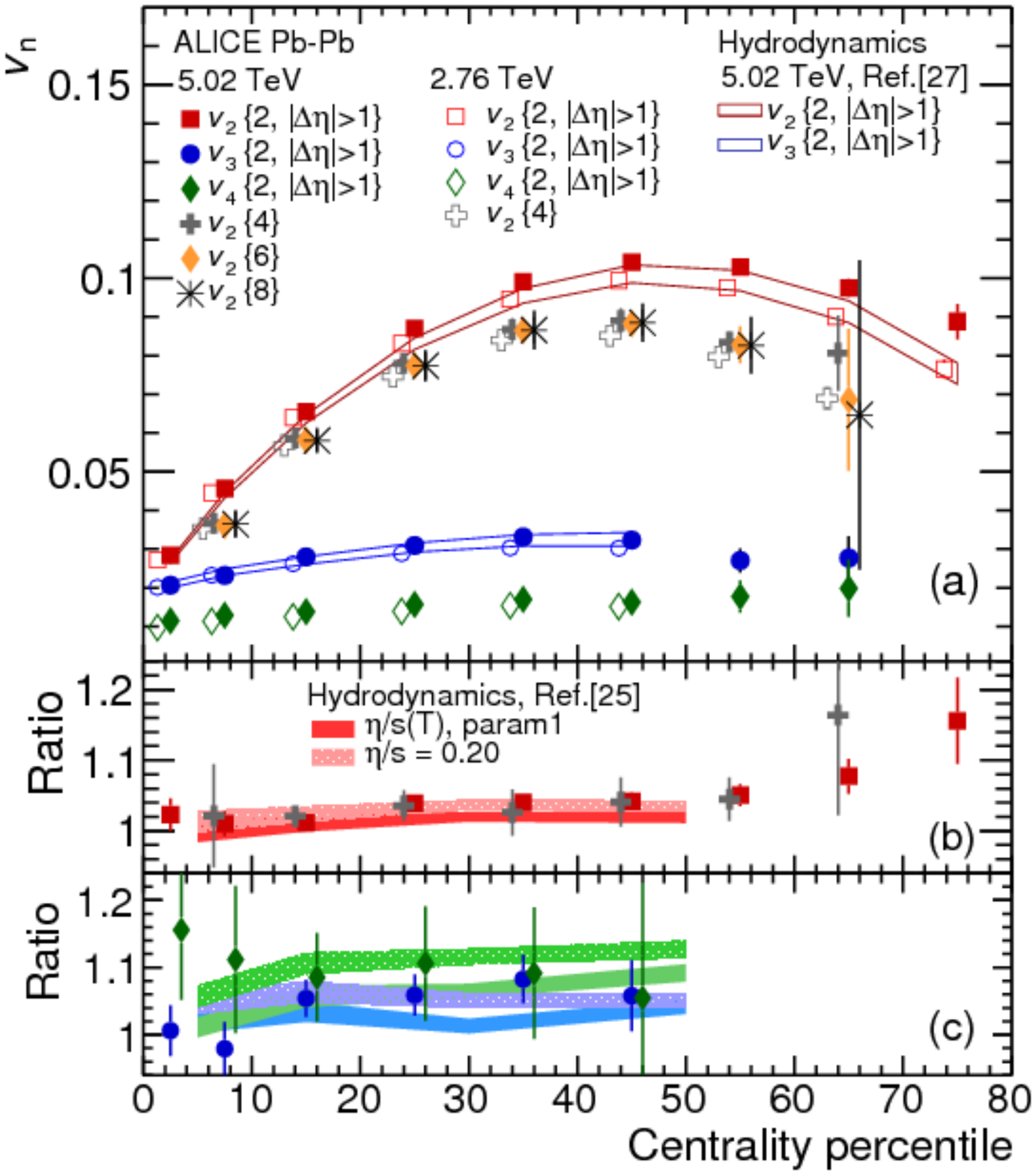}
\caption[Anisotropic flow for two-particle and multi-particle correlation]{Anisotropic flow integrated over $0.2< pT<5$ GeV for two-particle and multi-particle correlation as a function of centrality for PbPb collisions at $2.76$ and $5.02$ TeV. Comparing the hydrodynamical model prediction \cite{Noronha-Hostler:2015uye, Niemi:2015voa} with ALICE measurements \cite{ALICE:2011ab}.} \label{fig:alice_flow}
\end{figure}

People originally thought the QGP would behave like a gas due to the small coupling strength at high temperatures. However, collective flow data from RHIC has very good agreement with ideal hydrodynamic calculation. Collective flow means the final hadrons are moving collectively in a specific direction. Azimuthal anisotropic flow is related to particle motion in the transverse plane. As seen from Fig.~\ref{fig:reaction_plane}, the overlapping region is like an almond shape in the transverse plane, so the final hadrons' angular distribution is not expected to be uniform from such initial collision geometry. To quantify this non-uniformity, one can expand the final state particle azimuthal distribution as a Fourier series:
\begin{equation}
    \frac{d^3 N}{p_T dp_T dy d\phi}(p_T, y, \phi) = \frac{1}{2\pi} \frac{d^2N}{p_Tdp_Tdy}\left[1 + \sum_{n=1}^{\infty} 2v_n(p_T, y) \cos[n(\phi - \Psi^{\rm RP}_n)]\right],
\end{equation}
where $E, p_T, y, \phi$ are the energy, transverse momentum, rapidity, and azimuthal angle of the particle, $\Psi^{\rm RP}$ is the reaction plane angle associated with the initial density distribution. The Fourier coefficients $v_n(p_T, y)$, among which the first three are named as direct($v_1$), elliptic($v_2$) and triangular($v_3$) flow, characterize the geometric anisotropy of the system, they are given by:
\begin{equation}
v_n e^{in\Psi^{\rm EP}_n} = \frac{\int p_T dp_T dy d\phi e^{in\phi} \frac{dN}{dyp_Tdp_Td\phi}}{\int p_T dp_T dy d\phi \frac{dN}{dyp_Tdp_Td\phi}} = \left<e^{in\phi}\right>.
\end{equation}

\begin{figure}[!h]
\centering
\includegraphics[width=0.95\textwidth]{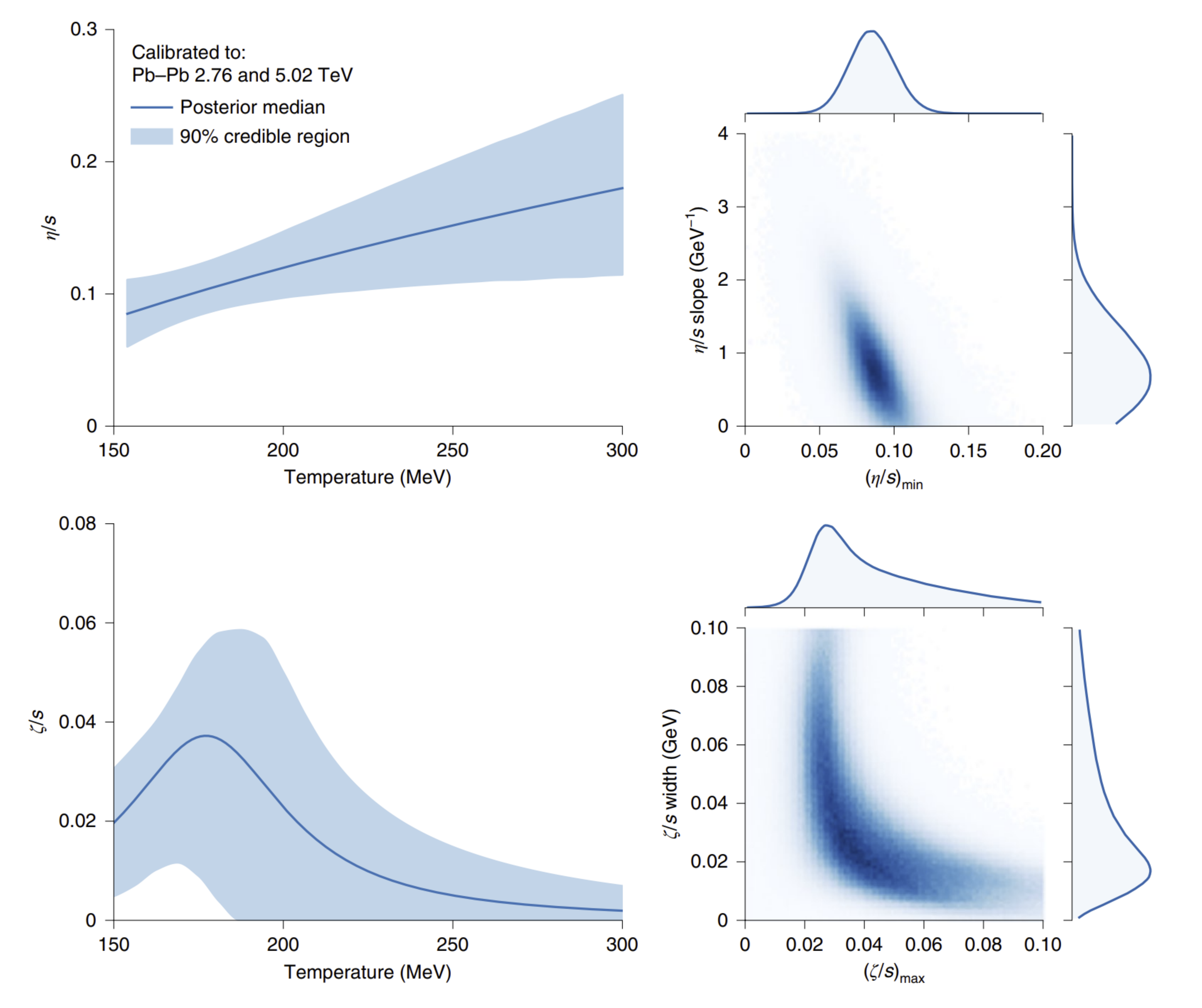}
\caption[Estimated temperature dependence of the specific shear and bulk viscosity]{Estimated temperature dependence of the specific shear and bulk viscosity. \textbf{Left column}: posterior medians and $90\%$ credible regions for $\eta/s(T)$ and $\varepsilon/s(T)$ estimated from PbPb collision data at $2.76$ TeV and $5.02$TeV. \textbf{Right column}: one dimensional (1D) histograms showing the marginal
distributions for the indicated parameters, along with 2D density histograms showing the joint distributions between the parameters. \textbf{Top row}: shear viscosity. \textbf{Bottom row}: bulk viscosity. \cite{bernhard2019bayesian}} \label{fig:hydro_viscocity_posterior}
\end{figure}

The angular bracket denotes the average over particles of interest in all selected events, $\Psi^{\rm EP}_n$ is the event plane angle that points to the direction where the $n^{\rm th}$ harmonic coefficient is the largest.

\begin{figure}[!h]
\centering
\includegraphics[width=0.99\textwidth]{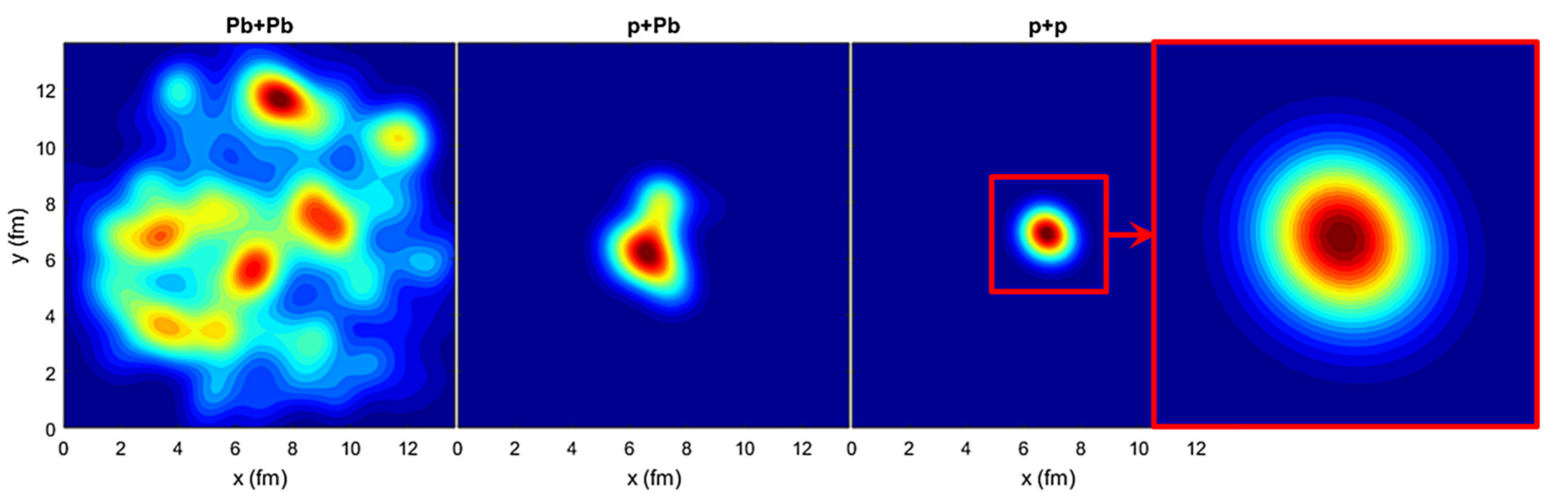}
\includegraphics[width=0.99\textwidth]{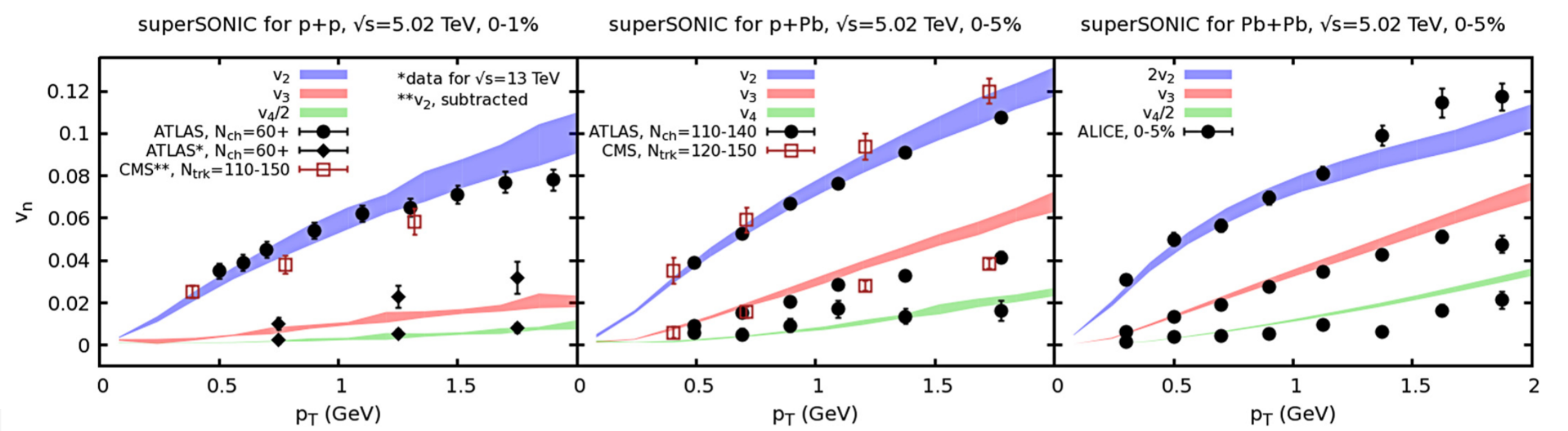}
\caption[Energy density profiles in the transverse plane and elliptic flow for PbPb, pPb and pp collisions]{\textbf{Top}: Snapshots of typical energy density profiles in the transverse plane for PbPb (left panel), pPb (center panel) and pp collisions (right panel, including zoom-in to enlarge system) at $\sqrt{s}=5.02$ TeV. \textbf{Bottom}: Elliptic ($v_2$), triangular ($v_3$) and quadrupolar ($v_4$) flow coefficients from superSONIC simulations (bands) compared to experimental data from ATLAS, CMS and ALICE (symbols) for pp (left panel), pPb (center panel) and PbPb (right panel) collisions at $\sqrt{s}=5.02$ TeV. Simulation parameters used were $\eta/s = 0.08$ and $\zeta/s = 0.01$ for all systems\cite{weller2017one}.} \label{fig:hydro_in_systems}
\end{figure}

Fig. \ref{fig:alice_flow} shows the comparison between experimental measurements and a hydrodynamic calculation of different orders of harmonic coefficients. The calculation matches the data with a small viscosity to entropy ratio $\eta/s$, which leads to the conclusion that the QGP behaves like a perfect fluid. In fact, from state-of-the-art Bayesian analysis, we see a $T$ dependence of $\eta/s$ and a minimum value around $0.1$ near $T_c$ (see Fig.~\ref{fig:hydro_viscocity_posterior}).

In fact, the collective flow has been even observed in proton-lead (pPb) and proton-proton (pp) collisions and hydrodynamic calculations are able to fit the flow coefficients in all three collision systems with the same parameters (see Fig.~\ref{fig:hydro_in_systems}), which indicates that the QGP droplet may exist even in smaller systems. 

\subsection{Jet quenching}

Jet quenching is another crucial evidence for the existence of the QGP medium, described as the suppression of high transverse momentum $p_T$ hadron spectra in heavy ion collisions compared to in pp collisions. A jet is a collimated ensemble of large $p_T$ hadrons that tries to probe partonic interactions. If no QGP medium is created, the $p_T$ spectra of individual hadrons or jets should be similar to what we see in pp collisions after normalizing it with the number of individual nucleon-nucleon collisions. The suppression is quantified by the nuclear modification factor:

\begin{equation}
    R_{AA}(p_T)=\frac{dN_{AA}/dp_T}{\langle N_{coll}\rangle dN_{pp}/dp_T}.
\end{equation}

\begin{figure}[!h]
\centering
\includegraphics[width=0.7\textwidth]{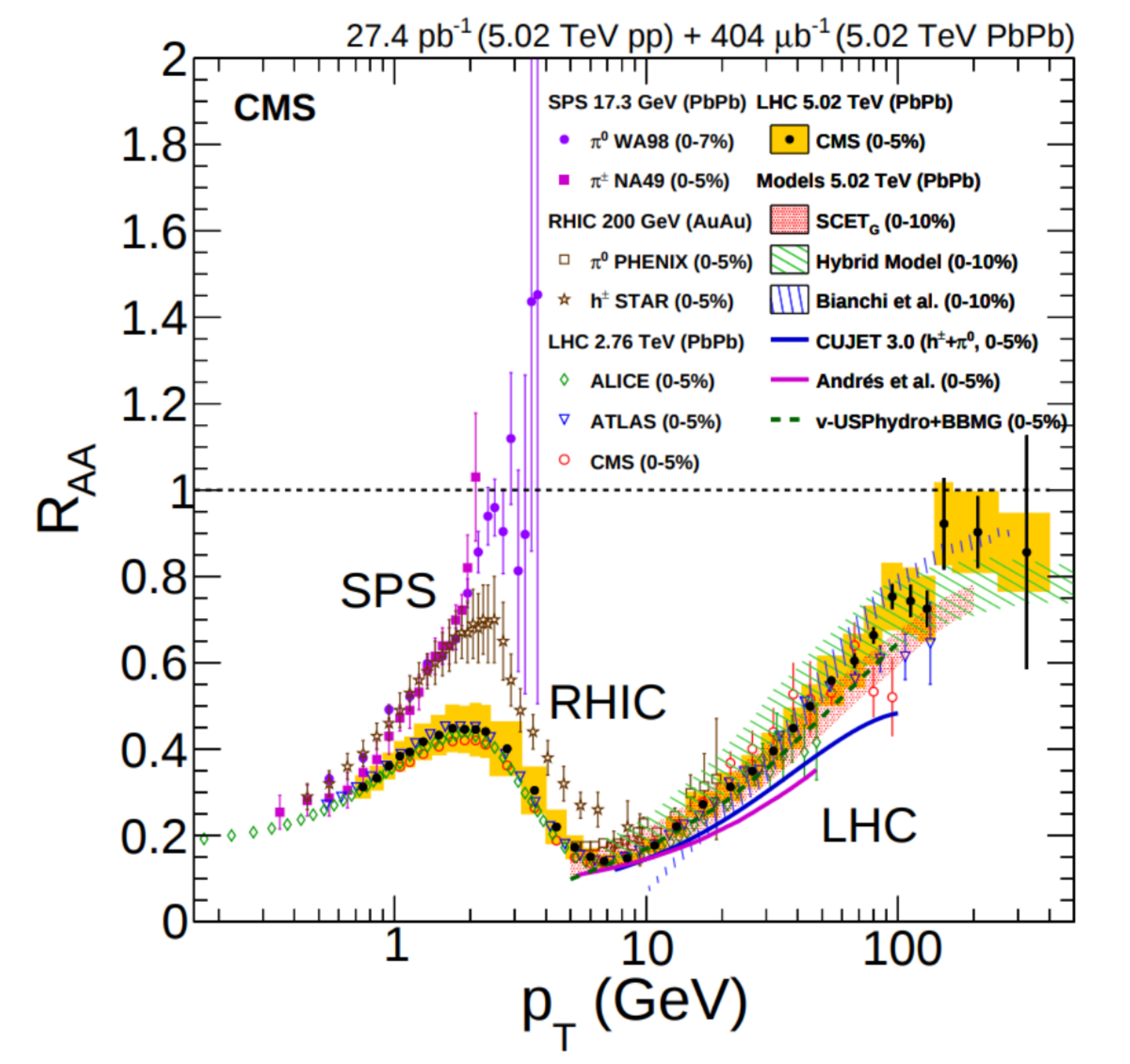}
\caption[Experimental measurements of nuclear modification factor $R_{AA}$ of neutral pions $\pi^0$, charged hadrons $h^{\pm}$ and charged particle]{Experimental measurements of nuclear modification factor $R_{AA}$ of neutral pions $\pi^0$, charged hadrons $h^{\pm}$ and charged particle at PbPb collisions at $17.3$ GeV, PbPb collisions at $2.76$ TeV and AuAu collisions at $200$ GeV, as a function of transverse momentum $p_T$, compared with several theoretical models \cite{CMS:2012aa}.} \label{fig:charged_RAA}
\end{figure}

\begin{figure}[!h]
\centering
\includegraphics[width=0.66\textwidth]{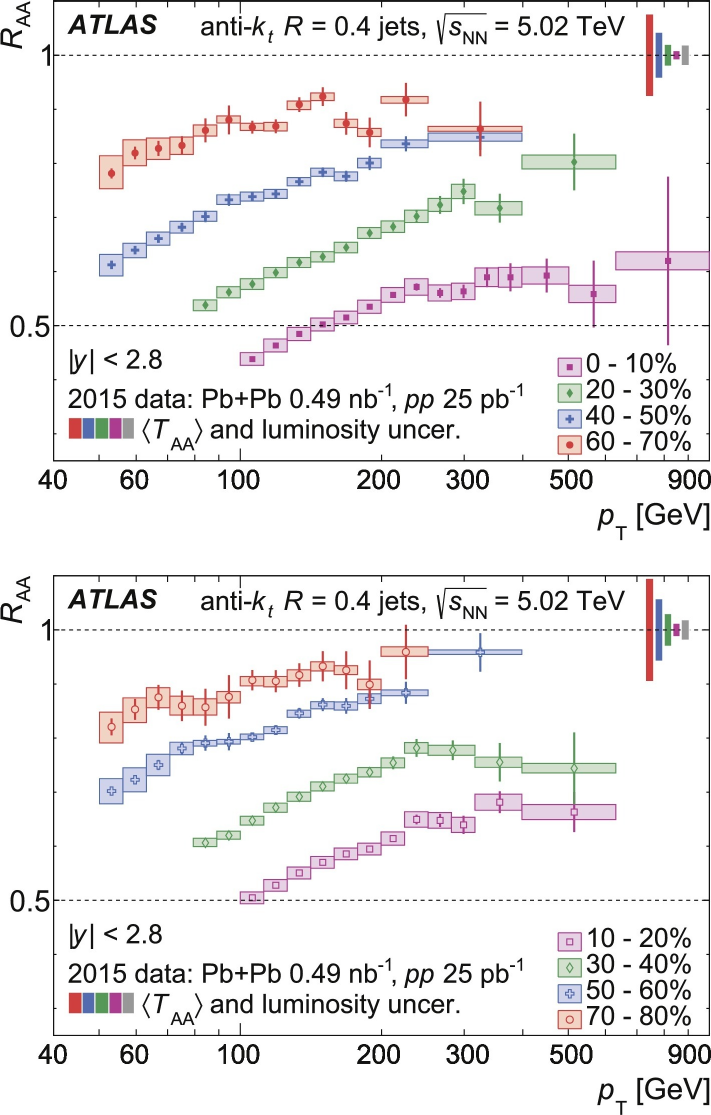}
\caption[The $R_{AA}$ values as a function of jet $p_T$ for jets]{Upper panel: The $R_{AA}$ values as a function of jet $p_T$ for jets with $|y|<2.8$ for different centrality intervals. Bottom panel: The $R_{AA}$ values as a function of jet $p_T$ for jets for four other centrality intervals \cite{aaboud2019measurement}.} \label{fig:jet_RAA}
\end{figure}

In reality, the $R_{AA}$ of both charged hadrons and jets are significantly smaller than $1$ over a wide range of $p_T$ (see Fig.~\ref{fig:charged_RAA} and Fig.~\ref{fig:jet_RAA}), indicating a strong suppression due to the interaction between high $p_T$ partons and the medium. 

\subsection{Heavy flavor probes}

\begin{figure}[!h]
\centering
\includegraphics[width=0.72\textwidth]{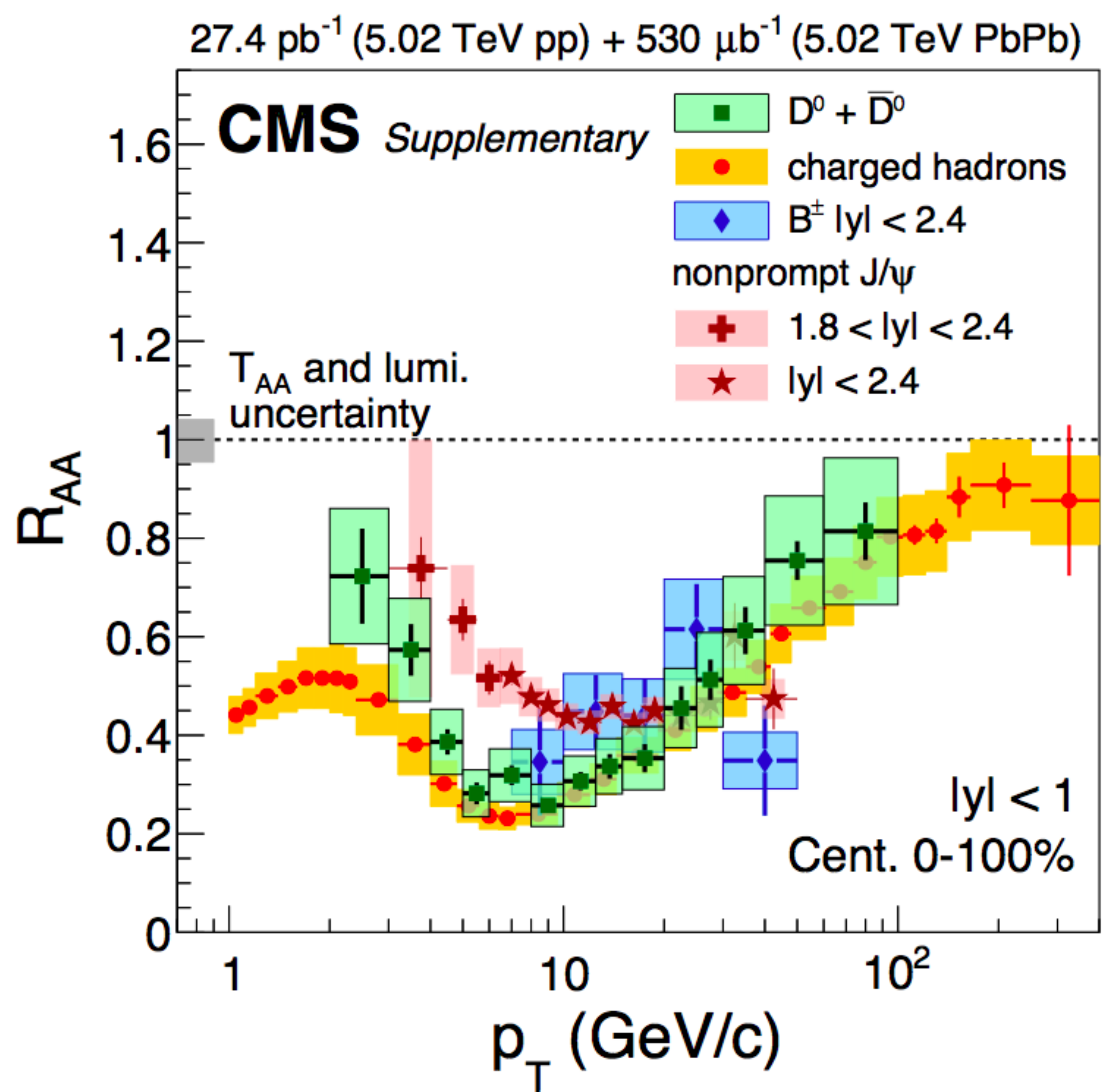}
\caption[Nuclear  modification  factor  of  charged particles, $D^0$ meson,non-prompt $J/\Psi$ and $B$ mesons]{Nuclear  modification  factor  of  charged particles, $D^0$ meson,non-prompt $J/\Psi$ and $B$ mesons performed by CMS at PbPb collisions at $5.02$ TeV \cite{Khachatryan:2016odn,Sirunyan:2017oug,Sirunyan:2017xss}.} \label{fig:D_RAA}
\end{figure}

\begin{table}[h!]
\centering
\caption[Properties of hadrons carrying open heavy flavor]{Properties of hadrons carrying open heavy flavor with charm or bottom quantum numbers $C = +1$ or $B = +1$ \cite{particle1998review}. }
\begin{tabular}{ |p{3cm}||p{3cm}|p{3cm}|p{3cm}|  }
 \hline
 \multicolumn{4}{|c|}{Open Heavy Flavor Mesons} \\
 \hline
 Name & Quark content & $I(J)^P$ & Mass ($GeV/c^2$)\\
 \hline
$D^+$   & $cd$    &$\frac{1}{2}(0^-)$ &  $1.8696\pm 0.0002$\\
$D^0$   & $c\bar{u}$    &$\frac{1}{2}(0^-)$ &  $1.8648\pm 0.0001$\\
$D^+_s$   & $c\bar{s}$    &$0(0^-)$ &  $1.9685\pm 0.0003$\\
$D^{*+}$   & $c\bar{d}$    &$\frac{1}{2}(0^-)$ &  $2.0102\pm 0.0001$\\
 \hline
 $B^+$   & $ub$    &$\frac{1}{2}(0^-)$ &  $5.2792\pm 0.0003$\\
 $B^0$   & $d\bar{b}$    &$\frac{1}{2}(0^-)$ &  $5.2795\pm 0.0003$\\
 $B^0_s$   & $s\bar{b}$    &$0(0^-)$ &  $5.3663\pm 0.0006$\\
  \hline
  
\end{tabular}
\end{table}

Heavy flavor (charm and bottom) quarks are good candidates for probing the QGP medium. Their mass ($m_c \approx 1.3-1.5$~GeV, $m_b \approx 4.2-4.5$~GeV) are much larger than the temperature of the QGP and the QCD scale ($\approx 200$~MeV). Because of this, heavy quarks are dominantly produced by hard scatterings at the beginning of the collision, before thermalization of the QGP medium. They participate in the full evolution of the QGP medium and can provide valuable information on the transport properties of the medium. A large mass also guarantees a negligible thermal production contribution. 

Since the gluon bremsstrahlung radiation of an accelerated heavy quark is suppressed within an angular cone of size $\theta < M/E$ (called the dead cone effect), one would expect that the heavy quarks will lose less energy in the medium compared to light quarks and gluons. The nuclear modification factor would show a mass-dependent hierarchy of $R^{h}_{AA}< R^c_{AA}< R^b_{AA}$ if the dead cone effect is the dominant contribution. However, there is also the collisional energy loss mechanism. One of the questions addressed in this thesis is to see whether both light and heavy flavor $R_{AA}$ can be described simultaneously with proper consideration of the quark mass. 

\begin{figure}[!h]
\centering
\includegraphics[width=0.75\textwidth]{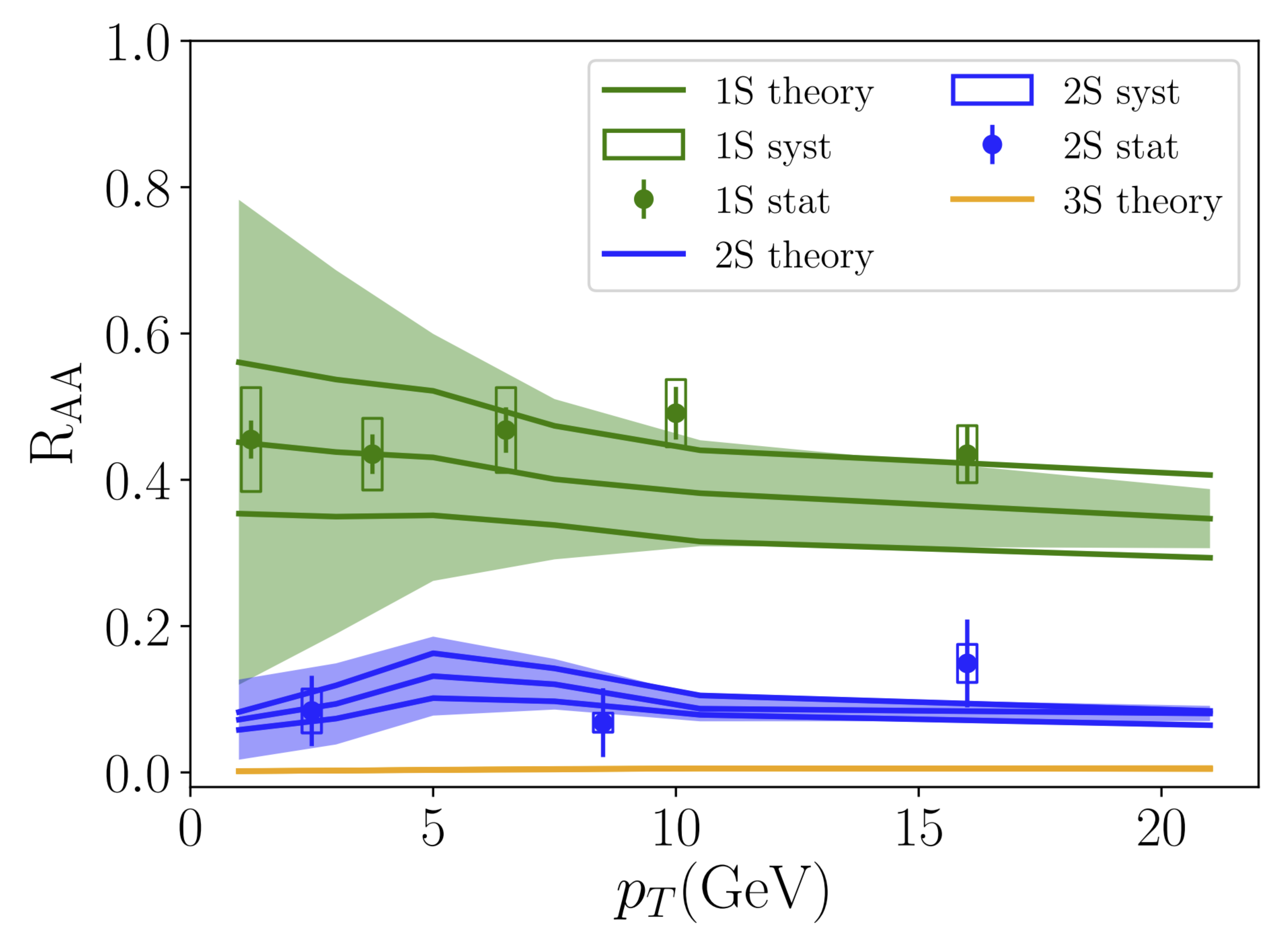}
\caption[Bottomonia $R_{AA}$ as functions of transverse momentum]{Bottomonia $R_{AA}$ as functions of transverse momentum at
$2.76$ TeV PbPb collision. The upper and lower curves correspond to calculations with parameters that differ by $\pm10\%$ respectively from the parameters used in the middle curve. The band indicates the nPDF uncertainty that is centered at the middle curve. \cite{yao2021coupled}} \label{fig:quarkonium_RAA}
\end{figure}

In this study, the focus is on open heavy mesons, such as D mesons. However, one can also study heavy quarkonium, which are bound states of $Q\bar{Q}$. $c\bar{c}$ is called charmonium and $b\bar{b}$ is called bottomonium. The ground state of charmonium and bottomonium are $J/\psi$ and $\Upsilon$. One of the most important features of quarkonium is its small size or large binding energy. Compared with the typical hadron radius $1$ fm, the radii of $J/\psi$ and $\Upsilon$ ground states are around $0.1$ and $0.2$ fm respectively (with binding energies around $0.6$ and $1.2$ GeV) \cite{muller1985physics}. This indicates that they can still survive in the QGP within a certain range of temperatures above the critical temperature $T_c$. The higher excited states are less stable due to their larger radius. Consequently, the production of different quarkonium states and extract thermal information of the QGP can be observed. A recent calculation of quarkonium $R_{AA}$ based on perturbative non-relativistic QCD (pNRQCD) is shown in Fig.~\ref{fig:quarkonium_RAA}.

\subsection{Small systems}

\begin{figure}[!h]
\centering
\includegraphics[width=0.9\textwidth]{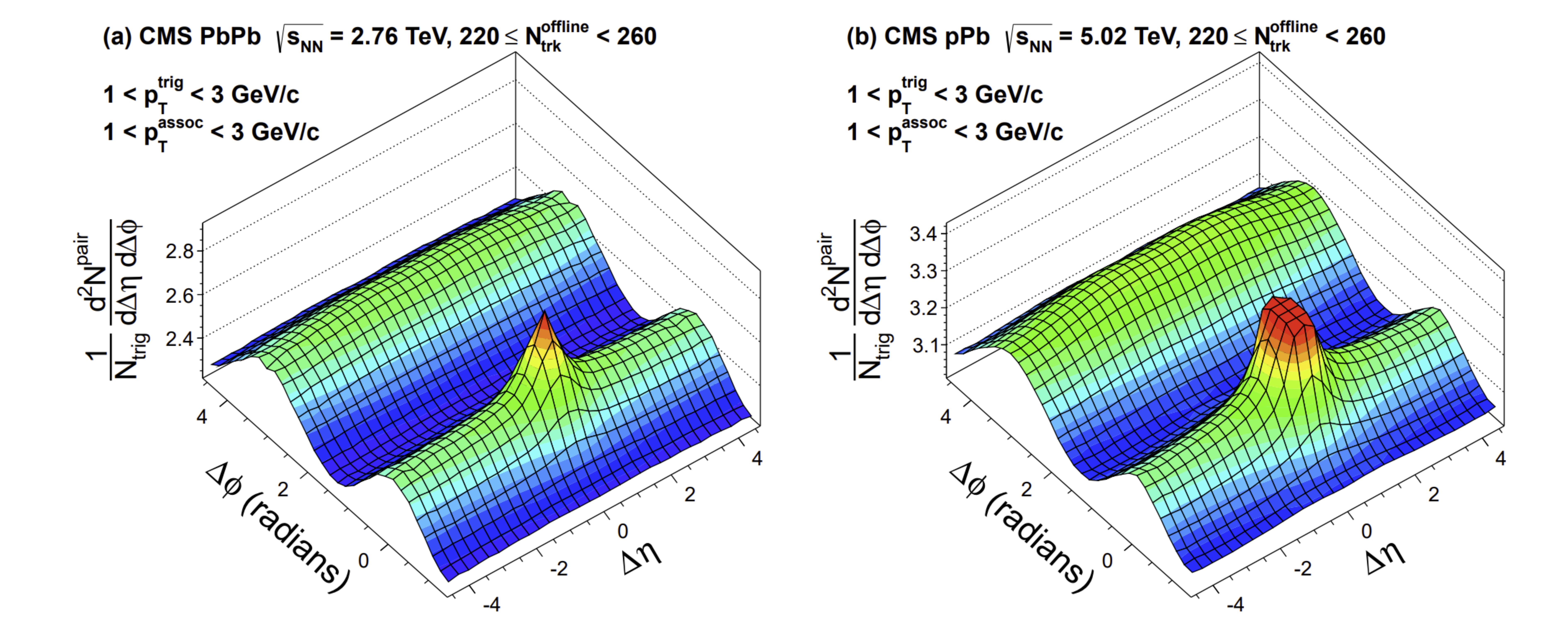}
\caption[The 2D two-particle correlation functions]{The 2D two-particle correlation functions for (a) PbPb 2.76 TeV  and (b) pPb 5.02 TeV collisions for pairs of charged particles with $1 < p^{
trig}_T < 3$ GeV/c and $1 < p^{assoc}_T < 3$ GeV/c within the $220 \le N^{offline}_{trk} < 260$ multiplicity bin. \cite{chatrchyan2013multiplicity}.} \label{fig:correlation-2d}
\end{figure}

\begin{figure}[!h]
\centering
\includegraphics[width=0.98\textwidth]{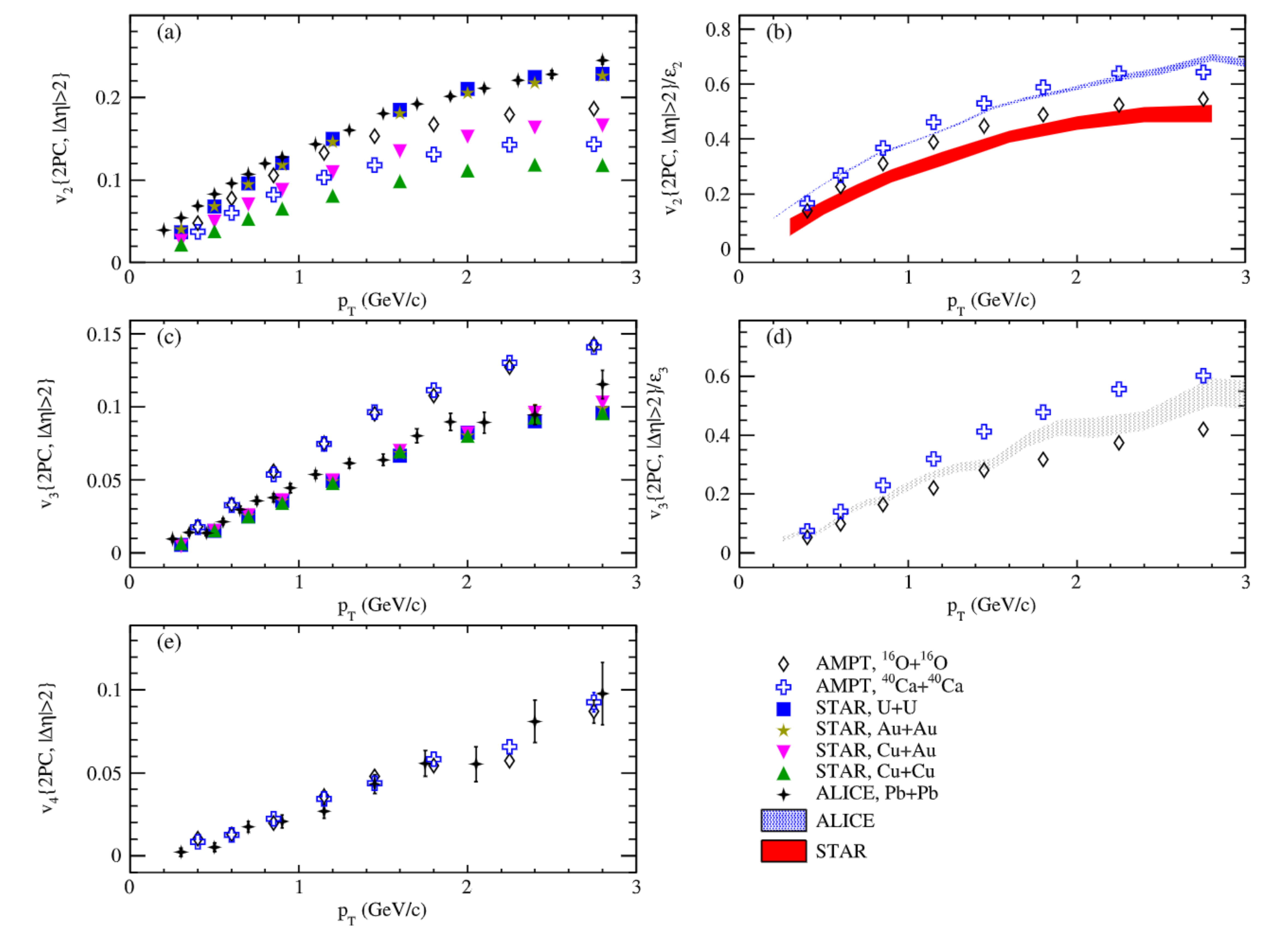}
\caption[Collective flows as a function of $p_T$ for the most central collision events in various symmetric collision systems]{\textbf{Left columns}: Collective flows $v_2$, $v_3$, and $v_4$ as a function of $p_T$ for the most central collision events in various symmetric collision systems at center of mass energy $\sqrt{s} =6.73$ TeV via two-particle correlation method. \textbf{Right columns}: the ratios of $v_n / \epsilon_n$ ($n=2,3$) as a function of $p_T$ \cite{zhang2020collision}.} \label{fig:system_size_scan_v2}
\end{figure}

Key evidence for the formation of a hot quark gluon plasma (QGP) in nucleus-nucleus (AA) collisions at high collision energies is the presence of jet quenching and collective behavior, along with their absence in smaller collision systems like proton–nucleus (pA) or deuteron–gold (dAu) \cite{gyulassy2004qgp}. The control measurements are needed to characterize the extent to which initial state effects can be differentiated from effects due to final state interactions in the QGP. Indeed, in the case of hard processes at mid-rapidity, control experiments, both at RHIC in dAu collisions at $\sqrt{s} = 200$ GeV \cite{back2003centrality, adler2003absence}, and at the LHC in pPb collisions at $\sqrt{s} = 5.02$ TeV \cite{abelev2013transverse,khachatryan2015nuclear,cms2016measurement,abelev2014j,abelev2014measurement,cms2015study}, demonstrated the absence of significant final state effects. In particular, the minimum bias pPb data can be well described by superimposing $N_{coll} = A\sigma_{pp}/\sigma_{pPb} \approx 7$ independent pp collisions with only small modifications induced by cold nuclear matter effect.

However, measurements of multi-particle correlations over large pseudorapidity range in high multiplicity pPb and pp collisions exhibit remarkable similarities with PbPb collisions \cite{chatrchyan2013multiplicity,chatrchyan2013observation}. The appearance of these ridge structures in high multiplicity pp and pA events caused speculation of similar physics being present in these small collision systems (see Fig.~\ref{fig:correlation-2d}). 
Another possible direction is to collide smaller nuclei in symmetric collisions. Collective flow in these smaller systems are also observed (see Fig.~\ref{fig:system_size_scan_v2}). It is therefore very interesting to study possible QGP effects in these small systems (asymmetric collision like pPb as well as symmetric collision like CC and OO) \cite{zhang2020collision}.

\section{Parameter inference in a complex system}\label{intro_bayes}

Since relativistic heavy ion collisions represent a many-body, multi-stage and multi-scale problem, there is no single analytical model that can describe the complete process. People have developed many models to describe different aspects of the problem: hydrodynamic models for evolving low $p_T$ (soft) partons, and transport models for evolving high $p_T$ (hard) partons, to name a few. Those models generally contain input parameters that are not calculable from first principle or are not measurable by experiments. Whether those parameters can be constrained given experimental data or whether those models, with appropriate values for those parameters, can describe data well are not easy questions to answer due to the following difficulties:

\begin{itemize}
\item The models are complex and expensive to compute. Some observables require very high statistics due to their small cross sections. It is computationally impossible to explore every combination of the model parameters, even when the number of model parameters is just more than a few.
\item The uncertainties needs to be taken into account. Some observables have huge uncertainties and yield almost no constraining power over the parameters. There's also uncertainties from model calculation that may be hard to reduce due to limited computing resources.
\end{itemize}

Bayesian analysis is the current best practice for inferring model parameters for such a complex problem. The key ingredients of such an analysis include Gaussian process emulators, Latin hypercube sampling, principal component analysis and so on. The end result is: the posterior distribution of the model parameters obtained at reasonable computation cost and with all the relevant uncertainties taken into account.

\section{Outline of the thesis}

In this thesis, I will focus on describing both light and heavy flavor observables within the JETSCAPE framework. I will also perform a Bayesian analysis that constrains the relevant energy loss parameters.

In Chapter~\ref{sec:multistage}, I will briefly introduce the various models used to describe the dynamics of heavy ion collisions. The models are categorized by different stages and scales of the collision. I will also introduce the JETSCAPE framework, a modular computational framework that tries to organize all these models systematically. 

In Chapter~\ref{sec:transport}, I will focus on the discussion of the transport models that are used in our calculation for studying both light and heavy flavor parton transport inside the QGP medium. 

In Chapter~\ref{sec:jetscape_results}, I will explore the effects of different energy loss formulations on the $R_{AA}$ of both light and heavy flavors. A multi-stage approach that combines the MATTER and LBT model for the energy loss with a virtuality dependent parameterization of the transport coefficient is found to give the best description of the experimental data. 

In Chapter~\ref{sec:bayesian}, a brief introduction to Bayesian model-to-data analysis technique is carried out. I will also apply this technique to an analytical bulk physics model and analyze the effect of varying uncertainties and different model assumptions. 

In Chapter~\ref{sec:bayesian_results}, I will put Bayesian analysis into action to try to constrain the parameters in our multi-stage approach. With the set of optimal values drawn from the posterior distribution of the parameters, I then show that our multi-stage approach can achieve a simultaneous description of charged hadron, D meson, and inclusive jet observables.

Finally, a summary of this thesis is given in Chapter~\ref{sec:summary}.

\chapter{A Multi-Stage Approach to Relativistic Heavy Ion Collision} \label{sec:multistage}

\vspace{1in}

Significant progress has been made over the past two decades regarding studying relativistic heavy ion collisions. A multi-stage, multi-scale description of the collision has been now proven successful for describing various observables across different collision systems and energies. 

\begin{figure}[!h]
\centering
\includegraphics[width=0.95\textwidth]{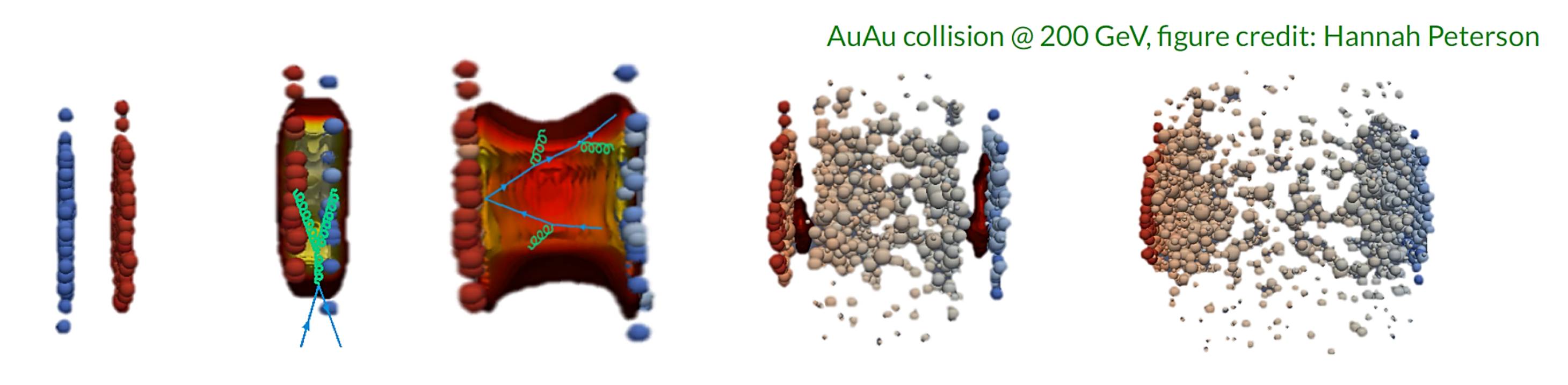}
\caption{Different stages of the heavy ion collision process.}\label{fig:RHIC_process_diagram}
\end{figure}


For describing the evolution of the bulk medium, some of the models are listed below:

\begin{itemize}
    \item \textbf{Initial condition}. Initial condition model describes the energy/entropy deposition of the collision, including fluctuation, into the QGP medium. Different models have been developed for this stage: the Glauber model \cite{Perdrisat:2006hj, Miller:2007ri}, the color glass condensate (CGC) inspired KLN model \cite{Kharzeev:2001yq,Drescher:2006pi}, and the IP-Glasma model \cite{Schenke:2012wb, Schenke:2012fw}. In this work, the \trento model which is a parametric model that maps the initial nuclear overlap density into an entropy density distribution is used.
    \item \textbf{Bulk QGP evolution}. Relativistic viscous hydrodynamics model is used for this stage in this study with the assumption that the medium is close to local thermal equilibrium. An event-by-event (2+1)-dimensional viscous hydrodynamical model called VISHNEW is employed in this work\cite{Song:2007ux, Bernhard:2016tnd}.
    \item \textbf{Particlization}. Particlization describes the switching from a hydrodynamical medium to individual particles as the medium cools down \cite{pratt2010coupling,Bernhard:2016tnd,mcnelis2021particlization}.
    \item \textbf{Hadronic rescattering}. The stage when hadrons keep interacting with each other until reaching kinetic freeze-out (the time when elastic scatterings stop) is described by hadronic rescattering models. There is also the chemical freeze out when inelastic scatterings cease. Hadrons are then detected by the surrounding detectors. UrQMD is a well known model for simulating this stage \cite{Bass:1998ca}. 
\end{itemize}

The above models are used to describe the bulk (soft final hadrons with $p_T \le 3$~GeV) observables such as charged particle/identified particle spectra, multiplicity, mean transverse momentum, mean energy, momentum anisotropy. By studying those observables, we can get a handle on the geometry, fluctuation, and transport properties like shear and bulk viscosity of the medium.

There are also the hard probes ($p_T \gtrsim 10$~GeV), including but not limited to charged hadrons, jets, heavy quarks, and photons for the collision. This work focuses on the first three and aims for a simultaneous description of these observables. A sequence of different models to describe the evolution of the hard probes during different stages of the collision are listed below:

\begin{itemize}
    \item \textbf{Initial condition}. Hard probes are produced via initial hard scatterings. Their momentum distribution can be calculated by either a Monte Carlo generator such as PYTHIA or sampled from a pQCD calculation like the fixed-order plus next-to-leading log formula (FONLL) \cite{Cacciari:1998it, Cacciari:2001td}. 
    \item \textbf{In medium evolution}. After the initial production, the hard probes/partons propagate through the QGP medium, losing energy as they interact with the medium. The interaction can be described by transport models with various assumptions on both the medium and the partons. The MATTER model is used specifically for describing the high virtuality showering in both the vacuum and the medium \cite{majumder2013incorporating}.
    \item \textbf{Hadronization}. Hadronization is usually done by Lund fragmentation \cite{fries2003hadron} in both the vacuum and in medium. For heavy quarks, there is also contribution from recombination with the medium at low $p_T$ that accounts for the observed large heavy meson flow \cite{fries2003hadron}. 
    \item \textbf{Hadronic interactions}. Hadronic interactions can again be taken into account with models like rQMD.
\end{itemize}

\begin{figure}[!h]
\centering
\includegraphics[width=0.95\textwidth]{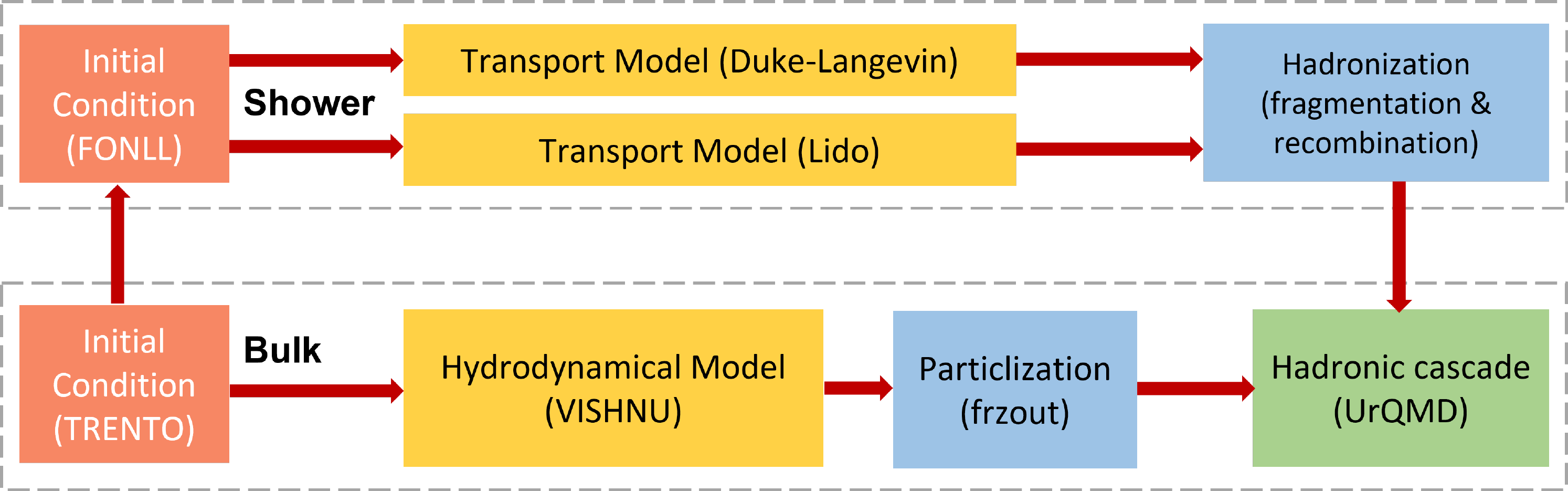}
\caption{Summary of different models used in the Duke multi-stage approach to describe heavy ion collision.} \label{fig:Duke_workflow}
\end{figure}

The workflow of the multi-stage approach the Duke group has developed is illustrated in Fig.~\ref{fig:Duke_workflow}. As we can see, the top panel shows the models that describe the hard parton evolution and the bottom panel shows the models used to describe the soft medium evolution.

This multi-stage approach works great except for the following few limitations:
\begin{enumerate}
    \item The models are developed by different people and use different coding languages (like FORTRAN, C++, and Python) and different interfaces. Maintaining and further development of the code become more and more complex over time. 
    \item Comparing results with experimental data and other people's result is challenging, as it is difficult to identify which model is causing the difference. Even two models based on the same theory can have slightly different implementations. In order to test which theory/model better describes the data, a controlled workflow that keeps all the other models used the same is needed. 
\end{enumerate}

These limitations are what the Jet Energy loss Tomography with a Statistically and Computationally Advanced program Envelope (JETSCAPE) collaboration is trying to overcome. The goal of this collaboration is to form an interdisciplinary team of physicists, computer scientists, and statisticians to develop a comprehensive software framework that will provide a systematic, rigorous approach to simulate the complex dynamical environment of relativistic heavy ion collisions \cite{putschke2019jetscape}. 

\begin{figure}[!h]
\centering
\includegraphics[width=0.95\textwidth]{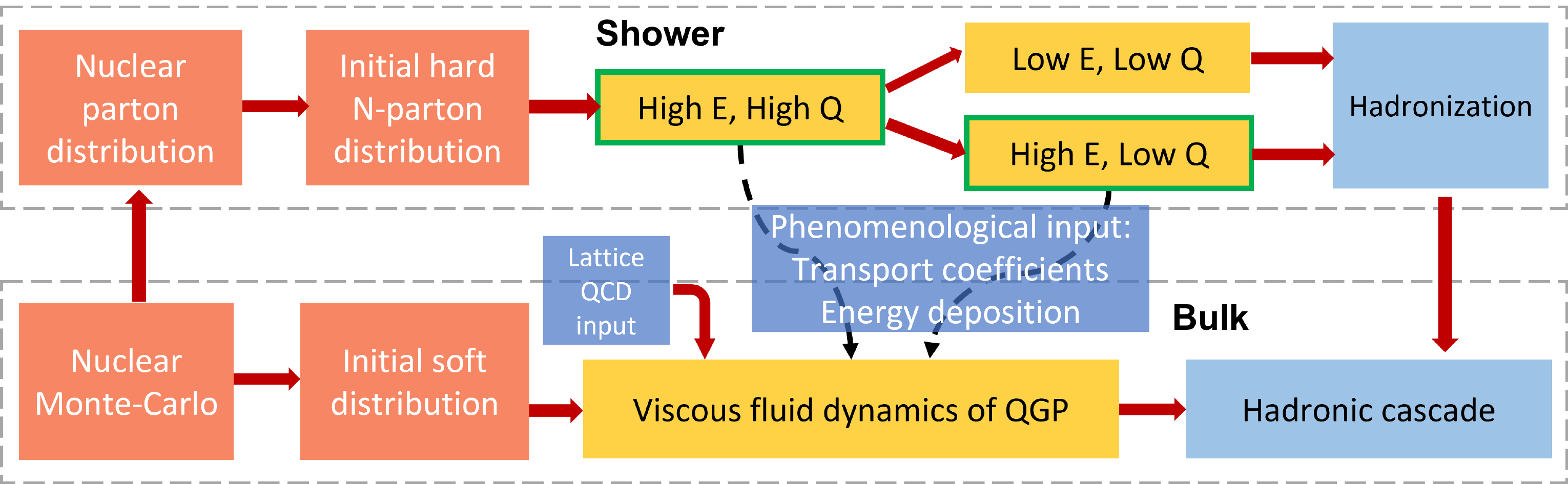}
\caption{The structure of the JETSCAPE event generator.} \label{fig:JETSCAPE_workflow}
\end{figure}

JETSCAPE is developed mainly in modern C++, with object-oriented programming (OOP) and modularity in mind. Below is the current structure of the JETSCAPE event generator. As can be seen in Fig.~\ref{fig:JETSCAPE_workflow}, the structure is very similar to what has been used at Duke. However, each slot may use different models that only need to share a common interface. It is also possible to change the structure of the workflow. For example, if one has developed an energy loss model that can handle both high energy and low energy partons, one can just use that single model for the entire shower evolution. With this fully fledged event generator, one may overcame the above limitations.

\begin{enumerate}
    \item JETSCAPE still depends on external libraries, but they are all well documented open source C++ libraries that are easy to compile. One can even use a Docker container environment to run JETSCAPE without having to worry about compilation issues. Developing custom modules is also relatively simple thanks to the modularity of JETSCAPE. One can also reuse many classes provided by JETSCAPE to save time and trouble. There are also online documentation and workshops held every year to help with the development. 
    \item Comparing between different models is now very easy to do. One just need to specify the relevant models in a XML document. 
\end{enumerate}

I will discuss more about JETSCAPE later in this chapter. 

\section{Initial condition}

As in proton-proton collisions, initial hard scatterings can be calculated via perturbation theory. The non-perturbative processes, which are directly involved in the initialization of the soft medium, are hard to calculate. In this work, a top-down approach is adopted that generates the initial condition with some parameterization and the relevant model is named \trento.

\subsection{The \trento model}

\trento is a parametric initial condition model that can generate initial entropy/energy profile for proton-proton (pp), proton-nucleus (pA), and nucleus-nucleus (AA) collisions. \trento does not assume a particular physical mechanism for the energy deposition in heavy ion collisions. However, it constructs an initial static profile in the transverse plane by mapping the nuclear density overlap function to the initial density via an effective function at a proper time $\tau$:
\begin{equation}
    \frac{ds}{dy}|_{\tau} = f(T_A(x,y), T_B(x,y)),
\end{equation}
where $ds/dy$ is assumed to be a fixed value at near rapidity (called the boost invariant assumption). $T_A(x,y), T_B(x,y)$ are the nuclear thickness function defined as:
\begin{equation}
    T_{A/B}(x,y) \propto \int dz \rho_{A/B} (x, y, z),
\end{equation}
where $\rho (x, y, z)$ is the nuclear matter density. $\int dz\rho(x,y,z)$ is assumed to be a Gaussian distribution in the transverse plane with an effective nucleon width $w$:
\begin{equation}
\int dz \rho(x,y,z) = \frac{1}{2\pi w^2} \exp\left(- \frac{x^2+y^2}{2w^2}\right).
\end{equation}

Therefore the nucleon thickness function defined before can be seen as a sum of individual Gaussian functions:
\begin{equation}
T_{\rm A/B}(x,y) = \sum_{i=0}^{N_{\rm part}} \gamma_i \frac{1}{2\pi w^2} \exp\left(- \frac{(x-x_i)^2+(y-y_i)^2}{2w^2}\right),
\end{equation}
where one needs to sum over all the participants centered at $(x_i, y_i)$ that at least collide once. This is effectively assuming that heavy ion collisions are superpositions of nucleon-nucleon collisions. 

The functional form of $f$ is difficult to get from first principle calculations. In \trento, it is proposed to be:
\begin{equation}
    f(T_A, T_B)  \equiv \left(\frac{T_A^p + T_B^p}{2}\right)^{1/p},
\end{equation}
where $p$ is an unknown parameter that needs to be determined from experiments. If $p$ is close to $1$, then $f$ becomes the general mean $(T_A+T_B)/2$ which is the Monte Carlo wounded nucleon model. If $p=0$, $f(T_A, T_B)=\sqrt{T_A T_B}$. It is found that by taking $p$ close to $0$, \trento is best at describing various soft observables in different collision systems and energies \cite{Bernhard:2016tnd,Moreland:2018gsh}. 

\begin{figure}
	\centering
	\includegraphics[width=1\textwidth]{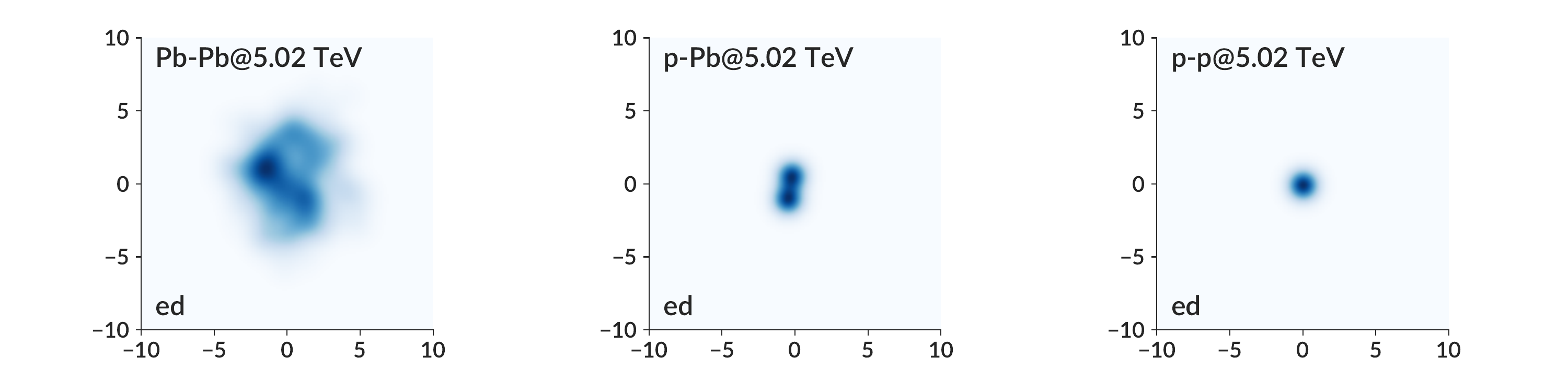}
	\caption[Examples of initial energy density generated by \trento]{Examples of initial energy density generated by \trento for a single collision in: PbPb, pPb, pp at $\sqrt{s_{\rm NN}}=5.02$ TeV.} \label{fig:trento_ed}
\end{figure}

Fig.~\ref{fig:trento_ed} shows the initial energy density in the transverse plane generated by \trento for three different collision systems at $5.02$~TeV for a single event. The geometric anisotropy in PbPb and pPb collisions are evident and should contribute to the final momentum anisotropy of the measured final hadrons.

\subsection{Hard parton initial momentum distribution}

The initial position distribution for the hard partons is sampled consistently from the energy density generated by \trento. The initial momentum distribution, on the other hand, are calculable using perturbative QCD. Specifically, for initial heavy quark generation, the leading order processes are gluon fusion $gg\rightarrow Q\bar{Q}$ and quark anti-quark annihilation  $q\bar{q}\rightarrow Q\bar{Q}$. 
In the Duke framework, the fixed-order plus next-to-leading log formula (FONLL) is adopted to calculate the heavy quark initial momentum distribution, which conveniently allows one to switch between different parton distribution function (PDF) parameterizations.
\begin{figure}
	\centering
	\includegraphics[width=0.6\textwidth]{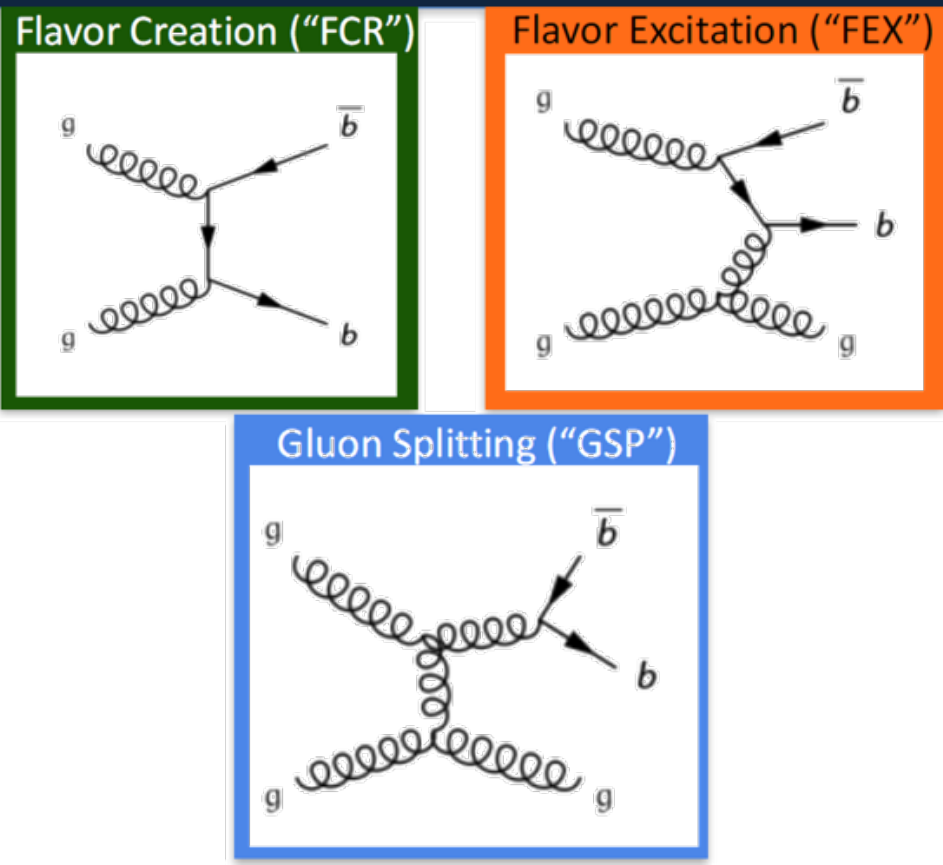}
	\caption[Dominant heavy flavor creation process from PYTHIA]{Dominant heavy flavor creation process from PYTHIA \cite{sjostrand2015introduction}.} \label{heavy flavor_creation}
\end{figure}

Apart from the two processes listed above, it is also possible to excite heavy quarks from the sea: $qQ\rightarrow qQ, \ gQ\rightarrow gQ$. The corresponding matrix elements can be found in \cite{Combridge:1978kx}. All these processes are included in PYTHIA and in the JETSCAPE framework.

\section{Relativistic viscous hydrodynamics}

\begin{figure}[h]
	\centering
	\includegraphics[width=1\textwidth]{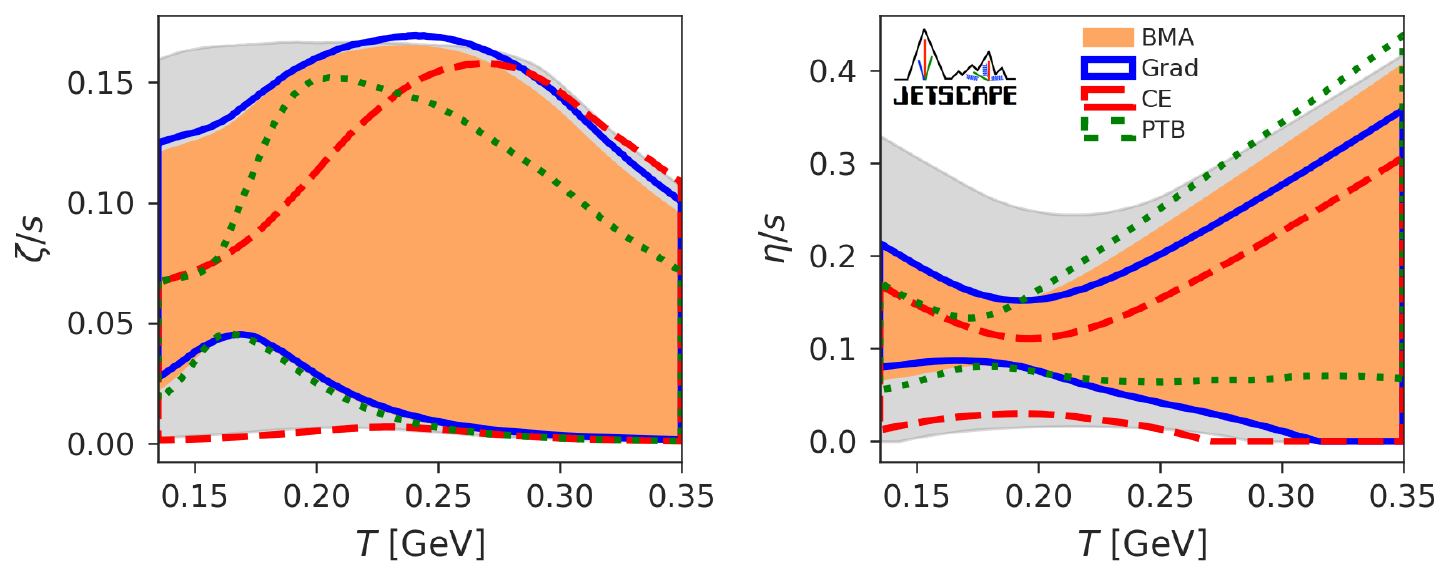}
	\caption[Temperature dependence of the shear and bulk viscosity]{A recent extraction of the temperature dependence of the specific bulk and shear viscosity \textbf{[left]} and \textbf{[right]}.  The 90\% credible intervals for the prior, the posteriors of the Grad, Chapman-Enskog and Pratt-Torrieri-Bernhard models, and their Bayesian model average are shown in gray, blue, red, and orange \cite{everett2021phenomenological}.} \label{fig:vishnu_shear}
\end{figure}

Relativistic viscous hydrodynamics is one of the most successful models for describing soft observables in heavy ion collisions. It is a macroscopic model based on the conservation of energy, momentum and charge current:
\begin{equation}
\partial_{\mu} T^{\mu\nu} = 0, \partial_{\mu} N^{\mu}  = 0,
\end{equation}
where $T^{\mu\nu} = (e+p) u^{\mu} u^{\nu} - p g^{\mu\nu} + \pi^{\mu\nu} - (g^{\mu\nu} - u^{\mu}u^{\nu}) \Pi$ is the energy momentum tensor, $N^{\mu} = n u^{\mu} + V^{\mu}$ is the net baryon charge current in the Landau frame, $e$ and $p$ are the energy density and pressure in the local fluid rest frame. $u^\mu =\gamma\ (1,\boldsymbol{\beta})$, with $\boldsymbol{\beta}=(\beta_x,\beta_y,\beta_z)$ being the 3-velocity of the considered fluid element and $\gamma = 1/\sqrt{1-\boldsymbol{\beta}^2}$ the corresponding Lorentz factor. $V^{\mu}$ is the baryon flow. $\pi^{\mu\nu}$ and $\Pi$ are the first order shear and bulk viscous corrections and can be further decomposed into:
\begin{equation}
\pi^{\mu\nu} = 2\eta \Delta^{\mu\nu\alpha\beta} \partial_{\alpha}u_{\beta}, \Pi = -\zeta \partial_{\mu}u^{\mu},
\end{equation}
where $
\Delta^{\mu\nu\alpha\beta} = \frac{1}{2} (\Delta^{\mu\alpha}\Delta^{\nu\beta} + \Delta_{\nu\alpha}\Delta_{\mu\beta}) - \frac{1}{3}\Delta^{\mu\nu}\Delta^{\alpha\beta}, \Delta^{\mu\nu} = g^{\mu\nu} - u^{\mu}u^{\nu}
$ are the projection operators. $\eta$ and $\zeta$ are the shear and bulk viscosities. One of the goals by studying the QGP with hydrodynamics is to determine the magnitude of those viscosities (a recent result using Bayesian analysis is shown in Fig.~\ref{fig:vishnu_shear}).

There are five equations above and six unknown variables ($e, p, n$ and three $ u^\mu$). To close the equations, one needs one more equation which is the equation of state (EOS) $p=p(n,e)$. The state-of-the-art result of the QCD EOS is calculated by the HotQCD collaboration \cite{Bazavov:2014pvz} using (2+1) flavor lattice QCD. A smooth interpolation is employed to connect between the lattice QCD EOS by the HotQCD collaboration and a hadron resonance gas EOS (in the interval between $110$ and $130$MeV) \cite{Moreland:2015dvc}.

The hydrodynamical implementation used in this study is called {\rm \tt VISHNU(2+1)}~\cite{Song:2007ux,shen2016iebe} which solves the boost invariant (2+1) dimensional viscous hydrodynamical equations event-by-event (EBE). It includes the shear and bulk viscosity corrections through the second-order Israel-Stewart equation in the 14-momentum approximation \cite{Israel:1979wp}. Their values are determined by a state-of-the-art Bayesian model-to-data comparison \cite{Bernhard:2016tnd}. In principle, the hydrodynamical model should be solved in (3+1) dimensions. However, if boost invariant symmetry is assumed to be true (which means the system behaves the same at different space-time rapidity up to a longitudinal boost) \cite{bjorken1983highly,Moreland:2014oya}, one can solve the hydro in (2+1) dimensions and boost to other rapidities. Experimentally, the event-averaged rapidity distribution of charged particles $dN_{ch}/dy$ in symmetric nuclei-nuclei collisions at the RHIC has a central plateau at least within $|y|<2$. If observables involving only mid rapidity particles are explored, using (2+1) dimension hydrodynamic simulation could be justified. However, the event-by-event particle production may break the boost invariance symmetry and asymmetric nuclear collisions such as pPb, pAu clearly don't even have boost invariance in the mid-rapidity region \cite{Adler:2006wg, abelev2014multiparticle,aad2013measurement,aad2014measurement}. Studying observables at large rapidity and in small collision systems requires the full (3+1) dimension hydro simulation.  

The stage between the initial condition and the time where the medium reaches local thermal equilibrium is called the pre-equilibrium stage. This stage is complicated to model since we are in the non-perturbative and non-equilibrium territory. The most straightforward modeling is to assume that partons free stream during this stage \cite{kurkela2019matching}. Other models, such as the Parton-Hadron-String dynamics (PHSD) model \cite{linnyk2011dilepton}, or the classic Yang-Mills equation \cite{Schenke:2012wb} can also be used to simulate this stage.



\section{Hadronization and hadronic transport}

Hydrodynamics is a macroscopic description of the medium in local thermal equilibrium. As the medium expands and cools down, the relaxation time becomes too long for the hydrodynamic approach to be applicable. A microscopic description with discrete particles is needed at this stage. This transition is called particlization and is assumed to happen near the pseudo-critical temperature $T_c$ calculated by lattice QCD when the $dp/dT$ reaches maximum. Since the phase transition is a smooth crossover, it is not a critical temperature. It is also in the range where the hadron resonance gas model converges with lattice calculation, indicating we can describe the medium with hadronic degrees of freedom. 

\subsection{Particlization}

Particlization is performed on a space-time hypersurface $\Sigma$ with a constant temperature using the Cooper-Frye formulation:
\begin{equation}
E \frac{dN_i}{d^3p} (x^{\mu}, p^{\mu})=  \frac{g_i}{(2\pi)^3} \int_{\Sigma} f_i(x^{\mu}, p) p^{\mu} d^3\sigma_{\mu},
\end{equation}
where $g_i$ is the spin degeneracy of particle species $i$, $f(x^\mu,p^\mu)=f_0(x^\mu,p^\mu)+\delta f(x^\mu,p^\mu)$ represents the phase space distribution of one particle species and is assumed to have small deviation from the thermal equilibrium distribution $f_0(x^\mu,p^\mu)$ due to viscous corrections $\delta f$. Different implementations of $\delta f$ can be found in Ref.~\ref{Bernhard:2016tnd,mcnelis2021particlization}. In Ref.~\cite{JETSCAPE:2020mzn}, Bayesian model selection are applied to select from four different implementations of $\delta f$ by comparing with experimental data.


\subsection{Hadronization}

Hadronization is the process when partons turn into hadrons. Hadronization for the medium happens implicitly during particlization. For the hard partons, the hadronization mechanism in proton-proton collisions is called fragmentation. The fragmentation process is non-perturbative but assumed to be universal. However, in heavy ion collisions where the system is dense, it is possible for several partons to combine into a hadron at the hadronization stage \cite{fries2003hadron}. 

We use PYTHIA for modeling the fragmentation mechanism and a sudden coalescence model for the recombination mechanism. For fragmentation, the probability distribution for a heavy quark producing a heavy hadron that carries $z=p_{H}/p_{Q}$ fraction of its momentum is known as the fragmentation function $D(z)$. There are various parameterizations for $D(z)$. For example, the Peterson fragmentation function is defined as:
\begin{equation}
    D(z)  \propto \frac{1}{z(1-\frac{1}{z}-\frac{\epsilon}{1-z})^2},
\end{equation}
where $\epsilon$ is a parameter that scales $m_Q^{-2}$ ($\epsilon_c \approx 0.05$,$\epsilon_b \approx 0.006$).

\begin{figure}[h]
	\centering
	\includegraphics[width=0.6\textwidth]{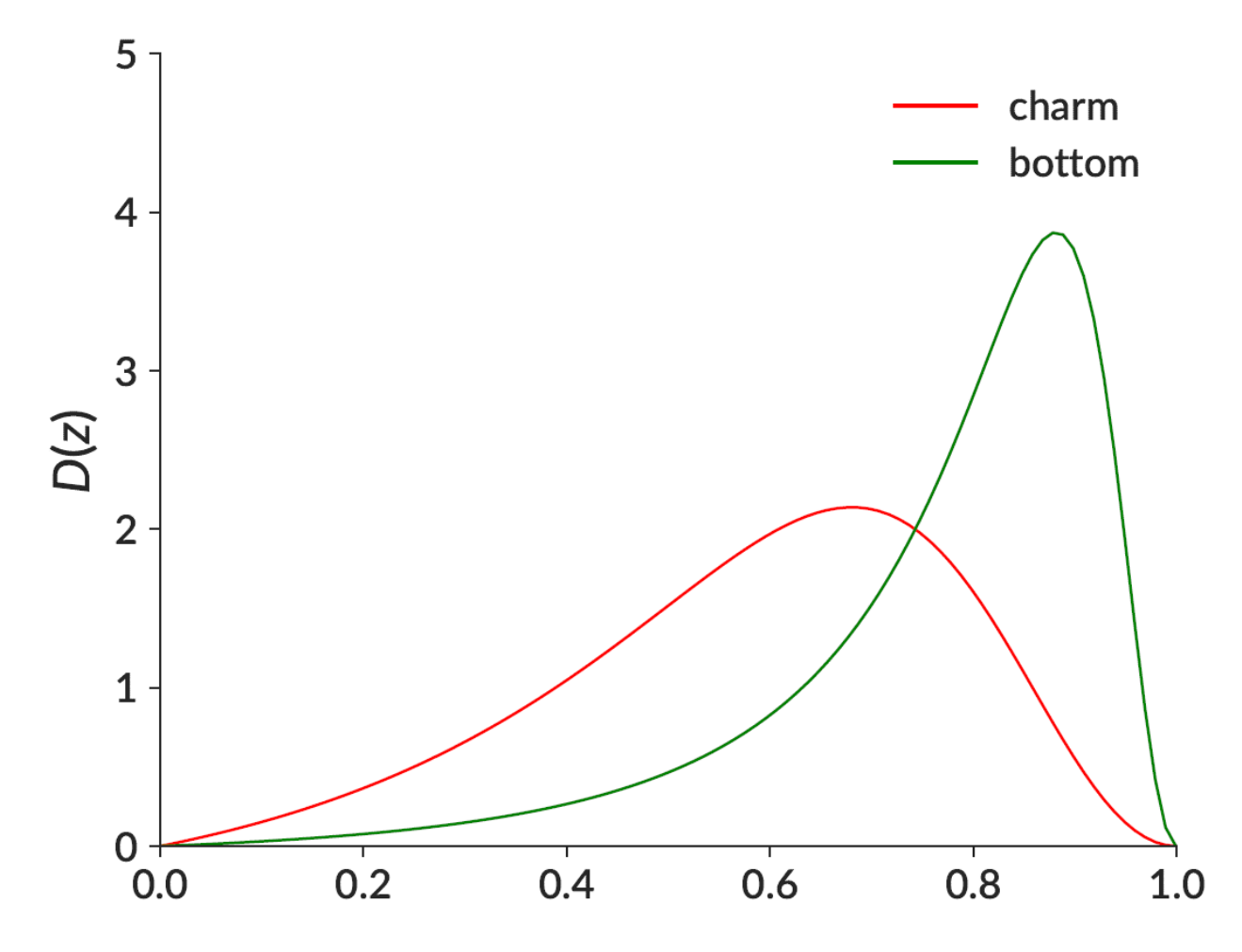}
	\caption[The Peterson fragmentation function for charm and bottom quarks]{\label{fig:peterson_fragmentation_function} The Peterson fragmentation function for charm and bottom quarks as a function of $z$ \cite{peterson1983scaling}.}
\end{figure}

For studying the recombination mechanism, the probability of recombination is determined by the overlapping between the initial and final state wave functions.

\begin{figure}[h]
	\centering
	\includegraphics[width=0.75\textwidth]{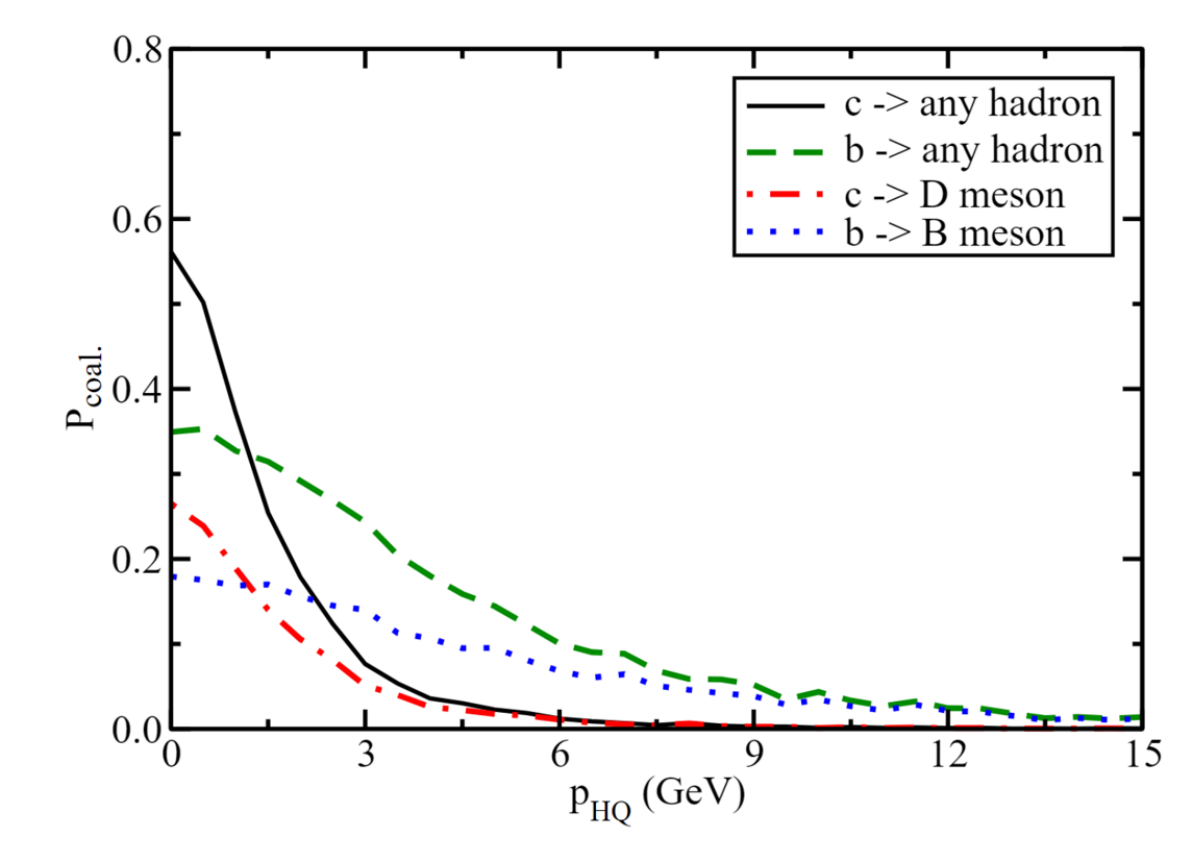}
	\caption[ The recombination probability for charm and bottom quarks]{\label{fig:recomb_probability} The recombination probability for charm and bottom quarks as a function of heavy quark momentum \cite{Cao:2016gvr}.}
\end{figure}

The momentum distribution of the recombined mesons and baryons are respectively:
\begin{equation}
    \frac{dN_M}{d^3 p_M}=\int d^3 p_1 d^3 p_2 \frac{dN_1}{d^3p_1} \frac{dN_2}{d^3p_2} \ f_M^W(\Vec{p}_1,\Vec{p}_2 )\delta(\Vec{p}_M-\Vec{p}_1-\Vec{p}_2),
\end{equation}
\begin{equation}
    \frac{dN_B}{d^3p_B}=\int d^3p_1d^3p_2d^3p_3 \frac{dN_1}{d^3p_1}\frac{dN_2}{d^3 p_2}\ \frac{dN_3}{d^3 p_3}\ f_B^W(\Vec{p}_1,\Vec{p}_2, \Vec{p}_3)\delta(\Vec{p}_M-\Vec{p}_1-\Vec{p}_2-\Vec{p}_3),
\end{equation}
where $\frac{dN_i}{d^3p_i}$ represents the momentum distribution of the $i^{th}$ valence parton in the recombined meson or baryon. The distribution for heavy quarks are obtained after their evolution inside the QGP medium whereas the distribution for light quarks are assumed to be the thermal distribution in the local cell frame. The Wigner function $f^W_M, f^W_B$ for meson and boson respectively are defined as:
\begin{equation}
    f^W_M(q^2)=\frac{g_M(2\sqrt{\pi}\sigma)^3}{V}e^{-q^2\sigma^2},
\end{equation}
\begin{equation}
    f^W_B(q_1^2,q_2^2)=\frac{g_B(2\sqrt{\pi}\sigma_1)^3(2\sqrt{\pi}\sigma_2)^3}{V}e^{-q_1^2\sigma_1^2-q_2^2\sigma_2^2},
\end{equation}
where $q$ is the magnitude of the momentum difference between two quarks (single $q$ value in the meson case and two $q$ values in the baryon case), $\sigma=1/\sqrt{\mu\omega}$ where $\mu$ is the reduced mass between two quarks (again, single $\mu$ for the meson case and two $\mu$ for the baryon case) and $\omega$ is called the oscillator frequency. The Wigner function is simplified to this form since the wave function of the quarks are assumed to be all $s$-wave of a harmonic oscillator and that's where $\omega$ comes from. $\omega$ is fitted to the charged radii of the charged hadrons:
\begin{equation}
    \langle r_M^2\rangle_{ch}=\frac{3}{2\omega}\frac{1}{(m_1+m_2)(Q_1+Q_2)}(\frac{m_2}{m_1}Q_1+\frac{m_1}{m_2}Q_2),
\end{equation}
\begin{equation}
    \langle r_B^2\rangle_{ch}=\frac{3}{2\omega}\frac{1}{(m_1+m_2+m_3)(Q_1+Q_2+Q_3)}(\frac{m_2+m_3}{m_1}Q_1+\frac{m_3+m_1}{m_2}Q_2+\frac{m_1+m_2}{m_3}Q_3),
\end{equation}

For the sake of simplicity, we assign $\omega=0.33$GeV for all charm and beauty mesons, $\omega=0.43$GeV for charm baryons and $\omega=0.41$GeV for beauty baryons by fitting to the charged radii of $0.43$fm of $D^+$, $0.62$fm of $B^+$, and $0.39$fm of both $\Lambda_c$ and $\Lambda_b$ \cite{oh2009heavy, hwang2002charge}.

The hybrid hadronization model containing both fragmentation and recombination works like the following. A random number is first generated from a uniform distribution between $0$ and $1$ and compared with the probability of the heavy quark recombining into any hadron. If the number is bigger than the probability, that heavy quark is sent to PYTHIA for fragmentation. Otherwise, light quarks are sampled from a thermal distribution in the local rest frame of the fluid cell and recombined with the heavy quark into either a meson or a baryon. Fig.~\ref{fig:recomb_probability} shows the recombination probabilities for a charm or bottom quark to all heavy flavor hadron channels and to only D or B meson. For the same $p_T$, bottom quarks have a larger recombination probability than charm quarks to produce heavy flavor hadrons due to their larger masses. 

\section{Hadronic stage interaction}

After all partons have turned into hadrons, they will keep decaying and scattering with each other until chemical freeze-out and kinetic freeze-out. The dynamics at this stage can be described by the Boltzmann transport equation:
\begin{equation}
\frac{df_i(x, p)}{dt} = {\cal{C}}_i(x,p),
\end{equation}
which states that the time evolution of phase space distribution $f_i(x,p)$ of species $i$ is determined by the collision terms ($\cal{C}$), including binary collisions, $2\rightarrow n$ inelastic process, annihilation, resonance formation and decays.

The Ultra-relativistic Quantum Molecule Dynamics (UrQMD) model \cite{Bass:1998ca} is one of the most widely used models to simulate such processes in the hadronic stage. It solves the Boltzmann equation by sampling the collision term stochastically and propagating the particles along a straight line trajectory. The inputs for the UrQMD model are the cross-section $\sigma_{\rm tot}$ between different species, which depend on the particle species and collision energies, and are tabulated from experimental data or parametrized according to the analytic calculations. In the semi-classical criterion, the cross-section between a pair of particles is approximated as $\sigma_{\rm tot}(\sqrt{s}) = \pi d_0^2$, which means that if the relative distance between the two particles $d_{\rm trans} < d_0$, the collision would happen. After hadrons cease interacting and reach kinetic freeze-out, the energy and momentum of light and heavy hadrons are collected to construct the final observables.

Currently, only the scatterings between D meson and $\pi,\rho$ mesons are implemented in UrQMD, using cross section calculated in Ref.~\cite{Lin:2000jp}. 

\section{The JETSCAPE framework}

The JETSCAPE simulation framework is an overarching computational envelope for developing complete evolution models for heavy ion collisions. It allows for modular incorporation of a wide variety of existing and future physics models that simulates different aspects of a heavy ion collision. The default JETSCAPE package contains both the framework and an entire set of indigenous and third-party routines that can be used to compare with experimental data directly \cite{putschke2019jetscape}. JETSCAPE is open source and its GitHub repository is hosted at github.com/JETSCAPE/JETSCAPE. In that repository, you can find the source code of the stable version of JETSCAPE, tools for statistical analysis, and past workshop material. You can find other information of JETSCAPE on its official website: jetscape.org.

Below is a typical workflow of one JETSCAPE event:

\begin{figure}[h]
	\centering
	\includegraphics[width=0.98\textwidth]{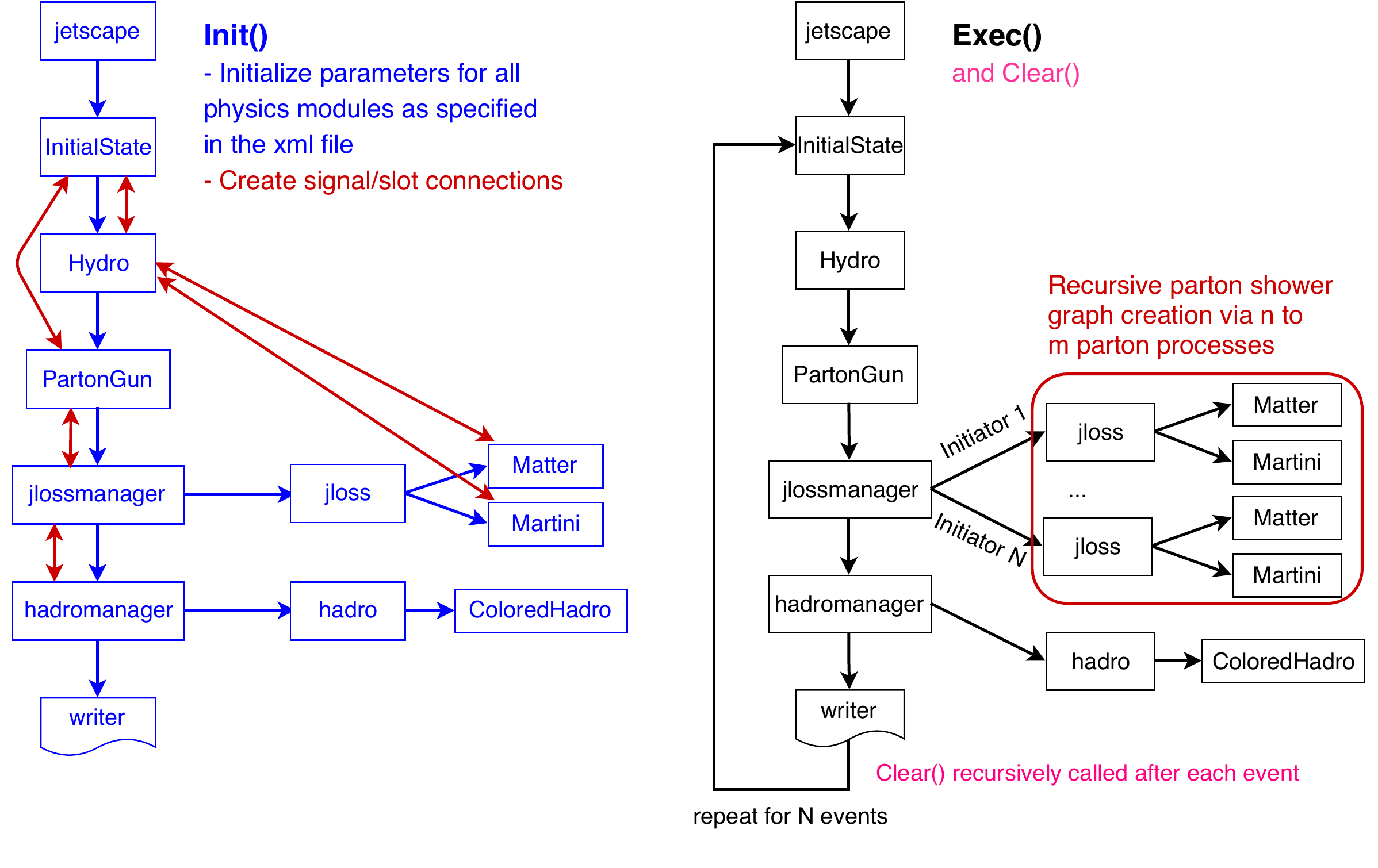}
	\caption[Typical workflow of JETSCAPE simulation]{\label{fig:jetscapeflow_extended} Typical workflow of JETSCAPE simulation. The modules are first initialized (left) and then executed (right). The modules can talk to each other during execution using the signal/slot mechanism \cite{putschke2019jetscape}.}
\end{figure}

\begin{figure}[h]
	\centering
	\includegraphics[width=0.65\textwidth]{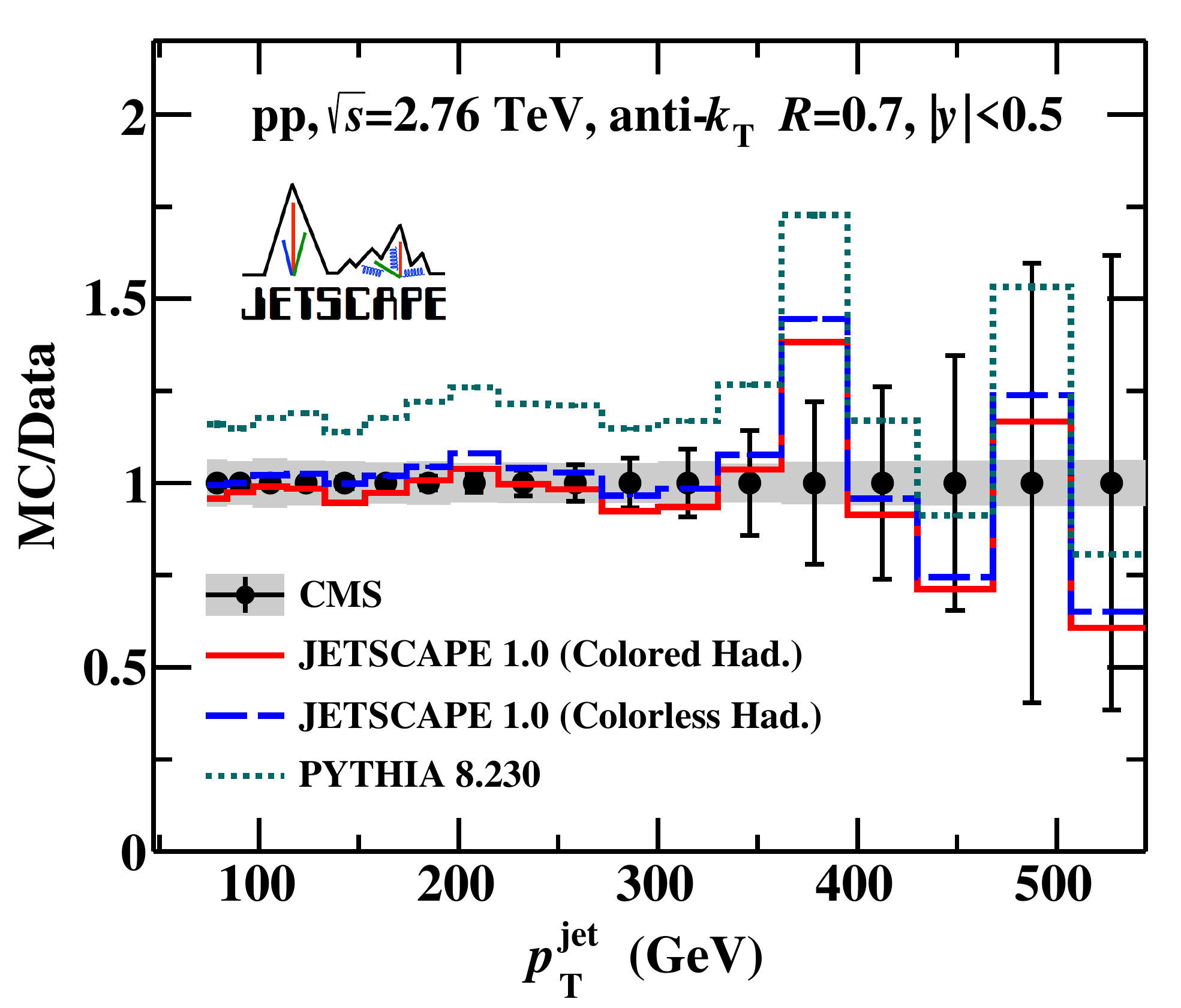}
	\caption[Ratio of inclusive jet yield between simulation and experiment data in pp collision at $2.76$TeV]{\label{fig:jetscape_pp_jet} Ratio of inclusive jet yield between simulation and experiment data in pp collision at $2.76$TeV \cite{putschke2019jetscape}.}
\end{figure}

\begin{figure}[!h]
\begin{center}
\begin{tabular}{cc}
\includegraphics[width=0.495\textwidth]{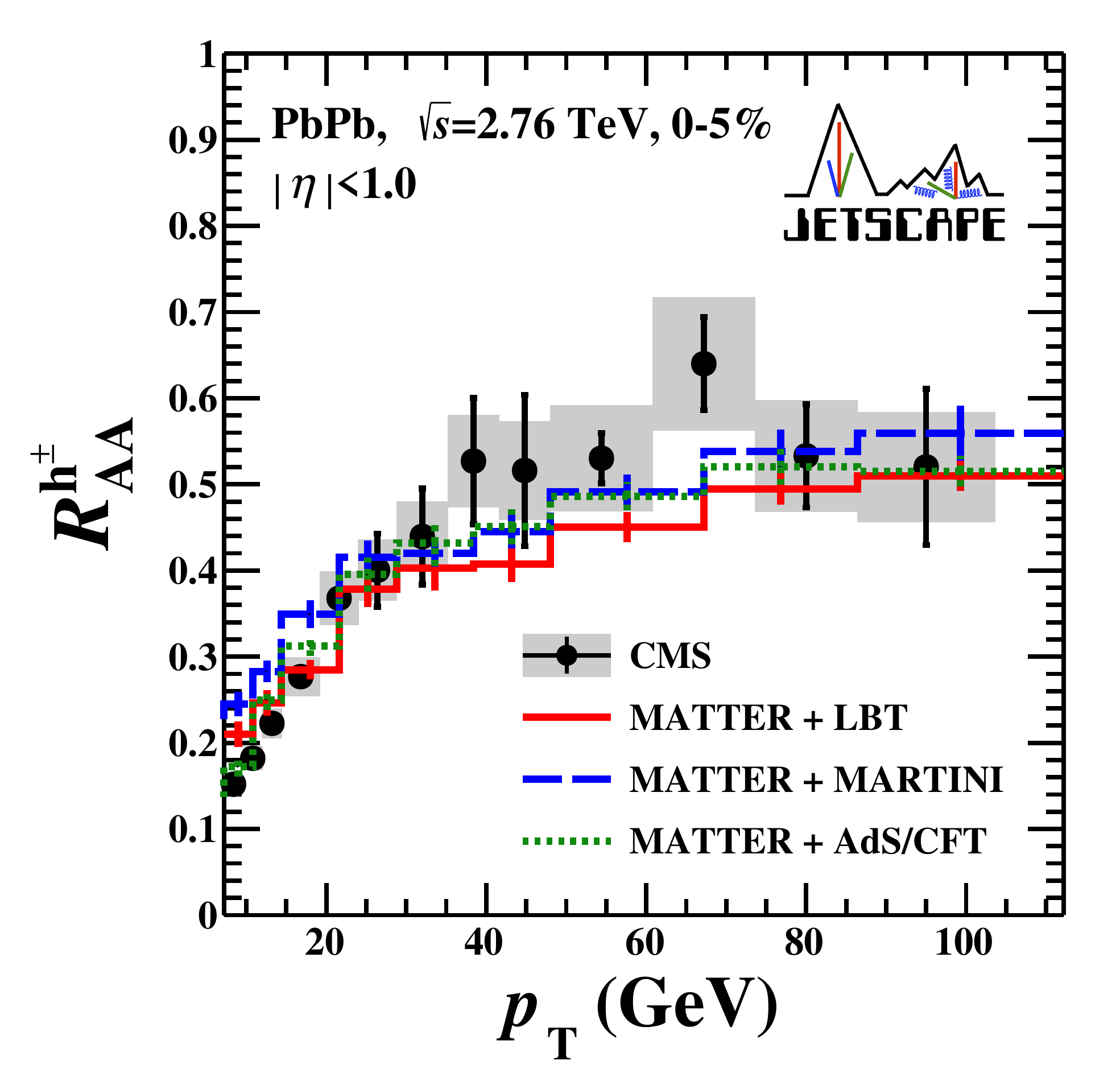} & \includegraphics[width=0.495\textwidth]{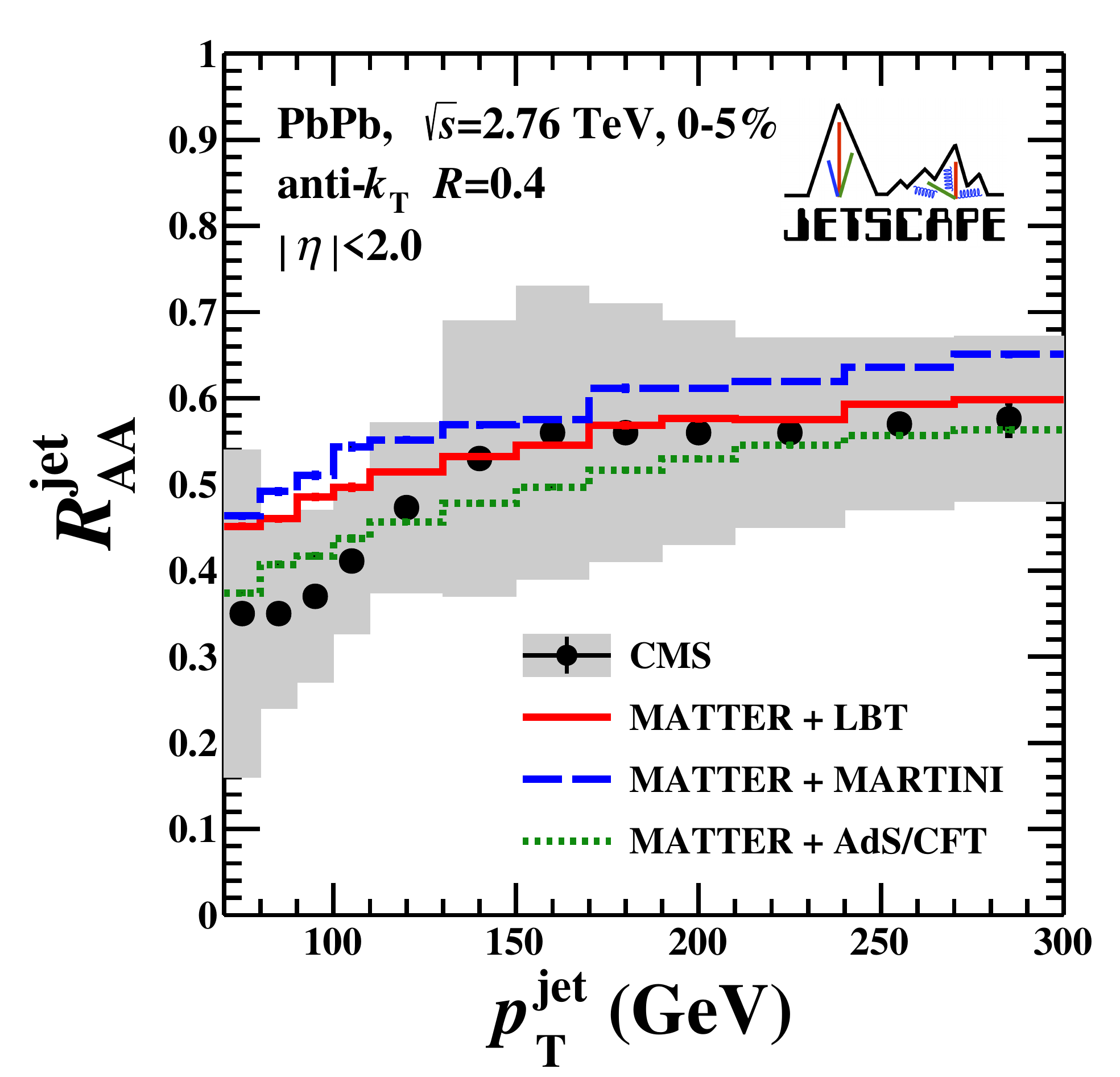}
\end{tabular}
\end{center}
\caption[Charged hadron and inclusive jet $R_{AA}$ in the most central PbPb collision at $2.76$TeV]{\textbf{Left:} Charged hadron $R_{AA}$ in the most central PbPb collision at $2.76$TeV. \textbf{Right:} Inclusive jet $R_{AA}$ in the most central PbPb collision at $2.76$TeV \cite{putschke2019jetscape}.}
\label{fig:jetscape_PbPb}
\end{figure}


Various calculation/benchmark have been done by the JETSCAPE collaboration (see Fig.~\ref{fig:jetscape_pp_jet} and Fig.~\ref{fig:jetscape_PbPb}). Different lines represent different models for hadronization (in Fig.~\ref{fig:jetscape_pp_jet}) or for parton energy loss in Fig.~\ref{fig:jetscape_PbPb}. This is one of the advantages of JETSCAPE that was mentioned before: being able to perform systematic comparisons between models.

\section{Summary}

I have listed the ``standard models'' in heavy ion collisions that I will utilize in this thesis. These models have successfully described various observables in both soft and hard sectors. Notice that they are not the only models/mechanisms that can describe the data, given the current varieties and uncertainties of the experimental data. We need to approach the RHIC problem in a systematic way. That is where the JETSCAPE framework comes into place. As a modular framework, JETSCAPE lets people develop and reuse code easily. It also allows researchers worldwide to perform full scale heavy ion collision simulations and compare them with experimental data, thanks to JETSCAPE being open source.

\chapter{Parton Energy Loss Inside the Medium} \label{sec:transport}

\vspace{1in}

During the evolution inside the QGP medium, hard partons will typically lose energy from the interactions with the medium. The interaction will not only depend on the momentum of the hard parton but also the temperature and flow of the medium. Different models have been formulated based on various assumptions, such as the degrees of freedom of the medium. Listed below are some of the examples:

\begin{enumerate}
    \item Langevin dynamics which makes no assumption on the constituents of the medium but focuses on the macroscopic properties of the medium such as the transport coefficients. Models following this assumptions include the Duke Langevin model\cite{Cao:2013ita,Xu:2017hgt} and the T-matrix model\cite{He:2012df,He:2011yi,He:2014cla}.
    \item Boltzmann dynamics which assumes the medium consists of quasi particles that interact with the hard partons via the Boltzmann equation. Models in this category includes the Lido model \cite{Ke:2018tsh}, the Catania-QPM model \cite{Plumari:2011mk,Scardina:2017ipo}, the BAMPS model\cite{Uphoff:2010sh,Uphoff:2011ad,Uphoff:2013rka,Uphoff:2014hza} and so on. These models differ by solving either the linearized or full Boltzmann transport equations, including or not including inelastic/radiative processes and the Landau-Pomeranchuk-Migda (LPM) effect, using different propagators, etc.
    \item The Parton-Hadron-String dynamics (PHSD) transport approach \cite{Song:2007fn,Song:2015sfa} is a microscopic covariant dynamical model which simulates the strongly interacting QGP medium based on Kadanoff-Baym equations. PHSD does not assume local equilibrium like hydrodynamics.
    \item The Ads/CFT model\cite{Horowitz:2007su,Horowitz:2011wm} which connects a field theory in $n$-dimensions to a string theory in $n+1$ dimensions. This correspondence can provide an upper limit for hard parton suppression in the strong coupling limit.
\end{enumerate}

In this chapter, the Boltzmann equation is first introduced. Then Langevin equation is derived from the Fokker-Plank equation which is a Boltzmann equation with the assumption of small momentum exchange. Next the radiation modifications to the aforementioned transport equations are discussed. Finally, the MATTER model is introduced which treats the in-medium Dokshitzer-Gribov-Lipatov-Altarelli-Parisi (DGLAP) evolution of a highly virtual parton. 

\section{Boltzmann dynamics}

The Boltzmann transport equation evolves the particle distribution in position and momentum space via localized collisions which occur at time scales much smaller than that of the mean free path $\tau \ll \lambda$. Then the collision probabilities can be evaluated using local particle distribution function and only include few-body collision processes. The Boltzmann equation reads:
\begin{equation}
\left(\frac{\partial}{\partial t} + \frac{\vec{p}}{E} \frac{\partial}{\partial \vec{x}}\right) f(t, \vec{x}, \vec{p}) = \mathcal{C}[f],
\label{eqn:Boltzmann_equation}
\end{equation}
where $f(t, \vec{x}, \vec{p})$ is the position and momentum distribution of the particle. $\mathcal{C}[f]$ represents the collision integral, including both elastic ($2\rightarrow2$) and inelastic ($2\rightarrow3$ and $3\rightarrow2$) processes. 

In the case of heavy ion collisions, the occupation number of hard particles (jet partons, heavy flavors) drops exponentially fast with the increase of $p_T$, so hard partons are very rare (occupation number $\ll 1$). Therefore, quantum statistical corrections to the hard parton distribution function and collision terms with more than one incoming hard partons can be neglected. The effect of hard partons on the distribution function of bulk particles can also be ignored. With these approximations, one can:
\begin{itemize}
    \item Approximate the distribution of the bulk particles to follow the local thermal distribution.
    \item Linearize the Boltzmann equation for the hard partons. 
    \begin{equation}
        \left(\frac{\partial}{\partial t} + \frac{\vec{p}}{E} \frac{\partial}{\partial \vec{x}}\right) f_H=-\mathcal{C}_H[f_H,f_{bulk}],
    \end{equation}
    where the collision integral $\mathcal{C}_H$ is a linear operator on $f_H$.
    
\end{itemize}

The collision integral $\mathcal{C}_H$ can be decomposed into a gain and a loss term:
\begin{equation}
\mathcal{C}_H[f(\vec{p})] = \int d^3k [\omega(\vec{p}+\vec{k}, \vec{k}) f(\vec{p}+\vec{k}) - \omega(\vec{p}, \vec{k}) f(\vec{p})],
\label{eqn:collision_integral}
\end{equation}
where $\omega(\vec{p}, \vec{k})$ denotes the collision rate for a parton changing momentum from $\vec{p}$ to $\vec{p}-\vec{k}$. In \ref{eqn:collision_integral}, the first term in the RHS represents the gain term while the second term is the loss term. 
Considering $2\rightarrow2$ elastic collisions between a heavy quark ($Q$) and light partons (light quark $q$ or gluon $g$), the collision rate can be written as:
\begin{equation}
\omega(\vec{p}_1, \vec{k}) = \sum_{2,3,4} d_{2} \int \frac{d^3p_2}{(2\pi)^3} f_2(\vec{p_2})[1\pm f_3(\vec{p}_1-\vec{k})][1\pm f_4(\vec{p}_2+\vec{k})] v_{\rm rel} d\sigma_{12\rightarrow 34}, 
\label{eqn:collision_rate}
\end{equation}
where index $1,2$ denote the incoming partons and $3,4$ denote the outgoing partons, $d$ is the spin-color degeneracy factor ($d_q=2 \times 3$ and $d_g=2 \times 8$), $v_{\rm rel}= \frac{\sqrt{(p_1^{\mu}p_{2\mu})^2 - m_1^2 m_2^2}}{E_1 E_2}$ is the relative velocity of the two incoming particles, and $d\sigma_{12\rightarrow 34}$ is the differential cross section. After all the different scattering channels are summed over, the elastic collision integral is derived as:
\begin{equation}
\begin{split}
\mathcal{C}^{12\rightarrow 34} (\vec{p}_1)= & \int \frac{d^3p_2}{(2\pi)^32E_2} \frac{d^3p_3}{(2\pi)^32E_3}  \frac{d^3p_4}{(2\pi)^32E_4} \sum \frac{d_2}{2}|\mathcal{M}|^2_{12\rightarrow 34} \left[f_3f_4 - f_1f_2\right]\\
&  \times (2\pi)^4  \delta^{(4)} (p_1+p_2 - p_3 - p_4),
\end{split}
\end{equation}

where $\mathcal{M}_{12\rightarrow 34}$ represents the matrix element for the scattering process $12\rightarrow 34$.

\subsection{Vacuum leading order matrix elements}

This section takes a look at the matrix elements of those elastic collision processes $|\mathcal{M}|^2_{12\rightarrow 34}$ in the vacuum. For the $Q+g \rightarrow Q+g$ process where $Q$ stands for a heavy quark:
\begin{equation}
\begin{split}
\left|\mathcal{M}_{Qg\rightarrow Qg}\right|^2 = & \pi^2 \alpha_s^2  [ \frac{32(s-M^2)(M^2 - u)}{t^2} + \frac{64}{9} \frac{(s-M^2)(M^2-u) + 2M^2(s+M^2)}{(s-M^2)^2}\\
& + \frac{64}{9} \frac{(s-M^2)(M^2-u) + 2M^2(u+M^2)}{(M^2-u)^2} + \frac{16}{9} \frac{M^2(4M^2-t)}{(s-M^2)(M^2-u)} \\
& + 16 \frac{(s-M^2)(M^2-u) + M^2(s-u)}{t (s-M^2)} - 16 \frac{(s-M^2)(M^2-u) - M^2(s-u)}{t (M^2 - u)} ],
\end{split}
\end{equation}
where $M$ is the quark mass, $s  = (p_1 + p_2)^2  = (p_3 + p_4)^2, t  = (p_1 - p_3)^2 = (p_2 - p_4)^2, u = (p_1 - p_4)^2 = (p_2 - p_3)^2$ are called the Mandelstam variables. 

The differential cross section in the center of mass frame reads:
\begin{equation}
\frac{d\sigma}{dt} = \frac{1}{16 \pi \sqrt{s} |\bf{p_1}|} \frac{1}{2E_1 2E_2 v_{\rm rel}}  \left|\mathcal{M}_{12\rightarrow 34}\right|^2,
\label{eqn: differential_sigma}
\end{equation}
where $v_{\rm rel}= \frac{\sqrt{(p_1^{\mu}p_{2\mu})^2 - m_1^2 m_2^2}}{E_1 E_2}$ is the relative velocity of the two incoming particles.

\begin{figure}
	\centering
	\includegraphics[width=0.3\textwidth]{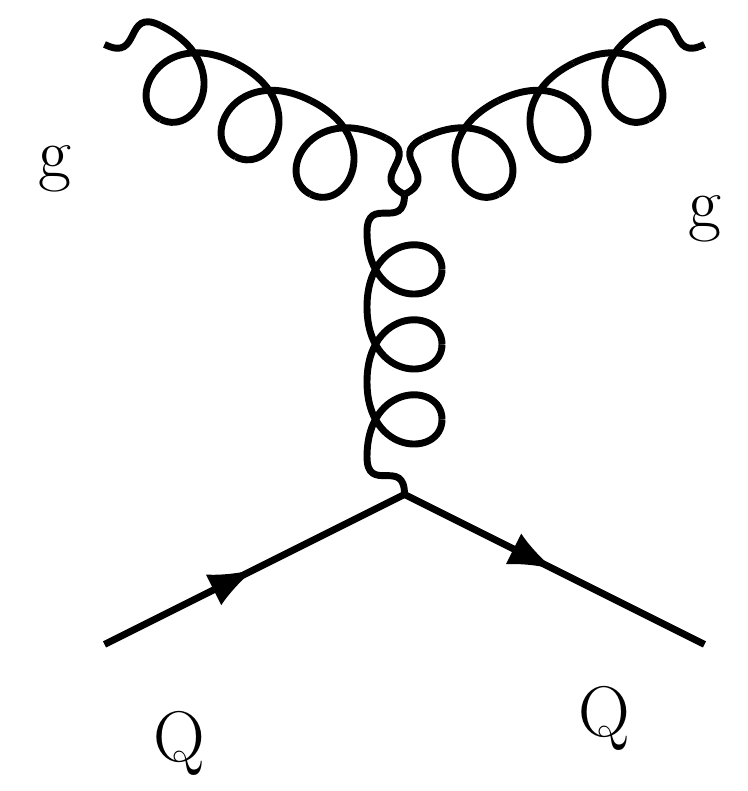}
	\includegraphics[width=0.3\textwidth]{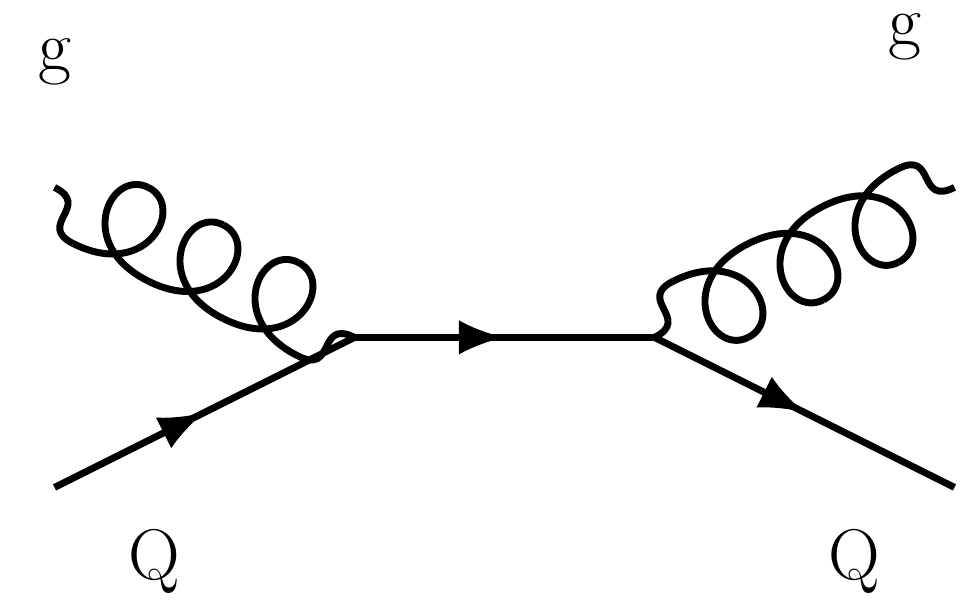}
	\includegraphics[width=0.3\textwidth]{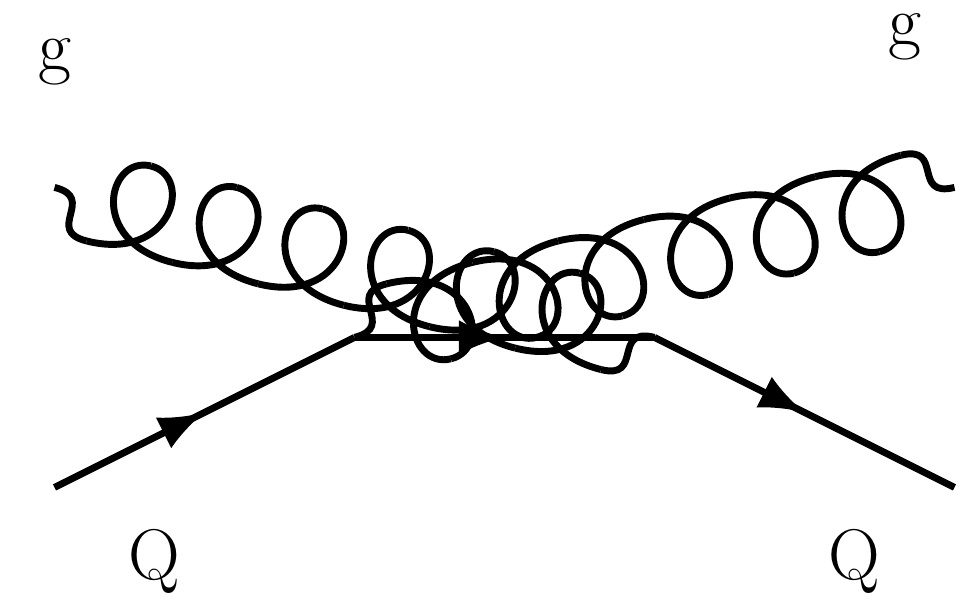}
	\caption[Leading order pQCD \textsc{Feynman} diagrams for heavy quark elastic scattering with gluons]{\label{fig:feynman diagram_Qg} Leading order pQCD \textsc{Feynman} diagrams for heavy quark elastic scattering with gluons $Q + g \rightarrow Q + g$ --- represents $t, s, u$ channels from left to right. Time line goes from left to right. }
\end{figure}
\begin{figure}
	\centering
	\includegraphics[width=0.3\textwidth]{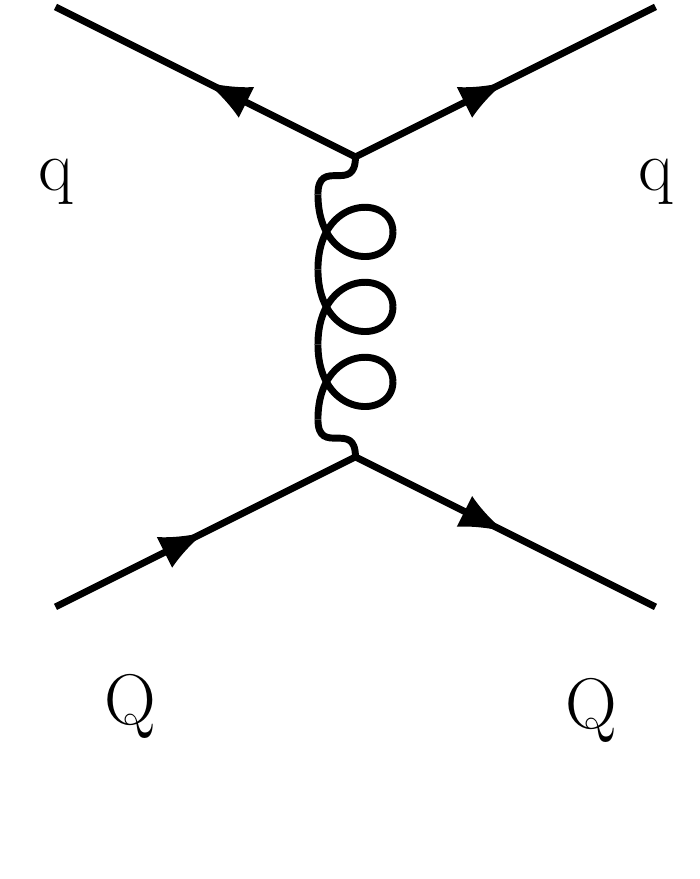}
	\caption[Leading order pQCD \textsc{Feynman} diagrams for heavy quark elastic scattering with light quarks]{\label{fig:feynman diagram_Qq} Leading order pQCD \textsc{Feynman} diagrams for heavy quark elastic scattering with light quarks $Q + g \rightarrow Q + g$ --- only $t$ channel survives. Time line goes from left to right. }
\end{figure}

For the  $Q+q \rightarrow Q+q$ where $q$ stands for a light quark:
\begin{equation}
\left|\mathcal{M}_{Qq\rightarrow Qq}\right|^2 = \frac{64}{9}\pi^2 \alpha_s^2 \frac{(M^2 -u)^2 + (s-M^2)^2 + 2M^2 t}{t^2}.
\end{equation}

And the differential cross section is given by:
\begin{equation}
\frac{d\sigma}{dt} = \frac{|\mathcal{M}_{12\rightarrow34}|^2}{16\pi (s-M^2)^2},
\end{equation}
where $v_{\rm rel}= \frac{\sqrt{(p_1^{\mu}p_{2\mu})^2 - m_1^2 m_2^2}}{E_1 E_2}$ is the relative velocity of the two incoming particles.

\subsection{In medium leading order matrix elements}

In the medium, two contributing factors will modify the matrix elements:

1. In a thermal medium with mobile charge, the scattering between a static charge and a fast moving charge are screened. The net effect is a modification to the parton propagator by a effective mass $\mu_t=\kappa_t m_D^2$:
\begin{equation}
\frac{1}{t} \rightarrow \frac{1}{t - \mu_t}.
\end{equation}

This so called Debye screening mass for quarks and gluons can be calculated as \cite{Xu:2004mz}:
\begin{equation}
m^2_{D} = \frac{8\alpha_s}{\pi} (N_c + n_f) T^2.
\label{eqn:debye_mass}
\end{equation}

2. The strong coupling constant $\alpha_s$ depends on the momentum transfer scale $Q^2$. In the case of $Q^2 \rightarrow 0$, $\alpha_s$ will diverge. When solving the Boltzmann equation, one need to either add a screening scale (for example, a scale proportional to the medium temperature) to regulate this divergence, or use an effective fixed value of $\alpha_s$ of the medium. 

Combining a Debye screening mass and a regulated coupling constant, the matrix elements of $2\rightarrow2$ scatterings in the medium can then be calculated. 

\subsection{Scattering rate}

In practice, when solving the linearized Boltzmann equation, the scattering rate is what is being calculated. For a time step of $\Delta t$, the average number of scattering for a particle with energy $E_1$ is:
\begin{equation}
    \begin{split}
\Gamma_{12\rightarrow 34}(E_1, T)  &=  \frac{d_{2}}{2E_1}  \int \frac{d^3p_2}{(2\pi)^32E_2} \frac{d^3p_3}{(2\pi)^32E_3}  \frac{d^3p_4}{(2\pi)^32E_4} \\
&f_2(E_2, T) \sum |\mathcal{M}|^2_{12\rightarrow 34} (2\pi)^4 \delta^{(4)}(p_1+p_2-p_3-p_4).
\label{eqn:gamma_2to2}
\end{split}
\end{equation}

The total scattering rate is summed over all possible channels. When propagating the hard parton through the medium, the average number of scattering during this $\Delta t$ is then $\left<n_s\right> = \Gamma \Delta t$. The actual distribution of the number of scatterings follows a Poisson distribution:

\begin{equation}
P(\rm{n_s\ge 1}) = e^{-\Gamma \Delta t} \frac{(\Gamma \Delta t)^{n_s}}{n_s !},
\end{equation}
and the probability for no collisions during $\Delta t$ is:
\begin{equation}
P(\rm{n_s=0}) = 1 - \sum_{n_s=1}^{\infty} P(n_s) = e^{-\Gamma \Delta t}.
\label{eqn:collision_probability}
\end{equation}

For solving the Boltzmann equation at a certain time step $t$, one must first sample a number from a uniform distribution between 0 and 1 to determine if a scattering happens (by comparing with $P(\Gamma(E_1, T))$). If no scattering happens, the hard parton will keep its momentum and propagate in a straight line. If a scattering does happen, the scattering channel is then determined by sampling according to the probability of each individual channel. Then the momentum of the final state particles are sampled from the differential cross sections. One thing to note is that the hard partons are propagated in the lab frame. But since the calculation of the scattering rate and sampling of the final state particles are most easily done in the center of mass frame of the scattering, one needs to perform Lorentz boost back and forth between the two frames. 

\section{Langevin dynamics}

\subsection{From Boltzmann equation to the Fokker-Plank equation}

If one assumes that the momentum exchange is small between the hard parton and the medium partons ($|\vec{k}| \ll |\vec{p}|$), the first term in the collision integral \ref{eqn:collision_integral} can be expanded with respect to $\vec{k}$ up to second order:
\begin{equation}
    \omega(\vec{p}+\vec{k}, \vec{k})f(\vec{p}+\vec{k}) \approx \omega(\vec{p}, \vec{k})f(\vec{p})+ k_i\frac{\partial}{\partial p_i}[\omega(\vec{p}, \vec{k})f(\vec{p})]+\frac{1}{2}k_i k_j \frac{\partial^2}{\partial p_i\partial p_j} [\omega(\vec{p}, \vec{k})f(\vec{p})].
\end{equation}

The collision kernel now becomes:
\begin{equation}
    \mathcal{C}_H[f(\vec{p})]\approx \int d^3k(k_i\frac{\partial}{\partial p_i} + \frac{1}{2} k_i k_j \frac{\partial^2}{\partial p_i \partial p_j}) \omega(\vec{p}, \vec{k})f_Q(\vec{p}),
\end{equation}
and the Boltzmann equation will reduce to the Fokker-Planck equation:
\begin{equation}
   \frac{\partial}{\partial t} f(t, \vec{p}) = \frac{\partial}{\partial p_i} \{ A_i(\vec{p}) f(t,\vec{p}) + \frac{\partial}{\partial p_j} [B_{ij}(\vec{p})f(t,\vec{p})]\},
   \label{eqn:fokker_plank_equation}
\end{equation}
where 
\begin{equation}
\begin{split}
A_i(\vec{p}) & =\int d^3k  k_i\omega(\vec{p},\vec{k}), \\
B_{ij}(\vec{p}) & =\frac{1}{2}\int d^3k  k_i k_j\omega(\vec{p},\vec{k}).
\end{split}
\end{equation}

If one defines the average operator over some quantity as:
\begin{equation}\label{eqn:average_operator}
\begin{split}
\left<X\right> &=  \frac{d_2}{2E_1}  \int \frac{d^3p_2}{(2\pi)^32E_2} \frac{d^3p_3}{(2\pi)^32E_3}  \frac{d^3p_4}{(2\pi)^32E_4} f_2(E_2, T) \\
&\sum |\mathcal{M}|^2_{12\rightarrow 34} (2\pi)^4 \delta^{(4)}(p_1+p_2-p_3-p_4) X \\
& = \int \frac{d^3 p_2}{(2\pi)^3} f_2(E_2, T) \Theta(s\ge 2m^2)  \frac{2s}{2E_1 2E_2} \sigma_{12\rightarrow 34}(s, T) X.
\end{split}
\end{equation}

The scattering rate $\Gamma=\left<1\right>$ by this definition. And one also gets:
\begin{equation}
\begin{split}
A_i(\vec{p}) & =\left<(p_i-p^{'}_i)\right>, \\
B_{ij}(\vec{p}) & =\frac{1}{2}\left<(p_i-p^{'}_i)(p_j-p^{'}_j)\right>.
\end{split}
\end{equation}

If one also assumes that the medium is in local equilibrium and rotational symmetry is preserved in the local rest frame of the scattering, $A_i, B_{ij}$ can be further decomposed into components that are in the longitudinal and transverse direction of $\vec{p}$:
\begin{equation}
\begin{split}
A_i(\vec{p}) & = A(\vec{p})p_i, \\
B_{ij}(\vec{p}) & = B_\parallel (\vec{p}) \frac{p_ip_j}{p^2}+B_\perp (\vec{p}) (\delta_{ij}-\frac{p_ip_j}{p^2}).
\end{split}
\end{equation}

\subsection{From Fokker-Plank equation to Langevin equation}
The classical Langevin equation which describes the Brownian motion of a single particle inside a thermal medium is :

\begin{equation}
\begin{split}
& \frac{dx_i}{dt} = \frac{p_i}{E}, \\
&  \frac{dp_i}{dt} = -\eta_D p_i + \xi_i(t).
\end{split}
\label{eqn:Langevin_equation}
\end{equation}

where $\eta_D$ is the drag coefficient and $\xi_i(t)$ describes the uncorrelated thermal random force exerted on the particle that has the following statistical properties:
\begin{equation}
\begin{split}
\left<\xi_i(t)\right> & = 0, \\
\left< \xi_i(t) \xi_j(t^{'}) \right> & = (\kappa_L \frac{p_i p_j}{p^2} + \kappa_T(\delta_{ij}-\frac{p_i p_j}{p^2}))\delta(t-t^{'}).
\end{split}
\end{equation}

Langevin equation can be derived from the Fokker-Planck equation as shown in Sec.~\ref{section:appendix_langevin_derivation}. Integrating over the position space in \ref{eqn:fokker_plank_equation_full} will yield \ref{eqn:fokker_plank_equation} with the following relations:
\begin{equation}
\begin{split}
& A_i(\vec{p}) = - \eta_D p_i,\\
& B_{ij}(\vec{p}) = \kappa_{ij} = \kappa_L \frac{p_i p_j}{p^2} + \kappa_L (\delta_{ij} - \frac{p_i p_j}{p^2}).
\end{split}
\end{equation}

When actually solving the Langevin equation, it remains ambiguous at which momentum the drag and random noise $\eta_D, \xi_{i}$ are evaluated. One can define a general momentum evaluation form:
\begin{equation}
\xi_{ij} = \xi(\vec{p} + C d\vec{p}),
\end{equation}
with $C\in [0,1]$. 

The momentum at two time steps are related by:
\begin{equation}
p_i^{t+\Delta t} - p_i^t = A_i (\vec{p}) \Delta t + \xi_i(\vec{p}) \Delta t.
\end{equation}

For the $C=0$ (called pre-point) scenario, the drag and diffusion terms are evaluated as:
\begin{equation}
\begin{split}
&A_i (\vec{p}) \equiv - \eta_D(p) p^i, \\
&B^{ij}(\vec{p}) \equiv \kappa_L(p) \frac{p_i p_j}{p^2} + \kappa_T(p) (\delta_{ij} - \frac{p_i p_j}{p^2}).
\end{split}
\end{equation}
where $p_i = p_i^{t}$,

The other choice of the discretization which has $C>0$, ($C=1/2$ refers to the mid-point scenario, and $C=1$ refers to the post-point scenario). The drag and diffusion terms is then evaluated as:
\begin{equation}
\begin{split}
&A_i (\vec{p}) \equiv - \eta_D(p) p^i - C \frac{\partial B_{ij}(\vec{p})}{\partial p_j}, \\
&B^{ij}(\vec{p}) \equiv \kappa_L(p) \frac{p_i p_j}{p^2} + \kappa_T(p) (\delta_{ij} - \frac{p_i p_j}{p^2}).
\end{split}
\end{equation}
where $p_i = (1-C)p_i^{t+\Delta t} + C p_i^{t}$.

In a large medium in thermal equilibrium, the hard partons will reach thermal equilibrium after evolving for sufficiently long time, meaning $f(\vec{p}) \propto e^{-E/T}$. The Fokker-Plank equation should still hold under this distribution and the time dependence is now zero. One then gets a constraint on the coefficients (dropping the dependence of the momentum):
\begin{equation}
A - \frac{B^{\parallel}}{2ET} + \frac{\partial B^{\parallel}}{\partial p^2} + \frac{B^{\parallel} - B^{\perp}}{p^2} = 0.
\end{equation}

This is referred to as the Einstein relationship, or the fluctuation dissipation relation. In terms of the drag and momentum transfer coefficients $\eta, \kappa_L, \kappa_T$, it can be written as:
\begin{equation}
\begin{split}
C = 0: & \eta_D = \frac{\kappa_L}{2ET} - \frac{\kappa_L - \kappa_T}{p^2} - \frac{\partial \kappa_L}{\partial p^2}, \\ 
C = 1: &\eta_D = \frac{\kappa_L}{2 ET}  - \frac{(\sqrt{\kappa_L} - \sqrt{\kappa_T})^2}{p^2}.
\end{split}
\label{eqn:Einstein_relation}
\end{equation}

\section{Radiation modification to the transport equations}

So far only consider elastic scatterings with the medium are considered in the Boltzmann equation and the Langevin equation. The next order correction for parton energy loss would be the $2\rightarrow3$ scattering with an additional gluon in the final state.  

High energy jet in-medium radiation is one of the most important topic in heavy ion physics. Four major phenomenological schemes that have been developed and widely used are:
\begin{itemize}
	\item Higher Twist ({\rm \tt HT})~\cite{Majumder:2009zu,Majumder:2007ne,Majumder:2007hx,Guo:2000nz,Majumder:2009ge}
	\item Path integral formalism of re-scattering summation on multiple static centers ({\rm \tt BDMPS-Z/ASW})~\cite{Baier:1996vi,Baier:1996kr,Baier:1996sk}
	\item Opacity expansion ({\rm \tt GLV})~\cite{Gyulassy:1999zd,Gyulassy:1999ig,Gyulassy:2000er,Gyulassy:2000fs,Gyulassy:2001nm}
	\item Finite temperature field theory approach ({\rm \tt AMY})~\cite{Arnold:2001ba, Arnold:2001ms, Arnold:2002ja, Arnold:2002zm}
\end{itemize}

The differences among those schemes lie in the different assumptions of: the nature of the medium, the virtuality of the energetic parton, and the kinetic approximations of the parton-medium interactions. Further comparisons among those approaches can be found in Ref.~\cite{Armesto:2011ht, Bass:2008rv}. In \cite{Bass:2008rv} those schemes are implemented with a 3-dimensional hydrodynamic approach, where the hard parton nuclear modification factor $R_{\rm AA}$ is compared with experimental data and a quantitative consistency of the momentum transport coefficients $\hat{q}$ is observed.

In this thesis, we adopted the Higher Twist formalism for radiative energy loss in the QGP medium. Under the assumption of collinear ($\omega \gg k_\perp$) and soft ($\omega \ll E$) radiation, the radiation rate of $Q\to Q+g$, where $Q$ is a heavy quark, is \cite{zhang2004heavy}:
\begin{equation}\label{eqn:heavy_quark_radiate_rate}
  \Gamma_{Q\rightarrow Q+g}=\int dydk_{\perp}^2\frac{dN_g}{dydk_{\perp}^2 dt} .   
\end{equation}

In Eq.~\ref{eqn:heavy_quark_radiate_rate}, $y$ is the fraction of the energy of the emitted gluon compared to the parent parton, $k_{\perp}$ is the gluon transverse momentum, and
\begin{equation}
    \frac{dN_g}{dy dk_{\perp}^2 dt}=\frac{2\alpha_s P(y)}{\pi k_{\perp}^4}\hat{q}(\frac{k_{\perp}^2}{k_{\perp}^2+y^2M^2})^4\sin^2(\frac{t-t_i}{2\tau_f}),
\end{equation}
where $P(y)$ is the splitting function, $\hat{q}$ is the transport coefficient and defined as $\left<\vec{p}_3-\hat{p}_1\cdot \vec{p}_3\right>$. $\tau_f=2Ex(1-x)/(k_\perp^2+x^2M^2)$ is the formation time of the radiated gluon and $t_i$ is the production time of the parent parton.

The average number of gluons emitted from a hard heavy quark, between $t$ and $t + \Delta t$, is:
\begin{equation}
    \bar{N}(t \to t + \Delta t) \approx\Delta t \Gamma_{Q\rightarrow Q+g}.
\end{equation}

As different successive emissions are independent, a Poisson distribution probability is employed, whereby the probability of emitting $n$ gluons is
\begin{equation}
\mathcal{P}(n)=\frac{\left(\bar{N}\right)^{n}}{n!}\exp\left[-\bar{N}\right],
\end{equation}
while the probability of a total inelastic process is $\mathcal{P}_{\rm inel.}=1-\exp\left[-\bar{N}\right]$. This procedure works for the linear Boltzmann equations. In the Langevin equation, the evolution of the radiated gluons are not recorded. The radiation will just modify the heavy quark's momentum through a recoil force term:
\begin{equation}
    f^g_i=-\frac{dp_i^g}{dt},
\end{equation}
where $p_i^g$ is the $i^{th}$ component of the momentum of the radiated gluons during $\delta t$. The Langevin equation is then updated as:

\begin{equation}
\frac{dp_i}{dt} = -\eta_D p_i + \xi_i(t) + f_i^{\rm gluon}.
\end{equation}

\section{Medium modified virtuality ordered parton showering}

So far the discussion has been focused on the evolution of on-shell hard particles inside the medium. In previous studies , these partons are generated by inclusive calculations like FONLL \cite{xu2018data} or Monte Carlo event generators, such as PYTHIA \cite{Ke:2018tsh,ke2021qgp}. In PYTHIA the parton showers are constructed first with some hard processes then recursively dressed up by emissions at successively “softer” (longer-wavelength) and/or more “co-linear” (smaller-angle) resolution scales. For example, the final state radiation (FSR), which generates time-like showers, will gradually reduce the virtuality of the partons inside the shower following the vacuum Dokshitzer-Gribov-Lipatov-Altarelli-Parisi (DGLAP) evolution. PYTHIA follows the QCD factorization theorem which separates the physics that live at different scales\cite{collins1989factorization} and is used to extract universal parton distribution functions (PDF) consistently from different processes and experiments \cite{thorne2019updates,dulat2016new,alekhin2017parton}. The parton distribution function $f(x_i,Q^2)$ represents the probability of finding a parton $i$ carrying $x_i$ fraction of the momentum of the incoming proton. It also depends on $Q^2$ which is the scale the proton is probed at. This scale is required to be large enough so that $\alpha_s(Q^2)$ is asymptotically small and the hard process can be calculated perturbatively. The parton fragmentation function $D^H(x_H,Q^2)$ is defined as the probability to find a certain hadron $H$ carrying a fraction $x_H$ of the parton’s momentum. The cross section for the inclusive production of the hadron $H$ can be written as \cite{field1989applications}:
\begin{equation}
    \frac{d\sigma_{p+p \rightarrow H+X}}{dydp_T^2}=\frac{1}{\pi} \int dx_idx_j f_i(x_i,Q^2)f_j(x_j,Q^2)\frac{d\sigma_{ij\rightarrow kl}}{d\hat{t}}\frac{1}{x_H}D^H(x_H,Q^2).
\end{equation}

\begin{figure}
	\centering
	\includegraphics[width=0.7\textwidth]{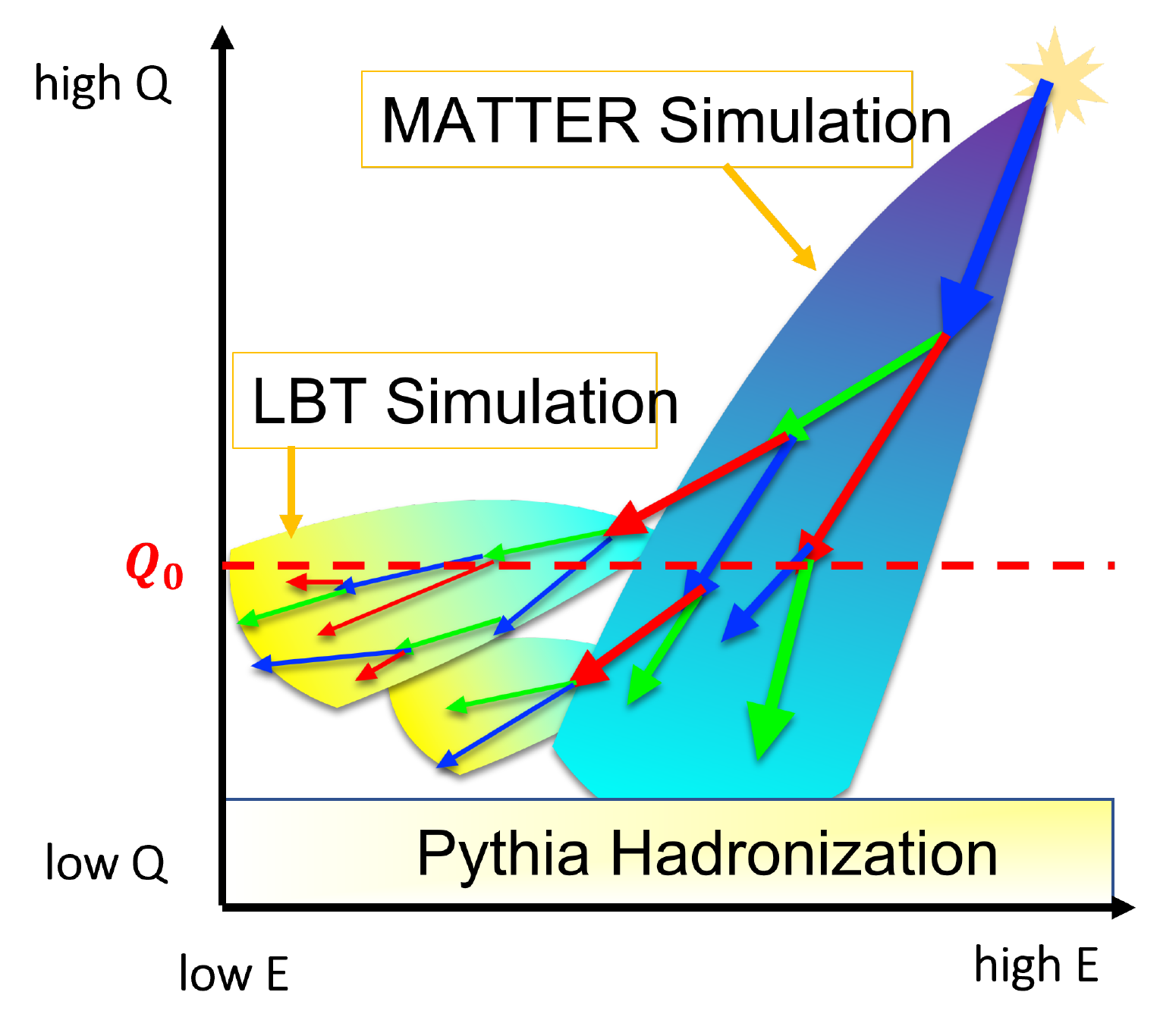}
	\caption[The phase space evolution of the parton shower]{\label{fig:parton_shower} The phase space evolution of the parton shower. Parton from hard scatterings generally start in the high $E$, high $Q$ region where the MATTER model is employed to describe the evolution. The parton will gradually lose virtuality via the medium-modified Sudakov factor. When the virtuality of the parton is below a switching virtuality $Q_0$, the LBT approach is used instead.}
\end{figure}

Although the parton distribution function and the parton fragmentation function are essentially non-perturbative objects, they parametrize universal long-distance physics and can be extracted from independent experiments at certain scales $Q^2_0$. Moreover, the evolution from the “definition”scale $Q^2_0$ to the process scale $Q^2$ can be described by the DGLAP evolution equations. 

The natural expectation in the case that a medium is formed in heavy ion collisions is that the DGLAP evolution is modified. Indeed, the formation time $\tau_f$ of a radiated gluon from a hard parton is $\approx 2\omega/k_\perp^2$ ($\omega, k^\perp$ are the energy and transverse momentum of the radiated gluon) and can be much longer than the lifetime of the medium. In this case, the radiation probability would be modified by scatterings with the medium. A more detailed discussion on the medium modification of the DGLAP evolution can be found in Ref.~\cite{cao2021medium}.

The Modular All Twist Transverse-scattering Elastic drag and Radiation (MATTER) model is a higher-twist formalism-based event generator that simulates the modification of the parton shower both in a vacuum and medium environment \cite{majumder2013incorporating}. It is primarily applicable to the high-virtuality, high-energy epoch of the parton shower, where the virtuality of the parton $t=Q^2\gg \sqrt{\hat{q}E}$ (see Fig.~\ref{fig:parton_shower}). In this phase, the medium-modified radiative processes are dominant, and the successive emissions from the parton are ordered in virtuality. A similar approach based on the multiple-soft scattering approximation (BDMPS) is called Q-PYTHIA \cite{armesto2009q}.

In MATTER, the distribution of the medium-modified radiated gluon from a single scattering with the medium is given as:
\begin{equation}
    \frac{dN_g}{dy dt}=\frac{\alpha_s}{2\pi} \frac{\Tilde{P}_a(y,t)}{t},
\end{equation}
where 
\begin{equation}
   \Tilde{P}_a(y,t)=P^{vac}_a(y)[1+\int_{\xi_0^{+}}^{\xi_0^{+}+\tau^{+}}d\xi^{+}K_a(\xi^{+}, \xi_0^{+},y,p^{+},t)].
\end{equation}

The index $a$ denotes the species of the parent parton. $P^{vac}_a(y)$ is the standard vacuum splitting function, $y$ is the momentum fraction carried by the emitted daughter parton, $p^+=(p^0+p^3)/\sqrt{2}$ is the light cone momentum for the parton in the z-direction, and $\tau^+=2p^+/t$ is the formation time of the radiated gluon. The parent parton started at $\xi_0^+$ and did the split at $\xi^+$ between $\xi_0^+$ and $\xi_0^{+}+\tau^+$. The quantity $K_a(\xi^{+}, \xi_0^{+},y,p^{+},t)$ is the single-emission-single-scattering kernel given as:
\begin{equation}
   K_a(\xi^{+}, \xi_0^{+},y,p^{+},t)=\frac{1}{y(1-y)t(1+\chi_a)^2}[2-2cos(\frac{\xi^+-\xi^+_0}{\tau^+})](C_1^a\hat{q}_a+C_2^a\hat{e}_a+C_3^a\hat{e}_{2,a}),
   \label{eqn:matter_scattering_kernel}
\end{equation}
where 
\begin{equation}
\begin{split}
& C_1^a=[1-\frac{y}{2}(\delta_{a,q}+\delta_{a,\overline{q}})]-\chi_a[1-(1-\frac{y}{2})\chi_a],\\
& C_2^a=\frac{2(1-y)}{\tau^+}\chi_a(1+\chi_a),\\
& C_3^a=\frac{4(1-y)}{yt(\tau^+)^2}\frac{\chi_a}{1+\chi_a}(\frac{1}{2}-\frac{11}{4}\chi_a).
\end{split}
\end{equation}

$\chi_a=(\delta_{a,q}+\delta_{a,\overline{q}})y^2m_a^2/(y(1-y)t-y^2m_a^2)$ with $m_a$ being the mass of the parent parton. The transport coefficients $\hat{q},\hat{e},\hat{e}_{2,a}$ encode the strength of parton-medium interactions. The coefficient $\hat{q}$ measures the average squared transverse momentum broadening per unit length of the medium, whereas $\hat{e}$ characterizes the average change in the longitudinal component of the parton momentum. $\hat{e}_{2,a}$ characterizes the average squared of change in the longitudinal momentum of the parton. If these transport coefficients are zero, the distribution of the emitted gluon in Eq.~\ref{eqn:matter_scattering_kernel} reduces to a vacuum-like distribution.

The virtuality ordered shower is generated based on the Sudakov formalism where one solves the in-medium DGLAP equation using Monte Carlo sampling. Given a maximum allowed virtuality $t_{max}$ and minimum virtuality $t_{min}$, one determines the virtuality of the parent parton $a$ by sampling the Sudakov form factor:
\begin{equation}
   S_a(t_{max},t)=\exp[-\int_{t}^{t_{max}}dt^{'}C_F\frac{\alpha_s(t^{'})}{2\pi t^{'}}\int_{y_{min}}^{y_{max}}dy\Tilde{P}_a(y,t^{'})].
   \label{eqn:matter_sudakov_factor}
\end{equation}

The Sudakov form factor represents the probability for a parton to transition from virtuality $t_{max}$ to $t$. The virtuality of the parent parton is determined by sampling a random number $R$ from the uniform distribution between $0$ and $1$. If $S_a(t_{max},t_{min})>R$, then the parton is assigned $t=t_{min}$ and propagates to the next time step without radiation (in the current simulations, $t_{min}$ is fixed to be $1GeV^2$). Otherwise the virtuality is determined by solving $S_a(t_{max},t)=R$. Then the splitting function $\Tilde{P}_a(y,t)$ is sampled to determine the momentum fraction $y$ shared by the two daughter partons (which equals to $(1-y)p^+$ and $yp^+$ respectively). The daughter partons' virtuality ($t_1$ and $t_2$) are then determined by again sampling the Sudakov form factor with $t_{max}=(1-y)^2t$ and $t_{max=y^2t}$. Their transverse momentum are:

\begin{equation}
    l^2_\perp=y(1-y)t-yt_1-(1-y)t_2.
\end{equation}

The $l^-$ component is then determined by the $l_1^2=t_1, l_2^2=t_2$ constraints. This procedure is repeated iteratively until the parton reaches a switching virtuality scale $t_s=Q_s^2$. Then the parton will be considered on-shell and propagated by other energy loss models like Boltzmann transport or Langevin dynamics. 

\subsection{Kinematic limits of the Sudakov form factor for heavy flavors}

For processes involving a heavy quark, the phase space in the Sudakov form factor is modified by the mass of the heavy quark. For a heavy quark radiating a gluon $Q\rightarrow Q+g$, the minimum and maximum momentum fraction allowed for this process, up to linear order in $t_{min}/t$ are:
\begin{equation}
\begin{split}
& y_{min}=\frac{t_{min}}{t}+\frac{M^2}{M^2+t}+\mathcal{O}((\frac{t_{min}}{t})^2),\\
& y_{max}=1-\frac{t_{min}}{t}+\mathcal{O}((\frac{t_{min}}{t})^2).
\end{split}
\end{equation}

Requiring $y_{max}>y_{min}$ implies that $t$ has a new lower bound $t_{min}(1+\sqrt{1+2M^2/t_{min}})$.

Heavy quarks can also be produced in the medium via $g\rightarrow Q + \overline{Q}$. The kinematics of this process again limits the available phase space. Assuming $M^2/t \ll 1$ and $t_{min}/t \ll 1$, we get:
\begin{equation}
\begin{split}
& y_{min}=\frac{t_{min}+M^2}{t}+\mathcal{O}((\frac{t_{min}+M^2}{t})^2),\\
& y_{max}=1-\frac{t_{min}+M^2}{t}+\mathcal{O}((\frac{t_{min}+M^2}{t})^2).
\end{split}
\end{equation}

Requiring $y_{max}>y_{min}$ implies that $t$ has a lower bound $2(t_{min}+M^2)$ for this process.

\subsection{Effective transport coefficient $\hat{q}$ in the high virtuality phase}

The jet transport coefficient $\hat{q}$ is defined as:
\begin{equation}
    \hat{q}=\frac{\left<p_T^2\right>}{L},
\end{equation}
where $\left<p_T^2\right>$ corresponds to the squared transverse momentum change of a parton as it traverses a distance $L$ through the QGP medium before splitting, and thus $\hat{q}$ is the average transverse momentum change per unit length. 

In the limit of high temperature and weak-coupling, the hard thermal loop (HTL) calculation gives:
\begin{equation}
    \hat{q}^{HTL}=C_a \frac{42\zeta(3)}{\pi} \alpha_s^2T^3\ln[\frac{cET}{4m_D^2}],
    \label{eqn:qhat_HTL}
\end{equation}
where $\zeta(3) \approx 1.202$ is the Ap\'{e}ry's constant, $C_a=3$ is the number of colors, and the Debye screening mass $m_D^2=\frac{4\pi \alpha_s T^2}{3}(N_c+N_f/2)$, $N_f=3$ and $c\approx5.7$ \cite{caron2009transverse}.

A first systematic extraction of $\hat{q}$ based on phenomenology was carried out by the JET collaboration \cite{burke2014extracting}. Extractions were based on a comparison of jet quenching model calculations to the experimental measurement of the hadron $R_{AA}$, in only the most central collisions at RHIC and LHC energies. These were performed independently, for five different parton energy loss approaches: GLV-CUJET \cite{gyulassy2001reaction}, HT-M \cite{majumder2012suppression}, HT-BW \cite{chen2011suppression}, MARTINI \cite{schenke2009martini}, and McGill-AMY \cite{qin2008radiative}. These calculations were run on identical (2+1)D
viscous hydrodynamical medium. The main result of this work was that the interaction strength $\hat{q}/T^3$ for the QGP at RHIC energy appeared to be up to twice as big compared to that at LHC energy, the so called "JET puzzle". A more data driven approach was carried out in recent study \cite{cao2021determining} in order to further constrain the dependence of $\hat{q}$ on $E$ and $T$.

\begin{figure}[h]
	\centering
	\includegraphics[width=0.7\textwidth]{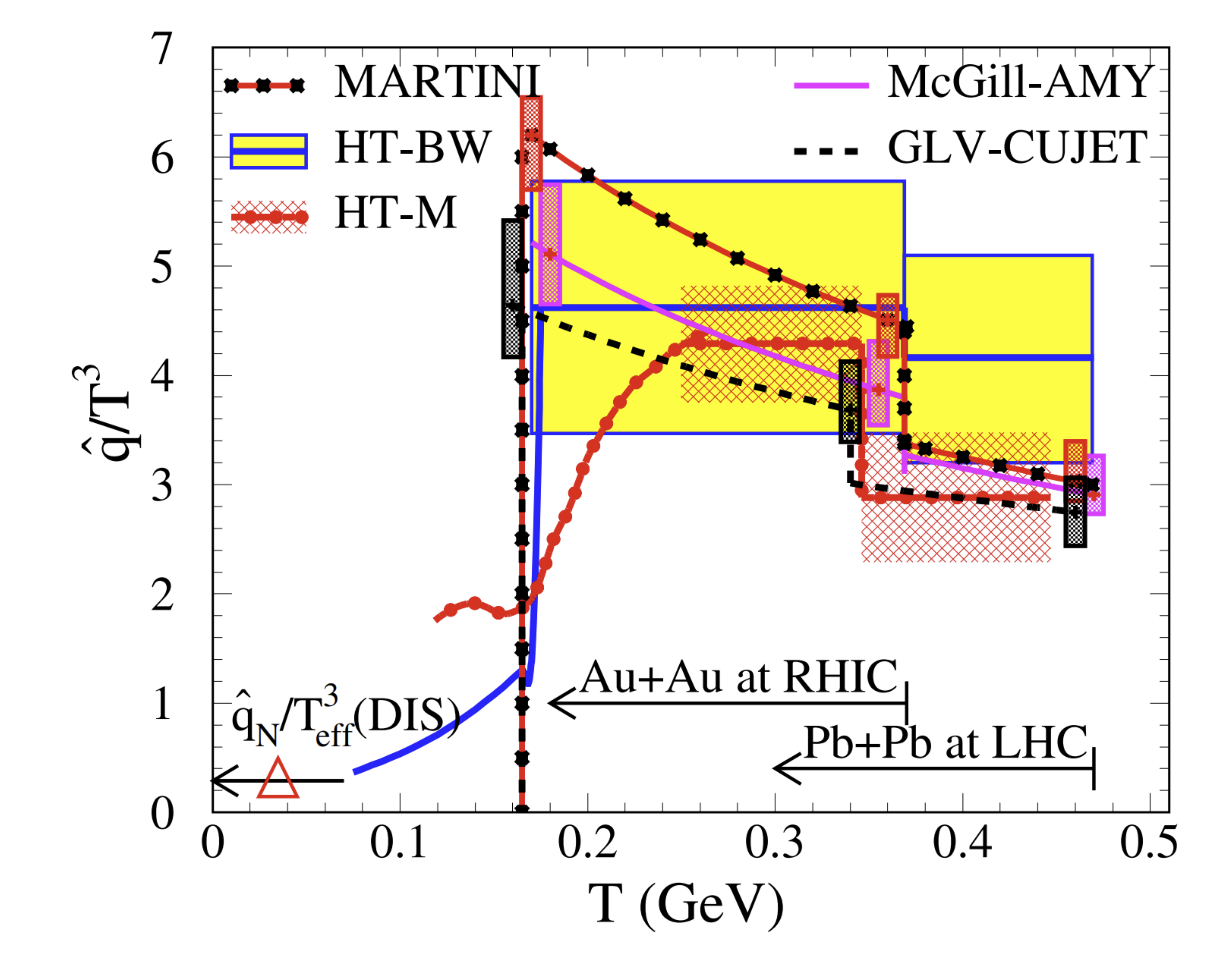}
	\caption[The assumed temperature dependence of the scaled jet transport parameter $\hat{q}/T^3$
in different jet quenching models]{\label{fig:qhat_jet_collab} The assumed temperature dependence of the scaled jet transport parameter $\hat{q}/T^3$
in different jet quenching models for an initial quark jet with energy $E = 10$ GeV. The values from different models are constrained by the charge hadron $R_{AA}$, in only the most central collisions at RHIC and LHC energies \cite{burke2014extracting}.}
\end{figure}

Up to this point, almost all attempts to extract the transport coefficient $\hat{q}$ have at most assumed dependence on $E$ and $T$, which are the only possibilities for an on-shell hard parton propagating through the plasma. This may not be the case for a highly virtual parton though. Several authors have argued that medium-induced radiation should depend on the resolution scale of the medium \cite{mehtar2012jets,casalderrey2011interference,mehtar2011antiangular}. The argument is that early in the history of the parton shower, the partons are very virtual and splittings involve large transverse momentum scales. The radiated gluons are unable to resolve the small transverse size of the dipole formed by the parton and the
emitted gluon, resulting in a reduction of the medium induced gluon radiation. This is called the coherence effect in jet propagation. 

Ref.~\cite{kumar2020energy} derived a more gradual reduction of medium induced emission in the high virtuality phase. The reduction in medium-induced emission is cast as a reduction in the effective value of $\hat{q}$ as a function of the parton virtuality $Q^2$. 

In this work I will employ a parameterization of the virtuality dependent $\hat{q}$ that reduces to the HTL formula at low virtuality but is suppressed at high virtuality:
\begin{eqnarray} 
\hat{q}(t) &=& \hat{q}^{HTL}\frac{c_0}{1+c_1\ln^2(t)+c_2\ln^4(t)}\nonumber\\
               &=& C_a \frac{42\zeta(3)}{\pi}\alpha_s(\mu^2) \alpha^{(\rm eff)}_s T^3 \ln \left[\frac{cET}{4m^2_D}\right]\frac{c_0}{1+c_1\ln^2(t)+c_2\ln^4(t)},
\label{eq:qhat_t}
\end{eqnarray}
where $c_1$ and $c_2$ are input parameters, $t$ is the virtuality of the parton, and $c_0$ is an overall normalization ensuring that the $t$-dependent contribution is unit-less and lies within 0 and 1. Another modification compared to the orignal HTL formula is that the $\alpha_s$ is now running with the scale $\mu^2=2ET$:
\begin{eqnarray}
\hat{q}^{HTL}=C_a \frac{42\zeta(3)}{\pi}\alpha_s(\mu^2) \alpha^{(\rm eff)}_s T^3 \ln \left[\frac{cET}{4m^2_D}\right],
\end{eqnarray}    
where
\begin{eqnarray}
\alpha_s(\mu^2)=\left\{
\begin{array}{rl}
\alpha^{(\rm eff)}_s & \mu^2 < \mu^2_0, \\
\frac{4\pi}{11-2N_f/3} \frac{1}{\ln\frac{\mu^2}{\Lambda^2}}& \mu^2 > \mu^2_0,\\
\end{array} \right.
\end{eqnarray}
with $\Lambda$ being chosen such that $\alpha_s(\mu^2)=\alpha^{(\rm eff)}_s$ at $\mu^2_0=1$ GeV$^2$. 

\section{Medium response in a weakly-coupled approach}

\begin{figure*}[htbp]
\centering
\includegraphics[width=0.45\textwidth]{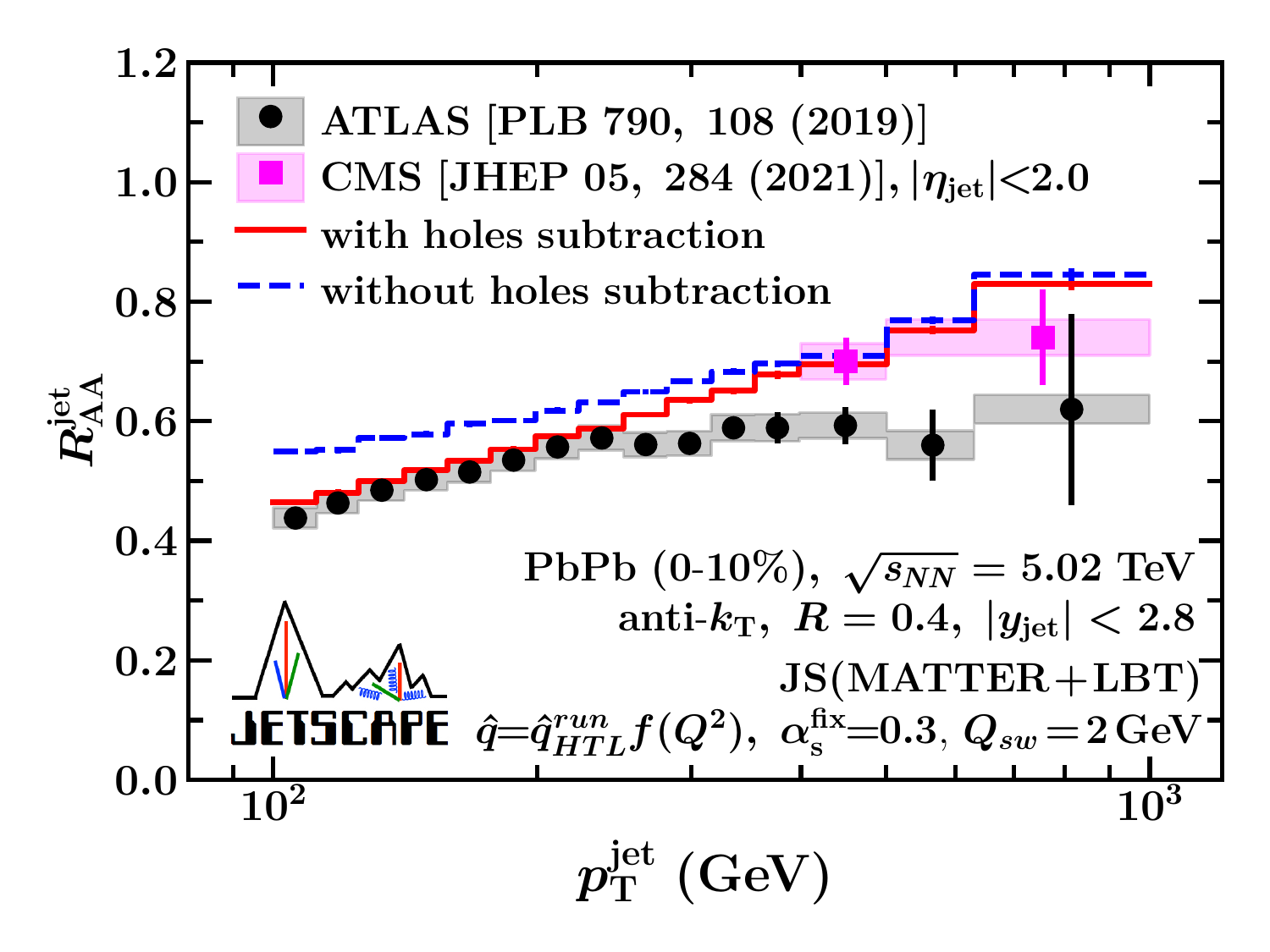}
\includegraphics[width=0.45\textwidth]{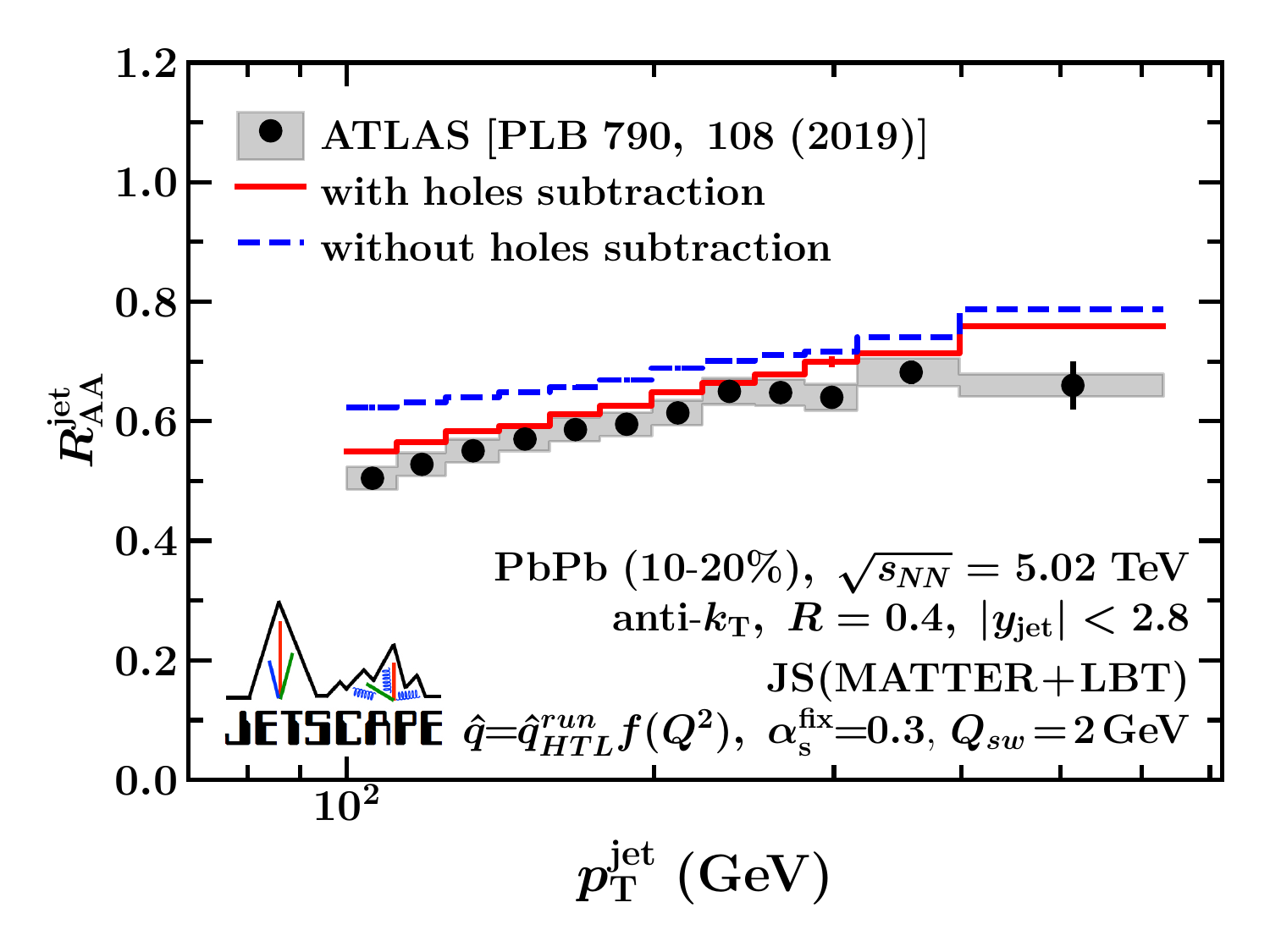}
\includegraphics[width=0.45\textwidth]{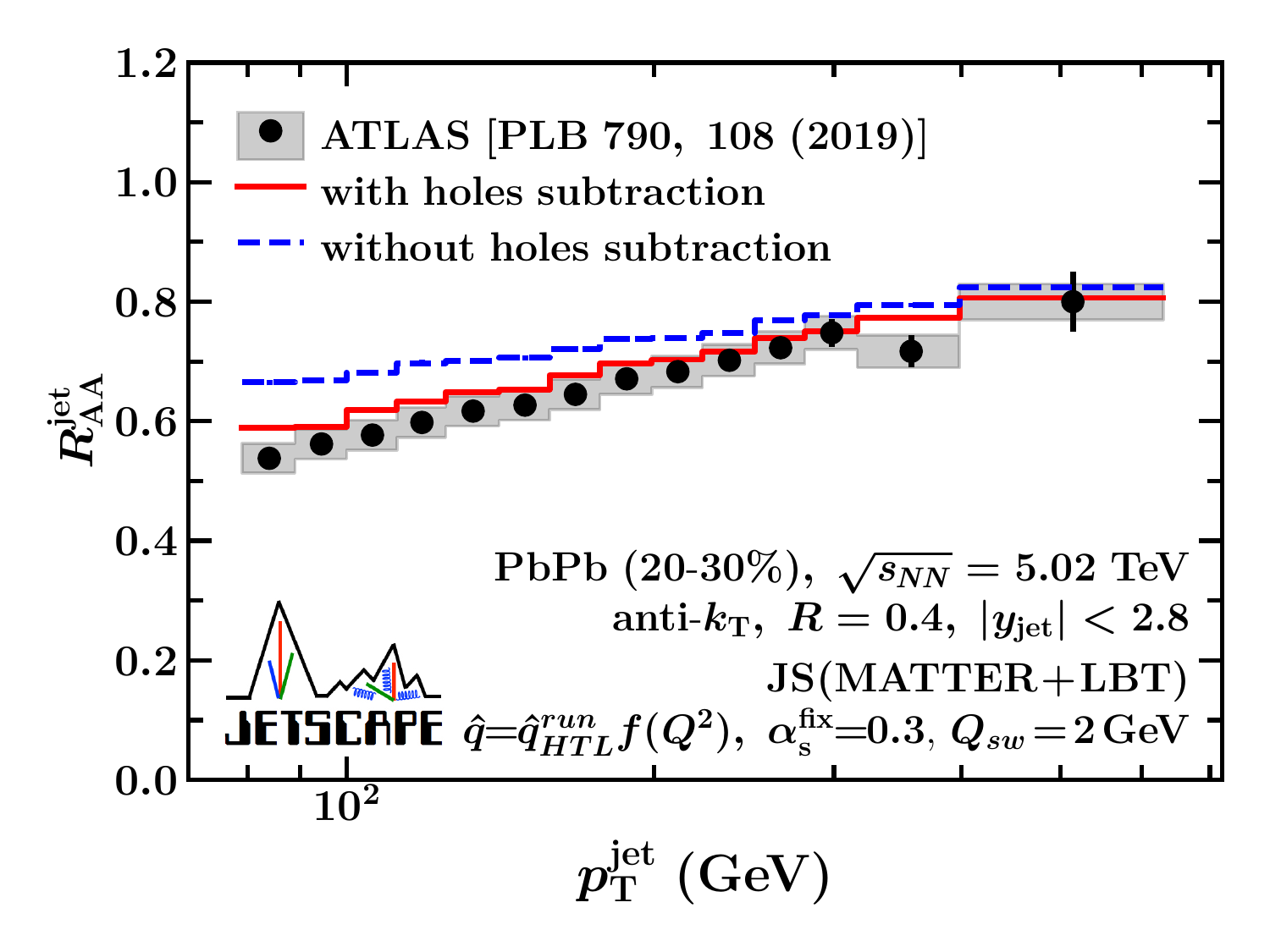}
\includegraphics[width=0.45\textwidth]{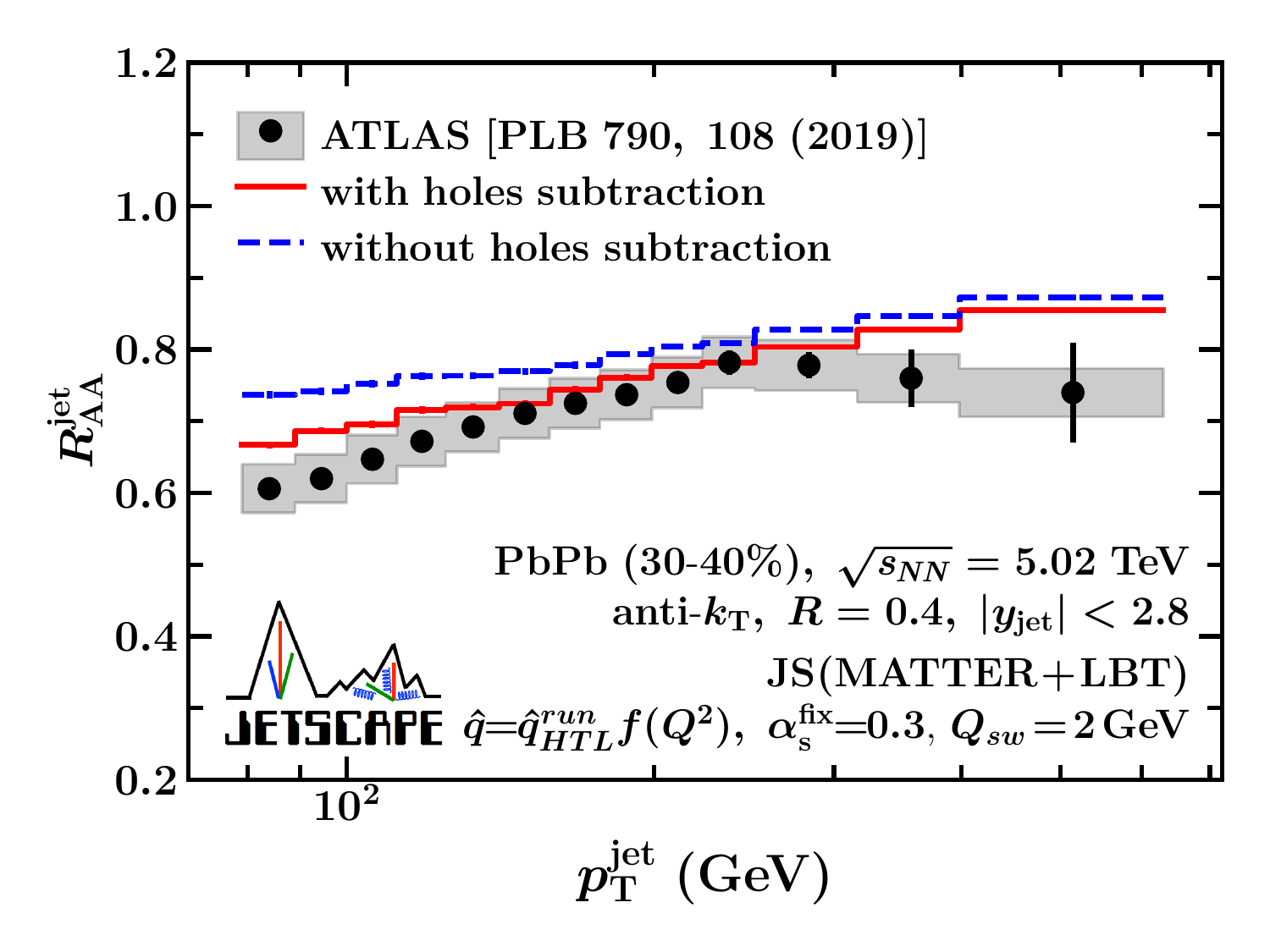}
\caption[Effect of hole subtraction on inclusive jet $R_{AA}$ at different centrality]{Centrality dependence of inclusive jet $R_{\mathrm{AA}}$ 
with $R=0.4$ and $y_{\mathrm{jet}} < 2.8$ 
at $\sqrt{s_\mathrm{NN}}$=5.02~TeV in PbPb collisions. The calculation is performed using a multi-stage jet quenching model (MATTER+LBT). The virtuality dependent $\hat{q}$ is used. Figures taken from Ref.~\cite{kumar2022inclusive}
}
\label{fig:semi_peripheral_collisions_5TeV_jets}
\end{figure*}

The parton showers exchange energy and momentum with the soft medium, during which they excite medium constituents. If we are just interested in leading parton observables, those relatively soft excitations can be ignored. When doing jet analysis, some of these excited partons are clustered within the jet which modify the structure of the reconstructed jets. In this study, the medium response is described as the propagation of recoil partons and their successive interactions with the medium in the JETSCAPE framework. In MATTER and LBT, the energy-momentum transfer between jets and the medium is executed via $2\rightarrow2$ scatterings. For each scattering, a medium parton is sampled from a thermal bath of a 3-flavor ideal QGP. After the scattering, the medium parton scattered by a hard parton, referred to as a recoil parton, is assumed to be on-shell, and its in-medium evolution is carried out by LBT, assuming weak coupling with the QGP medium. The parton showers including these recoil partons are hadronized together. On the other hand, the recoil parton leaves an energy-momentum deficit (hole) in the medium. We also keep track of these hole partons and subtract their contribution to ensure energy-momentum conservation. The hole partons are assumed to free-stream in the medium and are hadronized separately from other regular shower partons. The subtraction of the hole contribution for the final reconstructed jet momentum is performed as:
\begin{equation}
    p^{\mu}_{jet}=p^{\mu}_{shower}-\sum_{i \in holes} p^{\mu}_{i}
\end{equation}
where only holes inside the jet cone are considered. $p^{\mu}_{shower}$ denotes the four momentum of the jet reconstructed from all particles from the hadronization process including the recoils. The hole contribution to the inclusive jet $R_{AA}$ at different centralities is shown in Fig.~\ref{fig:semi_peripheral_collisions_5TeV_jets}.

This recoil prescription gives a reasonable description of the medium response as long as jet shower partons have sufficiently large energy and are far from thermalization, where their mean free paths are long enough to apply the kinetic theory. This recoil approximation breaks down when the showering partons' energy approaches the typical scale for the thermalized medium constituents \cite{tachibana2019medium, luo2021medium}. To extend this region of applicability, one needs to incorporate the hydrodynamic description for the soft modes of jets \cite{tachibana2017full, chen2018effects,tachibana2021hydrodynamic,casalderrey2021jet}. In this study, we do not include this hydrodynamic description for the medium response to jets, which requires a huge computational cost for the systematic studies of jets. Although they are essential for a more precise description of jet-correlated particle distribution and medium evolution, the recoil prescription is still a good approximation for the estimation of jet transverse momentum with typical jet cone sizes.

\section{Summary}

In this Chapter, I have first introduced the Boltzmann and Langevin equation, considering just elastic collisions. Then I discussed the radiative modification to these equations following the higher-twist formalism. Next, I introduced the in-medium DGLAP evolution described by the MATTER model. More importantly, a virtuality dependent parameterization for the transport coefficient $\hat{q}$ is proposed to try to explain the much smaller value of $\hat{q}$ extracted from collisions at LHC energy compared to at RHIC energy. Lastly, the treatment of recoil partons is laid out, which is important for the analysis of jet observables.

\chapter{Results of Parton Energy Loss in the JETSCAPE Framework} \label{sec:jetscape_results}

\vspace{1in}

In this chapter, the focus will be on calculating the $R_{AA}$ for charged hadron, D meson and inclusive jets with a multi-stage approach discussed in \ref{sec:transport}. Fig.~\ref{fig:parton_shower} shows the setup of our calculation. As discussed before, the MATTER model is suited for studying parton energy loss in the high virtuality, high energy regime. When the virtuality of the parton reaches some switching scale, one needs to switch to other appropriate energy loss models as the assumptions in MATTER no longer hold.

\section{The pp baseline}

\begin{figure}[htbp]
\centering
\includegraphics[width=0.49\textwidth]{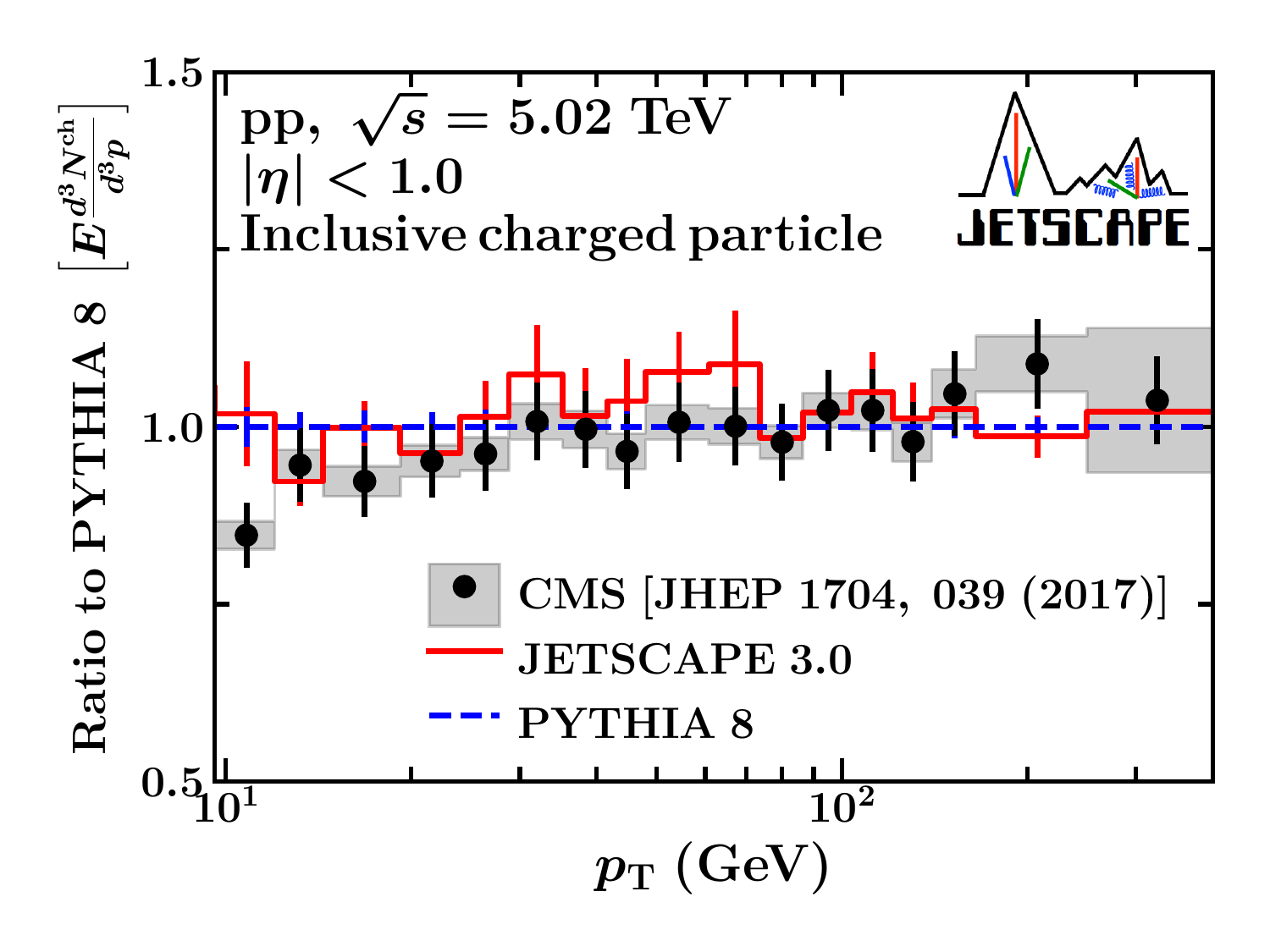}
\includegraphics[width=0.48\textwidth]{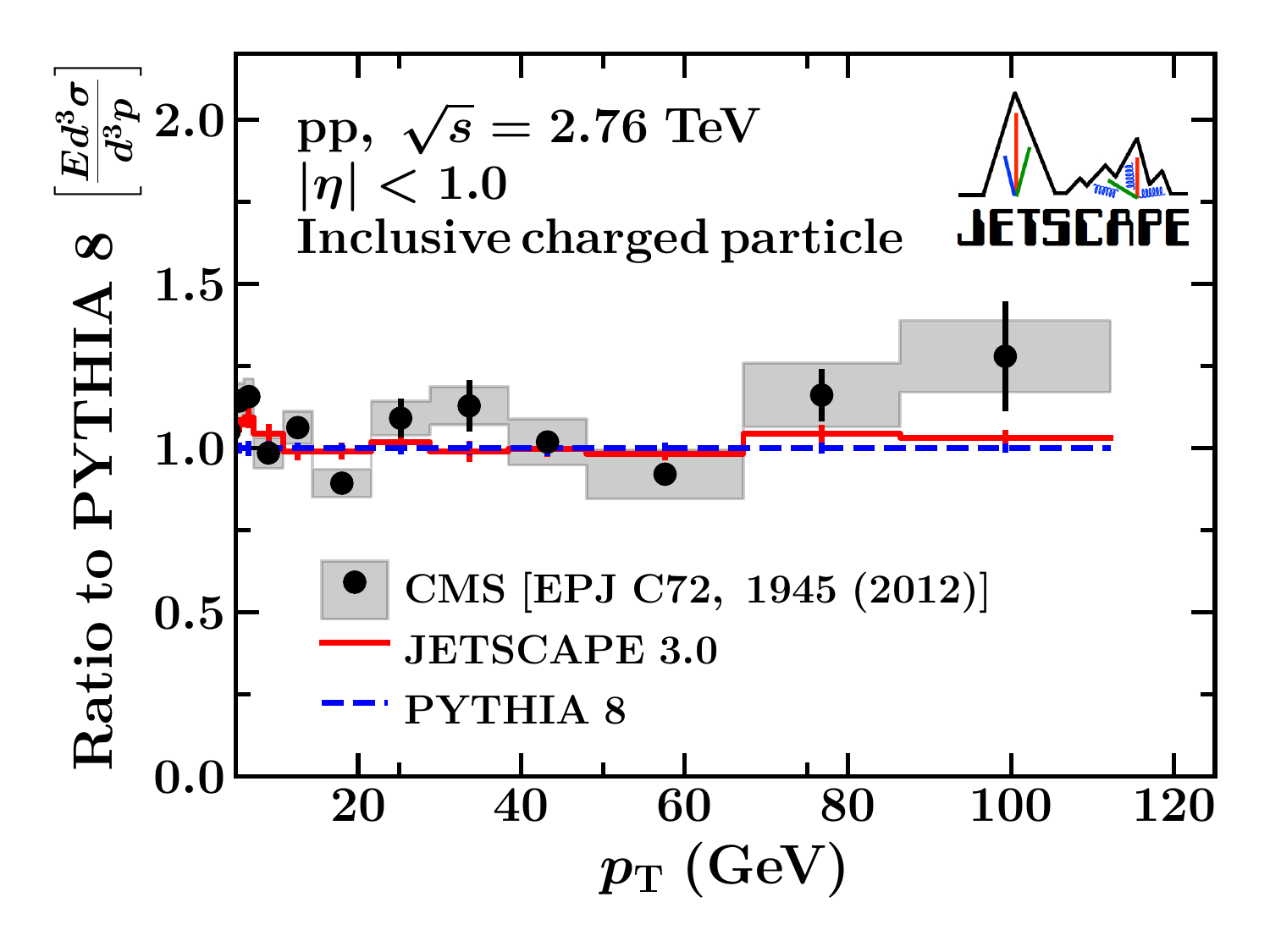}
\includegraphics[width=0.49\textwidth]{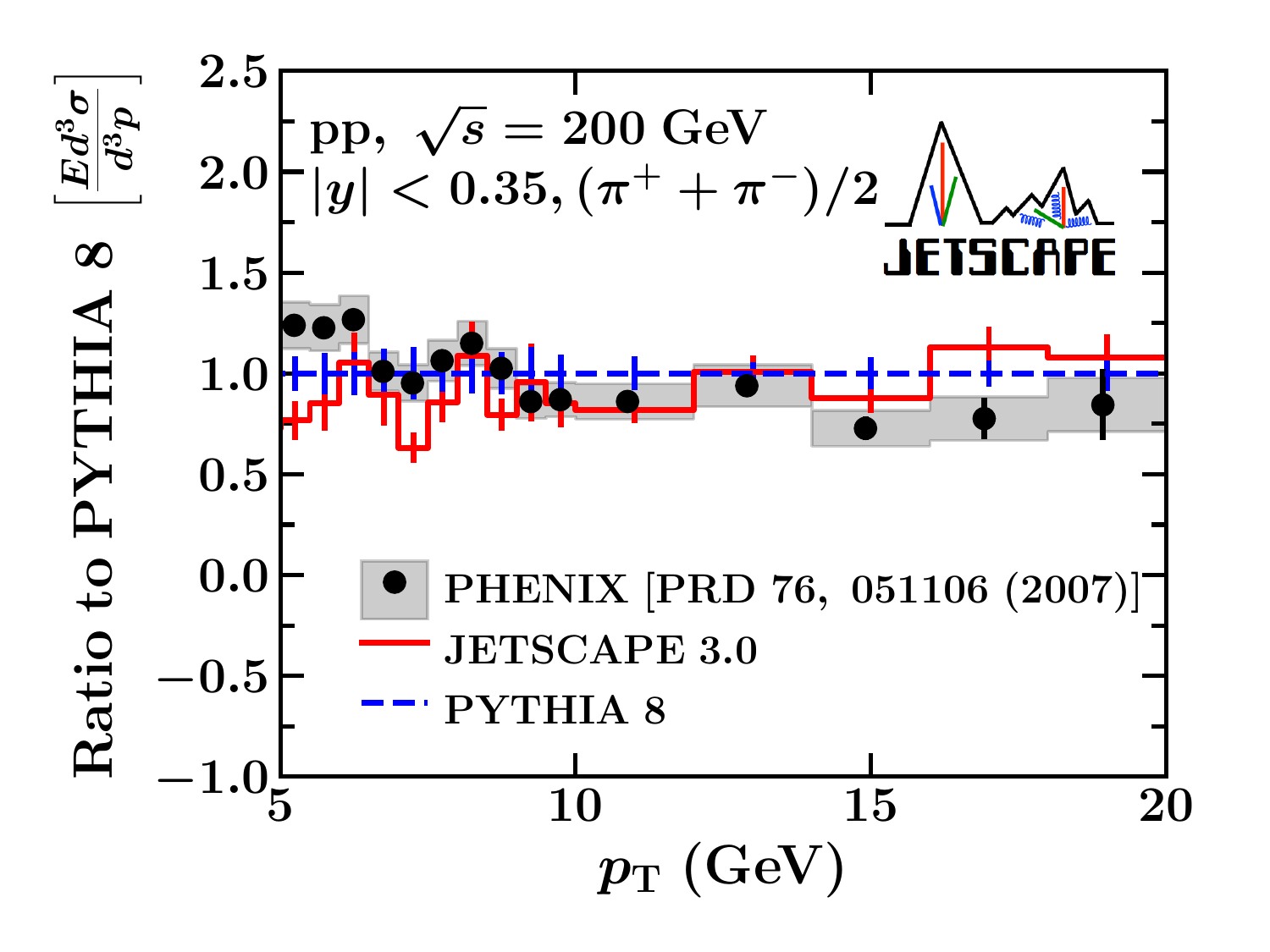}
\caption[Ratio of differential cross-section for inclusive charged-particle at mid-rapidity in pp collisions]{Ratio of differential cross-section for inclusive charged-particle at mid-rapidity in pp collisions. The ratio is taken w.r.t. the default PYTHIA. 
The solid red lines and dashed blue lines show the results from JETSCAPE and PYTHIA, respectively. 
Statistical errors (black error bars) and systematic uncertainties (grey bands) are plotted with the experimental data. 
\textbf{Top left}: 
Results for inclusive charged particle with $\eta < 1.0$ at $\sqrt{s}=5.02$~TeV. 
\textbf{Top right}: 
Results for inclusive charged particle with $\eta < 1.0$ at $\sqrt{s}=2.76$~TeV. 
\textbf{Bottom}: 
Results for charged pion with $y<0.35$ at $\sqrt{s}=200$~GeV. Figures taken from Ref.~\cite{kumar2022inclusive}.}
\label{fig:ratio_plot_chargedParticleYield_2760GeV_200GeV_CMS_RHIC}
\end{figure}

\begin{figure}[htbp]
\centering
\includegraphics[width=0.47\textwidth]{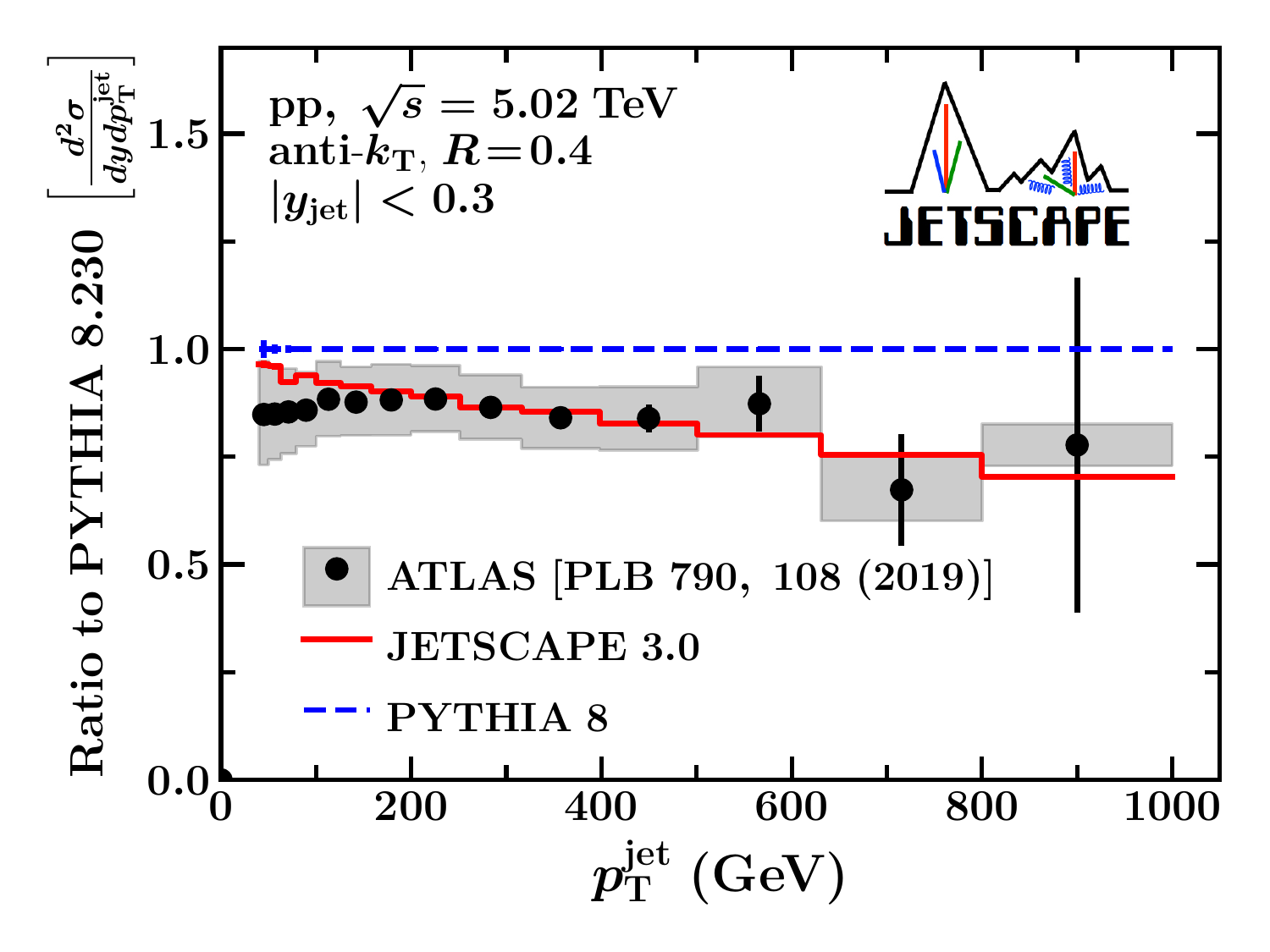}
\includegraphics[width=0.47\textwidth]{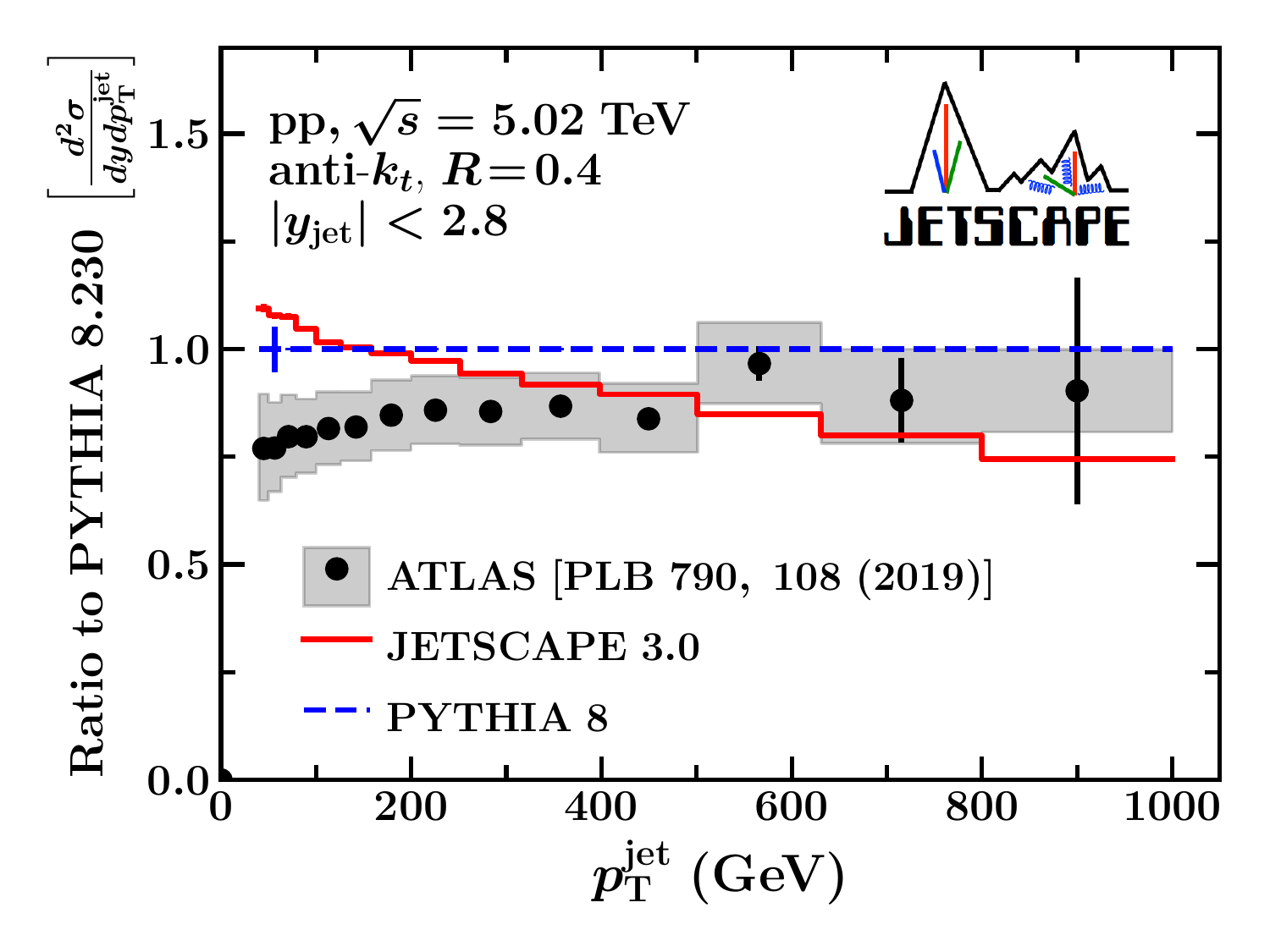}
\includegraphics[width=0.47\textwidth]{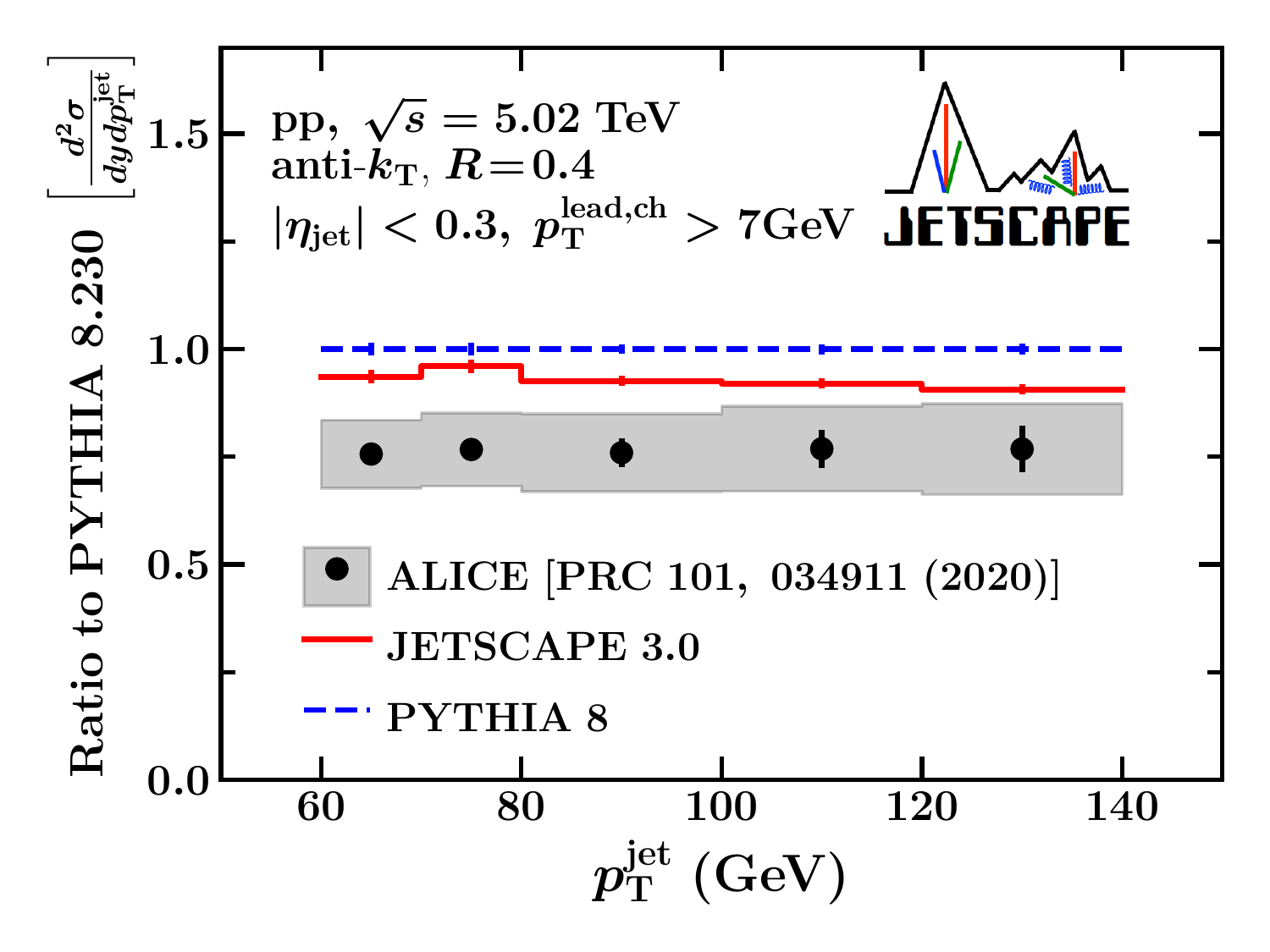}
\caption[Ratio of differential cross section for inclusive jets with cone size $R=0.4$ at mid-rapidity in pp collisions at $\sqrt{s}=5.02$~TeV]{Ratio of differential cross section for inclusive jets with cone size $R=0.4$ at mid-rapidity in pp collisions at $\sqrt{s}=5.02$~TeV. The ratio is taken w.r.t. the default PYTHIA MC. 
The solid red lines and dashed blue lines show the results from JETSCAPE and PYTHIA, respectively. 
Statistical errors (black error bars) and systematic uncertainties (grey bands) are plotted with the experimental data. 
\textbf{Top left}: 
Results for $y_{\mathrm{jet}} < 0.3$. 
\textbf{Top right}: 
Results for $y_{\mathrm{jet}} < 2.8$. 
\textbf{Bottom}: 
Results for $\eta_{\mathrm{jet}} < 0.3$ with $p^{\mathrm{lead,\,ch}}_{\mathrm{T}} > 7\,\mathrm{GeV}$. Figures taken from Ref.~\cite{kumar2022inclusive}.
}
\label{fig:ratio_plot_DJCS_5020GeV_ATLAS_ALICE}
\end{figure}

In order to isolate the effects of medium modification, the pp baseline needs to be checked first. A systematic study of inclusive jet, jet substructure and charged particle observables in pp collisions has been carried out using the JETSCAPE PP19 tune and presented in Ref.~\cite{kumar2020jetscape}. Here I only present the plots for charged hadron and inclusive jet $R_{AA}$. 
One can see that JETSCAPE yields similar results for charged hadron spectra compared to PYTHIA and are compatible with experimental data for three collision energies in pp collisions (see Fig.~\ref{fig:ratio_plot_chargedParticleYield_2760GeV_200GeV_CMS_RHIC}). For inclusive jet results, the story is a bit more complicated. JETSCAPE calculation is better than PYTHIA calculation for jets with smaller rapidity but overestimates the lower $p_T$ jet spectra for jets with a wider rapidity range (see Fig.~\ref{fig:ratio_plot_DJCS_5020GeV_ATLAS_ALICE}). Overall, JETSCAPE achieves a slightly better description of  observables in pp collisions compared to PYTHIA. 

\section{Comparison between different formulations in PbPb collisions at $5.02$~TeV and $0-10\%$ centrality}

In this section, I will try to identify which model or parameter contribute to the shape and magnitude of the $R_{AA}$. The first thing I want to explore is to compare between a single energy loss model and the multi-stage approach. If we only use the MATTER model, we will evolve the partons down to a fixed switching virtuality $\approx 1$GeV. If only the LBT model is used, which means MATTER is turned off, the final state radiation (FSR) in PYTHIA will be turned on. And the pp baseline calculation will not use MATTER for better consistency.

The evolution of the QCD medium used throughout this study is performed using a boost-invariant 2+1-dimensional hydrodynamic model which involves three stages: a pre-hydrodynamic, hydrodynamic and a hadronic transport stage~\cite{bernhard2019bayesian}. The pre-hydrodynamic stage is composed of the \trento{} model (initial condition for PbPb collisions), followed by free-streaming for a proper time of $\tau_{FS}=1.2$ fm/$c$. This generates a non-trivial initial condition for the hydrodynamical simulation to follow. We have generated in total 400 \trento{} initial PbPb configurations in the 0-10\% centrality class at $\sqrt{s_{NN}}=5.02$ TeV. 

The hydrodynamical simulation is performed until the the cross-over temperature of $T_c=154$ MeV is reached~\cite{Bazavov:2014pvz}, at which point fluid fields are converted into particles \cite{Bernhard:2018hnz} whose subsequent evolution is governed by hadronic Boltzmann transport \cite{bass1998microscopic}.

Hard probes do not interact during the pre-hydrodynamical evolution as it is given by free-streaming. Since we shall focus on momenta above $\approx 7$ GeV, we neglect hadronic final state interactions as well. Thus, charm quarks only interact during the hydrodynamical portion of the evolution, which is given by second order Israel-Stewart theory \cite{Israel:1979wp}. An estimation for the effect of hadronic interactions on $R_{AA}$ is given in Ref.~\ref{da2020studies}.

\subsection{$R_{AA}$ from LBT}
\label{sec:results_lbt}
To obtain $R_{AA}$ using LBT as the sole energy loss mechanism, an initial parton distribution needs to be provided. One way to obtain this distribution is using the PYTHIA vacuum shower mechanism. The latter is also used to provide the proton-proton baseline needed to calculate $R_{AA}$. Combining PYTHIA and LBT, two simulations were performed: one $\alpha^{\rm (eff)}_s=0.3$ serves as reference $R_{AA}$ calculation, while the other, using $\alpha_s(\mu^2)$, studies the effects of a running $\alpha_s$ on $R_{AA}$. 

\begin{figure}[!h]
\begin{center}
\begin{tabular}{cc}
\includegraphics[width=0.495\textwidth]{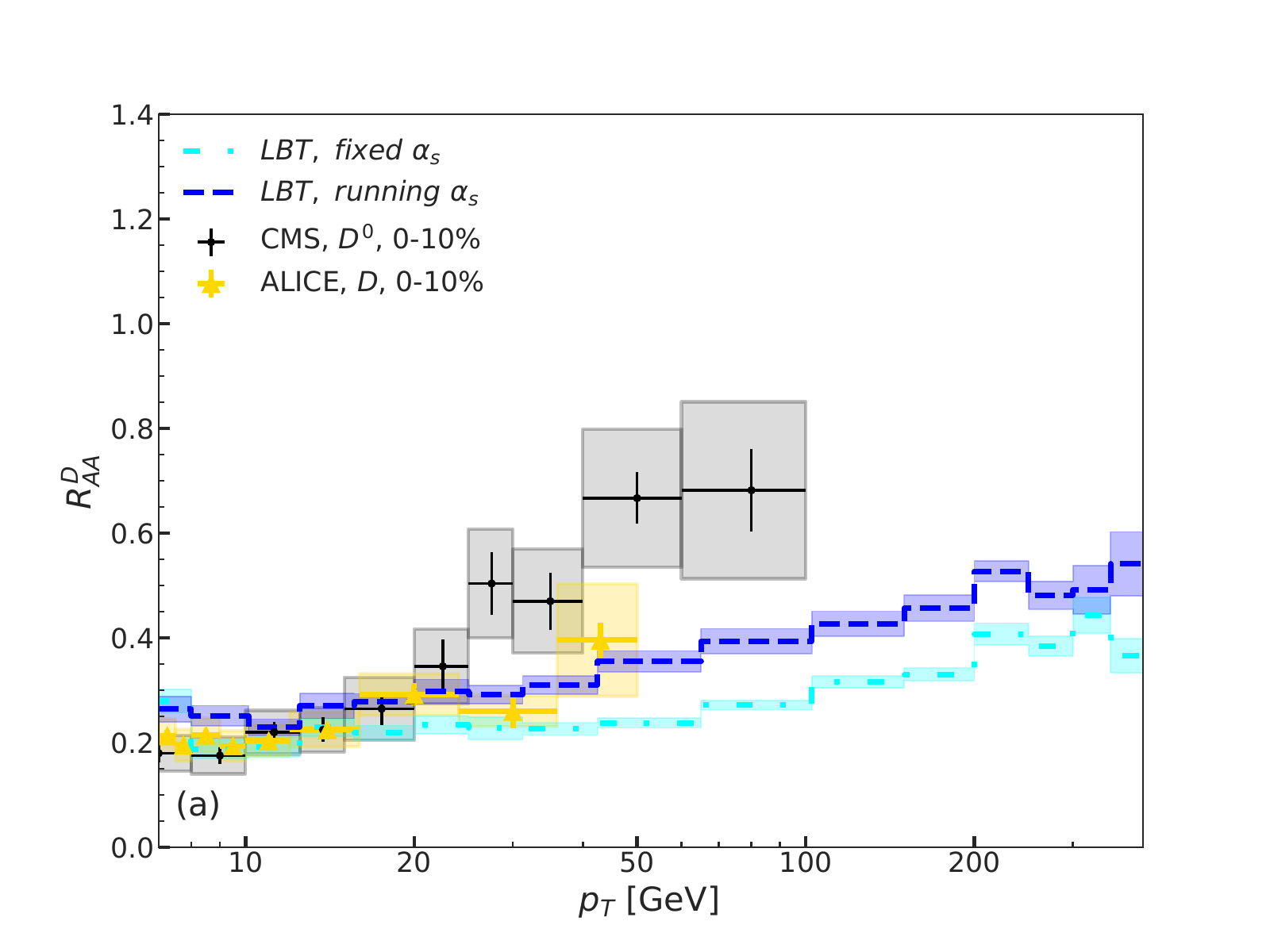} & \includegraphics[width=0.495\textwidth]{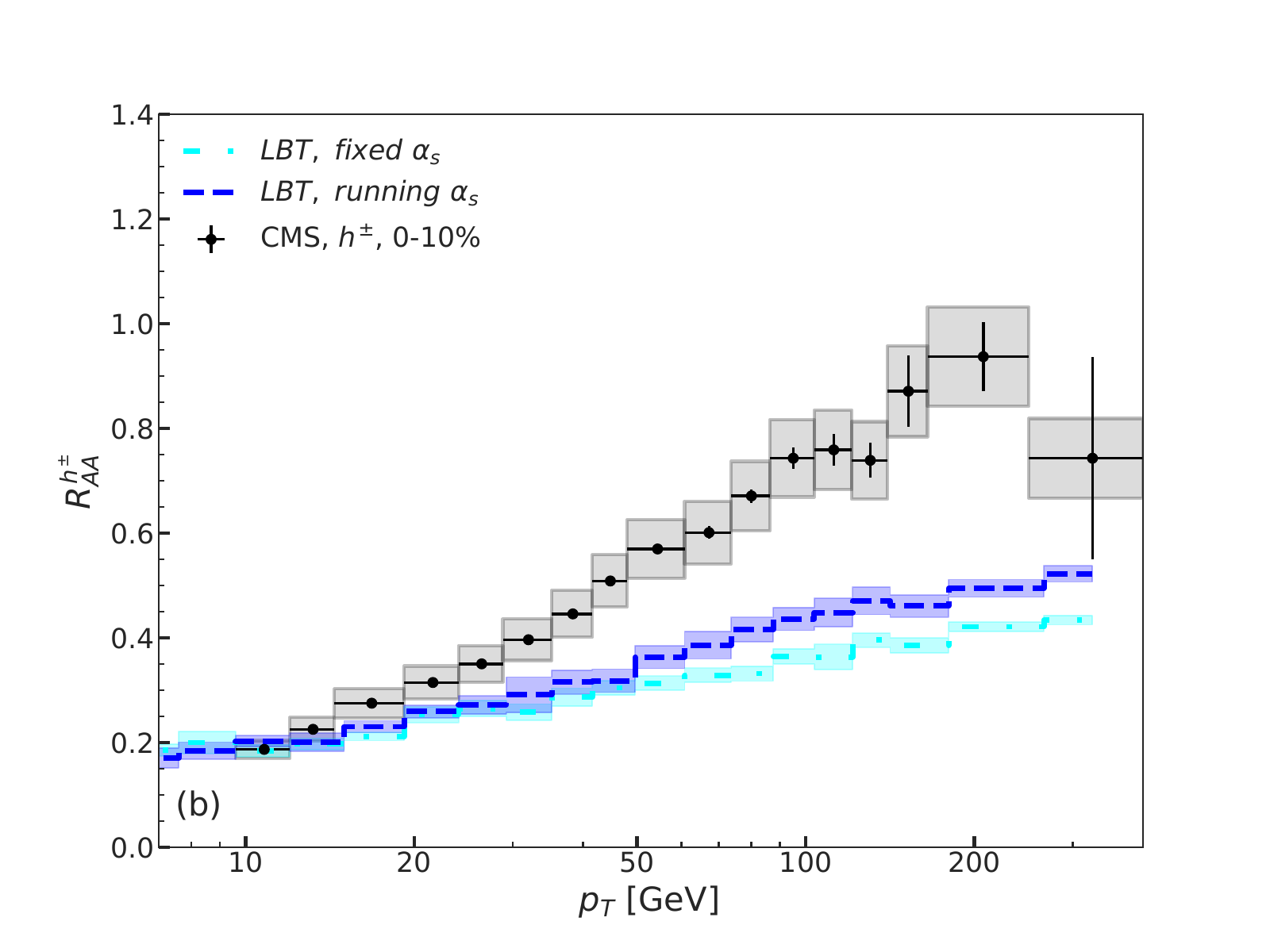}
\end{tabular}
\end{center}
\caption[Nuclear modification factor for D-mesons and charged hadrons]{Nuclear modification factor for D-mesons (a) and charged hadrons (b) in $\sqrt{s_{NN}}=5.02$ TeV PbPb collisions at the LHC at 0-10\% centrality. The pp baseline is calculated using PYTHIA. Data taken from Ref.~\cite{sirunyan2018nuclear,khachatryan2017charged,alice2018measurement}.}
\label{fig:LBT_comp}
\end{figure}

The results of these calculations are found in Fig. \ref{fig:LBT_comp}. Since these calculations rely on perturbation theory, we estimate them to be valid above a momentum of a few GeV. 
%

The calculations with constant $\alpha^{\rm (eff)}_s=0.3$ (dashed lines) generates too much energy loss at high $p_T$, producing an $R_{AA}$ slope that is inconsistent with data, for both charged hadrons and D-mesons. Including the effects of a running coupling $\alpha_s$ (dotted lines) reduces the amount of parton interactions at high $p_T$, which improves the overall $R_{AA}$ slope to better mimic what is seen in experimental data. 

Except for $D^0$-meson $R_{AA}$ at high $p_T$, assuming that no energy loss occurs during the high virtuality showering of partons in a jet is an approximation that doesn't provide a good description of the data. Thus, the goal of the next section is to investigate how energy loss affects the high-virtuality portion of the shower simulated via the higher twist formalism in MATTER.       

\subsection{$R_{AA}$ from MATTER}
\label{sec:results_matter}
\begin{figure}[!h]
\begin{center}
\begin{tabular}{cc}
\includegraphics[width=0.495\textwidth]{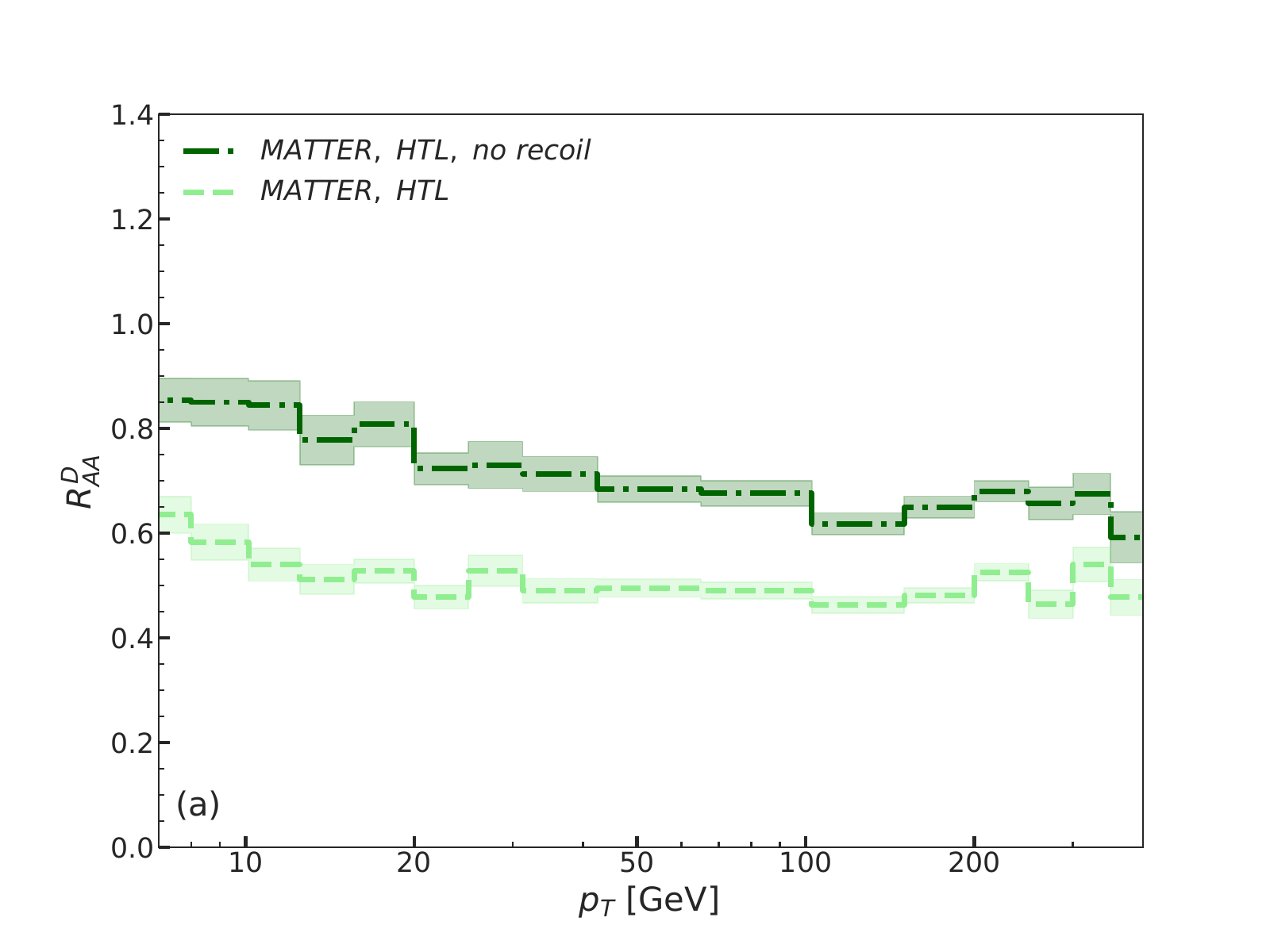} & \includegraphics[width=0.495\textwidth]{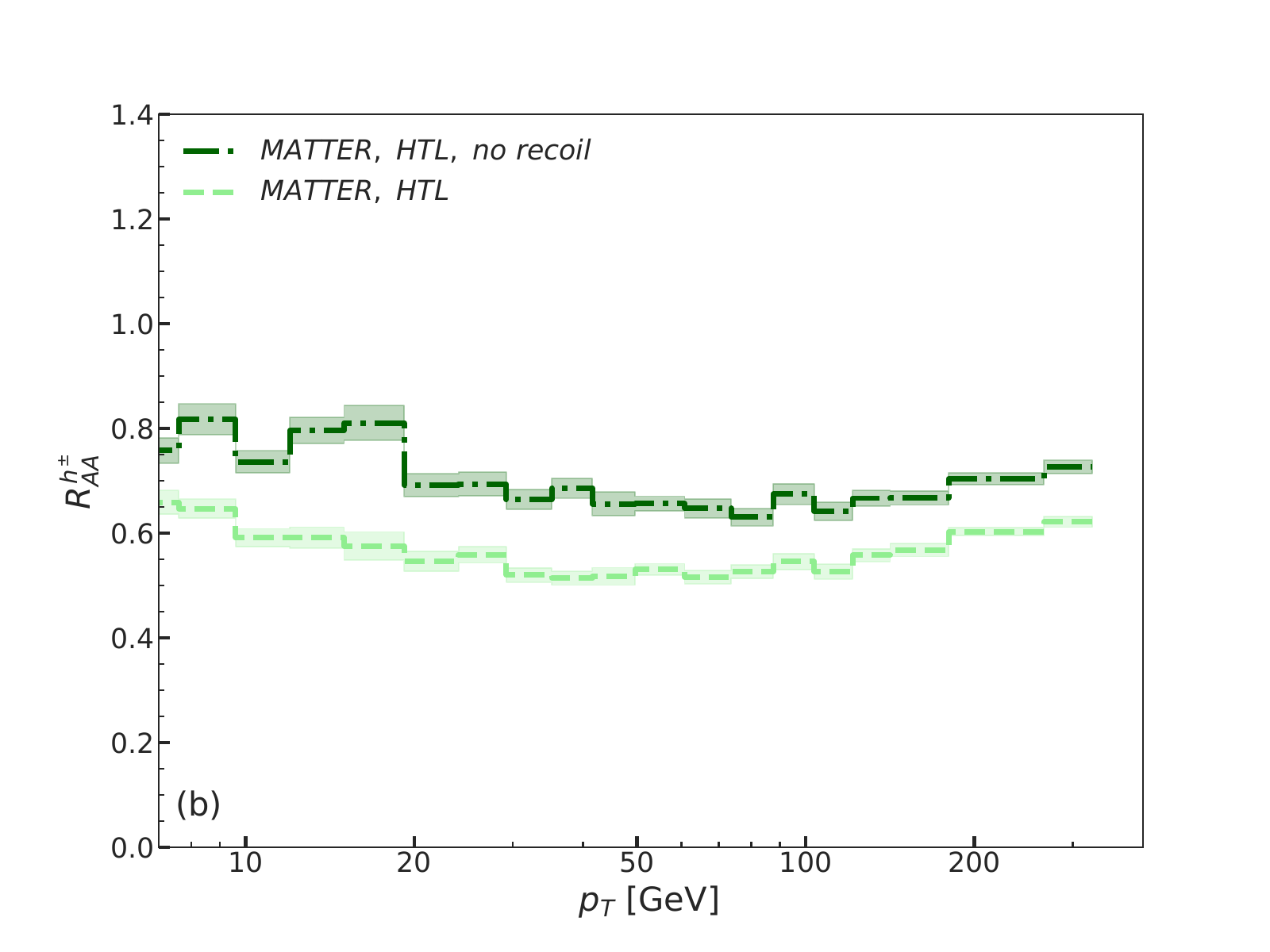}
\end{tabular}
\end{center}
\caption[Nuclear modification factor for D-mesons and charged hadrons]{Nuclear modification factor for D-mesons (a) and charged hadrons (b) in $\sqrt{s_{NN}}=5.02$ TeV PbPb collisions at the LHC at 0-10\% centrality. HTL denotes a calculation using $\hat{q}^{HTL}$ and no recoil refers to scattering processes being deactivated in MATTER.}
\label{fig:MATTER_comp_recoil}
\end{figure}

As MATTER is being used throughout the entire virtuality evolution herein, the higher twist formalism upon which it is based is employed until $t_s=1$ GeV$^2$. MATTER simulates the energy-momentum exchange between the partons of the medium and jet partons via two types of interactions. The first type of interaction is medium-induced inelastic radiation encapsulated in $\hat{q}$, a non-stochastic transport coefficient accounting for deviations  from vacuum splittings. Elastic  $2\to 2$ scatterings between jet and medium partons are treated stochastically. For each parton in the shower, the $2\to 2$ scattering rate is sampled. If a scattering occurs, the thermal parton involved can become part of the jet, leaving a negative contribution in the fluid, or become a source of energy-momentum to be deposited in the QGP. In Fig. \ref{fig:MATTER_comp_recoil}, the elastic and inelastic processes are studied in turn assuming a running $\alpha_s(\mu^2)$. 

Focusing on the result without $2\to2$ scatterings, labeled as no recoil in Fig. \ref{fig:MATTER_comp_recoil}, we can see that including elastic scatterings leads to additional energy loss compared to that incurred via radiative processes alone. One would also imagine these recoil partons would contribute to the final jet observables for both light and heavy flavor, which we shall study in the future. Unlike the LBT simulation where partons are long-lived and thus recoils are ever present, for a virtuality ordered shower like MATTER the importance of these elastic scatterings needs to be highlighted due to the highly variable lifetime of partons in the shower. Furthermore, our comparison between light and heavy flavor allows us to appreciate how much these recoils affect partons of different masses.

\begin{figure}[!h]
\begin{center}
\begin{tabular}{cc}
\includegraphics[width=0.495\textwidth]{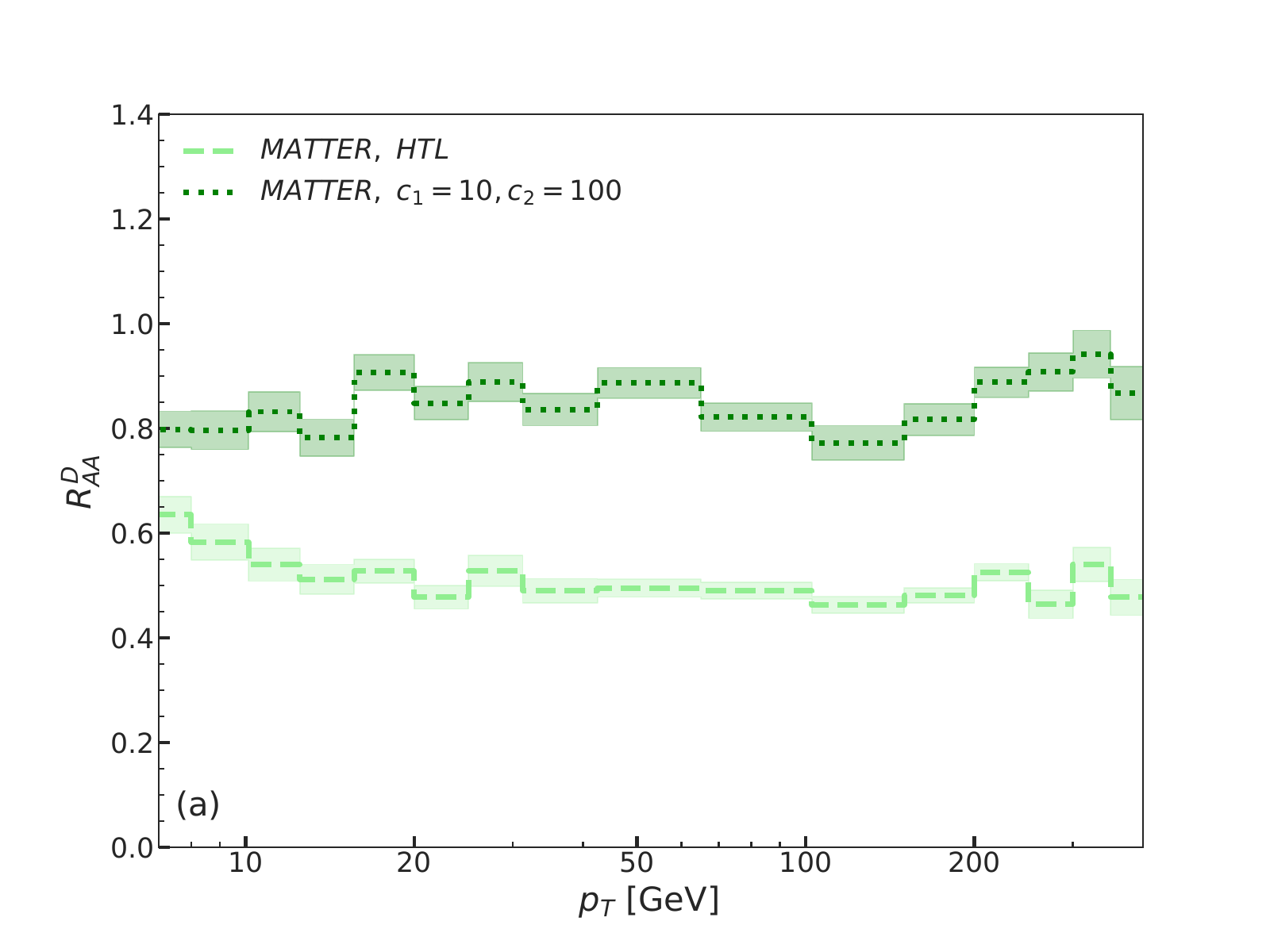} & \includegraphics[width=0.495\textwidth]{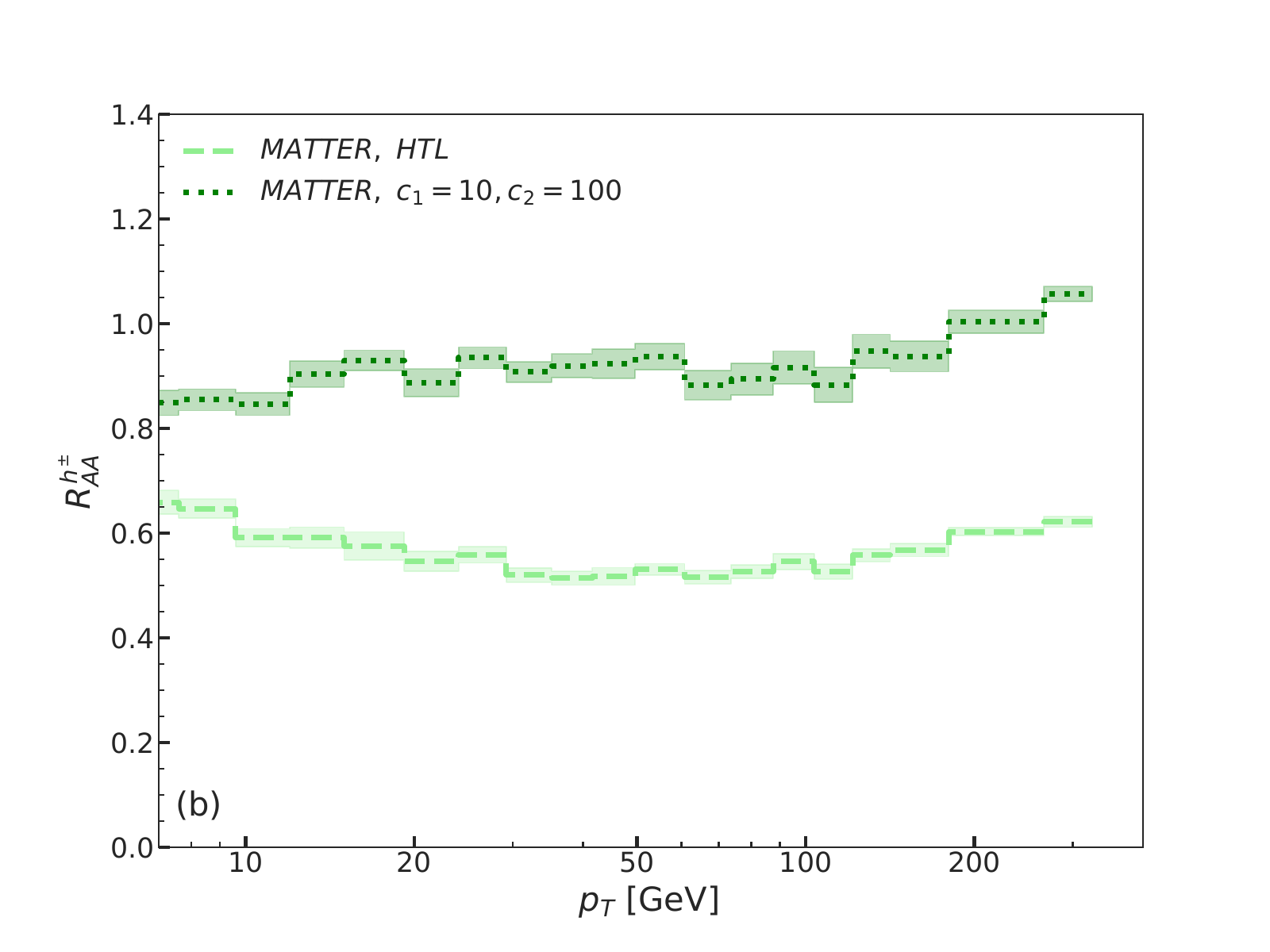}
\end{tabular}
\end{center}
\caption[Nuclear modification factor for D-mesons and charged hadrons]{Nuclear modification factor for D-mesons (a) and charged hadrons (b) in $\sqrt{s_{NN}}=5.02$ TeV PbPb collisions at the LHC at 0-10\% centrality. The difference between $\hat{q}^{HTL}$ and $\hat{q}(t)$ is significant especially at high $p_T$.}
\label{fig:MATTER_comp_qhat}
\end{figure}

As the virtuality dependent $\hat{q}(t)$ is smaller compared to the HTL result, the $R_{AA}$ tends to be much closer to $1$ for $\hat{q}(t)$ (dotted lines) compared to the one for $\hat{q}^{HTL}$ (dashed lines) as depicted in Fig. \ref{fig:MATTER_comp_qhat}. This effect is seen in both light and heavy flavor results at high $p_T$, as expected. If we were to turn off the scattering process for the virtuality dependent $\hat{q}$ case, $R_{AA}$ is even larger and almost consistent with $1$ across all $p_T$. The MATTER alone result is not supposed to be compared with data but to give us a sense what the MATTER+LBT $R_{AA}$ should look like when physics like scattering is not considered in MATTER. 

\subsection{$R_{AA}$ from the combined MATTER and LBT simulation}
\label{sec:results_matter_lbt}
The combination of MATTER and LBT simulations is done by separating, in virtuality, the parton evolution in MATTER from that in LBT. The virtuality at which the switch is performed is a parameter, which for light flavor was tuned to $t_s=4$ GeV$^2$. 

\begin{figure}[!h]
\begin{center}
\begin{tabular}{cc}
\includegraphics[width=0.495\textwidth]{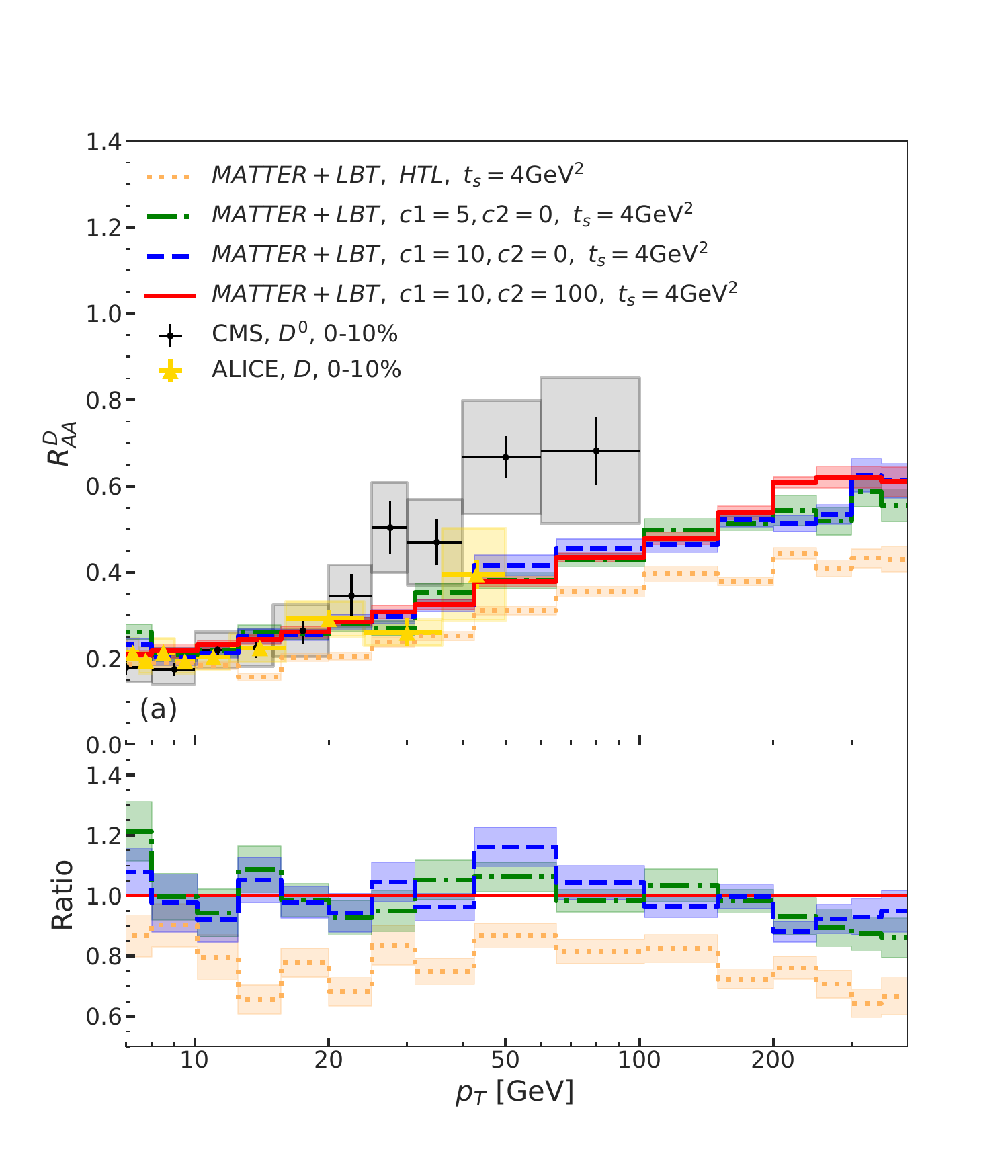} & \includegraphics[width=0.495\textwidth]{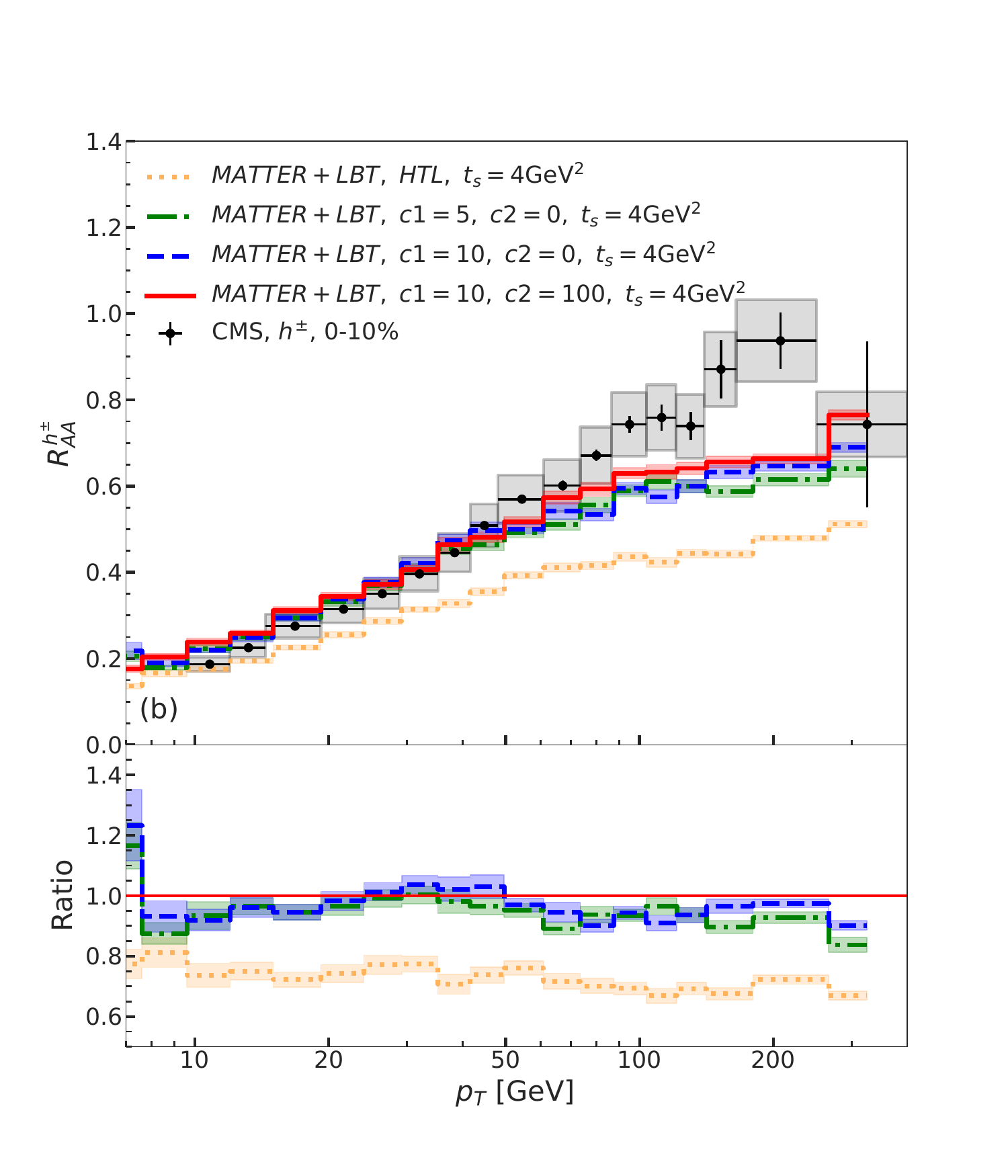}
\end{tabular}
\end{center}
\caption[Nuclear modification factor for D-mesons and charged hadrons]{Nuclear modification factor for D-mesons (a) and charged hadrons (b) in $\sqrt{s_{NN}}=5.02$ TeV PbPb collisions at the LHC at 0-10\% centrality. Here we are varying the parameterization of $\hat{q}(t)$ which is monotonically decreasing when $c_1$ and $c_2$ increase. The ratio in the bottom plots are taken with respect to the $c_1=10, c_2=100$ case with $\hat{q}(t)$ parameterization.}
\label{fig:MATTER_LBT_comp_qhat}
\end{figure}
\begin{figure}[!h]
\begin{center}
\begin{tabular}{cc}
\includegraphics[width=0.495\textwidth]{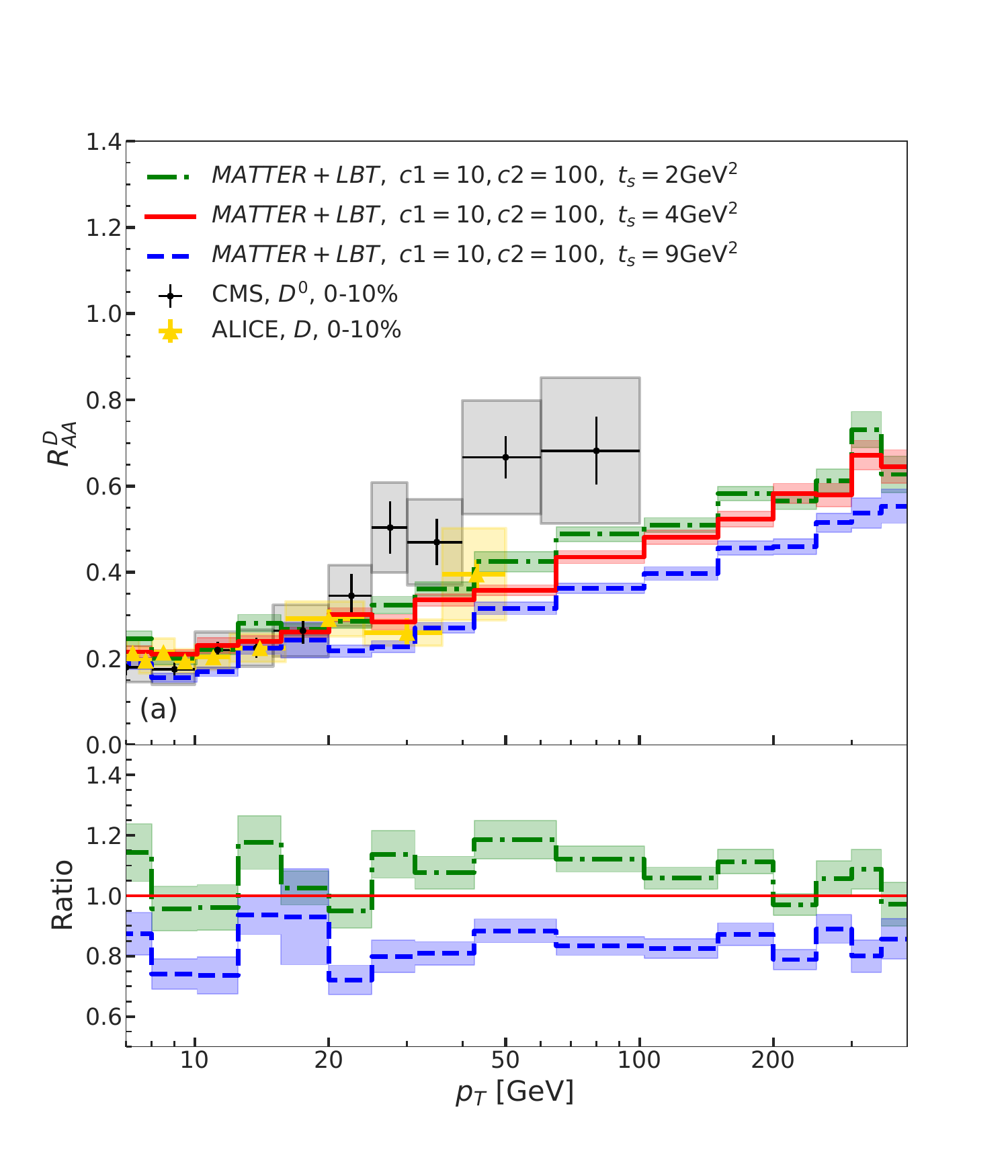} & \includegraphics[width=0.495\textwidth]{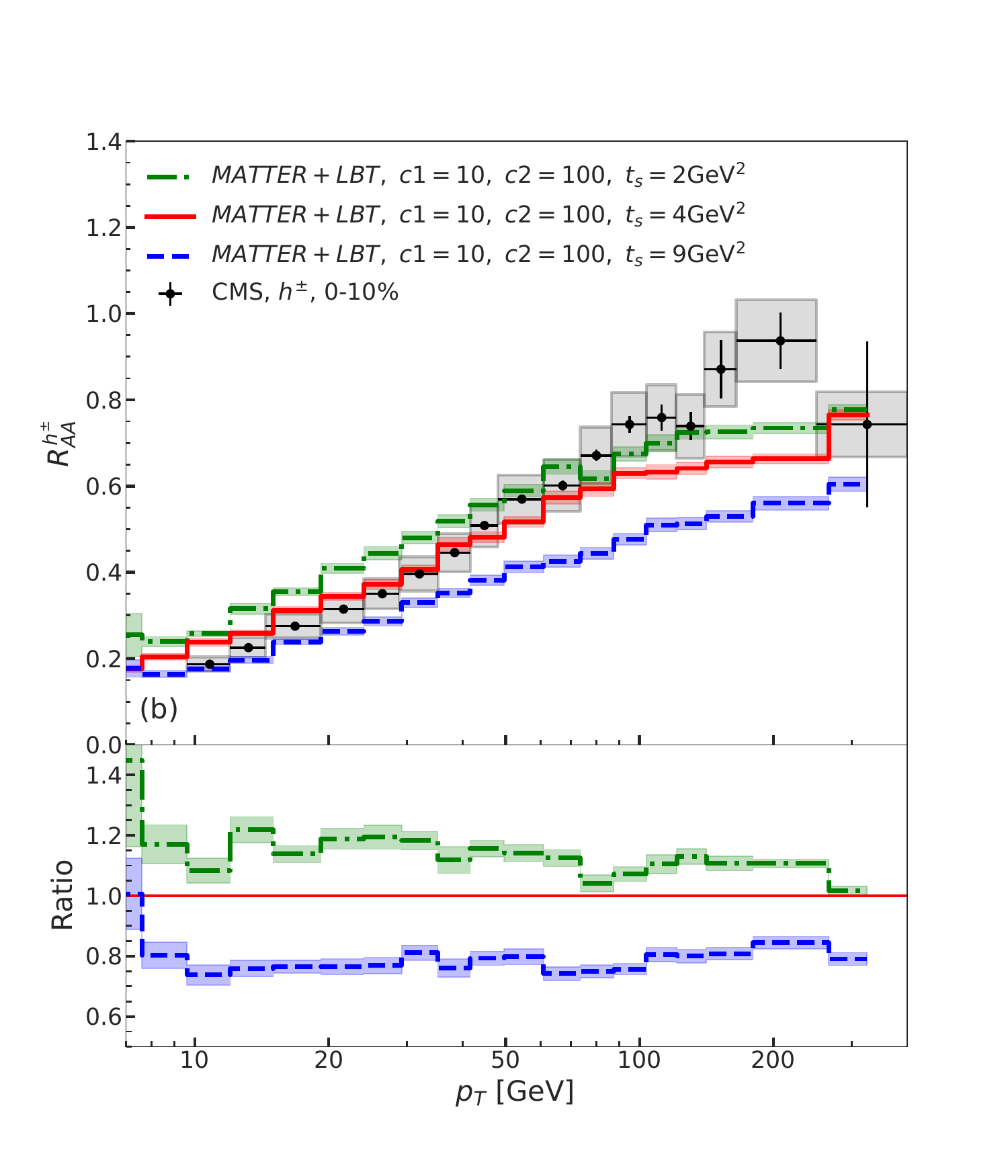}
\end{tabular}
\end{center}
\caption[Nuclear modification factor for D-mesons and charged hadrons]{Nuclear modification factor for D-mesons (a) and charged hadrons (b) in $\sqrt{s_{NN}}=5.02$ TeV PbPb collisions at the LHC at 0-10\% centrality. A bigger value of $t_s$ increases the effective length of LBT based energy-loss. The ratio in the bottom plots are taken with respect to the $c_1=10, c_2=100$ case with $\hat{q}(t)$ parameterization.}
\label{fig:MATTER_LBT_comp_Q0}
\end{figure}

A multi-stage $R_{AA}$ calculation using a virtuality-independent $\hat{q}^{HTL}$ alone shows an over suppression of $R_{AA}$ compared to data for both light and heavy flavors.  Additionally, the slope seen in the experimental data in the region $p_T\gtrsim 10$ GeV is steeper than what is obtained in our multi-stage $R_{AA}$ calculation using $\hat{q}^{HTL}$. A simple re-scaling of the overall normalization of $\hat{q}^{HTL}$ would not be enough to explain the slope seen in the data. In fact, a virtuality-dependent $\hat{q}$ whose value is suppressed as virtuality increases, such as found in this study and in Ref.~\cite{kumar2022inclusive}, helps in this regard. Employing a virtuality-dependent $\hat{q}$ indeed shows a significant effect on parton evolution not only in MATTER, but more importantly in the multi-stage MATTER+LBT evolution, affecting simultaneously light flavor and D-meson $R_{AA}$. It is the combination of a multi-stage simulation together with a virtuality-dependent $\hat{q}$ that is responsible for the agreement between the theoretical calculation and the data, in line with findings from the previous two sections. 

\subsubsection{Effects of $\hat{q}$ and $Q_s$}

Taking a closer look at Fig. \ref{fig:MATTER_LBT_comp_qhat}, we can see  how different parameterizations of $\hat{q}(t)$ affect the $R_{AA}$ especially at high $p_T$. This is of course due to the specific form of the parameterization of $\hat{q}(t)$. One can imagine that an even more aggressive reduction of $\hat{q}(t)$ at large $t$ shall further increase $R_{AA}$ at high $p_T$. We leave this to a future study, utilizing a Bayesian calibration to find the optimal values for the parameterization used here. 

Fig. \ref{fig:MATTER_LBT_comp_Q0} studies the effect of varying the switching scale $t_s$. A large $t_s$ implies that partons evolve longer in the LBT regime. This is why the $t_s=9$~GeV$^2$ curve looks very similar to the LBT alone curve, since the LBT mechanism generates significantly larger energy loss effects than the MATTER mechanism, especially at low $p_T$. Combining results from both Fig. \ref{fig:MATTER_LBT_comp_qhat} and \ref{fig:MATTER_LBT_comp_Q0}, we see that a parameter choice of $c_1=10, c_2=100, t_s=4$~GeV$^2$ provides the best simultaneous description of the charged hadron and $D^0$ meson $R_{AA}$ data.

\begin{figure}[!h]
\begin{center}
\begin{tabular}{cc}
\includegraphics[width=0.495\textwidth]{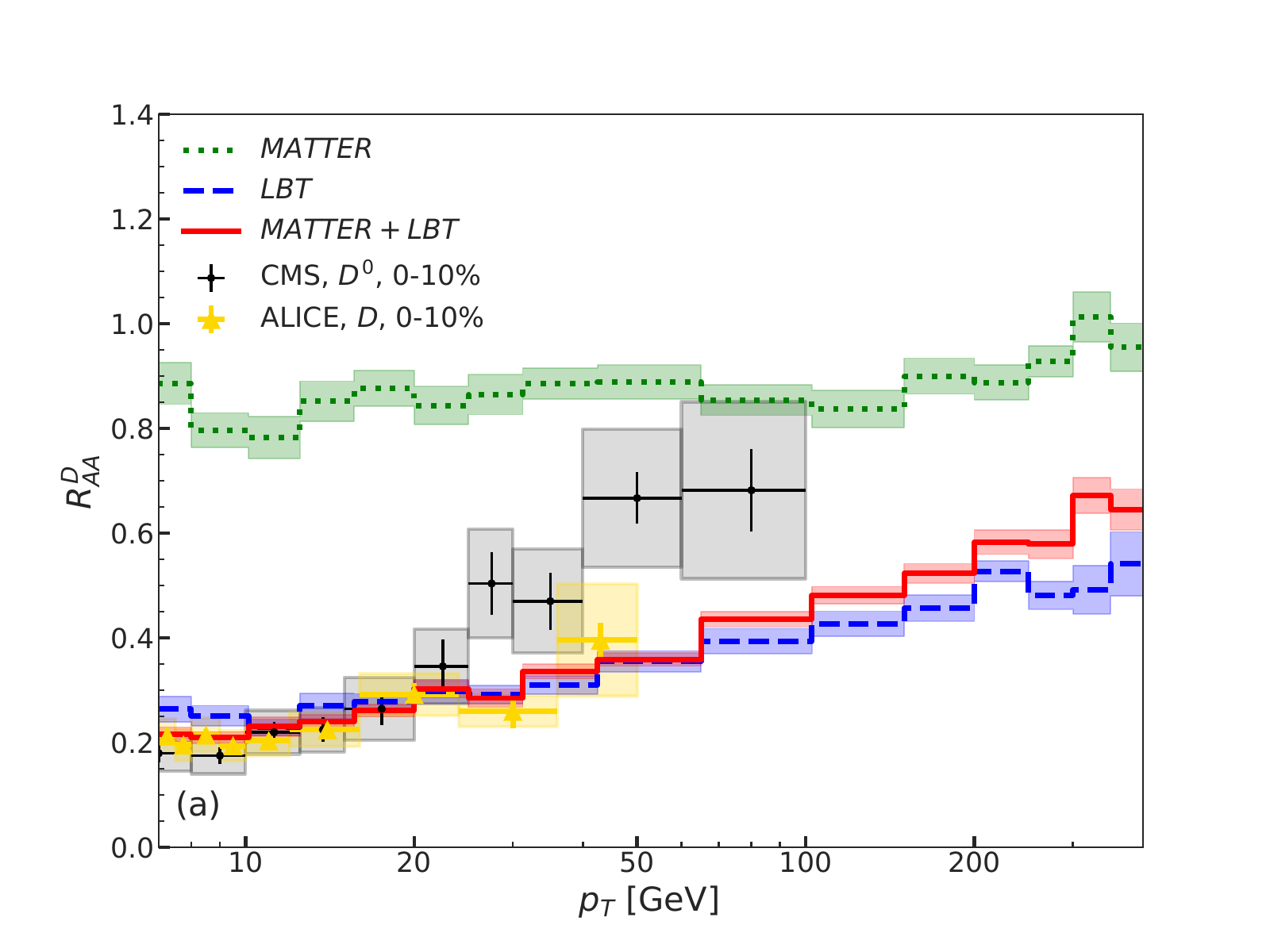} & \includegraphics[width=0.495\textwidth]{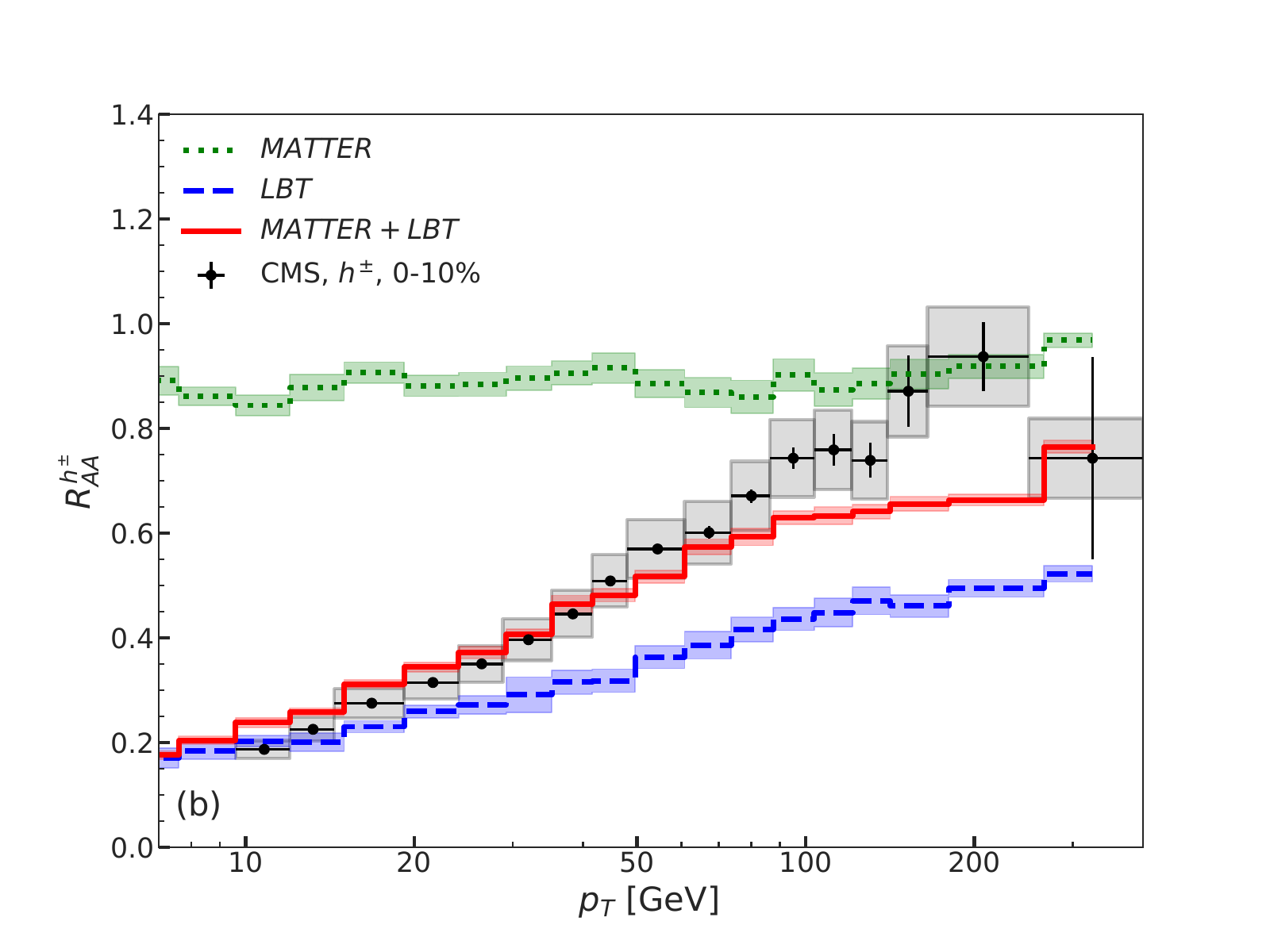}
\end{tabular}
\end{center}
\caption[Nuclear modification factor for D-mesons and charged hadrons]{Nuclear modification factor for D-mesons (a) and charged hadrons (b) at the $\sqrt{s_{NN}}=5.02$ TeV PbPb collisions at the LHC in the 0-10\% centrality. We set $c_1=10,\ c_2=100$ within the $\hat{q}(t)$ parameterization [see in Eq.~(\ref{eq:qhat_t})] for the MATTER alone and the MATTER+LBT curve. The other parameters for the MATTER+LBT curve is $t_s=4$~GeV$^2$ found to best describe the $R_{AA}$ data. The pp baseline for the LBT curve is calculated using PYTHIA whereas the pp baseline for the MATTER and MATTER+LBT case are calculated using MATTER vacuum \cite{JETSCAPE:2019udz}. Data taken from Ref.~\cite{sirunyan2018nuclear,khachatryan2017charged,alice2018measurement}.}
\label{fig:MAT_vs_LBT_all}
\end{figure}

With our simple parameter search, we compare MATTER+LBT result and LBT only result in Fig.~\ref{fig:MAT_vs_LBT_all}. A clear enhancement in $R_{AA}$ can be seen by employing the in medium DGLAP evolution (implemented by the MATTER model). This is because parton energy loss in the MATTER phase is now greatly suppressed with the virtuality dependent $\hat{q}$ and the time that a parton spent in the LBT phase is effectively reduced.

\subsubsection{Effects of gluon splitting to heavy quark pair}

\begin{figure}[!h]
\begin{center}
\begin{tabular}{cc}
\includegraphics[width=0.495\textwidth]{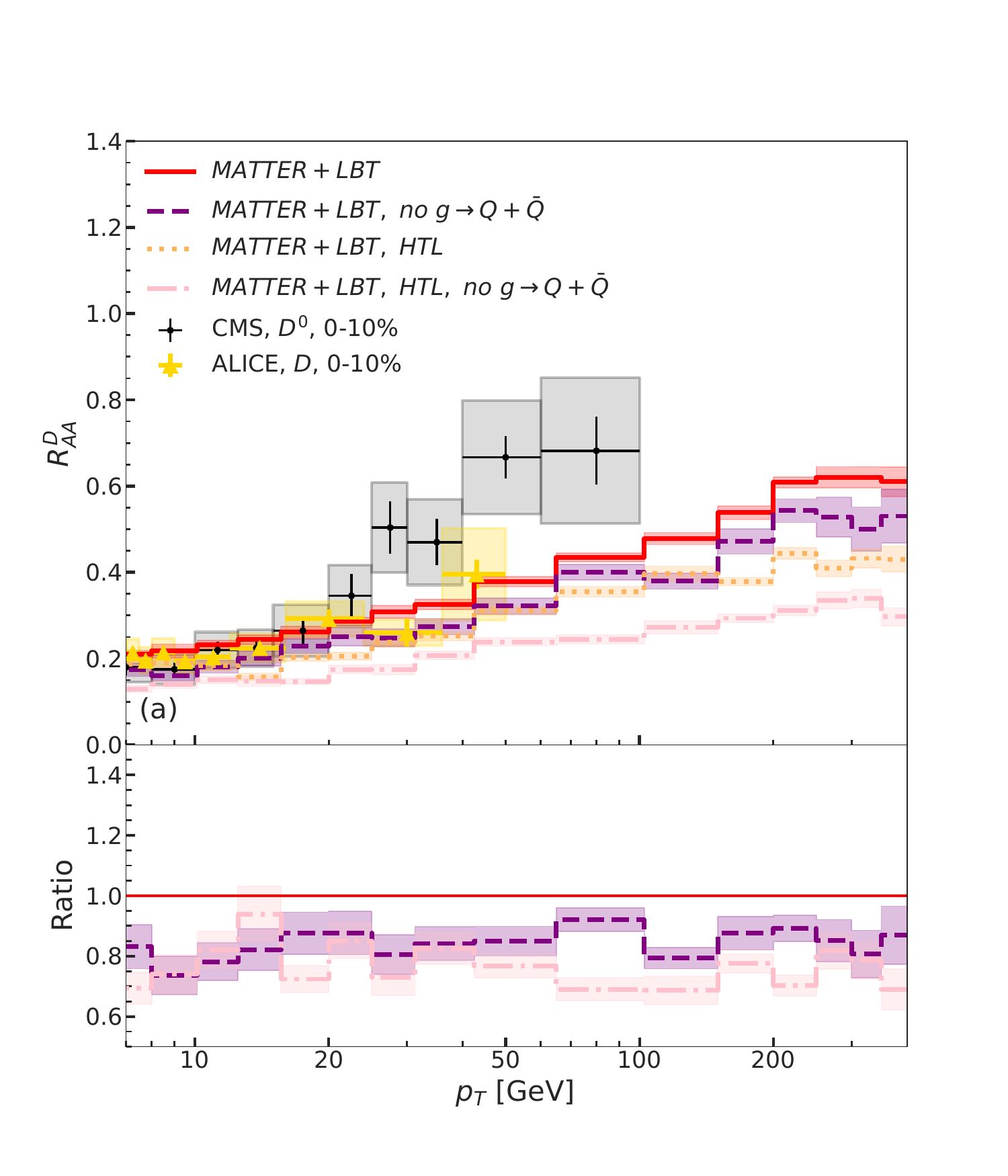} & \includegraphics[width=0.495\textwidth]{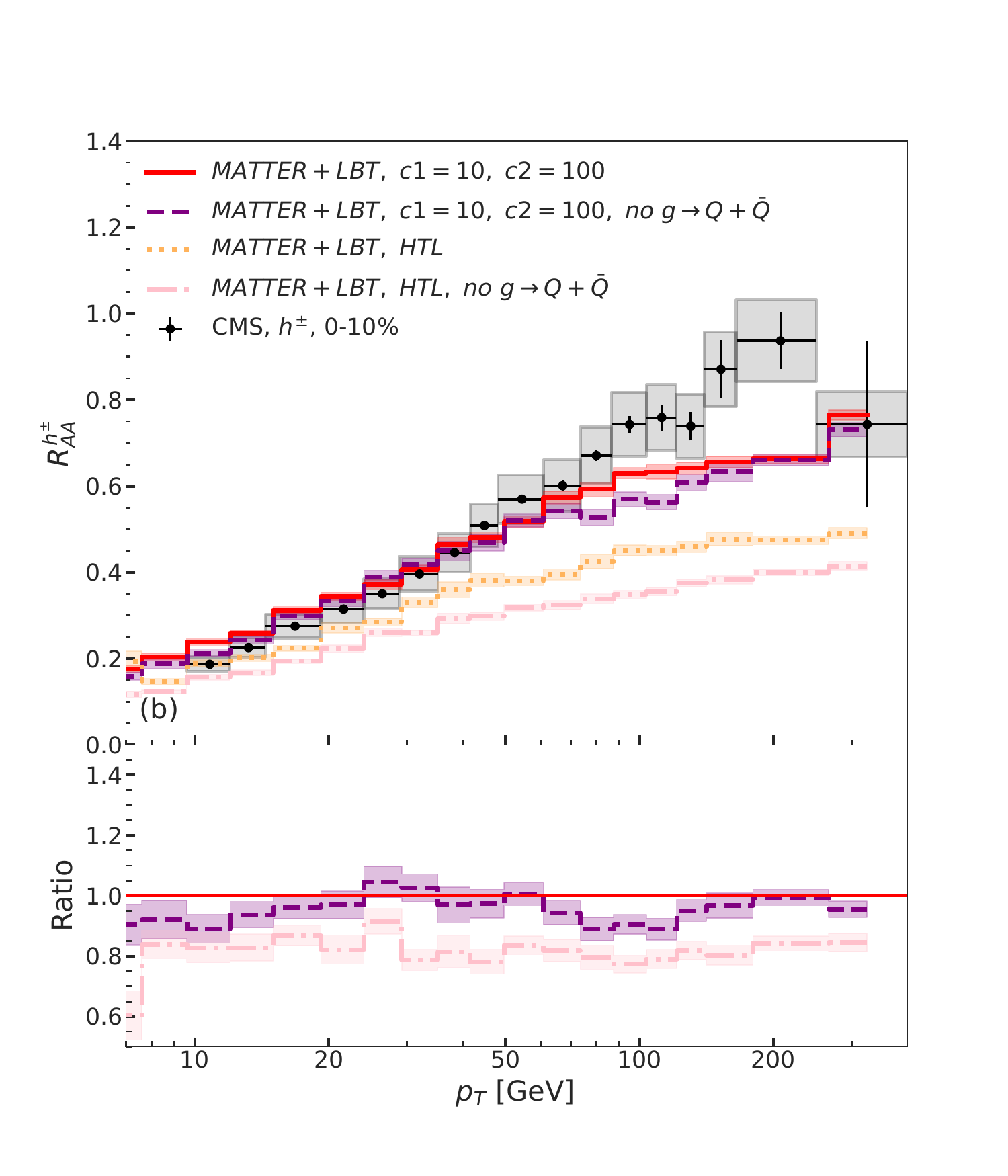}
\end{tabular}
\end{center}
\caption[Nuclear modification factor for D-mesons and charged hadrons]{Nuclear modification factor for D-mesons (a) and charged hadrons (b) in $\sqrt{s_{NN}}=5.02$ TeV PbPb collisions at the LHC at 0-10\% centrality. Here we choose $c_1=10, \ c_2=100$ for the $\hat{q}$ parameterization. Ignoring the $g \rightarrow Q+\bar{Q}$ process in MATTER impact the D meson $R_{AA}$ while has little effect on the charged hadron $R_{AA}$.}
\label{fig:MATTER_LBT_comp_gQQ}
\end{figure}

The novel physics ingredient that this study allows to explore is the creation of heavy flavor through $g \rightarrow Q+\bar{Q}$ in MATTER. To do so, both the $D$ meson $R_{AA}$ and the charged hadron $R_{AA}$ are explored. A combination using MATTER and LBT simulations is employed throughout as the goal is to investigate the effect of including the $g\to Q+\bar{Q}$ process in MATTER on the observed $R_{AA}$. As depicted in Fig.~\ref{fig:MATTER_LBT_comp_gQQ}, ignoring this process has a roughly $20\%$ impact on $D$ meson $R_{AA}$, while very little impact is seen for the charged hadron $R_{AA}$. A previous study using PYTHIA \cite{norrbin2000production} also reports non-negligible contribution from gluon splitting to the total charm cross section. 

\subsubsection{Effects of other parameters in the framework}

\begin{figure}[htbp]
\centering
\includegraphics[width=0.45\textwidth]{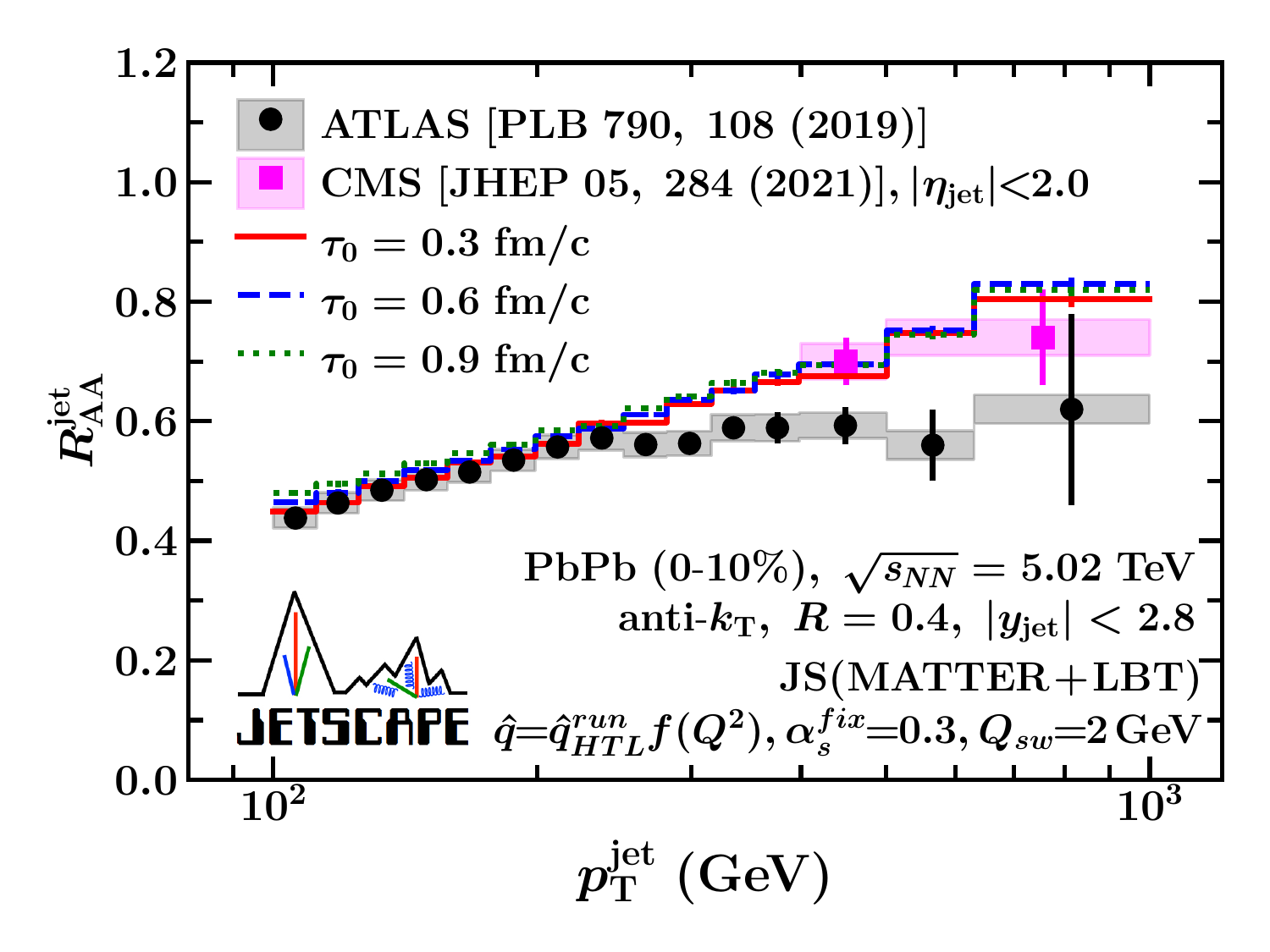}
\includegraphics[width=0.45\textwidth]{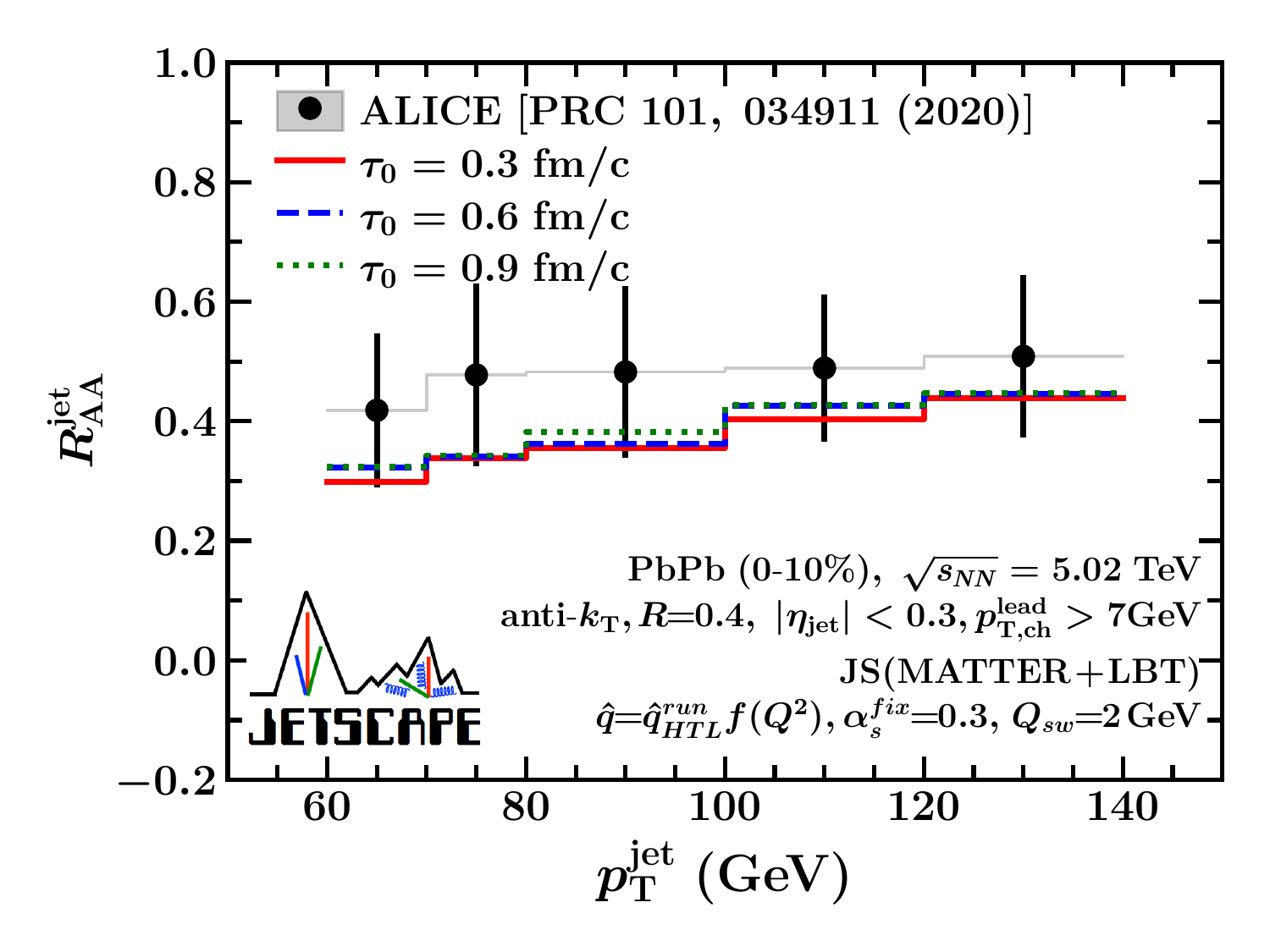}
\includegraphics[width=0.45\textwidth]{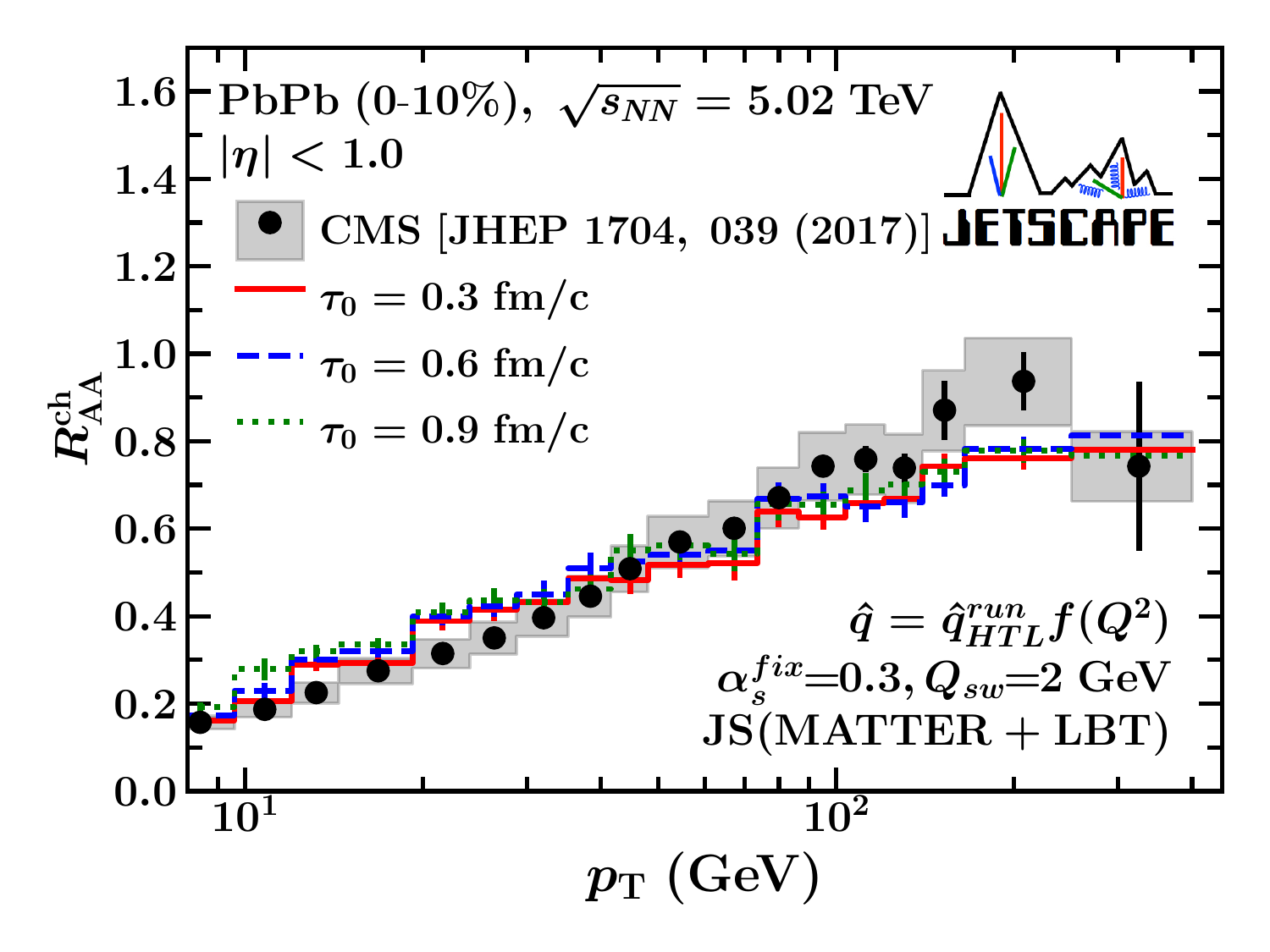}
\caption[Results for the virtuality dependent $\hat{q}(t)$ 
with starting longitudinal proper times for in-medium jet energy loss $\tau_0=0.3,\,0.6,\,\mbox{and }0.9$ fm]{The solid red, dashed blue, and dotted green lines show results for the virtuality dependent $\hat{q}(t)$ 
with starting longitudinal proper times for in-medium jet energy loss $\tau_0=0.3,\,0.6,\,\mbox{and }0.9$ fm, respectively \cite{kumar2022inclusive}. 
}
\label{fig:Type3-q-hat-Effect_of_tau0}
\end{figure}

\begin{figure}[htbp]
\centering
\includegraphics[width=0.45\textwidth]{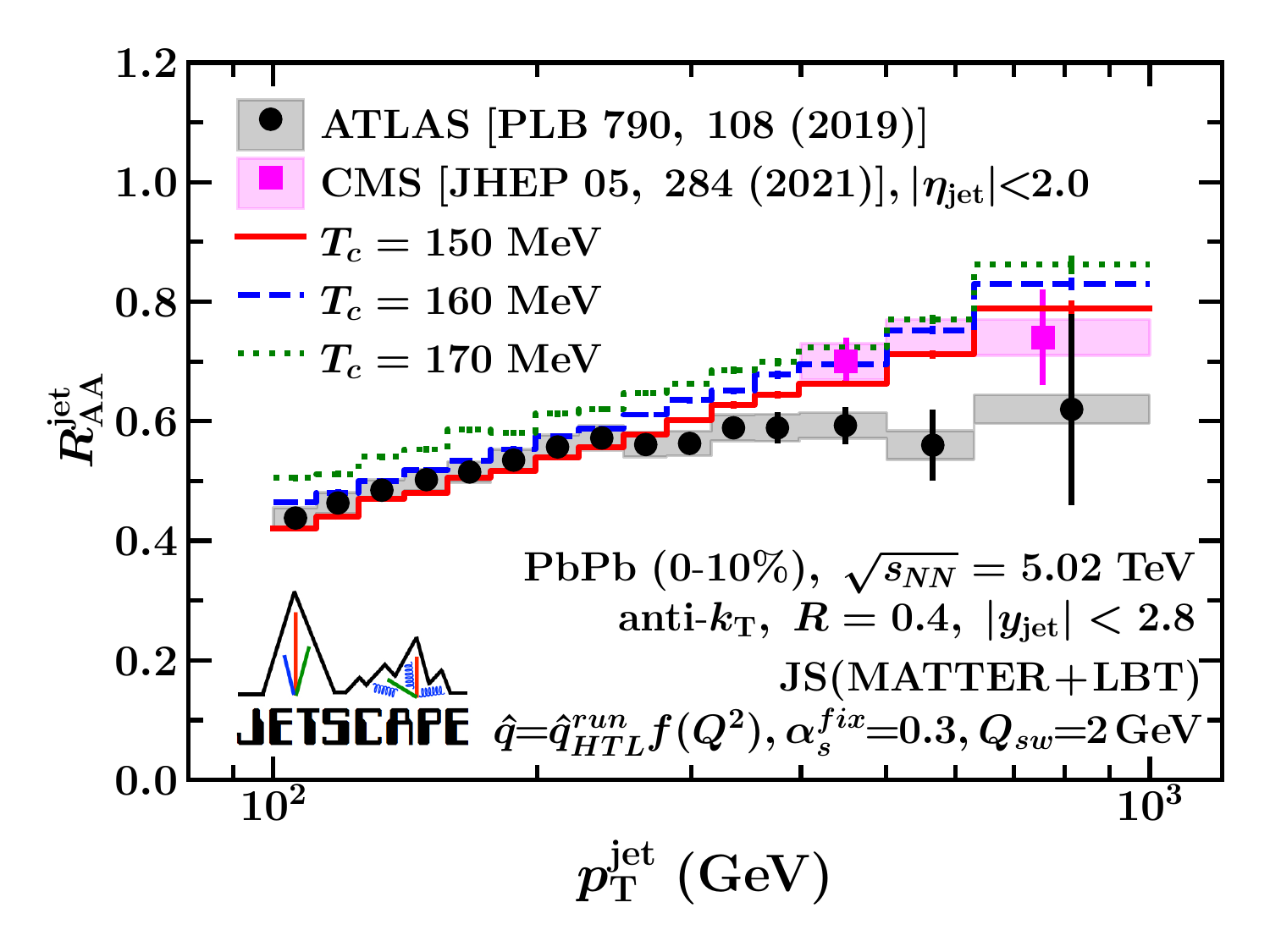}
\includegraphics[width=0.45\textwidth]{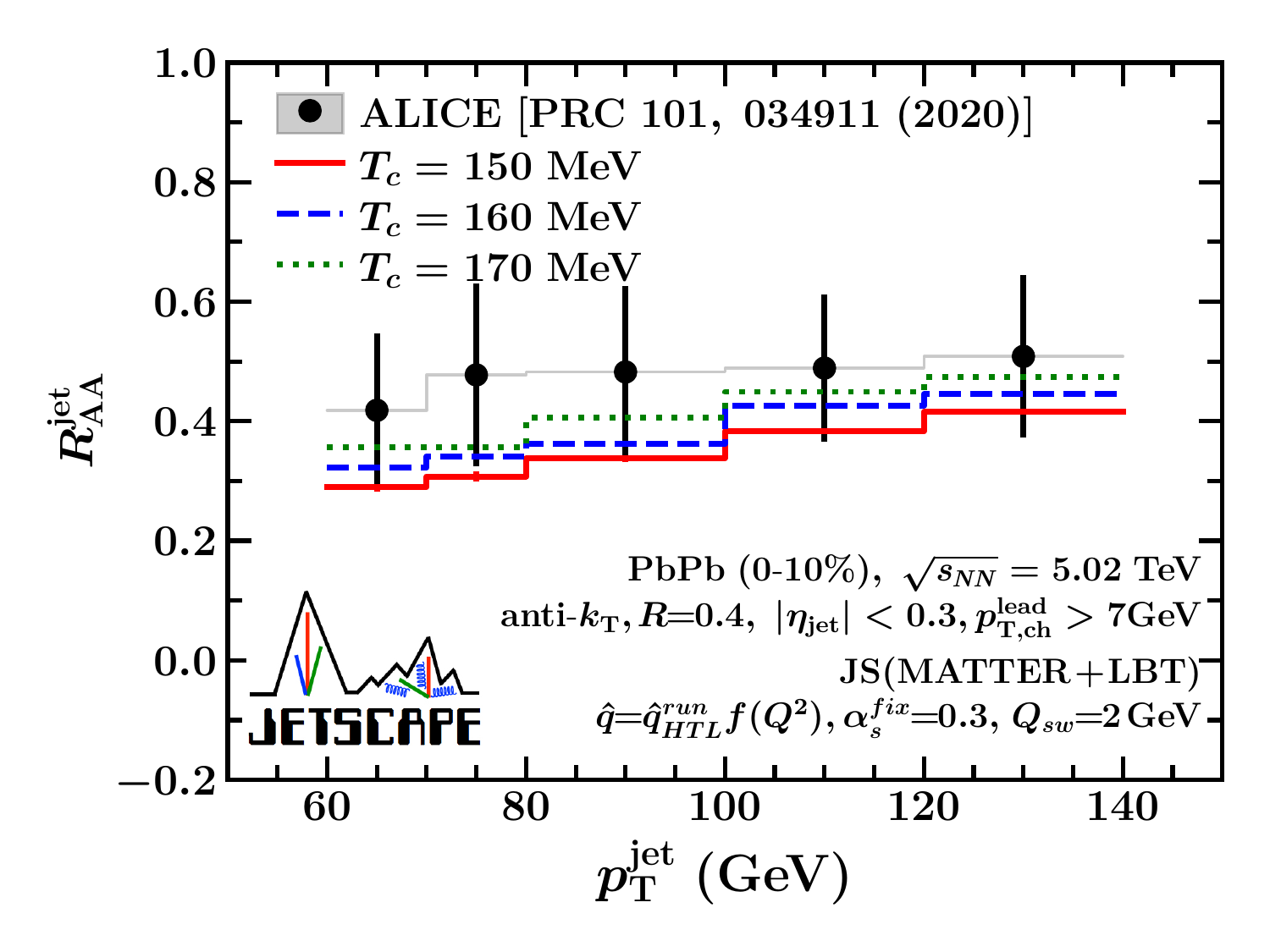}
\includegraphics[width=0.45\textwidth]{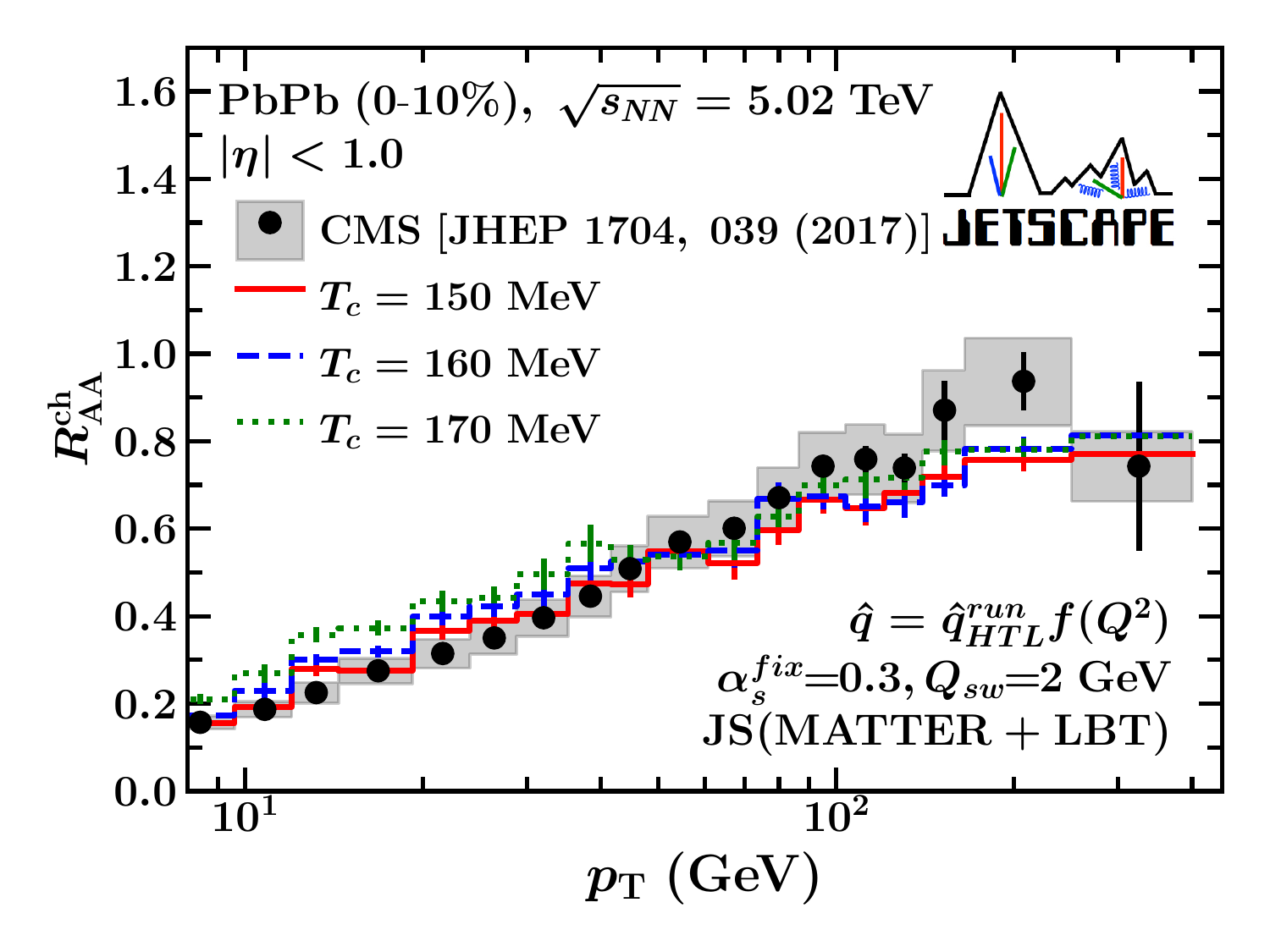}
\caption[Results for the virtuality dependent $\hat{q}(t)$ 
with the energy loss termination temperatures  
$T_{\mathrm{c}}=150,\,160,\,\mbox{and }170$~MeV]{
The solid red, dashed blue, and dotted green lines show results for the virtuality dependent $\hat{q}(t)$ 
with the energy loss termination temperatures  
$T_{\mathrm{c}}=150,\,160,\,\mbox{and }170$~MeV, respectively \cite{kumar2022inclusive}.
}
\label{fig:Type3-q-hat-Effect_of_Tc}
\end{figure}

There are other parameters in JETSCAPE that we can ``tune'' like the starting time of the energy loss $\tau_0$ and the stopping temperature of energy loss $T_c$. As we can see from Fig.~\ref{fig:Type3-q-hat-Effect_of_tau0} and Fig.~\ref{fig:Type3-q-hat-Effect_of_Tc}, varying these parameters have minor effects on the various $R_{AA}$. $T_c$ has a larger impact but basically just shifts the $R_{AA}$ curves up and down. It is possible to vary these parameters in order to achieve better description of the data. However, in our study, we will choose $\tau_0=0.6$ fm/c which is also when the hydro simulation begins, and $T_c=160$ MeV which is a little above the critical temperature our hydro uses ($154$ MeV). 

\subsection{$R_{AA}$ at $10-30\%$ and $30-50\%$ centrality}

\begin{figure}[!h]
\begin{center}
\begin{tabular}{cc}
\includegraphics[width=0.495\textwidth]{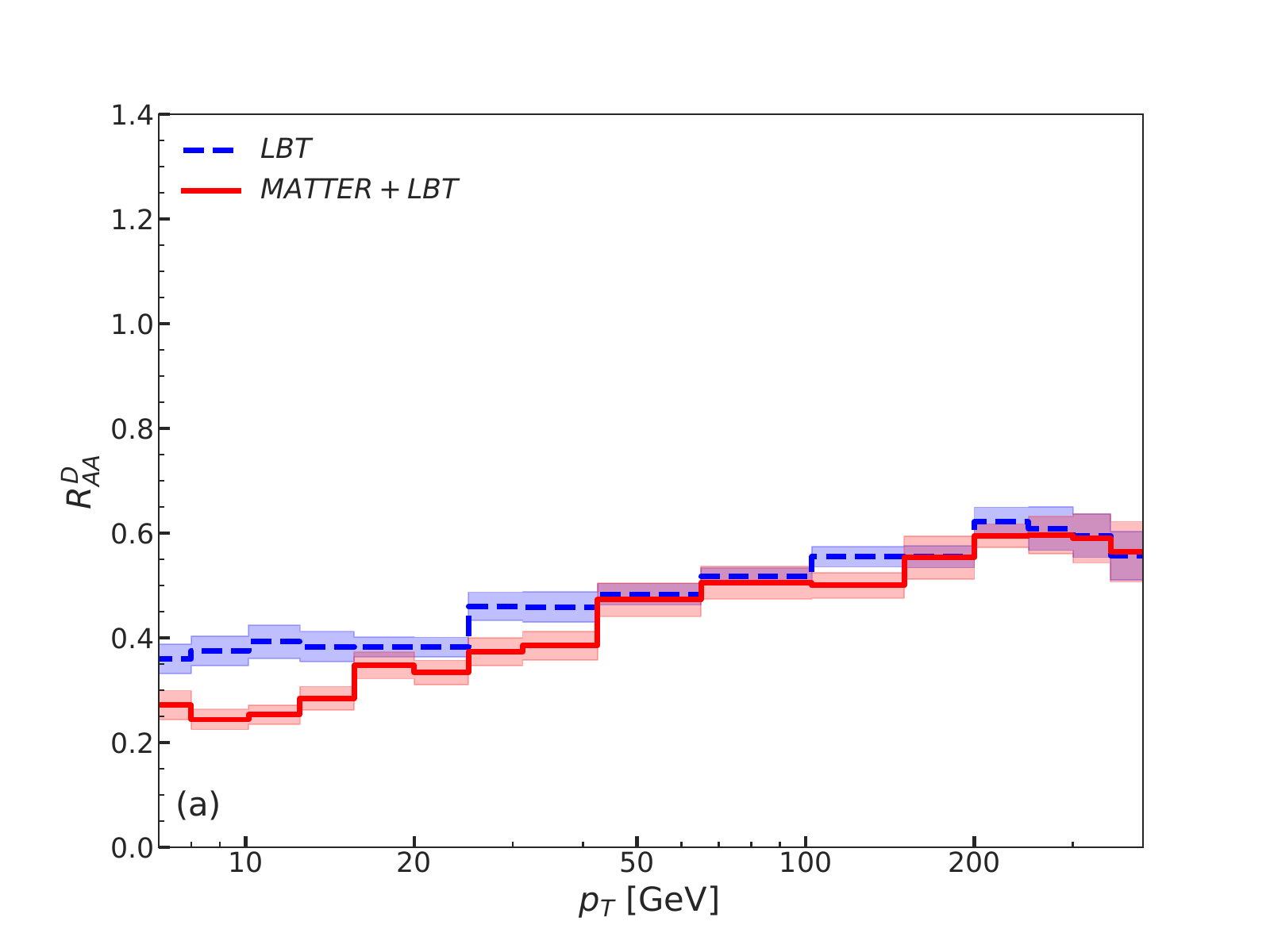} & \includegraphics[width=0.495\textwidth]{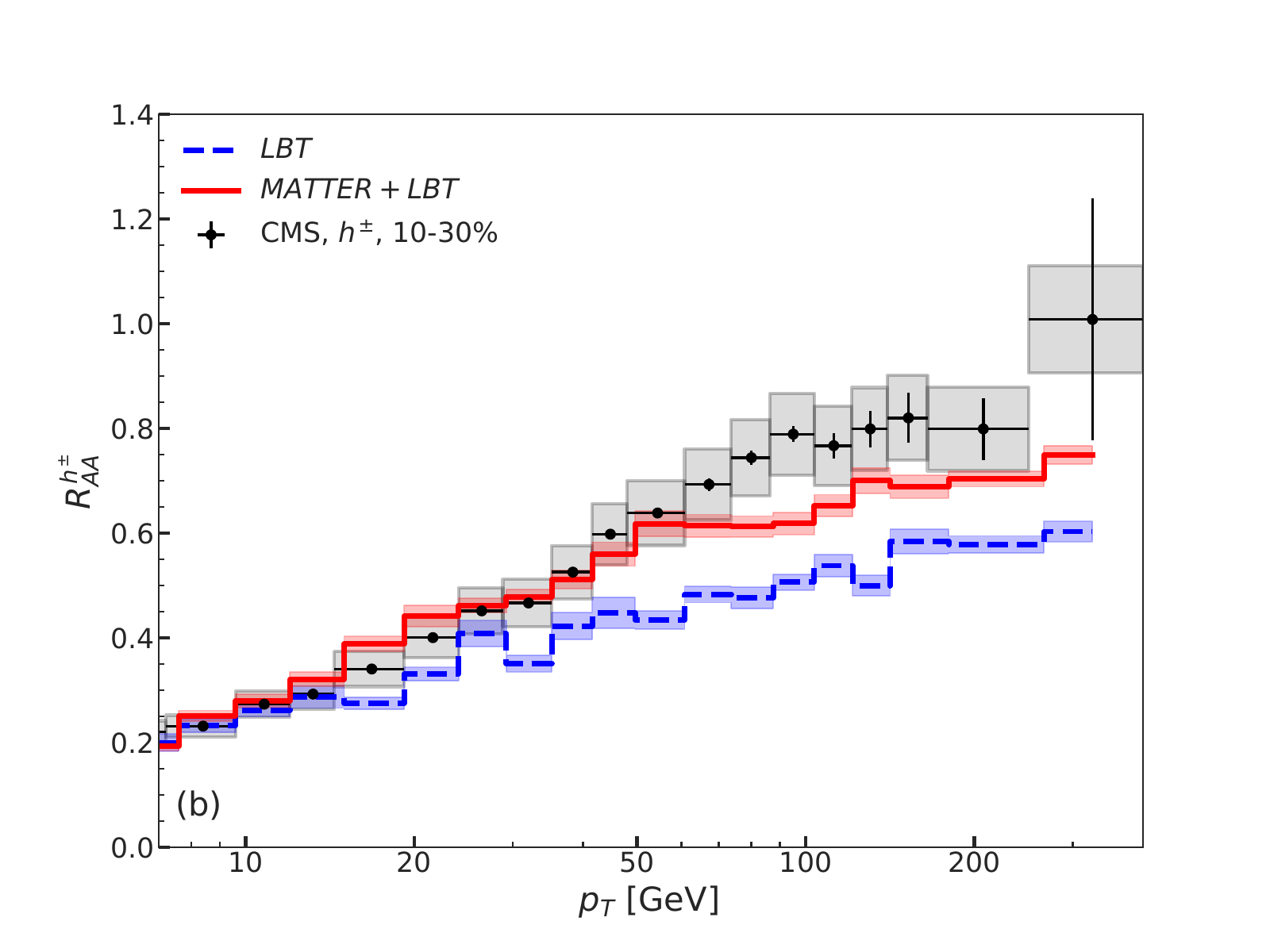}
\end{tabular}
\end{center}
\caption[Nuclear modification factor for D-mesons and charged hadrons]{Nuclear modification factor for D-mesons (a) and charged hadrons (b) in $\sqrt{s_{NN}}=5.02$ TeV PbPb collisions at the LHC at 10-30\% centrality. Here we choose $c_1=10, \ c_2=100$ for the $\hat{q}$ parameterization.}
\label{fig:MATTER_LBT_10-30}
\end{figure}

\begin{figure}[!h]
\begin{center}
\begin{tabular}{cc}
\includegraphics[width=0.495\textwidth]{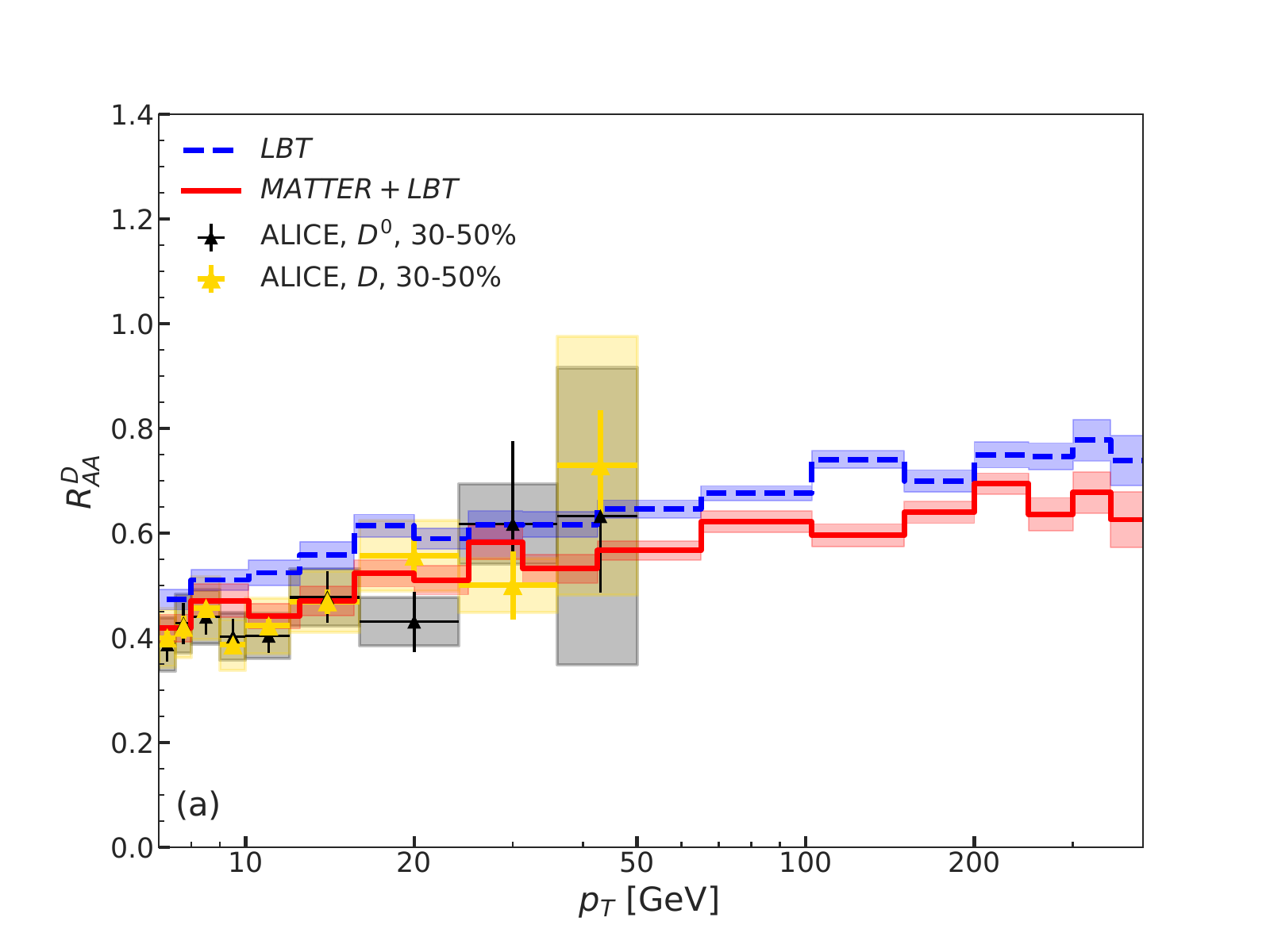} & \includegraphics[width=0.495\textwidth]{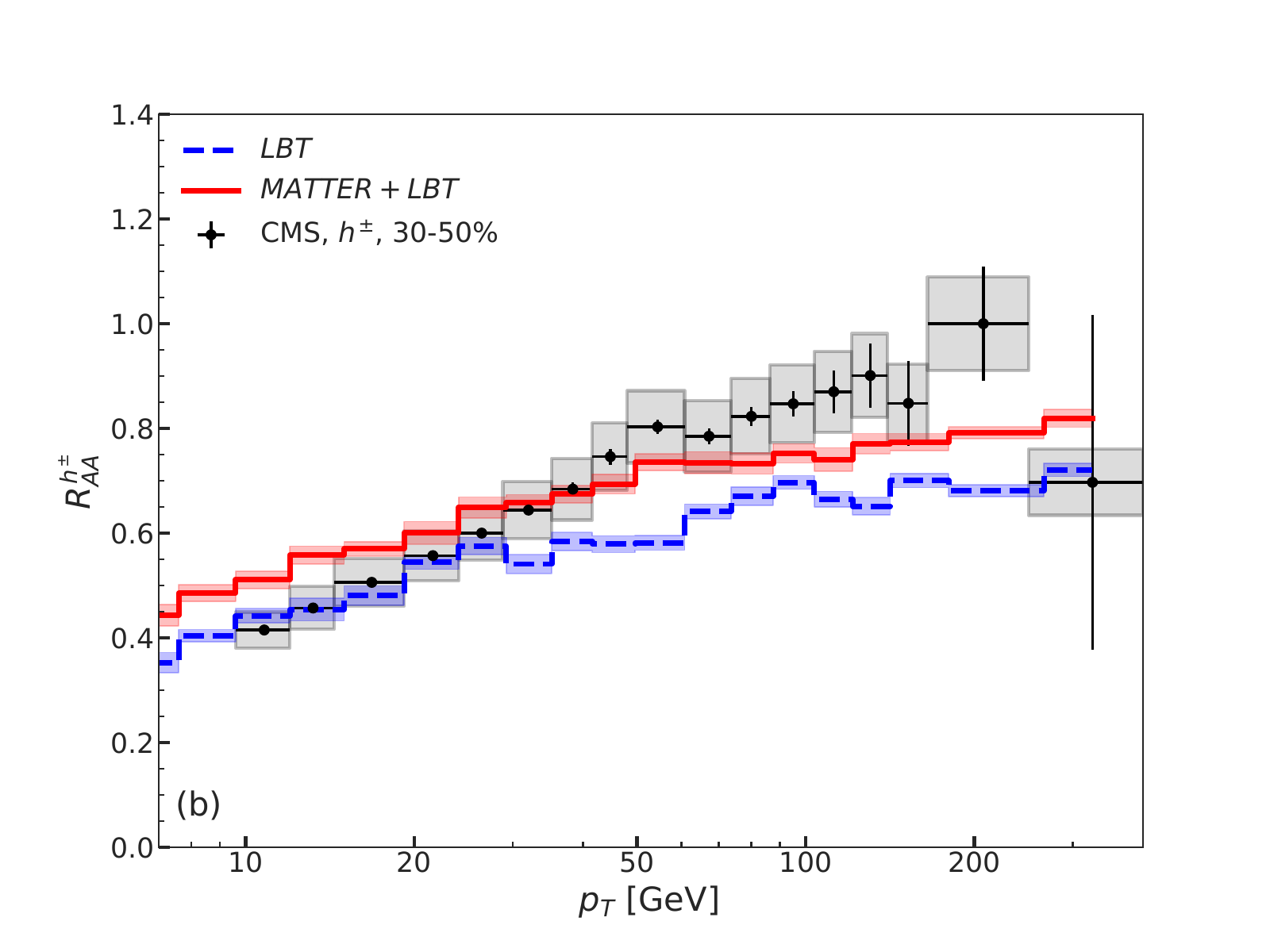}
\end{tabular}
\end{center}
\caption[Nuclear modification factor for D-mesons and charged hadrons]{Nuclear modification factor for D-mesons (a) and charged hadrons (b) in $\sqrt{s_{NN}}=5.02$ TeV PbPb collisions at the LHC at 30-50\% centrality. Here we choose $c_1=10, \ c_2=100$ for the $\hat{q}$ parameterization.}
\label{fig:MATTER_LBT_30-50}
\end{figure}

\begin{figure}[!h]
\begin{center}
\begin{tabular}{cc}
\includegraphics[width=0.495\textwidth]{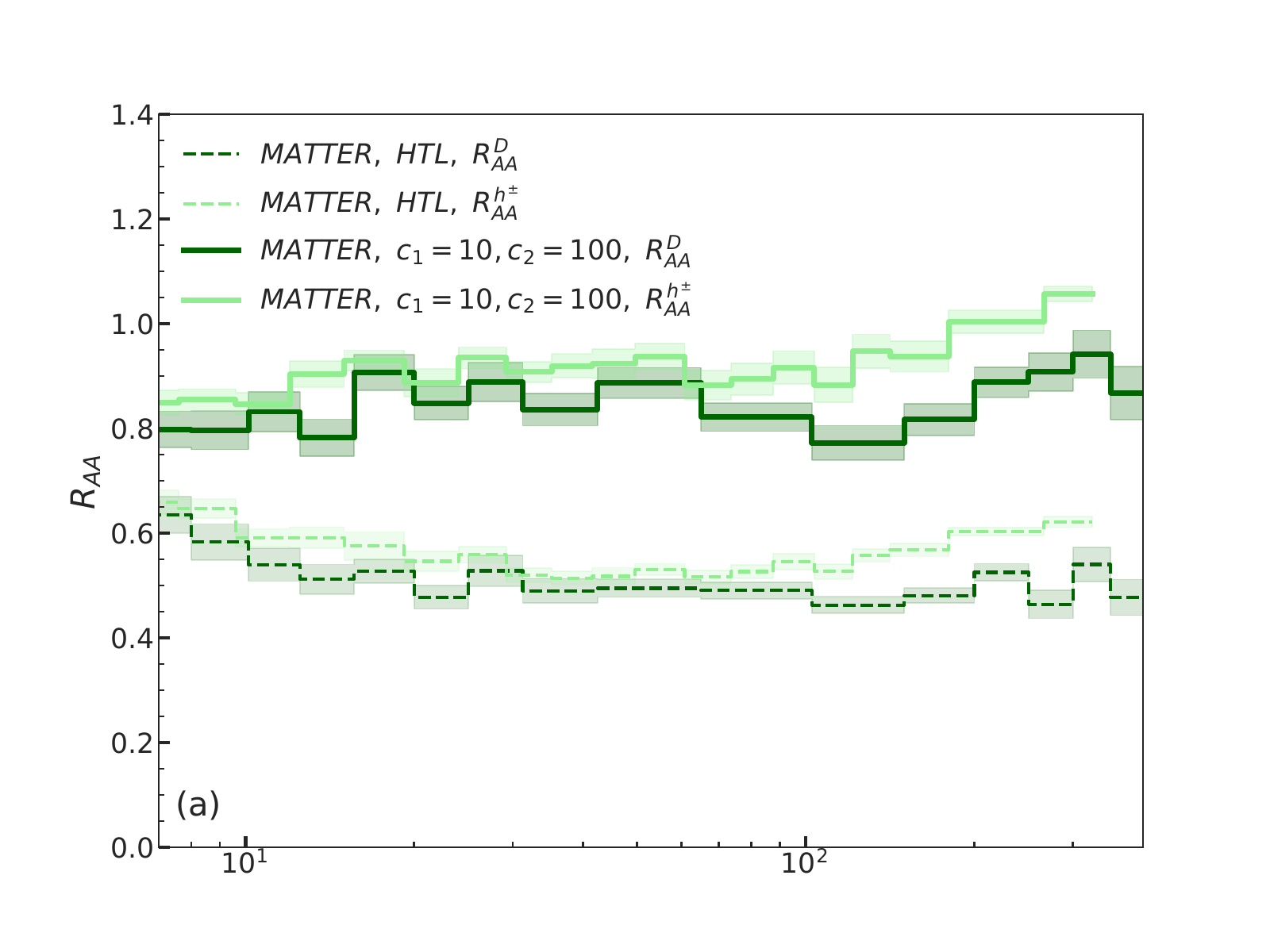} & \includegraphics[width=0.495\textwidth]{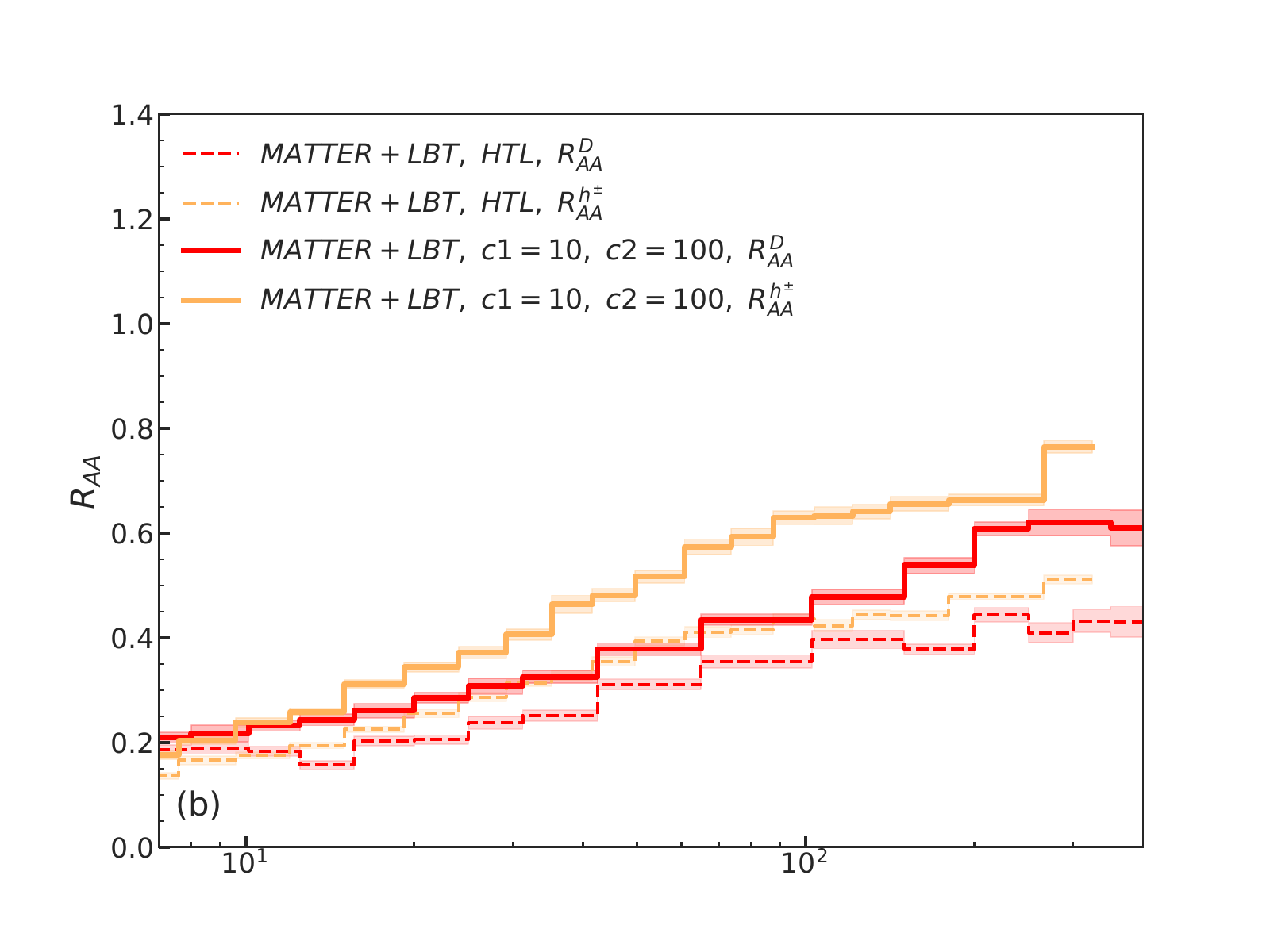}
\end{tabular}
\end{center}
\caption[Nuclear modification factor for D-mesons and charged hadrons]{Nuclear modification factor for MATTER only simulations (a) and for MATTER+LBT simulations (b) in $\sqrt{s_{NN}}=5.02$ TeV PbPb collisions at the LHC at 0-10\% centrality. $c_1=10, \ c_2=100$ parameters values are employed in Eq.~(\ref{eq:qhat_t}).}
\label{fig:MATTER_MATTER+LBT_compare}
\end{figure}

\begin{figure}[!h]
\begin{center}
\begin{tabular}{cc}
\includegraphics[width=0.495\textwidth]{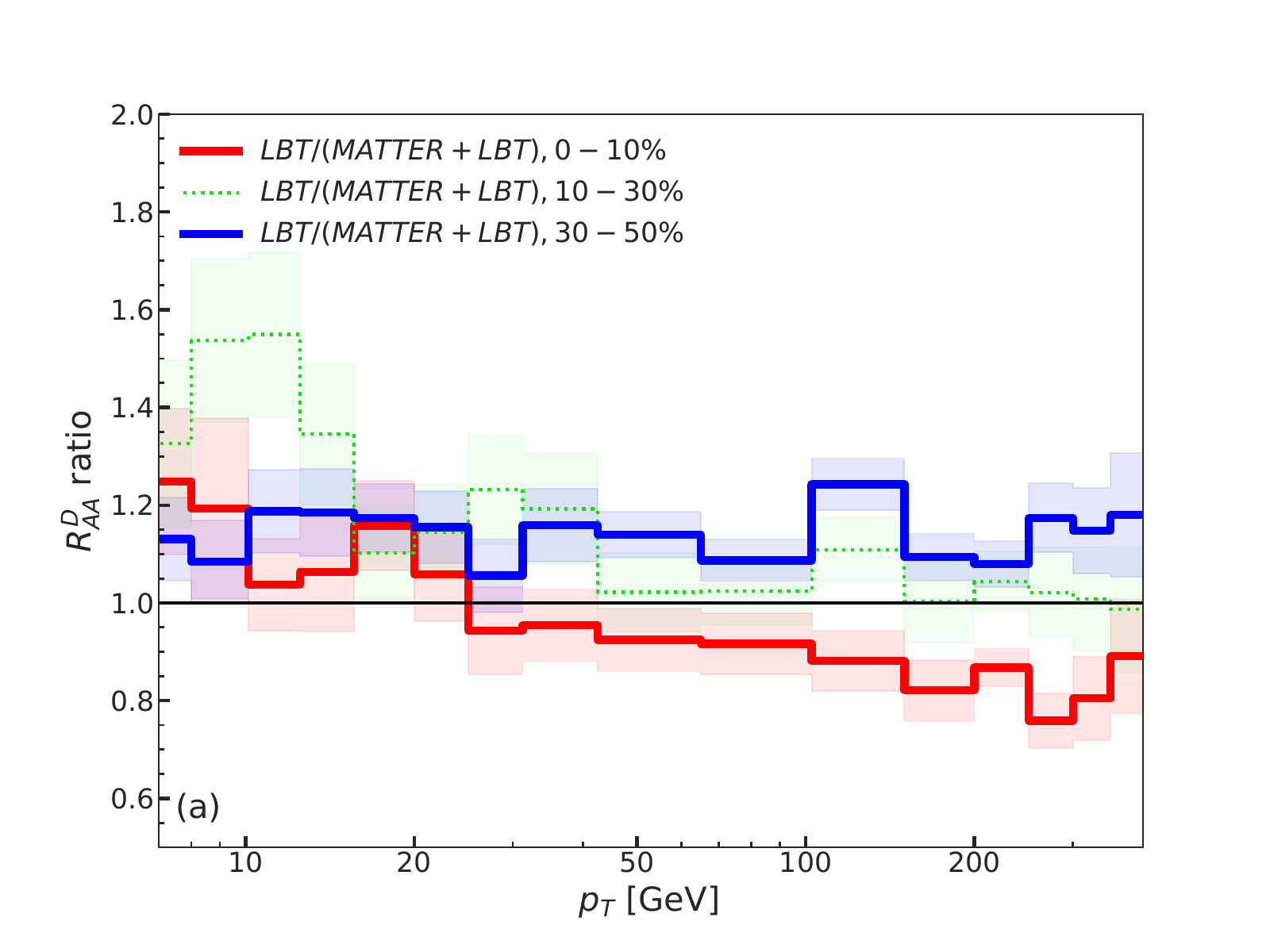} & \includegraphics[width=0.495\textwidth]{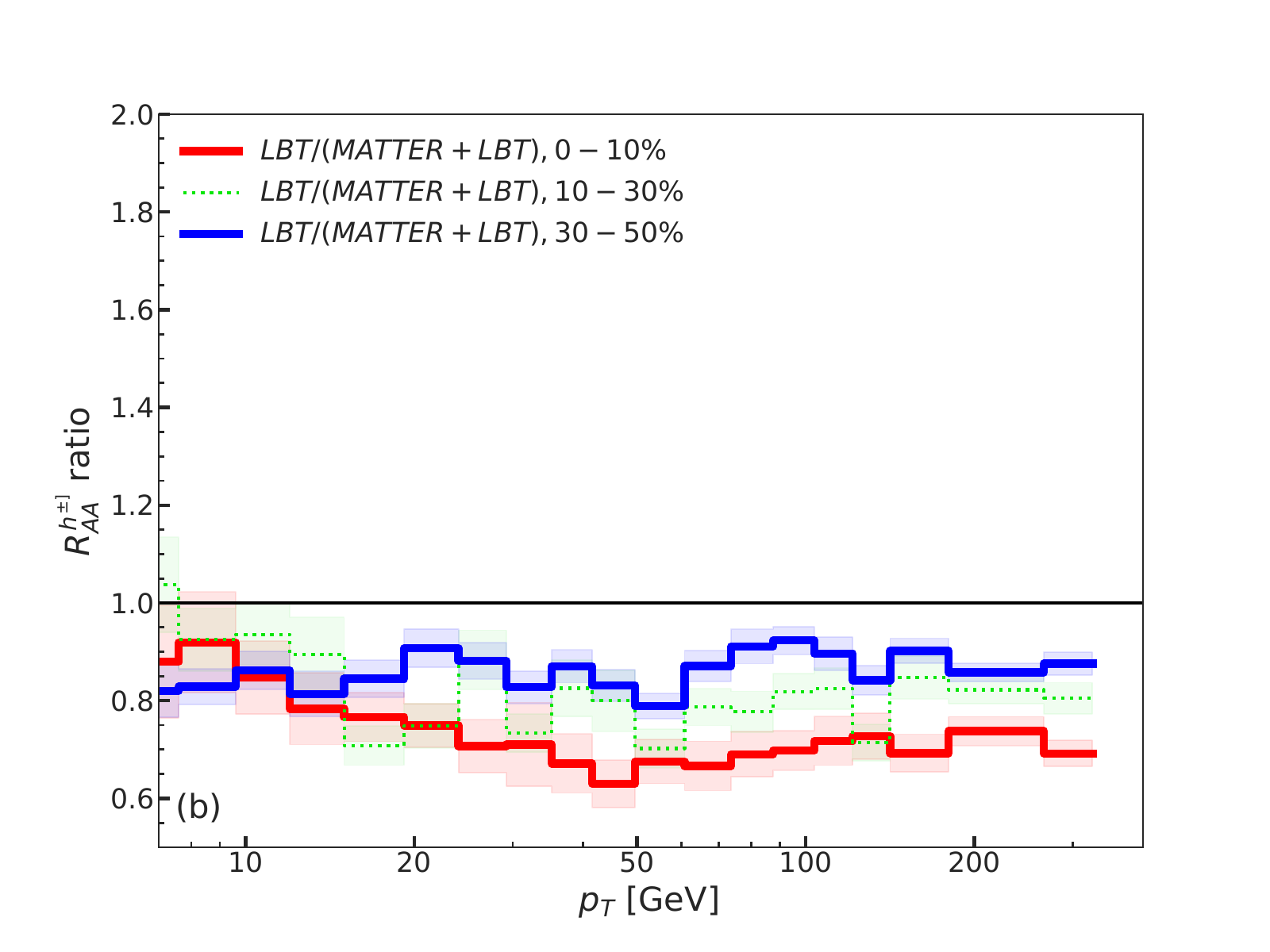}
\end{tabular}
\end{center}
\caption[Ratio of nuclear modification factor between LBT and MATTER+LBT for D-mesons and charged hadrons]{Ratio of nuclear modification factor between LBT and MATTER+LBT for D-mesons (a) and charged hadrons (b) in $\sqrt{s_{NN}}=5.02$ TeV PbPb collisions at the LHC at 0-10\%, 10-30\%, 30-50\% centrality. $c_1=10, \ c_2=100$ parameters values are employed in Eq.~(\ref{eq:qhat_t}).}
\label{fig:MATTER+LBT_to_LBT_ratio}
\end{figure}

Finally, the $R_{AA}$ result for the $10-30\%$ and $30-50\%$ centrality are studied (see Fig.~\ref{fig:MATTER_LBT_10-30} and Fig.~\ref{fig:MATTER_LBT_30-50}). Here the ``best'' fit parameters used in the $\hat{q}(t)$ parameterization (i.e. $c_1=10, \ c_2=100$) are employed. 
One interesting phenomenon to notice is the reversal in order between MATTER+LBT and LBT calculation at high $p_T$ for the $D$ meson $R_{AA}$ from the most central collisions to more peripheral collisions. Two important effects contribute to this observation. First, heavy quarks are more suppressed in the MATTER phase compared to light flavor partons at high $p_T$. This can be seen from both the MATTER only simulations  in Fig.~\ref{fig:MATTER_MATTER+LBT_compare} (a) as well as in MATTER+LBT simulations, see Fig.~\ref{fig:MATTER_MATTER+LBT_compare} (b). Second, going from central to more peripheral collisions, LBT simulations seem to be more affected by the amount of time partons spend interacting with the QGP, compared to MATTER+LBT simulations. Figure~\ref{fig:MATTER+LBT_to_LBT_ratio} shows that at higher $p_T$, the ratio of $R_{AA}$ between LBT and MATTER+LBT simulations increases as the centrality increases for both D meson and charged hadrons. The virtuality dependent $\hat{q}$ reduces the in-medium contribution to MATTER evolution, making it closer to a vacuum-like (DGLAP) evolution at high $p_T$, and thus the partons spend less time in the LBT phase for MATTER+LBT simulations compared to LBT-only simulations. It is important to recall that the same $\hat{q}(t)$ is used for both light and heavy quarks throughout this work, and thus the observation that parton evolution is more vacuum-like given our parameterization for $\hat{q}$ may not necessarily hold for heavy flavors. Whether a mass- and virtualty-dependent $\hat{q}(t,M)$ is needed when describing heavy flavor evolution, should be revisited in the future. In the present work, however, as centrality increases, simulations based solely on LBT evolution are more sensitive to the reduction in QGP space-time volume compared to MATTER+LBT simulations.

Please note that in all the $R_{AA}$ calculations involving MATTER, the pp baseline is using the MATTER vacuum results. You can find the differences between PYTHIA and MATTER vacuum in pp for charged hadrons and D mesons in \cite{JETSCAPE:2019udz, JETSCAPE:2020kwu}. This is essentially a comparison between PYTHIA, which generates an angular ordered shower, and MATTER which generates a virtuality ordered shower. If PYTHIA was used as the pp baseline calculation, then the $R_{AA}$ may further be improved in some $p_T$ ranges, at the expense of consistency. We choose to err on the side of a consistent calculation.  Another possibility is to use MATTER vacuum in PbPb calculations as well, see \cite{park2019multi}. 

\section{Summary}

In this chapter, I have explored how different physics entering a multi-stage description of jet partons interaction with the QGP affect both the D meson and the charged hadron $R_{AA}$. For the LBT regime, the effects of a running $\alpha_s(\mu^2)$ was studied. For the MATTER regime, we highlighted the effects of including scattering as well as considering a virtuality dependent $\hat{q}$ and found that both make a large contribution to the value of $R_{AA}$. The virtuality dependent $\hat{q}$ offers a possible explanation for the diminishing value of the interaction strength $\hat{q}/T^3$ at the LHC from previous extractions~\cite{Burke:2013yra}. However, neither of these two models alone is sufficient for describing the $R_{AA}$ at the $p_T$ range we are interested in. 

We find that the best simultaneous description of the D meson and charged hadron $R_{AA}$  requires the explicit inclusion of both, the high-energy and high-virtuality regime as well as the high energy and low virtuality regime of parton energy-loss. In this work, these have been modeled using the MATTER and the LBT schemes within the JETSCAPE framework. The specific form of the $\hat{q}(t)$ parameterization is still under investigation, yet we can already state that the suppression of $\hat{q}$ at higher virtuality mostly increases $R_{AA}$ at high $p_T$. We have also explored where, in virtuality, the transition point lies between these two regimes and how changing it affects the resulting $R_{AA}$. A higher switching scale $t_s$ implies that partons will evolve longer in the LBT regime and lose more energy. While we have found that our simple exploration of the parameter space already provides a decent simultaneous description of both light and heavy flavor $R_{AA}$, a Bayesian analysis could improve the description even further.

\chapter{Bayesian Analysis Overview} \label{sec:bayesian}

\vspace{0.1in}

\begin{figure}
	\centering
	\includegraphics[width=0.9\textwidth]{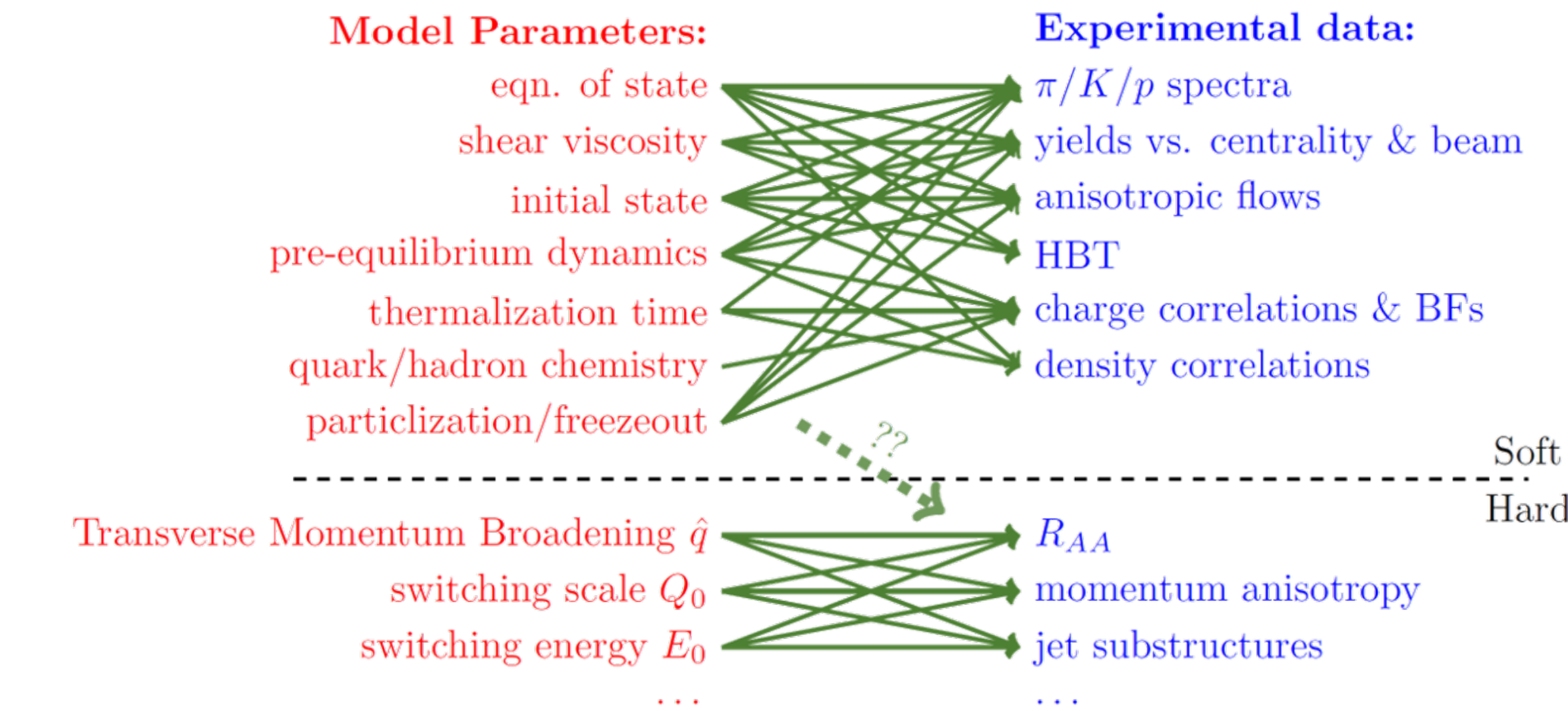}
	\caption{\label{fig:parameter_correlation} Model parameters in the simulation framework and their effects on experimental observables.}
\end{figure}

\begin{figure}
	\centering
	\includegraphics[width=0.8\textwidth]{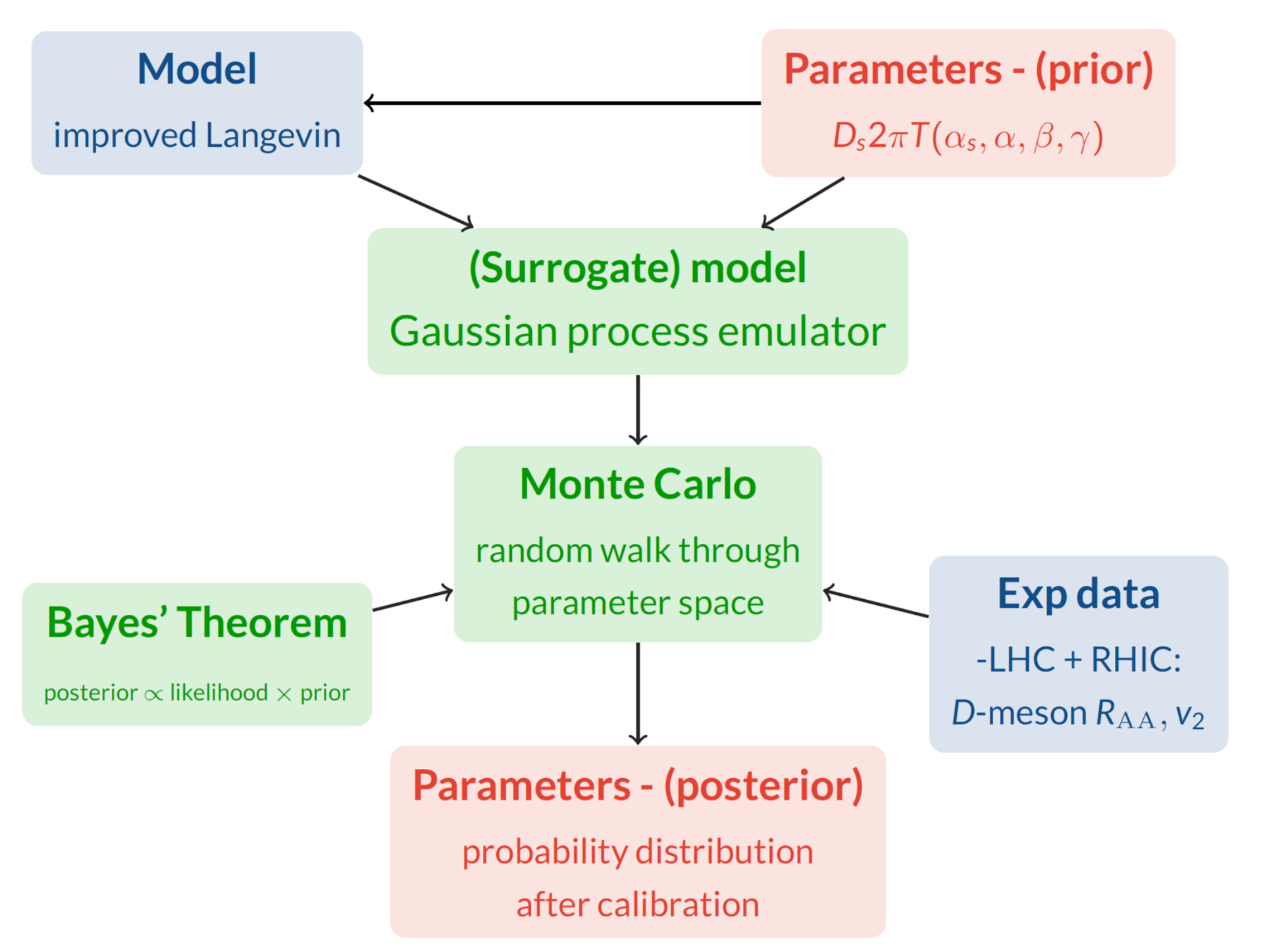}
	\caption{\label{fig:bayesian_framework} Workflow for the Bayesian model-to-data comparison framework.}
\end{figure}

So far, various models describing the evolution of both the soft medium and the hard parton shower in heavy ion collisions have been covered. Because of the many-body and multi-scale nature of heavy ion collision, many models will require input parameters that describe the properties of the system, but the values of which are not known a priori (see Fig.~\ref{fig:parameter_correlation}). It is therefore essential to study whether these models, with appropriate values of those parameters, are capable of describing the experimental data . Again, a modular framework has been developed to break this problem into smaller, easier-to-solve pieces (see Fig.~\ref{fig:bayesian_framework}).

\section{Bayes' theorem}

The Bayes' theorem states that the posterior distribution of the parameter set $\mathbf{q}$, given the experimental observation $\mathbf{y}_{\rm exp}$, is proportional to the product of the prior distribution $p(\mathbf{q})$ and the likelihood function $\mathcal{L}(\mathbf{y}_{\rm exp}|\mathbf{q})$:
\begin{equation}
    P(\mathbf{q}|\mathbf{y}) \propto \mathcal{L}(\mathbf{y}_{\rm exp}|\mathbf{q})p(\mathbf{q}).
\end{equation}

The prior $p(\mathbf{q})$ represents the prior knowledge of the parameter values. In this study, a joint uniform distribution within some reasonable ranges for the parameters will be used as the prior since little is known about what values the parameters should take except for some physical constraints (e.g., the switching scale $Q_s>1.5 ~\approx m_c$ where $m_c$ is the mass of the charm quark.). 

The likelihood function $\mathcal{L}(\mathbf{y_{\rm exp}}|\mathbf{q})$ is the probability of having an observation as $\mathbf{y}_{\rm exp}$ given a specific parameter set $\mathbf{q}$:
\begin{equation}
\label{eqn:likelihood_definition}
\mathcal{L}(\mathbf{y}|\mathbf{q}) = \frac{1}{\sqrt{(2\pi)^m \rm{det}\Sigma}} \exp\left[-\frac{1}{2} [\mathbf{f(\mathbf{q})} - \mathbf{y}_{\rm exp}]^T \Sigma^{-1} [\mathbf{f(\mathbf{q})} - \mathbf{y}_{\rm exp}]\right],
\end{equation}
where $m$ is the dimension of $\mathbf{y}_{\rm exp}$, $\Sigma = \Sigma_{M} + \Sigma_{\rm exp}$ is the uncertainty covariance matrix, which takes into account both model and experimental uncertainties. $f(\mathbf{q})$ is the model calculation giving the parameters $\mathbf{q}$. 


\section{Analytical solution for a simple linear model}

In some cases, the posterior distribution can be evaluated analytically and therefore the optimal values for the parameters can be determined by maximizing the posterior distribution.

For example, if the measurement $\mathbf{y}$ and the experiment condition $\mathbf{x}$ follow a simple linear relation:
\begin{equation}
    y_i=q_1 x_i+q_2.
\end{equation}

And if one assumes that the correlation matrix is diagonal and the prior distribution is uniform, the log of the posterior distribution can be simplified to:
\begin{equation}
\ln P(\mathbf{q}|y)= \rm{const} - \sum_{i=1}^N \frac{(y_i - q_1 x_i-q_2)^2}{2\sigma_{i}^2},
\label{eqn:log_post}
\end{equation}
where the parameters $\mathbf{q}=(q_1,q_2)$. 

Define the following matrices:
\begin{equation}
Y = \begin{pmatrix}
y_1\\
y_2\\
\vdots \\
y_N
\end{pmatrix} ,
X=  \begin{pmatrix}
1  & x_1\\
1 & x_2\\
\cdots & \cdots \\
1 & x_N
\end{pmatrix} ,
\Sigma =  \begin{pmatrix}
\sigma_1^2 & 0 & \cdots & 0\\
0 & \sigma_2^2 &  \cdots &0\\
\vdots & \cdots & \sigma_i^2  & \cdots \\
0 &  0 & \cdots & \sigma_N^2
\end{pmatrix} ,
\end{equation}
The best-fit values for the parameter $m$ and $b$ are then solved by:
\begin{equation}
Q=\begin{pmatrix}
q_1\\
q_2\\
\end{pmatrix} 
= [X^T\Sigma^{-1}X]^{-1} [X^T\Sigma^{-1}Y].
\label{eqn:analytic_Baye}
\end{equation}

\section{The Markov Chain Monte Carlo method}

In practice, analytical solutions are often difficult or even impossible to get. Markov Chain Monte Carlo (MCMC) sampling is an alternative approach to draw the posterior distribution in this case. The algorithm works as the following:

\begin{enumerate}
    \item Start with a point $\mathbf{q}_t$ in the parameter space at step $t$. Sample a new point $\mathbf{q}^{'}$ from some distribution $p(\mathbf{q}_t)$.
    \item Accept this new point based on the ratio of the posterior distribution of the two points:
    \begin{equation}
        a(\mathbf{q}_t,\mathbf{q}^{'})=min(1,\frac{\mathcal{L}(\mathbf{q}^{'} |y)p(\mathbf{q}^{'})}{\mathcal{L}(\mathbf{q}_t |y)p(\mathbf{q}_t)}).
    \end{equation}
    
    If $\mathbf{q}^{'}$ is accepted, $\mathbf{q}_{t+1}=\mathbf{q}^{'}$, otherwise  $\mathbf{q}_{t+1}=\mathbf{q}_t$.
    
    \item Repeat step 1 and 2 for a sufficient number of steps. The sample point set $\{\mathbf{q}_1,\mathbf{q}_2,...,\mathbf{q}_N\}$ is an approximation of the posterior distribution we want.
\end{enumerate}

In practise, a good idea is to run some steps first (called the burn-in stage) to explore the parameter space and try to get closer to the maximum posterior probability density region. One can also run multiple MCMC simulations in parallel and combine the sampled points together. A more detailed explanation of the mathematical derivation and implementation can be found in Ref.~\cite{FM:2013mc}.

\section{Gaussian process emulator}

The Markov Chain Monte Carlo (MCMC) method can draw the posterior distribution for models without an analytical solution. However, it still requires evaluating the model output at many different points in the parameter space. The number of evaluations is often at least $\mathcal{O}(10^4)$ to draw a stable posterior distribution and will grow as the number of parameters increases. The problem is that the model is often computationally slow for this number of evaluation to be feasible in terms of computational cost. 

For the event-by-event heavy ion collision simulation we are doing, the number of events needed to get good statistics on observables like the charged hadron $R_{AA}$ are on the order of $10^6$ and will be even several magnitudes larger for rare observables like $D$ meson $v_2$. Even if we use pre-computed $(2+1)D$ hydrodynamics, a single event may still take several seconds to compute. Thus the total run time for simulating a single point in the parameter space would be tens of thousands of CPU hours. We have to minimize the number of evaluations of the model while still being able to predict the model output at an arbitrary parameter point. It turns out that the Gaussian process emulator is just able to achieve this goal at the expense of introducing some additional uncertainties into the prediction. 

A GP essentially takes in the training data and interpolates them. Mathematically, one assumes that all the output $\mathbf{y}$ we want to predict at some input $Q$ and the known outputs $\mathbf{y}_{\rm train}$ at the training points $Q_{\rm train}$ ($Q_{\rm train}$ has dimension $m\times k$ where $m$ is the number of training data and $k$ is the dimension of the parameter set. $\mathbf{y}_{\rm train}$ is a $m \times 1$ vector as at each training point we are looking at just one dimension the model output) form a multivariate normal distribution:
\begin{equation}
\begin{pmatrix}
\mathbf{y} \\
\mathbf{y}_{\rm train} 
\end{pmatrix}  \sim 
\mathcal{N}  \left( \begin{pmatrix}
\mu\\
\mu_{\rm train}
\end{pmatrix}, 
\begin{pmatrix}
K(Q,Q) & K(Q, Q_{\rm train}) \\
K(Q_{\rm train}, Q) & K(Q_{\rm train},Q_{\rm train})
\end{pmatrix} \right),
\label{eqn:GP_predict}
\end{equation}
where $K$ denotes the covariance matrix. Each element in the covariance matrix $K$ is calculated with the kernel function $k(\mathbf{q},\mathbf{q}')$ that characterizes the correlation between two points in the parameter space. The kernel encodes our prior belief of the function that we want to mimic \cite{schulz2018tutorial} and one common choice of the kernel is the radial based function (RBF) kernel:
\begin{equation}\label{eqn:RBF_kernel}
    k(|x-x'|)=k(r)=\sigma^2\exp(-\frac{r^2}{2l^2}),
\end{equation}
where $\sigma^2$ is the auto correlation and $l$ is the correlation length. This form will make the covariance decay exponentially with the distance between two input points. So two outputs that are far away in their inputs are effectively uncorrelated. 

The distribution of $\mathbf{y}$ is then given by:
\begin{equation}
    \begin{split}
    \mathbf{y} \sim  &\mathcal{N}(K(Q, Q_{\rm train})K^{-1}(Q_{\rm train},Q_{\rm train})\mathbf{y}_{\rm train}, \\ 
    & K(Q,Q)-K(Q, Q_{\rm train})K^{-1}(Q_{\rm train},Q_{\rm train}) K(Q_{\rm train}, Q)).
\end{split}
\end{equation}

Unlike the polynomial interpolation, a Gaussian process does not provide a single estimation of the output but infers the probability distribution of the predicted outputs by predicting both the mean and the covariance matrix. It is a massive advantage of the Gaussian process to quantify its interpolation uncertainty.

\section{Principal component analysis}

The Gaussian process maps from a multi-dimensional input space to a one-dimensional space. In the case of multi-variate output (observables), we can train multiple Gaussian process emulators, one for each dimension of the output. However, the observables are not necessarily independent from each other. One can imagine that the $R_{AA}$ of charged hadrons will be highly correlated with the $R_{AA}$ of inclusive jet since they are both related to the evolution of hard probes. Therefore it is beneficial to reduce from high dimensional and correlated output to lower dimensional and orthogonal output of principal components(PCs), which are linear combinations of the original observables.

The output $Y$ can be seen as a $n\times m$ matrix where $m$ is the number of observables, and $n$ is the number of training data we have. This matrix can be decomposed by singular value decomposition (SVD):
\begin{equation}
    Y_{n\times m}=U_{n\times n}S_{n\times m}V_{m \times m}^T,
\end{equation}
where $S$ now only contains the variance of the principal components on its diagonal. The principal components are defined as:
\begin{equation}
    z=Vy.
\end{equation}

$S$ is typically chosen that its eigenvalues $\{\lambda_1,\lambda_2,...,\}$ are sorted in descending order. If the first $m^{'}$ principal components are able to take up the majority of the variance,

\begin{equation}
    CV(m^{'})=\frac{\sum_{i=1}^{m^{'}}\lambda_i}{\sum_{i=1}^{m}\lambda_i} \sim 1,
\end{equation}

we can therefore select only the first $m^{'}$ elements in the principal component space, and use the new training data set $\mathcal{D^{'}} = \{(\mathbf{x_i}, \mathbf{z_i})\}$ to train $m^{'}$ Gaussian Process emulators then make predictions of $\mathbf{z^{\*}}$ at each arbitrary parameter $\mathbf{x}$. The dimension of the output space is reduced by $m-m^{'}$.
Once the output $\mathbf{z^{\*}}$ of the principal components are predicted by the GP emulators, one would perform a inverse transformation to predict the original output $\mathbf{y^{\*}}$.
\begin{equation}
\mathbf{y^{\*}} = \frac{1}{\sqrt{n}} \mathbf{z^{\*}} V.
\end{equation}

\section{Latin hypercube sampling}

As discussed previously, the Gaussian process emulator serves as a fast surrogate model of the actual model. The idea is to use a few training data points while still being able to cover the parameter space with controlled uncertainty. The training data points play a vital role here. However, if a fixed number of values are sampled in each dimension, the number of total sampled points grows exponentially with the dimension of the input parameter space. 

\begin{figure}
	\centering
	\includegraphics[width=0.54\textwidth]{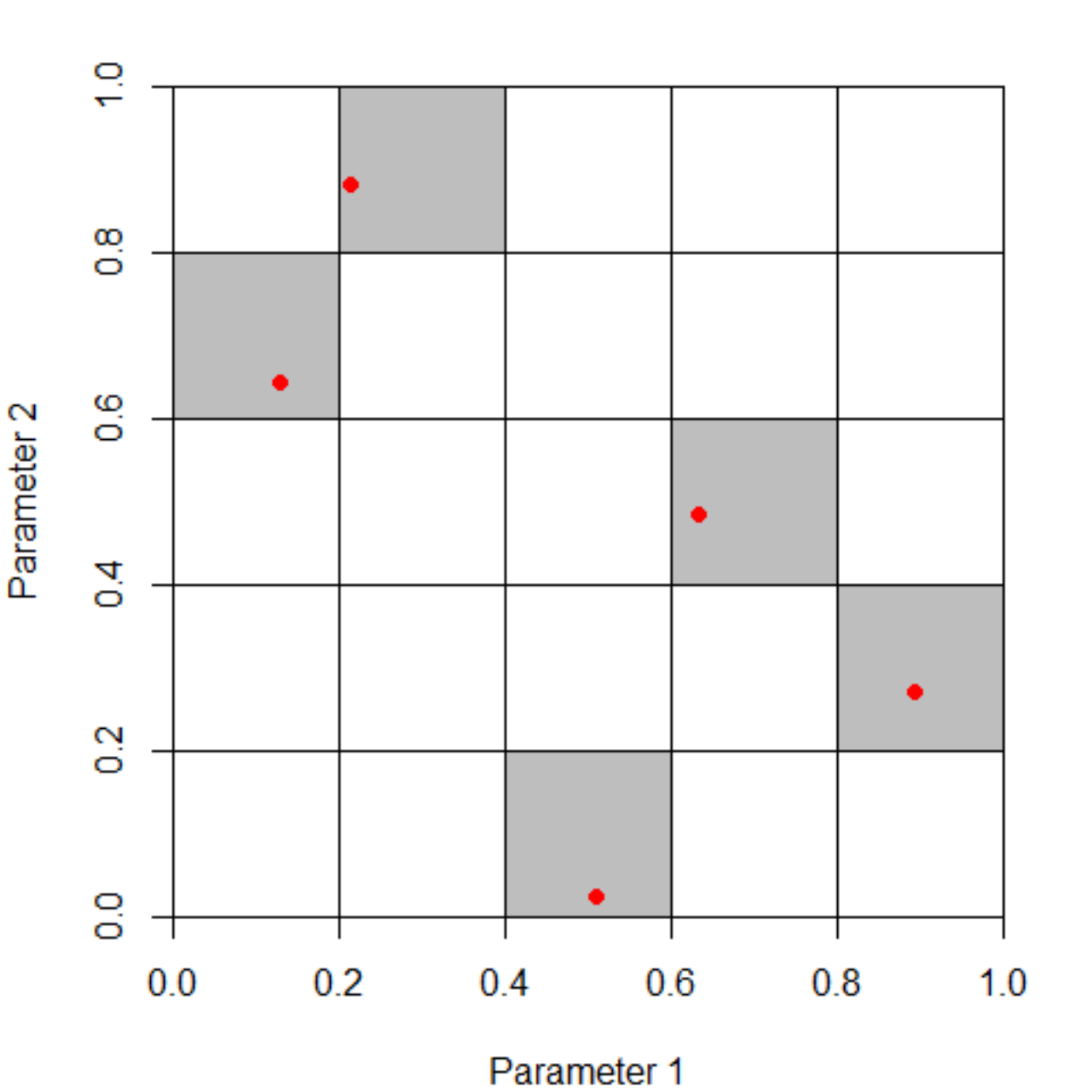}
	\caption[A 5-point latin hypercube sampling]{\label{fig:LHS_demo} A 5-point latin hypercube sampling. Each parameter is sampled only once in every interval.}
\end{figure}

On the other hand, Latin hypercube sampling (LHS) doesn't depend on the dimension of the input. Instead, one specifies the number of samples $N$ one would like to draw in advance. The range of each variable is partitioned into $N$ non-overlapping intervals based on equal probability size. One value from each interval is selected at random with respect to the probability density in the interval. The $N$ values obtained for $x_1$ are paired in a random manner with the $N$ values for $x_2$. These $N$ pairs of $(x_1, x_2)$ are combined in a random manner with the $N$ values of $x_3$, and so on, until a set of $N$ $n$-tuples is formed.  

\section{Bayesian parameter inference in action}\label{sec:simple_bulk_bayesian}

\subsection{A simple model for bulk physics}

In this section I would present the Bayesian analysis results for a simple bulk model \cite{paquet2020effective}. In this model, the final-state charged particle multiplicity and elliptic flow are related to an ``evolution history averaged'' viscosity:
\begin{equation}
    N_{ch}=N_0 \cdot \Tilde{E}_T \left[1+(\eta/s)_{eff}\right], \ v_2=\epsilon_2 \cdot \exp\left[-50 \frac{(\eta/s)_{eff}}{N^{1/3}_{ch}}\right],
\end{equation}
where $(\eta/s)_{eff}$ is the temperature average of $\eta/s$:
\begin{equation}\label{eqn:eta_s_def}
    (\eta/s)_{eff}=\frac{\int_{T_{min}}^{T_{max}}T^p(\eta/s) dT}{\int_{T_{min}}^{T_{max}}T^p dT}.
\end{equation}

The parameters $N_0=9, \ T_{min}=0.13, \ p=-1$ are assumed to be fixed. $T_{max}, \ \Tilde{E}_T, \ \epsilon_2$ are inputs depending on the centrality. 

We propose a simple parameterization of $\eta/s$ with respect to temperature $T$:
\begin{equation}
    \frac{\eta}{s}(T) = a + 
\begin{cases} 
b (T - T_s), & T>T_s,\\
c (T_s - T), & T<T_s,\\
\end{cases}
\end{equation}
$T_s=0.18$ is also fixed in this study but can be varied. And the parameters of our model are $a,b,c$.

One of the goals of this study is to go over the process of Bayesian analysis. Another goal is to test the validity of Bayesian analysis. The ``experimental data'' will be generated by the model, meaning we know the true value of the parameters and thus $\eta/s(T)$. Therefore we can also test whether we can constrain the model parameters given data with different levels of uncertainty. To achieve this goal, I need to minimize the uncertainties introduced by the model and the Gaussian process emulator. 

\subsection{Train surrogate model with Gaussian process}

The first step in the Bayesian analysis is to generate the Latin hypercube sampling. In order to reduce the emulator uncertainty, 1000 sampling points are generated:

\begin{figure}
	\centering
	\includegraphics[width=0.5\textwidth]{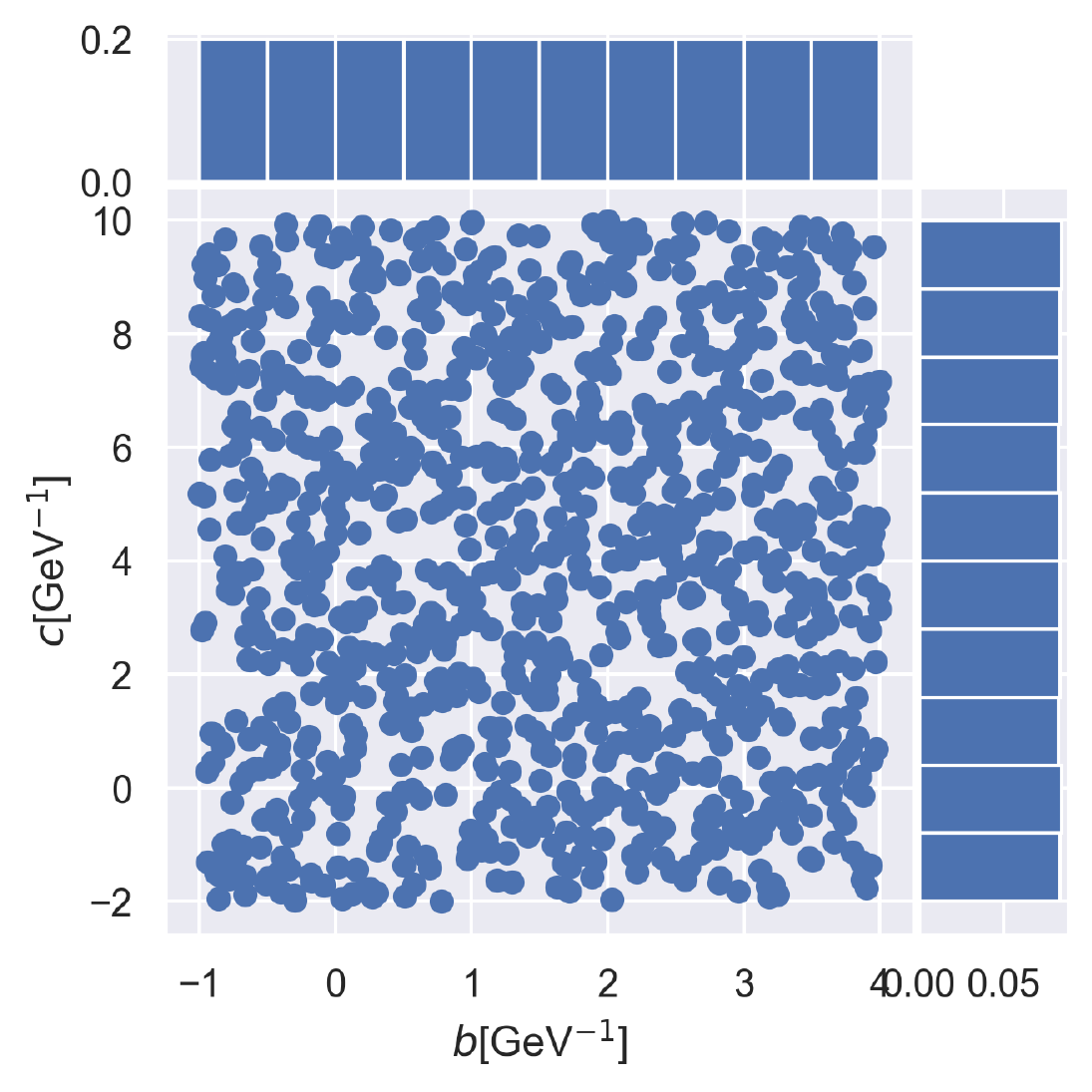}
	\caption{\label{fig:LHS_design} Latin hypercube sampling of 1000 design points. }
\end{figure}

As seen from Fig.~\ref{fig:LHS_design}, LHS yields almost uniform distribution on each of the parameters, as expected. 

\begin{figure}
	\centering
	\includegraphics[width=0.9\textwidth]{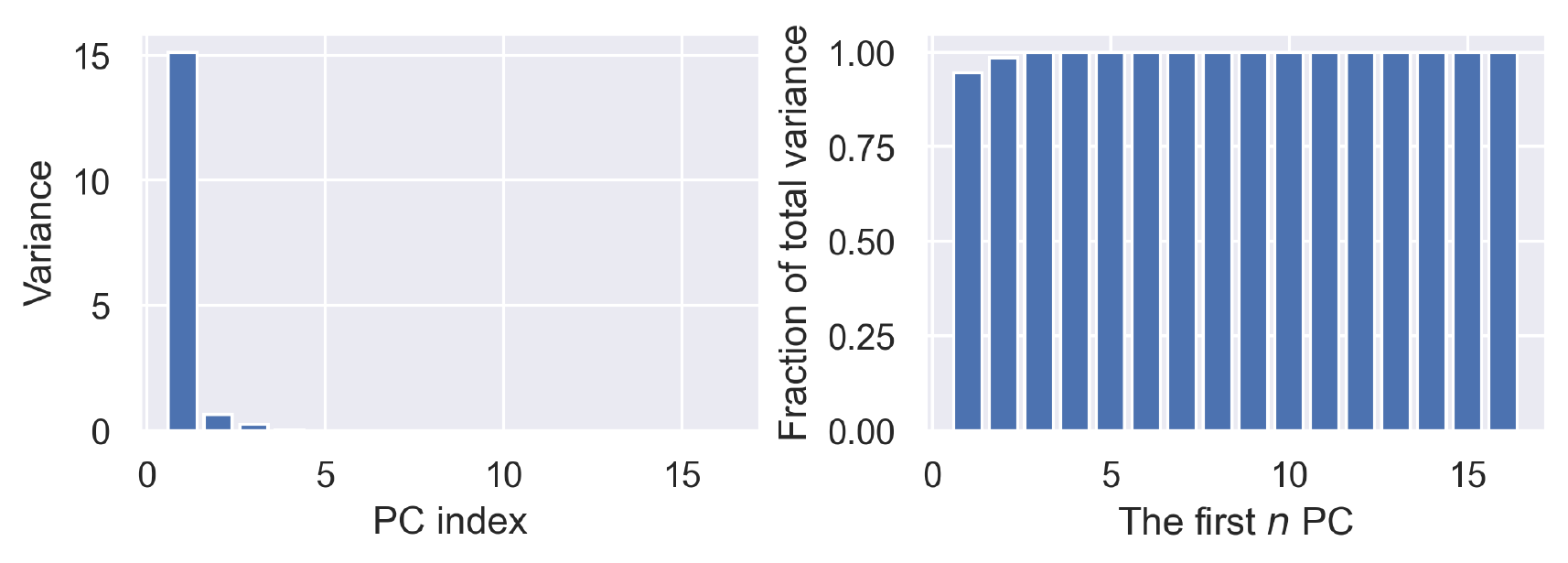}
	\caption[Variance of the principal components]{\label{fig:PC_importance} \textbf{Left}: Variance of the principal components by their index. \textbf{Right}: The ratio of the total variance of the first $n$ PC to the total variance of all the PC.}
\end{figure}

\begin{figure}
	\centering
	\includegraphics[width=0.96\textwidth]{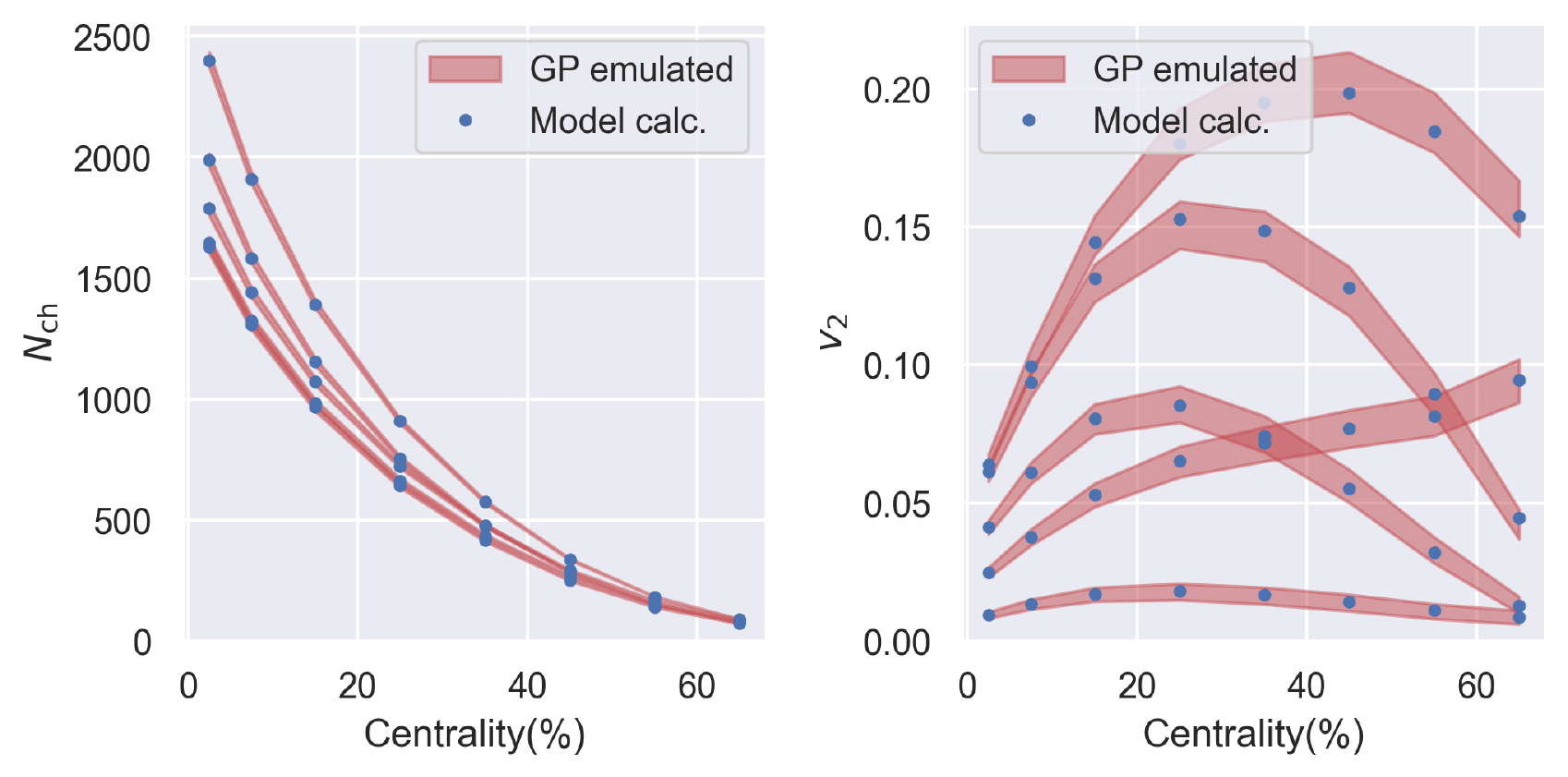}
	\caption[Comparison between model calculation and emulator prediction at 5 randomly sampled parameter points]{\label{fig:Emulator_validation_1} Comparison between model calculation and emulator prediction at 5 randomly sampled parameter points. The band of the GP represents $95\%$ CL.}
\end{figure}

The next step is to perform principal component analysis (PCA). From the model definition, one can see that $N_{ch}$ and $v_2$ are correlated. The total dimension of the ``experiment'' data $\mathbf{y}$ is $16$. But training 16 Gaussian process emulators would be redundant. In fact, after performing the PCA, one can see the variance is dominated by the first few components, and the first five will be used in this study. 

Now the Gaussian process emulators on the selected principal components can be trained. Before proceeding further, it is important to check the performance of these emulators (emulator validation). 

From Fig.~\ref{fig:Emulator_validation_1}, one can see the emulators can match with model calculations at those randomly generated new parameter points. 

\begin{figure}
	\centering
	\includegraphics[width=0.96\textwidth]{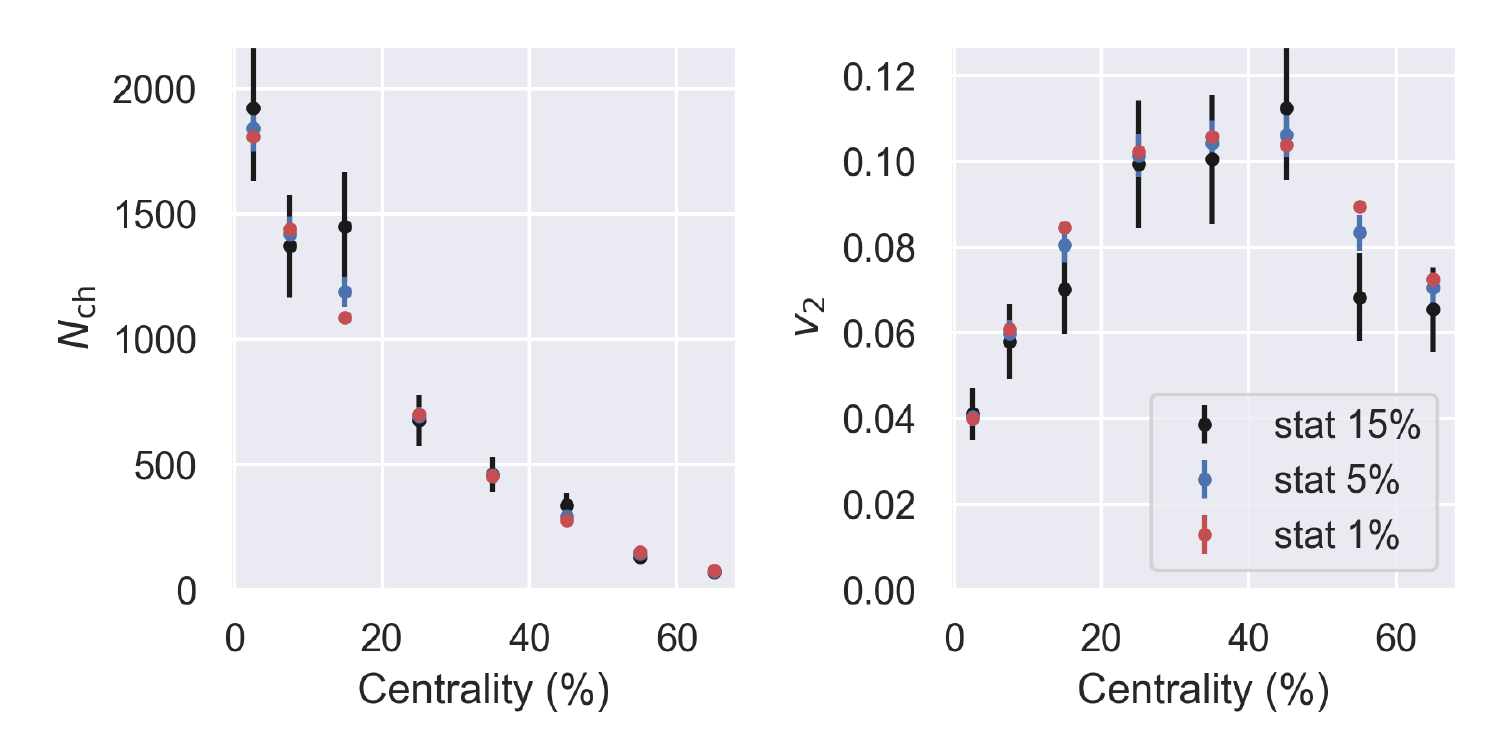}
	\caption[Pseudo data generated with $15\%$, $5\%$, and $1\%$ statistical uncertainty]{\label{fig:Pseudodata} Pseudo data generated with $15\%$, $5\%$, and $1\%$ statistical uncertainty.}
\end{figure}

Now I would like to revisit the definition of the covariance matrix in Eq.~\ref{eqn:likelihood_definition}. $\Sigma_{exp}$ can be further decomposed as $\Sigma_{exp}=\Sigma_{sys}+\Sigma_{stat}$. One often doesn't know the off-diagonal values in $\Sigma_{sys}$ from experiment publications but the diagonal values are generally available. As for the model covariance matrix $\Sigma_{M}$, the use of Gaussian process and principal component analysis both introduce uncertainties ($\Sigma_{emulator},\Sigma_{truncation}$) in addition to the statistical and systematic uncertainties from model calculations: $\Sigma_{M}=\Sigma_{emulator}+\Sigma_{truncation}+\Sigma_{model}$. In this example, the model calculation is exact. So only the experimental data uncertainty and emulator uncertainty are varied. 

\subsection{Inference with different levels of experimental data \mbox{uncertainty}}

The ``experimental data'' are prepared with three different levels of statistical uncertainties. If the data are sensitive to the parameters, we should be able to constrain the parameters even when sizable uncertainties are introduced. And the constraining power should increase as the uncertainties decrease. 

\begin{figure}
	\centering
	\includegraphics[width=0.9\textwidth]{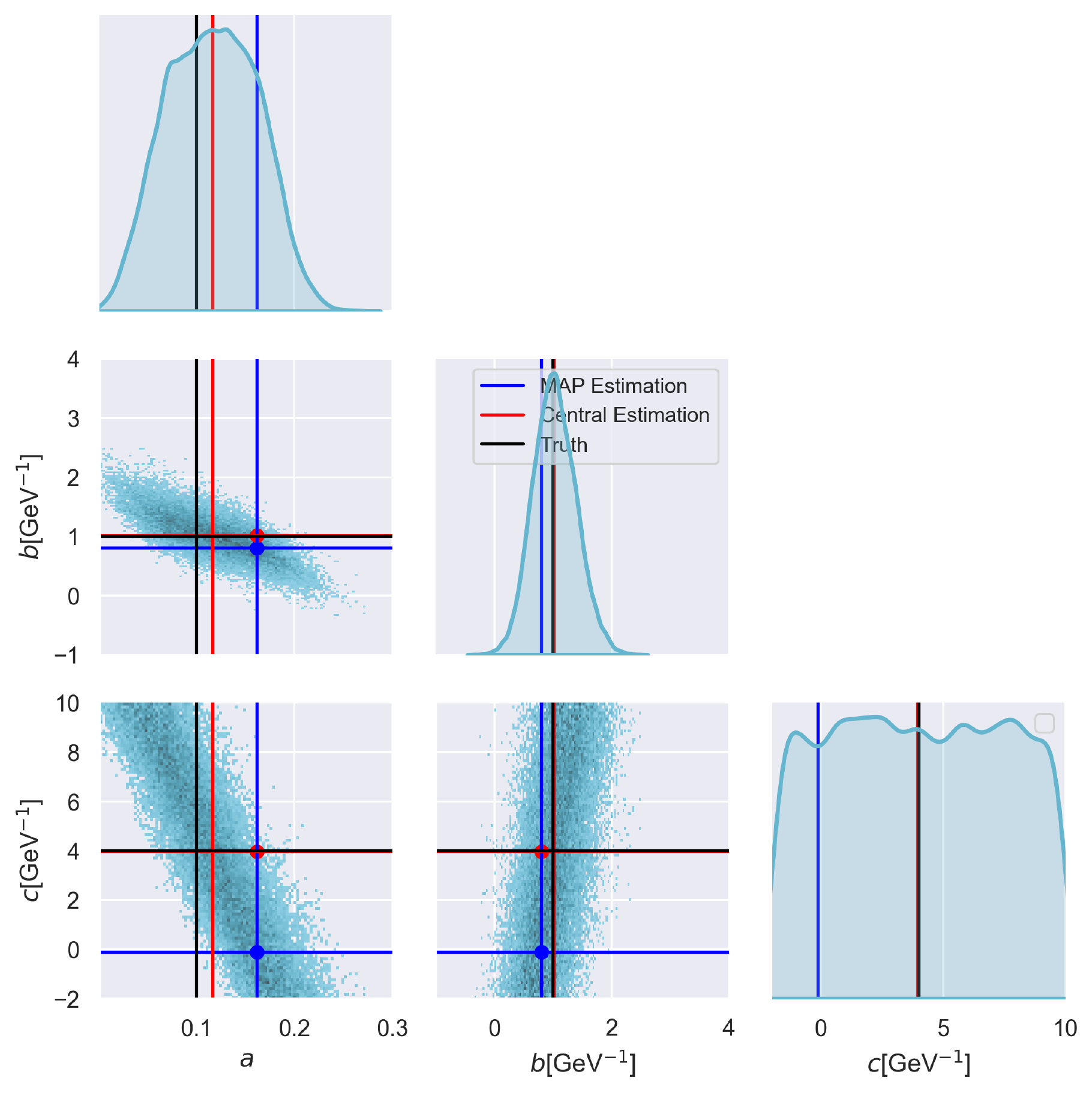}
	\caption{\label{fig:Posterior_of_parameters_15} Posterior distributions of the parameters $(a,b,c)$ inferred from data with $15\%$ of statistical uncertainties.}
\end{figure}

\begin{figure}
	\centering
	\includegraphics[width=0.9\textwidth]{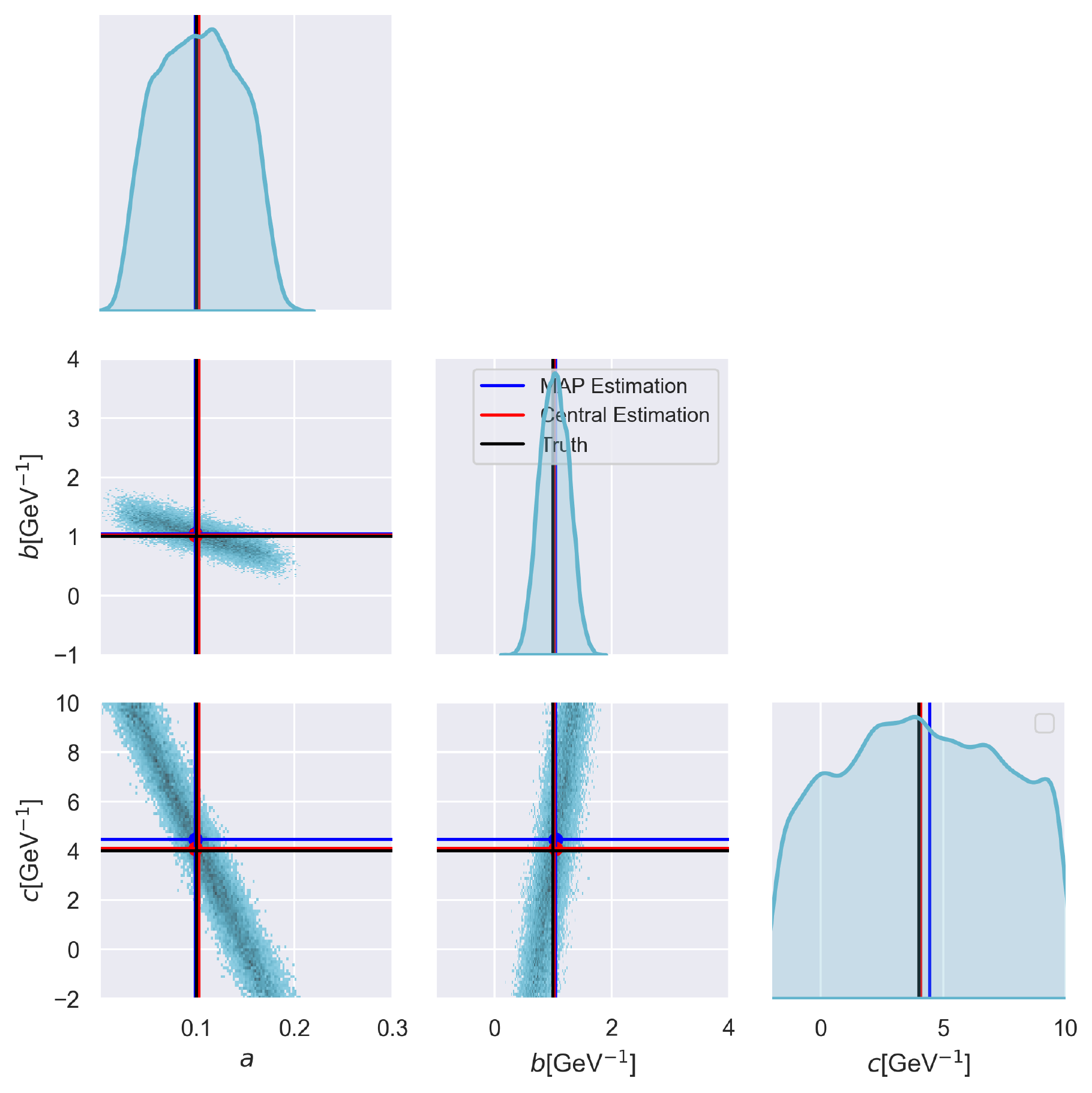}
	\caption{\label{fig:Posterior_of_parameters_5} Posterior distributions of the parameters $(a,b,c)$ inferred from data with $5\%$ of statistical uncertainties.}
\end{figure}

\begin{figure}
	\centering
	\includegraphics[width=0.9\textwidth]{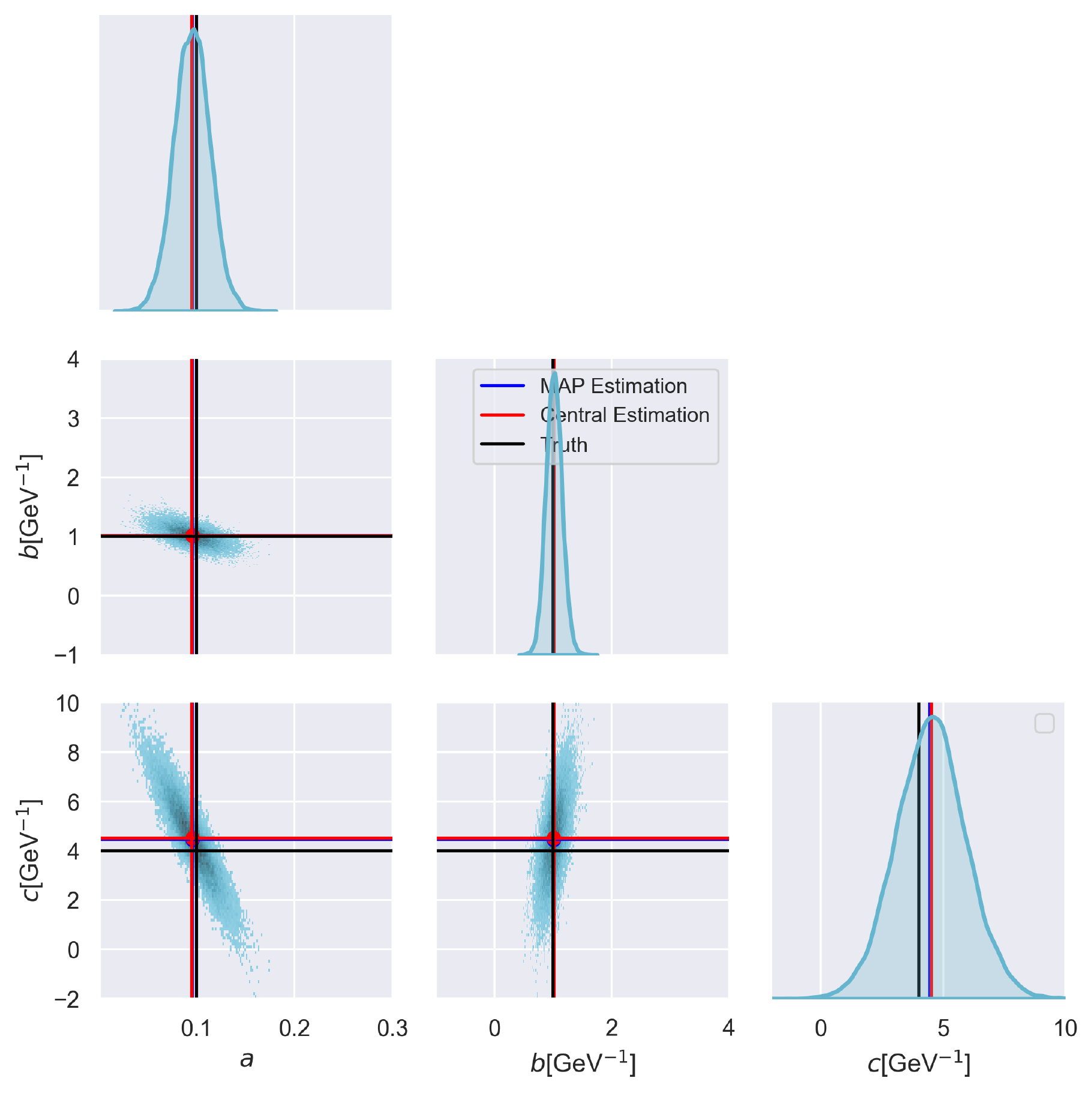}
	\caption{\label{fig:Posterior_of_parameters_1} Posterior distributions of the parameters $(a,b,c)$ inferred from data with $1\%$ of statistical uncertainties.}
\end{figure}

First, the posterior distribution of the parameters with $15\%$ of experimental data uncertainty is shown. In Fig.~\ref{fig:Posterior_of_parameters_15}, the diagonal plots are the posterior distribution of each of the parameters. The off-diagonal plots display the joint distributions between pairs of the parameters. The truth values are indicated by black lines, the center of the posterior distribution are indicated by red lines and the values that maximize the posterior distribution (also referred as the maximum a posteriori value) are labeled by the blue lines. In this case, only $b$ is constrained while the MAP estimation is still off. Next, the result with $5\%$ of data uncertainty is shown in Fig.~\ref{fig:Posterior_of_parameters_5}. $b$ is well constrained by the data and the prediction of $a$ and $c$ are much closer to their truth values. Lastly, the posterior distribution inferred from data with only $1\%$ of uncertainty is shown in Fig.~\ref{fig:Posterior_of_parameters_1}. In this case, both $a$ and $b$ are constrained very well. However, parameter $c$ shows some discrepancy between the truth value and the two estimations. 

\begin{figure}
	\centering
	\includegraphics[width=0.8\textwidth]{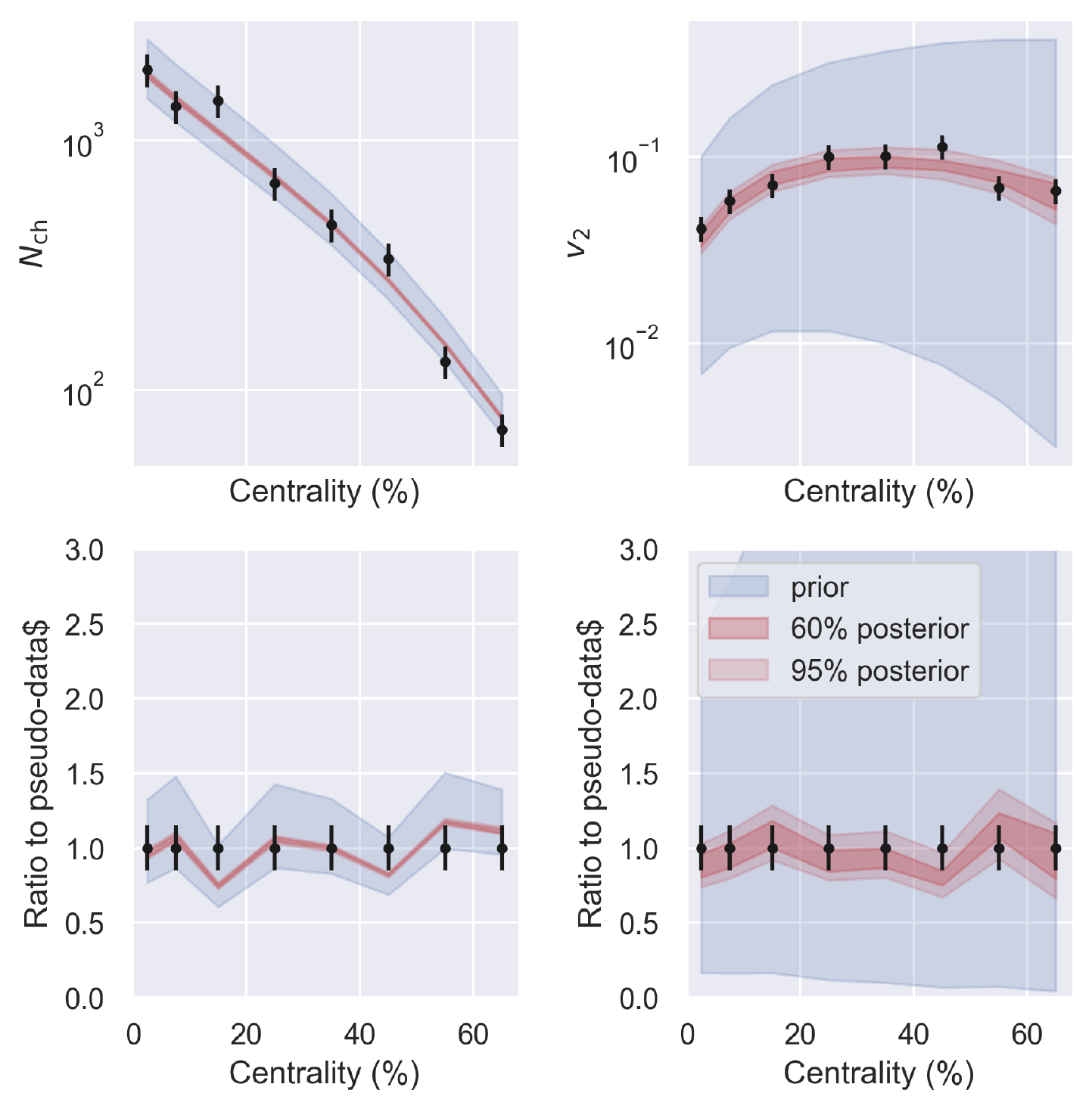}
	\caption[Comparison between data with $15\%$ uncertainty and emulator prediction with $95\%$ CL]{\label{fig:Posterior_validation_15} \textbf{Top row}: Comparison between data with $15\%$ uncertainty and emulator prediction with $95\%$ CL.  \textbf{Bottom row}: Ratio between emulator prediction and data.}
\end{figure}

Even though we failed to constrain the parameters for the case with $15\%$ of data uncertainty, the emulator prediction still matches the data pretty well (see Fig.~\ref{fig:Posterior_validation_15}).

\begin{figure}
     \centering
         \includegraphics[width=0.45\textwidth]{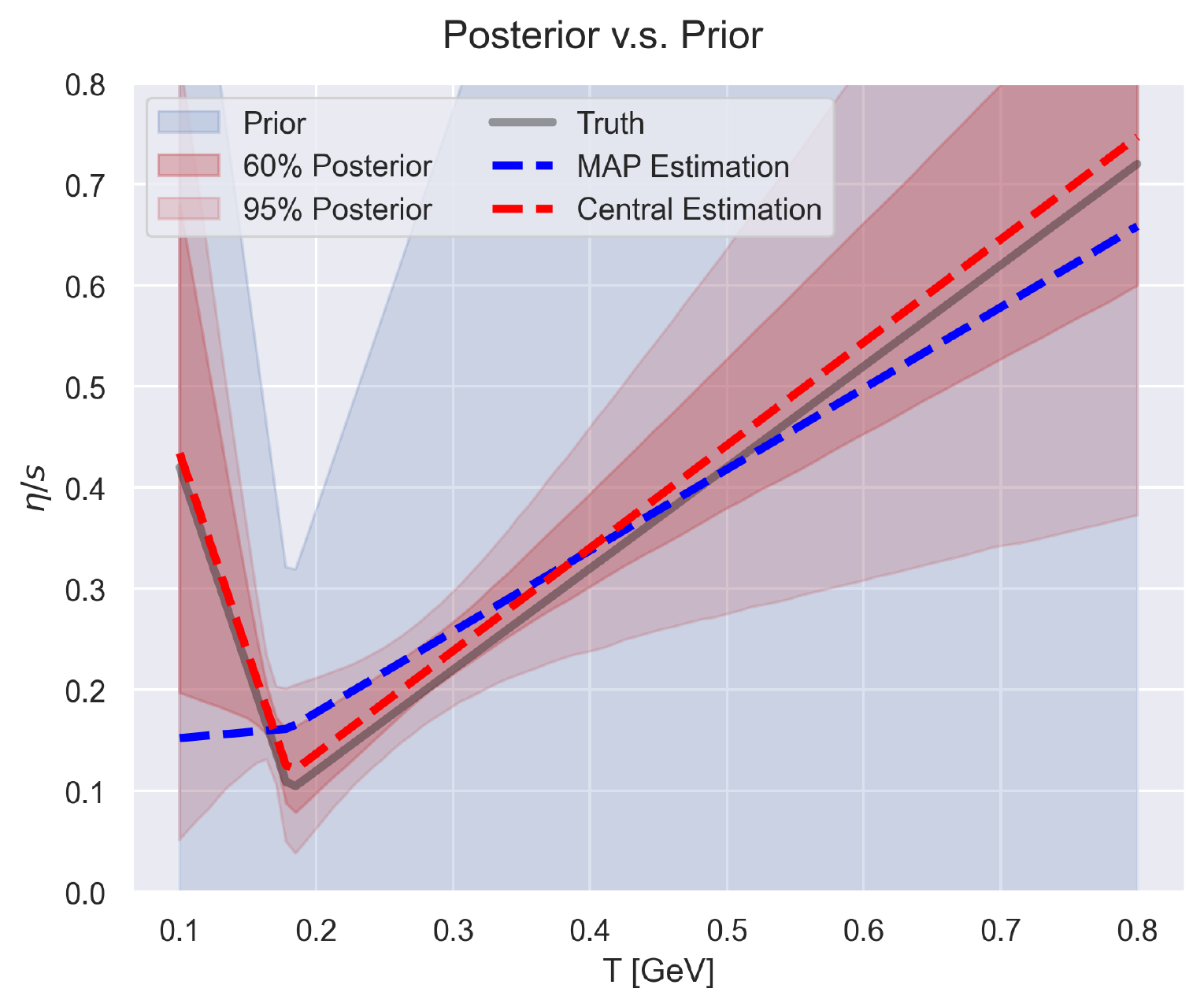}
         \includegraphics[width=0.45\textwidth]{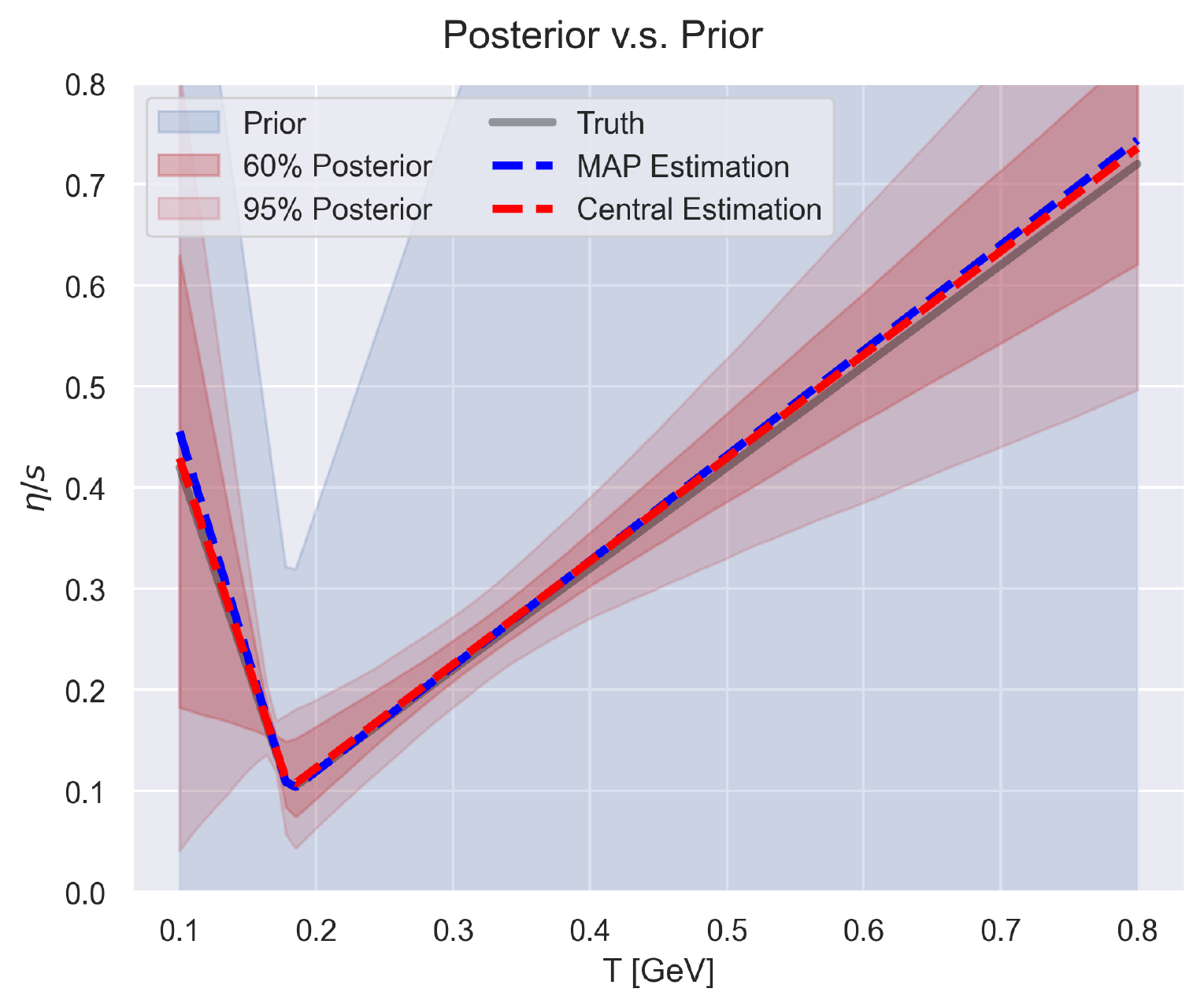}
         \includegraphics[width=0.45\textwidth]{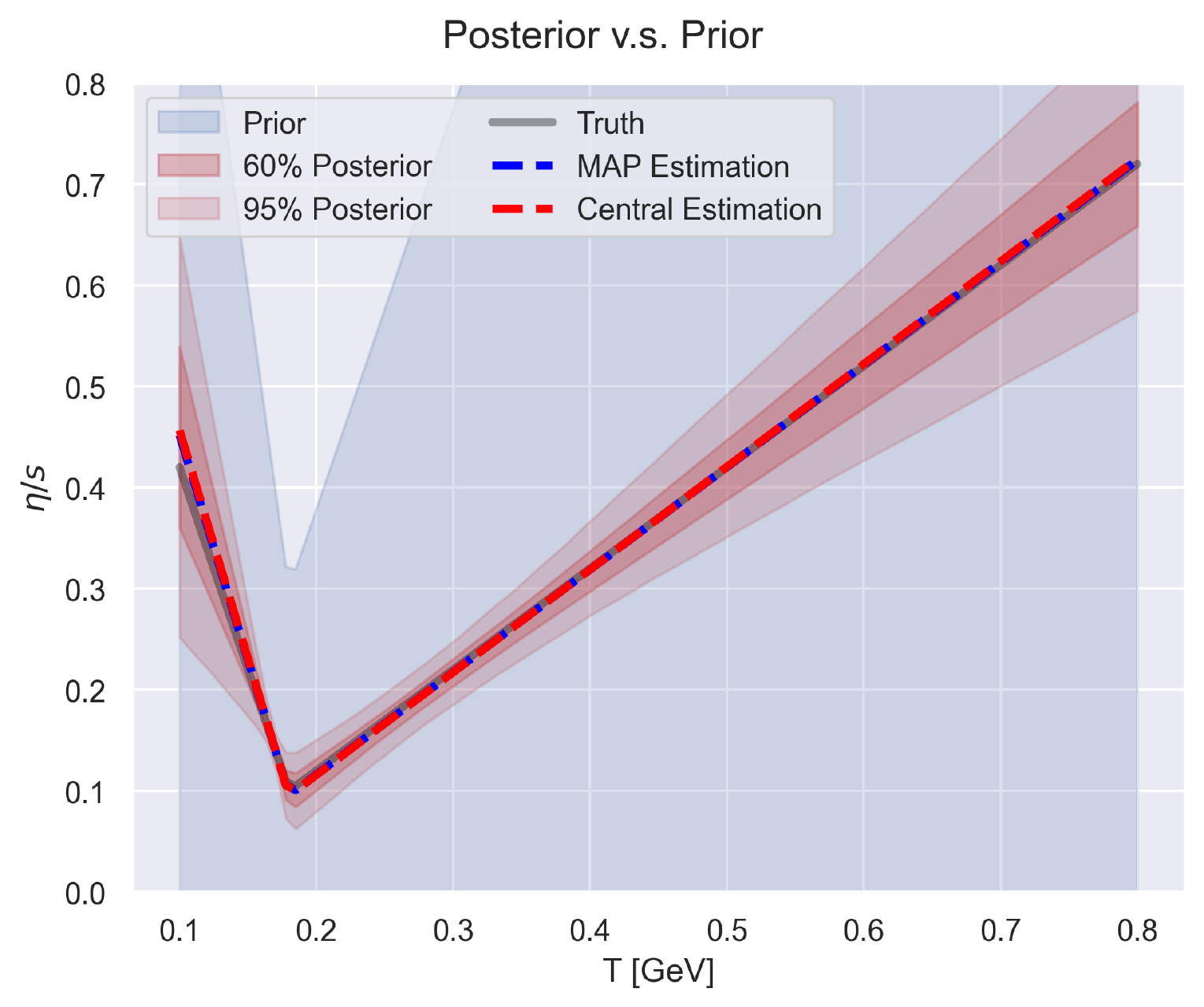}
        \caption[Comparison between the $95\%$ CL posterior distribution, the central estimation, the MAP estimation and the truth value of $\eta/s(T)$]{Comparison between the $95\%$ CL posterior distribution, the central estimation, the MAP estimation and the truth value of $\eta/s(T)$. \textbf{Top left}: Inferred from data with $15\%$ uncertainty. \textbf{Top right}: Inferred from data with $5\%$ uncertainty. \textbf{Bottom}: Inferred from data with $1\%$ uncertainty.}
        \label{fig:Posterior_eta_s_all_three}
\end{figure}

The posterior distribution of $\eta/s(T)$ is what people are really interested in. One can see that as the experimental uncertainty reduces, the posterior estimation gradually recovers the truth value (see Fig.~\ref{fig:Posterior_eta_s_all_three}). 

As seen from Fig.~\ref{fig:Posterior_of_parameters_15}, \ref{fig:Posterior_of_parameters_5}, \ref{fig:Posterior_of_parameters_1} and \ref{fig:Posterior_eta_s_all_three}, parameter $b$ is always the most constrained among the three parameters as varying this parameter will affect the slope of $\eta/s(T)$ curve above $T_s$. Parameter $a$ is also constrained for the last two cases because varying $a$ will shift the entire $\eta/s(T)$ curve up and down. $c$ is least constrained since the temperature range it affects ($T_s-T_{min}=0.05$ GeV) is much smaller than the temperature range $a$ affects ($\approx 0.5$ GeV). 

\subsection{Inference with different levels of emulator uncertainty}

\begin{figure}
	\centering
	\includegraphics[width=0.96\textwidth]{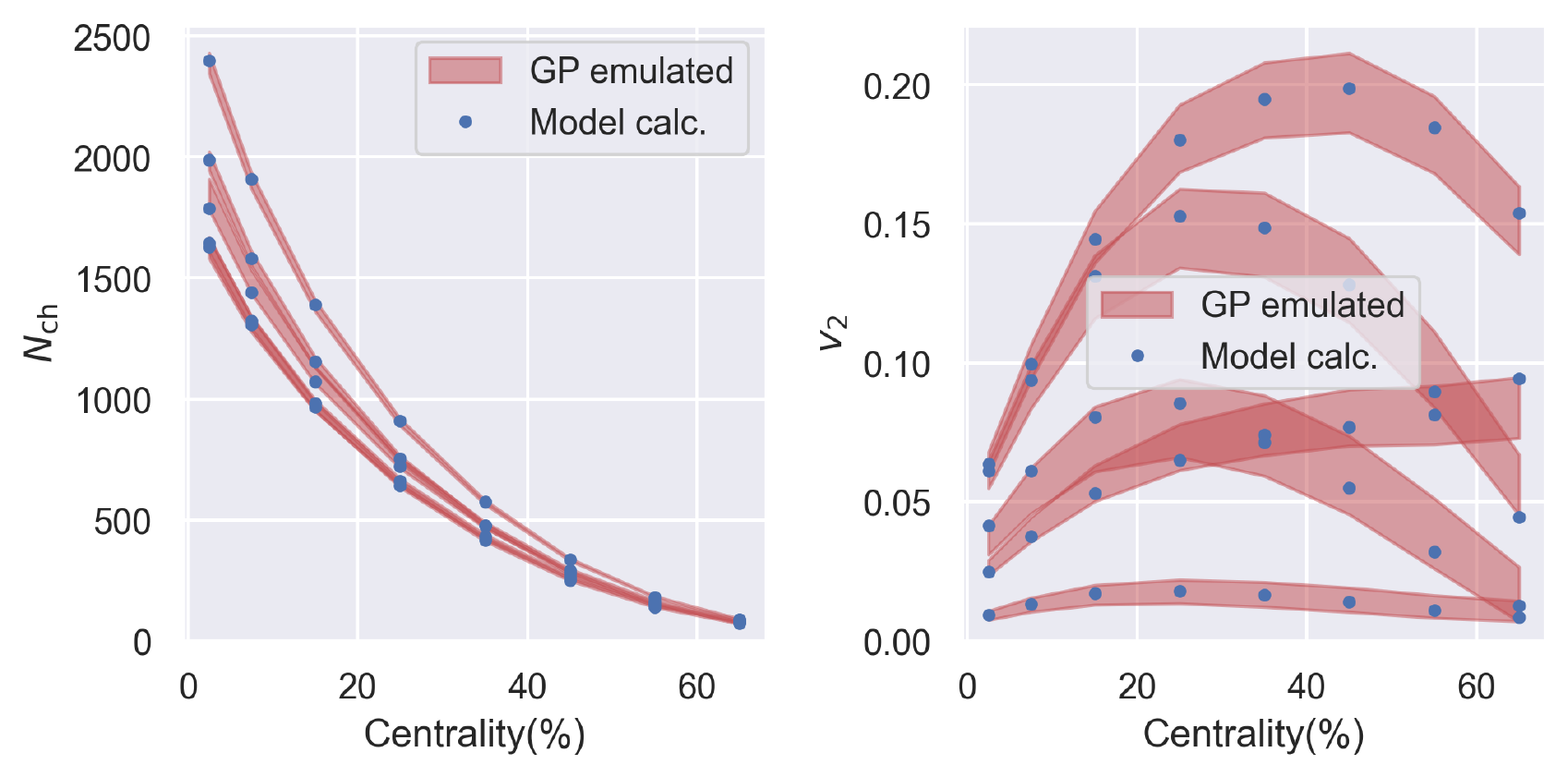}
	\caption[Comparison between model calculation and emulator prediction at 5 randomly sampled parameter points]{\label{fig:Emulator_validation_1_bad_gp} Comparison between model calculation and emulator prediction at 5 randomly sampled parameter points. The band of the GP represents $95\%$ CL. The emulator only uses 20 design points and 3 principal components. }
\end{figure}

\begin{figure}
	\centering
	\includegraphics[width=0.9\textwidth]{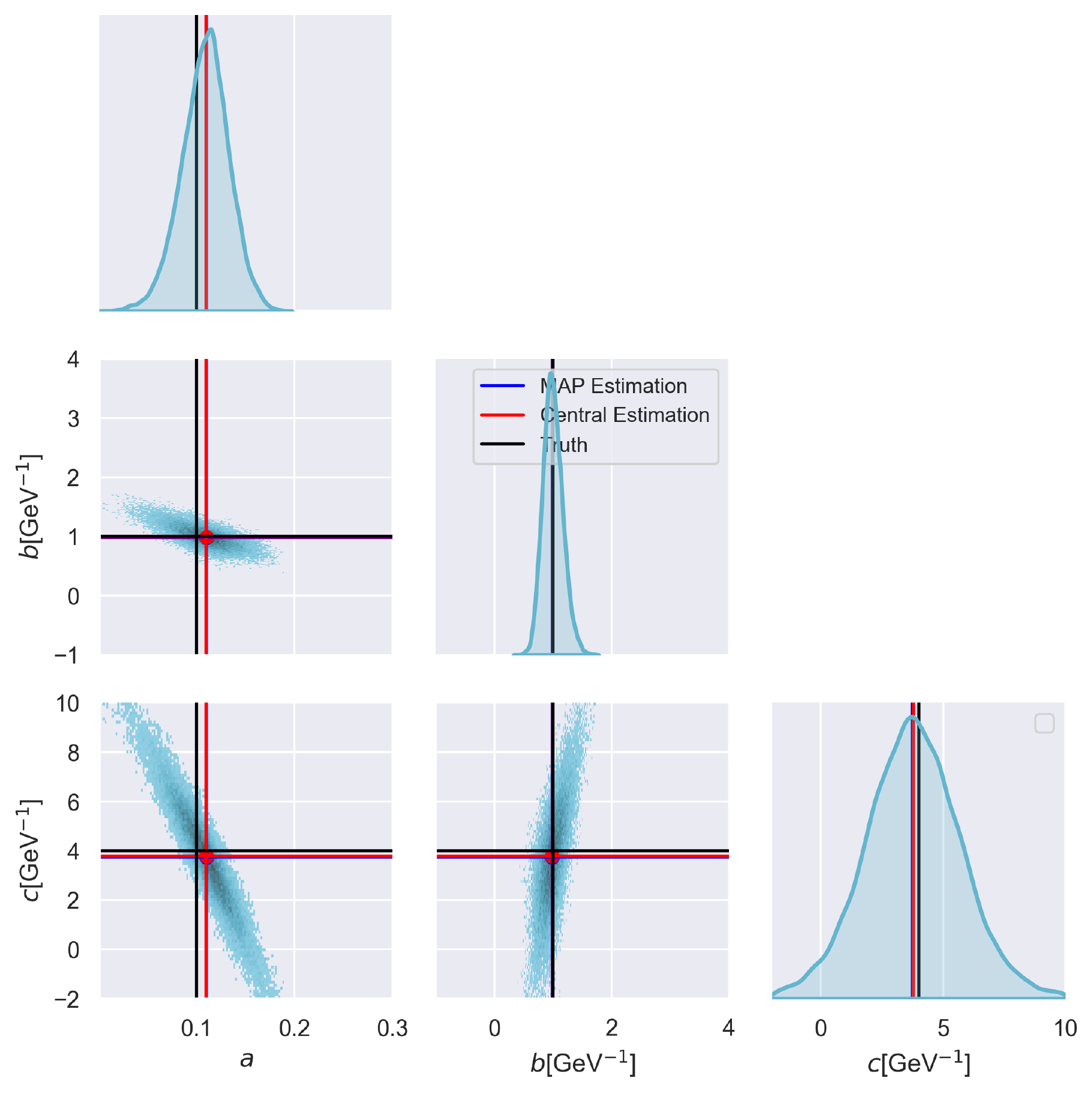}
	\caption[Posterior distributions of the parameters $(a,b,c)$ inferred from data with $1\%$ of statistical uncertainties]{\label{fig:Posterior_of_parameters_1_bad_gp_2} Posterior distributions of the parameters $(a,b,c)$ inferred from data with $1\%$ of statistical uncertainties. The Gaussian process emulator is only using 50 design points and 5 principal components in this case.}
\end{figure}

\begin{figure}
	\centering
	\includegraphics[width=0.9\textwidth]{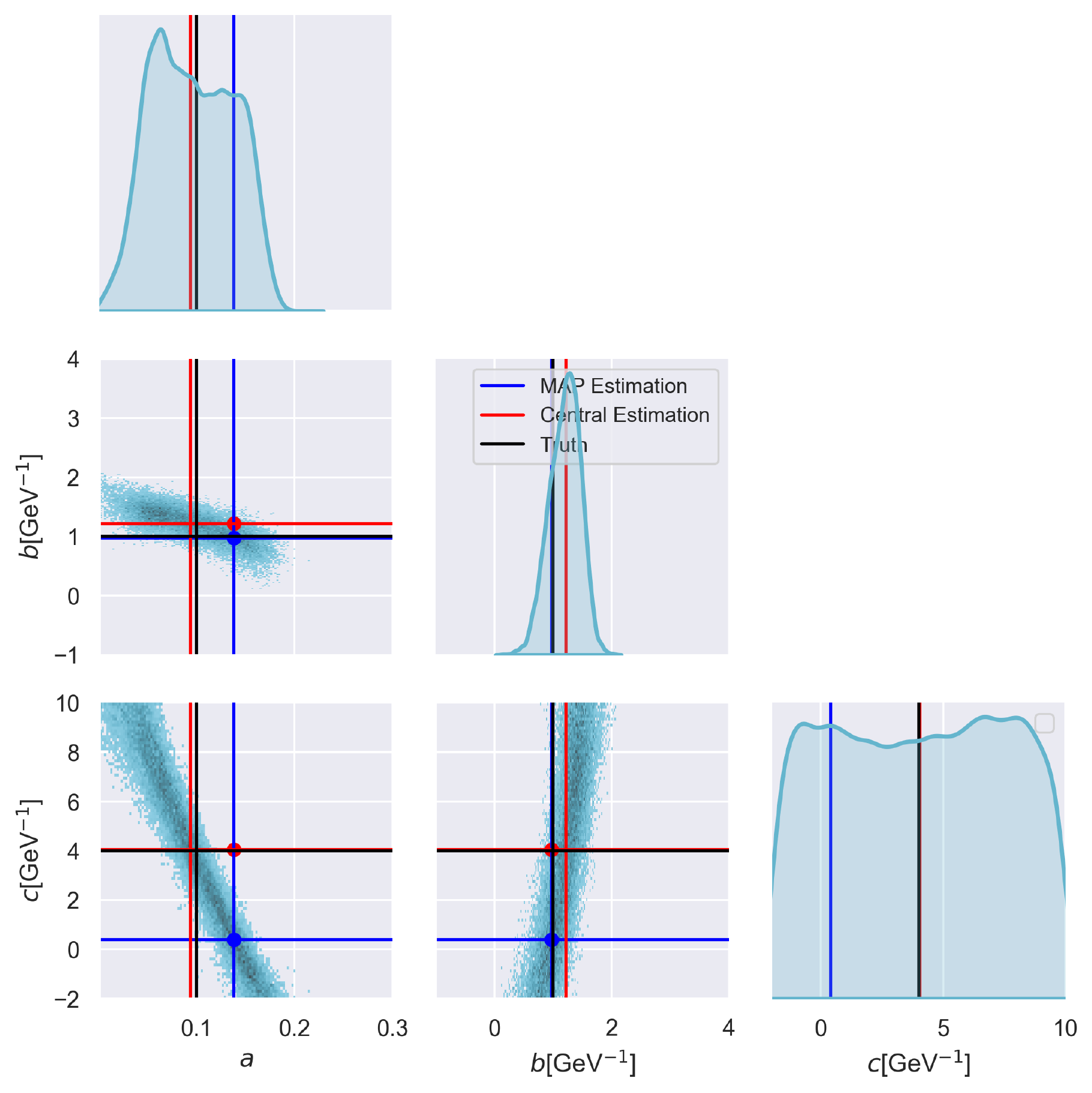}
	\caption[Posterior distributions of the parameters $(a,b,c)$ inferred from data with $1\%$ of statistical uncertainties]{\label{fig:Posterior_of_parameters_1_bad_gp} Posterior distributions of the parameters $(a,b,c)$ inferred from data with $1\%$ of statistical uncertainties. The Gaussian process emulator is only using 20 design points and 3 principal components in this case.}
\end{figure}

\begin{figure}
	\centering
	\includegraphics[width=0.8\textwidth]{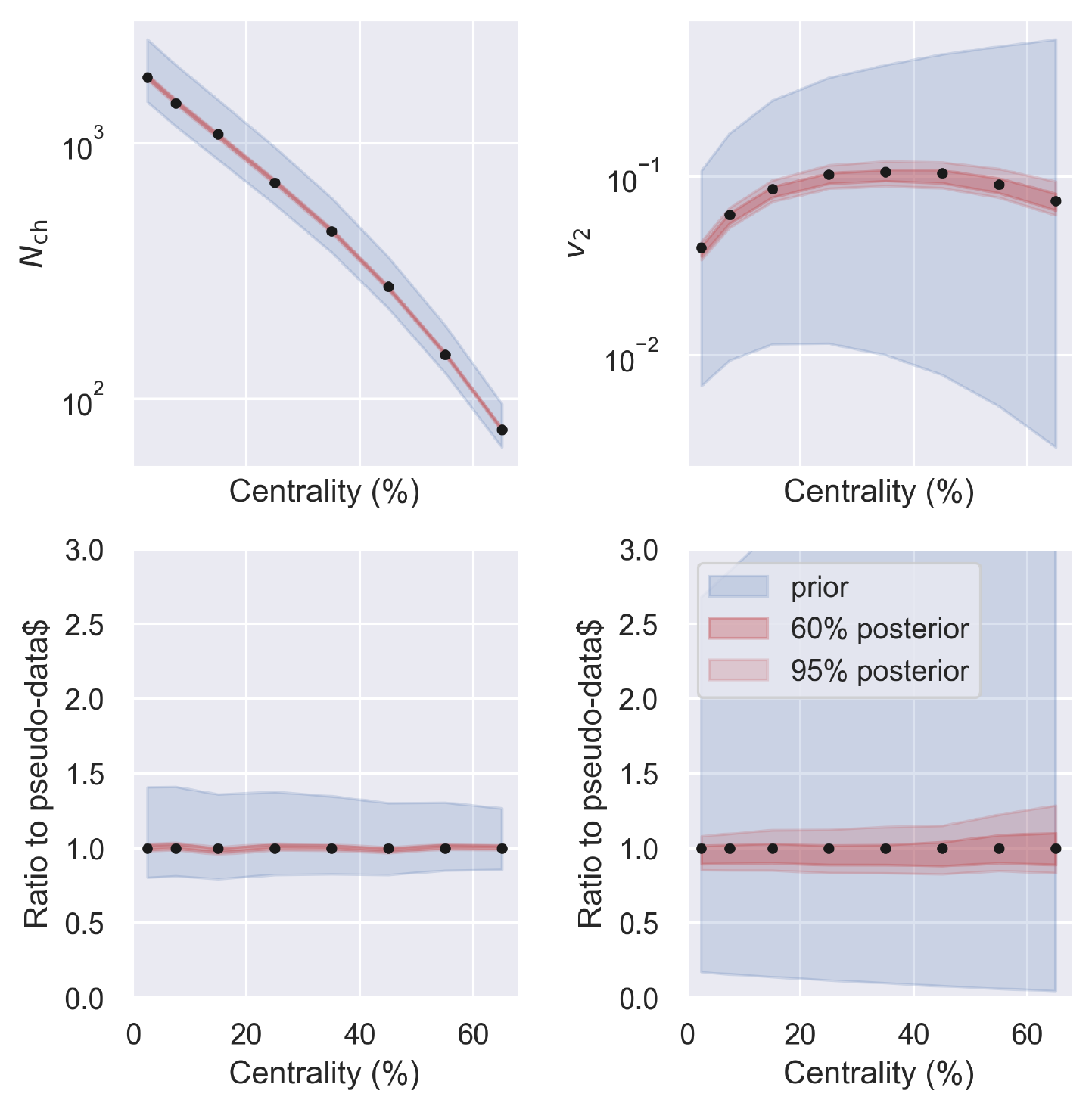}
	\caption[Comparison between data with $1\%$ uncertainty and emulator prediction with $95\%$ CL]{\label{fig:Posterior_validation_bad_gp} \textbf{Top row}: Comparison between data with $1\%$ uncertainty and emulator prediction with $95\%$ CL. The emulator is just using 20 design points and 3 principal components. \textbf{Bottom row}: Ratio between emulator prediction and data.}
\end{figure}

Next, I want to check the performance of Bayesian analysis when the uncertainties of the Gaussian process emulator are much larger than the experimental data uncertainties. The data uncertainty is fixed to be $1\%$ from now on. First, I only use $20$ design points to train the emulator and pick just the first $3$ of the principal components (as opposed to $1000$ and $5$ in the previous setup). While it is quite remarkable that Gaussian process emulator can still mimic the model (see Fig.~\ref{fig:Emulator_validation_1_bad_gp}), we are no longer able to constrain the parameters as shown in Fig.~\ref{fig:Posterior_of_parameters_1_bad_gp}. However, the emulator is still able to predict the data (see Fig.~\ref{fig:Posterior_validation_bad_gp}). If we increase the number of design points to $50$ and pick the first $5$ principal components, the performance is surprisingly almost the same as the setup in the previous section (see Fig.~\ref{fig:Posterior_of_parameters_1_bad_gp_2} and Fig.~\ref{fig:Posterior_validation_bad_gp}).  

\begin{figure}
     \centering
         \includegraphics[width=0.45\textwidth]{images/Posterior_of_eta_s_1.pdf}
         \includegraphics[width=0.45\textwidth]{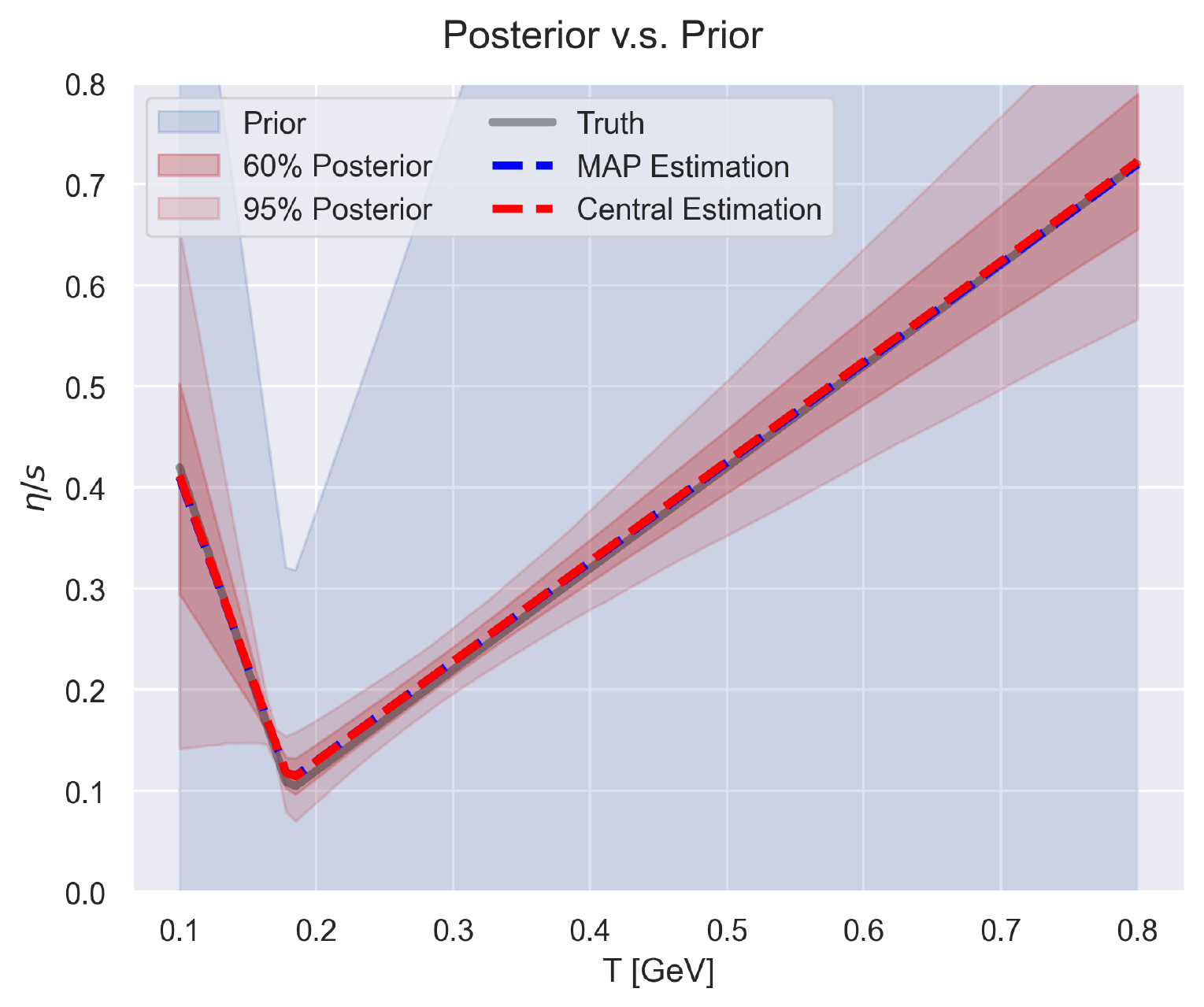}
         \includegraphics[width=0.45\textwidth]{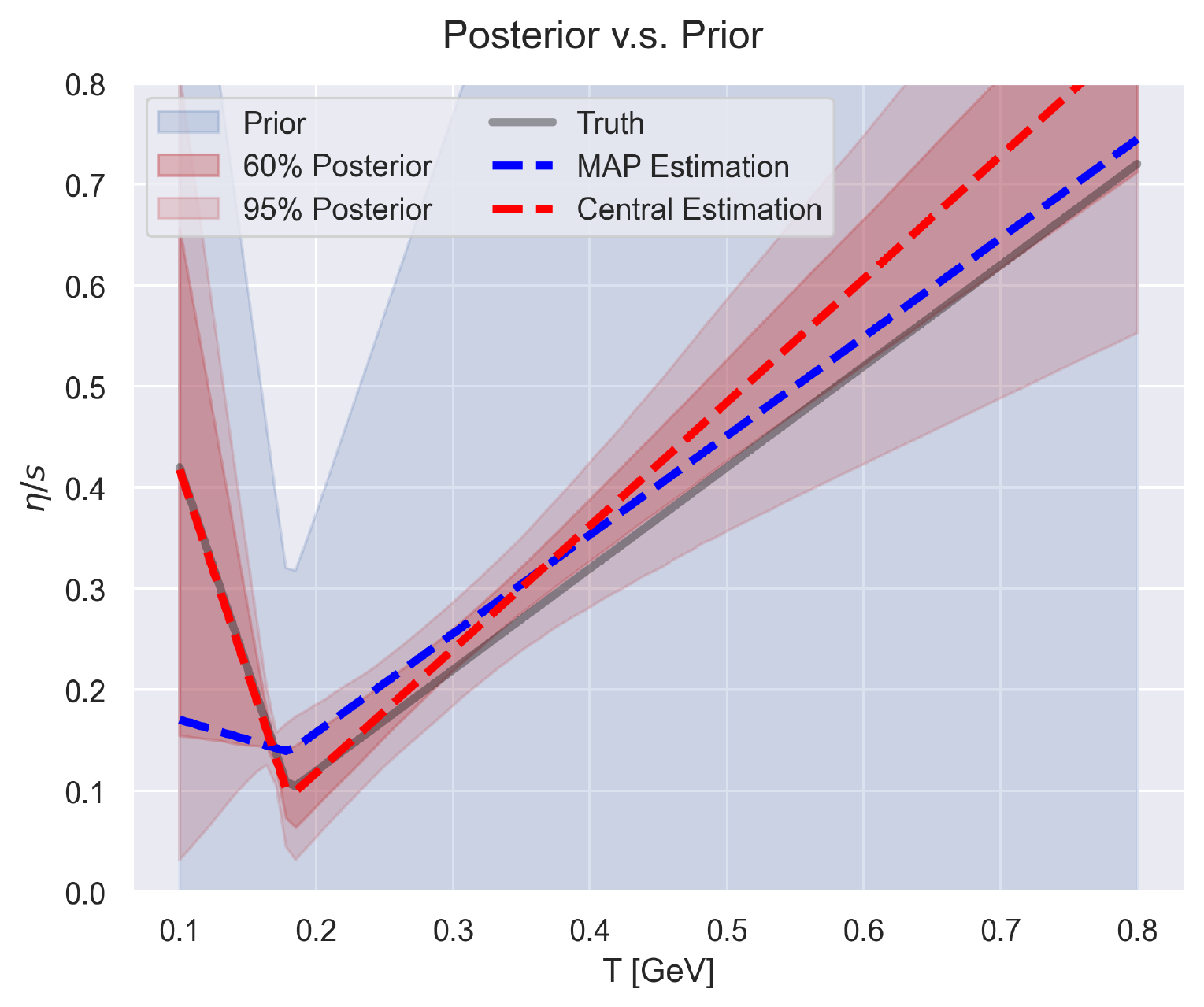}
        \caption[Comparison between the $95\%$ CL posterior distribution, the central estimation, the MAP estimation and the truth value of $\eta/s(T)$]{Comparison between the $95\%$ CL posterior distribution, the central estimation, the MAP estimation and the truth value of $\eta/s(T)$. \textbf{Top Left}: Inferred from data with $1\%$ uncertainty and Gaussian process emulators with minimal uncertainty (1000 design points and 5 principal components). \textbf{Top Right}: Inferred from data with $1\%$ uncertainty and Gaussian process emulators with only 50 design points and 5 principal components. \textbf{Bottom}: Inferred from data with $1\%$ uncertainty and Gaussian process emulators with only 20 design points and 3 principal components.}
        \label{fig:Posterior_eta_s_bad_gp}
\end{figure}

\subsection{Inference with systematic model uncertainty}

\begin{figure}
	\centering
	\includegraphics[width=0.98\textwidth]{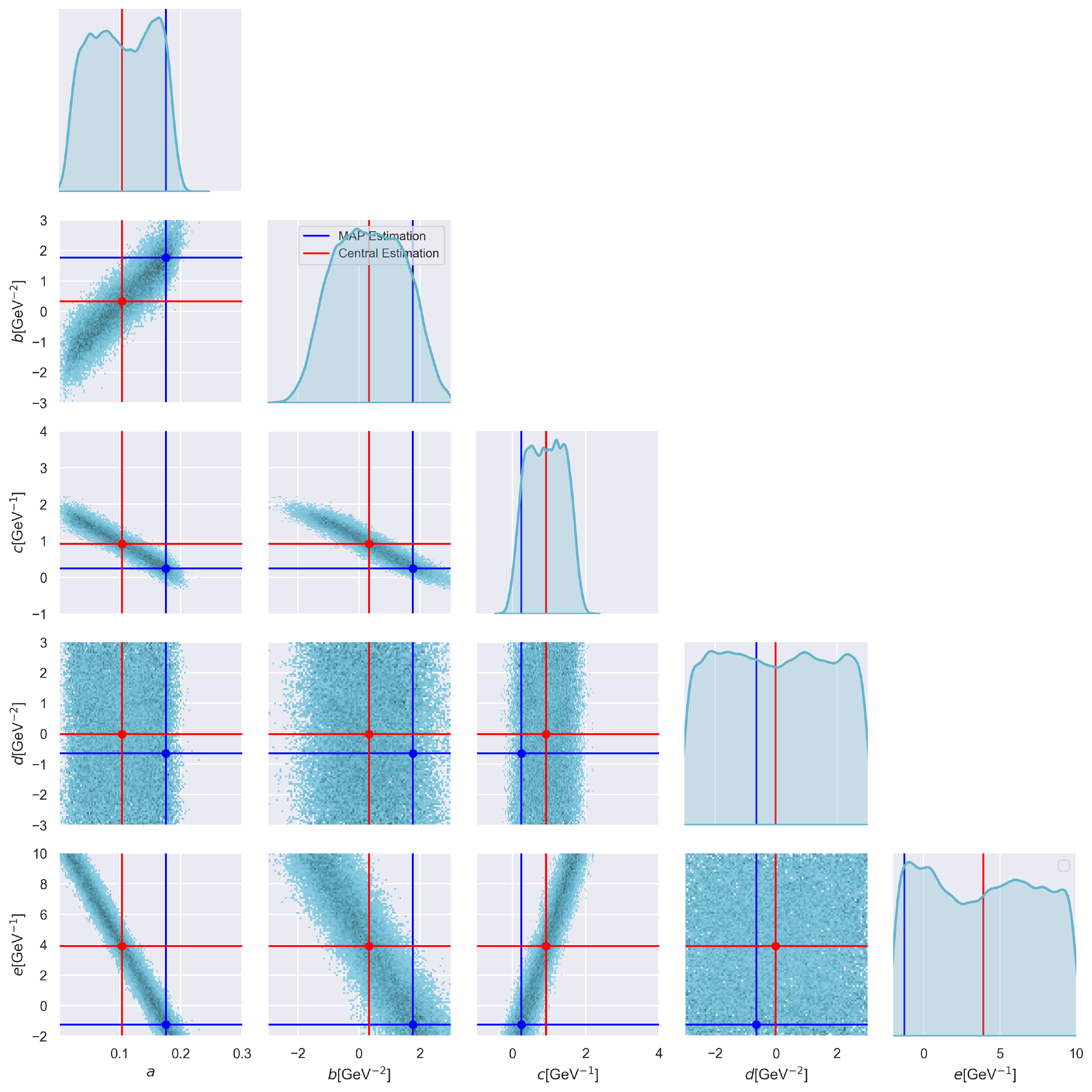}
	\caption[Posterior distributions of the parameters $(a,b,c)$ inferred from data with $1\%$ of statistical uncertainties]{\label{fig:Posterior_of_parameters_1_bad_model} Posterior distributions of the parameters $(a,b,c)$ inferred from data with $1\%$ of statistical uncertainties. The data are generated with the linear parameterization of $\eta/s(T)$ while the model is assuming a quadratic form.}
\end{figure}

\begin{figure}
     \centering
         \includegraphics[width=0.45\textwidth]{images/Posterior_of_eta_s_1.pdf}
         \includegraphics[width=0.45\textwidth]{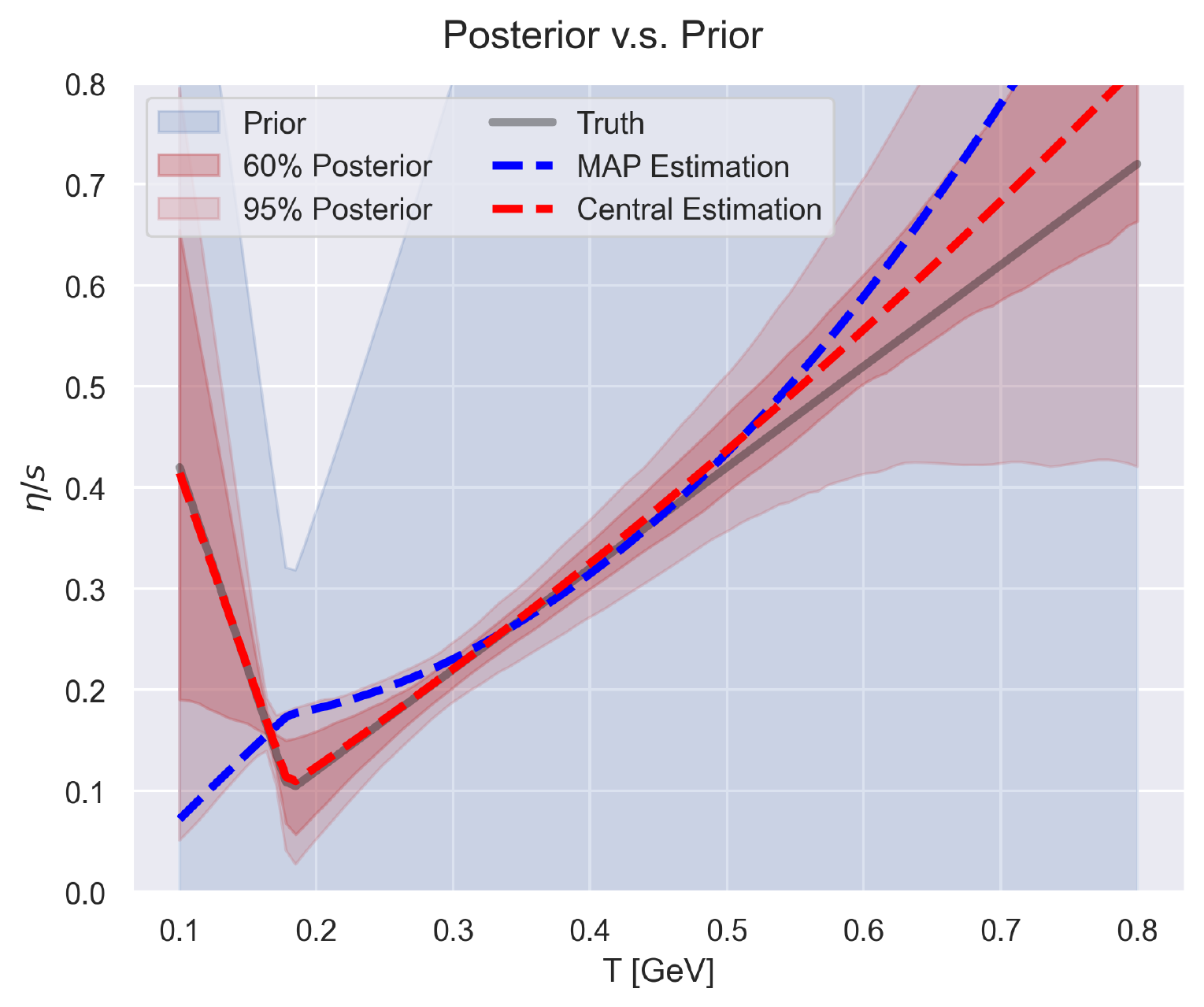}
        \caption[Comparison between the $95\%$ CL posterior distribution, the central estimation, the MAP estimation and the truth value of $\eta/s(T)$]{Comparison between the $95\%$ CL posterior distribution, the central estimation, the MAP estimation and the truth value of $\eta/s(T)$. \textbf{Left}: Inferred from data with $1\%$ uncertainty and Gaussian process emulators with minimal uncertainty. \textbf{Right}: Inferred from data generated with a linear form and Gaussian process emulators trained with a quadratic form.}
        \label{fig:Posterior_eta_s_bad_model}
\end{figure}

\begin{figure}
	\centering
	\includegraphics[width=0.8\textwidth]{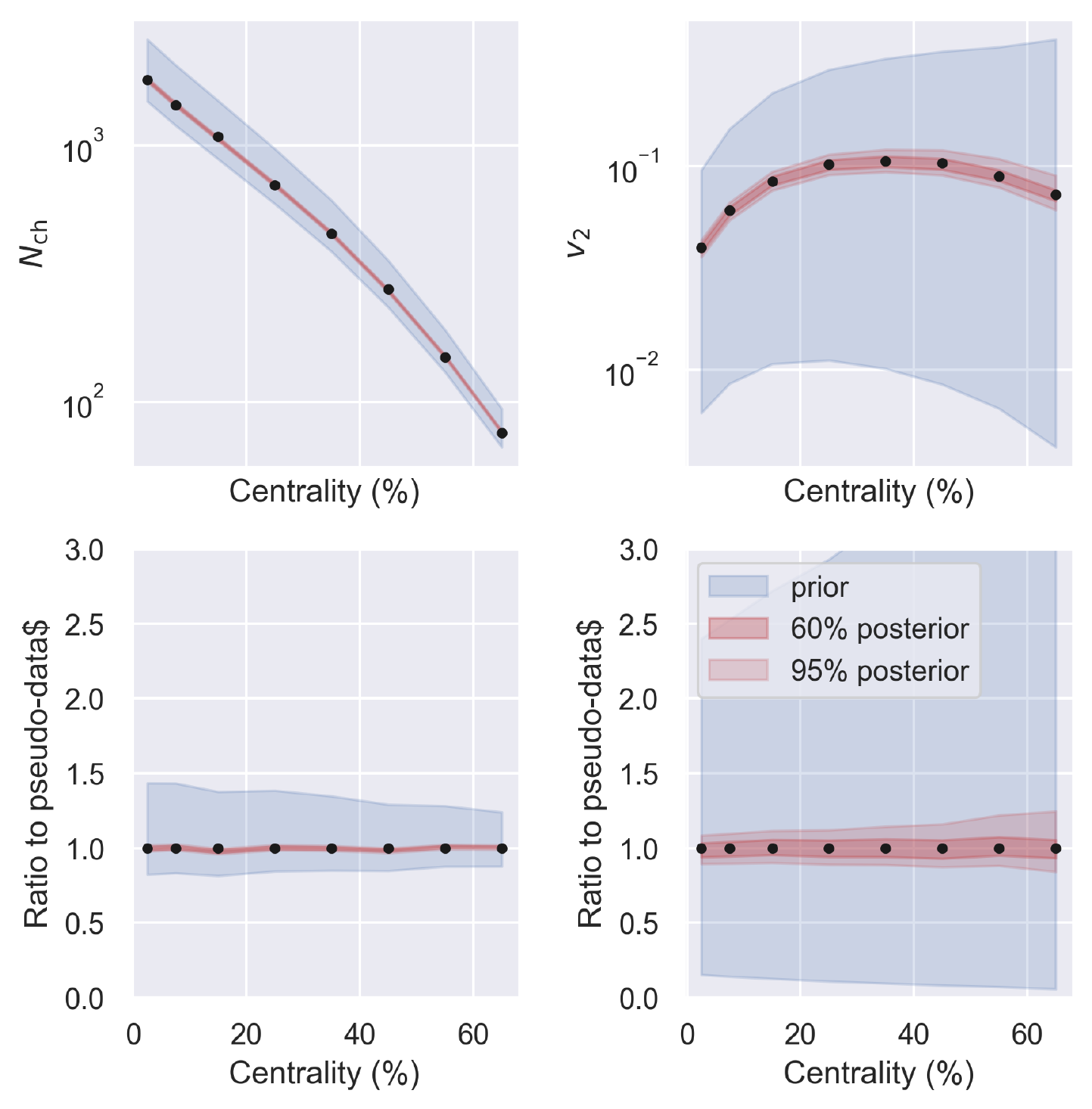}
	\caption[Comparison between data with $1\%$ uncertainty and emulator prediction with $95\%$ CL]{\label{fig:Posterior_validation_bad_model} \textbf{Top row}: Comparison between data with $1\%$ uncertainty and emulator prediction with $95\%$ CL. The emulator is trained with a different model. \textbf{Bottom row}: Ratio between emulator prediction and data.}
\end{figure}

\begin{figure}
	\centering
	\includegraphics[width=0.9\textwidth]{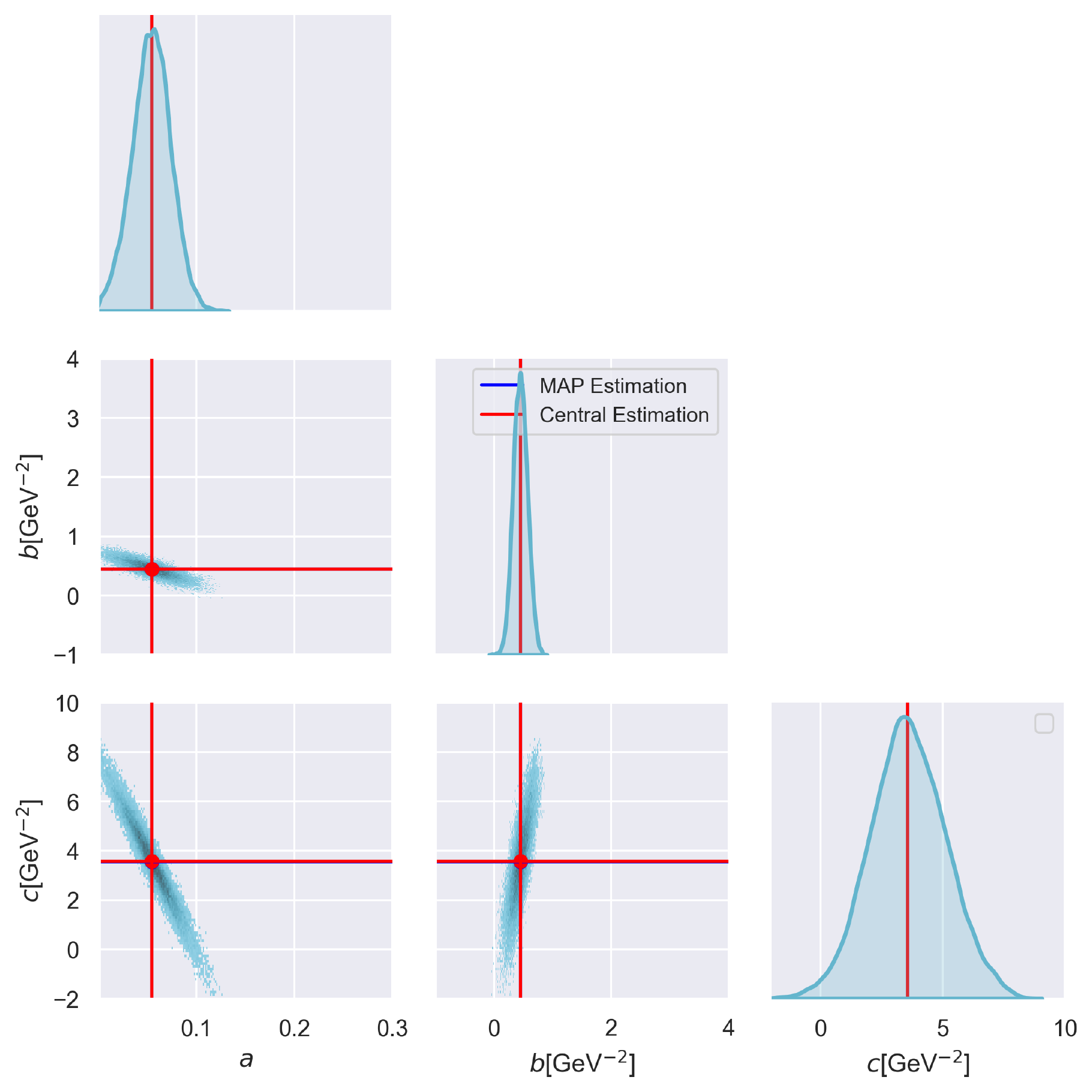}
	\caption[Posterior distributions of the parameters $(a,b,c)$ inferred from data with $1\%$ of statistical uncertainties]{\label{fig:Posterior_of_parameters_1_bad_model_2} Posterior distributions of the parameters $(a,b,c)$ inferred from data with $1\%$ of statistical uncertainties. The data are generated with the quadratic parameterization of $\eta/s(T)$ while the model is assuming a linear form.}
\end{figure}

\begin{figure}
     \centering
         \includegraphics[width=0.45\textwidth]{images/Posterior_of_eta_s_1.pdf}
         \includegraphics[width=0.45\textwidth]{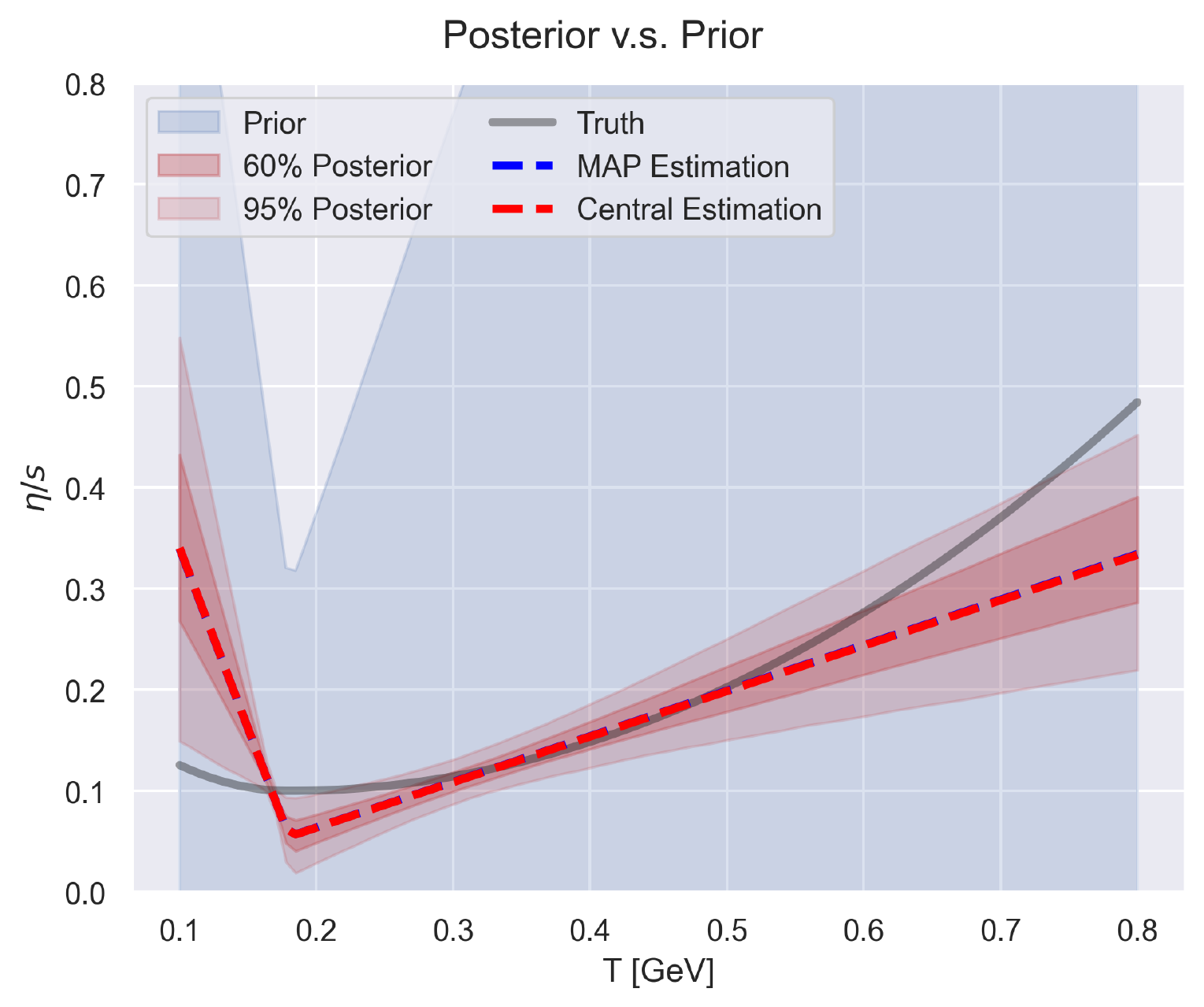}
        \caption[Comparison between the $95\%$ CL posterior distribution, the central estimation, the MAP estimation and the truth value of $\eta/s(T)$]{Comparison between the $95\%$ CL posterior distribution, the central estimation, the MAP estimation and the truth value of $\eta/s(T)$. \textbf{Left}: Inferred from data with $1\%$ uncertainty and Gaussian process emulators with minimal uncertainty. \textbf{Right}: Inferred from data generated with a quadratic form and Gaussian process emulators trained with a linear form.}
        \label{fig:Posterior_eta_s_bad_model_2}
\end{figure}

\begin{figure}
	\centering
	\includegraphics[width=0.8\textwidth]{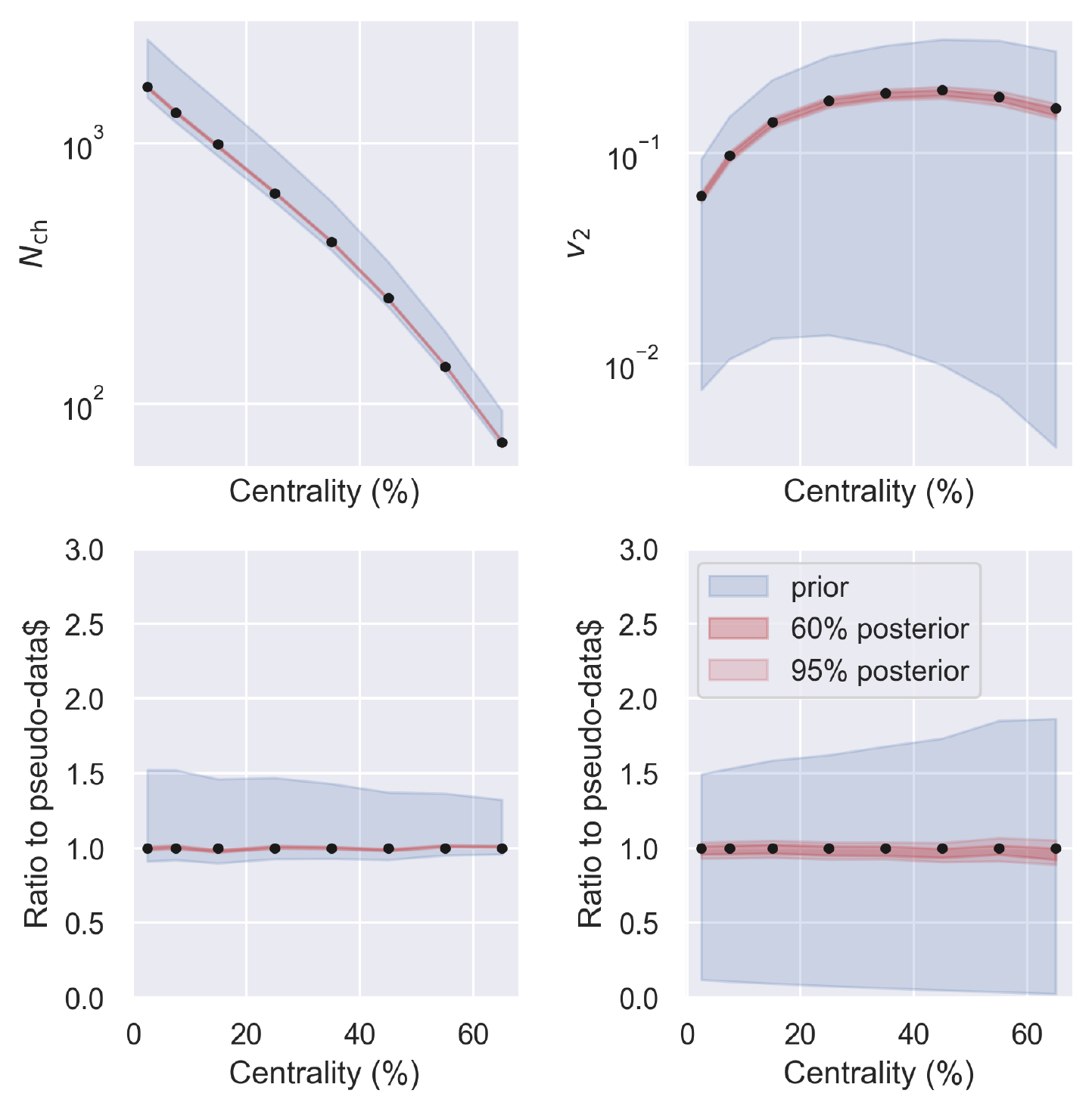}
	\caption[Comparison between data with $1\%$ uncertainty and emulator prediction with $95\%$ CL]{\label{fig:Posterior_validation_bad_model_2} \textbf{Top row}: Comparison between data with $1\%$ uncertainty and emulator prediction with $95\%$ CL. The emulator is trained with a different model. \textbf{Bottom row}: Ratio between emulator prediction and data.}
\end{figure}

In the previous two sections, I explored the scenarios with large data uncertainty or emulator uncertainty. The model, on the other hand, is used to generate the mock experimental data. In reality, our current modeling of nature is most likely incomplete or not precise, which translates to systematic model uncertainty. In this section, I introduce a different parameterization of $\eta/s(T)$ by changing its linear form into a quadratic form:

\begin{equation}
    \frac{\eta}{s}(T) = a + 
\begin{cases} 
b (T - T_s)^2+c (T-T_s), & T>T_s\\
d (T_s - T)^2+e(T_s - T), & T<T_s\\
\end{cases}
\end{equation}

The mock data are still generated using the linear parameterization of $\eta/s(T)$ and with $1\%$ uncertainty. There are no truth values in this case, as we are using different models for training the emulator and generating the data. The quadratic parameterization will fall back to the linear one if we set the two second-order coefficients to zero. And this is what we see in the posterior distribution (see Fig.~\ref{fig:Posterior_of_parameters_1_bad_model}). The central values for the second-order coefficients are close to $0$, which is what the data suggests. The reason that we are not able to constrain the two second-order coefficients is probably due to their small effect on the final observables. Since $|T-T_s|<0.5$, the second-order term has a smaller impact on $\eta/s_{eff}$ compared to the linear term. 

It is, however, still meaningful to compare the posterior distribution of $\eta/s(T)$ (see Fig.~\ref{fig:Posterior_eta_s_bad_model}). We can see that by introducing two extra parameters, the posterior distribution with $95\%$ CL is much wider at low temperature and at higher temperature since we are constraining the two second order coefficients. There is no systematic model uncertainty in this case as the parameterization in the model can reduce to the one used for generating the data. But still much constraining power is lost.

The other direction can be explored as well, meaning the linear parameterization of $\eta/s$ is used in the model, but the data are generated with a quadratic parameterization. The posterior distribution is shown in Fig.~\ref{fig:Posterior_of_parameters_1_bad_model_2} and Fig.~\ref{fig:Posterior_eta_s_bad_model_2}. The parameters seem to be well constrained although the parameterization used in the model is inherently wrong. The model with the linear parameterization is still able to describe the experiment data with very little statistical uncertainty (see Fig.~\ref{fig:Posterior_validation_bad_model_2}). Because it is really difficult to constrain $\eta/s(T)$ if only integrated quantities over $\eta/s(T)$ are observed. If one probes even higher temperatures or more differential observables, maybe the systematic bias can show up in Fig.~\ref{fig:Posterior_validation_bad_model_2}.

\section{Summary}

This Chapter laid out the fundamentals for performing Bayesian analysis. We utilize techniques like Latin hypercube sampling, Gaussian process emulator, and principal component analysis to be able to quickly predict model calculations at arbitrary parameter points. Markov Chain Monte Carlo is then used to draw the posterior distributions of the parameters. 

I then performed Bayesian analysis using a simple model for the bulk medium evolution with different scenarios for the covariance matrix and the $\eta/s(T)$ parameterization. The lessons learned from this exploration are the following:

\begin{enumerate}
    \item Reducing experimental data and emulator uncertainty can improve the constraining power. Even if the model has no systematic uncertainty, one may be unable to constrain the parameters if too much experimental or emulator uncertainty is introduced. The improvement of experimental uncertainty may require years in the field of heavy ion collision. The emulator uncertainty can be reduced by using more design points which also requires more computation time (by the scale of millions of CPU hours for our current project). Nevertheless, even with significant data uncertainty or emulator uncertainty, we are still able to recover the truth values of the parameters within our posterior distribution, meaning we are gaining knowledge about the truth values of the parameters by doing Bayesian analysis. 
    
    \item Some parameters may be difficult to constrain if their impact on the final observables is small. In the model we have used, the slope for the lower part of the $\eta/s(T)$ curve (parameter $c$) is very challenging to constrain because it only governs a small range of temperature, and the observables we looked at are both integrated over $T$. If we want to constrain it, we should look at more sensitive observables (i.e., differential observables measured at low temperatures). In practice, we should include as many as experimental observables in our calibration to help constrain the parameters. 
    
    \item Using a more complicated parameterization will likely reduce the constraining power while using a parameterization with insufficient flexibility will introduce systematic model uncertainty. It is essential to assume a reasonable parameterization with our prior knowledge about the underlying physics. A very good example is for the \trento model where we know that the density deposition function $f(T_A, T_B)$ is scale invariant ($f(cT_A, cT_B)=cf(T_A, T_B)$). It is therefore natural to assume that the parameterization to be a generalized mean $f(T_A, T_B)=(\frac{T_A^p+T_B^p}{2})^{1/p}$\cite{Moreland:2014oya}. If the form of the function is completely unknown, we can assume it to be in the form of a Taylor expansion or Gaussian random field \cite{xie2022information}.
\end{enumerate}

\chapter{Bayesian Model-to-Data Comparison Results} \label{sec:bayesian_results}

\vspace{1in}

In this chapter, I will apply Bayesian analysis along with other techniques discussed in Chapter~\ref{sec:bayesian} to estimate the various parameters explored in Chapter~\ref{sec:jetscape_results}. 

\section{Previous energy loss studies using Bayesian analysis}

\begin{figure}
	\centering
	\includegraphics[width=0.96\textwidth]{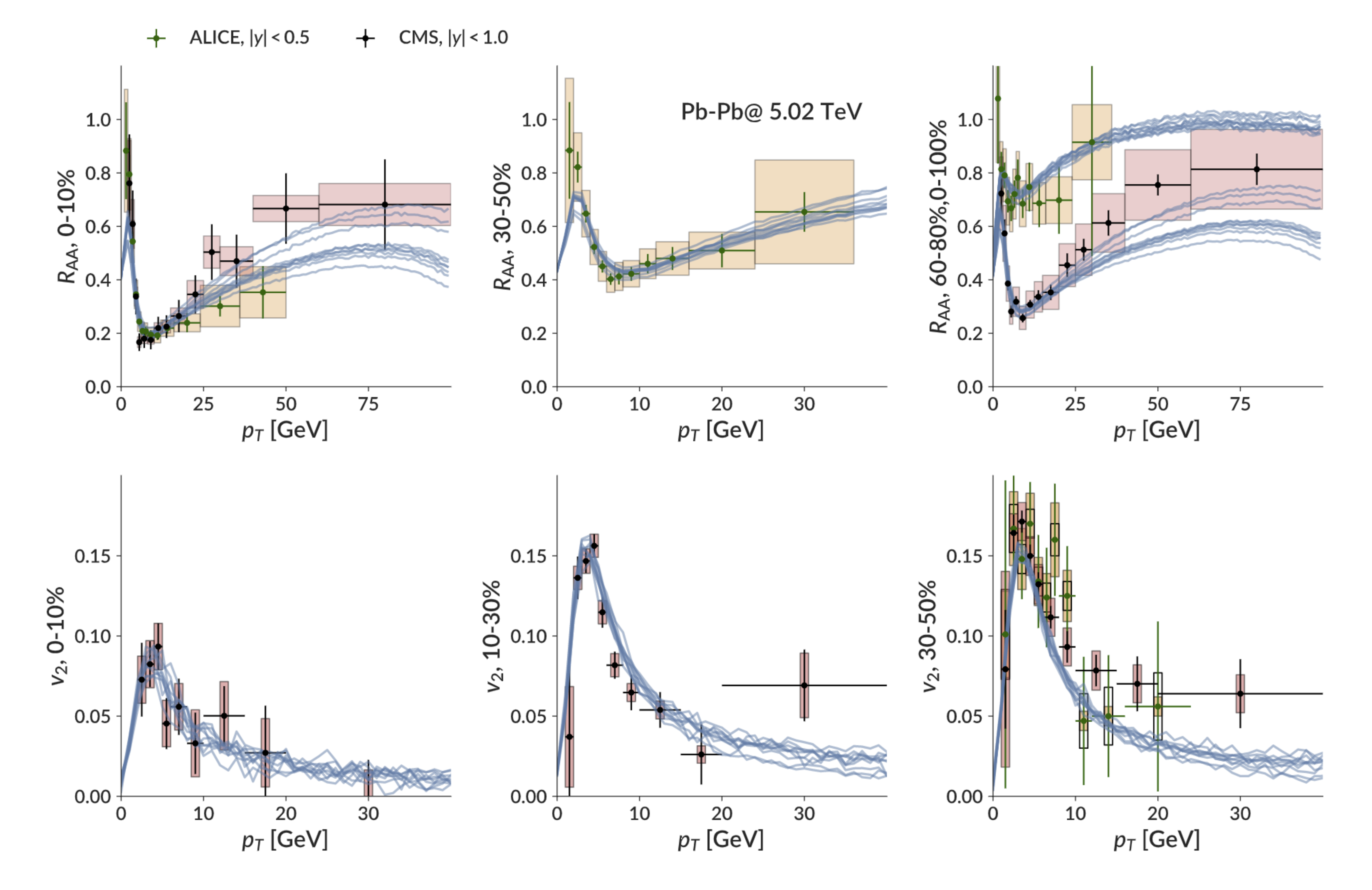}
	\caption[Calculation of the D-meson $R_{AA}$ and $v_2$ in PbPb collisions at 5.02 TeV]{\label{fig:yingru_calibration} Calculation of the D-meson $R_{AA}$ and $v_2$ in PbPb collisions at 5.02 TeV, taking the parameters randomly drawn from the posterior distributions. The improved Langevin model is used for parton energy loss. \cite{xu2019data}}
\end{figure}

\begin{figure}
	\centering
	\includegraphics[width=0.96\textwidth]{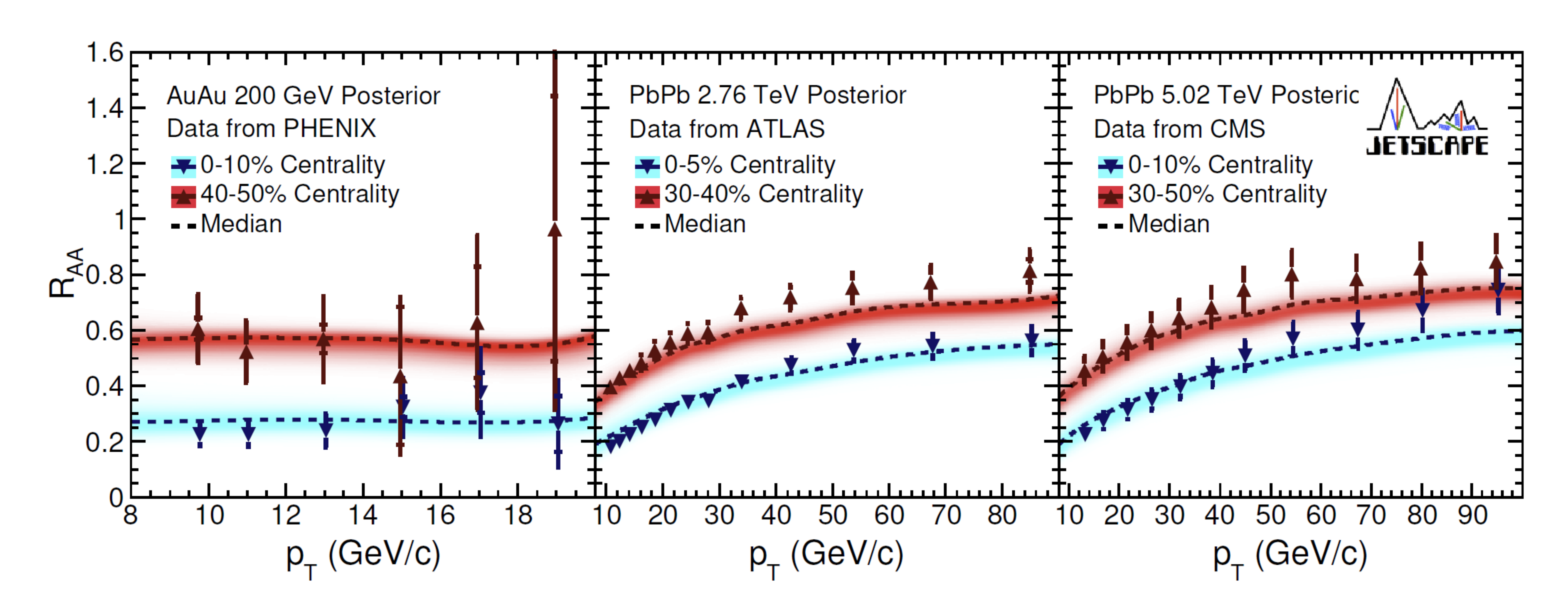}
	\caption[Posterior predictive distributions of $R_{AA}$ using MATTER+LBT]{\label{fig:shanshan_calibration} Posterior predictive distributions of $R_{AA}$ using MATTER+LBT compared to data. Dashed lines show model calculation using median values of parameters. A virtuality dependent parameterization of $\hat{q}$ is used. \cite{cao2021determining}}
\end{figure}

\begin{figure}
	\centering
	\includegraphics[width=0.9\textwidth]{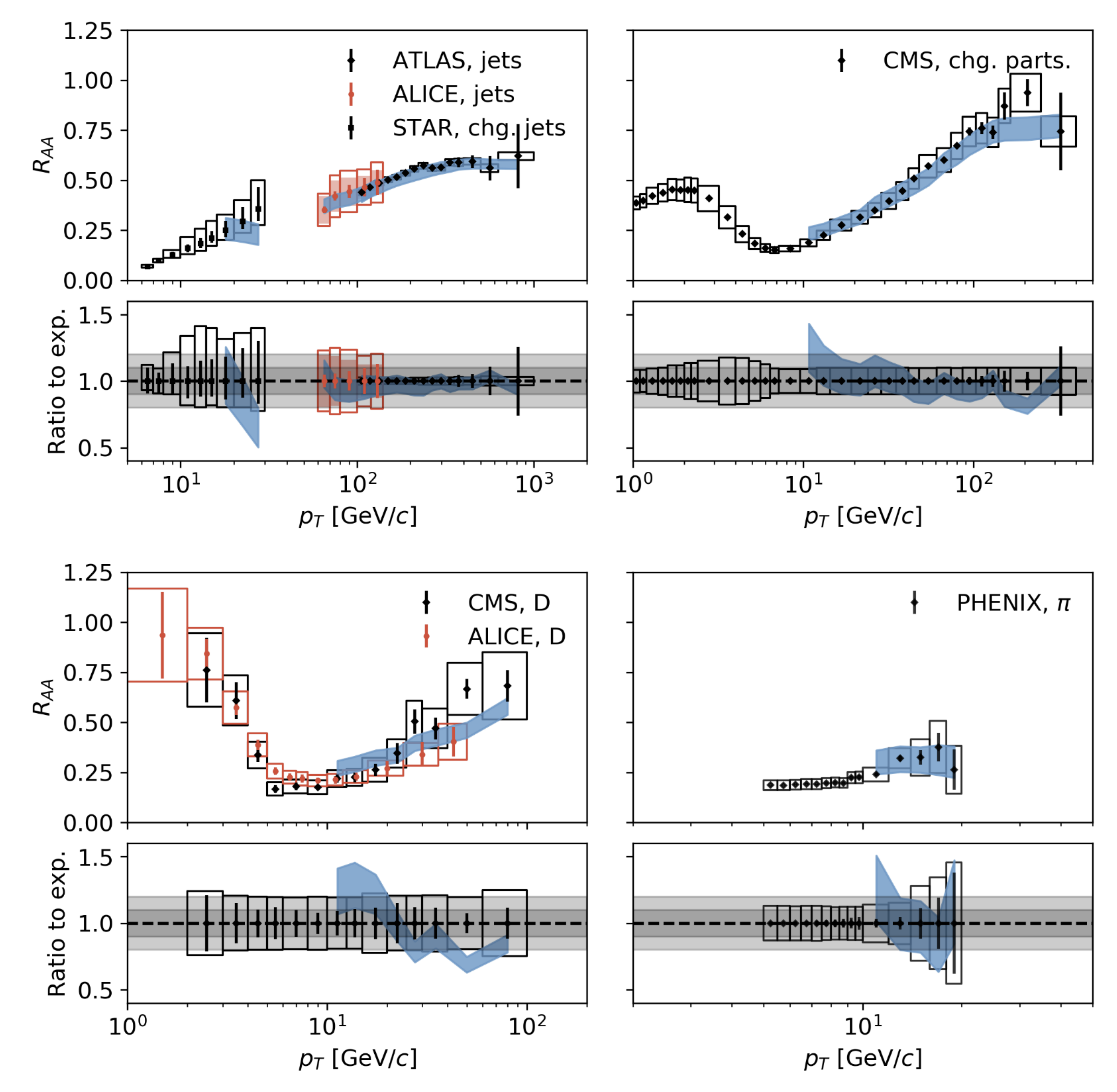}
	\caption[Emulated calculations of inclusive jet and single hadron suppression]{\label{fig:weiyao_calibration} Emulated calculations of inclusive jet and single hadron suppression after model parameters are systematically calibrated to these observables. In each set of plots, the $R_{AA}$ is shown in the upper panel, and the ratio to experiment is shown in the lower panel. The blue bands correspond to 95\% credible limit given by the model emulator. The gray bands in the ratio plot denote $\pm10\%$ and $\pm20\%$ level of discrepancy respectively. \cite{ke2021qgp}}
\end{figure}

The application of Bayesian analysis to jet energy loss in heavy ion collisions is inspired by the application of the same technique to soft observables in heavy ion collisions \cite{Bernhard:2016tnd,Bernhard:2018hnz}. Started in 2018, early attempts focused on open heavy flavor observables (see Fig.~\ref{fig:yingru_calibration})\cite{xu2018data,Ke:2018tsh}. As only the leading heavy parton was tracked, oversampling (sample an ensemble of heavy flavor partons in one event) was possible. Therefore enough statistics could be achieved for calculating even the flow coefficients with high precision. Next, people started to study charged hadron and jet observables with Bayesian analysis \cite{liu2021qlbt, cao2021determining}. Specifically, Ref.~\cite{cao2021determining} uses the JETSCAPE framework with the multi-stage MATTER+LBT approach for parton energy loss (see Fig.~\ref{fig:shanshan_calibration}). More recently, Ref.~\cite{ke2021qgp} performed the calibration on charged hadron, D meson and inclusive jet $R_{AA}$ (see Fig.~\ref{fig:weiyao_calibration}).

\section{Calibration setup}

\subsection{Parameter design}

As we have seen in Chapter~\ref{sec:jetscape_results}, parameters like $\tau_0, T_c$ have minor influence on $R_{AA}$ so they will be fixed in this analysis. What will be explored are the coupling constant $\alpha_s$, the switching scale $Q_s$ and $c_1,c_2$ in the parameterization of $\hat{q}$. The prior range for these parameters are considered to be:

\begin{table}[h!]
\centering
\caption{Prior ranges for the parameters in our Bayesian calibration.}
\begin{tabular}{ |p{3cm}|p{3cm}|  }
 \hline
 Parameter & Range\\
 \hline
$\alpha_s$  & 0.1 - 0.5 \\
 \hline
 $Q_s$  & 1.5 - 4 \\
 \hline
  $c_1$   & 1 - 10\\
 \hline
  $c_2$   & 50 - 300\\
 \hline
 \end{tabular}
\end{table}

\begin{figure}
	\centering
	\includegraphics[width=0.65\textwidth]{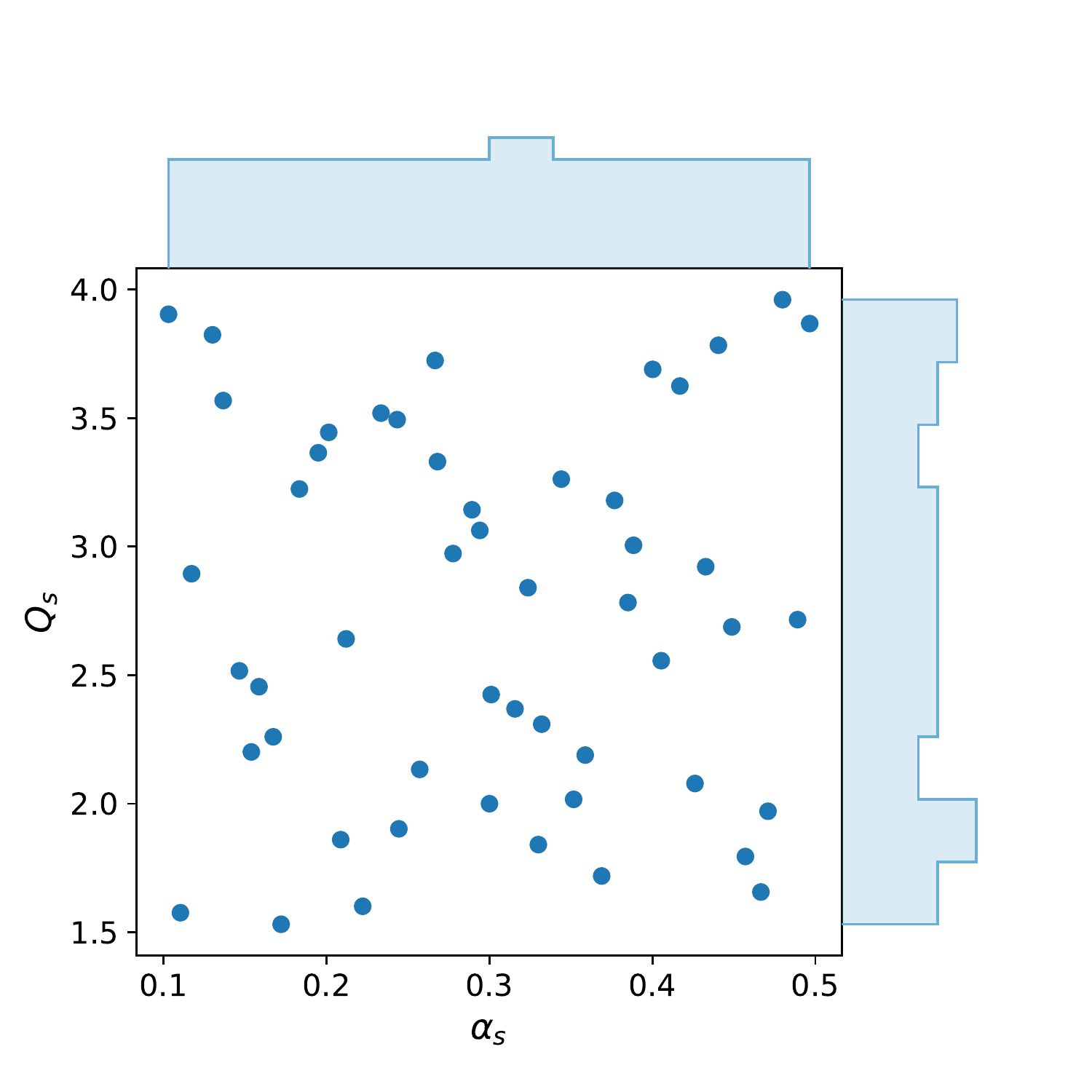}
	\caption{\label{fig:Design14} Distribution of input parameter $\alpha_s$ and $Q_s$ from 50 sampled design points within the prior range.}
\end{figure}

Those ranges are selected based on our simple exploration in Chapter~\ref{sec:jetscape_results} and also in Ref.~\cite{kumar2022inclusive}. Due to constrained computation budget, I use 50 design points drawn from Latin hypercube sampling (see Fig.~\ref{fig:Design14}). 

\subsection{Experimental observables}

There exist a wide range of experimental observables from different collision systems and at different collision energies and centralities. Due to a constrained computation budget, I will focus only on PbPb collision at 5.02~TeV and 0-10\% centrality. A previous study \cite{xu2018data} shows that calibrating to different collision energies independently or at the same time may yield slightly different posteriors or one may need to use different values for the same parameter in other collision systems \cite{ke2021qgp}. I will leave such an exploration to a future study. 

As for the experimental observables that will be in this calibration, I choose $R_{AA}$ for charged hadrons, D mesons, and inclusive jets. Calibrating to these observables together should give a better constraint on the transport properties of the QGP. One thing to notice is that data points for the charged hadron and D meson $R_{AA}$ below $7$ GeV are not considered in the calibration as they are affected by non-perturbative effects, medium response, recombination contributions to hadronization, etc. We could also consider flow observables like $v_2,v_3$ for the D mesons, but that would require around $10$ times more statistics to get meaningful results in central collisions. Another class of observables is the jet substructure observable which will be ignored as well since contributions from medium response and background subtraction are still being investigated.

\subsection{Model calculation at the design points}

\begin{figure}
	\centering
	\includegraphics[width=0.96\textwidth]{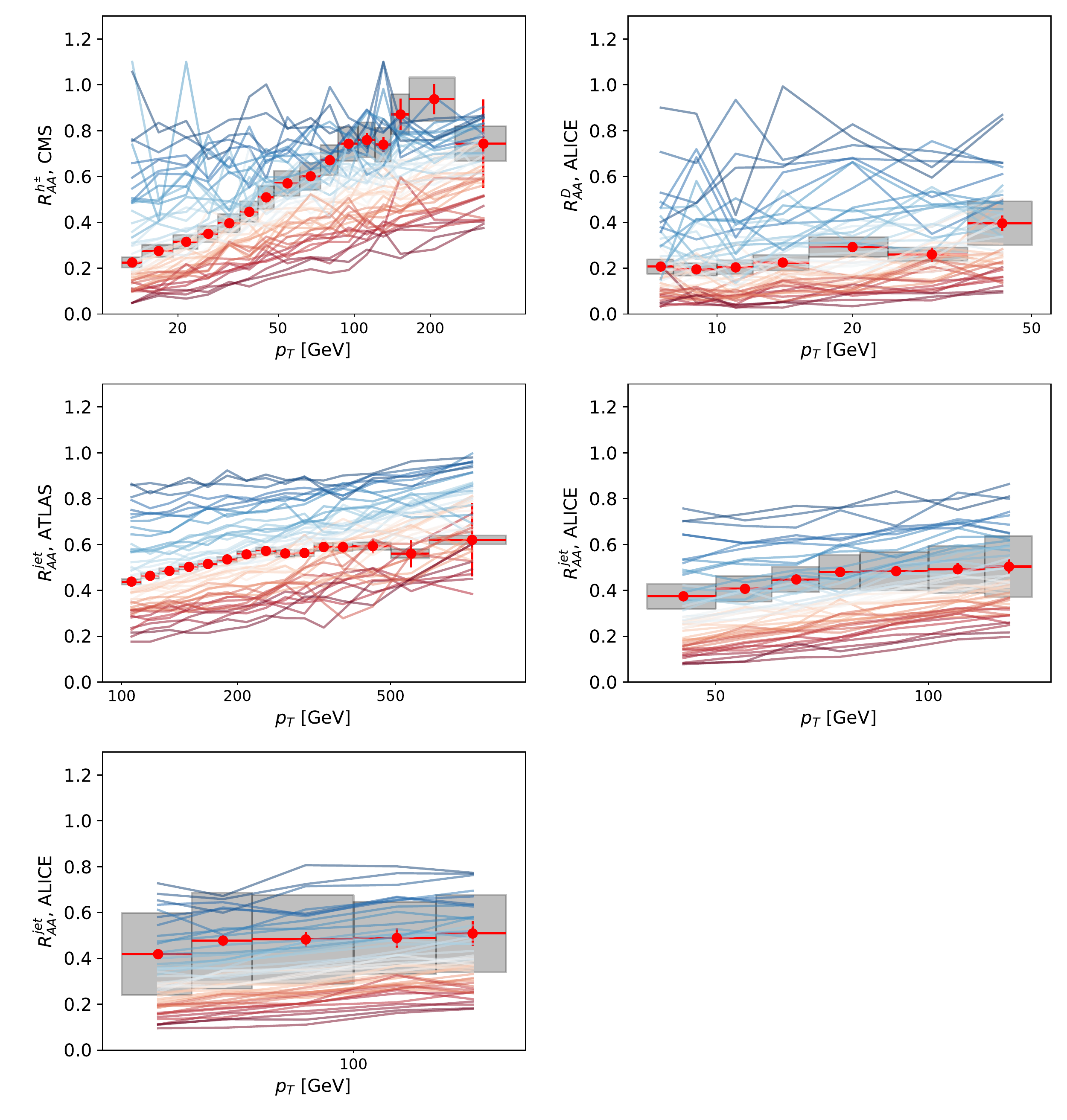}
	\caption[Model calculation using parameters from all the design points]{\label{fig:PredictedDesign} Model calculation using parameters from all the design points. From top to bottom, left to right, the measurements are charged hadron $R_{AA}$ from CMS \cite{khachatryan2017charged}, D meson $R_{AA}$ from ALICE \cite{alice2018measurement}, inclusive jet $R_{AA}$ from ATLAS \cite{acharya2020measurements} and two inclusive jet $R_{AA}$ with different jet radius from ALICE \cite{acharya2020measurements}. To better see how different calculations are distributed among these observables, each design point is assigned with a different color.}
\end{figure}

Now I want to show the model calculation from all the design points as an indication of the prior range of our prediction in the observable space. If the model calculations from all design points are all above or below the data, it is unlikely we will be able to describe that observable. Fortunately, the model calculations span a wide range in $R_{AA}$ and can cover the data for all five observables (see Fig.~\ref{fig:PredictedDesign}). 

For each design point, roughly 400k events are generated and are distributed among 400 fluid simulations which give rise to about 1000 events per fluid simulation. This number is only around $4\%$ of what was used in Chapter~\ref{sec:jetscape_results}, so we observe pretty big fluctuations in the calculations. Those fluctuations will impact our calibration. I will verify the validity of our Bayesian analysis against model calculation fluctuations in Chapter~\ref{sec:quantitative_closure_test} and Chapter~\ref{section:appendix_bayesian}.

\begin{figure}
	\centering
	\includegraphics[width=0.96\textwidth]{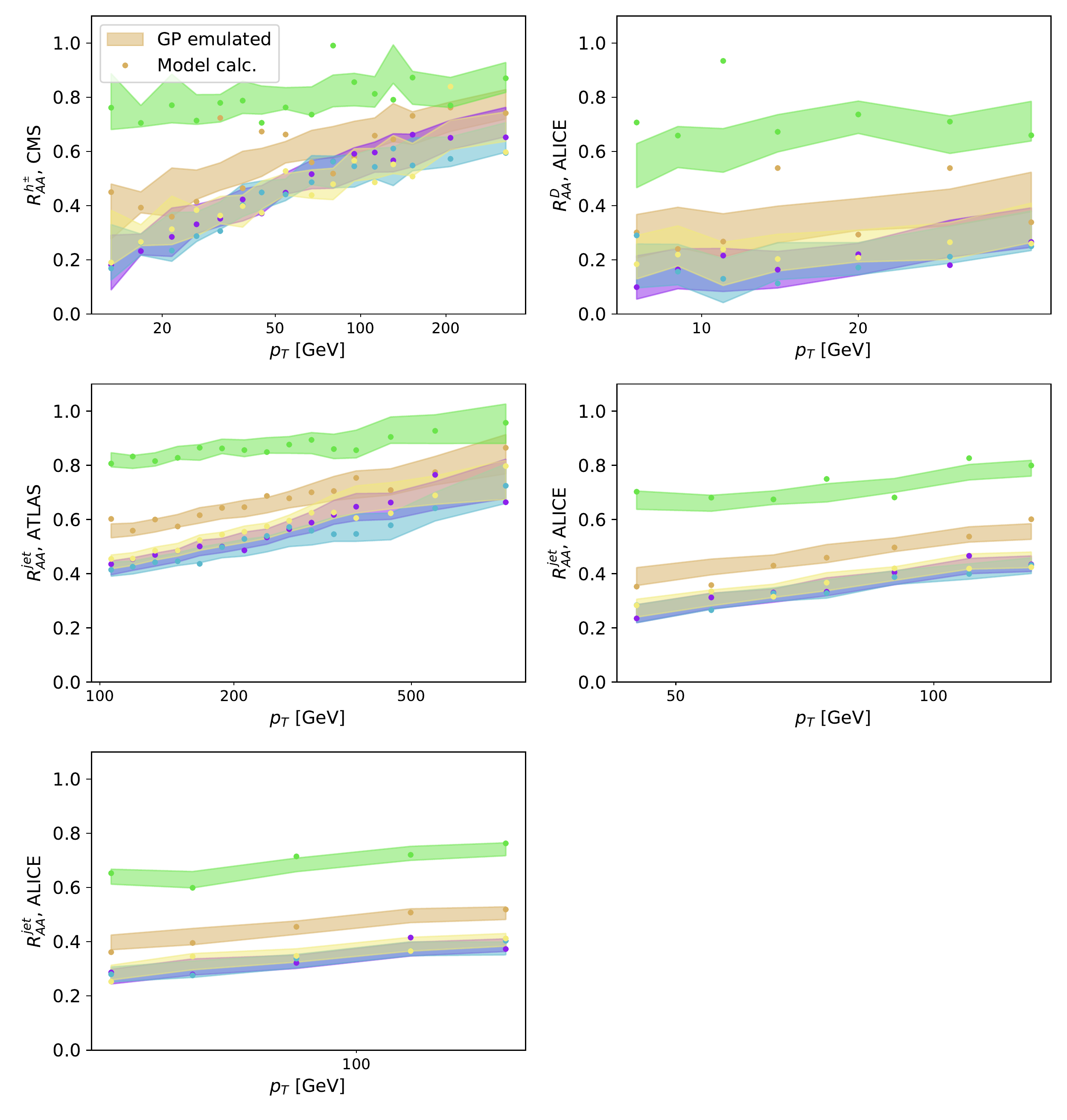}
	\caption{\label{fig:EmulatorValidation} Comparison between emulator predictions and model calculations at 5 random design points.}
\end{figure}

\begin{figure}
	\centering
	\includegraphics[width=0.8\textwidth]{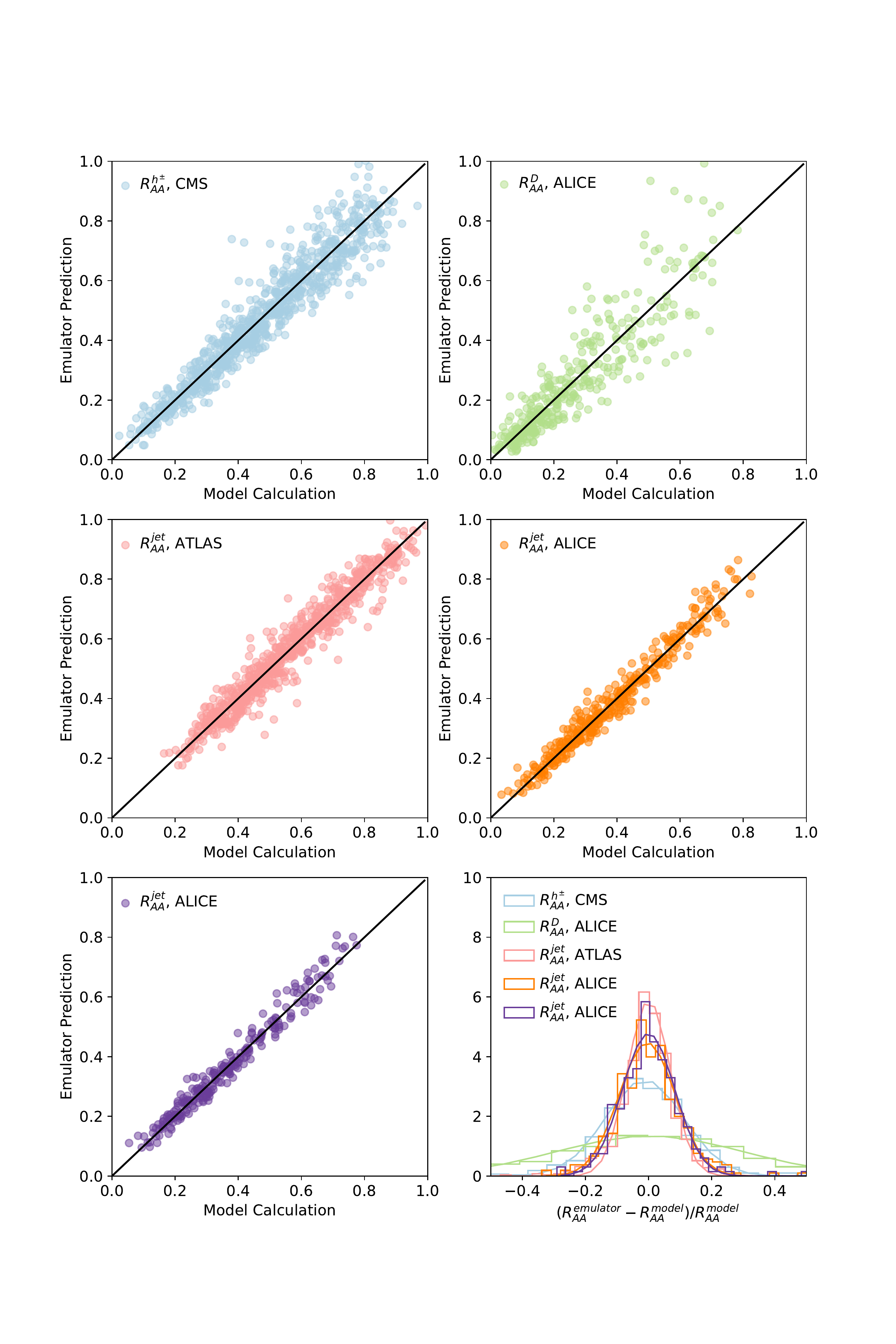}
	\caption[Comparison between emulator predictions and model calculations at all design points]{\label{fig:EmulatorValidationScatter} Comparison between emulator predictions and model calculations at all design points. The first five plots are scatter plots that plots the model calculation and emulator prediction for each observable in pairs. If the model calculation and emulator prediction are the same, they should be positioned on the $y=x$ line (the black solid line). The last plot shows the histograms of the relative difference between model calculation and emulator prediction for different observables.}
\end{figure}

\begin{figure}
	\centering
	\includegraphics[width=0.96\textwidth]{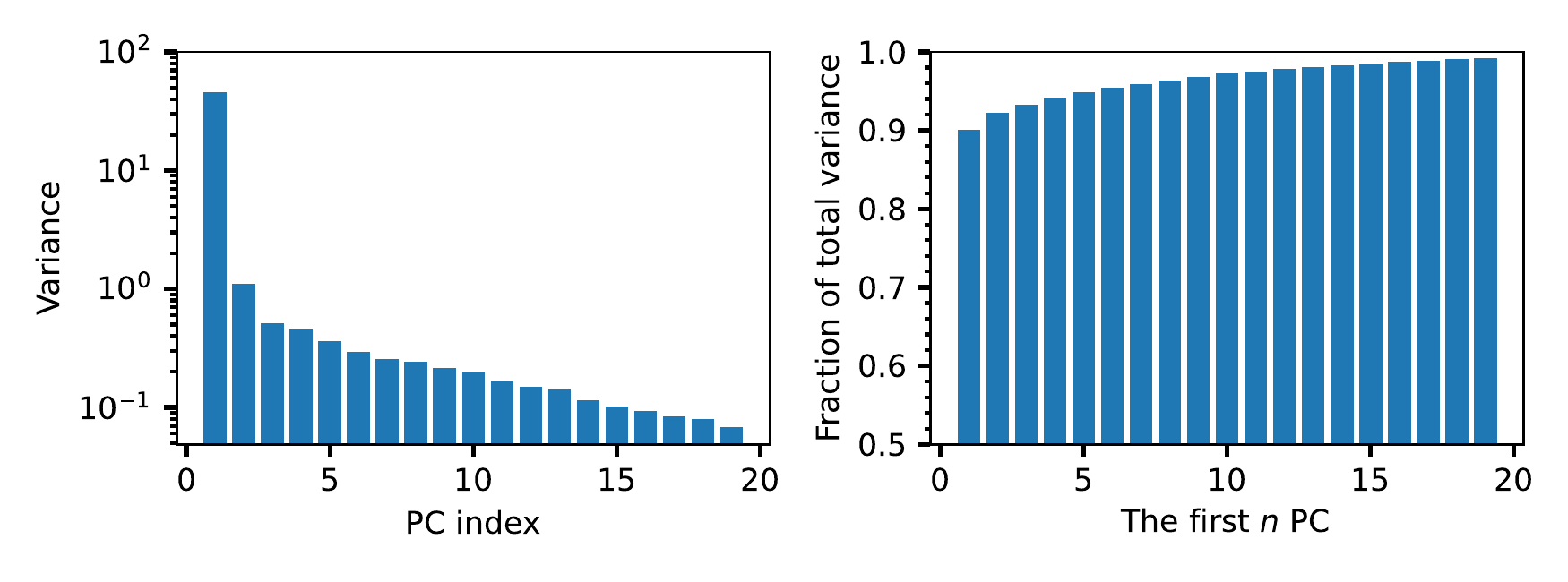}
	\caption[Relation between variances and principal components]{\label{fig:PC_importance} \textbf{Left}: The variance explained by each principal component. \textbf{Right}: The cumulative variance explained by the first $n$ principal components.}
\end{figure}

\section{Emulator training and validation}\label{sec:emulator_validation}

The emulator is the essential part that makes the Bayesian analysis possible. One can view it as a black box that accepts arbitrary input parameters and outputs predictions of the observables. One should first ensure that the emulator is working correctly before running the MCMC to get the posterior distribution of the parameters. Otherwise, one might be interpreting a biased posterior distribution. The key ingredients when training the emulator are the kernel to use, the number of principal components, and the treatment of uncertainties (covariance matrix). Below are the reasonings for choosing the specific settings for the emulator in this work:

\begin{itemize}
    \item The kernel I use is a linear combination of the radial based function (RBF) kernel (Eq.~\ref{eqn:RBF_kernel}) and the white noise kernel. The rationale for this particular choice is that the various $R_{AA}$'s are generally smooth and monotonic increasing functions (at least in the $p_T$ range we are interested in). So the smooth RBF kernel should be able to fit them well. The fluctuations in our simulation can be absorbed by the white noise kernel. I have also tried the Mat\'{e}rn kernel which is a generalization of the RBF kernel \cite{schulz2018tutorial}. It has an additional variable $\nu$ that controls its smoothness (Mat\'{e}rn kernel with $\nu\rightarrow \infty$ is equivalent to the RBF kernel). A more detailed comparison between different kernels will be carried out in Section~\ref{sec:quantitative_closure_test} and Appendix.~\ref{section:appendix_matern kernel}. Unless specifically stated, the kernel used in the emulator will be the RBF+white noise kernel.
    \item The number of principal components can be determined by examining the hyper parameters of the emulator for each PC (like the length scale $l$ in the RBF kernel or the noise level in the white noise kernel). One can expect that using too few PC will under-fit the training data. Using more PC can capture finer details of the model calculations but also risk of treating statistical fluctuations as real input. In this study, 5 PCs that explain around $95\%$ of the total variance show the best performance for closure tests and will be used hereafter (see Fig.~\ref{fig:PC_importance}). The effects of the number of PCs on closure tests will be studied in more detail in Section~\ref{sec:quantitative_closure_test}.
    \item As discussed in Chapter~\ref{sec:bayesian}, the covariance matrix takes the form:
    \begin{equation}
        \Sigma=\Sigma_{sys}+\Sigma_{stat}+\Sigma_{emulator}+\Sigma_{truncation}+\Sigma_{model}.
    \end{equation}
    
    The $\Sigma_{emulator}$ contains the total covariance of all the Gaussian processes. The $\Sigma_{truncation}$ contains the total covariance of all the remaining principal components not considered by the emulator. $\Sigma_{model}$ is not considered in this analysis as it is challenging to quantify. As for the experimental covariance $\Sigma_{sys}+\Sigma_{stat}$, it is assumed to be diagonal in this analysis. The effect of off-diagonal terms in $\Sigma_{sys}$ will be discussed in Appendix.~\ref{section:appendix_experiment_covariance}.
    
\end{itemize}

Some of the arguments listed above are more or less qualitative. In the next section, I will introduce a quantitative way to determine the optimal settings for the Gaussian process emulator.

\subsection{Emulator validation}

\begin{table}[h!] 
\centering
\caption{The mean $\mu$ standard deviation $\sigma$ when fitting Gaussian to the distribution of the relative difference between model calculation and emulator prediction.}\label{tab:gaussian_fit_emulator_difference}
\begin{tabular}{ |p{3cm}|p{2cm}|p{4cm}|  }
 \hline
  Gaussian fit & Mean $\mu$ & Standard deviation $\sigma$\\
 \hline
 $R_{AA}^{h^\pm}$, CMS  & -0.0126 & 0.124\\
 \hline
 $R_{AA}^{D}$, ALICE  & 0.00360 & 0.297\\
 \hline
 $R_{AA}^{jet}$, ATLAS  & 0.000912 & 0.0677\\
 \hline
 $R_{AA}^{jet}$, ALICE  & -0.00433 & 0.0892\\
  \hline
 $R_{AA}^{jet}$, ALICE  & 0.0000105 & 0.0835\\
 \hline
 \end{tabular}
\end{table}

Before running MCMC with the emulator, the emulator's performance on the training data should be validated. If the emulator can not reproduce the training data, it can not serve as a surrogate for the model. A direct comparison between emulator predictions and model calculations at $5$ random design points can be seen in Fig.~\ref{fig:EmulatorValidation}. The emulator predictions fit the model calculations pretty well and cut off some statistical fluctuations in our model calculations. One can also notice that the emulator prediction gives larger uncertainty bands for model calculations with larger fluctuations.

To see the performance of the emulator on all design points, the model calculation and emulator prediction for each observable from all design points are plotted in pairs in the first five plots in Fig.~\ref{fig:EmulatorValidationScatter}. The emulator seems to perform the best predicting inclusive jet $R_{AA}$, followed by charged hadron $R_{AA}$ and finally the $D$ meson $R_{AA}$. In the last plot in Fig.~\ref{fig:EmulatorValidationScatter}, the histograms of the relative difference between model calculations and emulator predictions for different observables are shown. The distributions are all be fitted by a Gaussian distribution centered near the origin, meaning there is little systematic bias introduced into the emulator. The mean and standard deviation are listed in Table.~\ref{tab:gaussian_fit_emulator_difference}. The uncertainty is around $6-9\%$ when predicting inclusive jet $R_{AA}$, $12\%$ when predicting charged hadron $R_{AA}$, and $30\%$ when predicting D meson $R_{AA}$. When calculating the emulator prediction at one design point, the training data will exclude data from that specific design point. This means we are training different emulators for each design point and the emulator does not know the truth values when making predictions. If all the design points are used to train a single emulator, the uncertainty of the relative difference will be slightly reduced (around $5-7\%$ when predicting inclusive jet $R_{AA}$, $10\%$ when predicting charged hadron $R_{AA}$, and $25\%$ when predicting D meson $R_{AA}$).

\subsection{Closure test}

\begin{figure}
	\centering
	\includegraphics[width=0.49\textwidth]{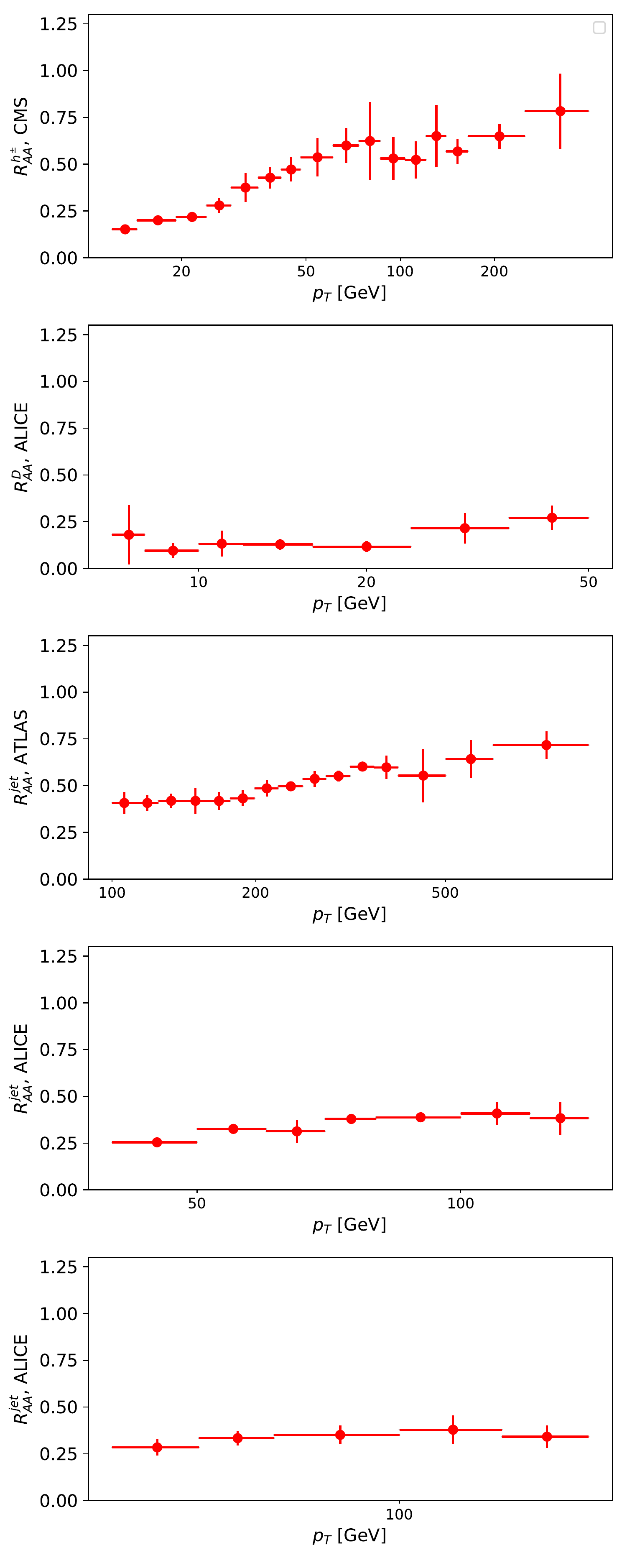}
	\includegraphics[width=0.49\textwidth]{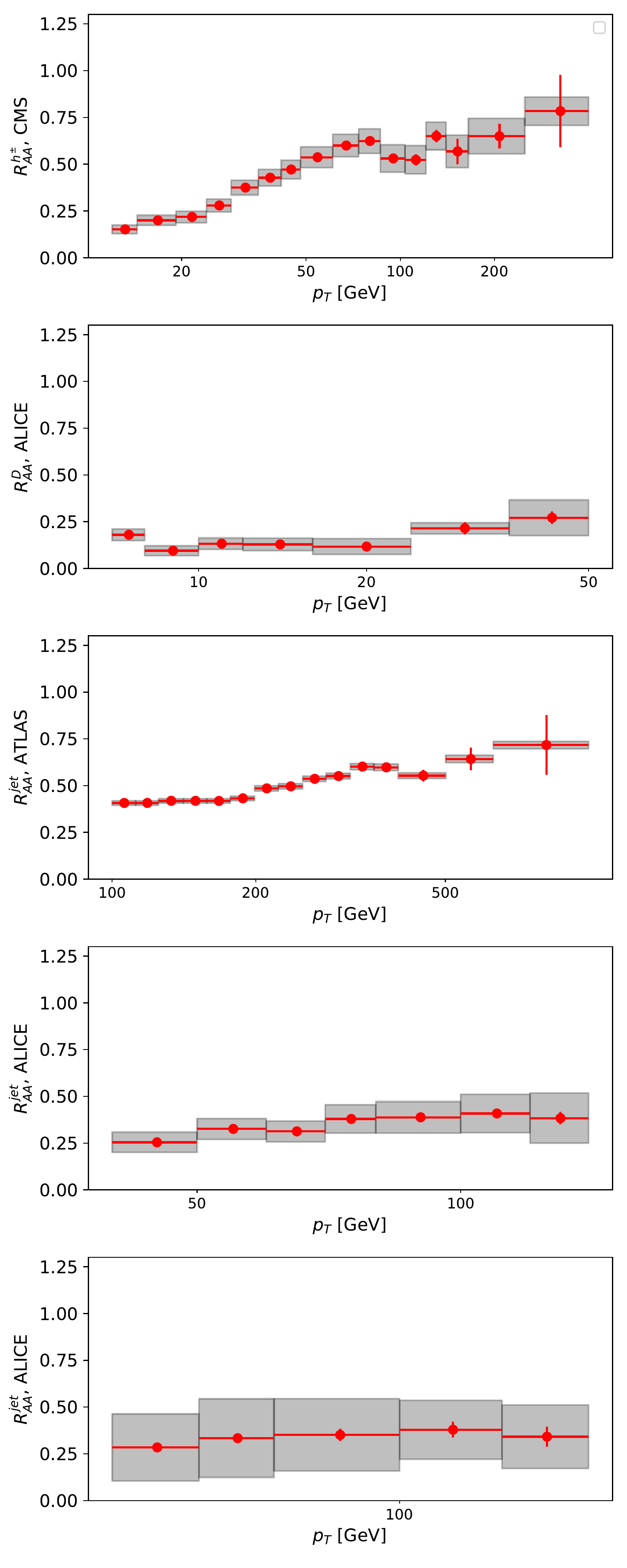}
	\caption[Mock data using simulation statistical uncertainty and using experimental uncertainty]{\label{fig:comparison_exp_own_fluc} \textbf{Left Column}: mock data using statistical fluctuations from simulation and no systematic fluctuations. \textbf{Right Column}: mock data using statistical fluctuations and systematic fluctuations from experiments.}
\end{figure}

\begin{figure}
	\centering
	\includegraphics[width=0.32\textwidth]{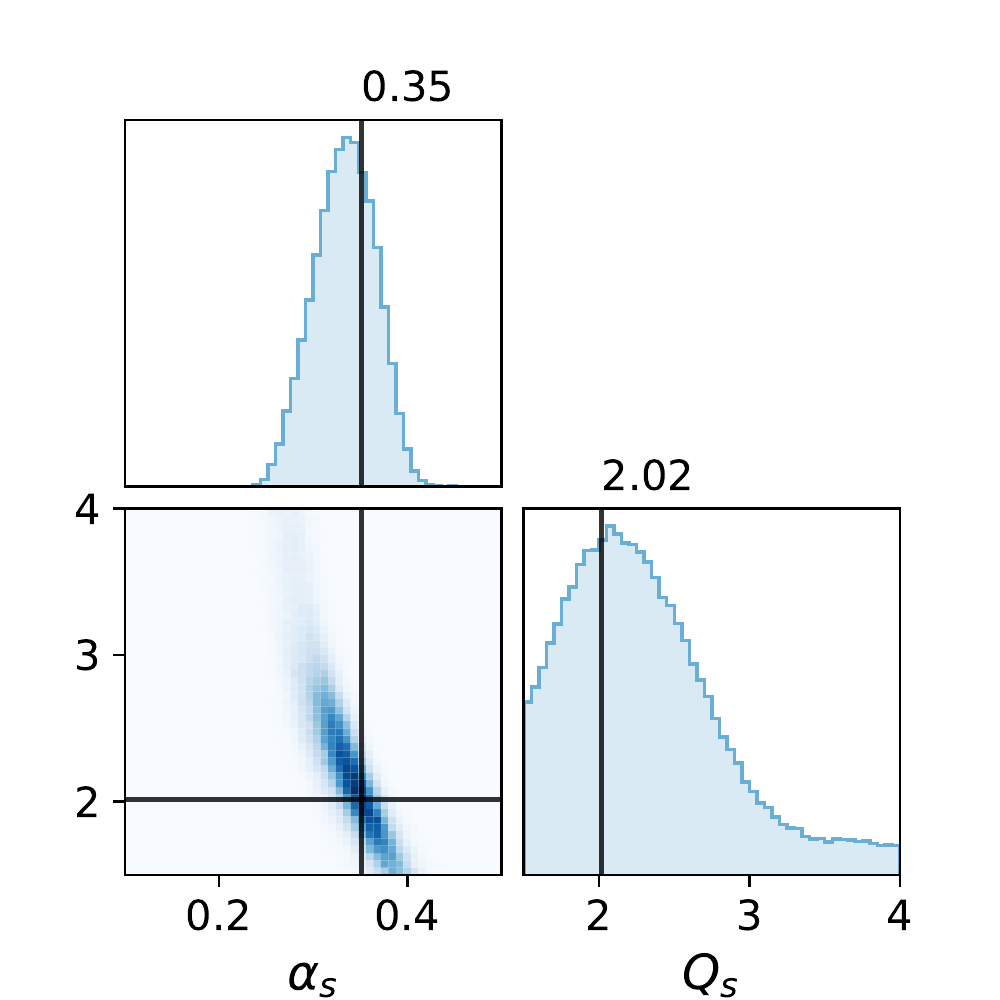}
	\includegraphics[width=0.32\textwidth]{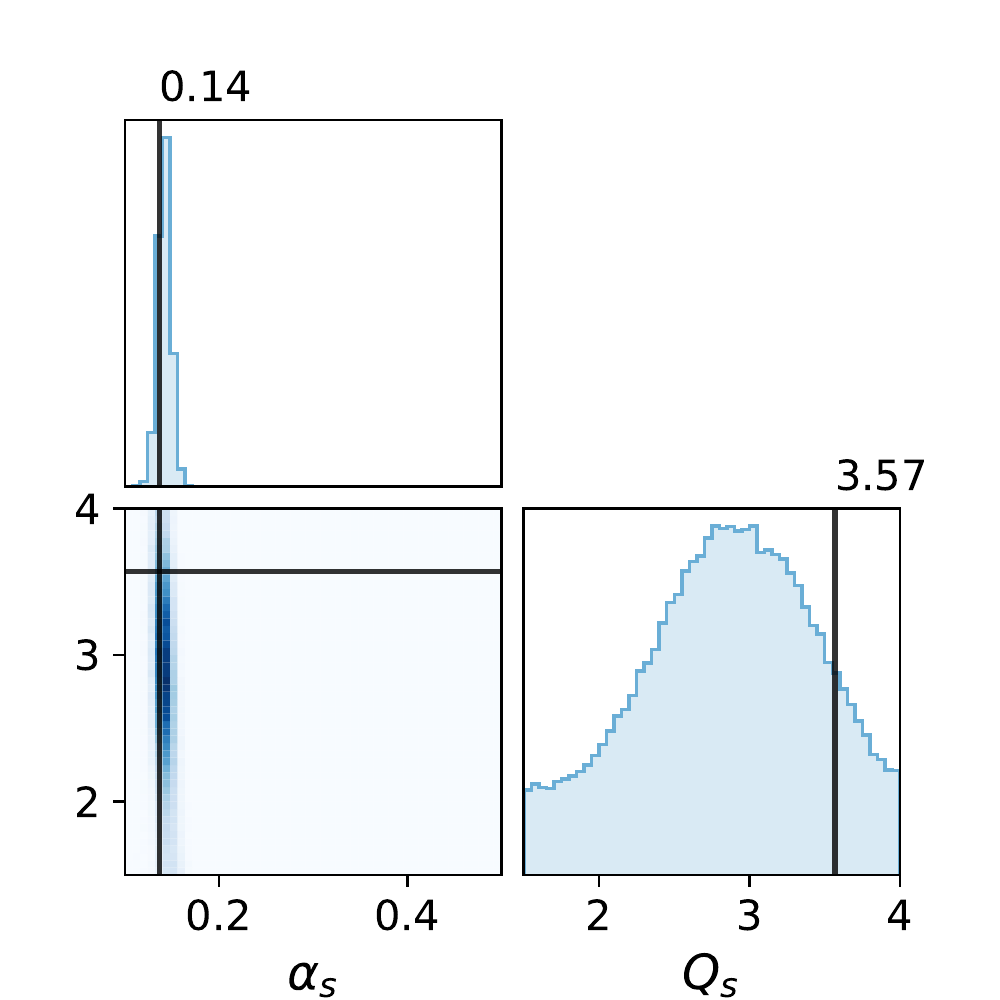}
	\includegraphics[width=0.32\textwidth]{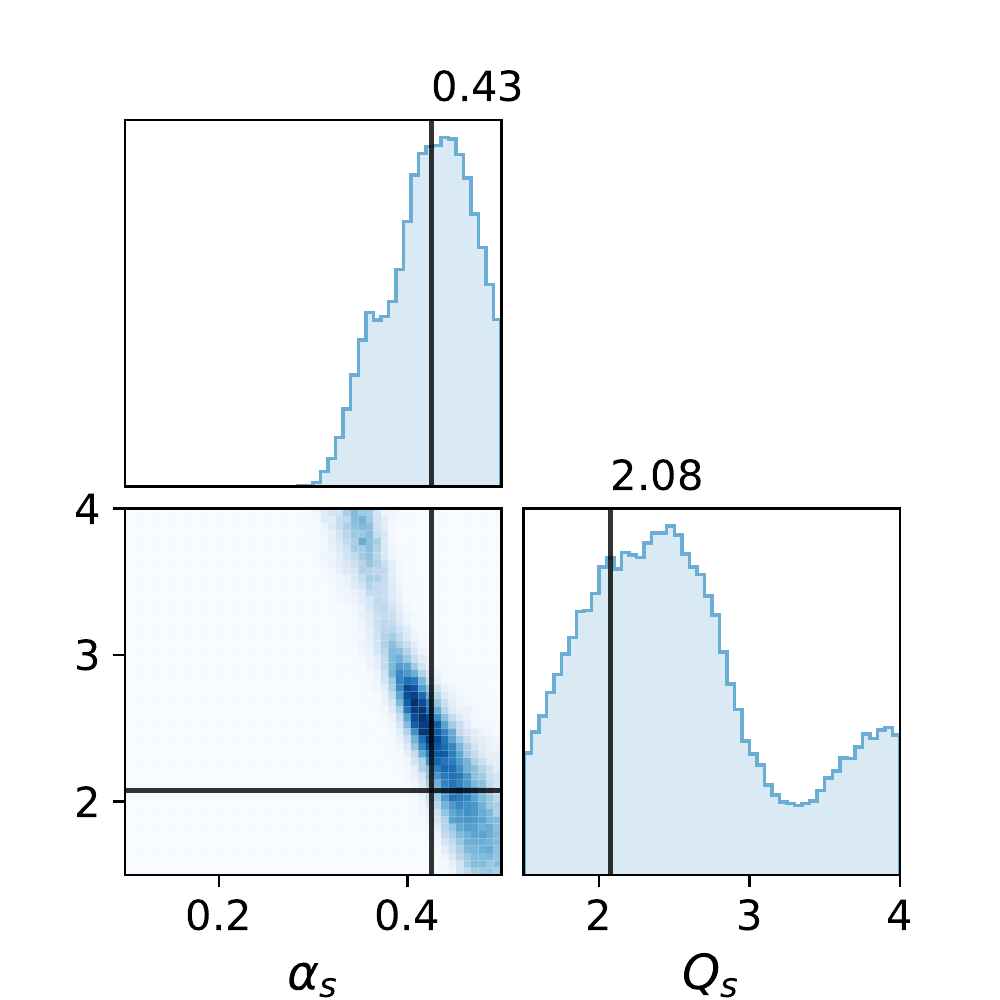}
	\includegraphics[width=0.32\textwidth]{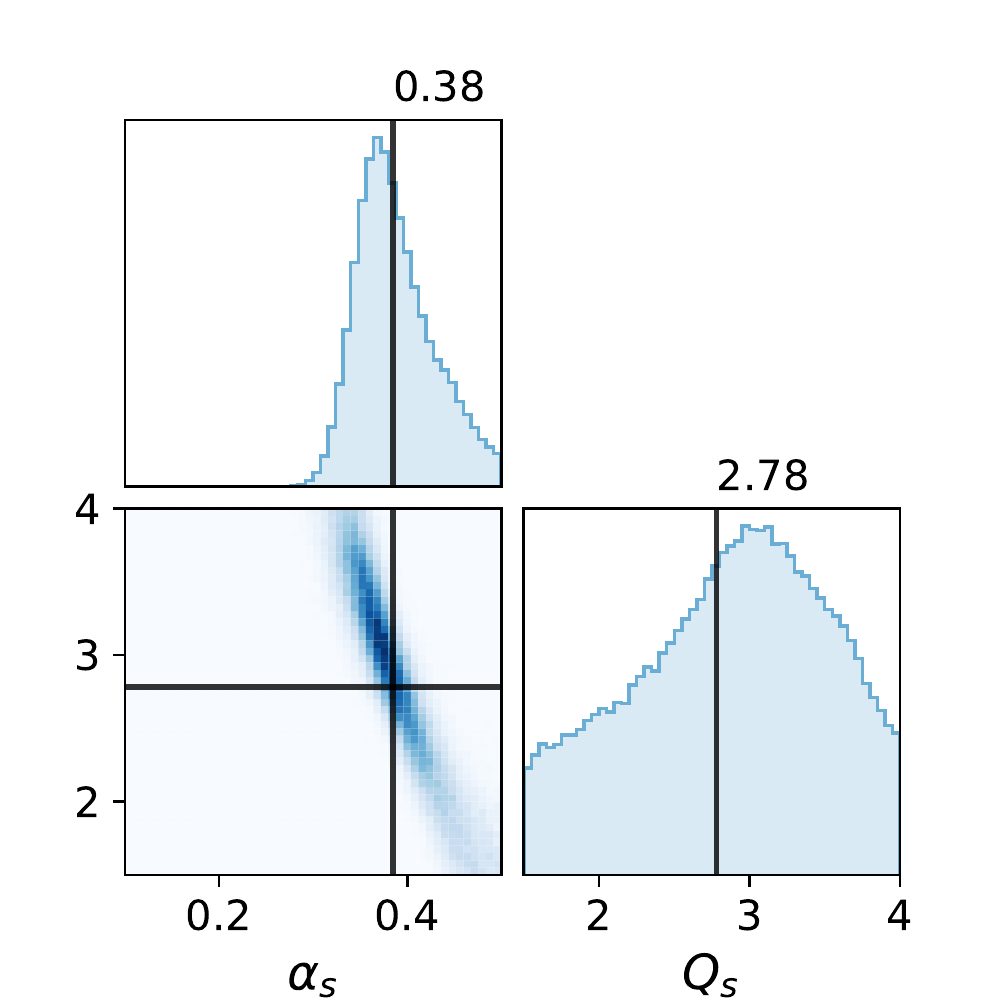}
	\includegraphics[width=0.32\textwidth]{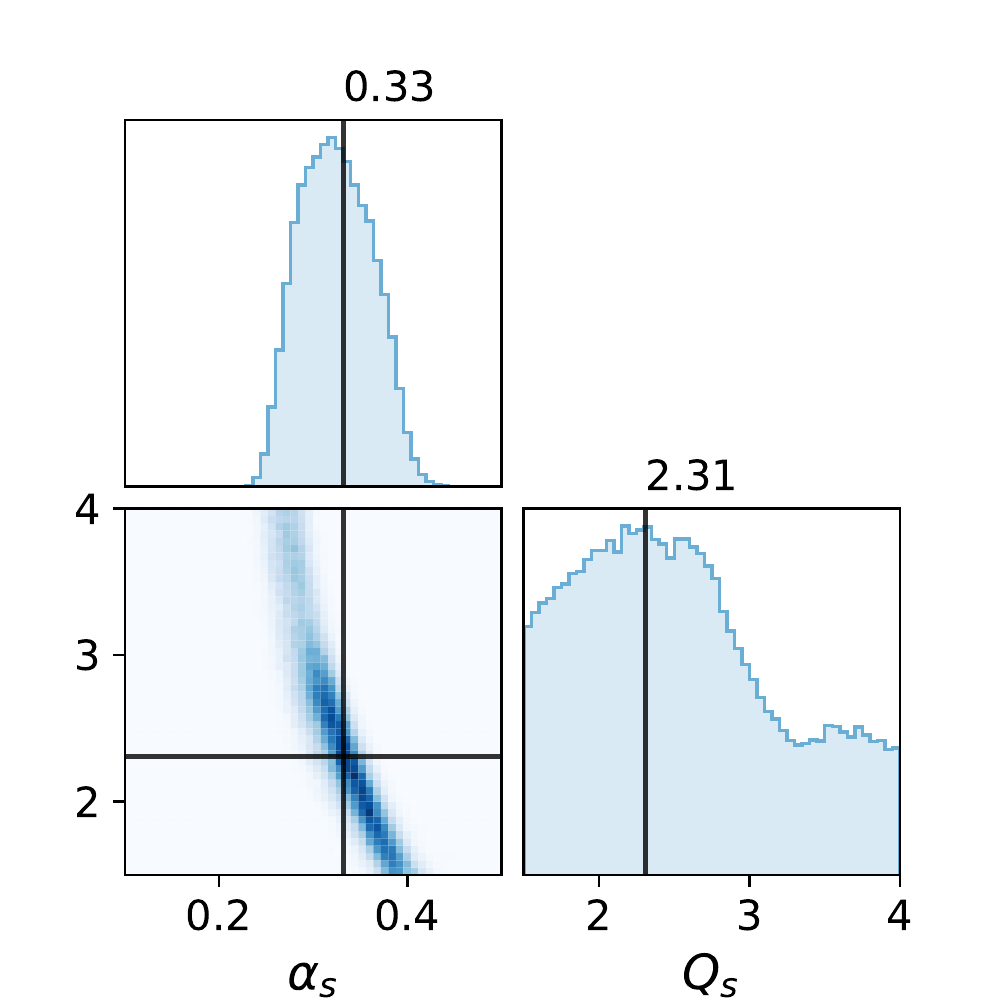}
	\includegraphics[width=0.32\textwidth]{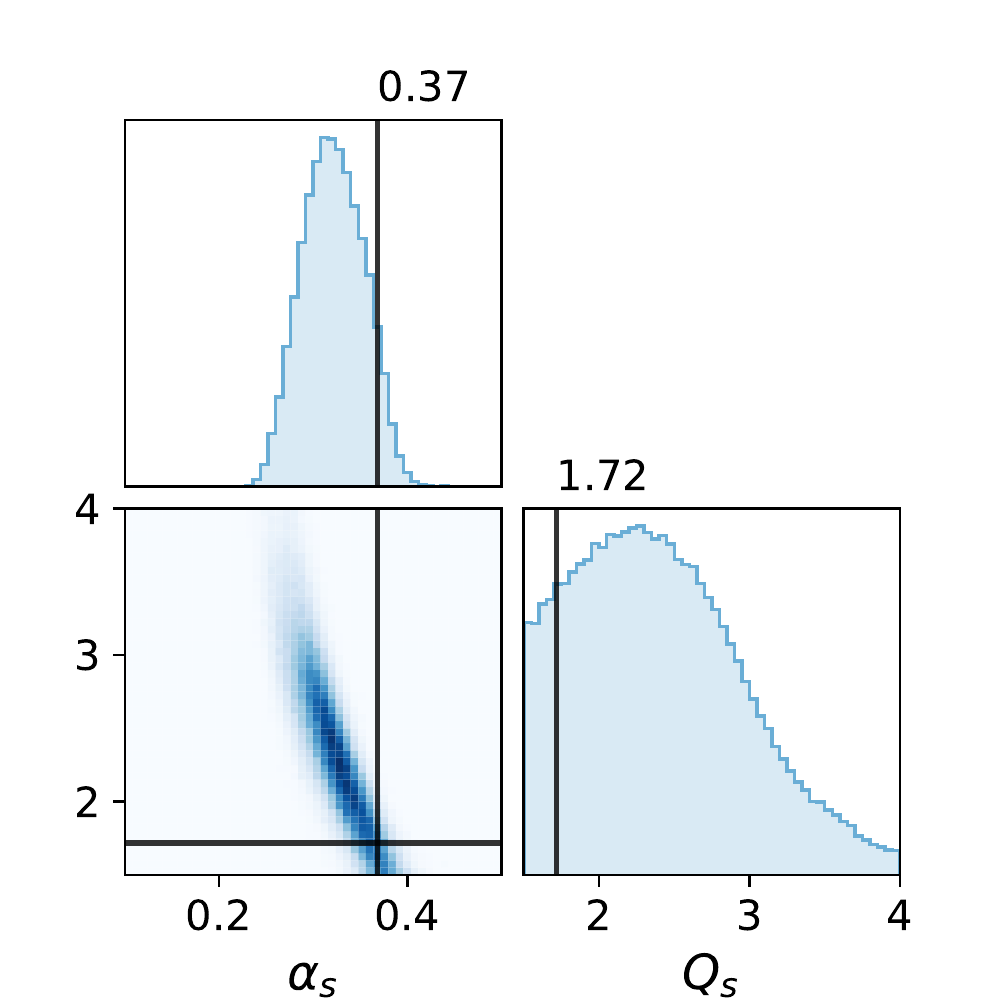}
	\includegraphics[width=0.32\textwidth]{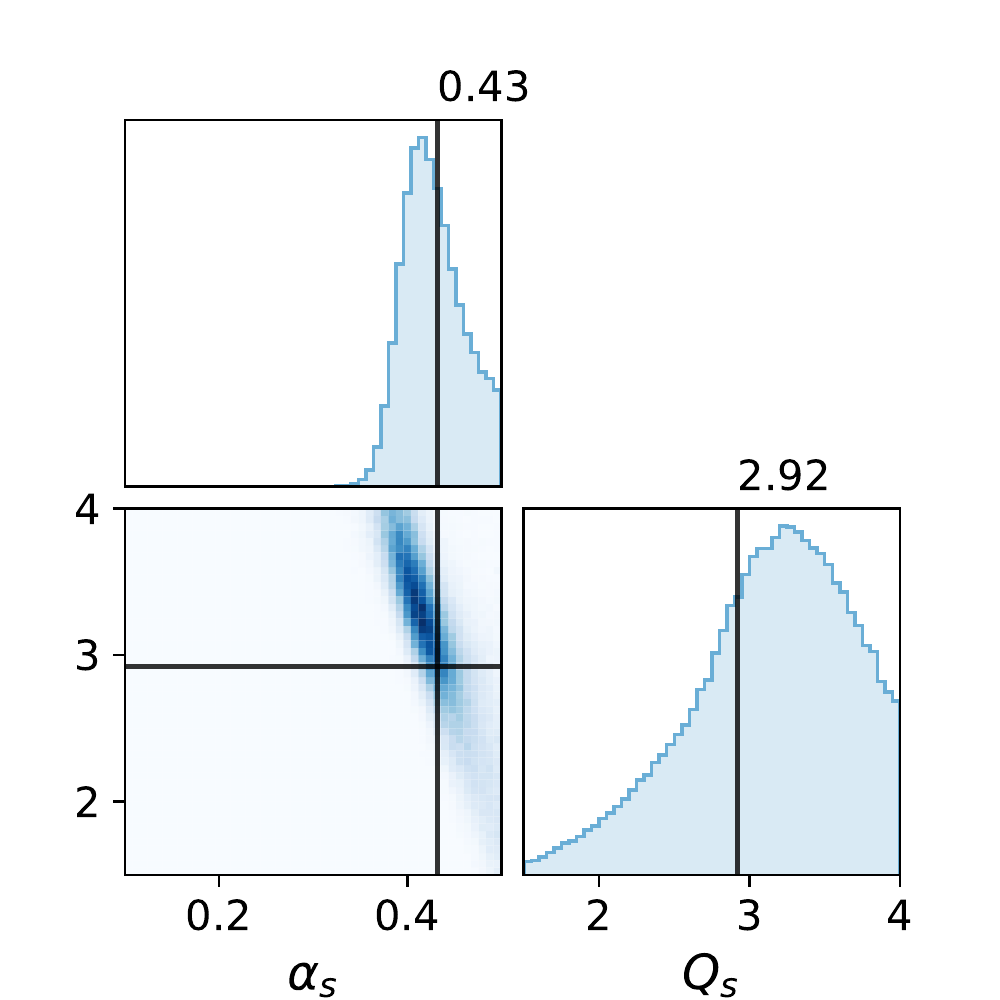}
	\includegraphics[width=0.32\textwidth]{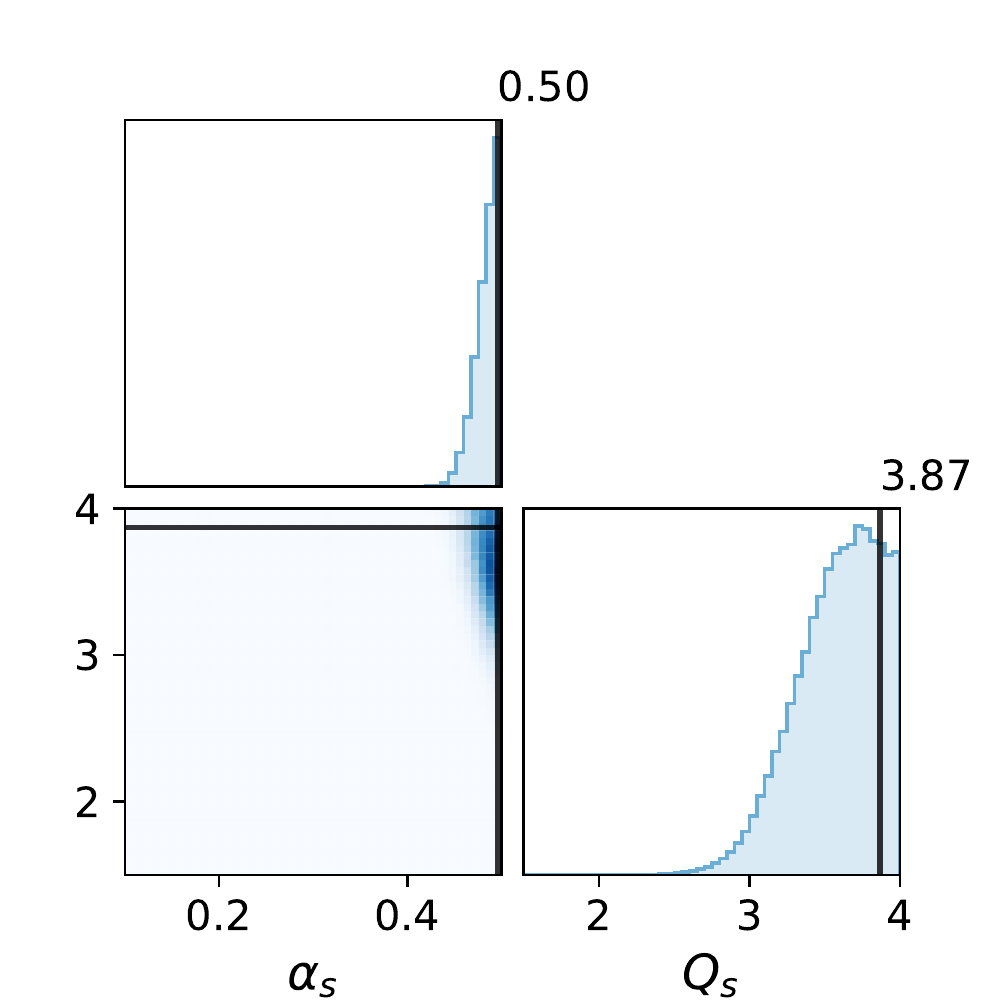}
	\caption{\label{fig:closure_own_fluc_14} Closure tests of $\alpha_s$ and $Q_s$ using 8 sets of mock data with statistical fluctuations from model calculations.}
\end{figure}

\begin{figure}
	\centering
	\includegraphics[width=0.48\textwidth]{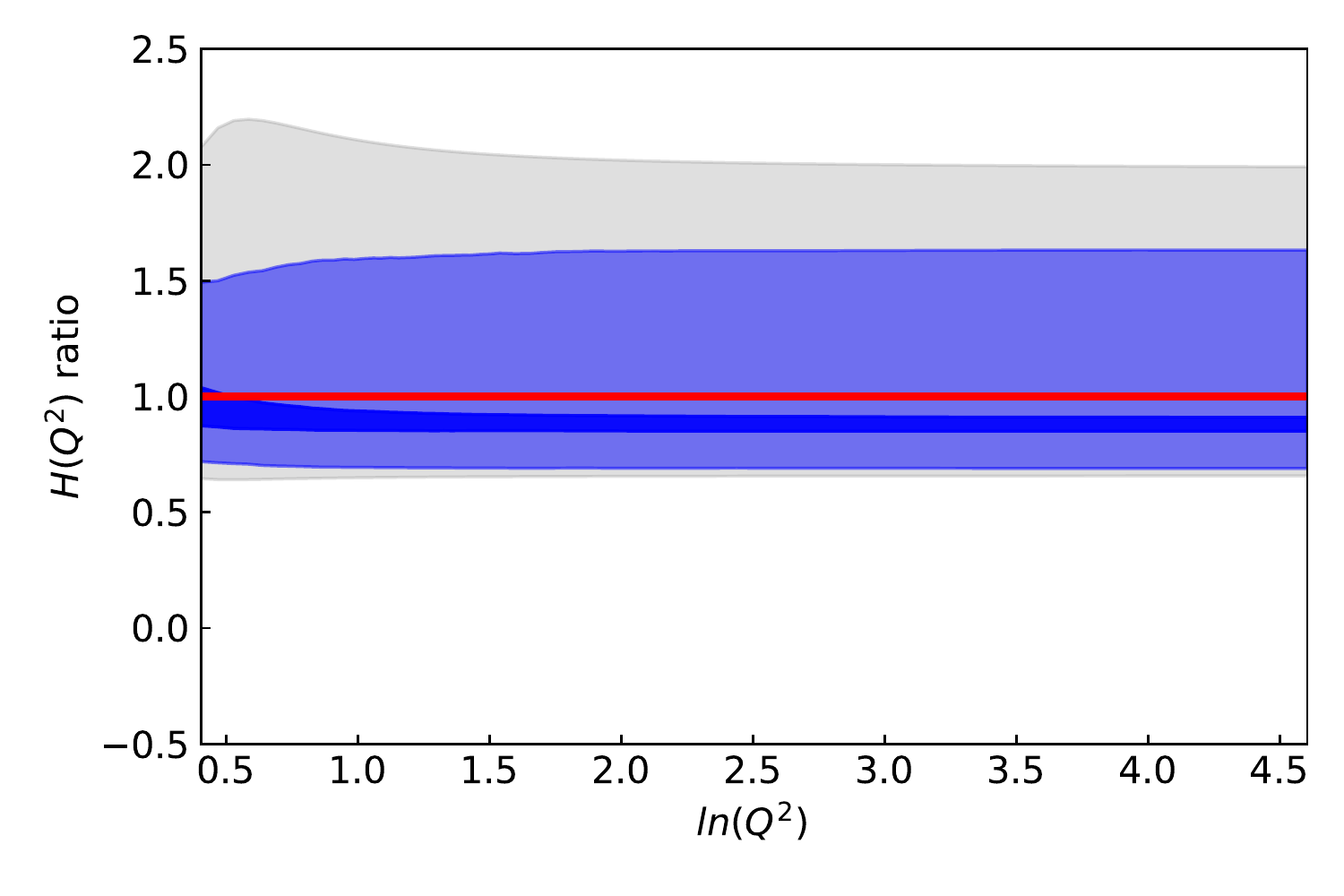}
	\includegraphics[width=0.48\textwidth]{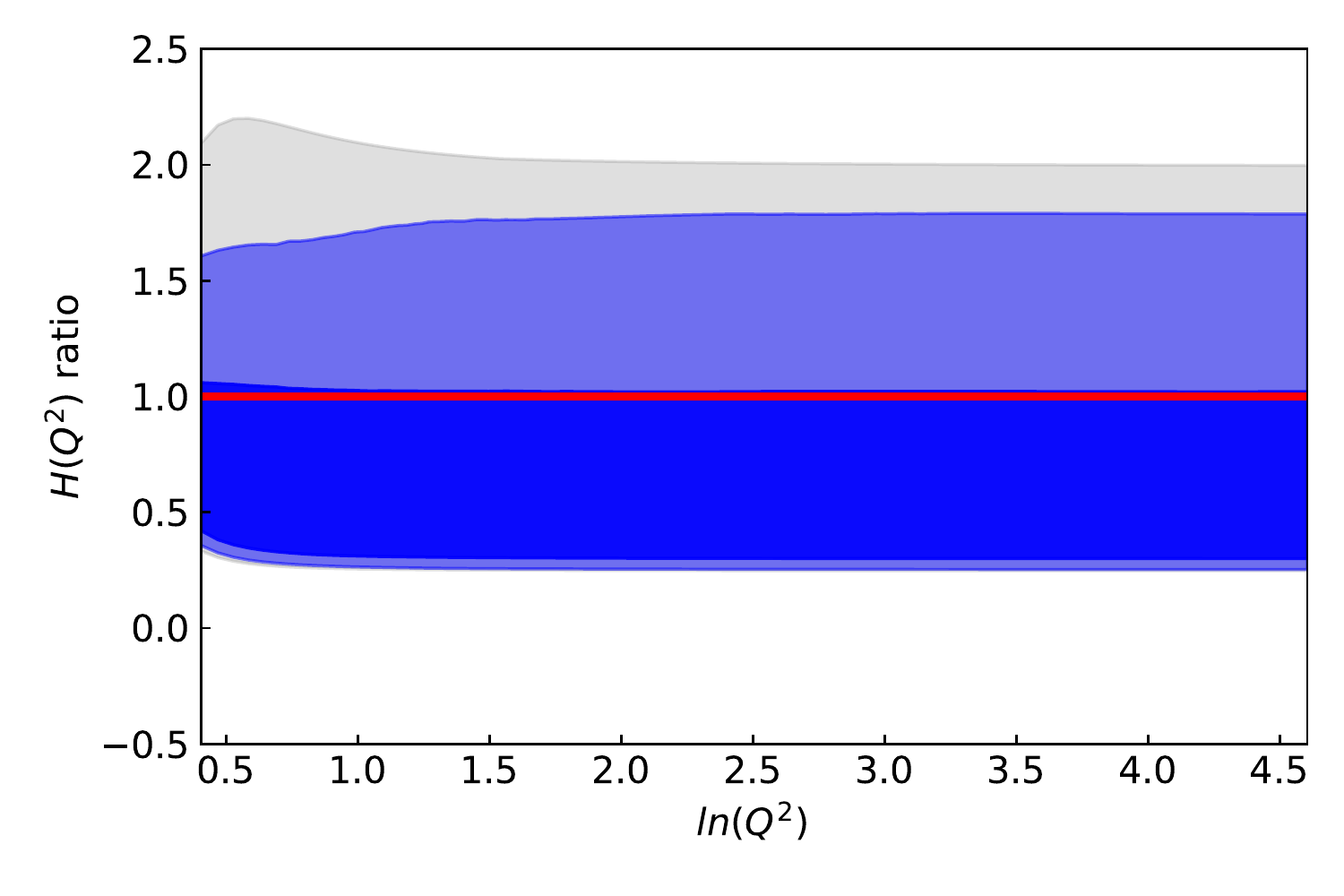}
	\includegraphics[width=0.48\textwidth]{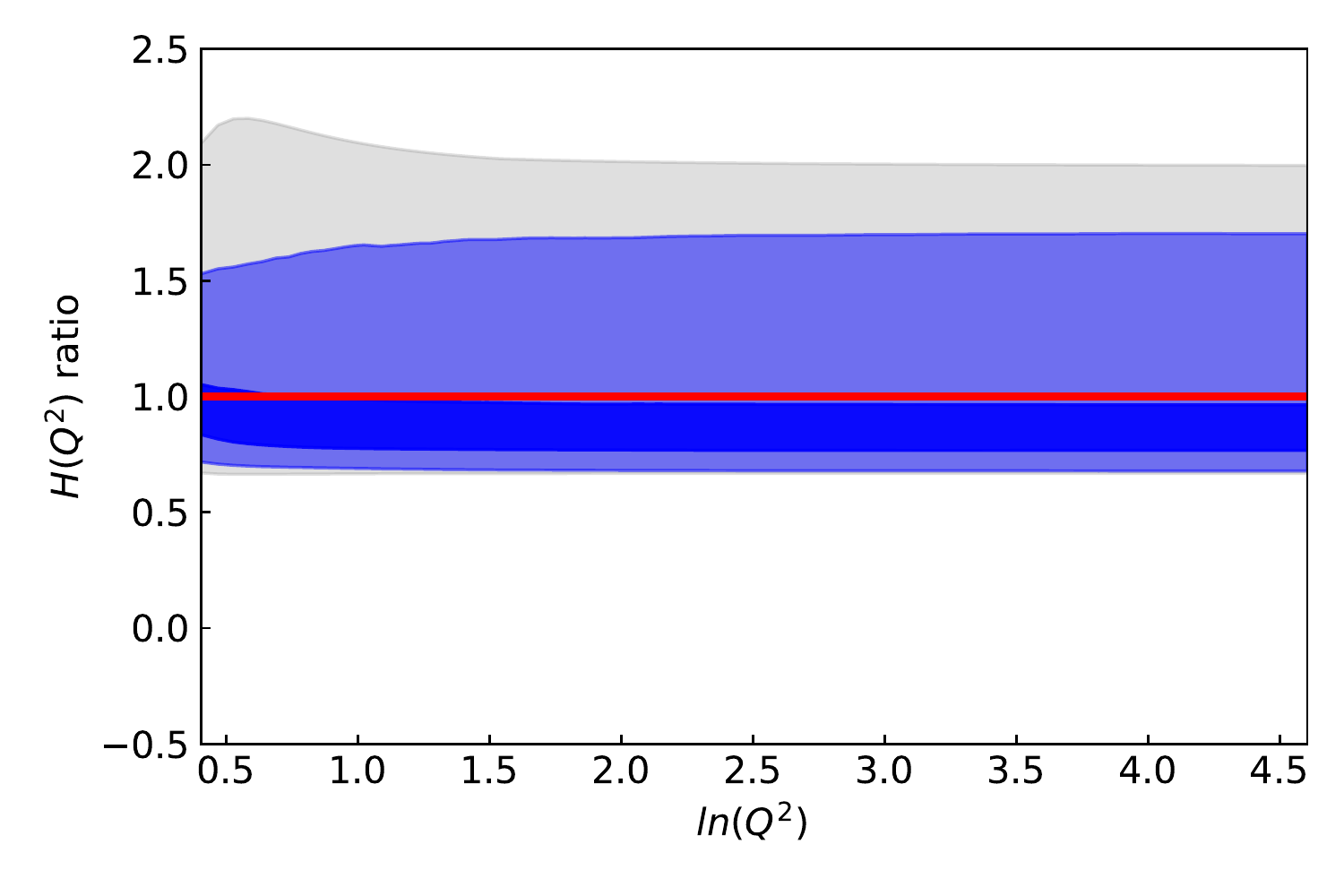}
	\includegraphics[width=0.48\textwidth]{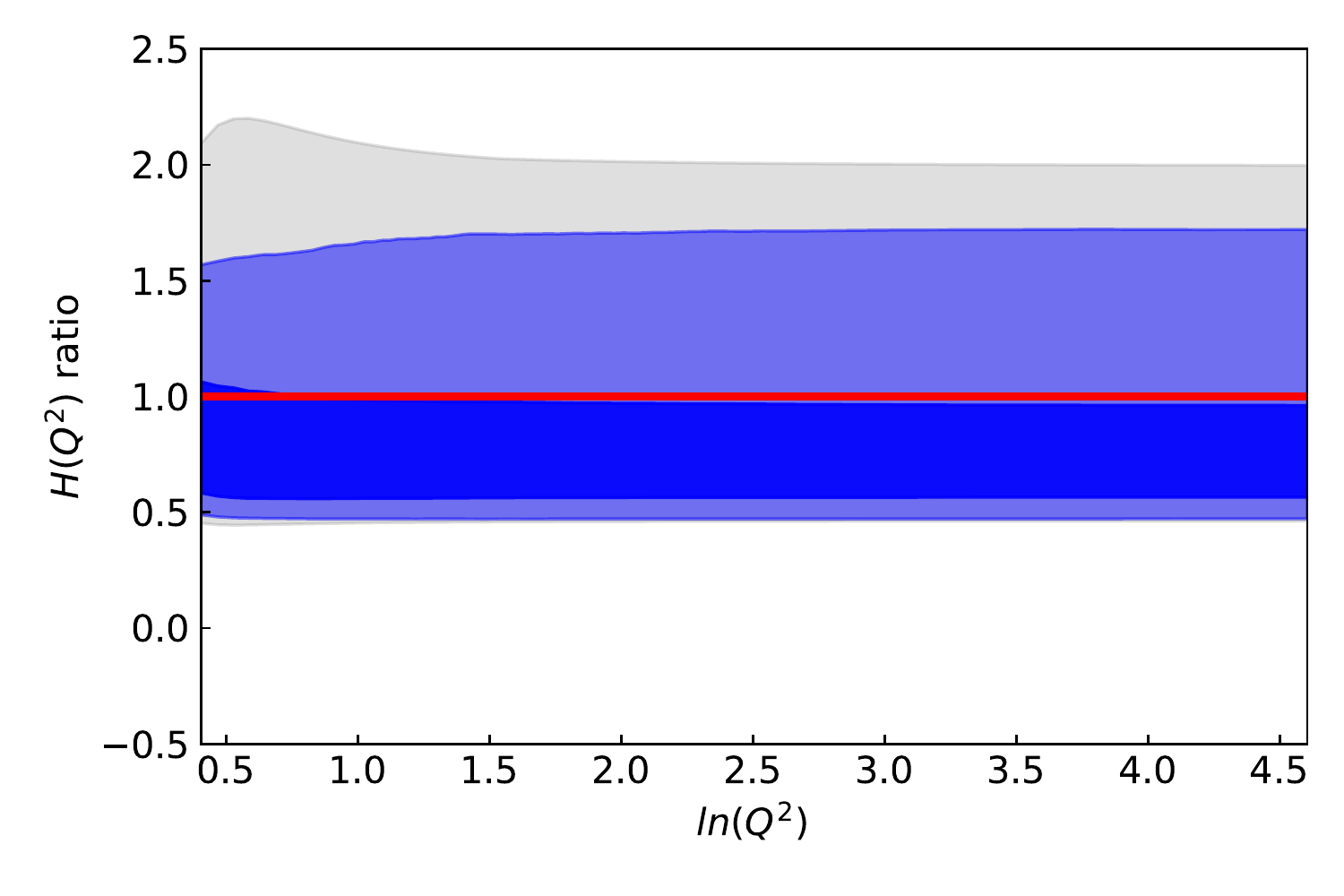}
	\includegraphics[width=0.48\textwidth]{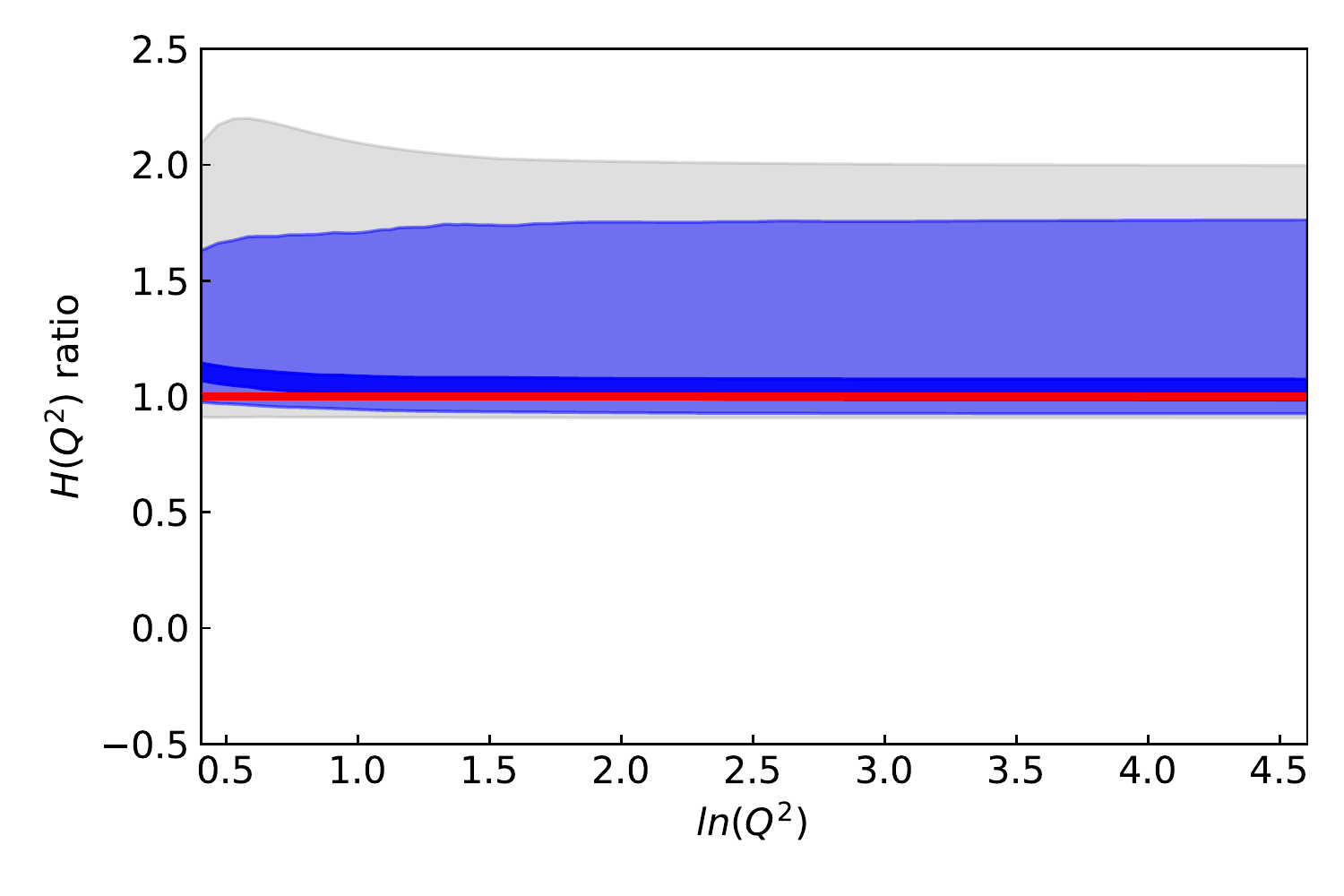}
	\includegraphics[width=0.48\textwidth]{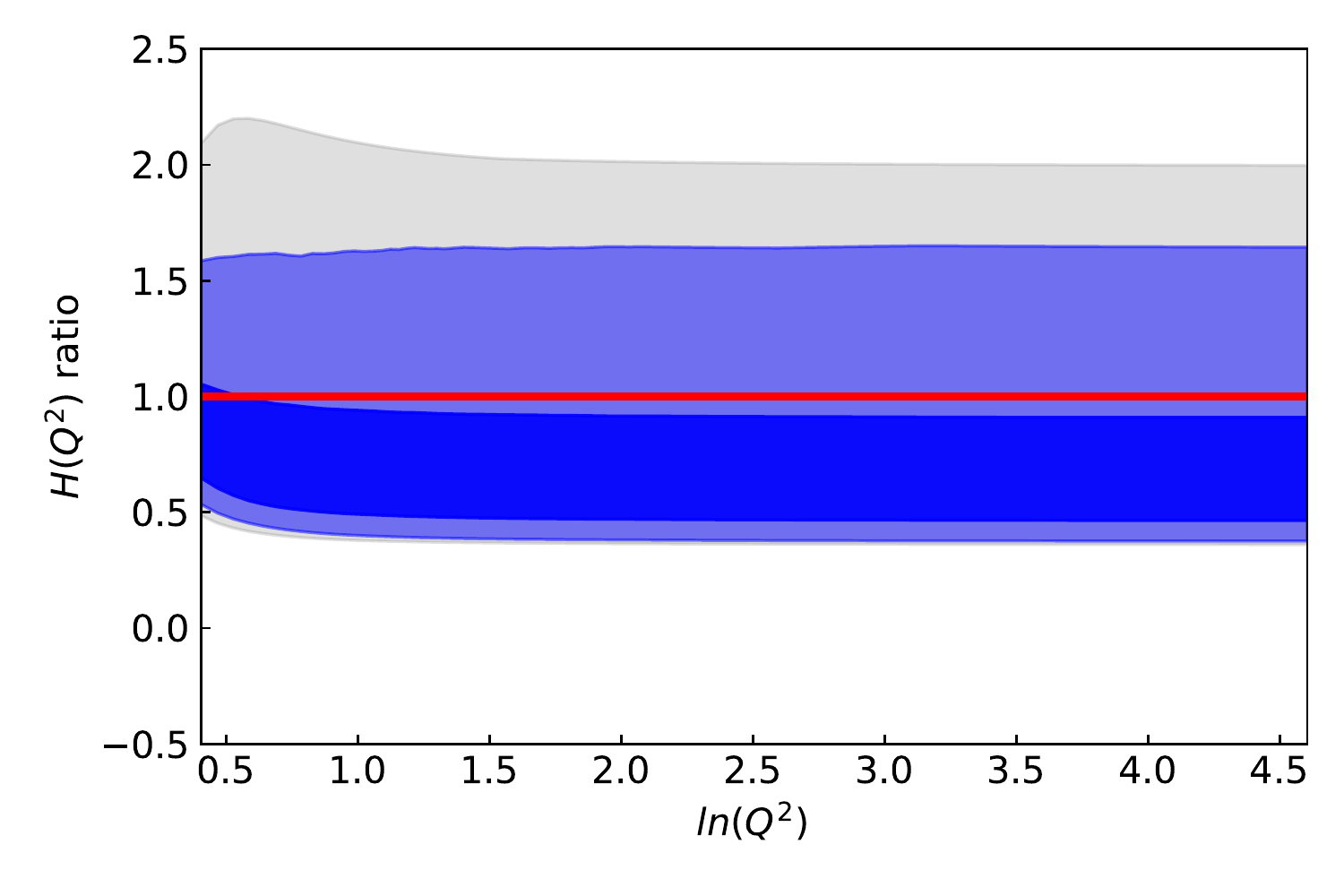}
	\includegraphics[width=0.48\textwidth]{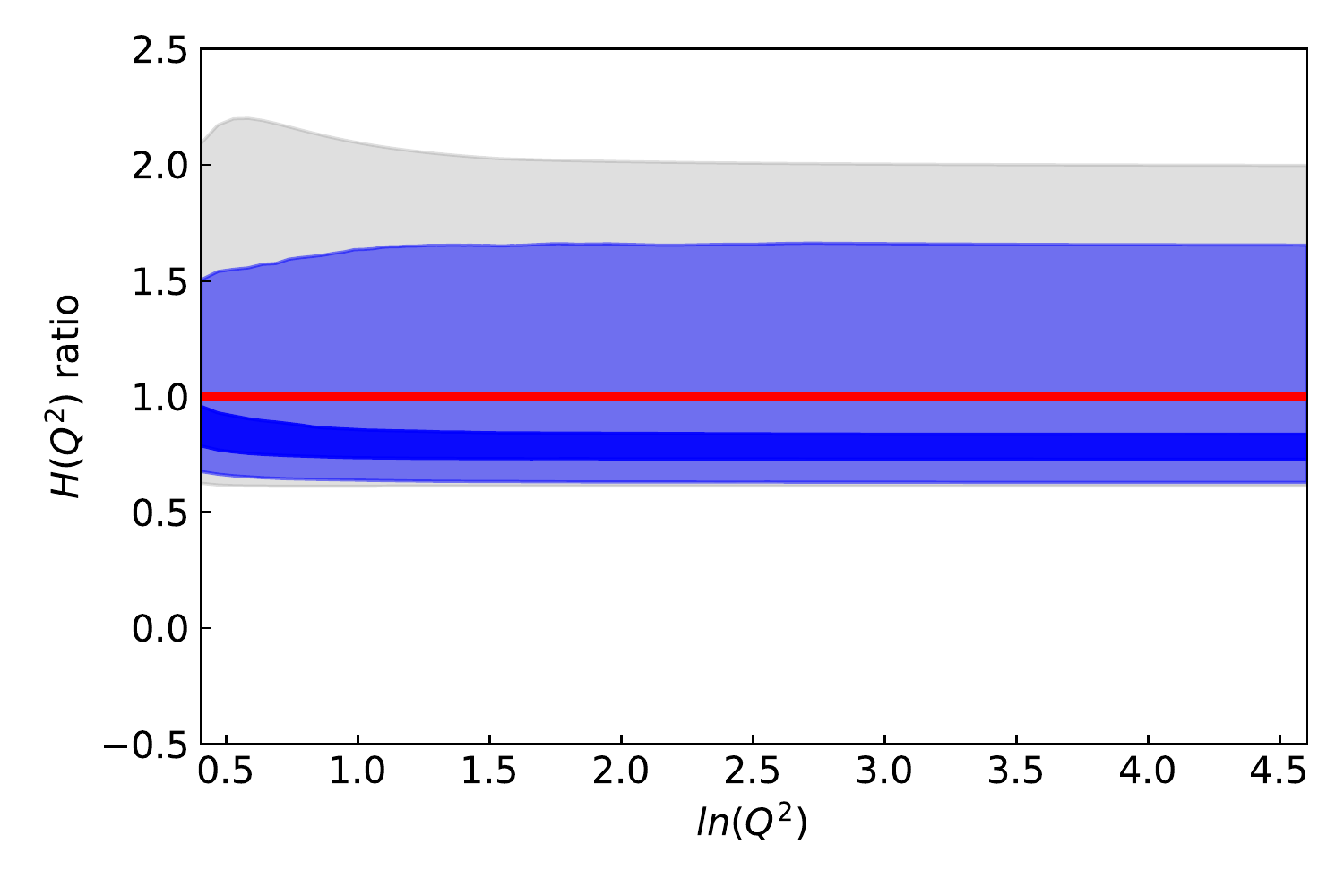}
	\includegraphics[width=0.48\textwidth]{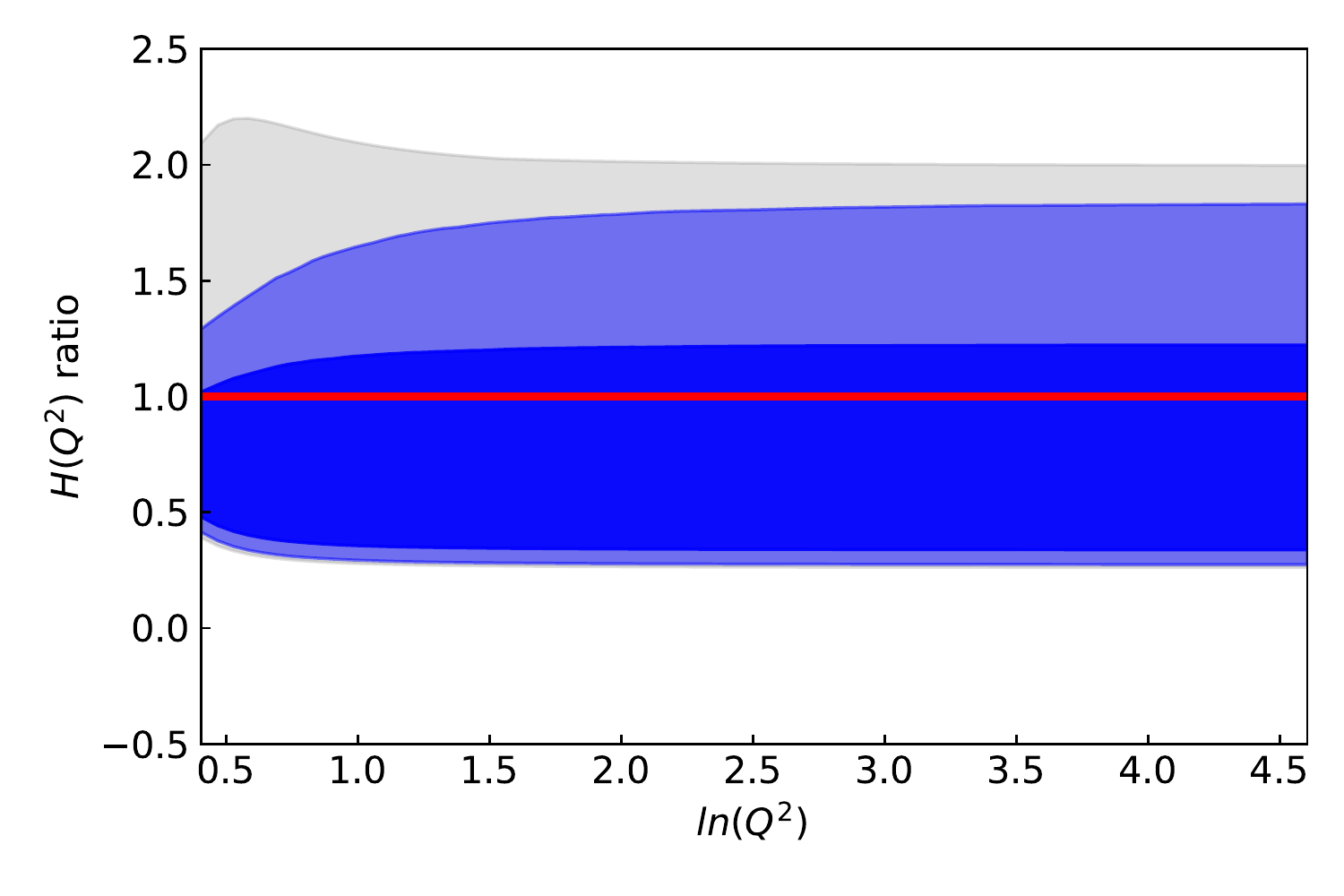}
	\caption{\label{fig:closure_own_fluc_23} Closure tests of $c_1$ and $c_2$ plotted as the ratio of $H(Q^2)$ using 8 sets of mock data with statistical fluctuations from model calculations.}
\end{figure}

\begin{figure}
	\centering
	\includegraphics[width=0.32\textwidth]{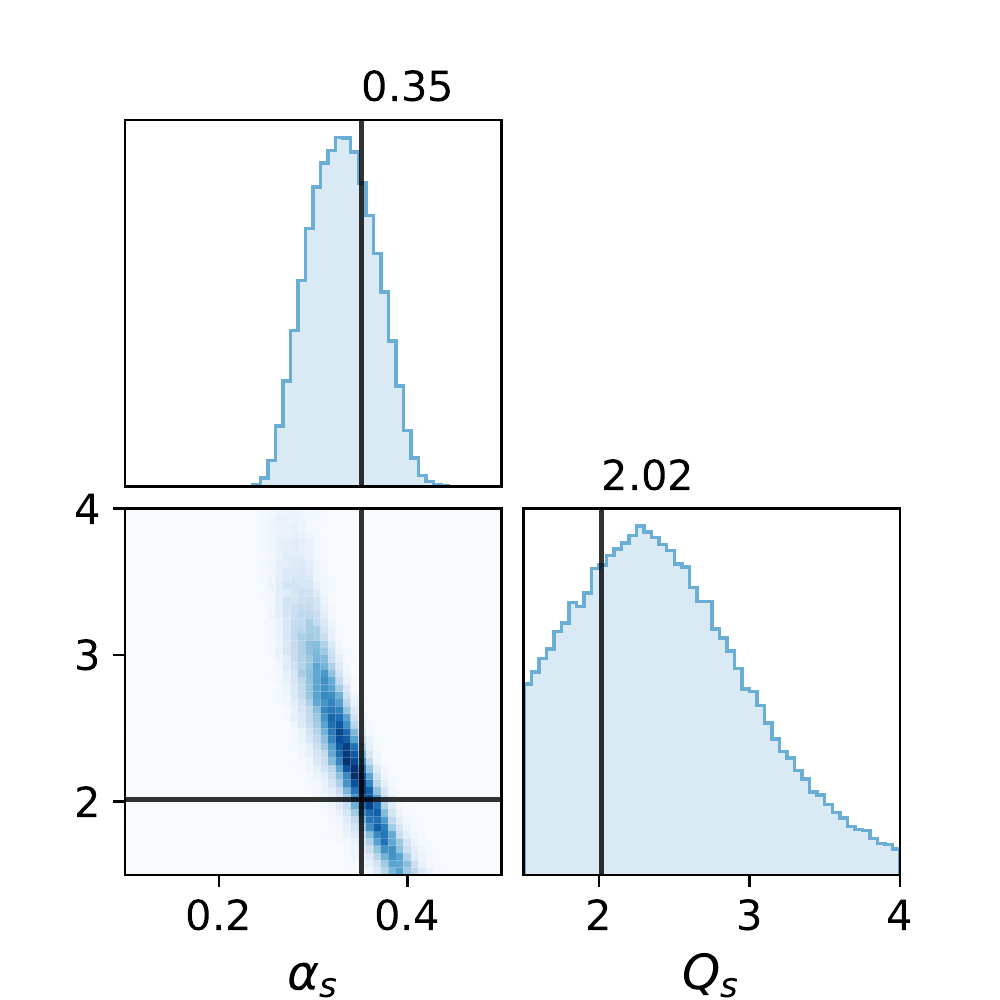}
	\includegraphics[width=0.32\textwidth]{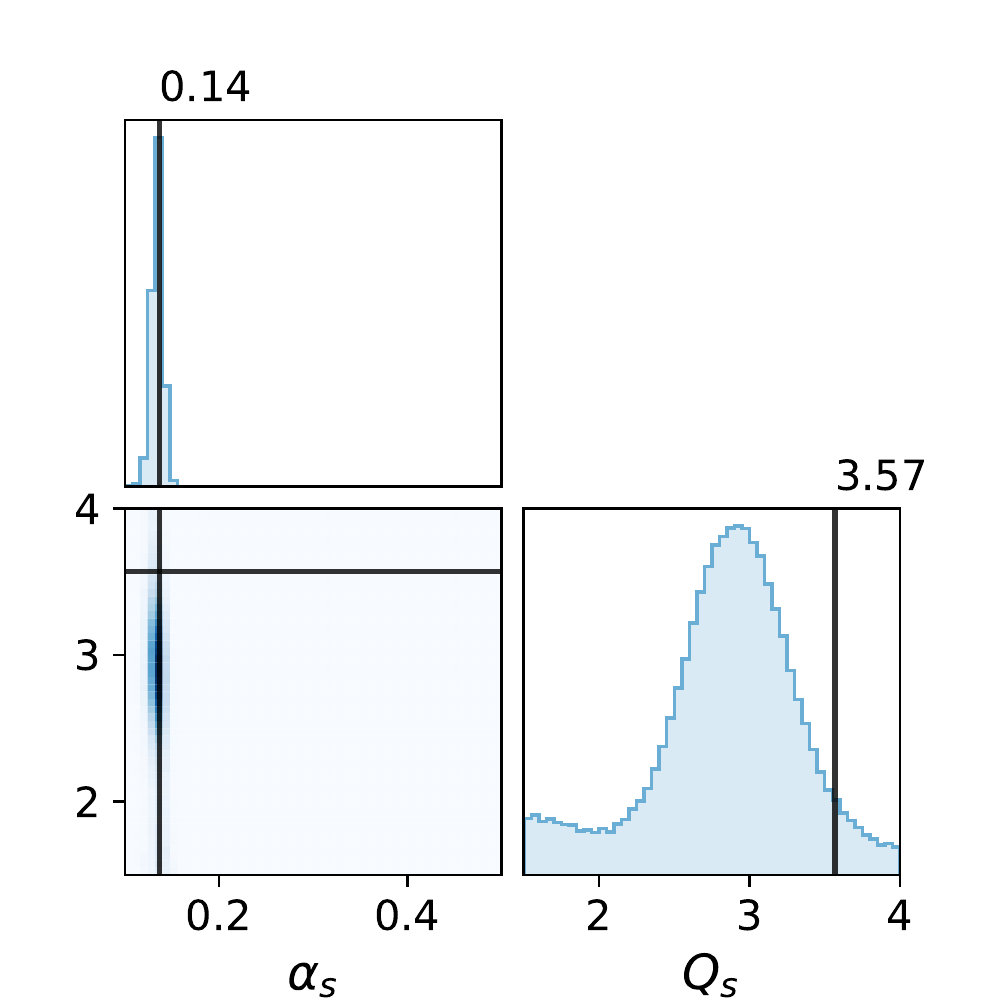}
	\includegraphics[width=0.32\textwidth]{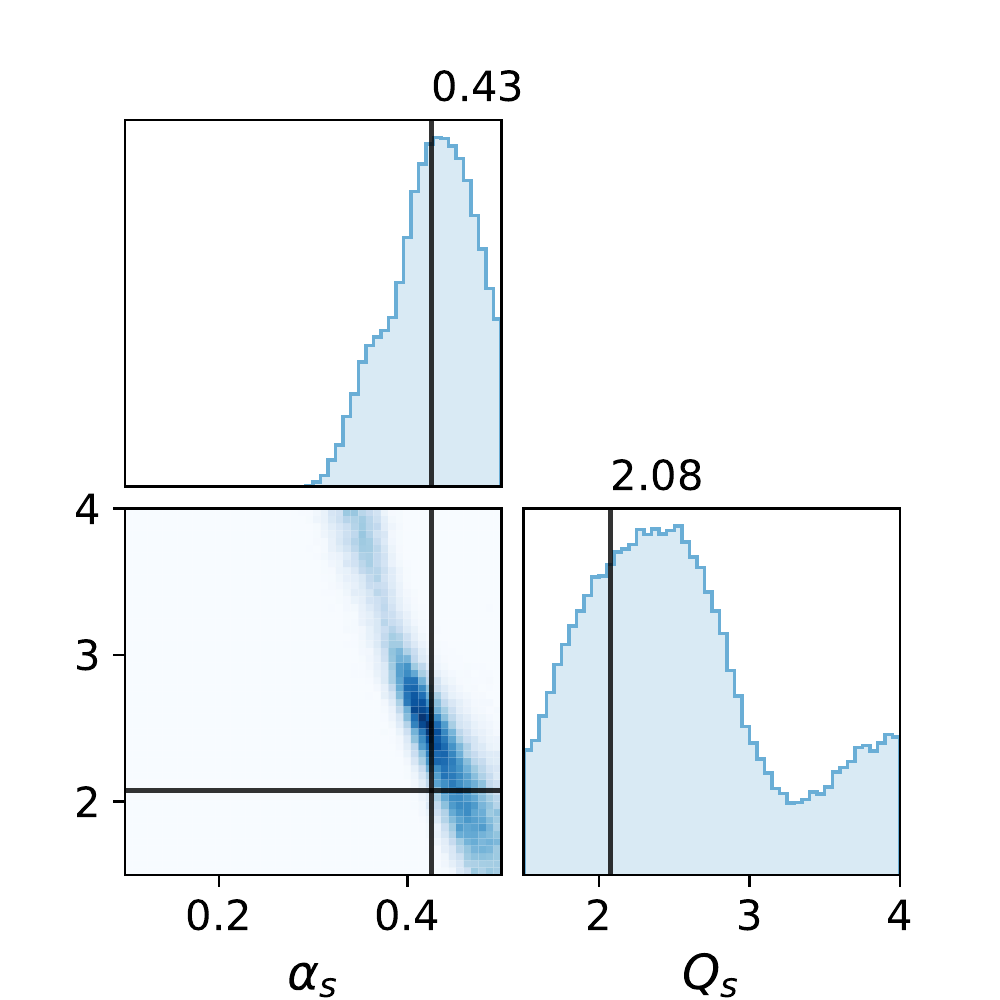}
	\includegraphics[width=0.32\textwidth]{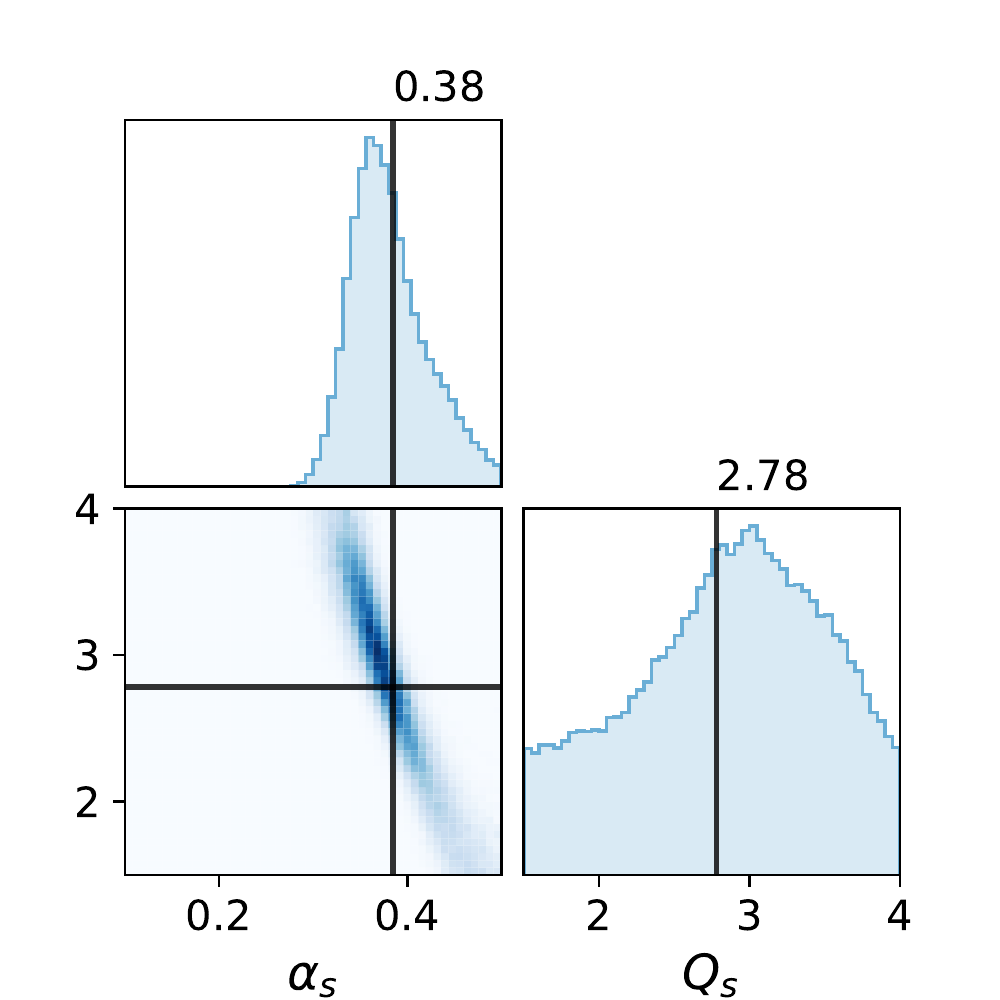}
	\includegraphics[width=0.32\textwidth]{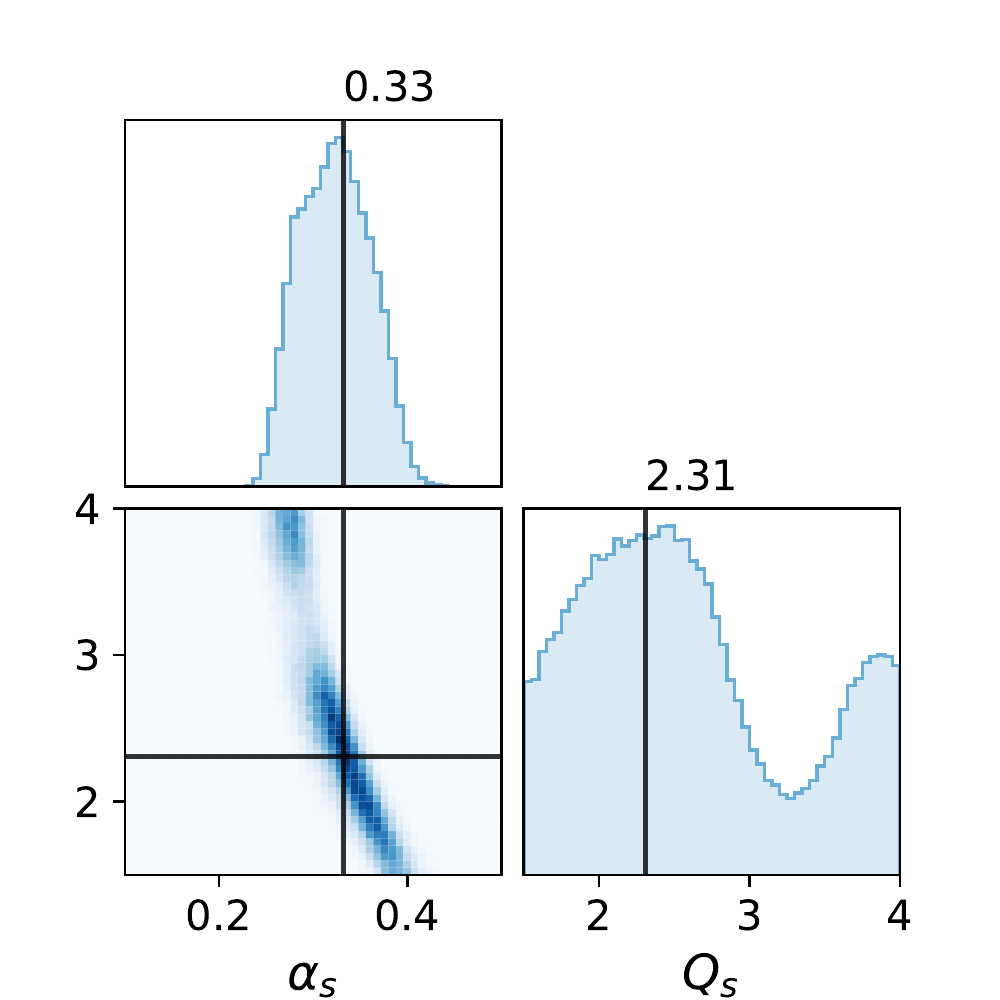}
	\includegraphics[width=0.32\textwidth]{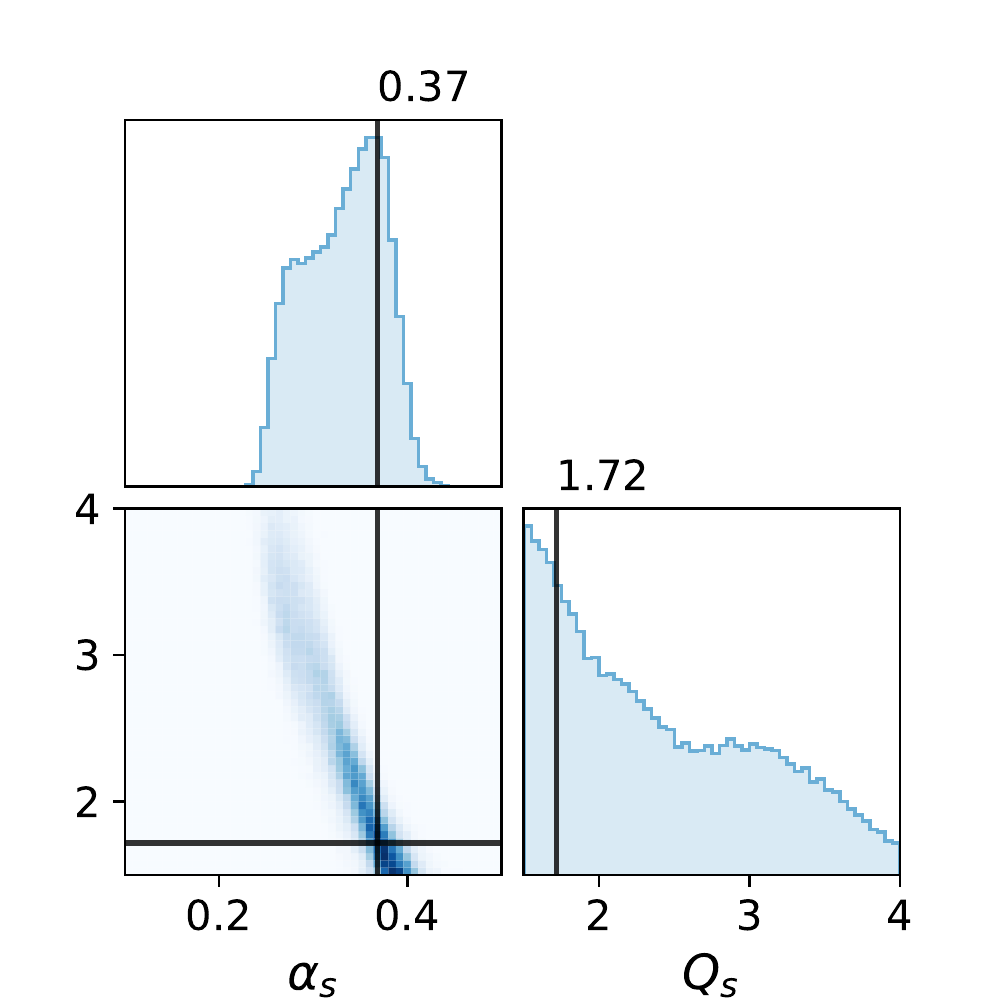}
	\includegraphics[width=0.32\textwidth]{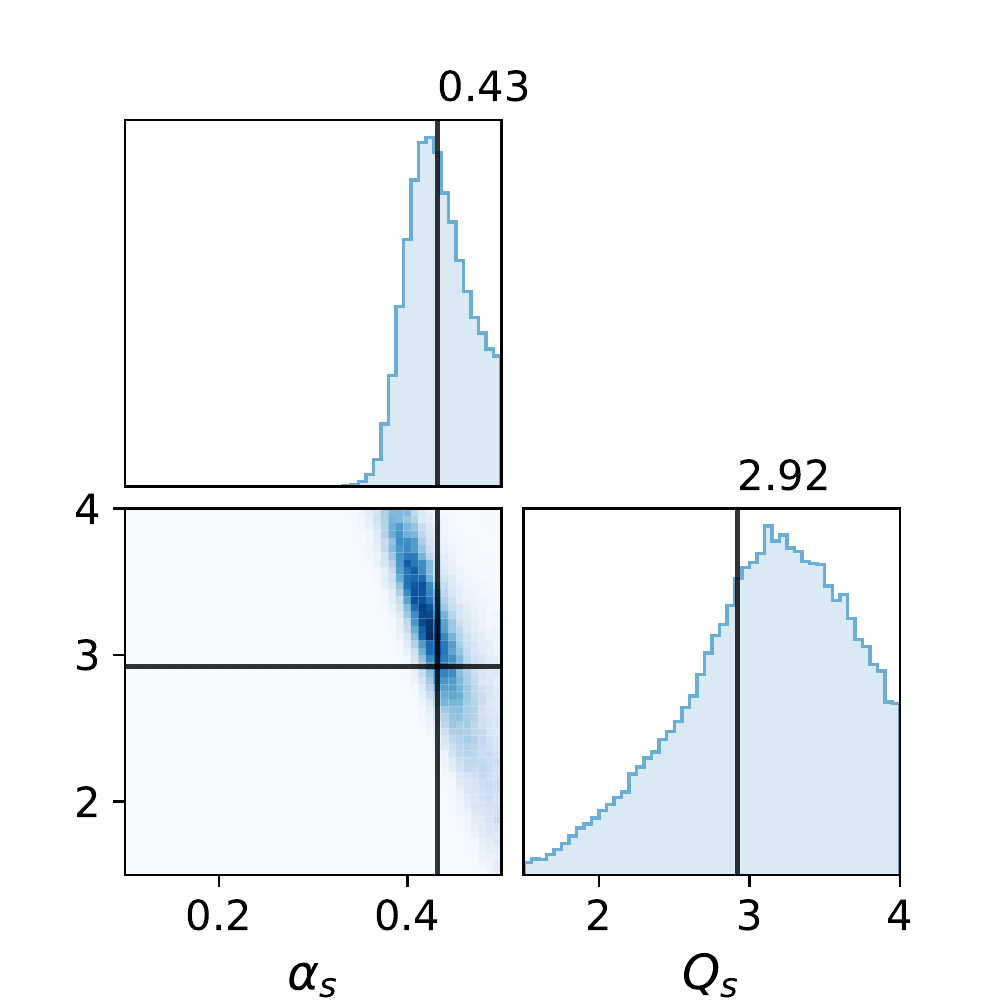}
	\includegraphics[width=0.32\textwidth]{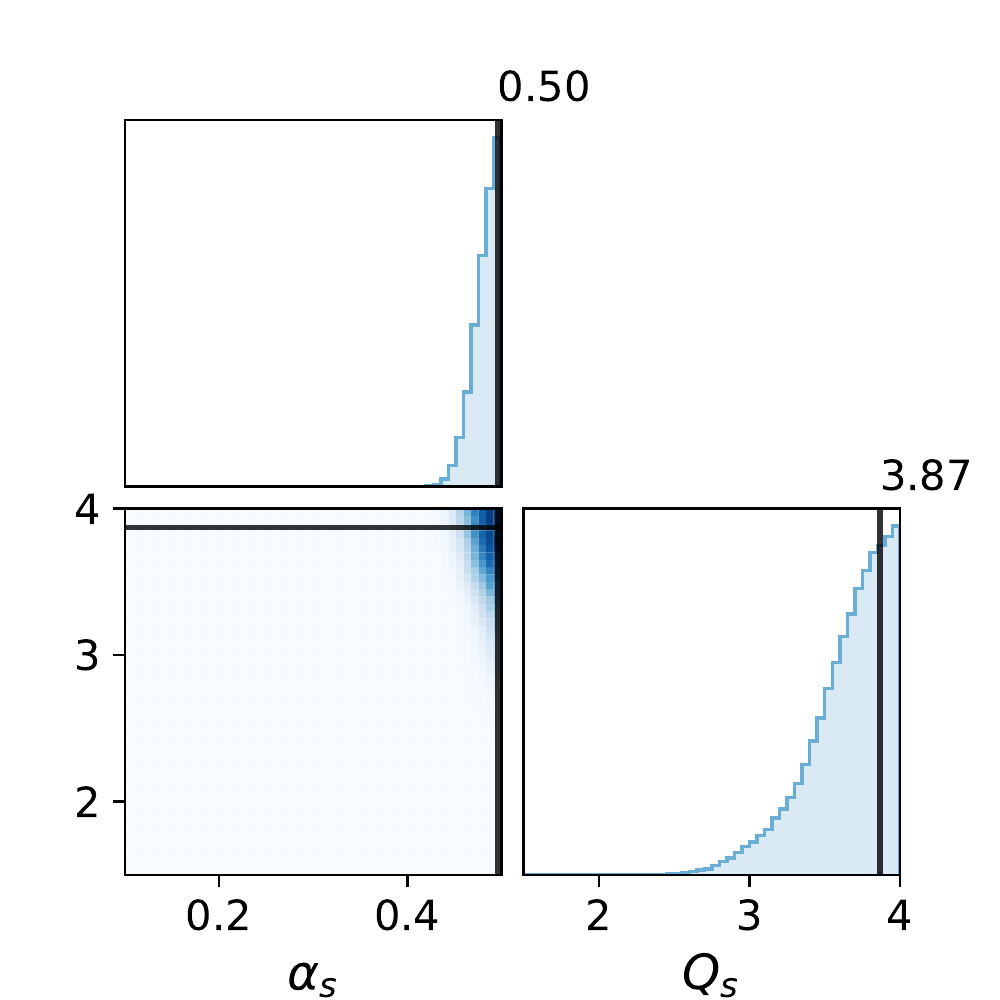}
	\caption{\label{fig:closure_exp_fluc_14} Closure tests of $\alpha_s$ and $Q_s$ using 8 sets of mock data with statistical and systematic fluctuations from experiments.}
\end{figure}

\begin{figure}
	\centering
	\includegraphics[width=0.48\textwidth]{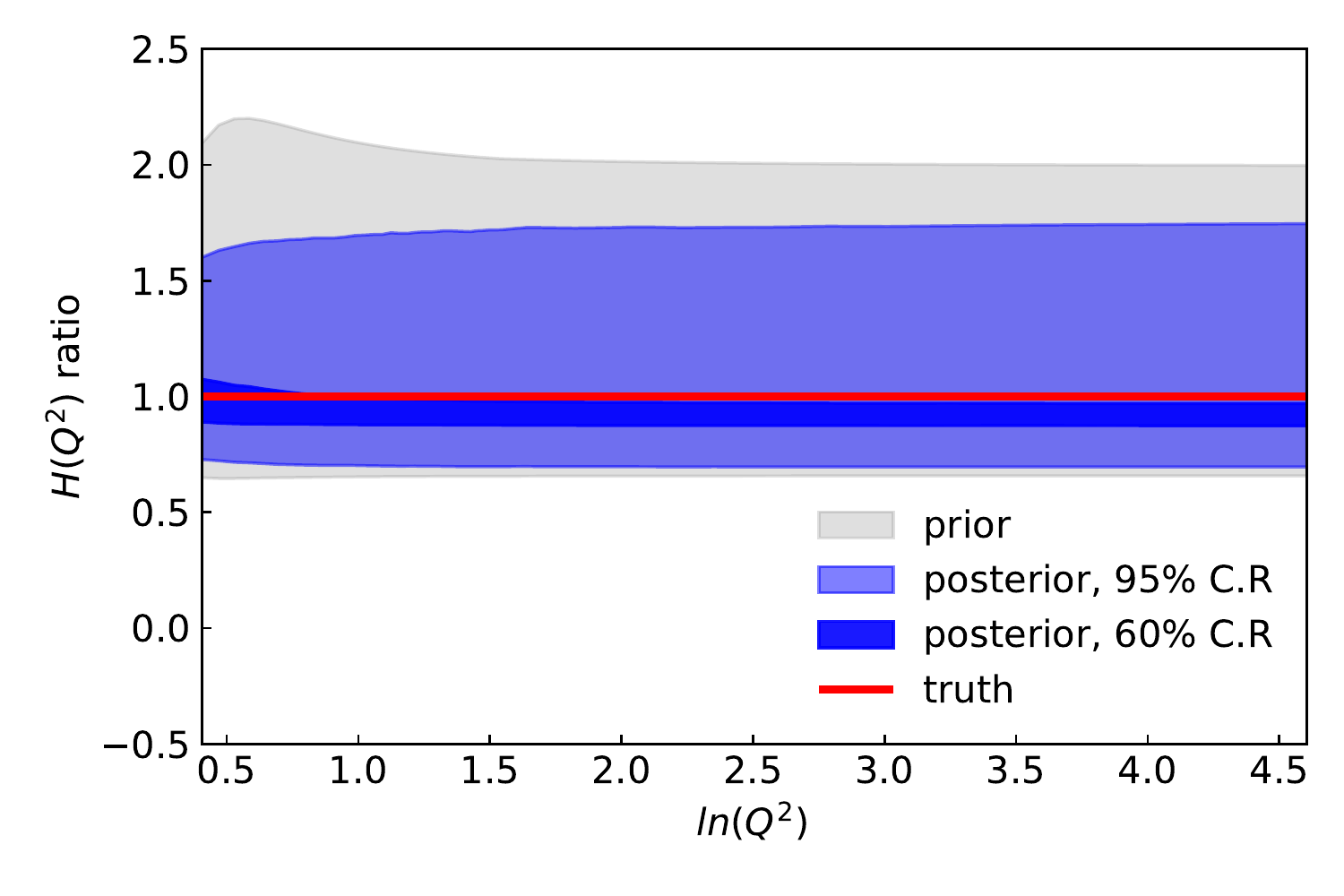}
	\includegraphics[width=0.48\textwidth]{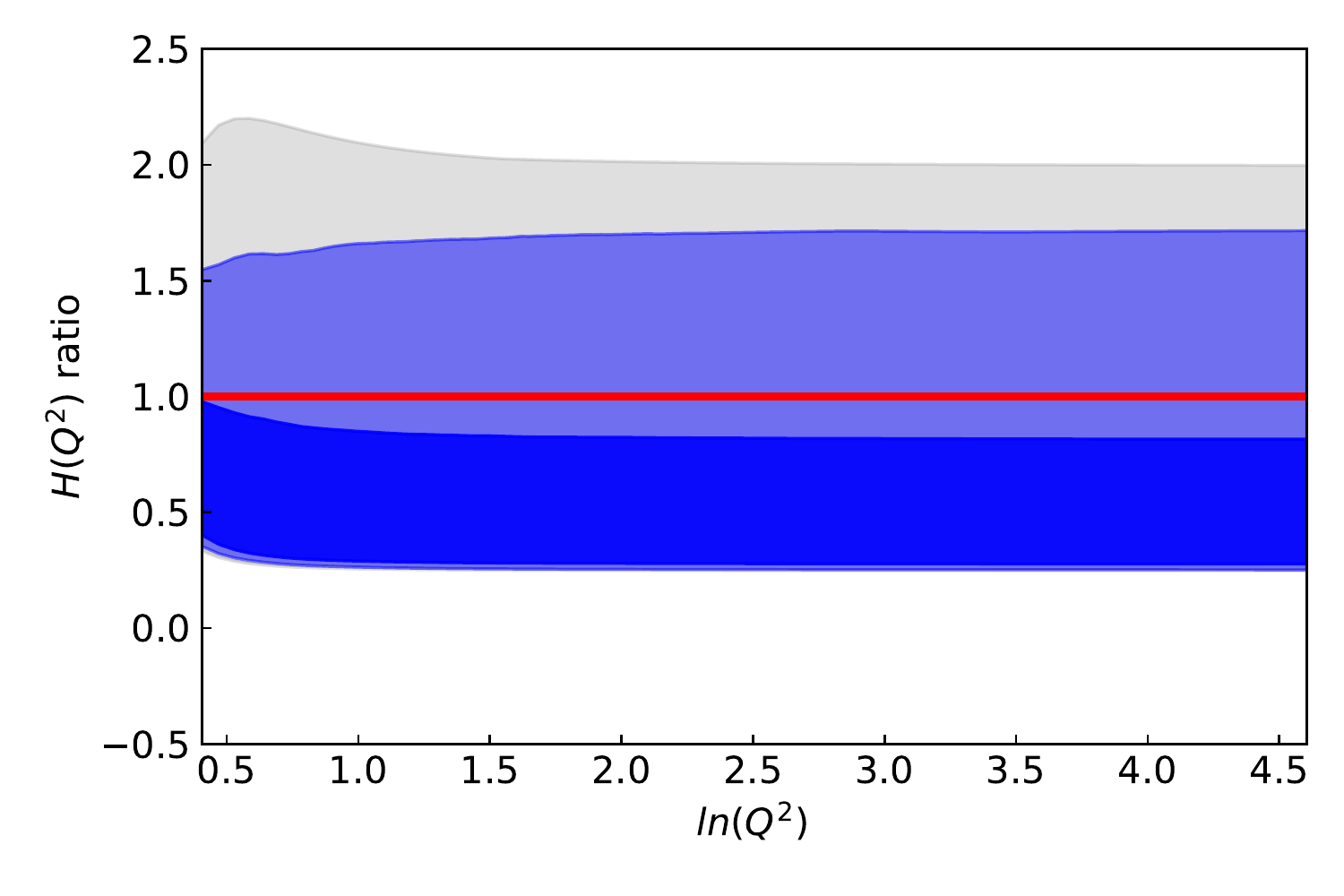}
	\includegraphics[width=0.48\textwidth]{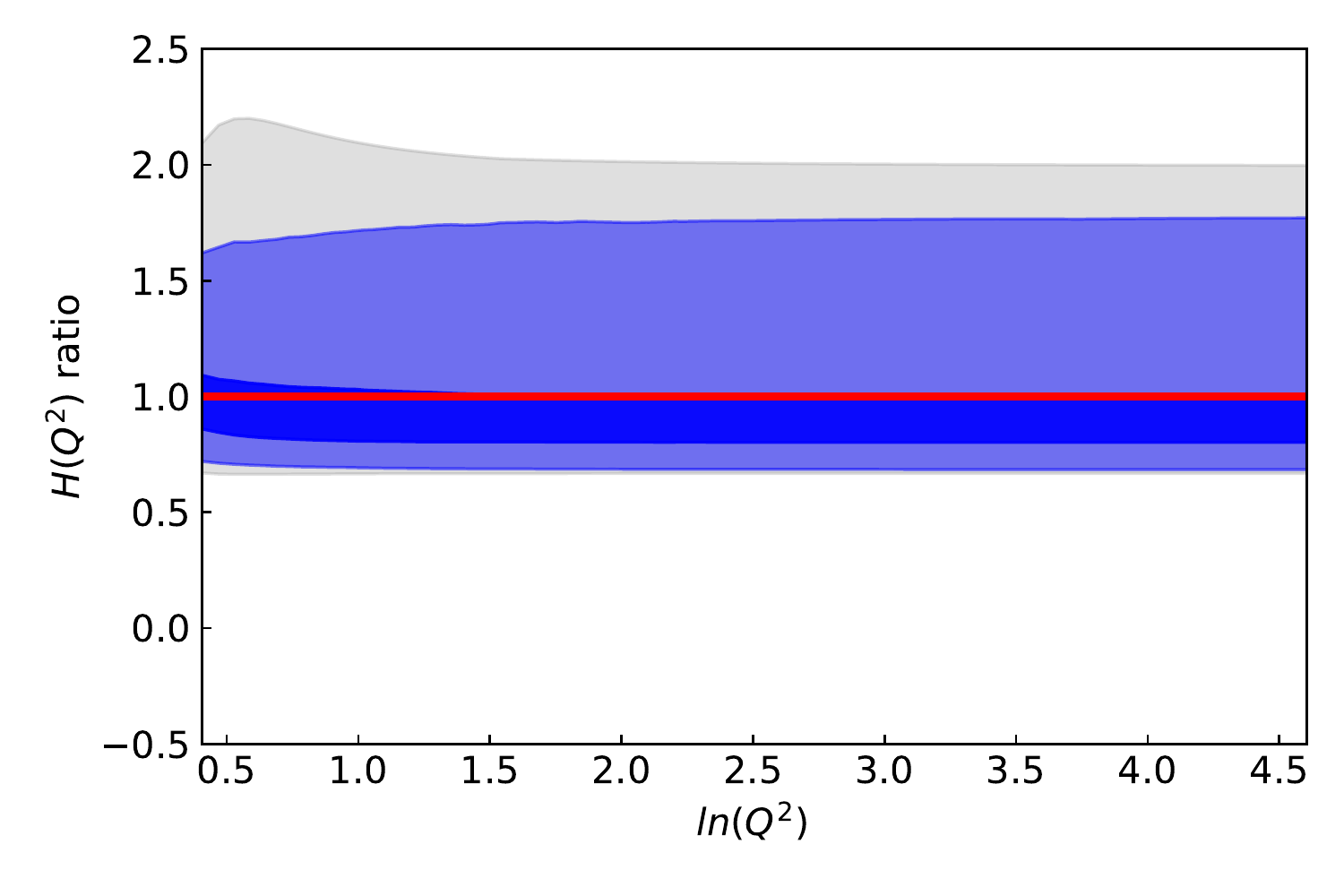}
	\includegraphics[width=0.48\textwidth]{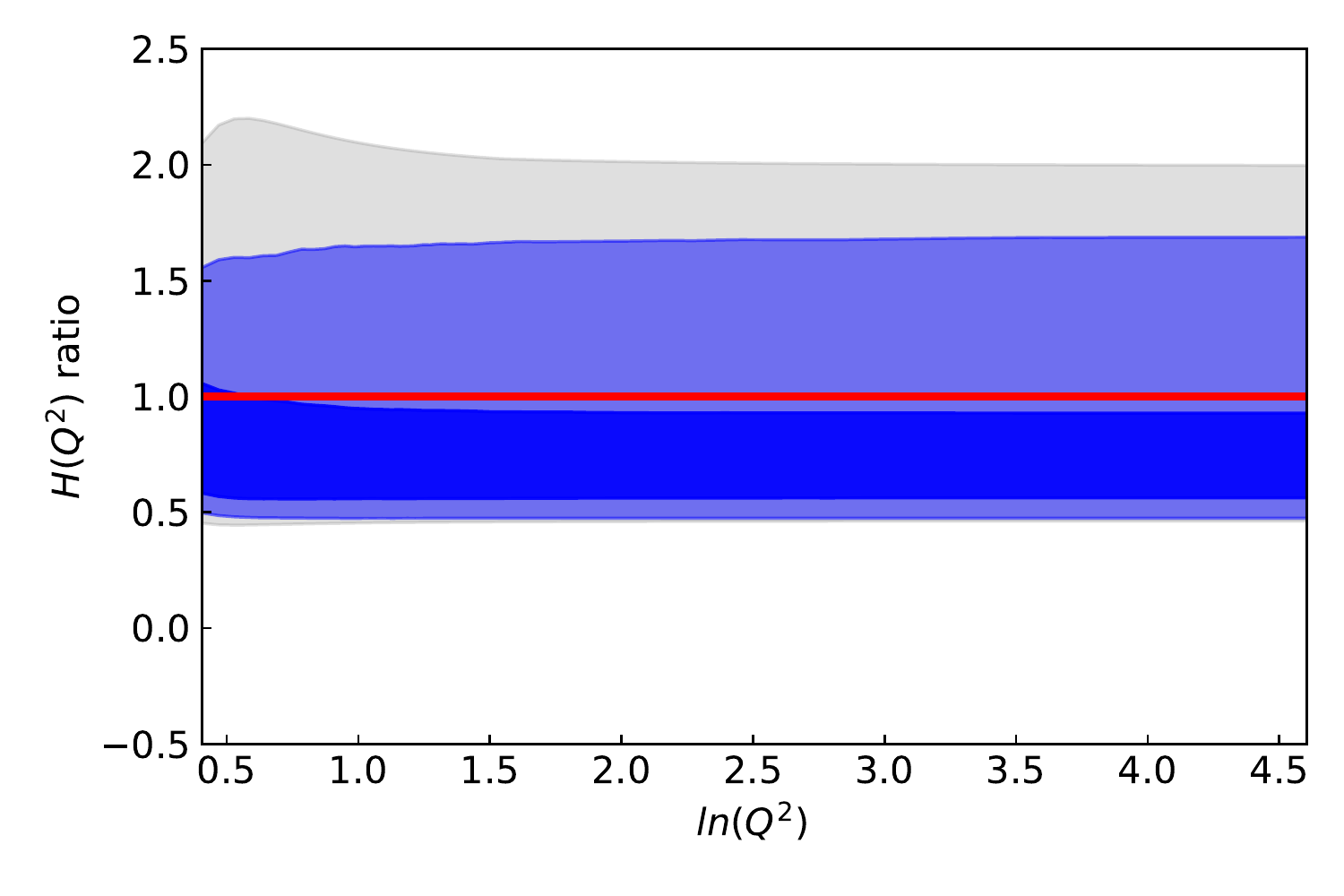}
	\includegraphics[width=0.48\textwidth]{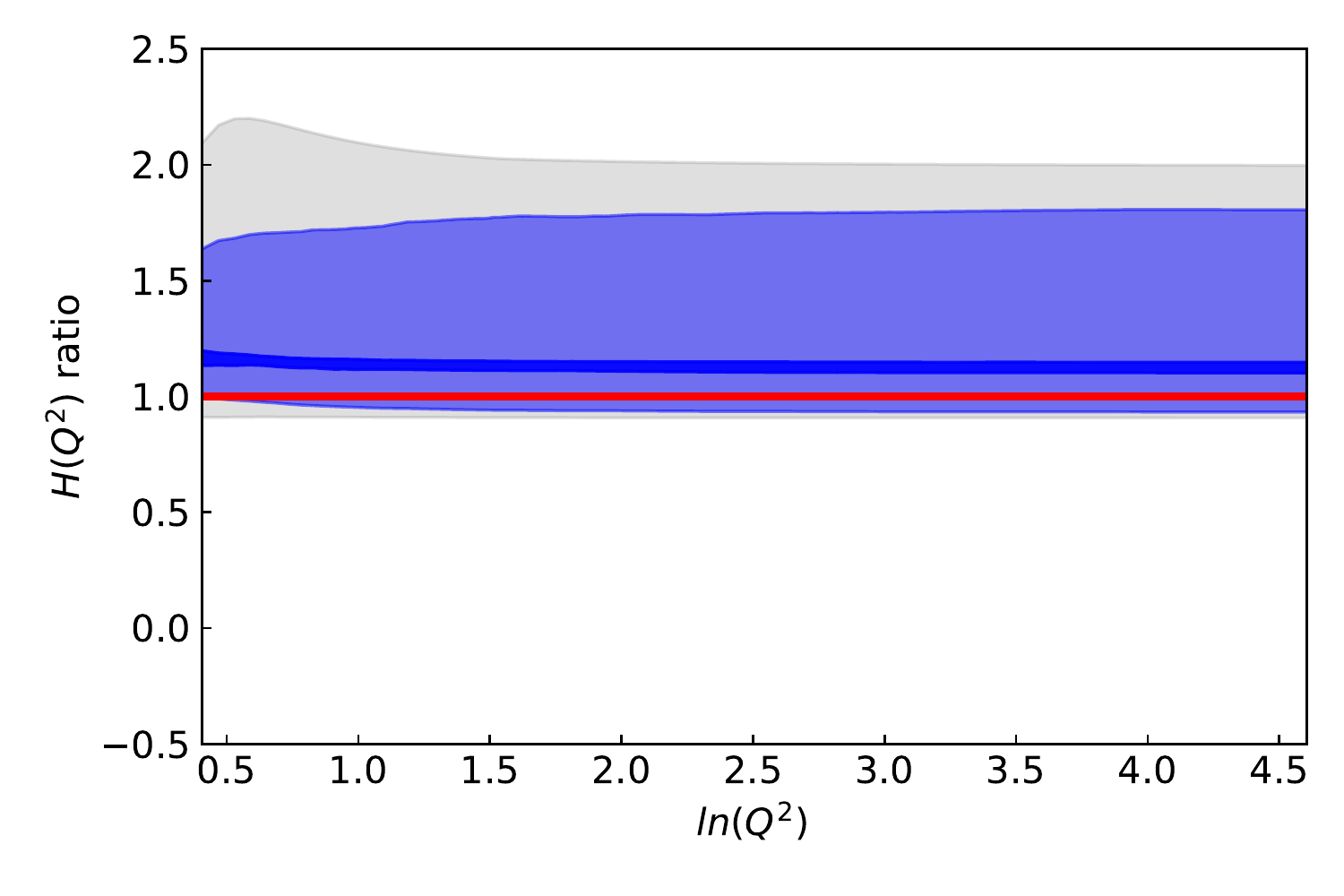}
	\includegraphics[width=0.48\textwidth]{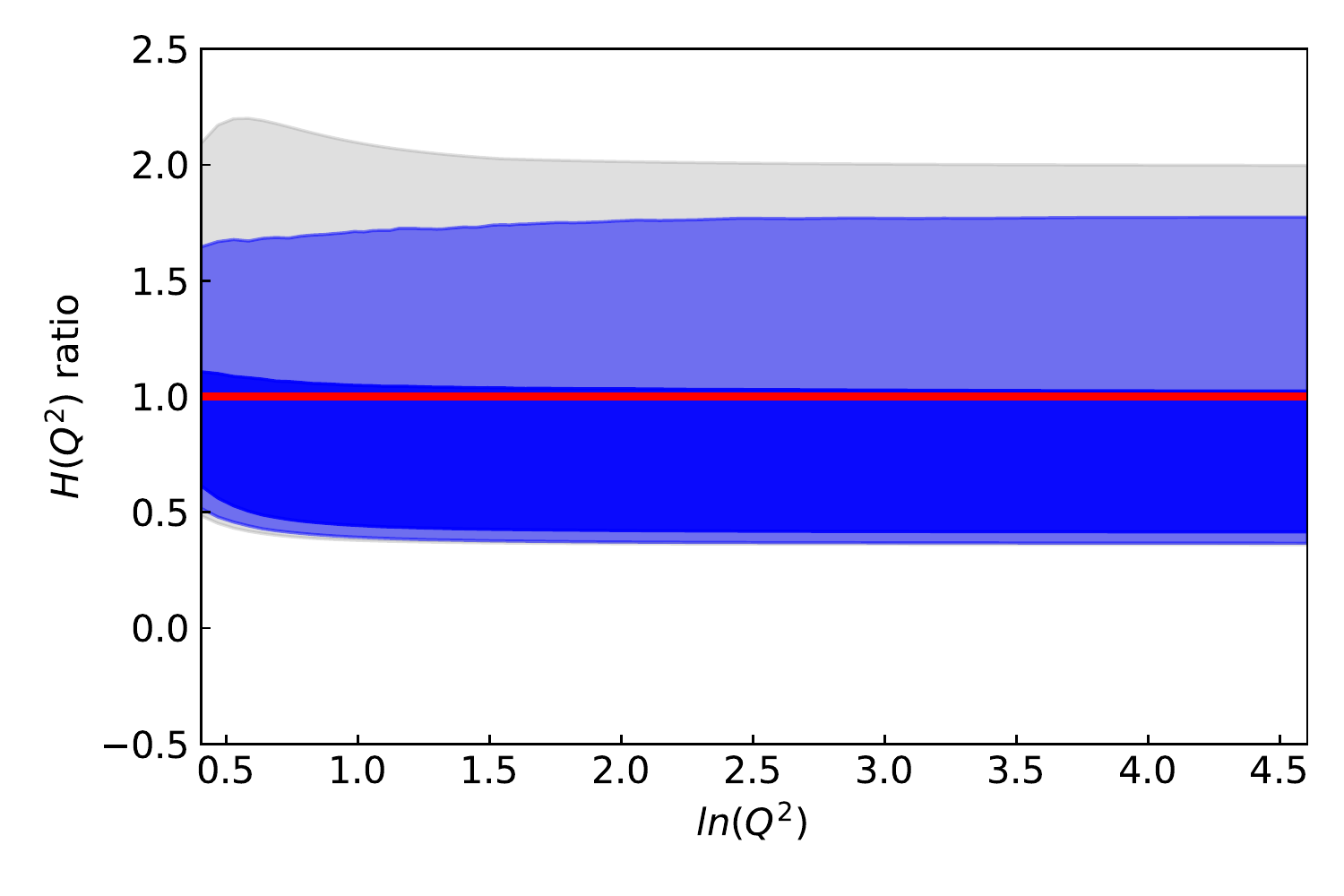}
	\includegraphics[width=0.48\textwidth]{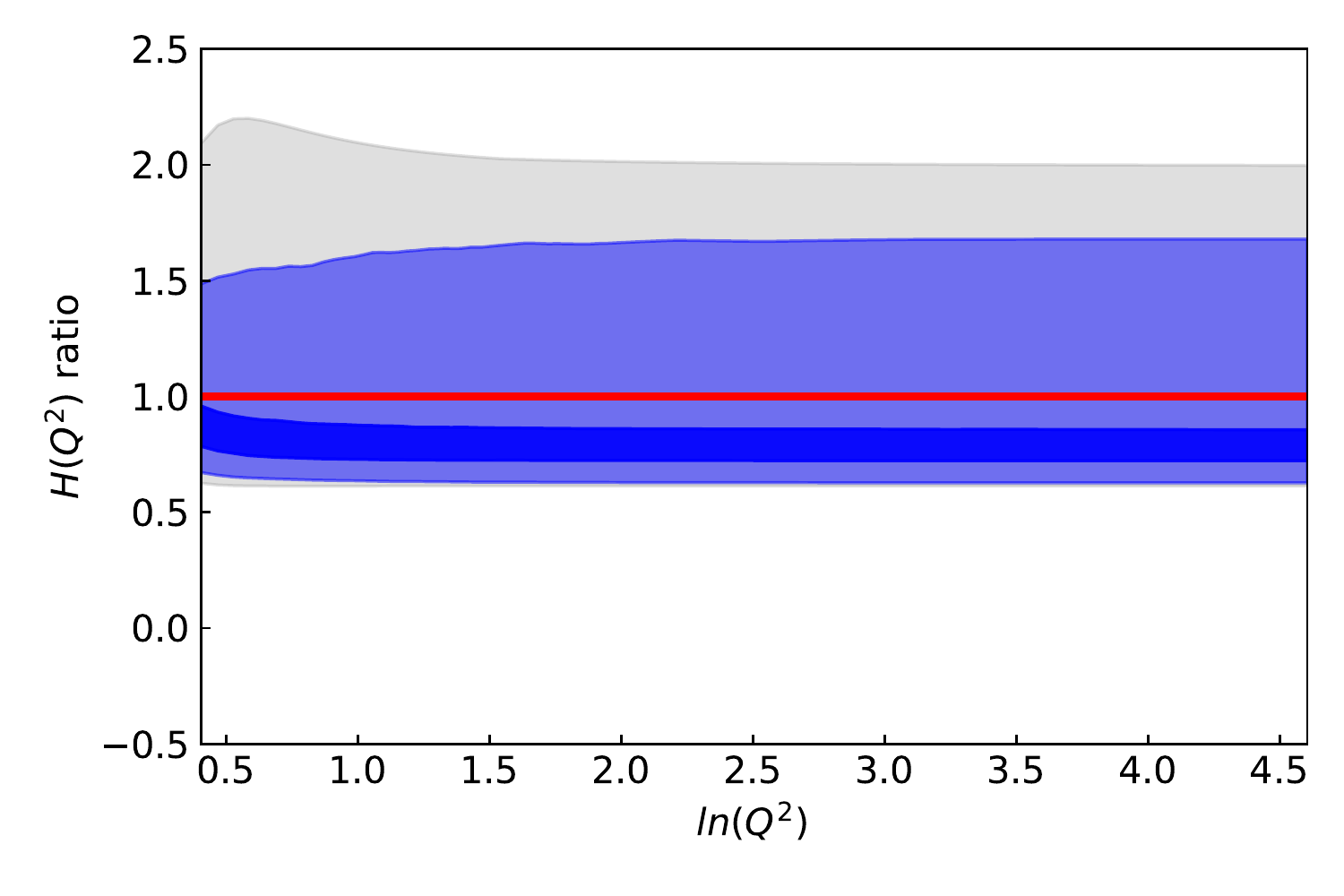}
	\includegraphics[width=0.48\textwidth]{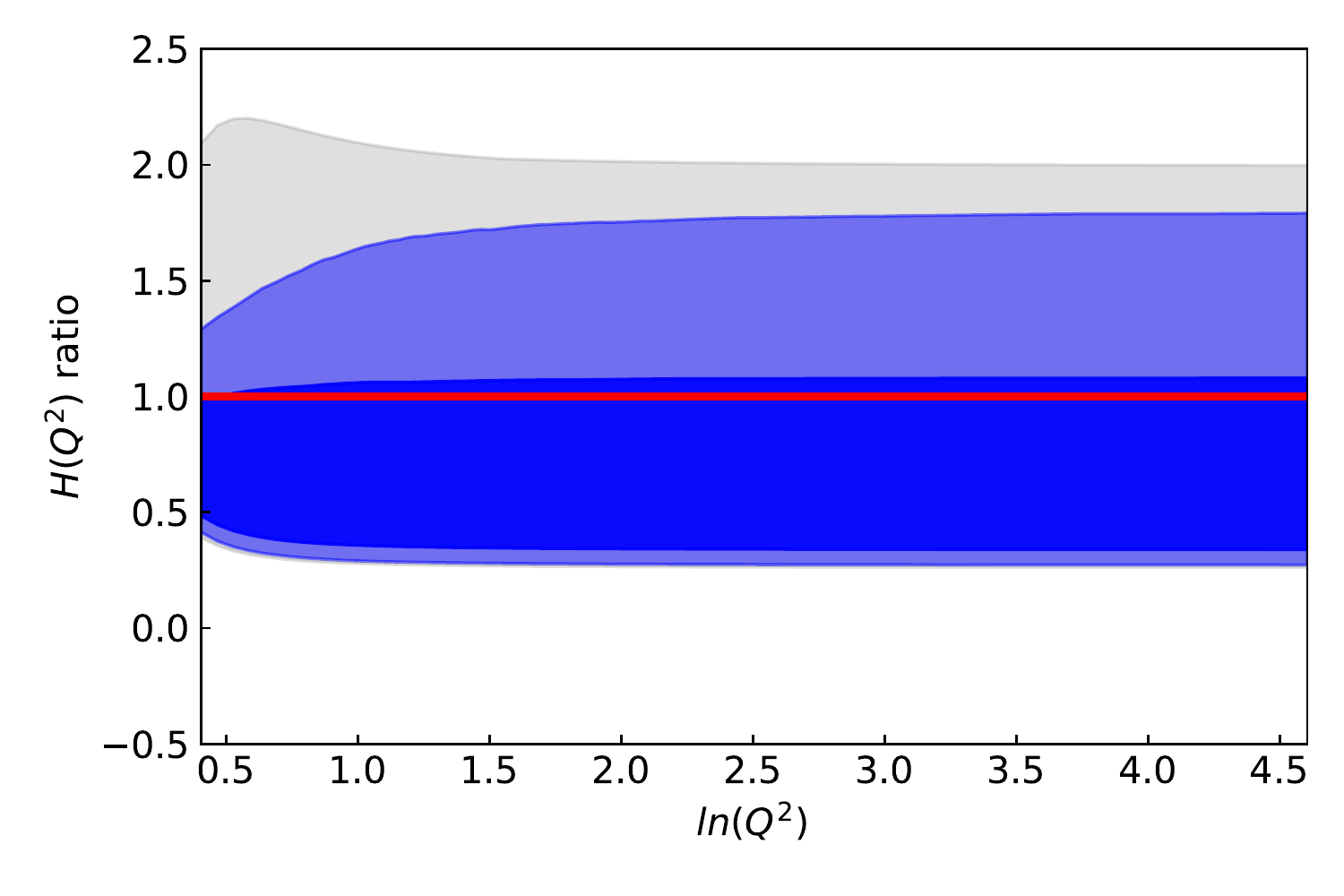}
	\caption{\label{fig:closure_exp_fluc_23} Closure tests of $c_1$ and $c_2$ plotted as the ratio of $H(Q^2)$ using 8 sets of mock data with statistical and systematic fluctuations from experiments.}
\end{figure}

Being able to predict the training data does not guarantee that our emulator can constrain the model parameters well. If the data are not sensitive to some parameters, those parameters may end up with a wide posterior. If the data are degenerate (can be described with multiple sets of parameters), the posterior distributions may end up with multiple peaks. If the emulator captures too much fluctuation in the model calculation, the distance in the parameter space may be distorted in the observable space. These scenarios can be checked by performing a cross-validation closure test, where one design point is taken out from the training process and treated as the truth. The emulator then calibrates to model calculations (mock data) at that design point. Since we know the truth values for the parameters in this case, we can make a comparison between the posterior distribution of the parameters and the truth. 

There are two cases to consider for the fluctuations of the mock data. First, the statistical fluctuations from the model simulation can be used. In our study, this means small fluctuations for inclusive jet $R_{AA}$ and charged hadron $R_{AA}$ and large fluctuations for D meson $R_{AA}$. Second, we can use the statistical and systematic fluctuations from experiments. This amounts to large fluctuations for the inclusive jet $R_{AA}$ data at ALICE, and large fluctuations at high $p_T$ for the rest of the three $R_{AA}$ results. These two cases are shown in Fig.~\ref{fig:comparison_exp_own_fluc} for a single design point. For each case, I will show the results for the closure test at $8$ random design points. Specifically, when showing the closure test results for $c_1$ and $c_2$, their posterior distributions are not directly shown. Instead, I introduce the modification to $\hat{q}^{HTL}$ in Eq.~\ref{eq:qhat_t} as:
\begin{equation} 
H(Q^2, c_1, c_2) = H(t, c_1, c_2) =\frac{c_0}{1+c_1\ln^2(t)+c_2\ln^4(t)}=\frac{1+c_1\ln^2(t_0)+c_2\ln^4(t_0)}{1+c_1\ln^2(t)+c_2\ln^4(t)}.
\label{eq:Hq}
\end{equation}

The ratio between the posterior distributions and the truth value of $H(Q^2)$ is plotted. This is because $c_1$ and $c_2$ does not affect the $\hat{q}$ independently. The posterior of $H(Q^2)$ is what we are actually looking for.

In Fig.~\ref{fig:closure_own_fluc_14}, the posterior distribution of $\alpha_s$ and $Q_s$ using $8$ random sets of mock data are shown. The truth values of $\alpha_s$ and $Q_s$ are shown by black lines. In Fig.~\ref{fig:closure_own_fluc_23}, the ratio of $H(Q^2)$ between the posterior distribution and the truth are plotted. The results demonstrate that the $H(Q^2)$ are well enclosed by the inferred 60\% and
95\% confidence regions. 

Next, the closure tests are done with experimental uncertainties as shown in Fig.~\ref{fig:closure_exp_fluc_14} and Fig.~\ref{fig:closure_own_fluc_23}. A similar agreement with the truth values is observed. The two closure tests indicate that the emulator is stable with both simulation and experimental level of uncertainty and can generally infer the parameters from mock data. 

The closure tests, however, rely on the fact that the mock data and the training data use the same underlying physics model. We still don't know if our model can describe the experiment observables given the appropriate choice of parameters.

\section{A quantitative method to determine the optimal settings for the emulator} \label{sec:quantitative_closure_test}

In the previous section, closure tests at $8$ random design points were performed with model statistical uncertainty or real experimental uncertainty for the covariance matrix $\Sigma_{exp}$. However, it is difficult to tell how well the emulator is doing by just looking at those posteriors, let alone compare the performance of one type of kernel with another type. I would like to find a quantitative measure of the emulator's performance. By performance, I mean how well the emulator can recover the truth from the mock data. Naturally, one can define this quantity $\Delta_d$, which measures the deviation from the truth value for a specific parameter while using the $d$-th design point as the truth:
\begin{equation}
    \Delta_d= \int \left(\frac{|p-p_{truth}|}{|p_{max}-p_{min}|}\right)^l P(p)dp = \frac{1}{N}\sum_{j=1}^N \left(\frac{|p_j-p_{truth}|}{|p_{max}-p_{min}|}\right)^l,
\end{equation}
where the summation denotes a Monte Carlo sampling of the integration, and $l$ can be $1,2,3$ or just any positive real number. The larger $l$ is, the more penalty is put for values further away from the truth. 

$\Delta_d$ is for a cross-validation test at one design point, we can then average over it over all design points and get the overall performance of the emulator:
\begin{equation}
    \langle\Delta\rangle=\frac{1}{N_{design}}\sum_d \Delta_d.
\end{equation}

\begin{figure}
	\centering
	\includegraphics[width=0.7\textwidth]{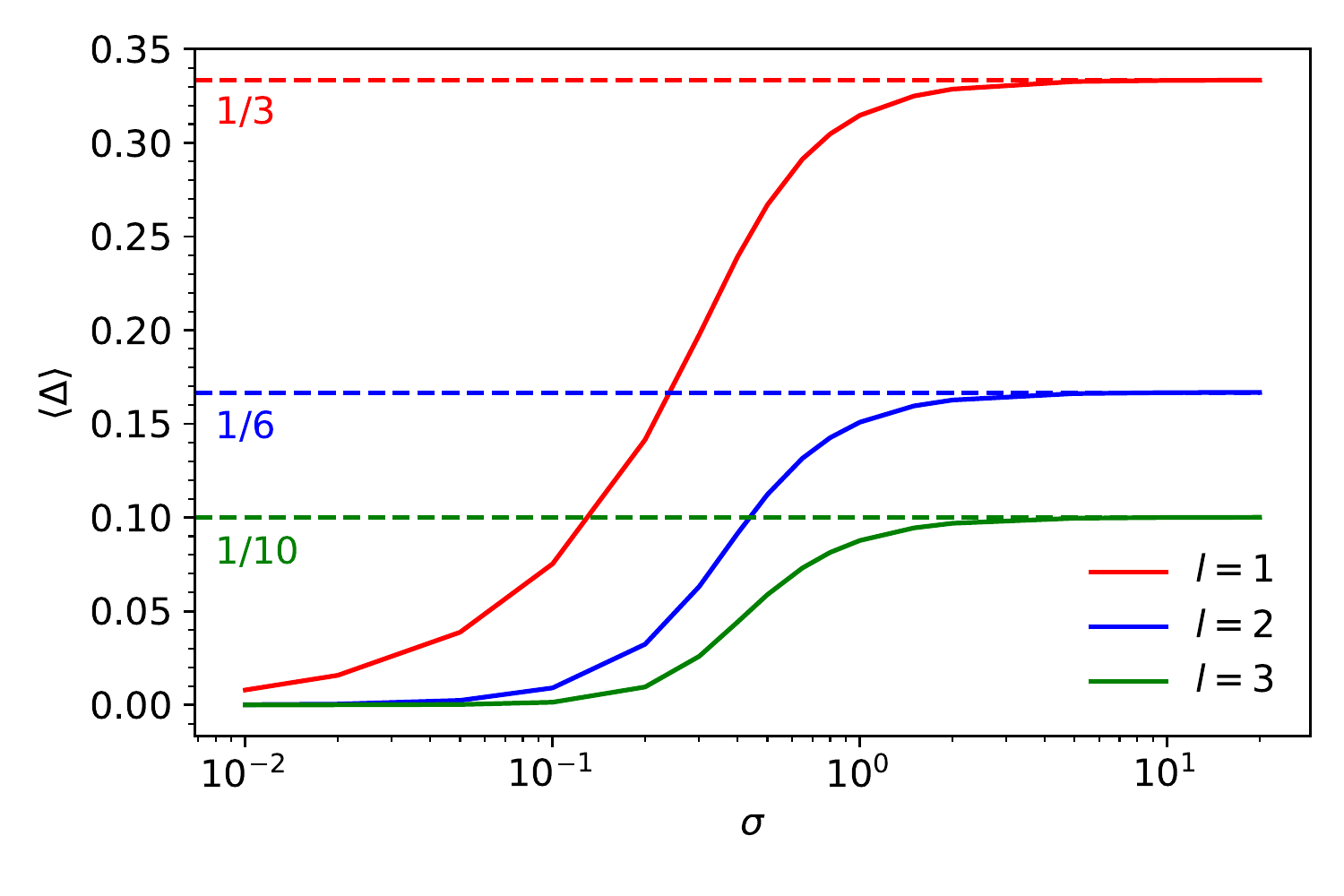}
	\caption[$\langle\Delta\rangle$ calculated with Gaussian posterior with different $l$ and $\sigma$]{\label{fig:Delta_analytical} $\langle\Delta\rangle$ calculated with Gaussian posterior centered at the truth versus the variance $\sigma$ of the Gaussian posterior. The analytical result with a uniform posterior are shown with dashed lines. }
\end{figure}

Suppose there are infinitely many design points uniformly distributed among the prior range, one can calculate the values of $\langle\Delta\rangle$ analytically with a uniform posterior or a Gaussian posterior distribution centered at the truth with different variances (see Table.~\ref{tab:delta_analytical}). As one would expect, $\langle\Delta\rangle$ gets smaller when $l$ increases or the posterior centers more around the truth. Fig.~\ref{fig:Delta_analytical} shows the change in $\langle\Delta\rangle$ versus $\sigma$ in a Gaussian if a Gaussian posterior centered at the truth is used. $\langle\Delta\rangle$ approaches the value calculated with a uniform posterior when $\sigma\rightarrow\infty$, as expected. 

\begin{table}[h!] 
\centering
\caption{Values of $\langle\Delta\rangle$ assuming a uniform posterior or a Gaussian posterior and infinite number of uniformly distributed design points.} \label{tab:delta_analytical}

\begin{tabular}{|p{1.2cm}|p{2.1cm}|p{2.1cm}|p{2.1cm}|p{2.1cm}|}
\hline
$\langle\Delta\rangle$ & uniform posterior & Gaussian ($\sigma=0.5$) & Gaussian ($\sigma=0.2$) & Gaussian ($\sigma=0.1$) \\
\hline 
$l=1$        & 0.333                & 0.267                      & 0.141                      & 0.0753                    \\
$l=2$        & 0.167                & 0.112                     & 0.0323                     & 0.00904                    \\
$l=3$        & 0.100                & 0.0588                    & 0.00954                   & 0.00139    \\        
\hline
\end{tabular}
\end{table}

When calculated with the $50$ design points used in this analysis, the $\langle\Delta\rangle$ values are shown in Table.~\ref{tab:delta_uniform_posterior}. The results are very close to those in Table.~\ref{tab:delta_analytical}, which confirms that Latin hypercube sampling (LHS) indeed samples the design points uniformly from the assumed uniform prior.

\begin{table}[h!] 
\centering
\caption{Values of $\langle\Delta\rangle$ for different parameters using a uniform posterior and the sampled design points.} \label{tab:delta_uniform_posterior}
\begin{tabular}{|l|l|l|l|l|}
\hline
 & $\langle\Delta\rangle(\alpha_s)$ & $\langle\Delta\rangle(c_1)$ & $\langle\Delta\rangle(c_2)$ & $\langle\Delta\rangle(Q_s)$ \\
\hline
$l=1$      & 0.333      & 0.333 & 0.333 & 0.333 \\
$l=2$      & 0.167      & 0.166 & 0.166 & 0.167 \\
$l=3$      & 0.100      & 0.0996 & 0.0994 & 0.100 \\
\hline
\end{tabular}
\end{table}

\subsection{Comparison with other performance metrics}

\begin{figure}
	\centering
	\includegraphics[width=0.48\textwidth]{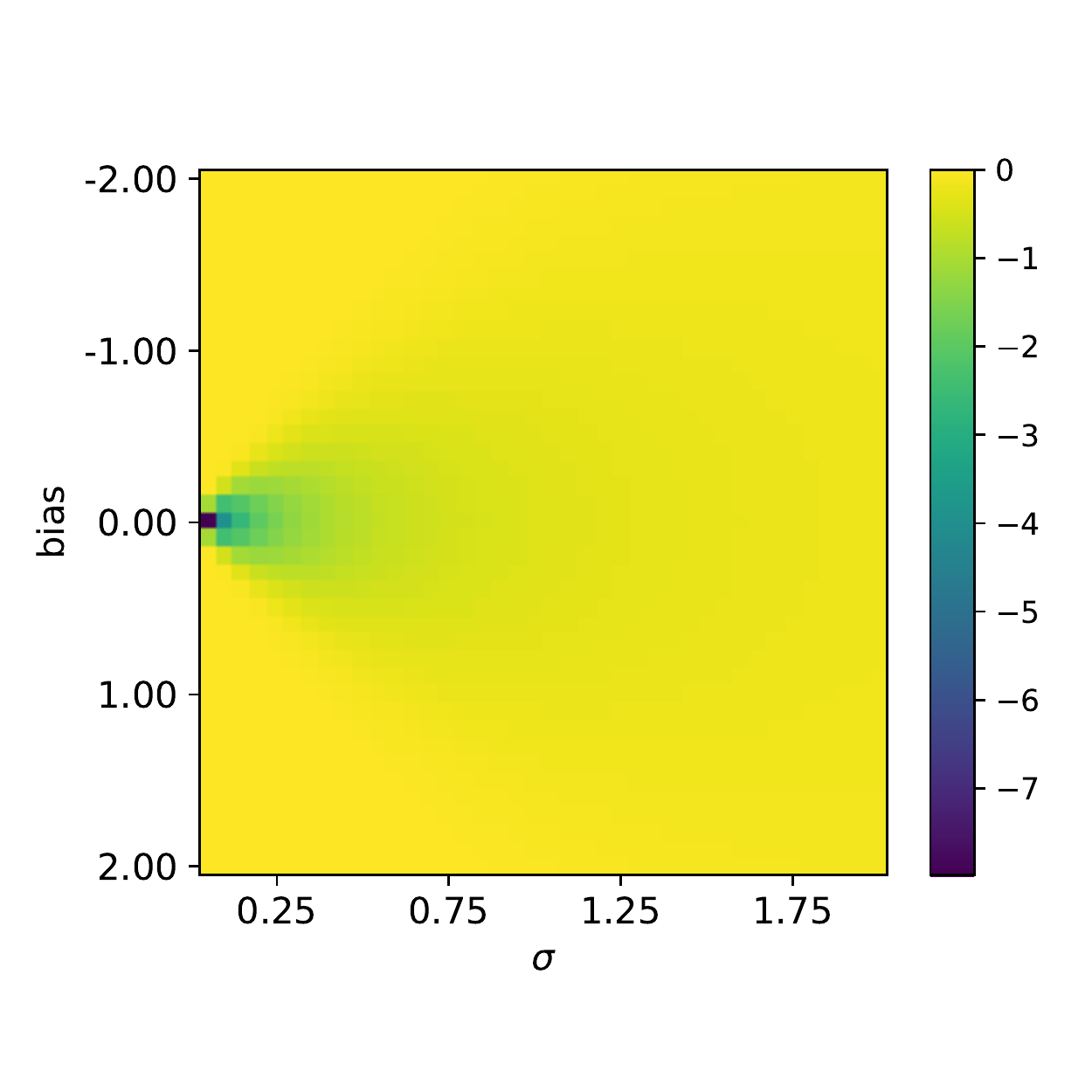}
	\includegraphics[width=0.48\textwidth]{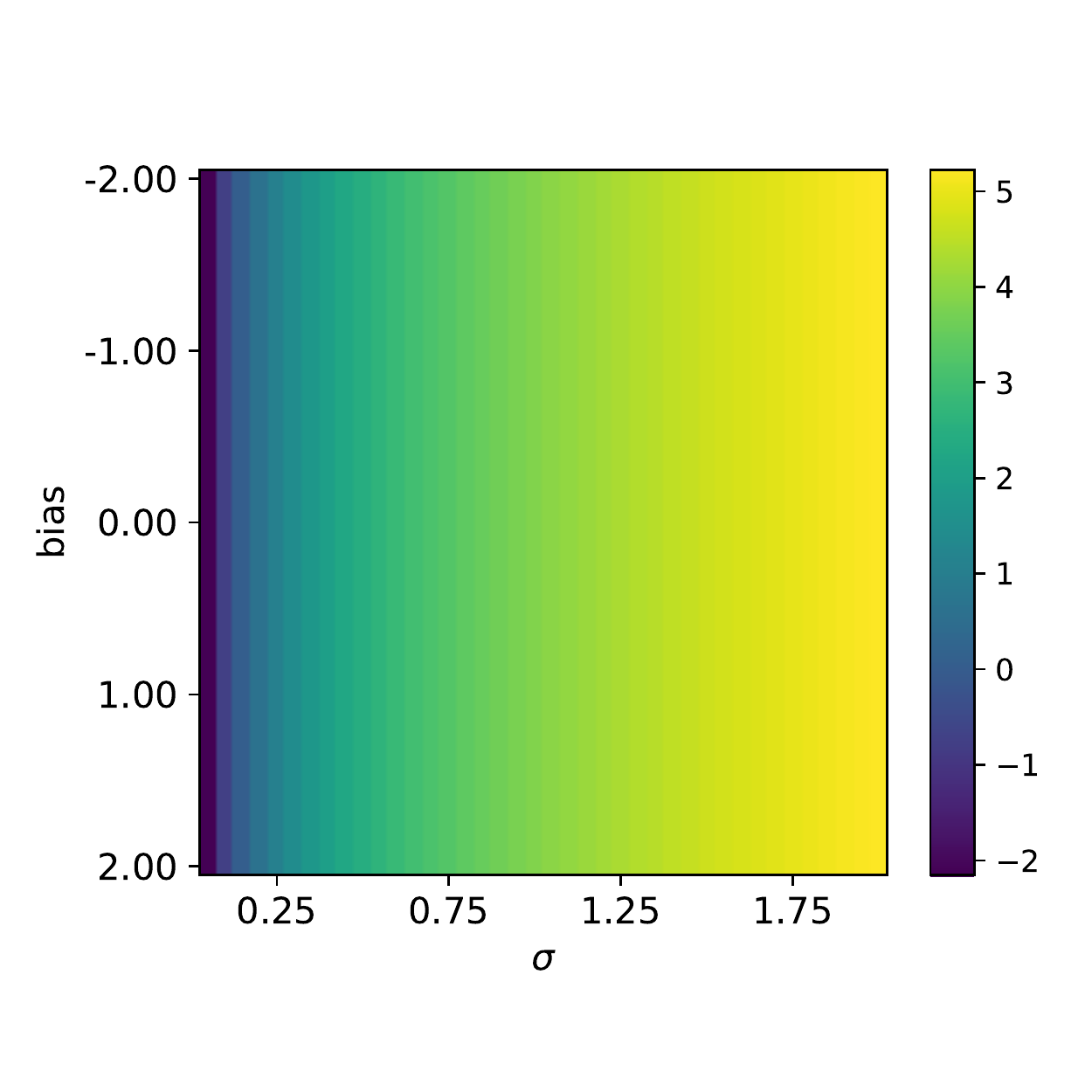}
	\includegraphics[width=0.48\textwidth]{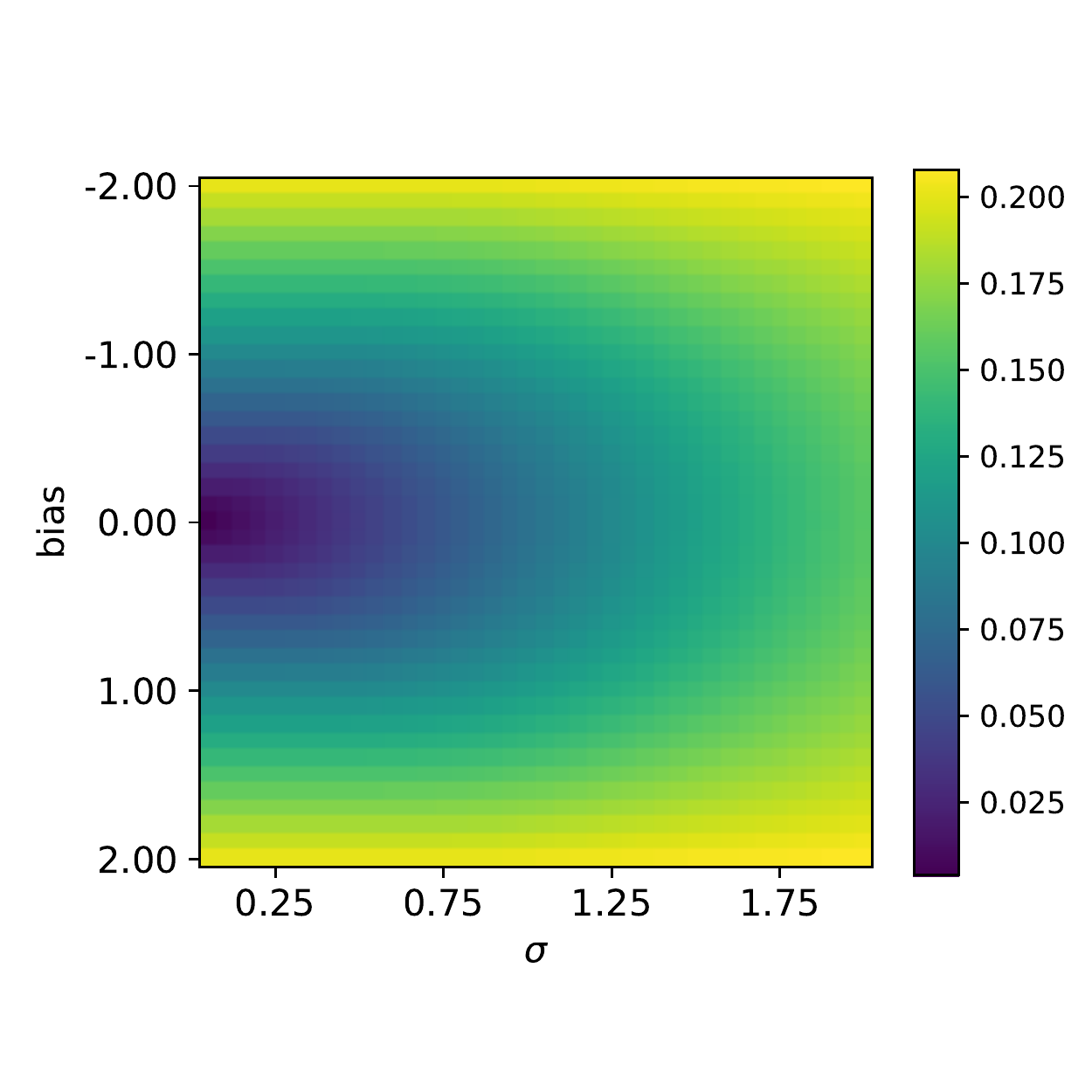}
	\includegraphics[width=0.48\textwidth]{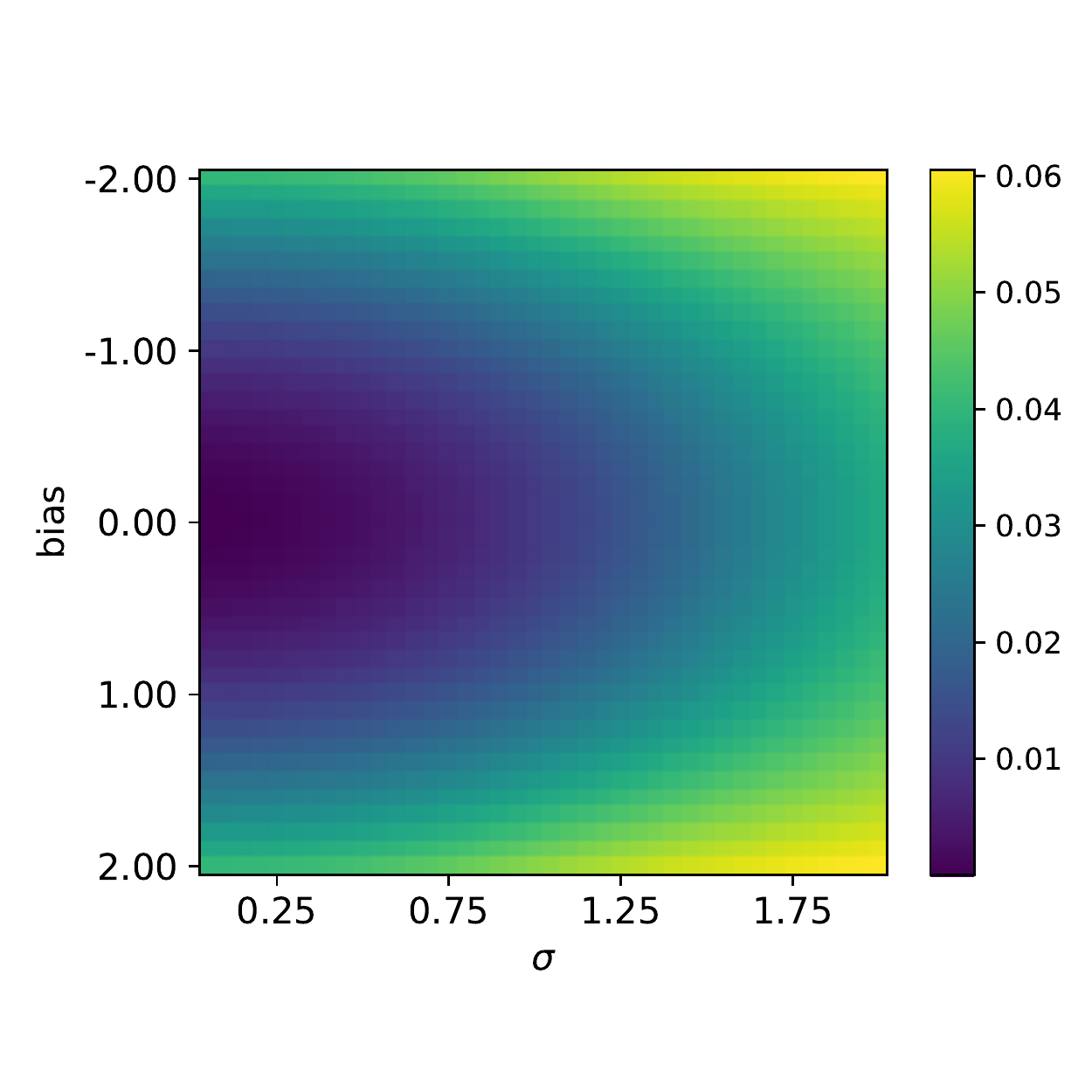}
	\includegraphics[width=0.48\textwidth]{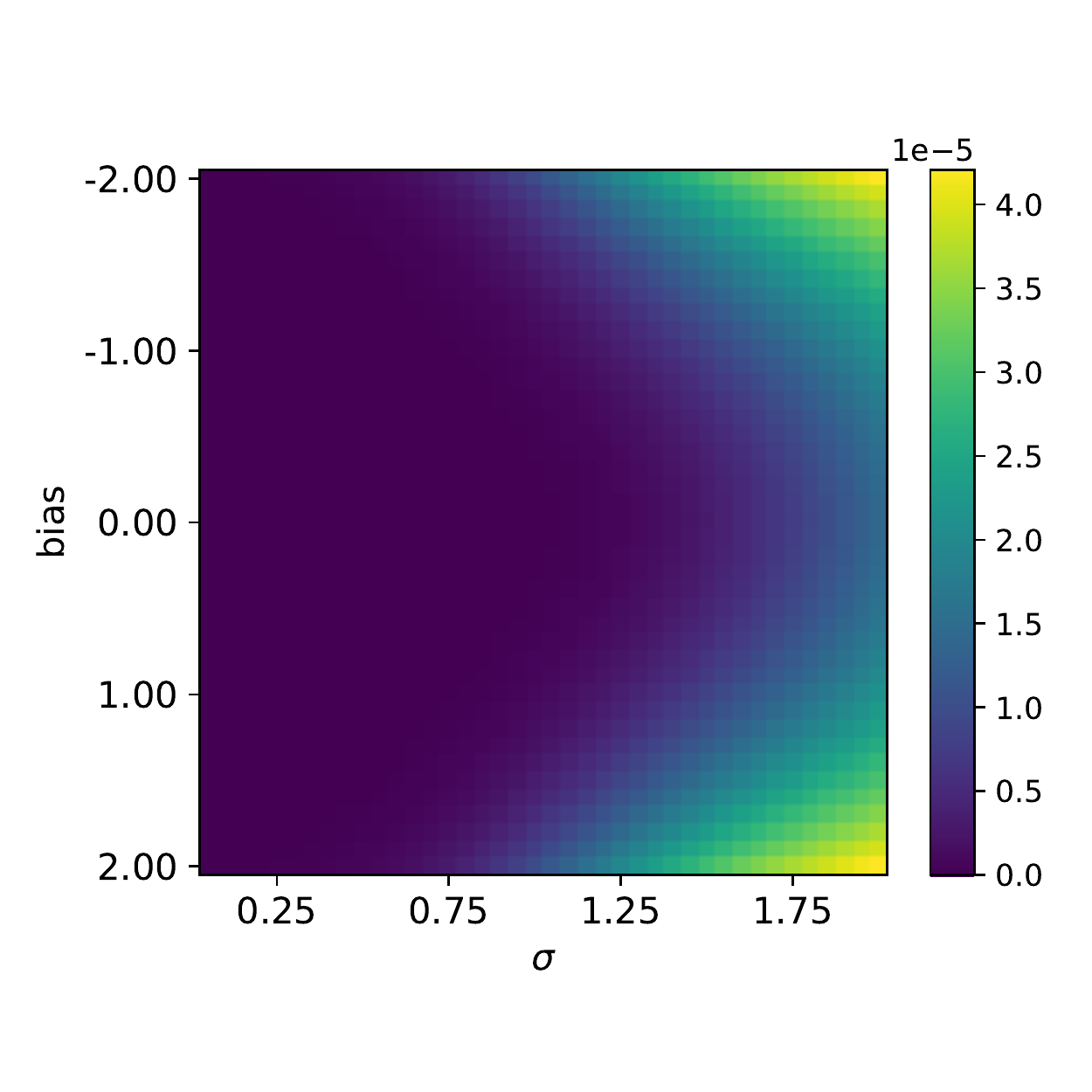}
	\includegraphics[width=0.48\textwidth]{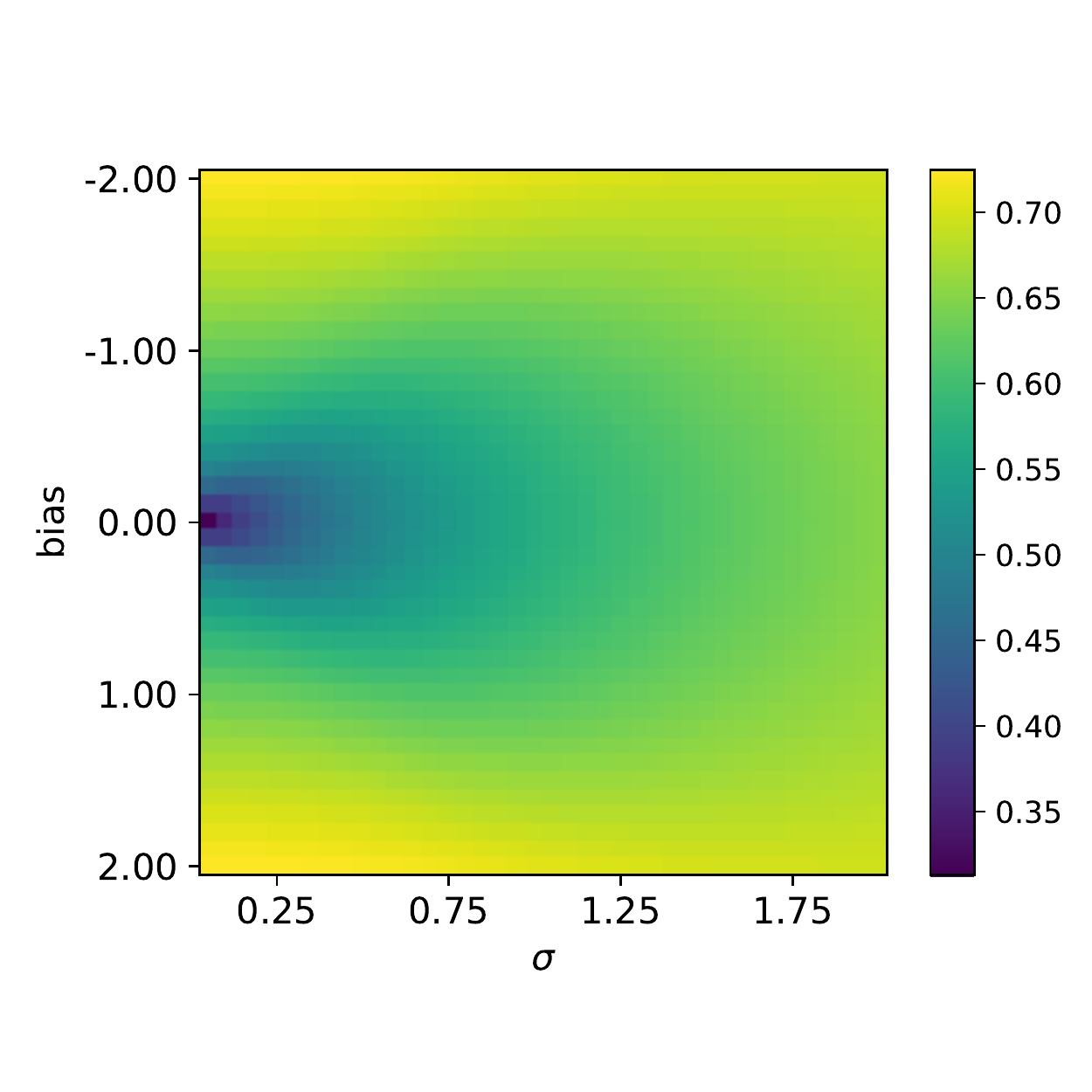}
	\caption[Comparison between different closure test metrics with a Gaussian posterior]{\label{fig:Metric_comparison} Change in $P(p_{truth})$ (top left), $AIC$ (top right), $\langle\Delta\rangle \ (l=1)$ (middle left),  $\langle\Delta\rangle \ (l=2)$ (middle right),  $\langle\Delta\rangle \ (l=10)$ (bottom left) and  $\langle\Delta\rangle \ (l=1/5)$ (bottom right) with different bias and $\sigma$ of the Gaussian posterior.}
\end{figure}

There are other metrics for measuring the performance of the closure test, namely the posterior at the truth $P(p_{truth})$ and the Akaike Information Criterion (AIC) \cite{akaike1974new} which is defined as:
\begin{equation}
    AIC=-2ln(\mathcal{L}_{max})+2k,
\end{equation}
where $\mathcal{L}_{max}$ is the maximum likelihood in the parameter space and $k$ is the number of parameters.

We can compare the variation of these metrics with a simple $1$-dimensional Gaussian posterior distribution centered away from the truth value $p_{truth}$ and some finite variance as shown in Fig.~\ref{fig:Metric_comparison}. The posterior at the truth and the $AIC$ flatten in some regions of the plane, which does not seem ideal. $AIC$ behaves particularly bad in this example as it is entirely indifferent to the change in the bias. $\langle\Delta\rangle$ looks more reasonable, yielding a larger penalty to both larger bias and larger $\sigma$. When the posterior becomes more complex, it becomes more complicated to compare one metric with another. I believe the story here is similar to the choice of the kernel in the Gaussian process emulator: there is no single kernel that can handle every possible situation. In this study, I will use $\langle\Delta\rangle$ with $l=2$ as the metric for calculating the deviation of the posterior from the truth in closure tests.

\subsection{Connection to the Kullback-Leibler Divergence}\label{sec:connection_DKL}

The Kullback-Leibler Divergence is defined as:
\begin{equation}
    D_{KL}(P|Q)=\int p(x)\ln\left(\frac{p(x)}{q(x)}\right)dx
\end{equation}
between two probability distribution $p(x)$ and $q(x)$. It is a measure of how one probability distribution $P$ is different from a second, reference probability distribution, $Q$. In Bayesian statistics, it can be seen as a measure of the information gain in moving from a prior distribution to a posterior distribution. We will calculate the $D_{KL}$ with the extracted posterior distribution in Section \ref{sec:bayesian_results}.

However, $D_{KL}$ is ill-defined (diverges) when the prior distribution $q(x)$ is a Dirac delta function centered at $x_0$ (here we use the limit representation $\delta(x-x_0)=\lim_{\sigma\rightarrow 0} \frac{1}{\sqrt{\pi}\sigma}e^{-(x-x_0)^2/\sigma^2}$): 
\begin{equation}\label{eqn:DKL_equivalence}
\begin{split}
    D_{KL}(P|Q)&=\int \lim_{\sigma\rightarrow 0} p(x)\ln\left(\frac{p(x)}{\frac{1}{\sqrt{\pi}\sigma}e^{-(x-x_0)^2/\sigma^2}}\right)dx \\
    &=\int p(x)\ln p(x) dx + \lim_{\sigma\rightarrow 0} \ln{\sqrt{\pi}\sigma} + \lim_{\sigma\rightarrow 0}\frac{1}{\sigma^2} \int  p(x)(x-x_0)^2dx \\
    &=\lim_{\sigma\rightarrow 0}\frac{1}{\sigma^2} \int  p(x)(x-x_0)^2dx + \mathcal{O}(\frac{1}{\sigma}).
\end{split}
\end{equation}

However, when we take the ratio between two $D_{KL}$, or argue that the divergence caused by $\frac{1}{\sigma^2}$ is independent from the posterior $p(x)$, the relevant part is:
\begin{equation}
    \int  p(x)(x-x_0)^2dx,
\end{equation}
which is equivalent to the definition of $\Delta$ when $l=2$. If we use other limit representations of the Dirac delta function, for example:
\begin{equation}
    \delta(x-x_0)=\lim_{\sigma\rightarrow 0} \frac{1}{2\sigma\Gamma(1+\frac{1}{l})}e^{-|x-x_0|^l/\sigma^l},
\end{equation}
where $l\ge 1$, we should get to the same equivalence relation with the specific $l$. When $l<1$, the derivation would be the same except that the relevant part is no longer the dominant contribution in Eq.~\ref{eqn:DKL_equivalence}.

The above derivations tell us that $\Delta$ is really measuring the finite relevant part in $D_{KL}$ when the prior is a Dirac delta function. $D_{KL}$ was not used before as a metric in closure test because of the divergence, now we have ``regulated'' it. 

One should notice that we are actually measuring information loss in closure tests, since the prior is a Dirac delta function which is the most certain probability distribution. In this case, we want the $D_{KL}$ (information loss) to be as small as possible. When we employ a uniform prior distribution in Bayesian analysis, we are calculating information gain because a uniform prior means as least information as  possible. This time larger $D_{KL}$ means we are constraining the parameter better.

\subsection{Comparison between different kernels}

Six types of kernel are compared in this section: 
\begin{enumerate}

    \item the RBF kernel (equivalent to the Mat\'{e}rn kernel with $\nu \rightarrow \infty$):
    \begin{equation}
        k(r)=\sigma^2\exp(-\frac{r^2}{2l^2}).
    \end{equation}
    \item the Mat\'{e}rn ($\nu=3/2$) kernel:
    \begin{equation}
        k(r)=\sigma^2(1+\frac{\sqrt{3}r}{l})\exp(-\frac{\sqrt{3}r}{l}).
    \end{equation}
    \item the Mat\'{e}rn ($\nu=5/2$) kernel:
    \begin{equation}
        k(r)=\sigma^2(1+\frac{\sqrt{5}r}{l}+\frac{5r^2}{3l^2})\exp(-\frac{\sqrt{5}r}{l}).
    \end{equation}
    \item the RBF + white noise kernel.
    \item the Mat\'{e}rn ($\nu=3/2$) + white noise kernel.
    \item the Mat\'{e}rn ($\nu=5/2$) + white noise kernel.

\end{enumerate}

A lesser common choice of $\nu$ in the Mat\'{e}rn kernel is $\nu=1/2$ which gives the following kernel function:
\begin{equation}
    k(r)=\sigma^2\exp(-\frac{r}{l}).
\end{equation}

\begin{figure}
	\centering
	\includegraphics[width=0.6\textwidth]{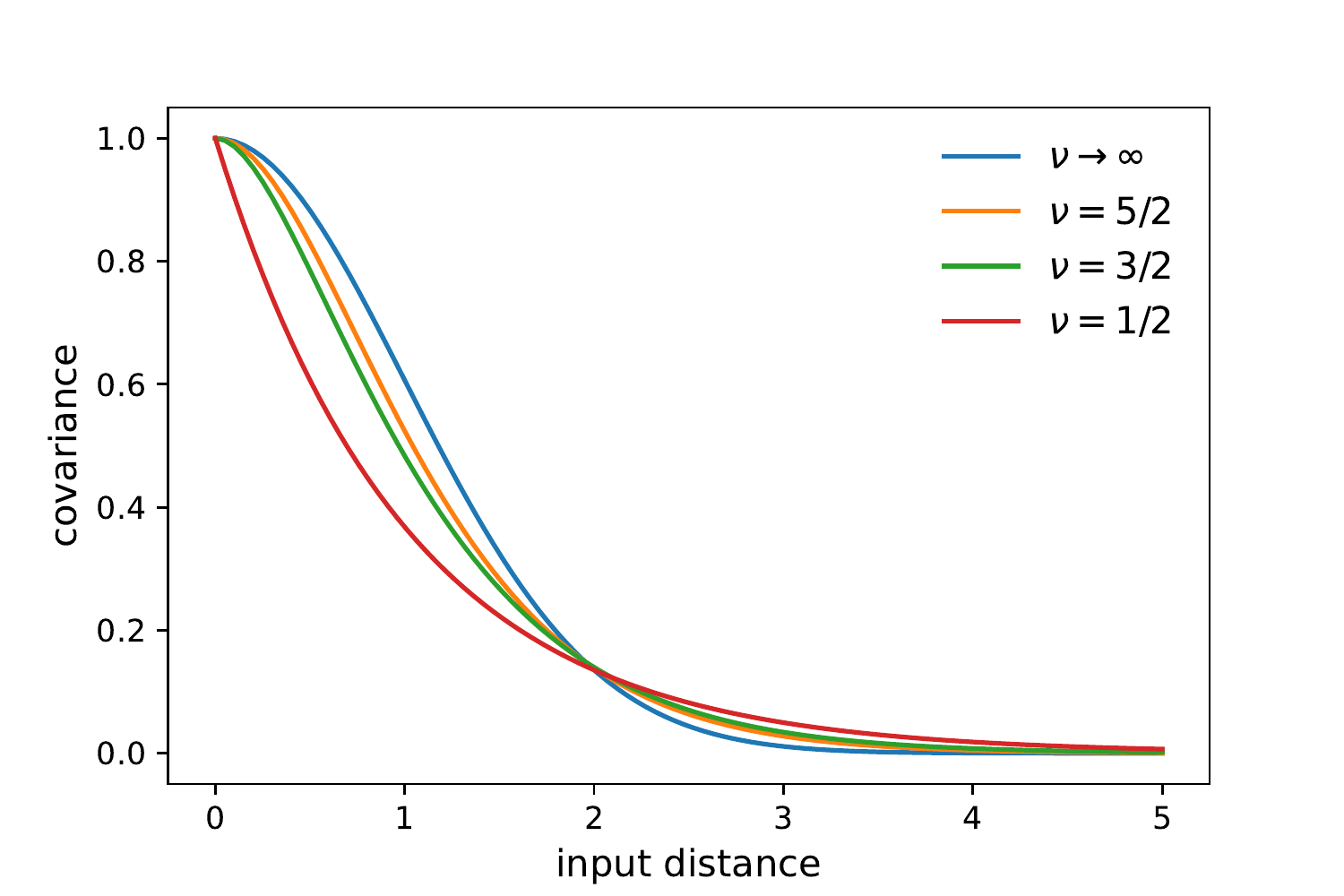}
	\caption{\label{fig:Matern_comparison} Comparison between Mat\'{e}rn kernels with different $\nu$.}
\end{figure}

\begin{figure}
	\centering
	\includegraphics[width=0.48\textwidth]{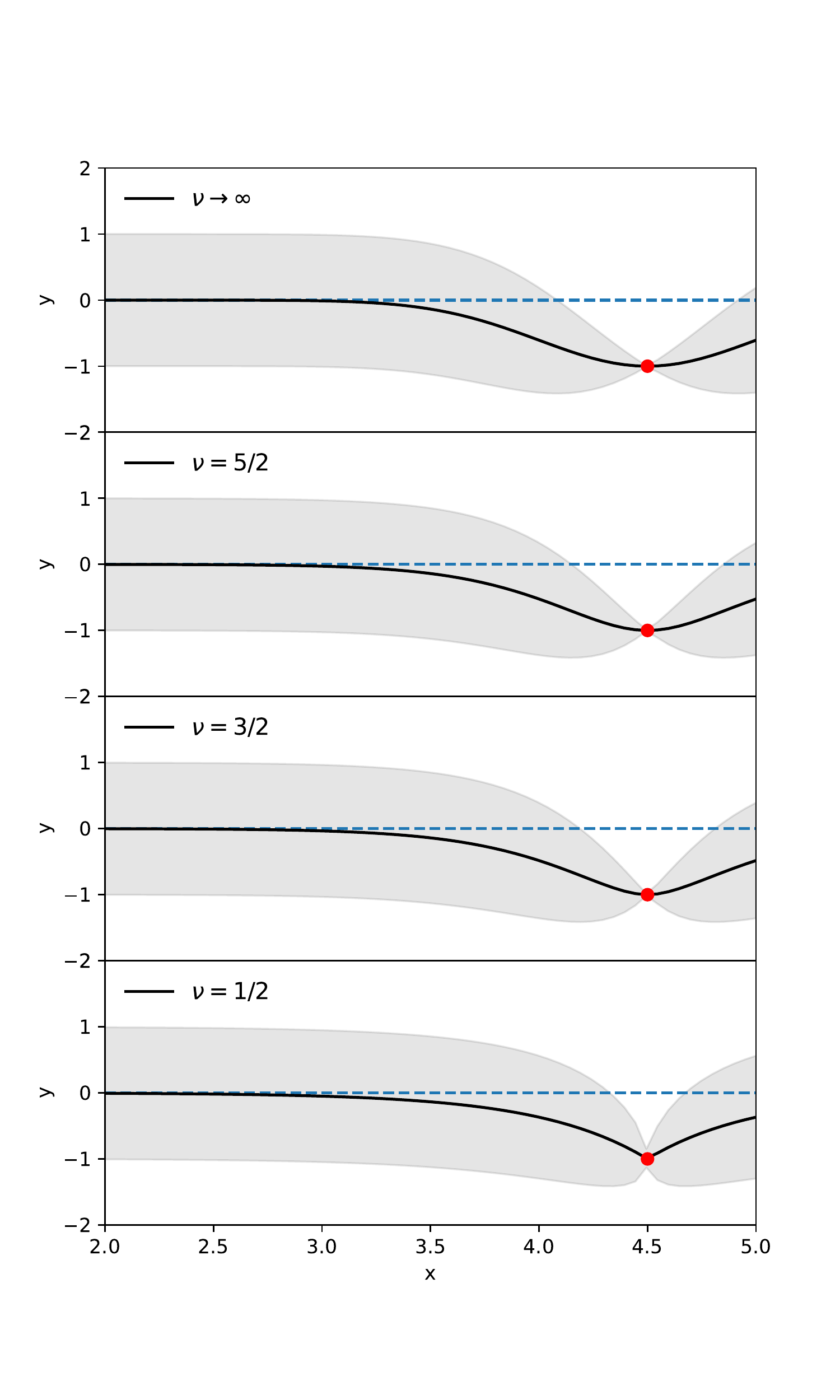}
	\includegraphics[width=0.48\textwidth]{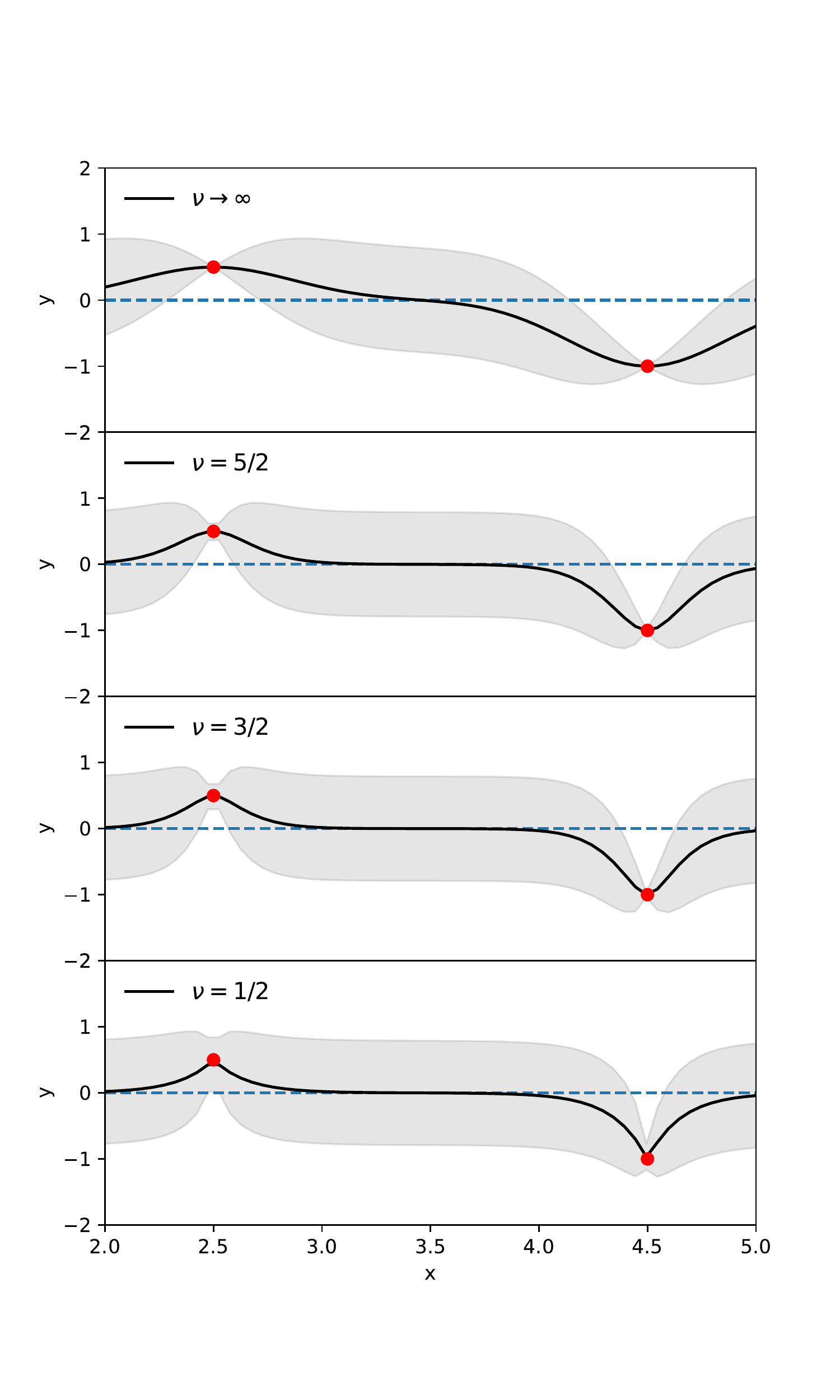}
	\caption[The posterior function fitted by  Mat\'{e}rn kernels with different $\nu$.]{\label{fig:Matern_comparison_manual} The posterior function fitted by  Mat\'{e}rn kernels with different $\nu$. Red points represent the training data, the slid black line is the mean of the posterior, and the grey band represents the $1\sigma$ credible region.}
\end{figure}

The reason for choosing these half-integer values $p+1/2$ for $\nu$ is that the Mat\'{e}rn kernel can be simplified into a product of an exponential and a polynomial of order $p$. The Mat\'{e}rn kernels with different $\nu$ are plotted in Fig.~\ref{fig:Matern_comparison}. As we can see, a larger $\nu$ means a stronger correlation at small distances and a weaker correlation at large distances. This is also observed in Fig.~\ref{fig:Matern_comparison_manual} where the posterior is trained with only one or two data point(s). The solid line is the mean and the grey band represents the $1\sigma$ credible region. The Mat\'{e}rn ($\nu=1/2$) kernel is used only in Fig.~\ref{fig:Matern_comparison} and Fig.~\ref{fig:Matern_comparison_manual} for comparison purposes, since it predicts non differentiable functions which do not fit in our use case. In our use case where we use the Gaussian process to train a model that maps input parameters to observables, it is not clear how this mapping would look like for an arbitrary parameterization. In our brief parameter exploration in Section~\ref{sec:jetscape_results}, it seems like the observables have monotonic relations with the model parameters. However, no conclusive observation can be drawn. This is another motivation for why we need to compare the performance of different kernels.

\begin{figure}
	\centering
	\includegraphics[width=0.48\textwidth]{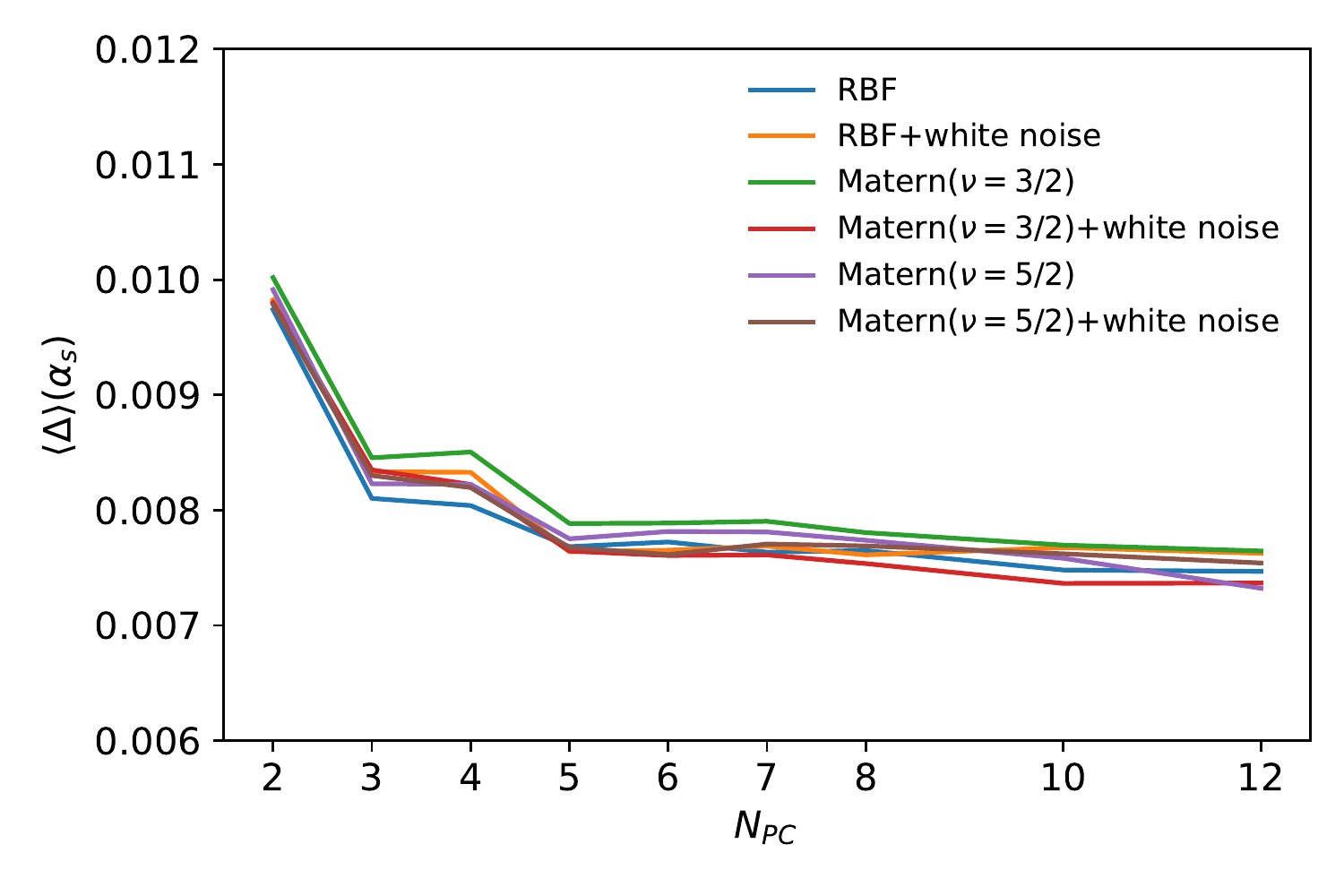}
	\includegraphics[width=0.48\textwidth]{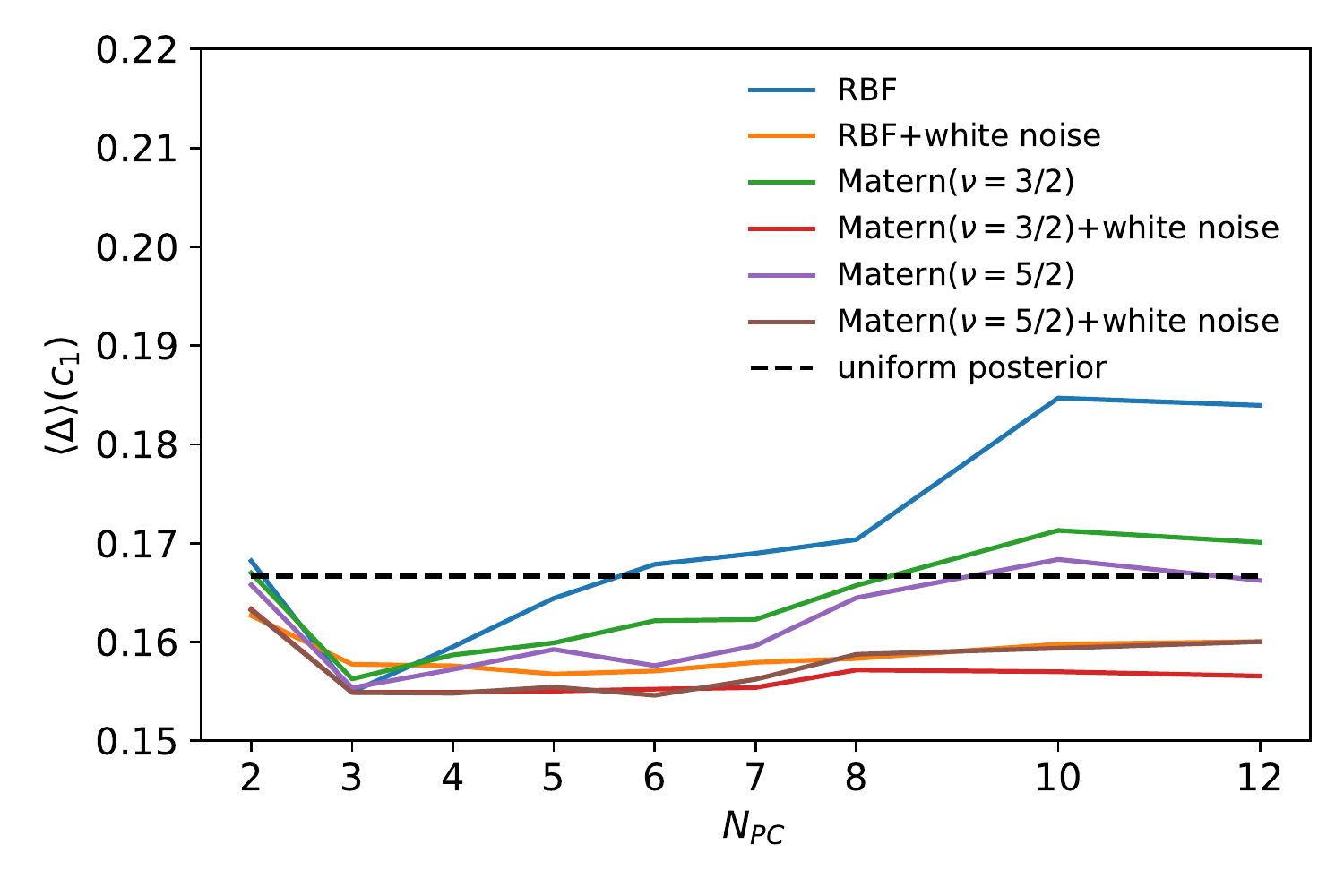}
	\includegraphics[width=0.48\textwidth]{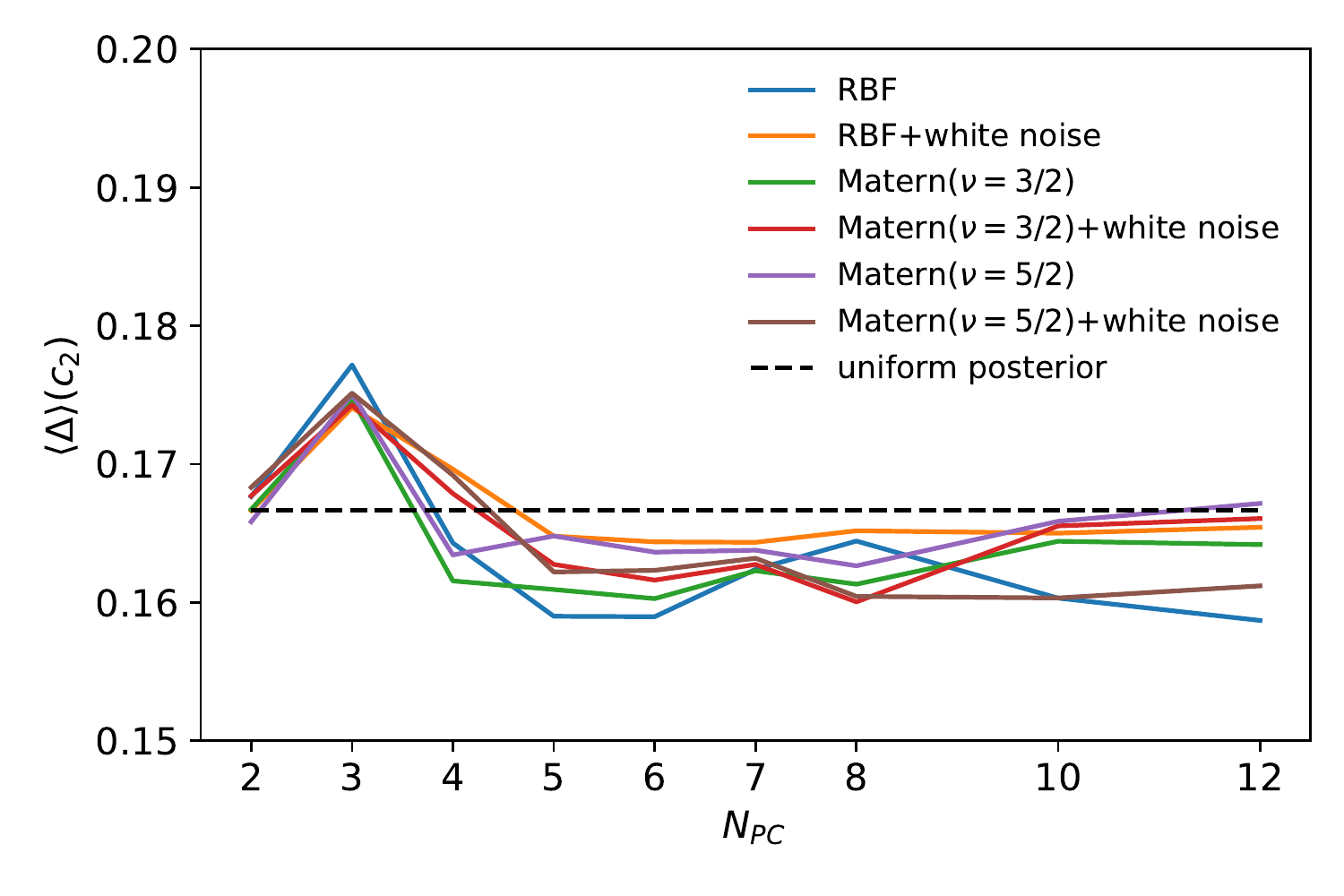}
	\includegraphics[width=0.48\textwidth]{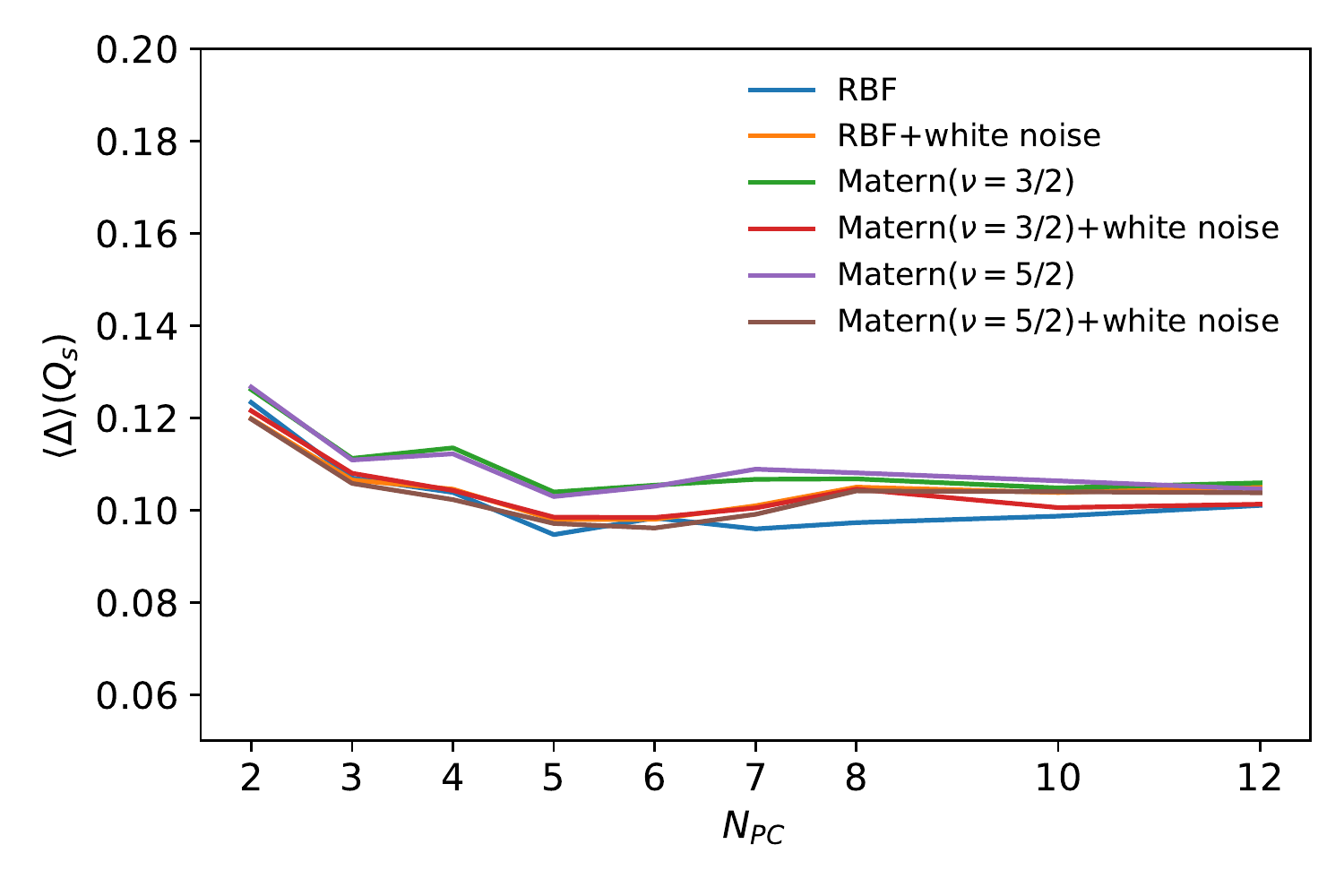}
	\caption{\label{fig:Delta_comparison} Comparison of $\langle\Delta\rangle$ for different parameters with different kernels and number of principal components.}
\end{figure}

Experimental uncertainty is used in the covariance matrix $\Sigma_{exp}$. $\langle\Delta\rangle$ are calculated with different kernels and varied number of principal components. The results are shown in Fig.~\ref{fig:Delta_comparison}. Reading from the numbers in the plots, $\langle\Delta\rangle(\alpha_s)$ are generally the smallest, followed by $\langle\Delta\rangle(Q_s)$. $\langle\Delta\rangle(c_1)$ and $\langle\Delta\rangle(c_2)$ fluctuates around $0.167$ (indicated by the black dashed line) which is the analytical result assuming a uniform posterior. This means the emulators are having trouble recovering these two parameters given the current level of uncertainties.

\begin{figure}
	\centering
	\includegraphics[width=0.78\textwidth]{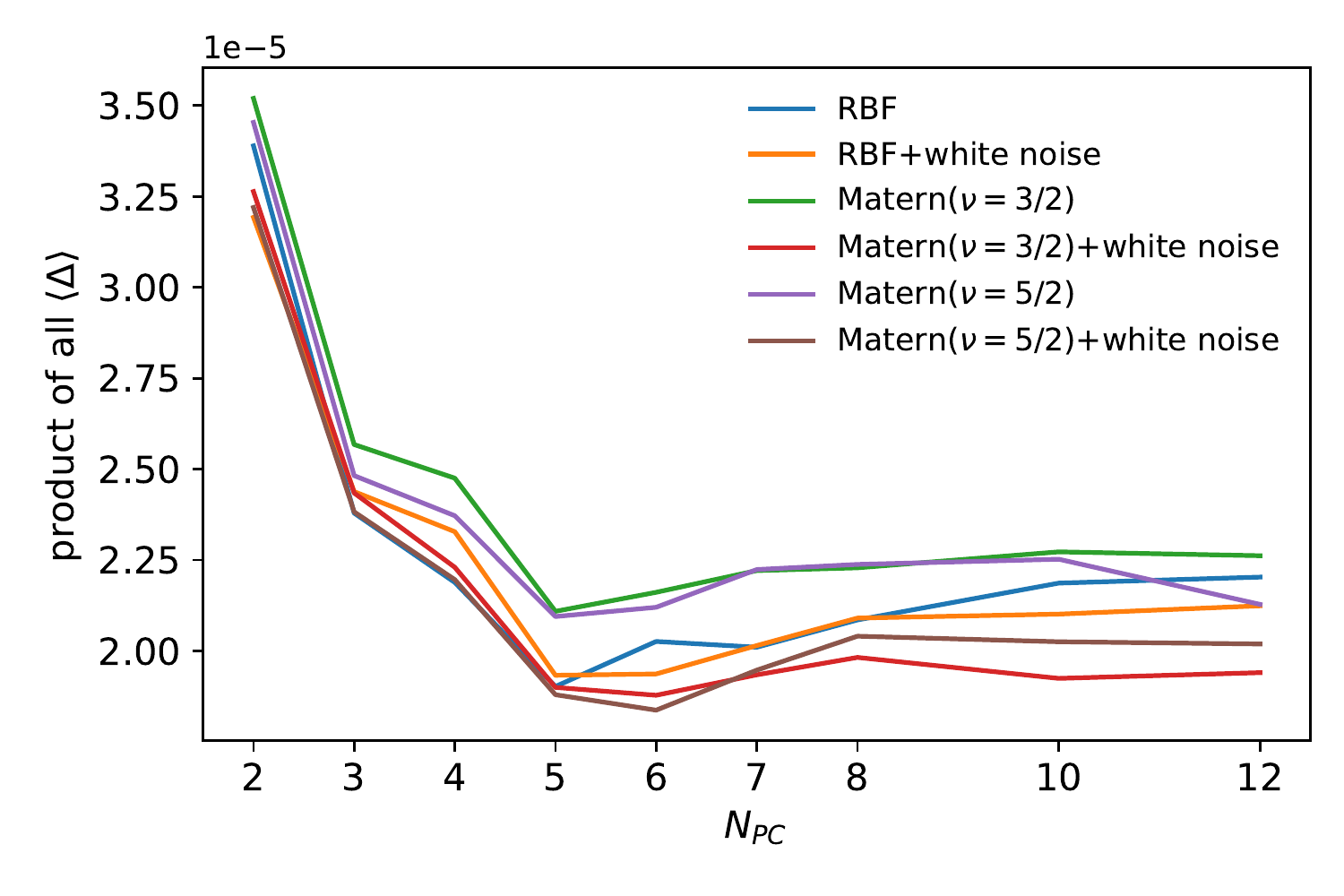}
	\caption{\label{fig:Delta_comparison_all} Comparison of the product of all $\langle\Delta\rangle$ for different parameters with different kernels, training data selection, and number of principal components.}
\end{figure}

In order to pick the optimal settings for the emulator, the product of all the $\langle\Delta\rangle$ will be compared (see Fig.~\ref{fig:Delta_comparison_all}). In this test, the Mat\'{e}rn ($\nu=5/2$) + white noise kernel with $6$ principal components performs the best. However, the RBF + white noise kernel with $5$ principal components performs at a similar level, and we will stick to it for consistency with previous studies. Closure test result using the Mat\'{e}rn ($\nu=5/2$) + white noise kernel with $6$ principal components is shown in Appendix.~\ref{section:appendix_matern kernel}. Another observation is that the white noise kernel improves the overall performance, especially for the Mat\'{e}rn kernel. 

\begin{figure}
	\centering
	\includegraphics[width=0.78\textwidth]{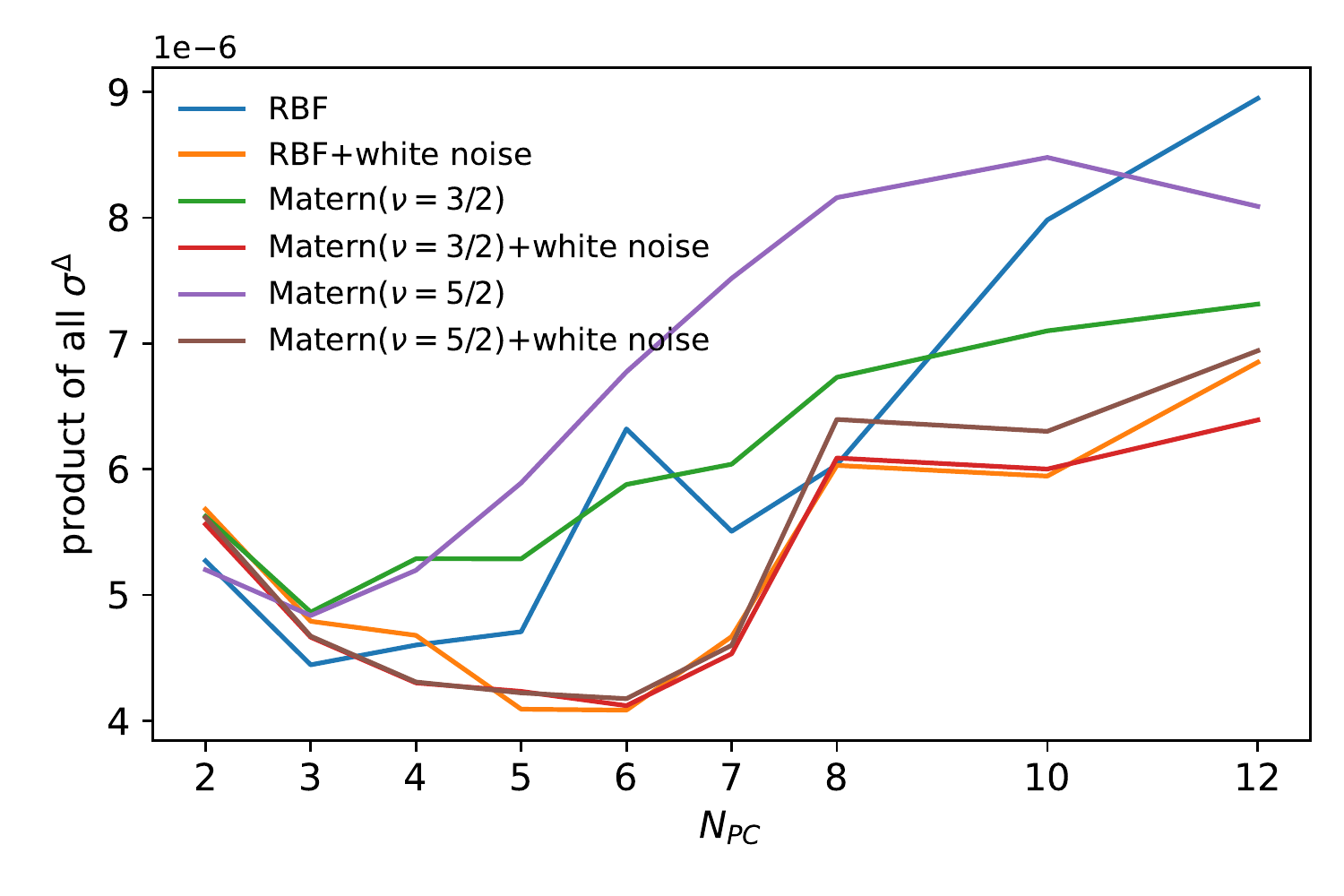}
	\caption{\label{fig:Delta_std_comparison_all} Comparison of the product of all $\langle\Delta\rangle$ for different parameters with different kernels, training data selection, and number of principal components.}
\end{figure}

Another quantity we can look at is the variance of $\Delta$ ($\sigma^\Delta$). $\sigma^\Delta$ is not an indication of the emulator's performance, as it is possible to have large mean and small variance (or small mean and large variance) at the same time. Nevertheless, the variance limits the distribution of $\Delta$ to the right side of the mean $\langle\Delta\rangle$. In Fig.~\ref{fig:Delta_std_comparison_all}, the product of all the variance of $\Delta$ are shown. Interestingly, the smallest values are achieved at around $5$ or $6$ of PC and with the kernels that includes the white noise kernel. 

The method discussed in this section is not limited to the particular Bayesian analysis in this work. It can be easily applied to other Bayesian analysis projects utilizing the Gaussian process emulator as a surrogate model. Key ingredients for training the emulator, including choice of the kernel and the number of principal components can all be determined via this method. The level of constraint on each parameter, can also be reflected by the magnitude of $\langle\Delta\rangle$. The caveat, however, is that these observations assume that we are using the correct model. No systematic uncertainty of the model is taken into account here. 

\section{Bayesian analysis results}\label{sec:bayesian_results}

\begin{figure}
	\centering
	\includegraphics[width=0.96\textwidth]{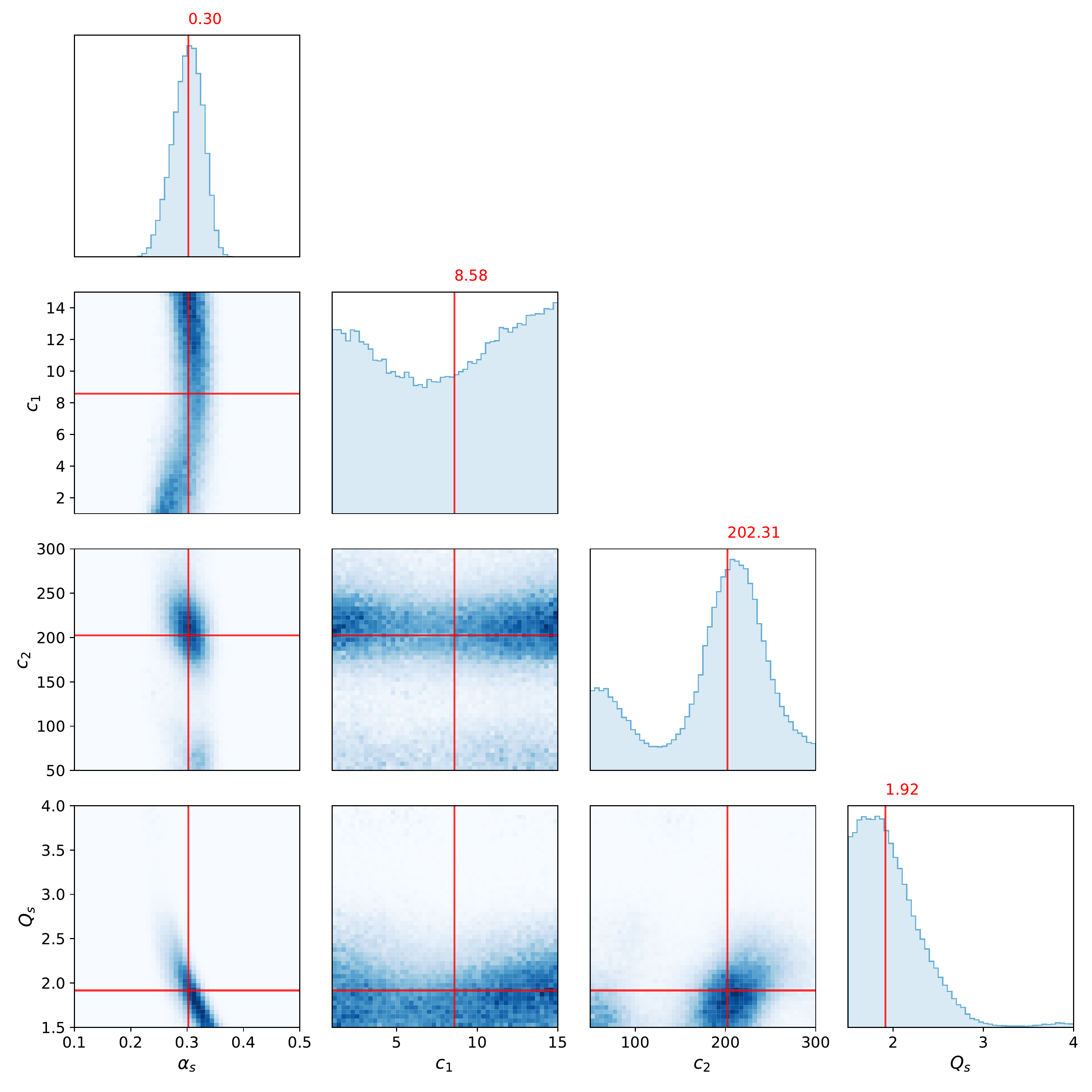}
	\caption{\label{fig:BestPosteriorEstimation} The posterior distribution of the model parameters. The emulator is using 5 PC and the RBF + white noise kernel and trained from 50 design points.}
\end{figure}

\begin{figure}
	\centering
	\includegraphics[width=0.75\textwidth]{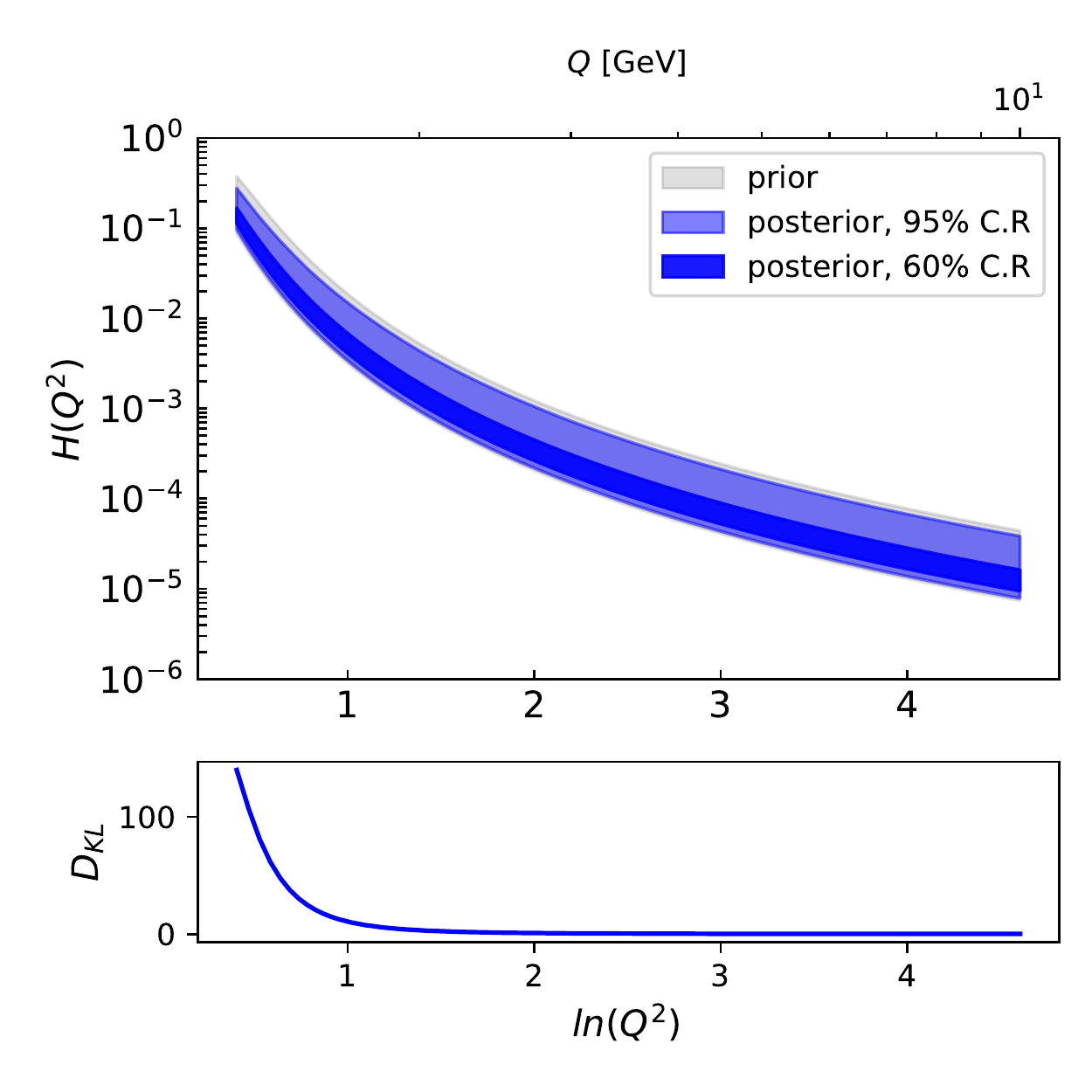}
	\caption[The prior and posterior distribution of $H(Q^2)$]{\label{fig:BestPosteriorHq} \textbf{Top}: The prior (grey), 95\% credible region of the posterior (blue), and 60\% credible region of the posterior (deep blue) of $H(Q^2)$ define in Eq.~\ref{eq:Hq}. \textbf{Bottom}: The corresponding information gain (Kullback-Leibler divergence $D_{KL}$).}
\end{figure}

\begin{figure}
	\centering
	\includegraphics[width=0.96\textwidth]{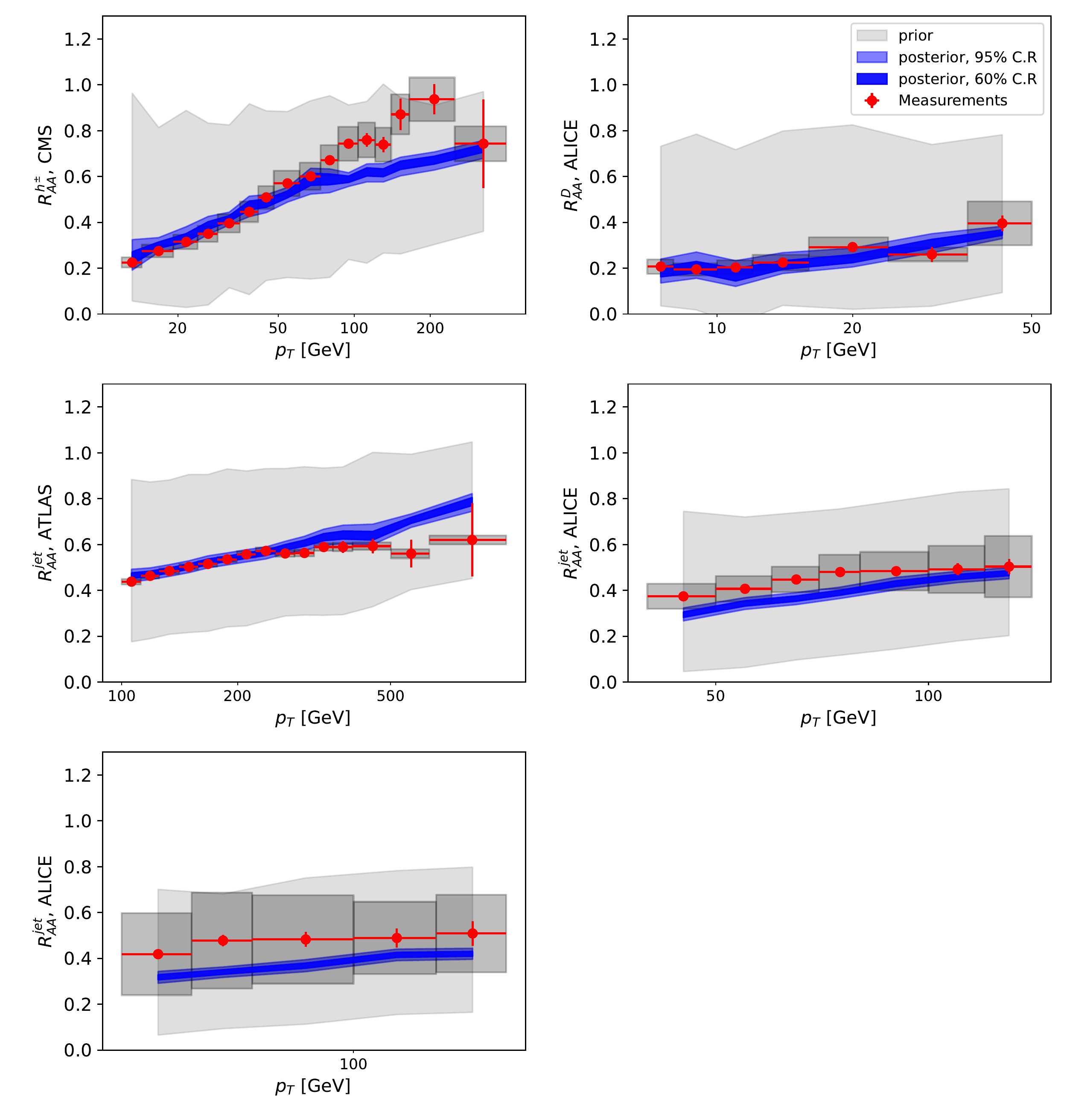}
	\caption{\label{fig:BestObservablePosterior} Comparison between the posterior distribution of the observables and experiment data.}
\end{figure}

In this section, I will first show the Bayesian analysis result calibrating to all five experiment data sets. This is the best estimation of the model parameters we can get at the current stage (see Fig.~\ref{fig:BestPosteriorEstimation}, Fig.~\ref{fig:BestPosteriorHq}). The posterior distribution of the observables compared to data are shown in Fig.~\ref{fig:BestObservablePosterior}. Compared to the parameters we used in Chapter~\ref{sec:jetscape_results} (where $\alpha_s=0.3, \ c_1=10, \ c_2=100, \  Q_s=2$), the posterior distribution in Fig.~\ref{fig:BestPosteriorEstimation} suggests similar value for $\alpha_s$ but a slightly smaller $Q_s$. From previous variation of $Q_s$ we know a smaller $Q_s$ would shift the charged hadron and D meson $R_{AA}$ up, possibly making up for the disagreement in charged hadron $R_{AA}$ at high $p_T$ seen in Fig.~\ref{fig:BestObservablePosterior}. Meanwhile, the constraint on $c_1$ is much wider and a peak over large $c_2$ is observed. However, from the tests done in Sec.~\ref{sec:quantitative_closure_test}, the constraints on $c_1$ and $c_2$ should be not be taken too seriously as on average the emulator's performance is just slightly better than using a uniform posterior. One can also see why $c_1$ and $c_2$ are hard to constrain by looking at Fig.~\ref{fig:MATTER_LBT_comp_qhat}. Even with much higher precision, the D meson and charged hadron $R_{AA}$ results are difficult to distinguish between different $\hat{q}$ parameterization except at high $p_T$ where the experimental data uncertainty are still large. 

In Fig.~\ref{fig:BestPosteriorHq}, the top plot shows the prior, 95\% credible region, and 60\% credible region of $H(Q^2)$ as a function of $\ln(Q^2)$. The Kullback-Leibler divergence (defined in Section \ref{sec:connection_DKL}) peaked near the lower limit of the prior range of $Q_s$ which is $1.5$. This is likely due to the fact that the second-order term and the fourth-order term in $H(Q^2)$ are comparable in magnitude in this region. This shows that the constraint is more on the joint distribution of $c_1$ and $c_2$ at low $Q$ than their individual values. At large $Q$, the fourth-order term dominates $H(Q^2)$ and the small $D_{KL}$ values here suggest that not much information is gained for the posterior of $c_2$.

\subsection{Sensitivity to different observables}

\begin{figure}
	\centering
	\includegraphics[width=0.96\textwidth]{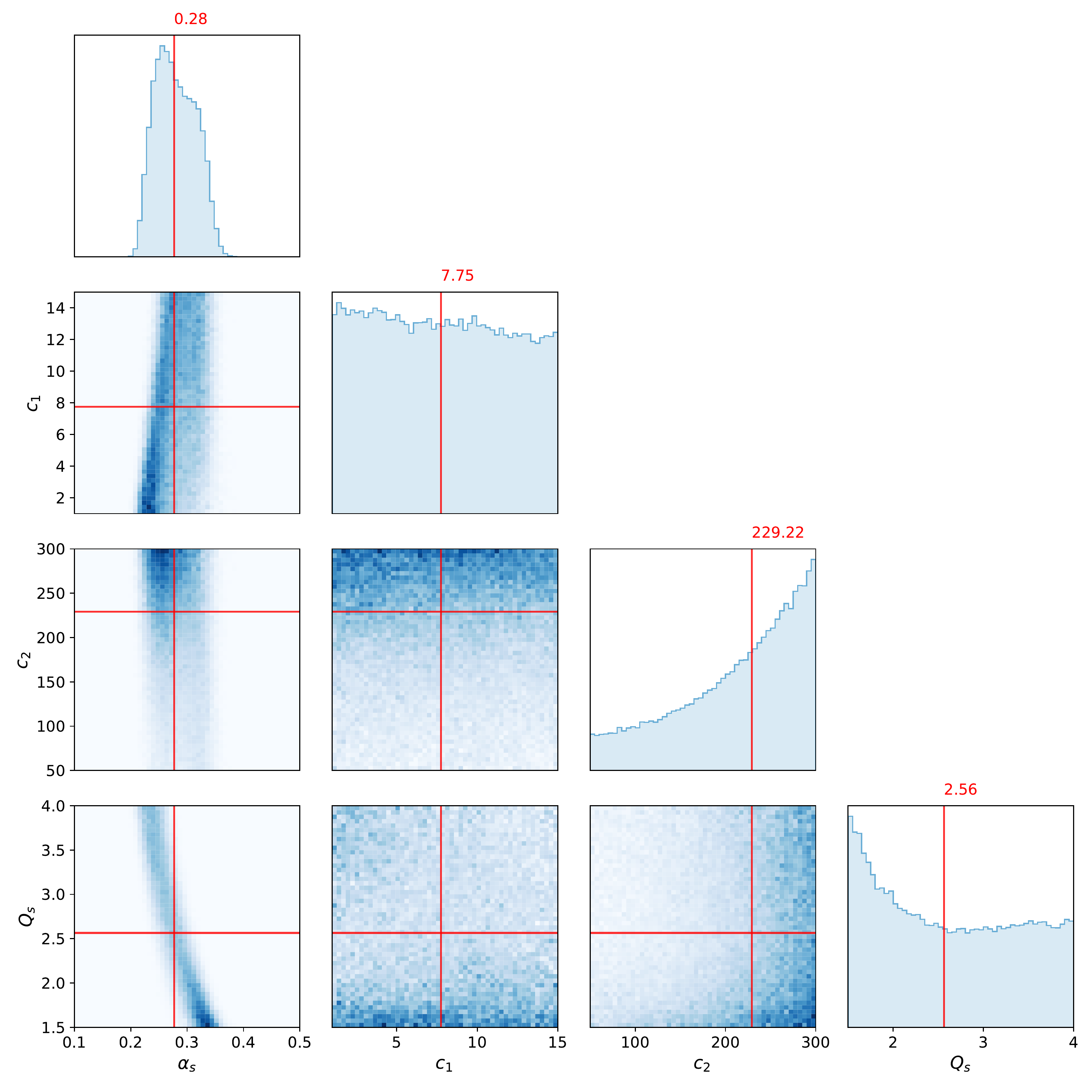}
	\caption{\label{fig:PosteriorEstimation3Obs} The posterior distribution of the model parameters. The emulator is only calibrating to the three inclusive jet $R_{AA}$ results.}
\end{figure}

\begin{figure}
	\centering
	\includegraphics[width=0.96\textwidth]{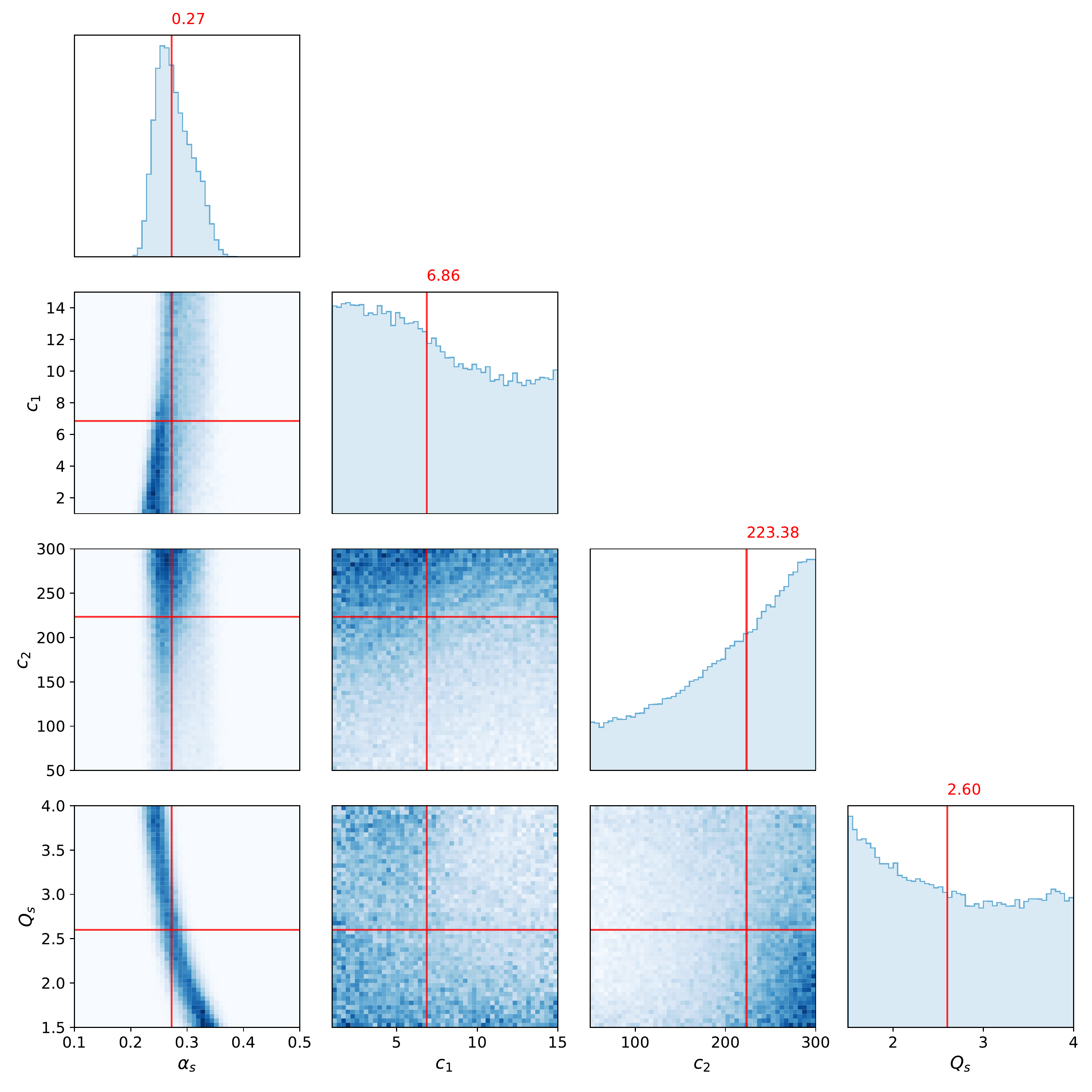}
	\caption{\label{fig:PosteriorEstimation4Obs} The posterior distribution of the model parameters. The emulator is calibrating to the three inclusive jet $R_{AA}$ results and the $D$ meson $R_{AA}$ result.}
\end{figure}

How different observables help with constraining the parameters is another interesting topic. There are essentially three types of observables we are calibrating on (charged hadron, D meson, and inclusive jet $R_{AA}$). In this section, I start with only calibrating to the three inclusive jet $R_{AA}$ results (see Fig.~\ref{fig:PosteriorEstimation3Obs}). Compared to Fig.~\ref{fig:BestPosteriorEstimation}, the major difference is a strong preference of large $c_2$ and weaker constraint on $Q_s$. Next, the calibration is done with both the inclusive jet $R_{AA}$ results and the $D$ meson $R_{AA}$ result as shown in Fig.~\ref{fig:PosteriorEstimation4Obs}. This time a combination of small $c_1$ and large $c_2$ is slightly preferred. The constraint on $Q_s$ is still weaker compared to Fig.~\ref{fig:BestPosteriorEstimation}. We can further divide the data into different $p_T$ ranges and test the sensitivity of the posterior to different combinations of the segmented data. I'll leave that to a future study. 

\section{Summary}

In this chapter, I have applied Bayesian analysis to the multi-stage energy loss model discussed in Chapter~\ref{sec:transport}, and Chapter~\ref{sec:jetscape_results}. I first performed closure tests on the emulator to validate its performance. With a quantitative measure $\langle\Delta\rangle$, the optimal settings for the Gaussian process emulator are determined, and the emulator's constraining power on each parameter is studied. Then I showed the sensitivity of the posterior distribution to different observables. In the end, I have successfully constrained the $\alpha_s$ and $Q_s$ and explained why $c_1$ and $c_2$ are difficult to constrain within the current setup. Continued discussion of the stability of the posterior distribution to additional fluctuations and the effect of correlated experimental systematic uncertainty or using the Mat\'{e}rn kernel in the emulator can be found in Appendix.~\ref{section:appendix_bayesian}. This analysis is the first result of a simultaneous description of $R_{AA}$ data for multiple hadron species and inclusive jets with a multi-stage energy loss description. 

However, there are still many things to improve in our analysis:

\begin{itemize}
    \item Include more realistic modeling into the JETSCAPE framework. For example, the MATTER model could use different $Q_s$ and $\hat{q}$ for different quark species. The $\hat{q}$ parameterization could be more flexible and dependent on quark mass. We should also apply a hybrid hadronization model for the heavy quarks and hadronic rescattering which are absent in current simulations. A more complex pre-equilibrium modeling and a (3+1)D hydrodynamic background are also things to improve upon.
    \item Include more observables into our calibration. Due to the current limitations, the flow coefficients are hard to simulate to a reasonable precision, but are helpful for studying the geometry and flow of the medium. Considering data from more centrality and collision systems is also important.
    \item Increase the statistics for each design point and the number of design points. The current level of model statistical uncertainty is not enough to constrain $c_1$ and $c_2$ (see Fig.~\ref{fig:Delta_comparison}).
\end{itemize}

\chapter{Conclusion} \label{sec:summary}

\vspace{1in}

Ultra-relativistic heavy ion collisions provide the opportunity for studying the extremely hot and dense form of matter, namely the quark gluon plasma (QGP). The degrees of freedom in the QGP, as its name suggests, are quarks and gluons. The dynamics of the QGP is believed to be dominated by quantum chromodynamics (QCD). 

In this thesis, I have adopted a multi-stage approach for simulating the evolution of the QGP during the collision. Specifically, the description is divided into two parts depending on the energy scale we are looking at: the bulk medium and the hard probes. The evolution of the bulk medium which produces large number of soft hadrons in the final state has been successfully described by relativistic viscous hydrodynamics with phenomenology modeling of the initial condition and stopping condition. 
As for the hard probes (leading partons with $p_T>10$ GeV, jets), they are created at the early stage of the collision and acting as probes to the evolution of the medium. In this study, their creation is simulated by PYTHIA and their interaction with the QGP medium described by transport models like MATTER and LBT. The MATTER model takes care of the in-medium DGLAP evolution of the highly virtual partons. LBT describes the evolution of on-shell partons inside the medium and contains both the elastic and inelastic scattering kernel. A virtuality dependent parameterization of the transport coefficient $\hat{q}$ is proposed in order to explain the smaller $\hat{q}$ extracted from experiments with higher collision energy. 

All these evolution models have been integrated into the JETSCAPE framework. This allows a systematic comparison between different energy loss mechanisms which is carried out in Section~\ref{sec:jetscape_results}. The effects of employing MATTER, LBT, or the multi-stage MATTER+LBT approach on the charged hadron and D meson $R_{AA}$ have been studied. We found that the MATTER model with the virtuality dependent $\hat{q}$, is approaching the vacuum DGLAP at high $p_T$. However, it effectively reduces the path length the hard parton spent in the LBT phase, thus enhances the $R_{AA}$. Using the current ``optimal'' parameters and the MATTER+LBT approach, we have predicted a slightly lower $R_{AA}$ for D meson compared to charged hadron at high $p_T$ across multiple centralities. The experimental data is yet to confirm this observation due to the lack of D meson $R_{AA}$ beyond $100$~GeV. 

Despite a good description of the experimental data with the MATTER+LBT approach, it is difficult to determine the optimal values for the model parameters and how far we can get in terms of describing the experimental data. The uncertainty in experimental data and simulation should also be taken into account. For that, I have utilized a state-of-the-art Bayesian model-to-data comparison framework which has been proven successful when applied to the soft sector of heavy ion collisions. In fact, the bulk medium evolution is simulated with the optimal extracted parameters from a previous Bayesian model-to-data comparison. 
The effects of emulator uncertainty, experiment uncertainty and model uncertainty, have been investigated in Section~\ref{sec:bayesian} and Section~\ref{sec:bayesian_results}. 
I have proposed a new metric for measuring the performance of the Gaussian process emulator against the choice of kernel and number of principal components in closure tests. The sensitivity of the posterior distribution of the parameters to different observables and uncertainty levels are investigated before we draw a conclusion from the final posterior distribution calibrated to charged hadron, D meson and inclusive jet $R_{AA}$ simultaneously. We found a similar peak of $\alpha_s$ and a slightly lower peak of $Q_s$ compared to our previous manual parameter fitting. And the current level of uncertainty from both experiment and simulation failed to constrain the $c_1$ and $c_2$ parameter in the parametrization of $\hat{q}$. 

Our study is a step forward towards a simultaneous description of energy loss for multiple hadron species and jet types in heavy ion collisions, with proper treatment of the uncertainties from different sources and a qualitative method for choosing the optimal Gaussian process emulator. However, there are still many places that can be improved. On the theoretical side, the recombination mechanism and hadronic rescattering should be included in the energy loss calculation. A more flexible parameterization of the transport coefficient could also be considered. When calibrating to data, the Bayesian analysis should include more observables, and the simulation statistics for a single design point as well as the total number of design points could be increased.

With better and expanded experimental measurements from LHC (Run 3) and upcoming electron ion collider (EIC), and advancement in both theoretical modeling and computing power, we are looking forward to a better understanding of the properties of QGP in the coming years.

\appendix
\chapter{Deriving the Langevin equation from the Fokker-Planck equation} \label{section:appendix_langevin_derivation}

One way to show the equivalence between the Fokker-Planck and the Langevin equation is to covert the Langevin equation to a path integral expression and recognize the Fokker-Planck equation as an Euclidean Schrodinger equation, which also has a path integral representation \cite{zinn2021quantum}. Another way is first to derive the equation of motion for the probability distribution $\rho(\vec{x}, \vec{p}, t)$ of finding a particle in the interval $(x,x+dx), (p,p+dp)$ at time $t$ for one realization of the random kick $\xi(t)$, then average $\rho(\vec{x}, \vec{p}, t)$ over a ensemble of $\xi(t)$. 

The normalization constraint on $\rho(\vec{x}, \vec{p}, t)$ reads:
\begin{equation}
\int_{-\infty}^{\infty} d^3x_i \int_{-\infty}^{\infty} d^3p_i \rho(\vec{x}, \vec{p}, t)  = 1,
\end{equation}
which gives the continuity equation in the phase space:
\begin{equation}
\begin{split}
\frac{\partial}{\partial t} \rho(\vec{x}, \vec{p}, t) & = - \frac{\partial }{\partial x_i} (\frac{\partial x_i}{\partial t} \cdot \rho(\vec{x}, \vec{p}, t)) - \frac{\partial }{\partial p_i} (\frac{\partial p_i}{\partial t} \rho(\vec{x}, \vec{p}, t))\\
& = - \frac{\partial }{\partial x_i} \left(\frac{p_i}{E} \cdot \rho(\vec{x}, \vec{p}, t)\right) - \frac{\partial}{\partial p_i} \left((-\eta_D p_i + \xi_i))\cdot \rho(\vec{x}, \vec{p}, t)\right) \\
& = \left(-\frac{p_i}{E} \frac{\partial}{\partial x_i} + p_i \frac{\partial}{\partial p_i} \eta_D p_i\right) \rho(\vec{x}, \vec{p}, t) - \xi_i(t) \frac{\partial}{\partial p_i} \rho(\vec{x}, \vec{p}, t) \\
& = - L_0 \rho(\vec{x}, \vec{p}, t) - L_1(t) \rho(\vec{x}, \vec{p}, t),
\label{eqn:probability_flow_equation}
\end{split}
\end{equation}
where:
\begin{align}
& L_0 = \frac{p_i}{E} \frac{\partial}{\partial x_i} - \frac{\partial}{\partial p_i} \eta_D p_i\\
& L_1 = \xi_i(t) \frac{\partial}{\partial p_i}.
\end{align}

Performing a change of variables $\rho(\vec{x}, \vec{p}, t) = e^{-L_0 t} \sigma(\vec{x}, \vec{p}, t)$ will transform Eq.~\ref{eqn:probability_flow_equation} into:
\begin{equation}
\frac{\partial}{\partial t} \sigma(\vec{x}, \vec{p}, t) = - e^{L_0 t} L_1(t) e^{-L_0 t} \sigma(\vec{x}, \vec{p}, t) = - V(t) \sigma(\vec{x}, \vec{p}, t),
\end{equation}
which is now a first order linear partial differential equation (PDE) and has the following solution:
\begin{equation}
\sigma(t) = \exp \left[- \int_0^t dt' V(t')\right]\sigma(0).
\end{equation}

Performing an ensemble average over the Gaussian random kick $\xi(t)$ leads to:
\begin{equation}
\left<\sigma(t)\right>_{\xi} = \exp\left[\frac{1}{2} \int_0^t dt_1 \int_0^t dt_2 \left<V(t_1) V(t_2)\right>_{\xi}\right] \sigma(0).
\label{eqn:characteristic function}
\end{equation}

And the integral in Eqn.~\ref{eqn:characteristic function} can be calculated as:
\begin{align*}
\frac{1}{2} \int_0^t dt_1 \int_0^t dt_2 \left<V(t_1) V(t_2)\right>_{\xi} &= \frac{1}{2} \int_0^t dt_1 \int_0^t dt_2 \left<e^{L_0 t_1} L_1(t_1) e^{-L_0 t_1} e^{L_0 t_2} L_1(t_1) e^{-L_0 t_1}\right> \\
& = \frac{1}{2} \int_0^t dt_1 \int_0^t d t_2 \left<e^{L_0 t_1} \xi_i(t_1) \frac{\partial}{\partial p_i} e^{-L_0 t_1} e^{L_0 t_2} \xi_j(t_2) \frac{\partial}{\partial p_j} e^{-L_0 t_1}\right> \\
& = \frac{1}{2} \int_0^t dt_1 \int_0^t d t_2 \left<e^{L_0 t_1}  \frac{\partial}{\partial p_i} e^{-L_0(t_1-t_2)}  \xi_i(t_1) \xi_j(t_2) \frac{\partial}{\partial p_j} e^{-L_0 t_1}\right> \\
& = \frac{1 }{2} \int_0^t dt_1 e^{L_0t_1}\frac{\partial}{\partial p_i} \kappa^{ij} \frac{\partial}{\partial p_j} e^{-L_0 t_1},
\end{align*}
which means:
\begin{equation}
\frac{\partial}{\partial t} \left<\sigma(t)\right>_{\xi} = \frac{1}{2} e^{L_0 t} \frac{\partial}{\partial p_i} \kappa^{ij} \frac{\partial}{\partial p_j} \left<\sigma(t)\right>_{\xi},
\end{equation}

If one defines $p(\vec{x}, \vec{p}, t)=\left<\rho(\vec{x}, \vec{p}, t)\right>_{\xi}=e^{-L_0 t}\left<\sigma(\vec{x}, \vec{p}, t)\right>_{\xi} $, Eq.~\ref{eqn:probability_flow_equation} becomes:
\begin{equation}\label{eqn:fokker_plank_equation_full}
\frac{\partial}{\partial t} p(\vec{x}, \vec{p}, t) =  - \frac{p_i}{E} \frac{\partial}{\partial x_i} p(\vec{x}, \vec{p}, t) - \frac{\partial}{\partial p_i}(\eta_D p_i p(\vec{x}, \vec{p}, t)) + \frac{1}{2} \frac{\partial}{\partial p_i} \kappa^{ij} \frac{\partial}{\partial p_j} p(\vec{x}, \vec{p}, t).
\end{equation}

This is the Fokker-Planck equation for the full phase distribution $p(\vec{x}, \vec{p}, t)$. The equivalence between the Fokker-Plank and Langevin equation under a Gaussian random force is now proved.

\chapter{More on Bayesian analysis}\label{section:appendix_bayesian}



\section{Closure test results with a simple analytical bulk model}


\begin{figure}
	\centering
	\includegraphics[width=0.48\textwidth]{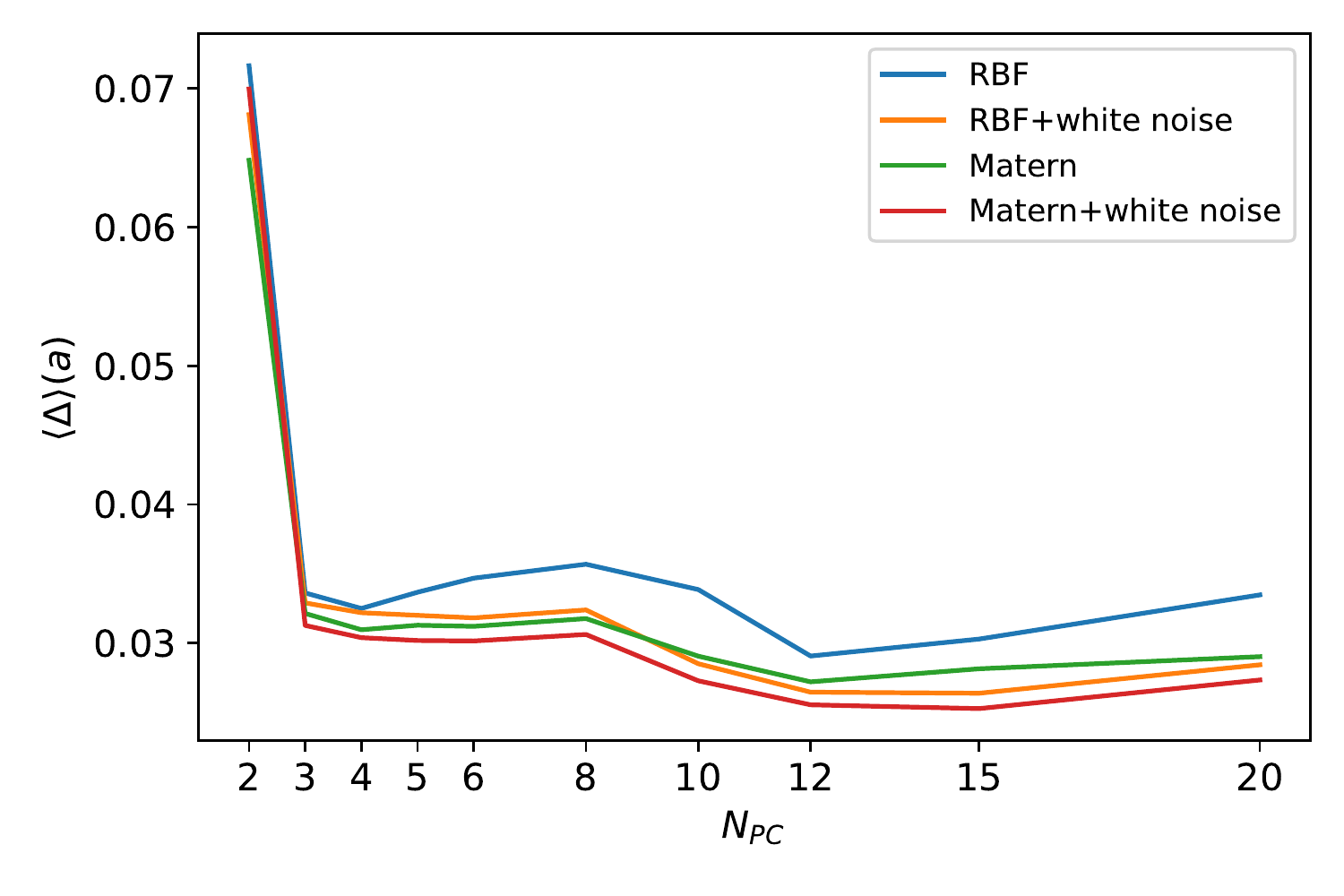}
	\includegraphics[width=0.48\textwidth]{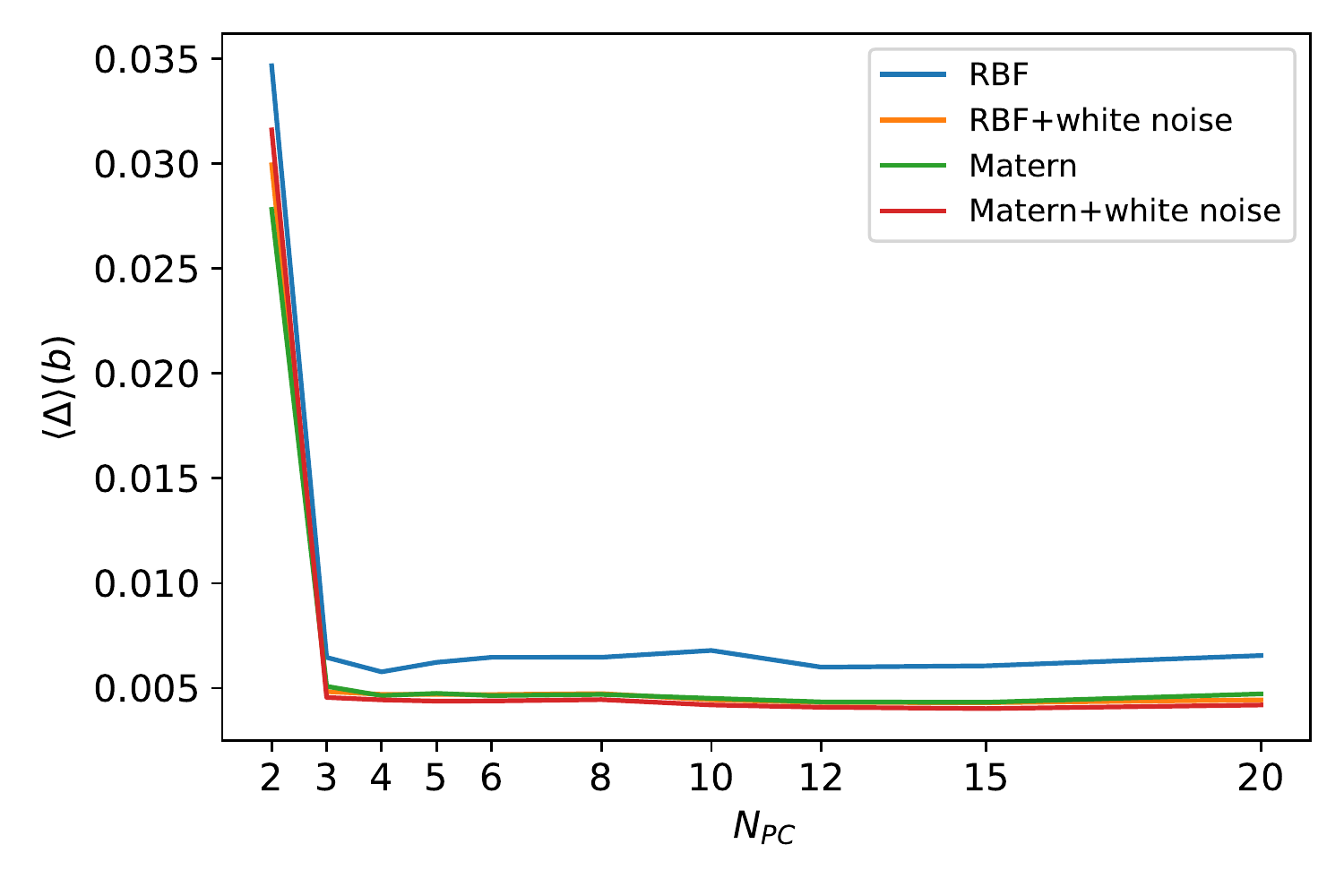}
	\includegraphics[width=0.48\textwidth]{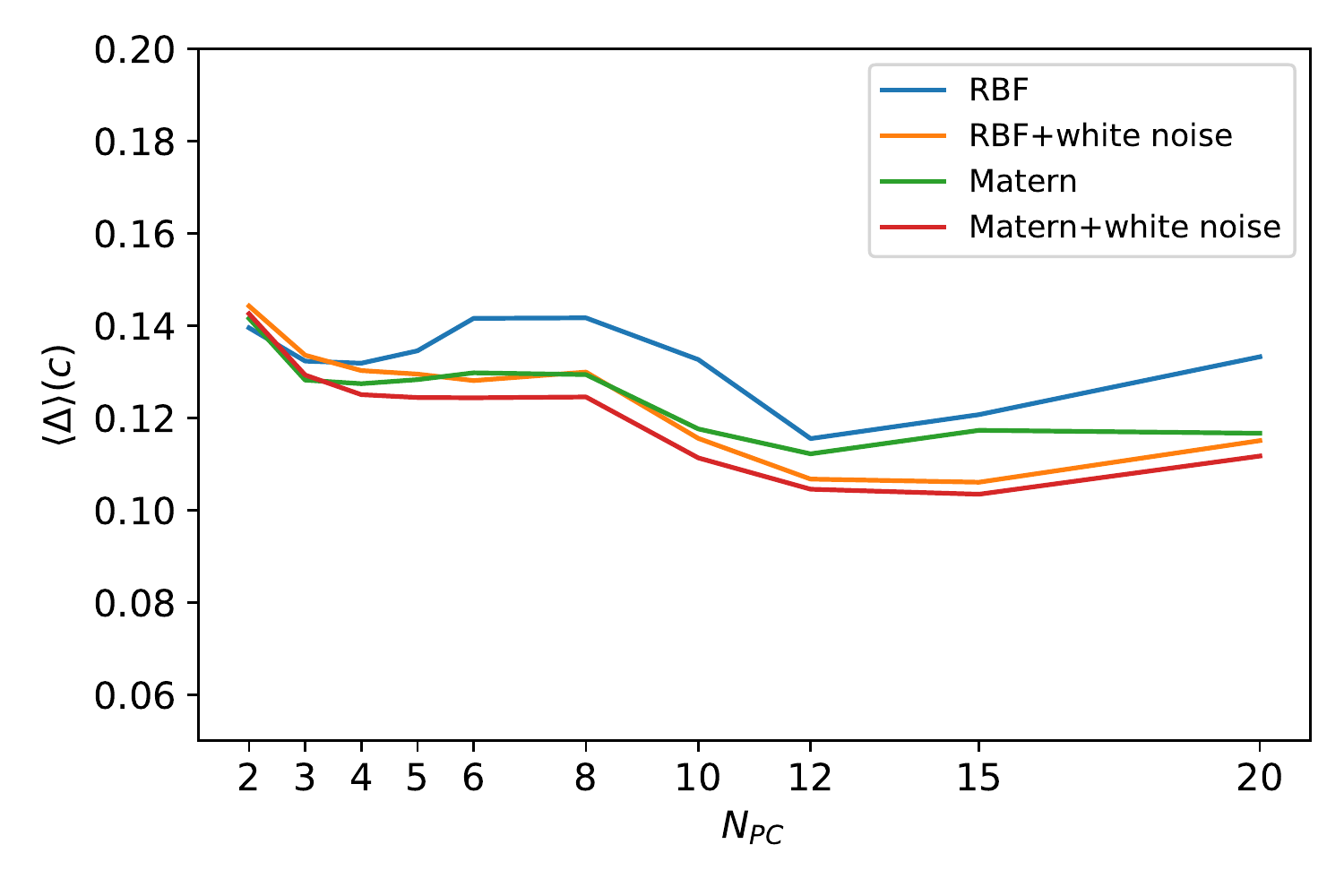}
	\includegraphics[width=0.48\textwidth]{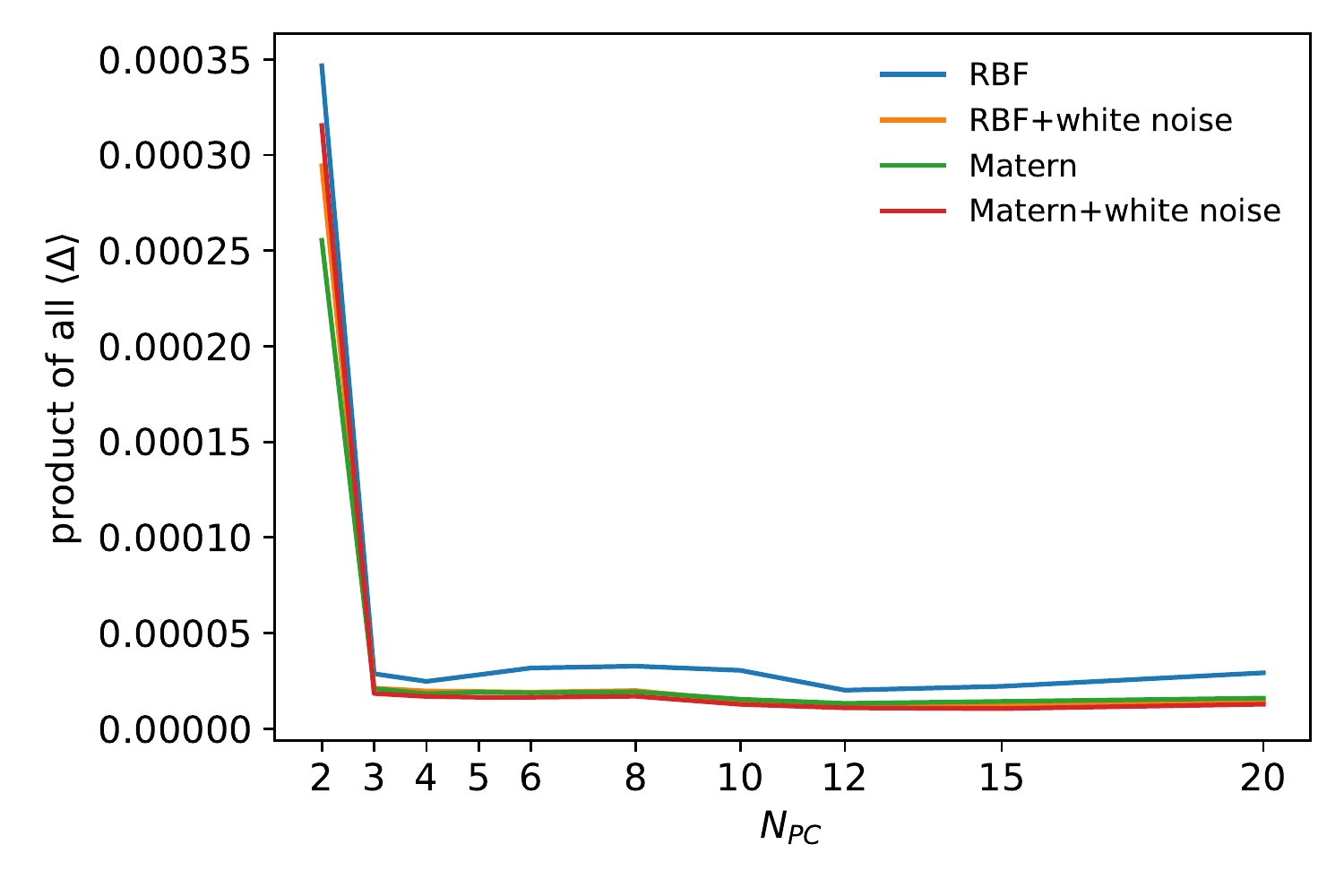}
	\caption[Comparison of $\langle\Delta\rangle$ for different parameters with different kernels and number of principal components]{\label{fig:Delta_comparison_simple_50_dp_1_stat} Comparison of $\langle\Delta\rangle$ for different parameters with different kernels and number of principal components. The last plot shows the product of all $\langle\Delta\rangle$. $1\%$ of statistical model uncertainty is introduced. The Mat\'{e}rn kernel uses $\nu=3/2$.}
\end{figure}

\begin{figure}
	\centering
	\includegraphics[width=0.48\textwidth]{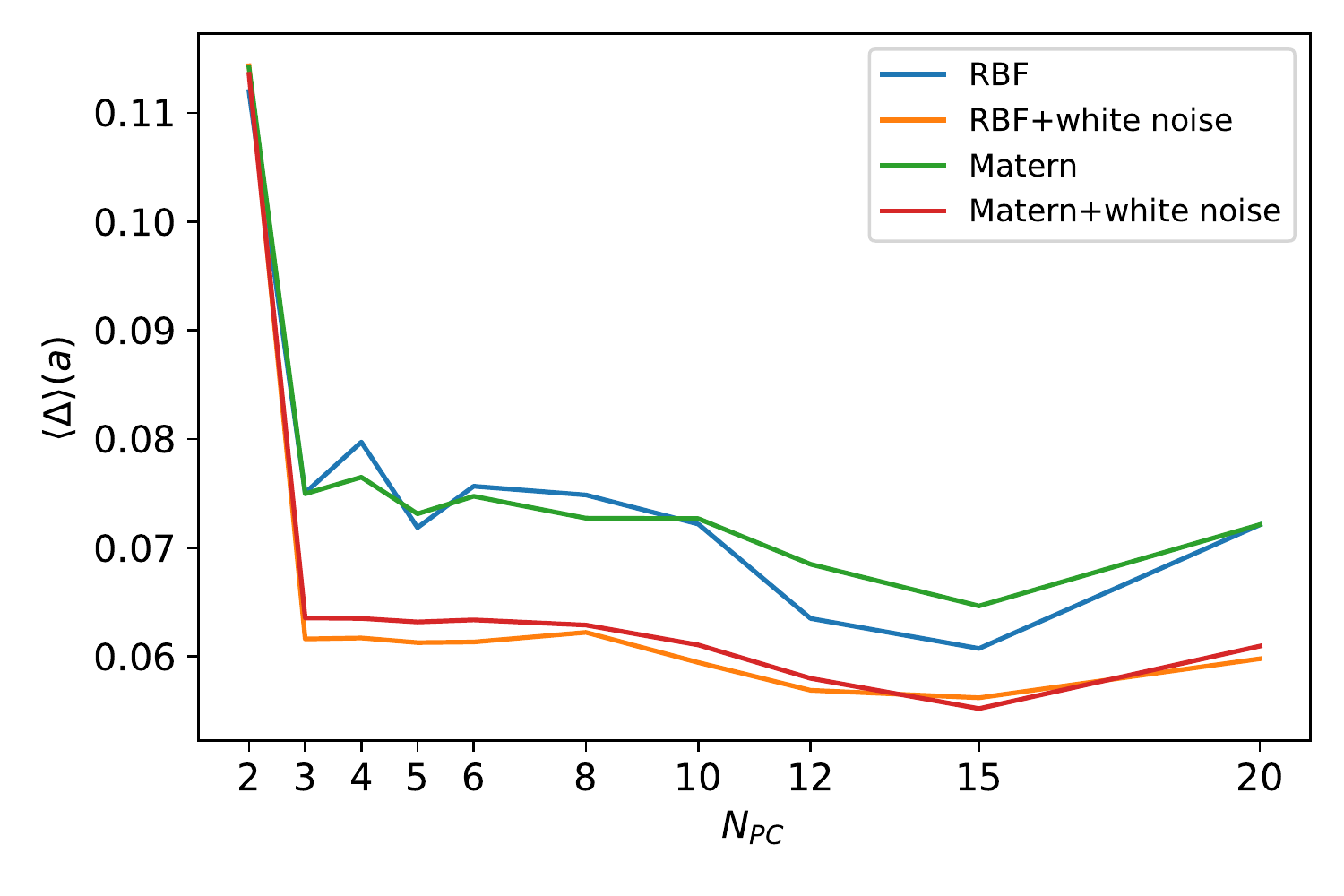}
	\includegraphics[width=0.48\textwidth]{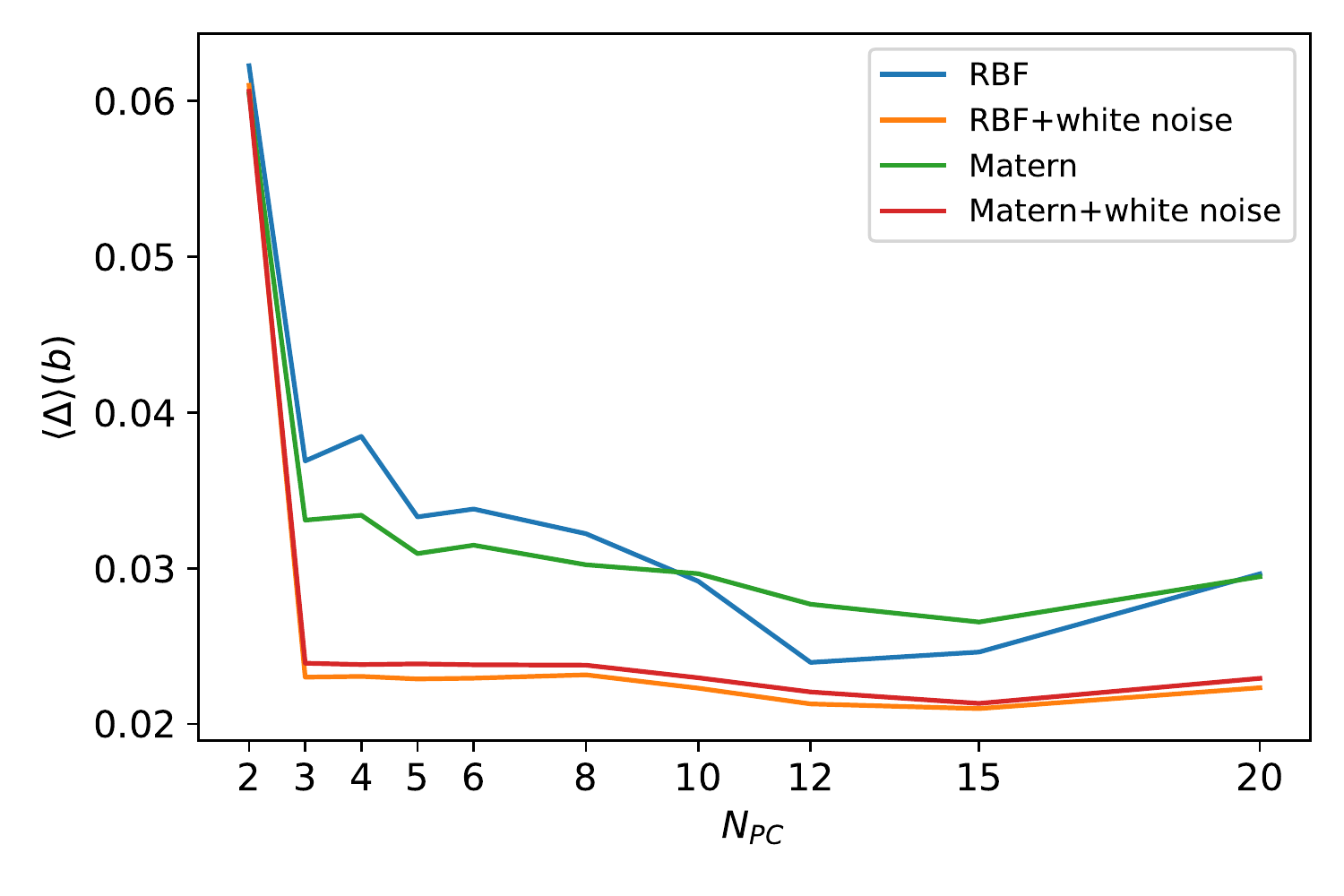}
	\includegraphics[width=0.48\textwidth]{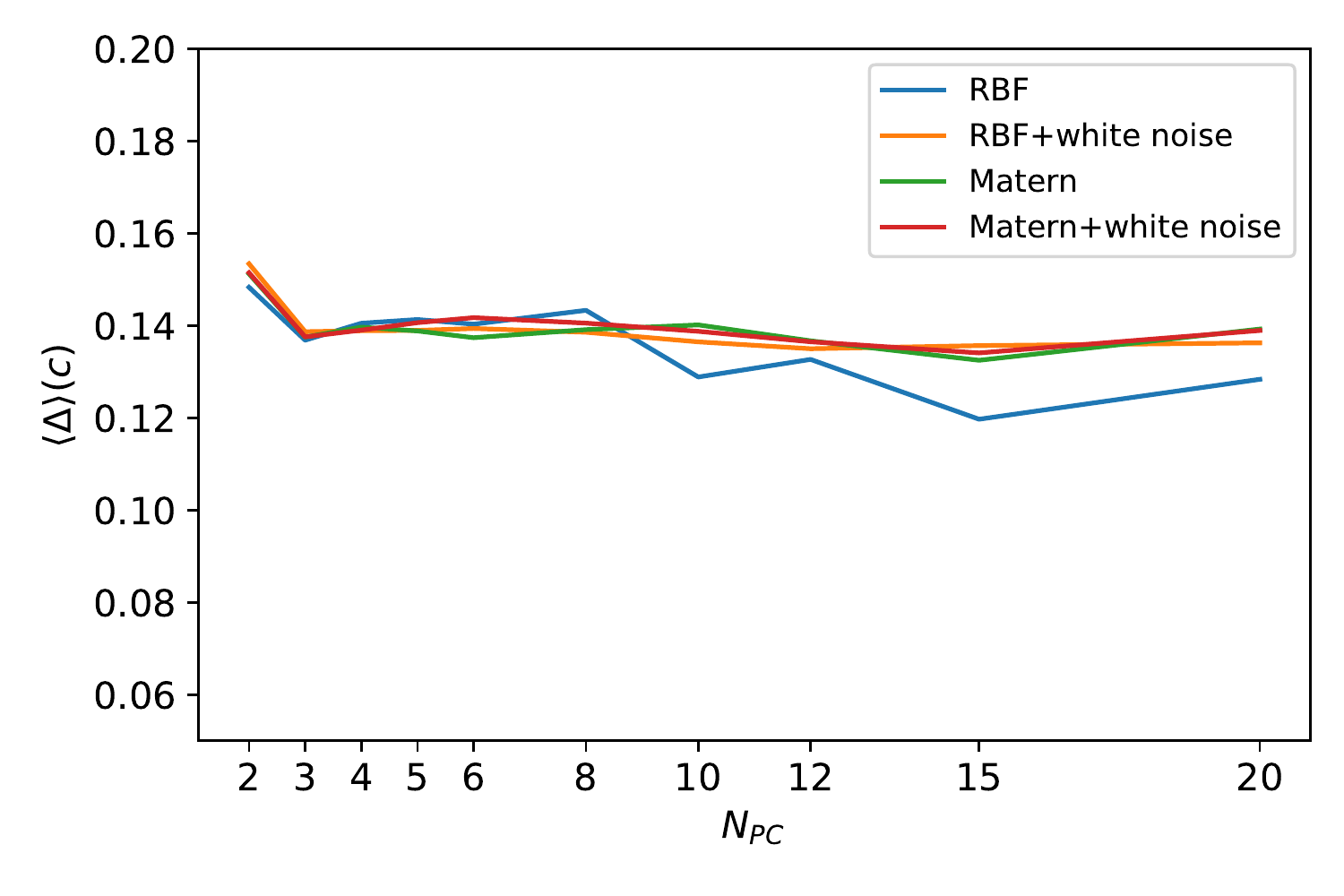}
	\includegraphics[width=0.48\textwidth]{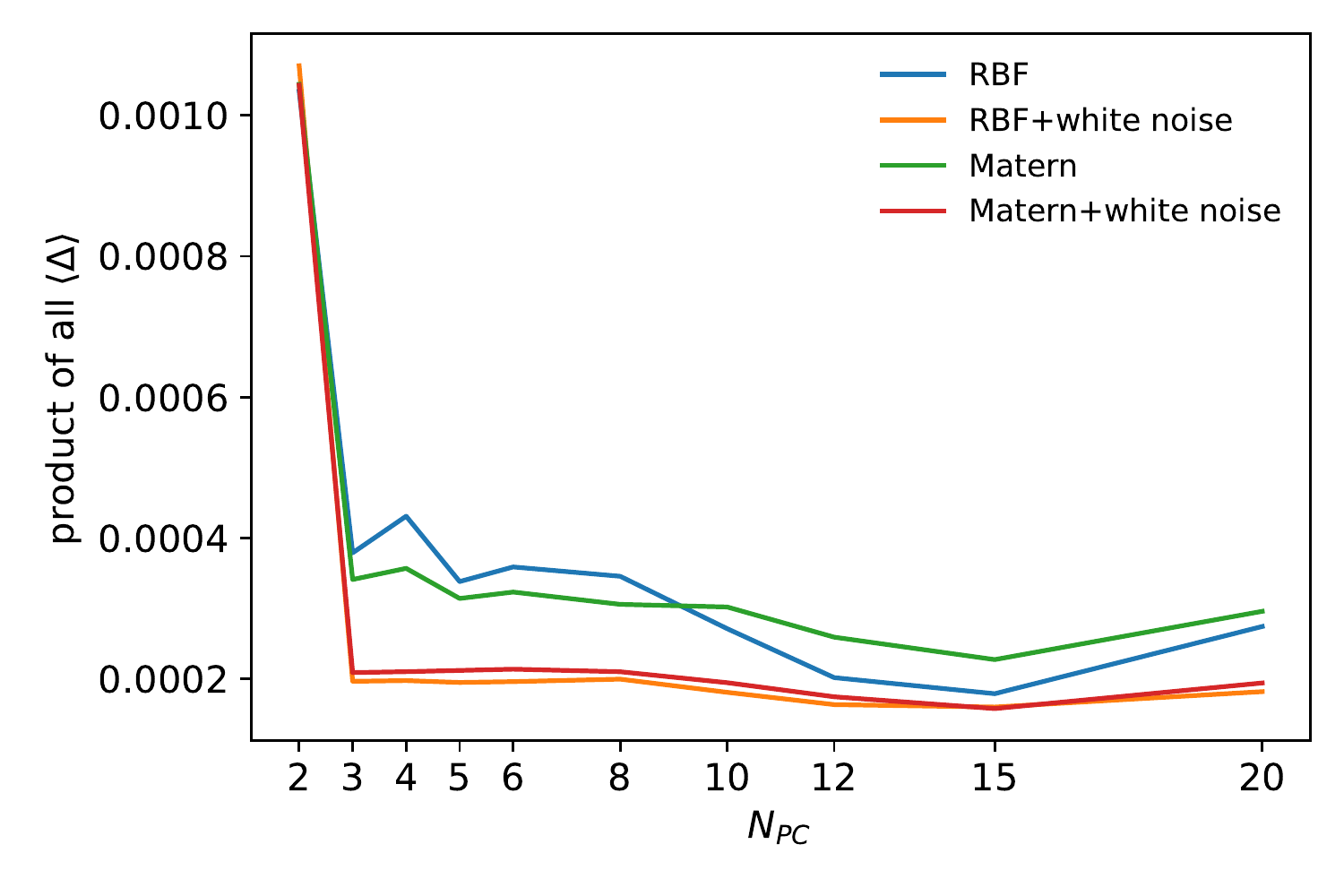}
	\caption[Comparison of $\langle\Delta\rangle$ for different parameters with different kernels and number of principal components]{\label{fig:Delta_comparison_simple_50_dp_5_stat} Comparison of $\langle\Delta\rangle$ for different parameters with different kernels and number of principal components. The last plot shows the product of all $\langle\Delta\rangle$. $5\%$ of statistical model uncertainty is introduced. The Mat\'{e}rn kernel uses $\nu=3/2$.}
\end{figure}

\begin{figure}
	\centering
	\includegraphics[width=0.48\textwidth]{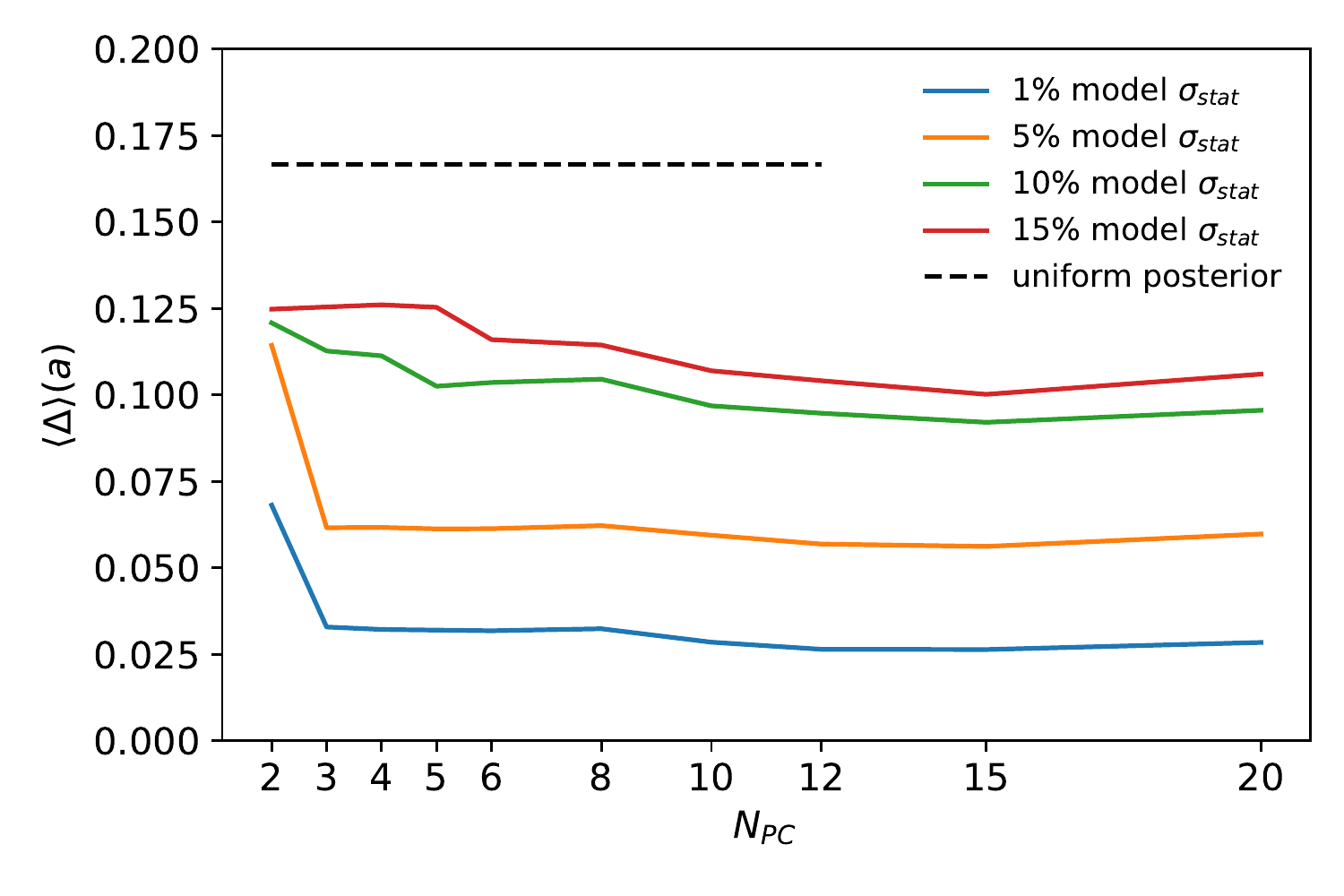}
	\includegraphics[width=0.48\textwidth]{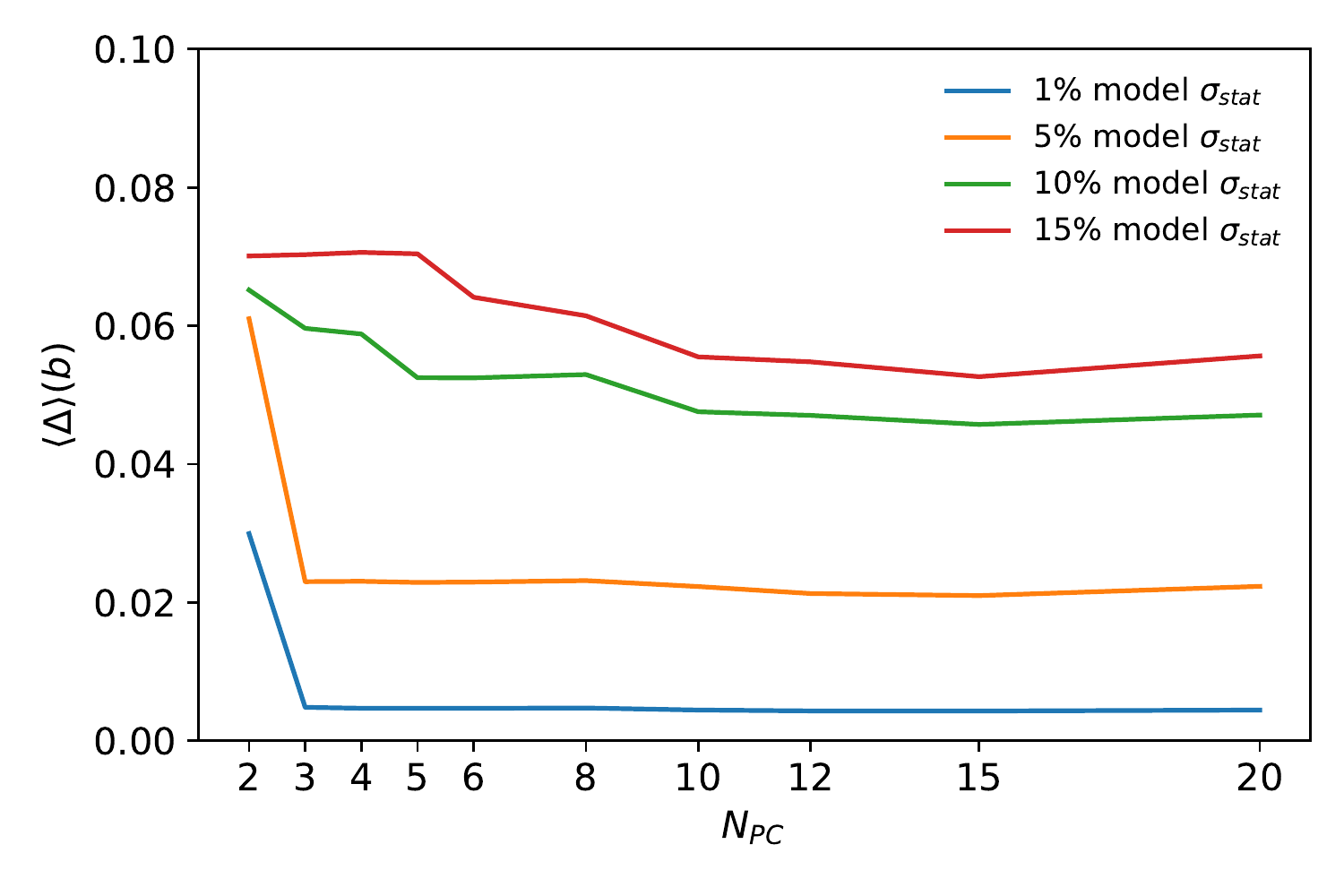}
	\includegraphics[width=0.48\textwidth]{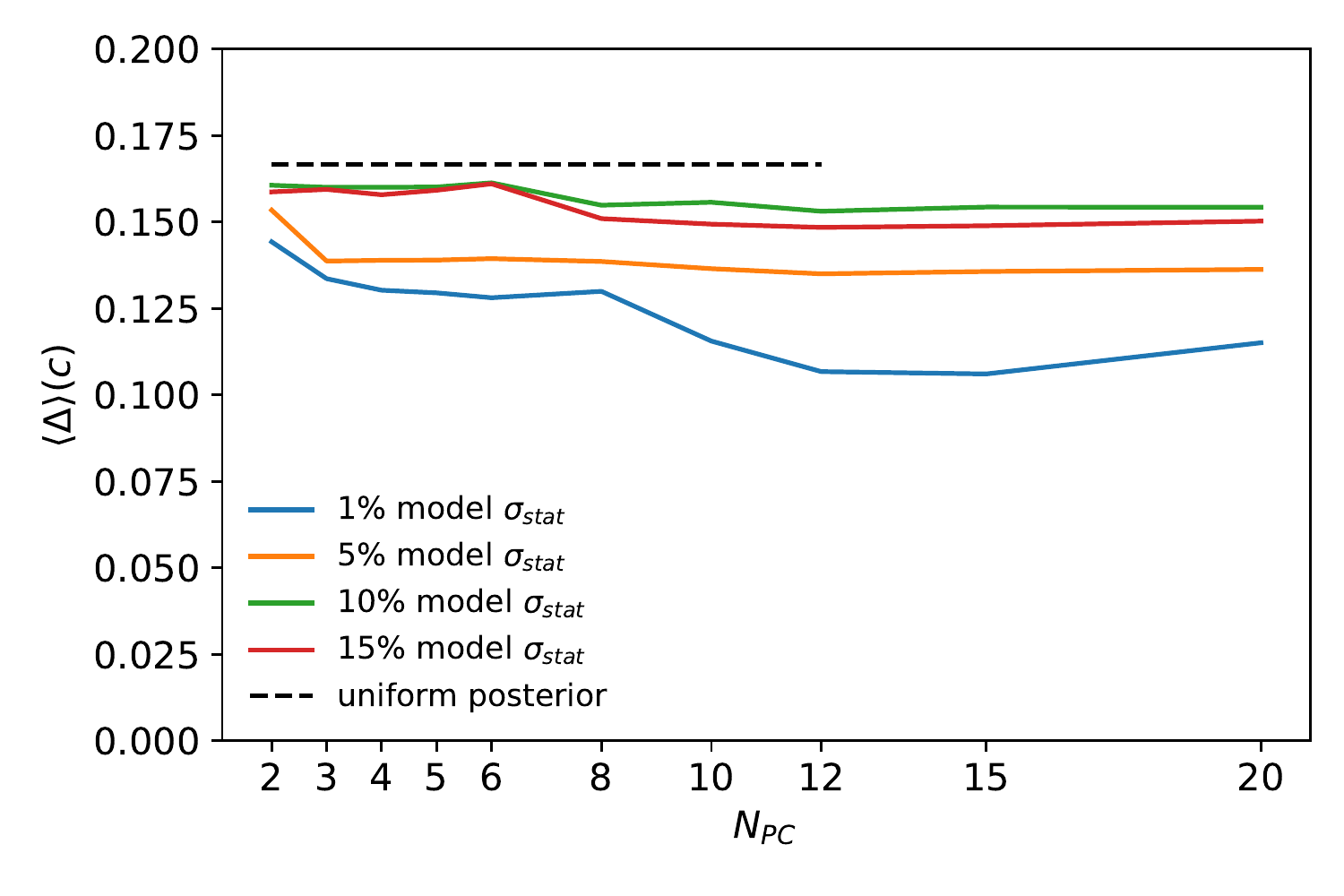}
	\includegraphics[width=0.48\textwidth]{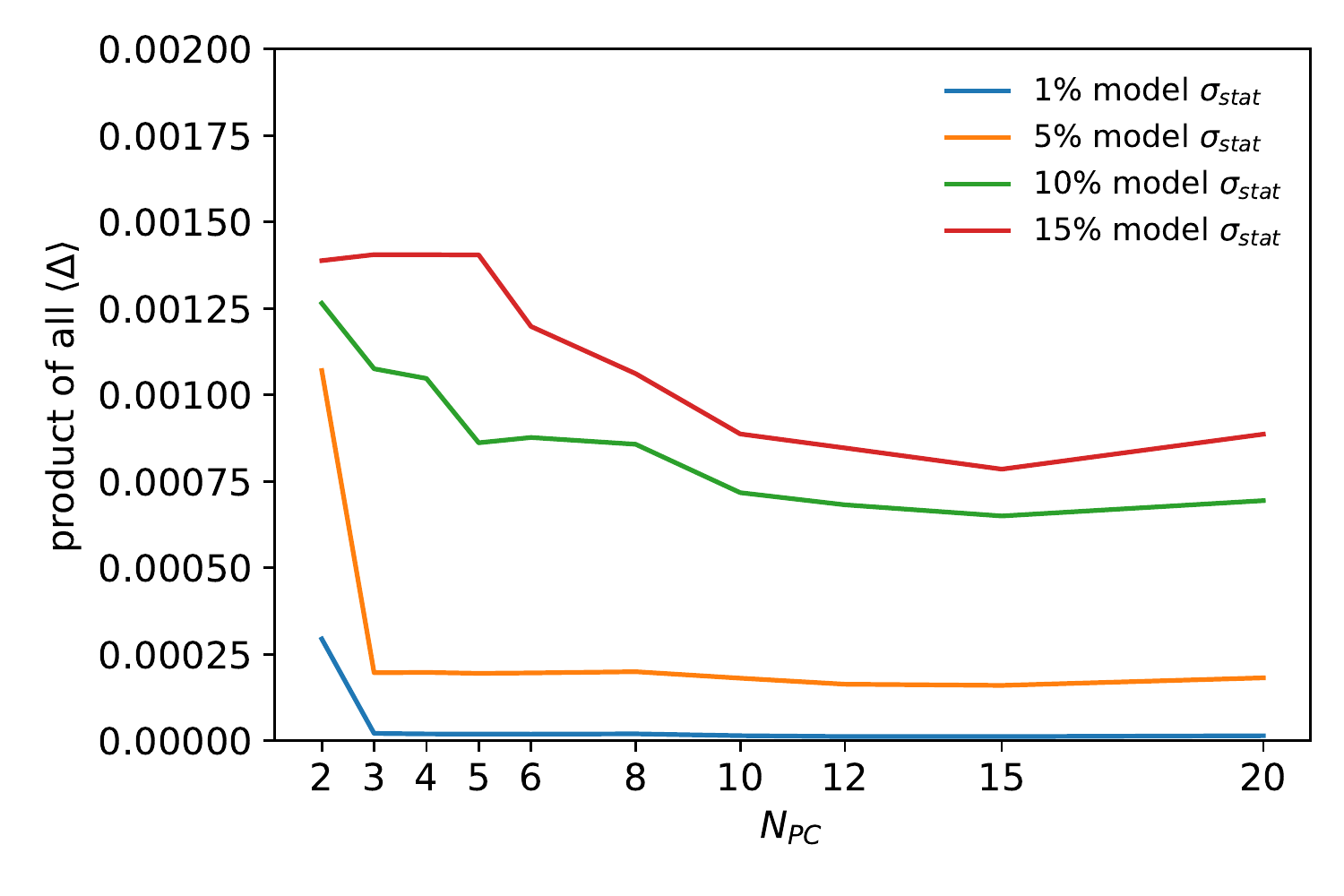}
	\caption[Comparison of $\langle\Delta\rangle$ for different parameters with different kernels, number of principal components and model statistical uncertainties]{\label{fig:Delta_comparison_simple} Comparison of $\langle\Delta\rangle$ for different parameters with different kernels, number of principal components and model statistical uncertainties. The last plot shows the product of all $\langle\Delta\rangle$. The Mat\'{e}rn kernel uses $\nu=3/2$.}
\end{figure}

We can also calculate $\langle\Delta\rangle$ using the analytical bulk model discussed in Section~\ref{sec:simple_bulk_bayesian} which has three parameters $a$, $b$, and $c$. The benefit of using this simple model is that the level of statistical model uncertainty can be tuned very easily. $50$ design points are sampled and four levels of statistical model uncertainty $1\%,5\%,10\%,15\%$ are investigated. 
In Fig.~\ref{fig:Delta_comparison_simple_50_dp_1_stat} and Fig.~\ref{fig:Delta_comparison_simple_50_dp_5_stat}, the effect of additional statistical model uncertainty on the choice of kernel can be seen. The inclusion of the white noise kernel helps reducing $\langle\Delta\rangle$. 

Fig.~\ref{fig:Delta_comparison_simple} compares the $\langle\Delta\rangle$ results between different levels of statistical model uncertainty. As expected, the smaller the statistical model uncertainty, the smaller $\langle\Delta\rangle$ for all parameters. However, one should notice that even with just $1\%$ of statistical model uncertainty, $\langle\Delta\rangle(c)$ is around $0.13$ which translates to a Gaussian posterior with $\sigma\approx 0.2$ on average. This is consistent with the findings from Section \ref{sec:simple_bulk_bayesian} that the parameter $c$ is the most difficult to constrain as it only affects a small temperature range in the $\eta/s$ parameterization.



\section{Stability of the posterior to fluctuations} \label{section:appendix_stability_fluctuations}

The closure test offers a excellent check that the emulator can reproduce mock data (model calculation) at many design points. But the relation between the posterior distribution and the uncertainty level of the training data is still not explored. It is difficult to reduce the fluctuations of the training data as it requires running more events. The other direction is easier to explore and can be investigated in two ways.

\begin{figure}
	\centering
	\includegraphics[width=0.96\textwidth]{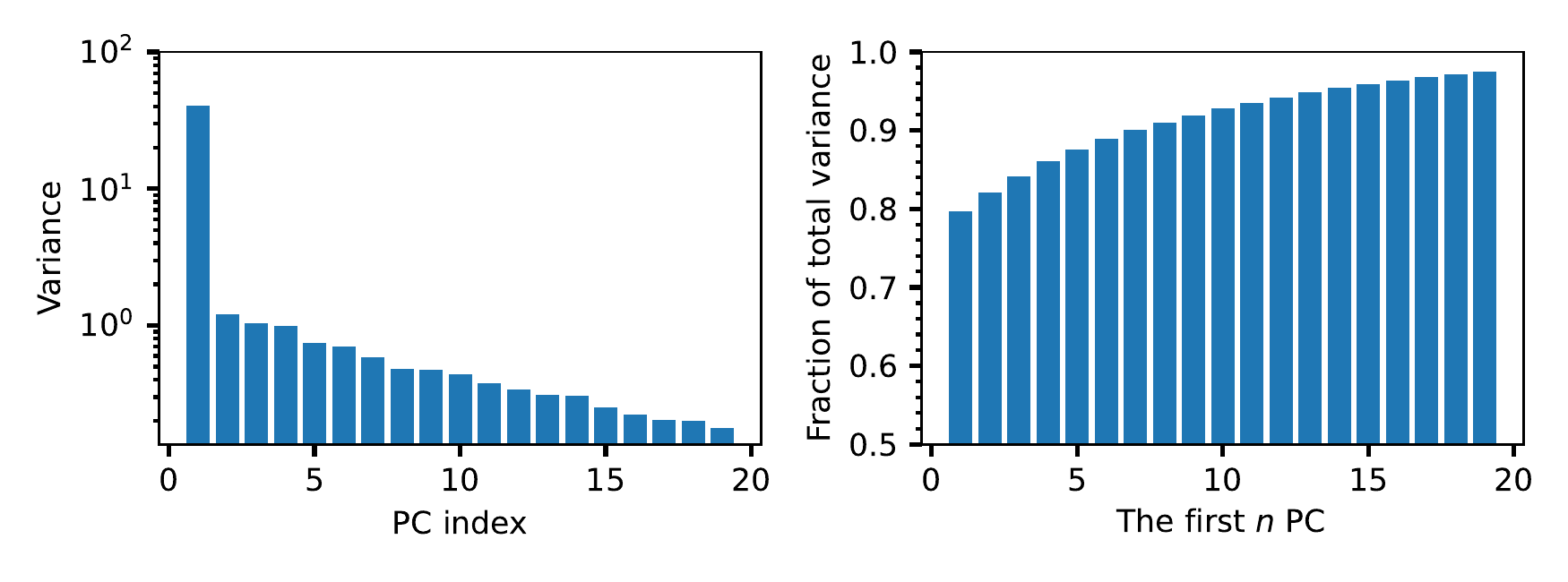}
	\caption[Relation between variances and principal components]{\label{fig:PC_importance_reduce_stat} \textbf{Left}: The variance explained by each principal component. \textbf{Right}: The cumulative variance explained by the first $n$ principal components. Only $1/3$}
\end{figure}

\begin{figure}
	\centering
	\includegraphics[width=0.96\textwidth]{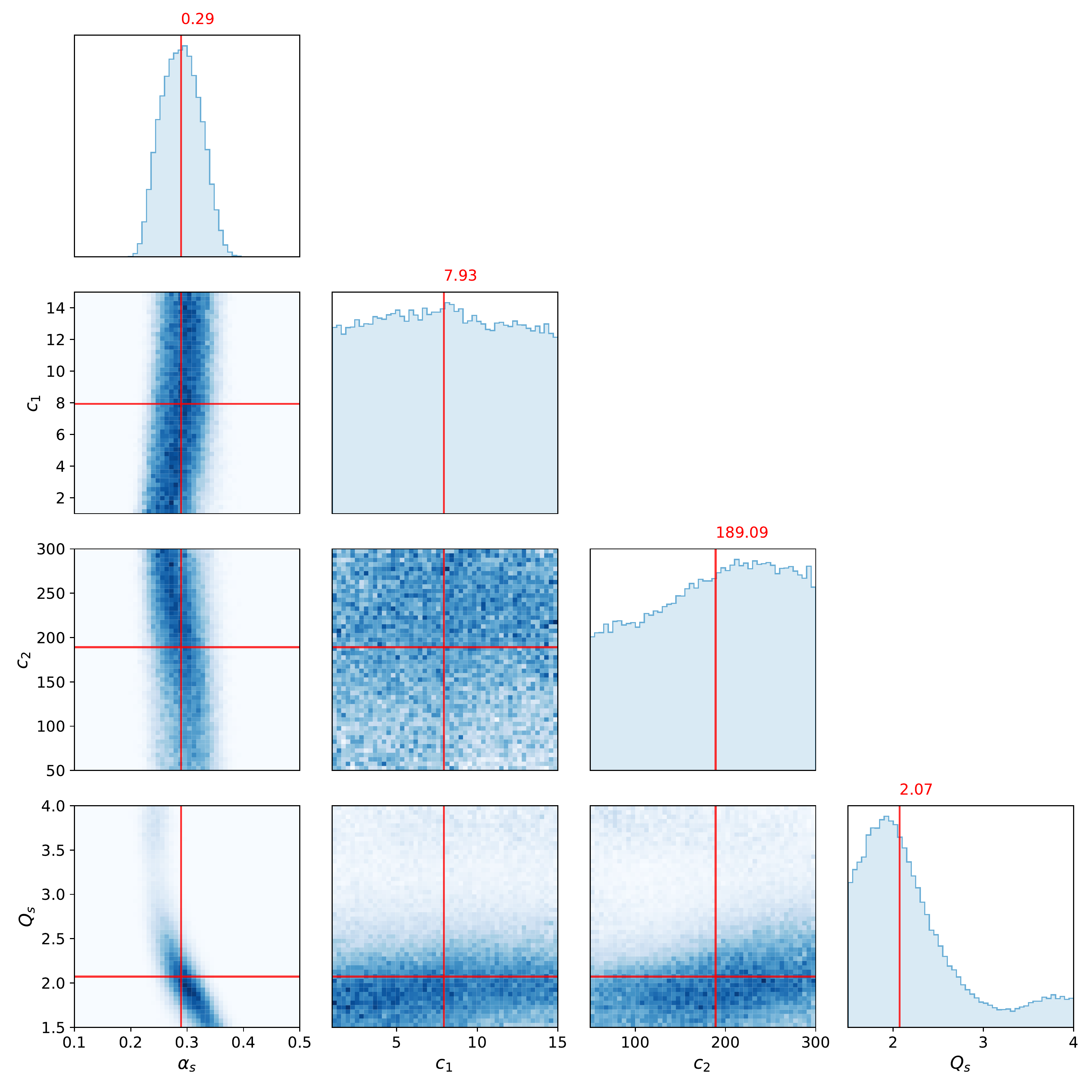}
	\caption{\label{fig:PosteriorEstimation_reduce_stat} The posterior distribution of the model parameters. Only $1/3$ of events are used for each design point.}
\end{figure}

First, we can reduce the statistics when generating the training data. Only $1/3$ of events are now used for each design point. This way the statistical fluctuations for each observable are scaled up proportionally. In Fig.~\ref{fig:PC_importance_reduce_stat}, we can see that now $5$ principal components only explains around $86\%$ of the total variance compared to $95\%$ before.  Fig.~\ref{fig:PosteriorEstimation_reduce_stat} shows the posterior distribution of the parameters. Compared to whats shown in Fig.~\ref{fig:BestPosteriorEstimation}, the posterior distribution is not altered much in this case, except for less constraint on $c_1$ and $c_2$. 

\begin{figure}
	\centering
	\includegraphics[width=0.48\textwidth]{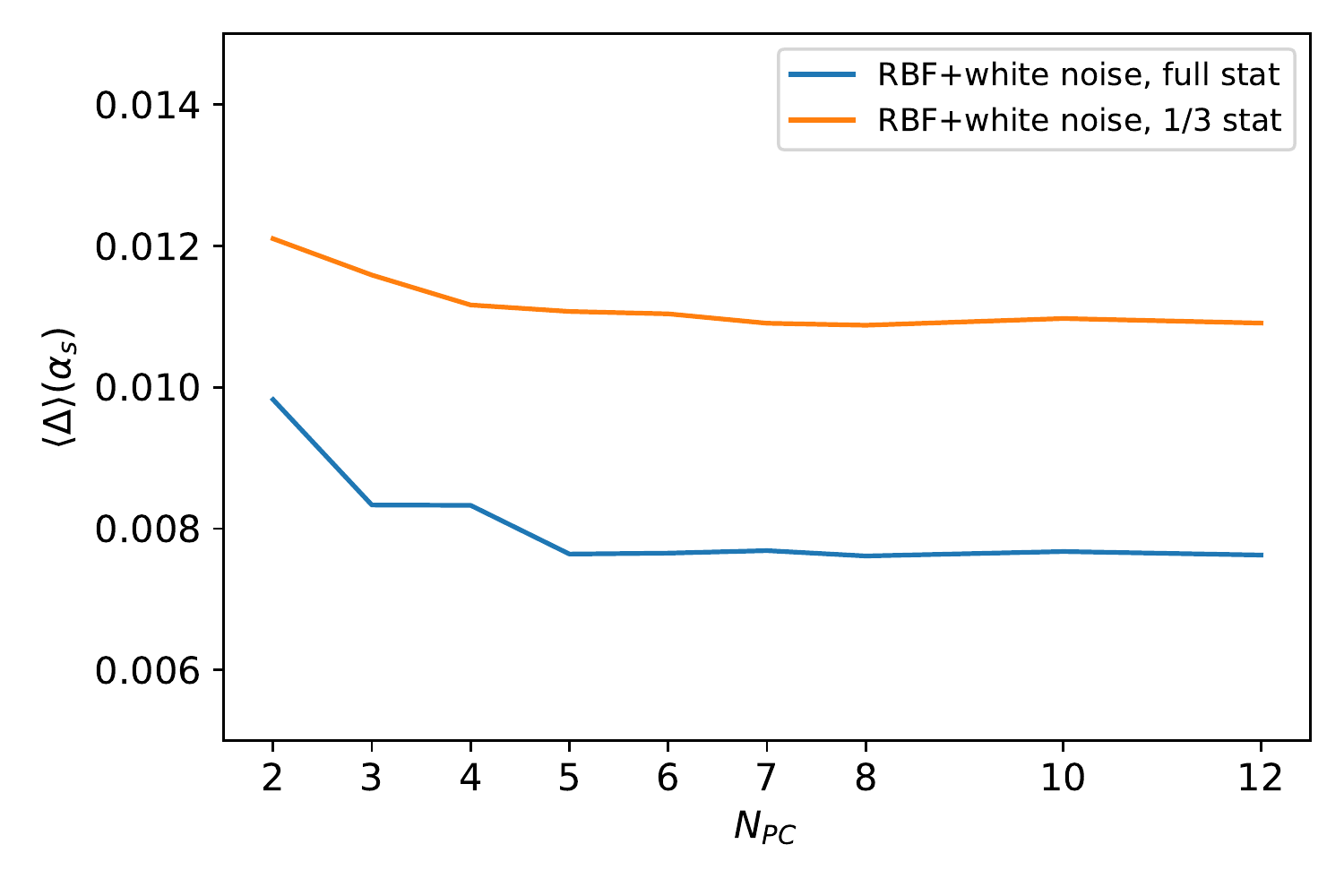}
	\includegraphics[width=0.48\textwidth]{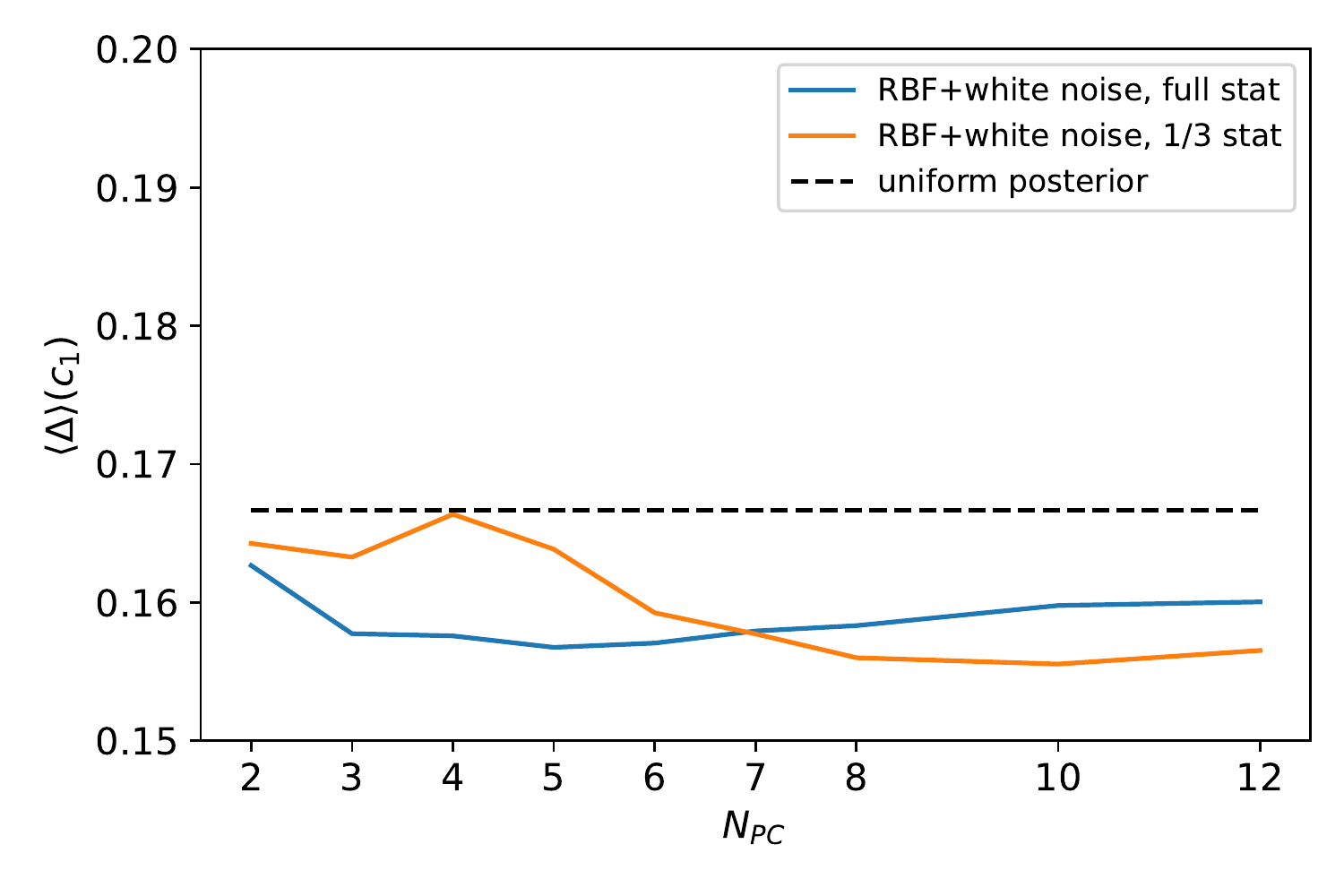}
	\includegraphics[width=0.48\textwidth]{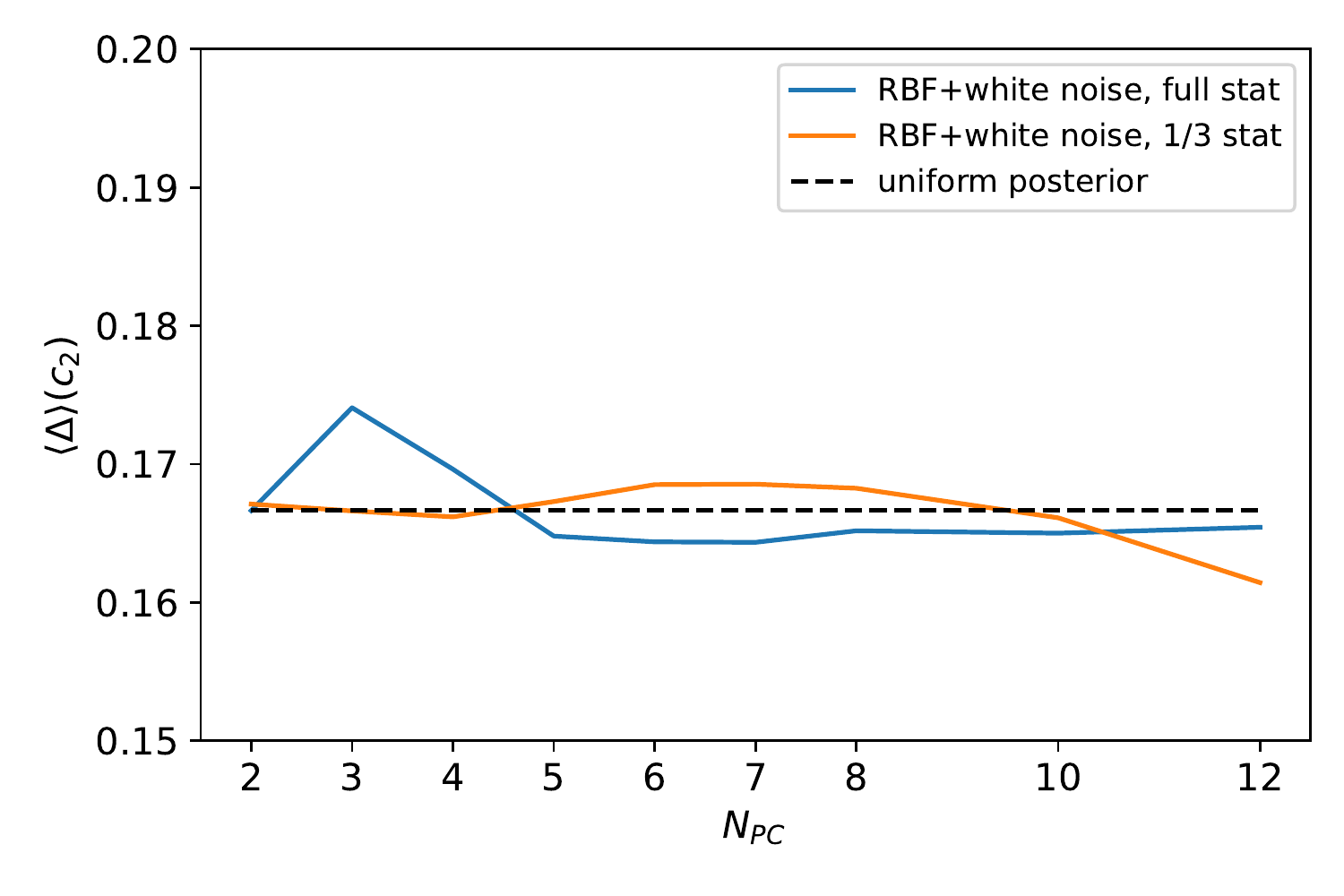}
	\includegraphics[width=0.48\textwidth]{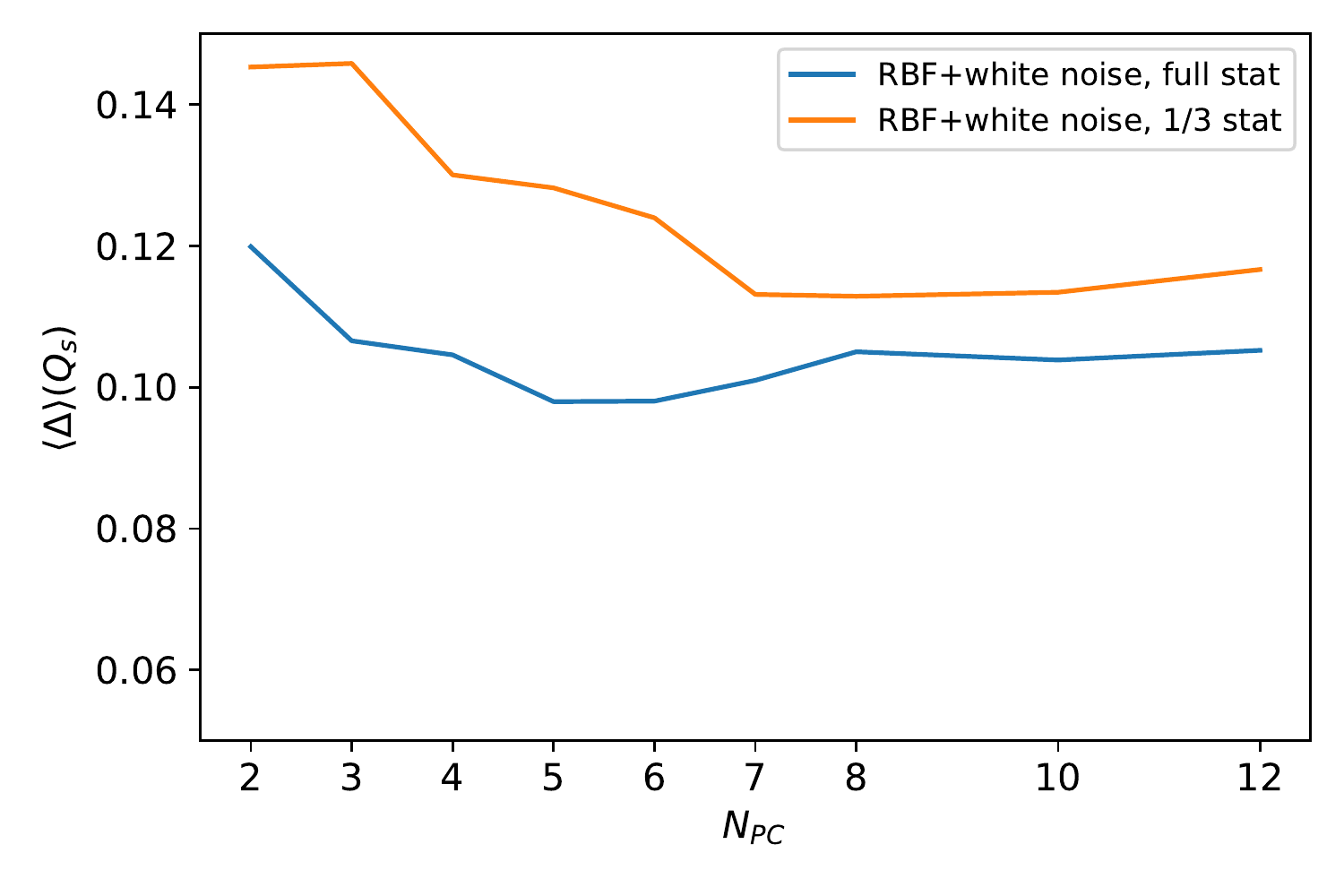}
	\caption{\label{fig:Delta_comparison_stat} Comparison of $\langle\Delta\rangle$ for different parameters with the same settings for the emulator but different statistics for the training data.}
\end{figure}

One question can be answered is how much does going from using $1/3$ of the statistics to using all the statistics improve the constraint on the parameters. In Fig.~\ref{fig:Delta_comparison_stat}, we can see that $\langle\Delta\rangle(\alpha_s)$ and $\langle\Delta\rangle(Q_s)$ does improve. However, $\langle\Delta\rangle$ for $c_1$ and $c_2$ still float around the dashed line representing result calculated with a uniform posterior, showing that $c_1$ and $c_2$ are indeed the most difficult to constrain. It is still unclear how much using more design points and more statistics for each design point can help constrain these two parameters at this moment. 

\begin{figure}
	\centering
	\includegraphics[width=0.96\textwidth]{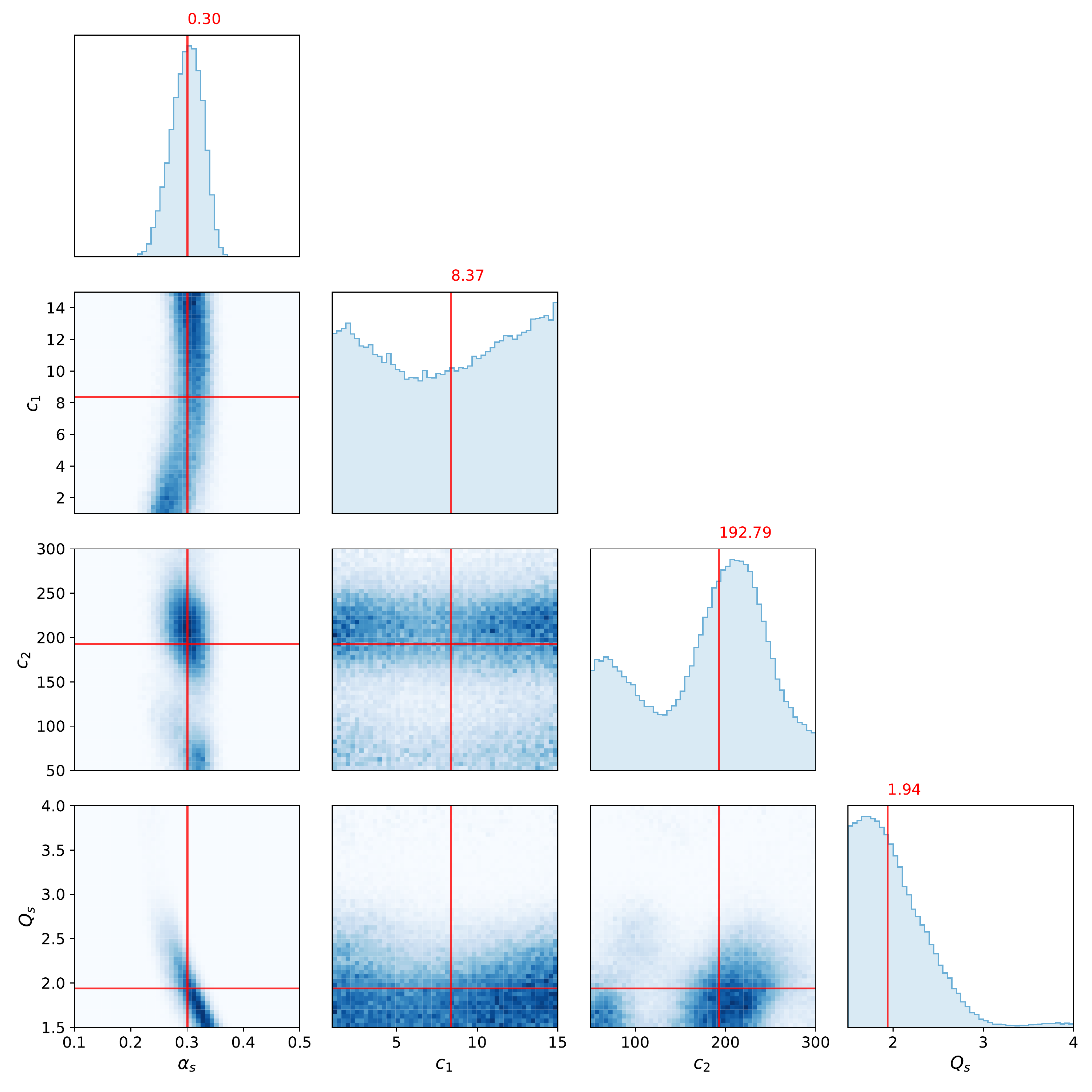}
	\caption{\label{fig:PosteriorEstimation_more_fluc_0.02} The posterior distribution of the model parameters. A Gaussian noise with zero $\mu=0$ and $\sigma=0.02$ is added to all the training data.}
\end{figure}

\begin{figure}
	\centering
	\includegraphics[width=0.96\textwidth]{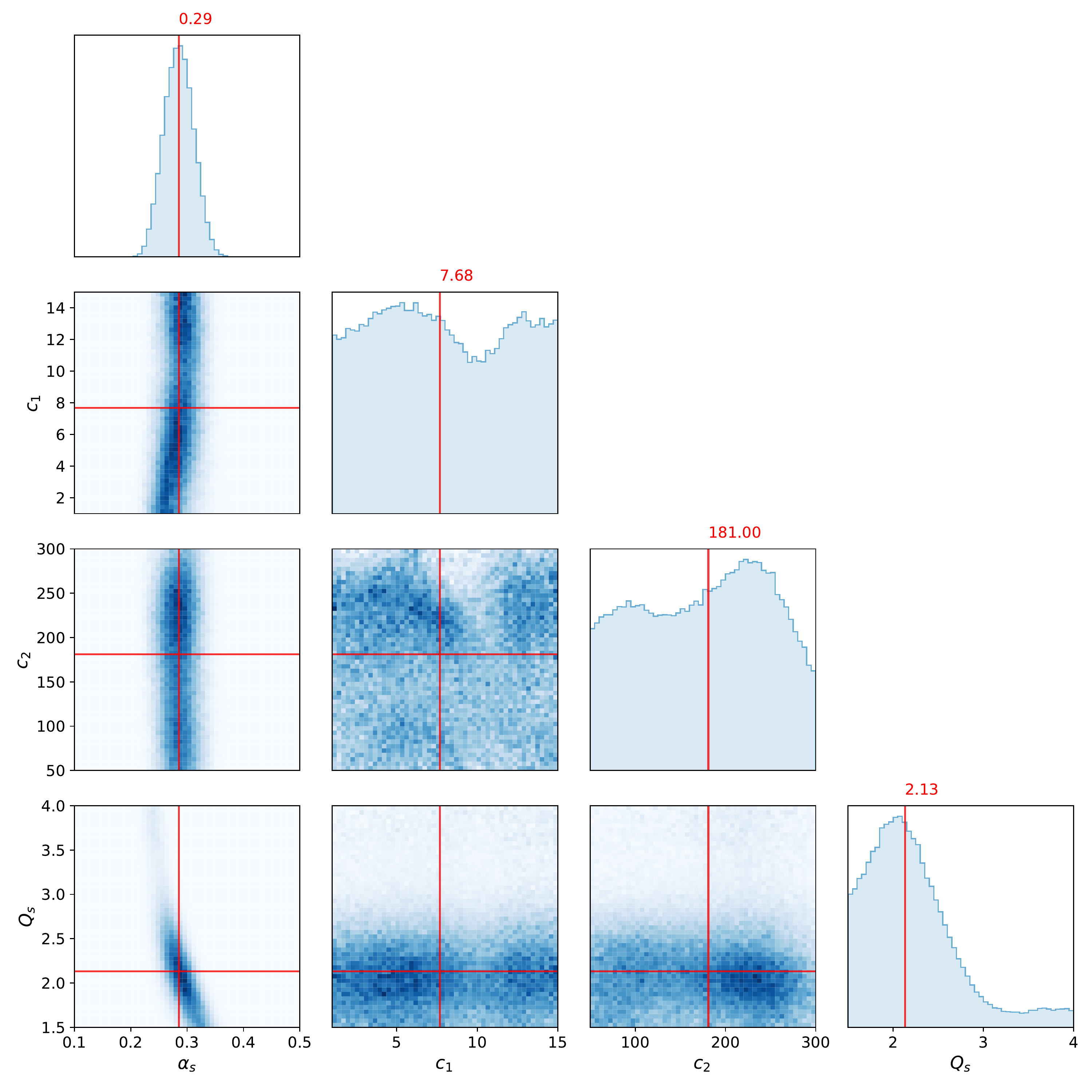}
	\caption{\label{fig:PosteriorEstimation_more_fluc_0.05} The posterior distribution of the model parameters. A Gaussian noise with zero $\mu=0$ and $\sigma=0.05$ is added to all the training data.}
\end{figure}

\begin{figure}
	\centering
	\includegraphics[width=0.96\textwidth]{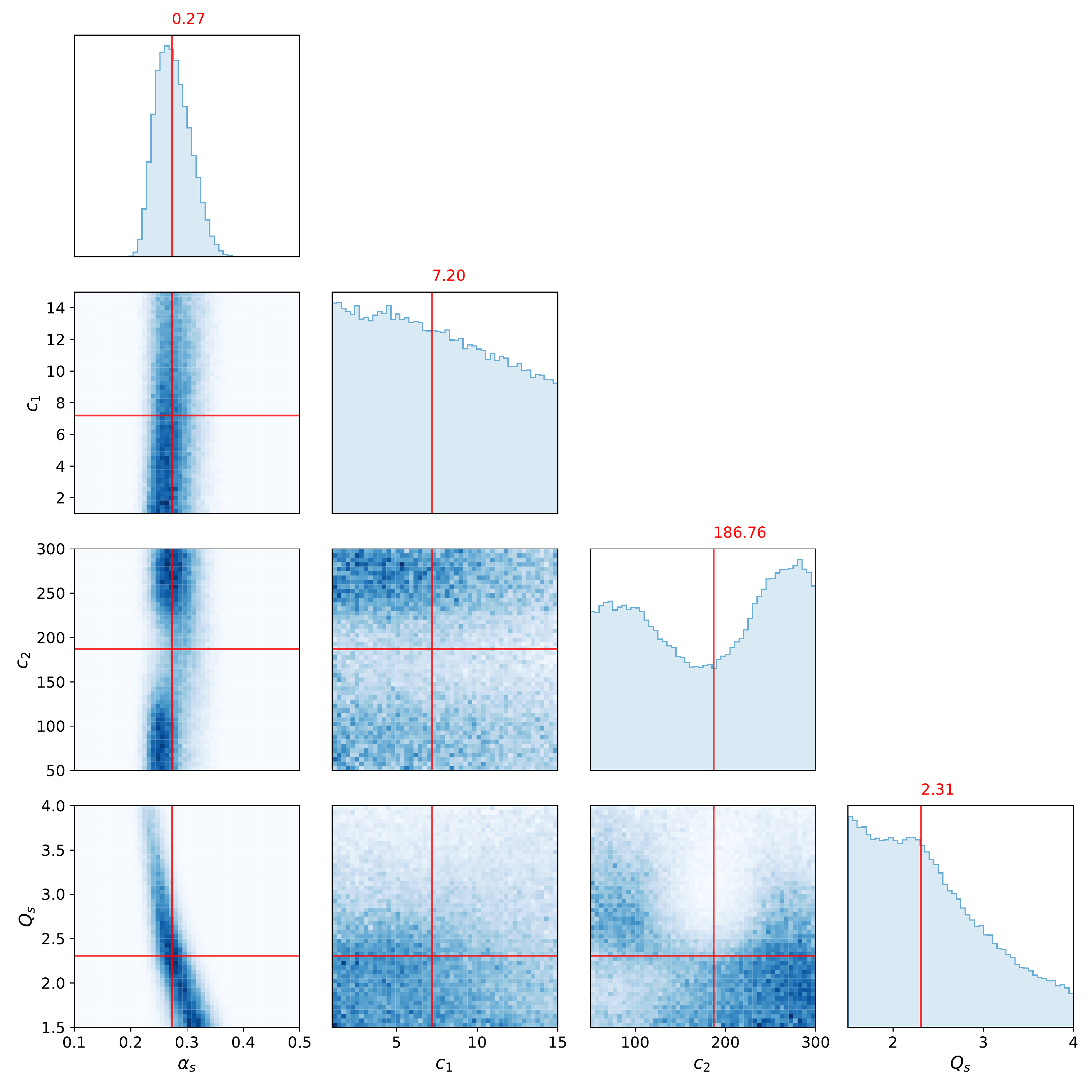}
	\caption{\label{fig:PosteriorEstimation_more_fluc_0.1} The posterior distribution of the model parameters. A Gaussian noise with $\mu=0$ and $\sigma=0.1$ is added to all the training data.}
\end{figure}

\begin{figure}
	\centering
	\includegraphics[width=0.96\textwidth]{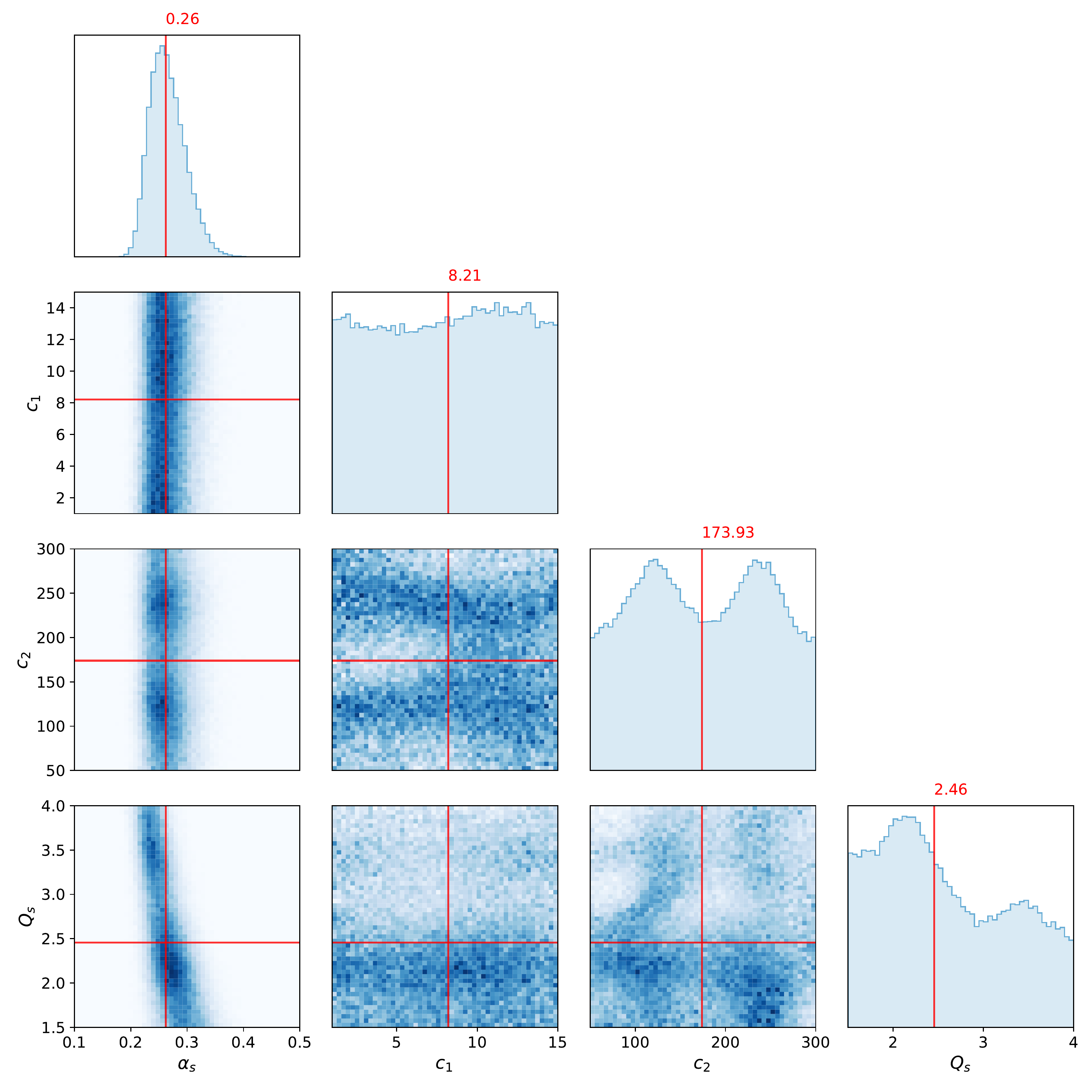}
	\caption{\label{fig:PosteriorEstimation_more_fluc_0.2} The posterior distribution of the model parameters. A Gaussian noise with $\mu=0$ and $\sigma=0.2$ is added to all the training data.}
\end{figure}

Another way is to add Gaussian noise to all the model calculations for each design point. This time every observable gets an equal amount of additional statistical fluctuation. In Fig.~\ref{fig:PosteriorEstimation_more_fluc_0.02}, Fig.~\ref{fig:PosteriorEstimation_more_fluc_0.05}, Fig.~\ref{fig:PosteriorEstimation_more_fluc_0.1}, and Fig.~\ref{fig:PosteriorEstimation_more_fluc_0.2}, we can see the constraints on each observable gradually reduce as more fluctuations are introduced. One should notice that additional noise with $0.02$ standard deviation is already larger than the model statistical fluctuation for low $p_T$ $R_{AA}$ data.

\section{Experimental covariance}\label{section:appendix_experiment_covariance}

\begin{figure}
	\centering
	\includegraphics[width=0.96\textwidth]{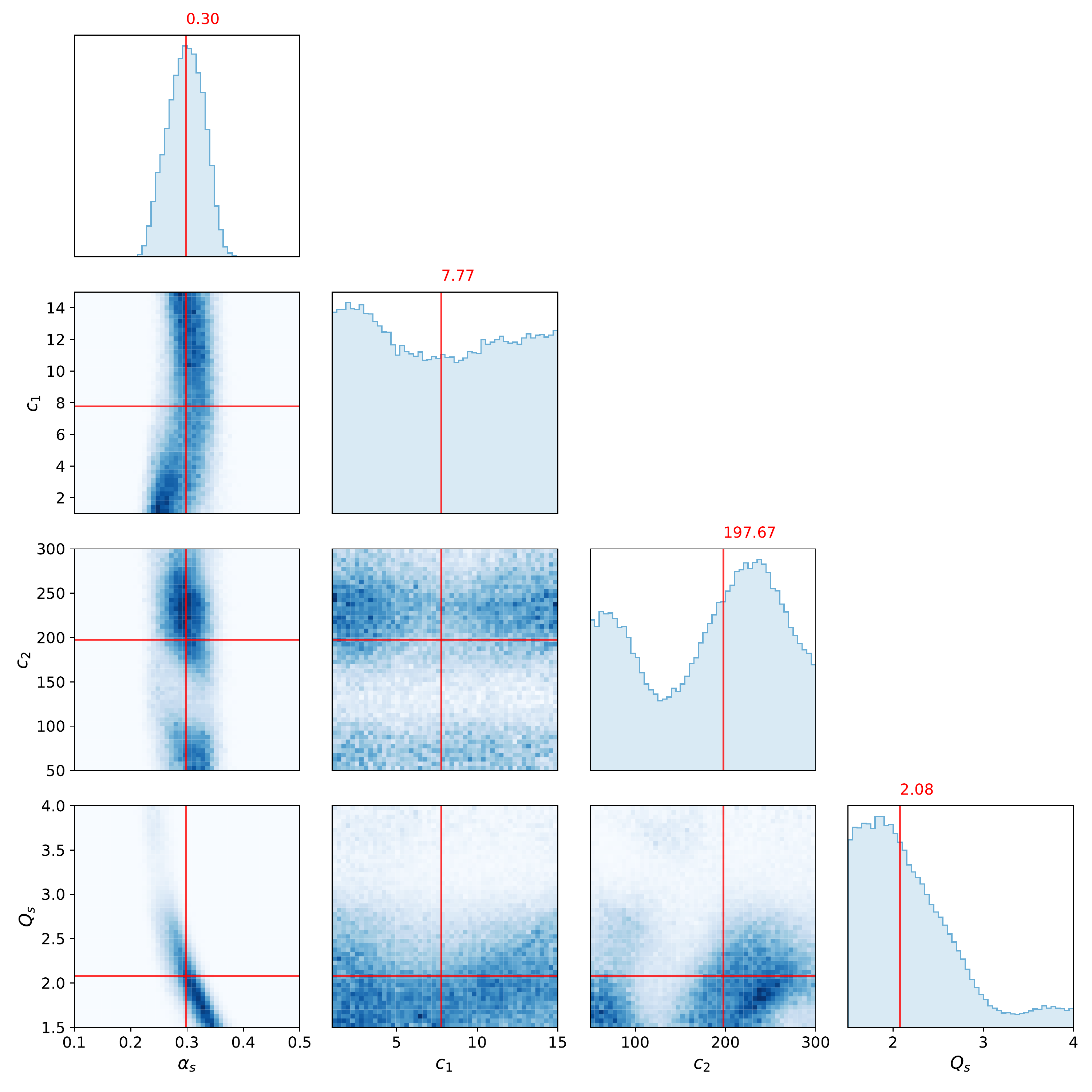}
	\caption[The posterior distribution of the model parameters]{\label{fig:PosteriorEstimation_Cor_Sys} The posterior distribution of the model parameters. A off diagonal systematic correlation with $l=0.1(p_{T,max}-p_{T,min})$ is assumed to be present within each observed $R_{AA}$. }
\end{figure}

At the moment the experimental covariance matrix is assumed to be diagonal since the off diagonal correlations are not reported by experiments. However there are possible systematic correlations among different observables. A simple way to account for such correlation is to assume a Pearson correlation only among observables in the same class:
\begin{equation}
    \Sigma_{exp}^{sys}=\frac{1}{\sigma_i \sigma_j} exp[-\frac{1}{2}(\frac{x_{i}-x_j}{l})^2]
\end{equation}

where $\sigma_i, \sigma_j$ are the standard deviation of the two observable, $x_i-x_j$ measures the distance between the two observable in the space they are defined (for example, the distance in $p_T$ when measuring $R_{AA}$). The hyper-parameter $l$ controls the correlation length and can be varied. In principle, a larger $l$ means more correlation across the measurements and less constraint on the parameters. The posterior distribution using this experimental covariance matrix is shown in Fig.~\ref{fig:PosteriorEstimation_Cor_Sys} and similar qualitative behavior is observed as in Fig.~\ref{fig:BestPosteriorEstimation}.

\section{Posterior distribution using the Mat\'{e}rn kernel}\label{section:appendix_matern kernel}

\begin{figure}
	\centering
	\includegraphics[width=0.96\textwidth]{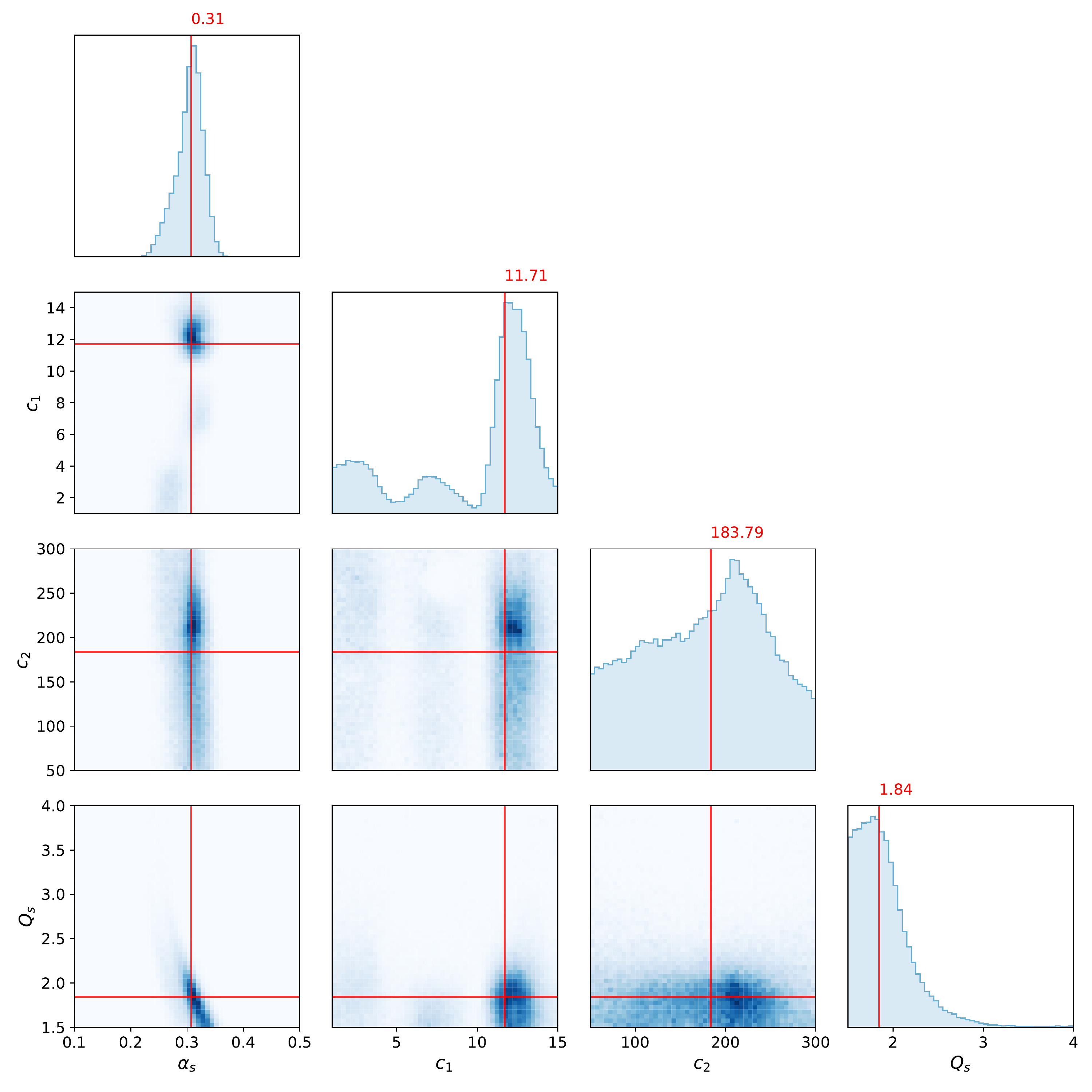}
	\caption[The posterior distribution of the model parameters]{\label{fig:PosteriorEstimation_Matern_kernel} The posterior distribution of the model parameters. The emulator is using the Mat\'{e}rn kernel with $\nu=5/2$ and $N_{PC}=6$.}
\end{figure}

The choice of kernel in the emulator also plays a important role in Bayesian analysis. The Mat\'{e}rn kernel \cite{schulz2018tutorial} encodes prior assumption of the smoothness of the underlying function. The posterior distribution using the Mat\'{e}rn($\nu=5/2$)+white noise kernel with $6$ principal components is shown in Fig.~\ref{fig:PosteriorEstimation_Matern_kernel}. This is the settings found to give the smallest product of all $\langle\Delta\rangle$ in Section~\ref{sec:quantitative_closure_test}. Similar posterior distribution is seen for $\alpha_s$ and $Q_s$ compared to the RBF + white noise kernel in Fig.~\ref{fig:BestPosteriorEstimation}. The main difference is that now there is a also strong peak for $c_1$. As mentioned in Section~\ref{sec:quantitative_closure_test}, we shouldn't expect tight constraint on $c_1$ and $c_2$ as the average $\langle\Delta\rangle$ for these two parameters is close to the one calculated with a uniform posterior. 

\begin{figure}
	\centering
	\includegraphics[width=0.32\textwidth]{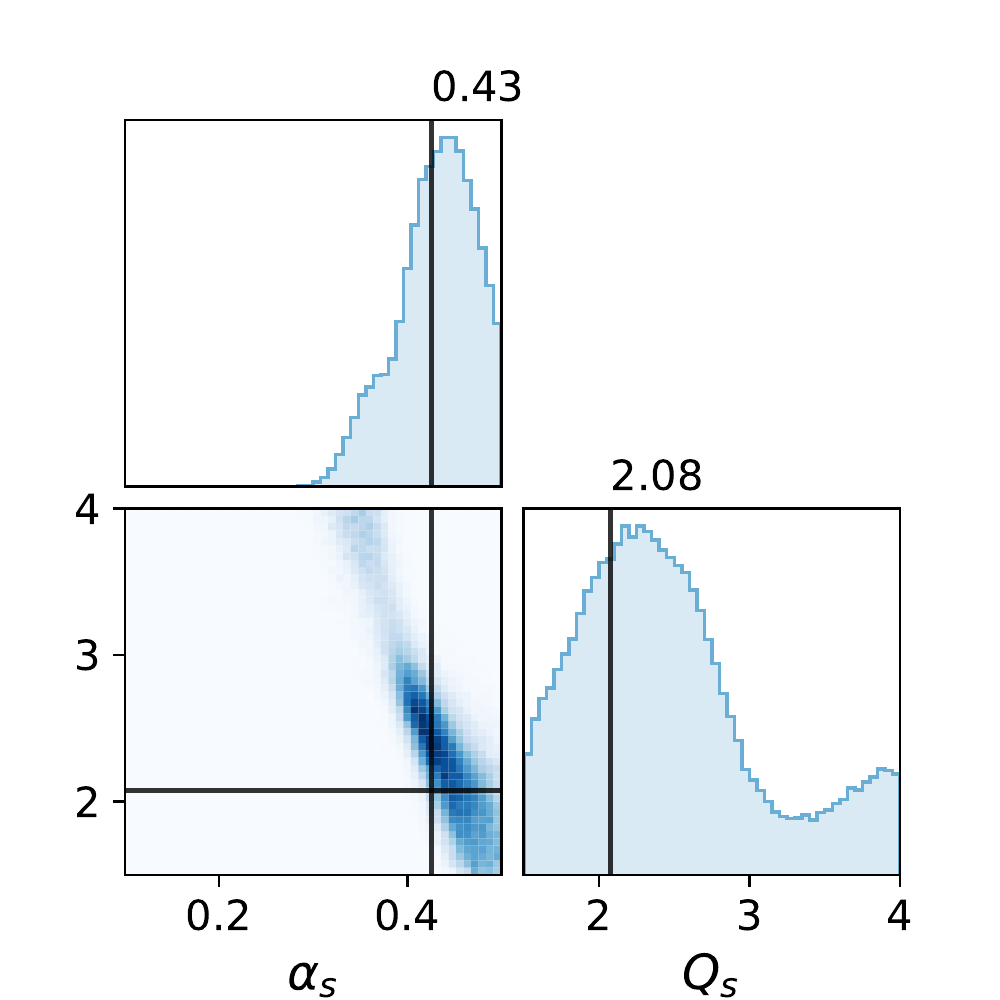}
	\includegraphics[width=0.32\textwidth]{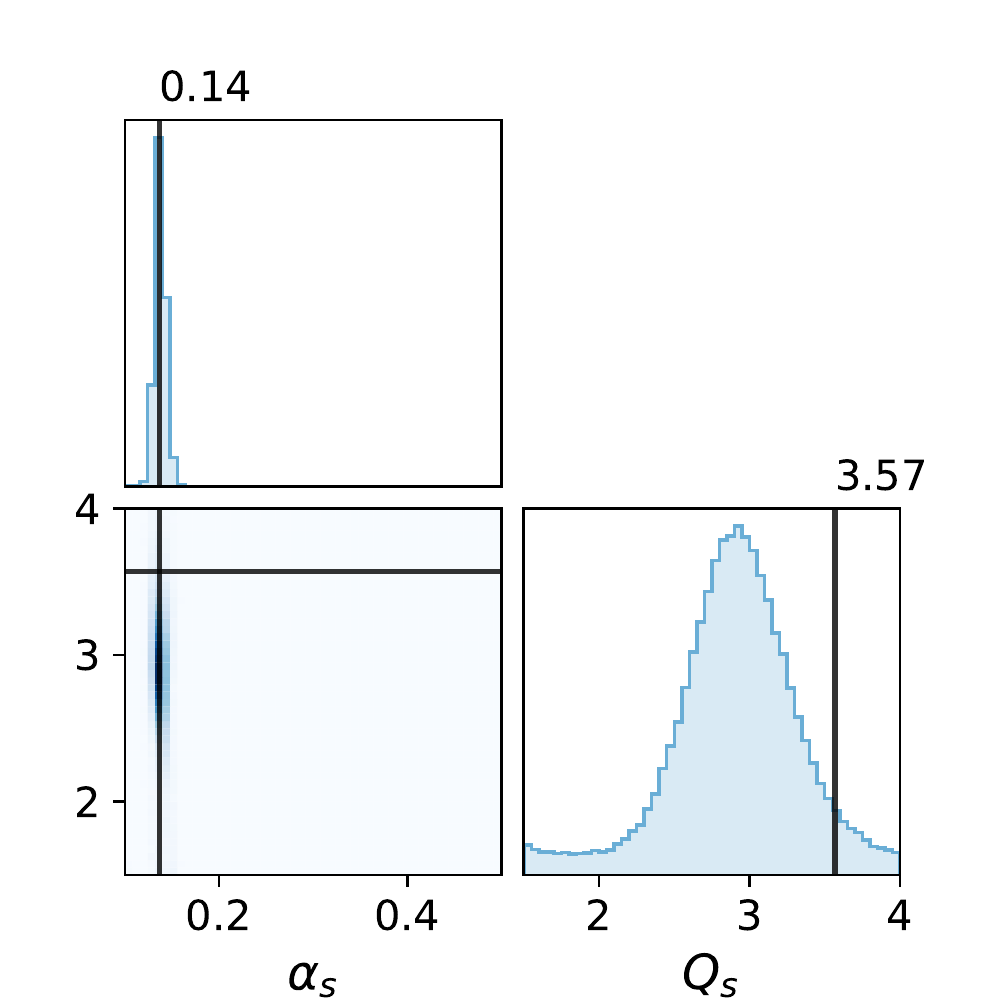}
	\includegraphics[width=0.32\textwidth]{images/Bayesian_Results/Matern_kernel/10/Correlation2.pdf}
	\includegraphics[width=0.32\textwidth]{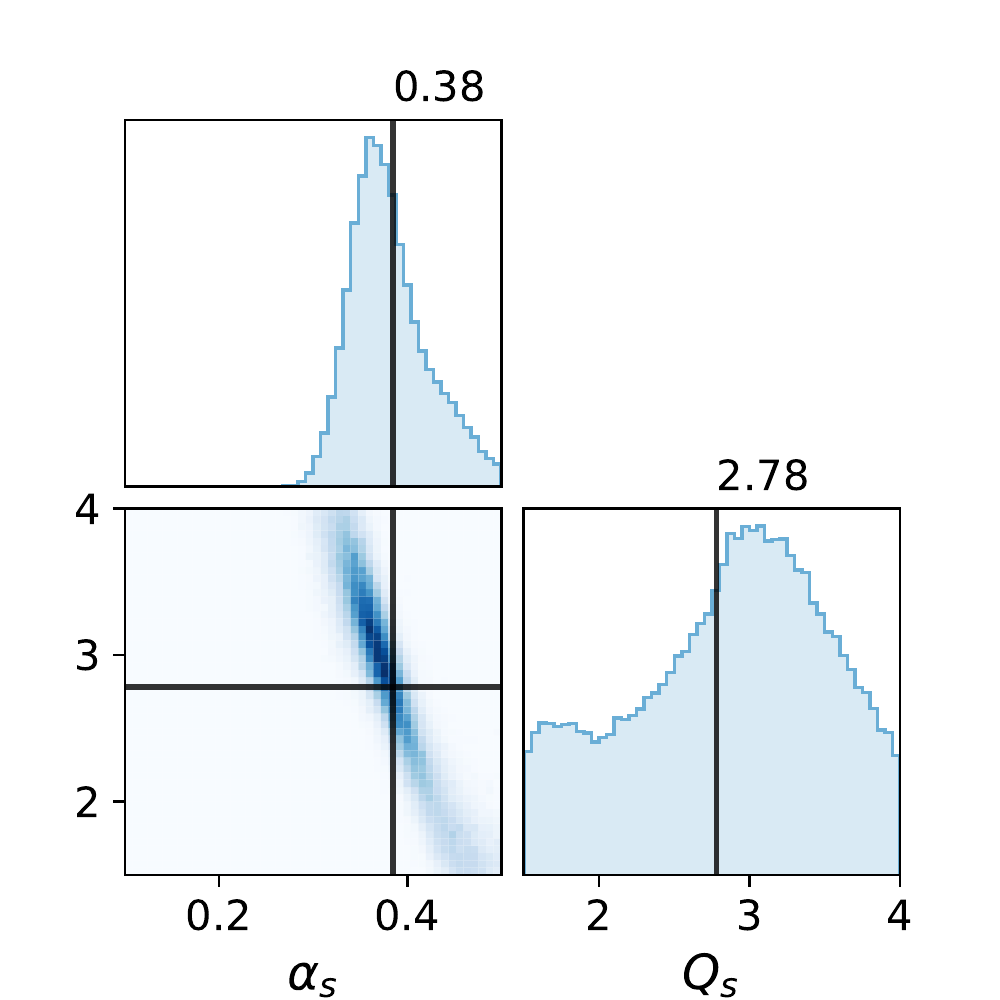}
	\includegraphics[width=0.32\textwidth]{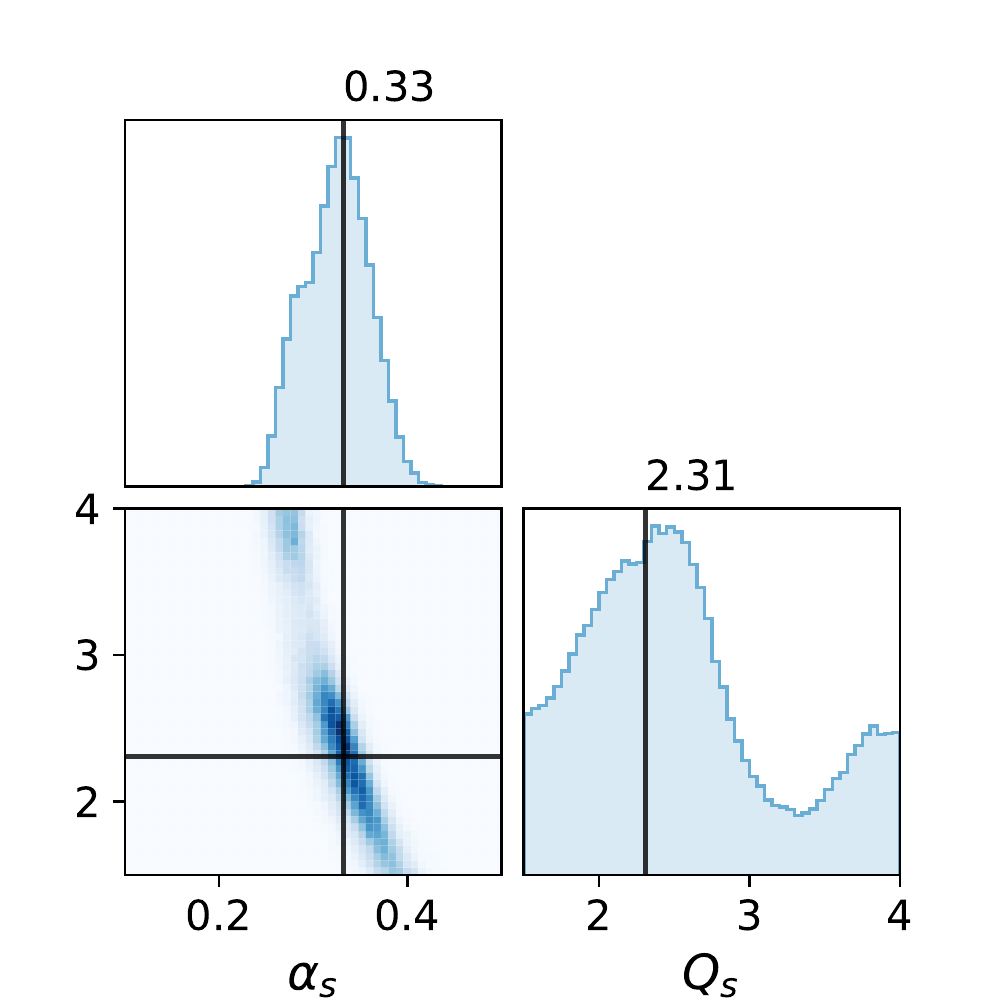}
	\includegraphics[width=0.32\textwidth]{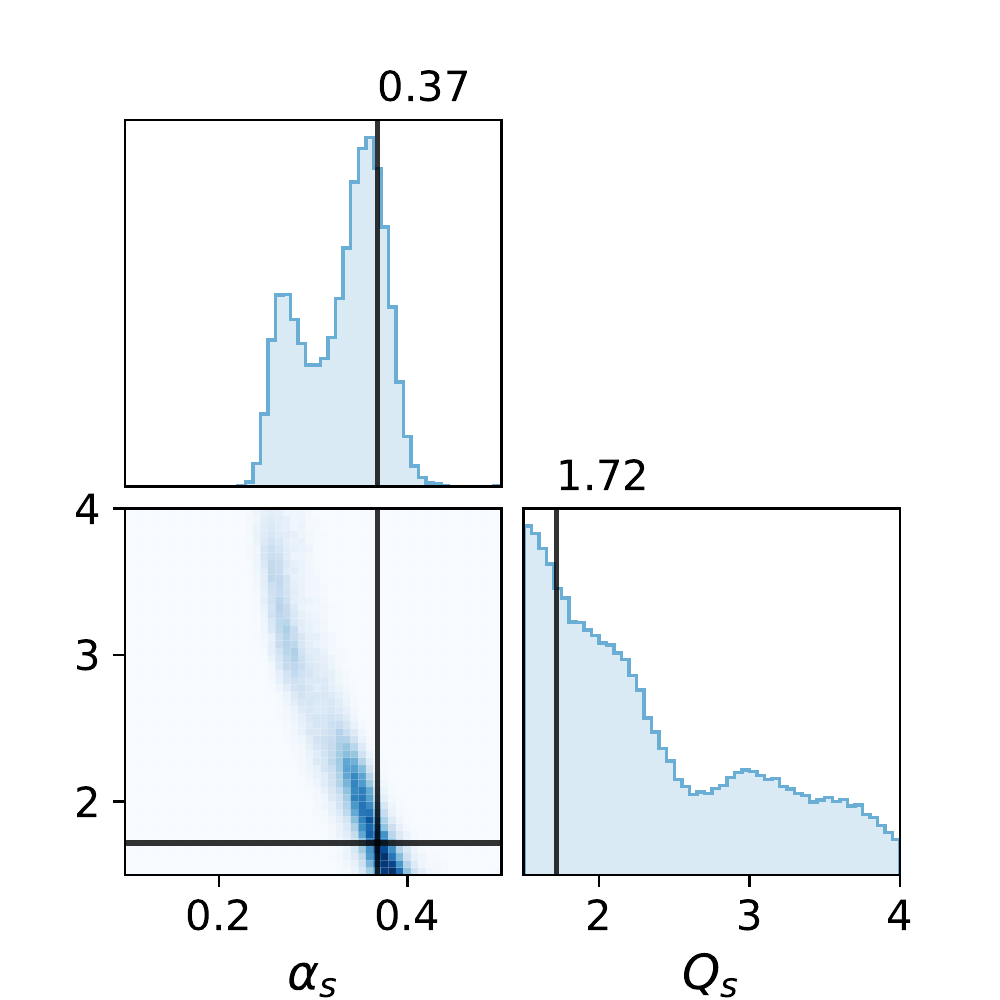}
	\includegraphics[width=0.32\textwidth]{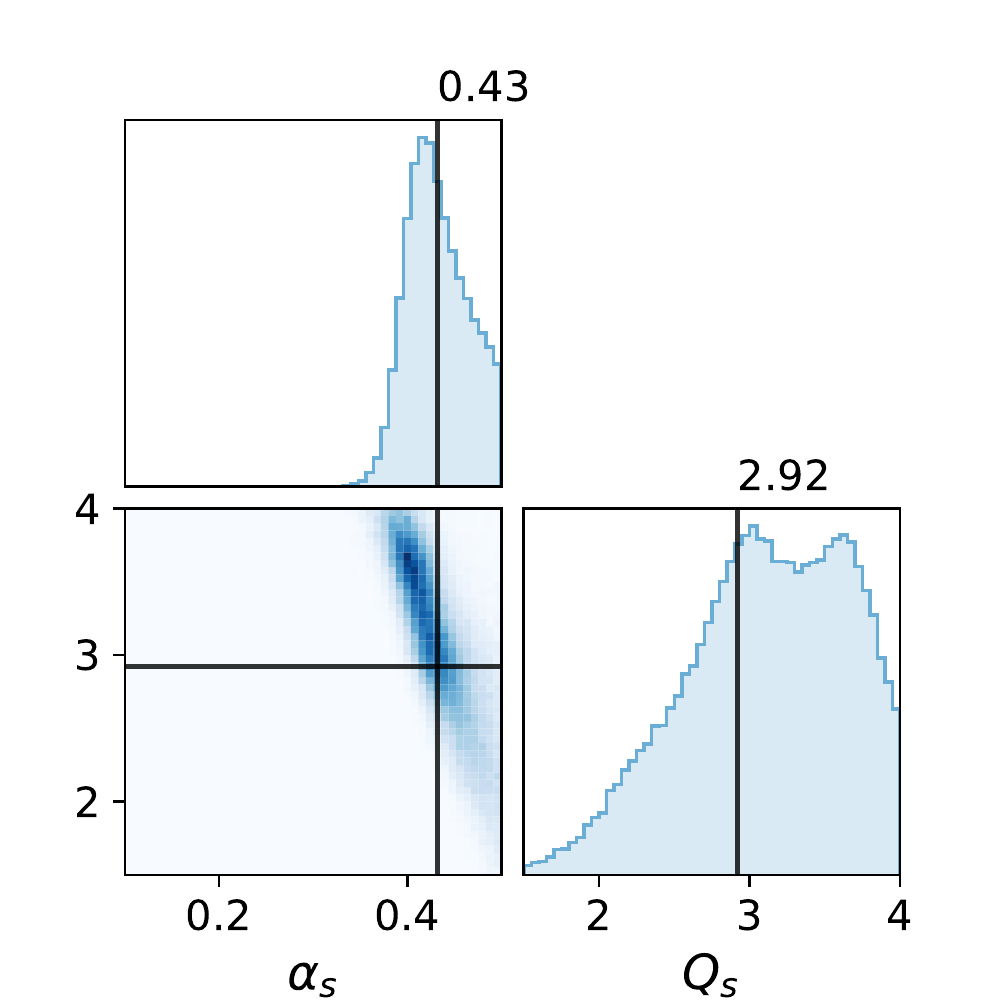}
	\includegraphics[width=0.32\textwidth]{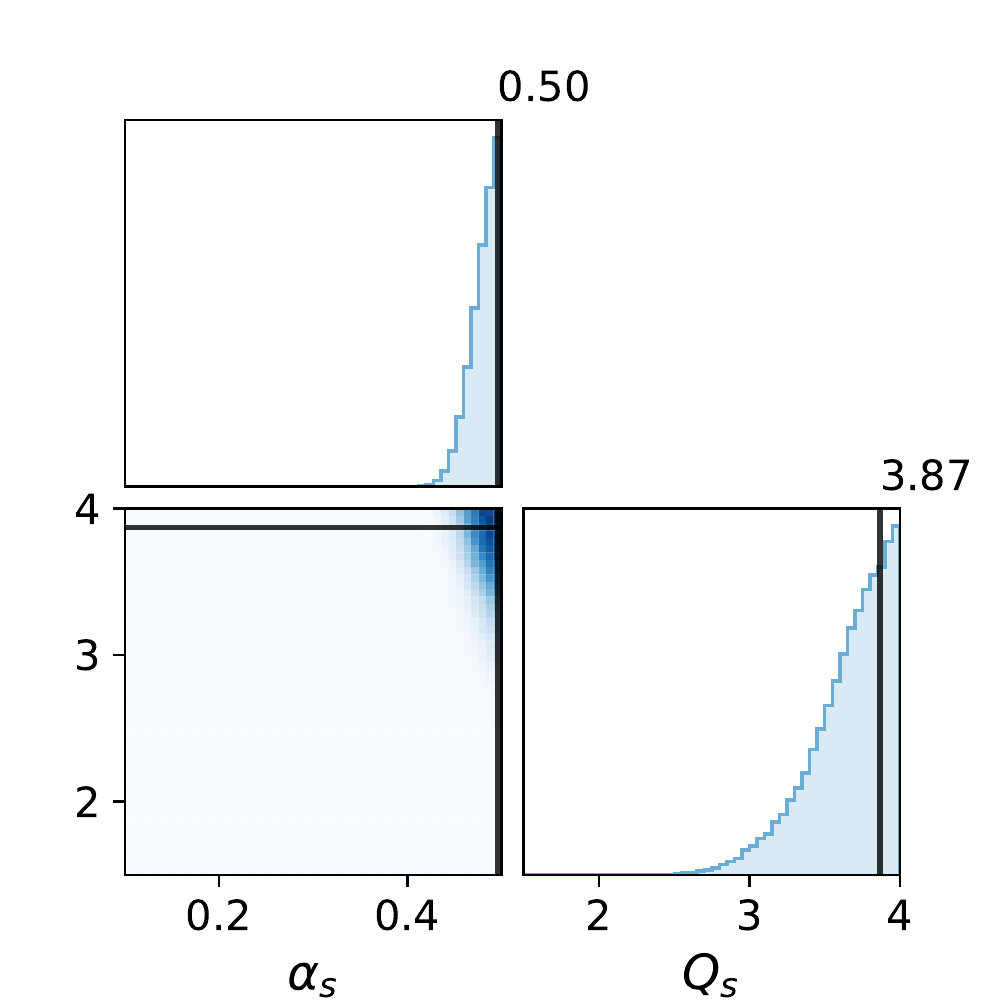}
	\caption[Closure tests of $\alpha_s$ and $Q_s$ using 8 sets of mock data]{\label{fig:closure_matern_14} Closure tests of $\alpha_s$ and $Q_s$ using 8 sets of mock data. The emulator is using the Mat\'{e}rn kernel with $\nu=5/2$ and $N_{PC}=6$.}
\end{figure}

\begin{figure}
	\centering
	\includegraphics[width=0.48\textwidth]{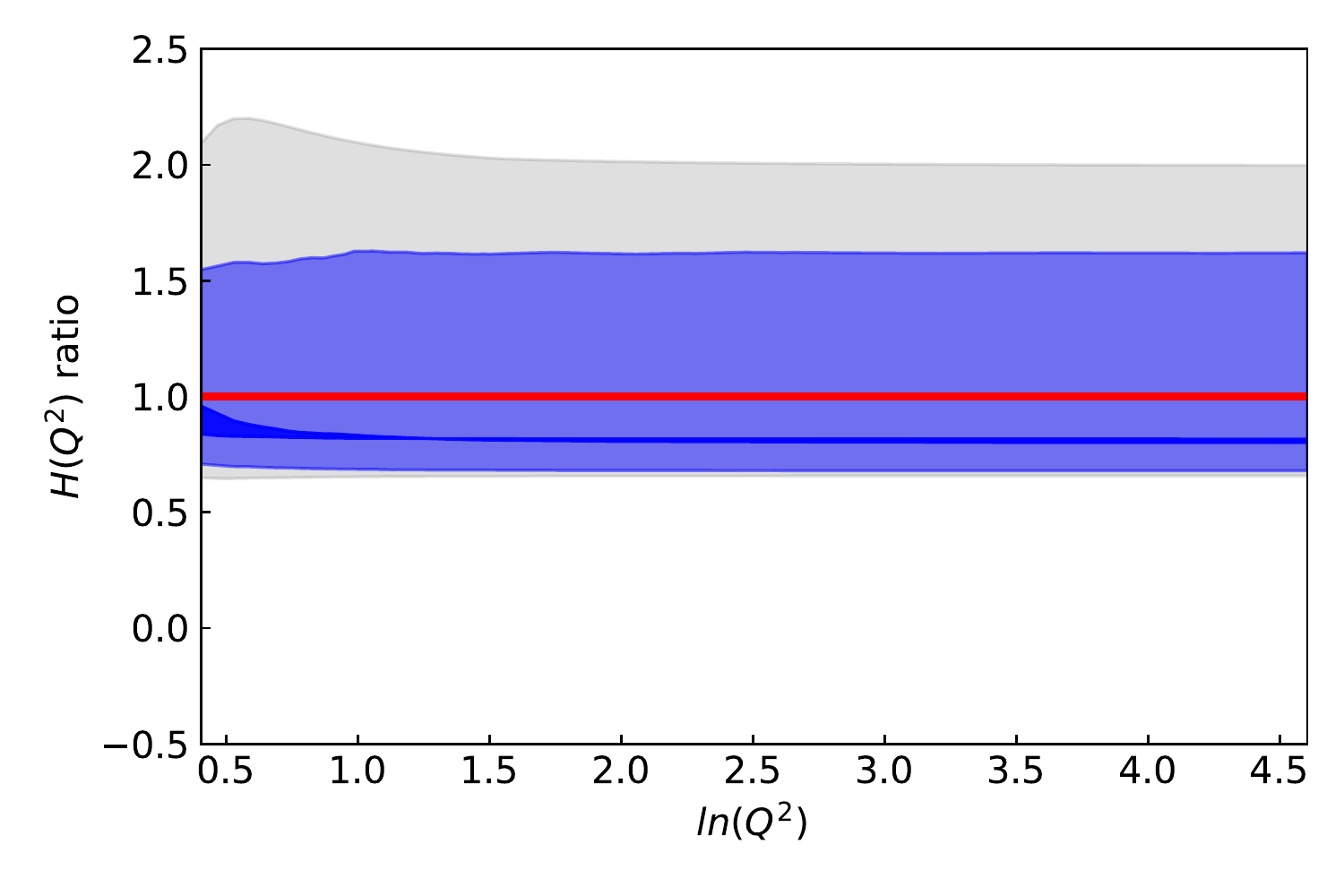}
	\includegraphics[width=0.48\textwidth]{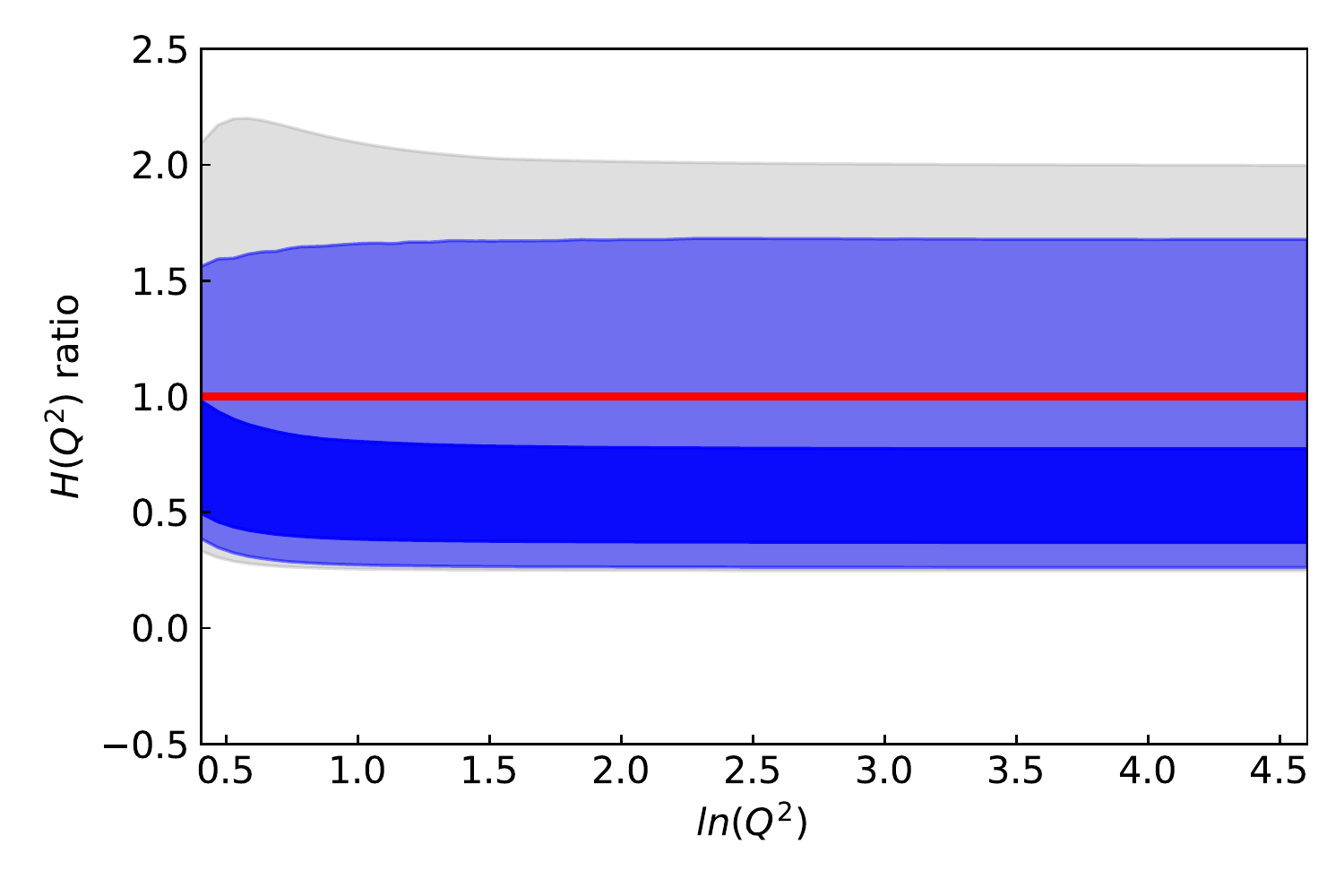}
	\includegraphics[width=0.48\textwidth]{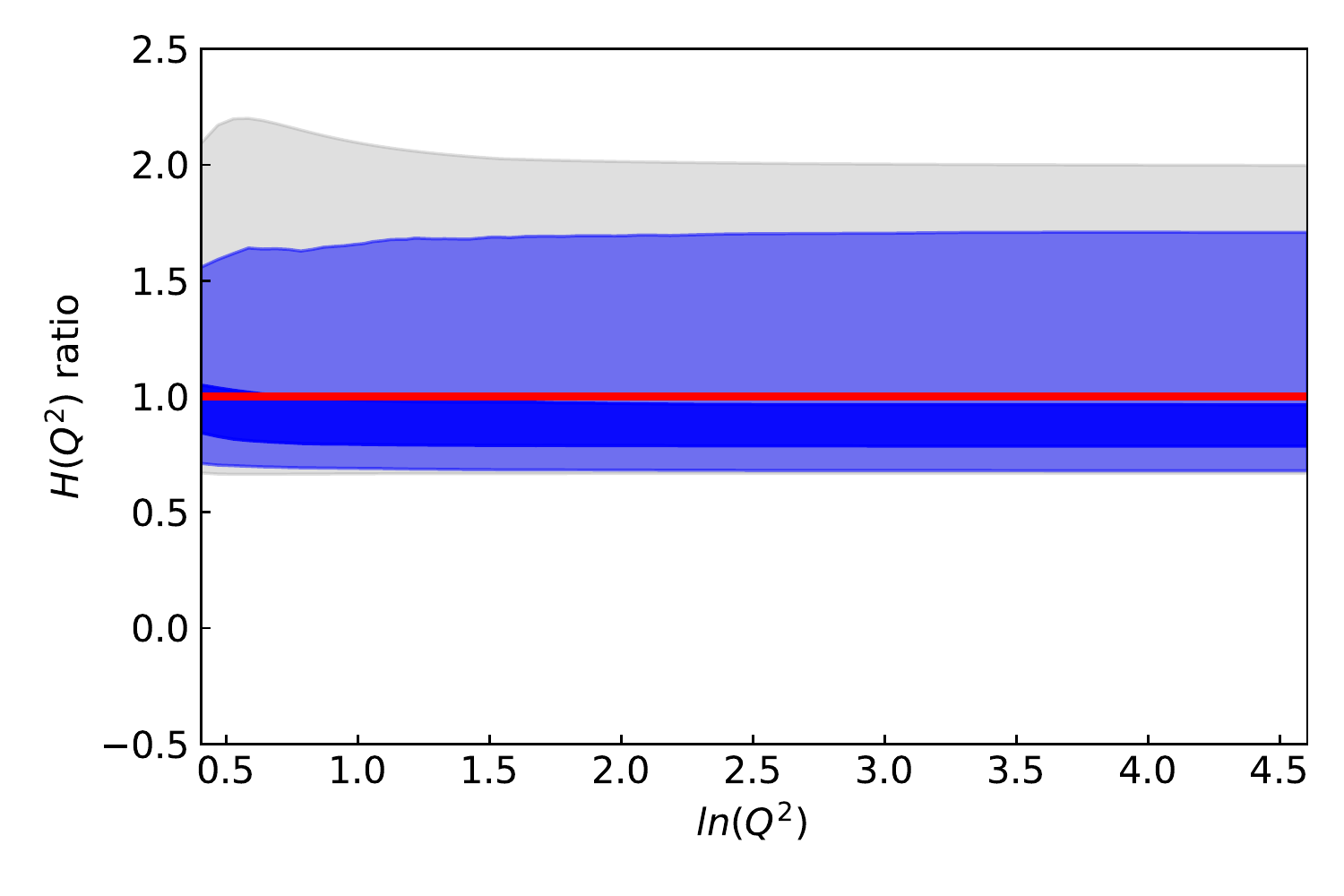}
	\includegraphics[width=0.48\textwidth]{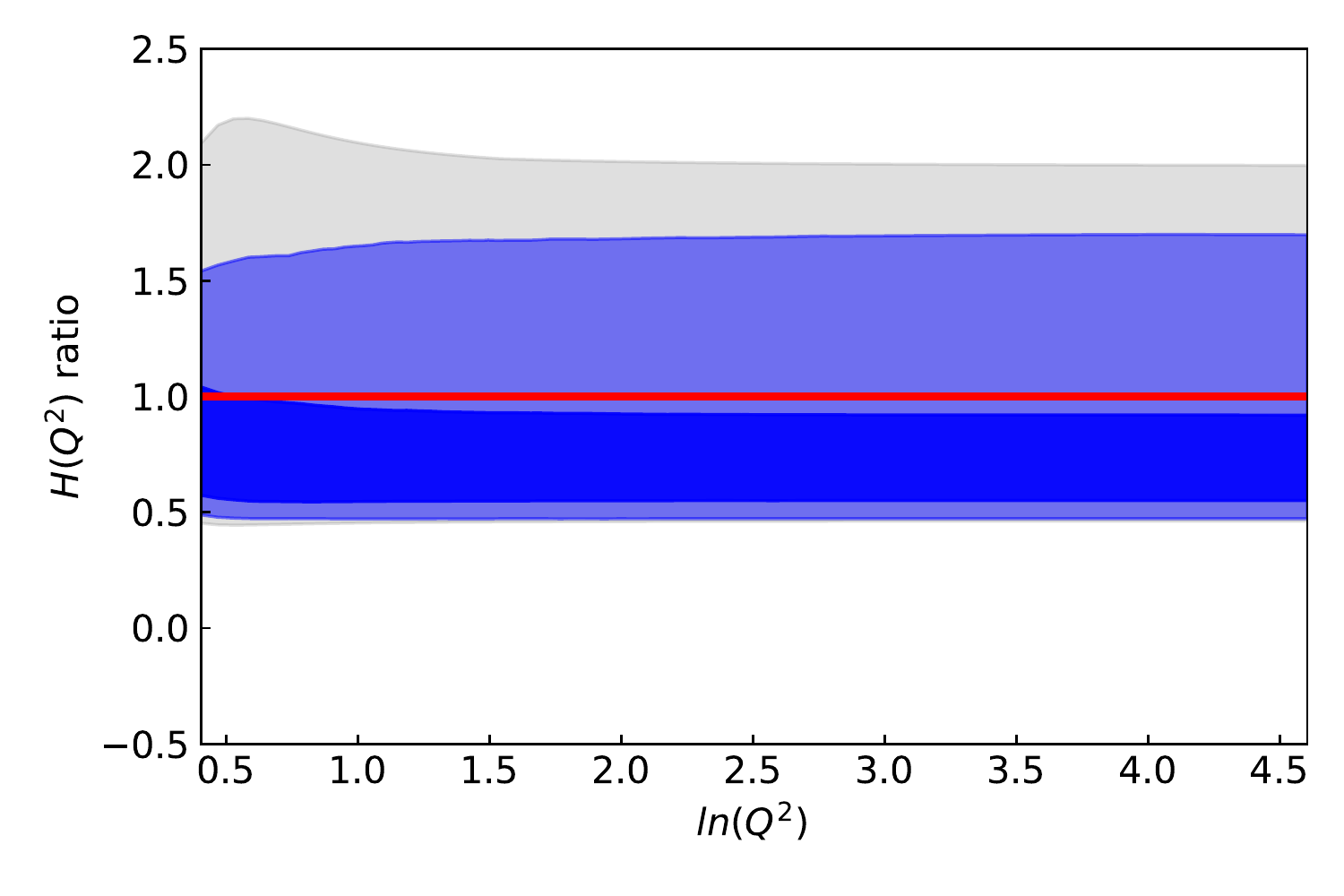}
	\includegraphics[width=0.48\textwidth]{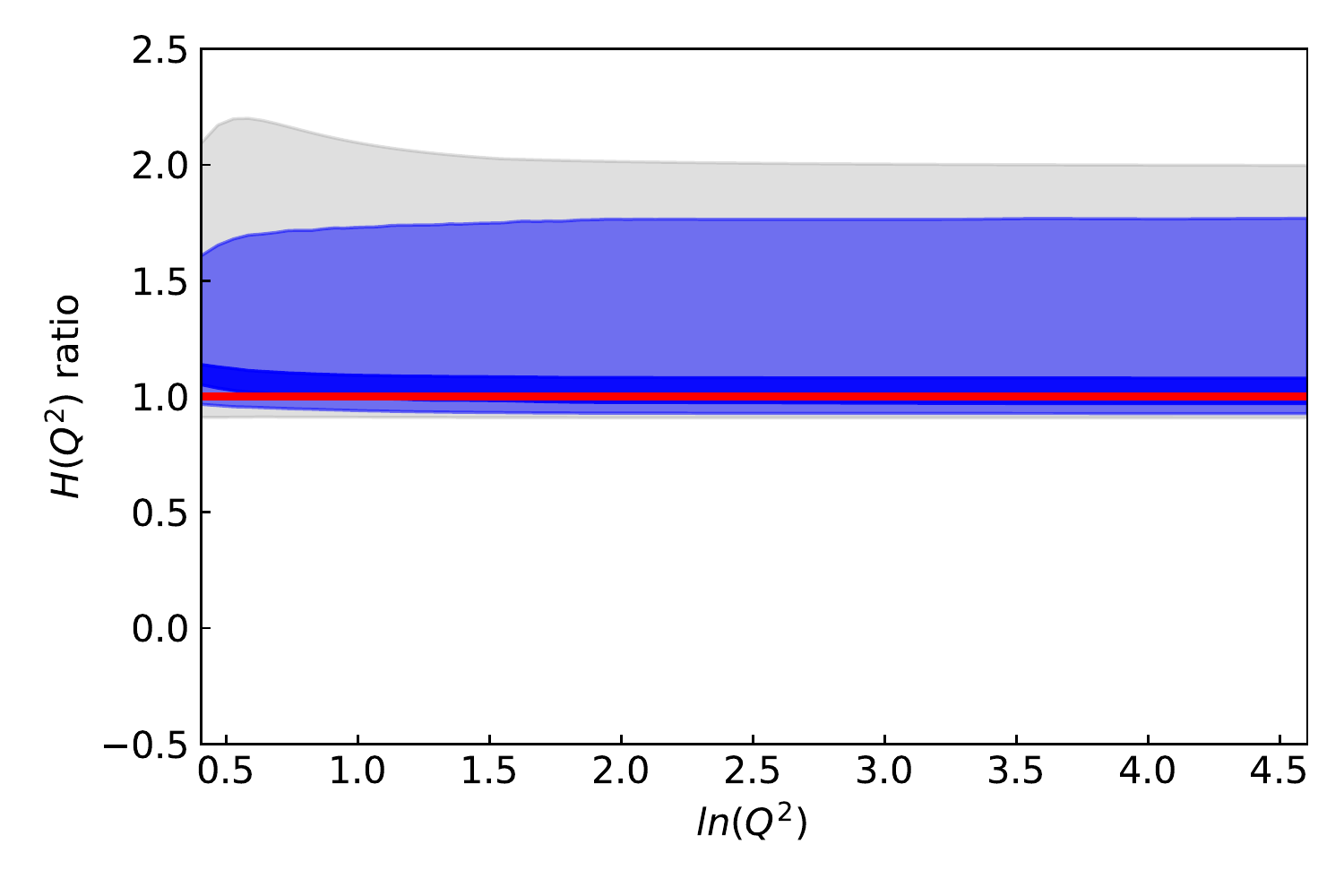}
	\includegraphics[width=0.48\textwidth]{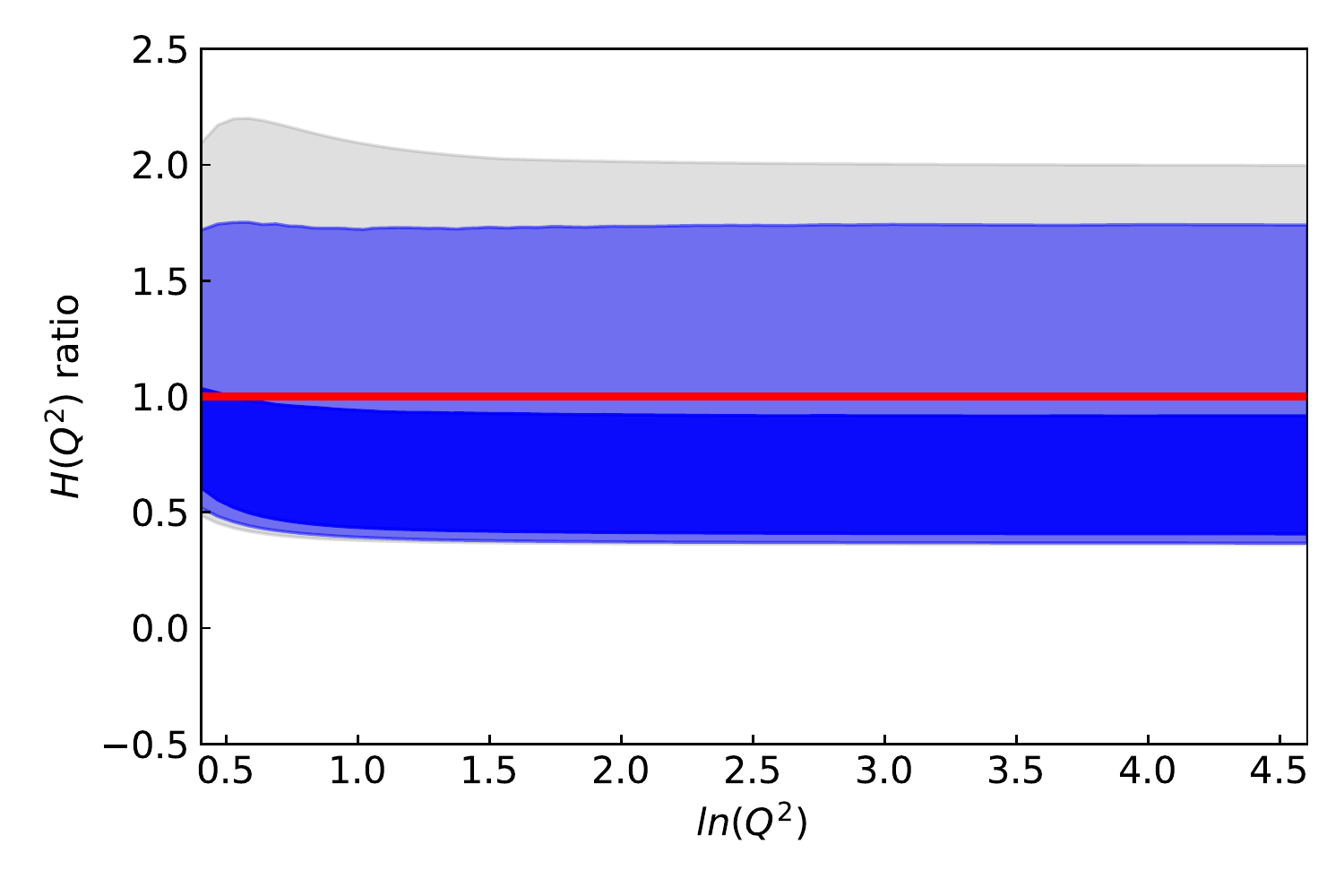}
	\includegraphics[width=0.48\textwidth]{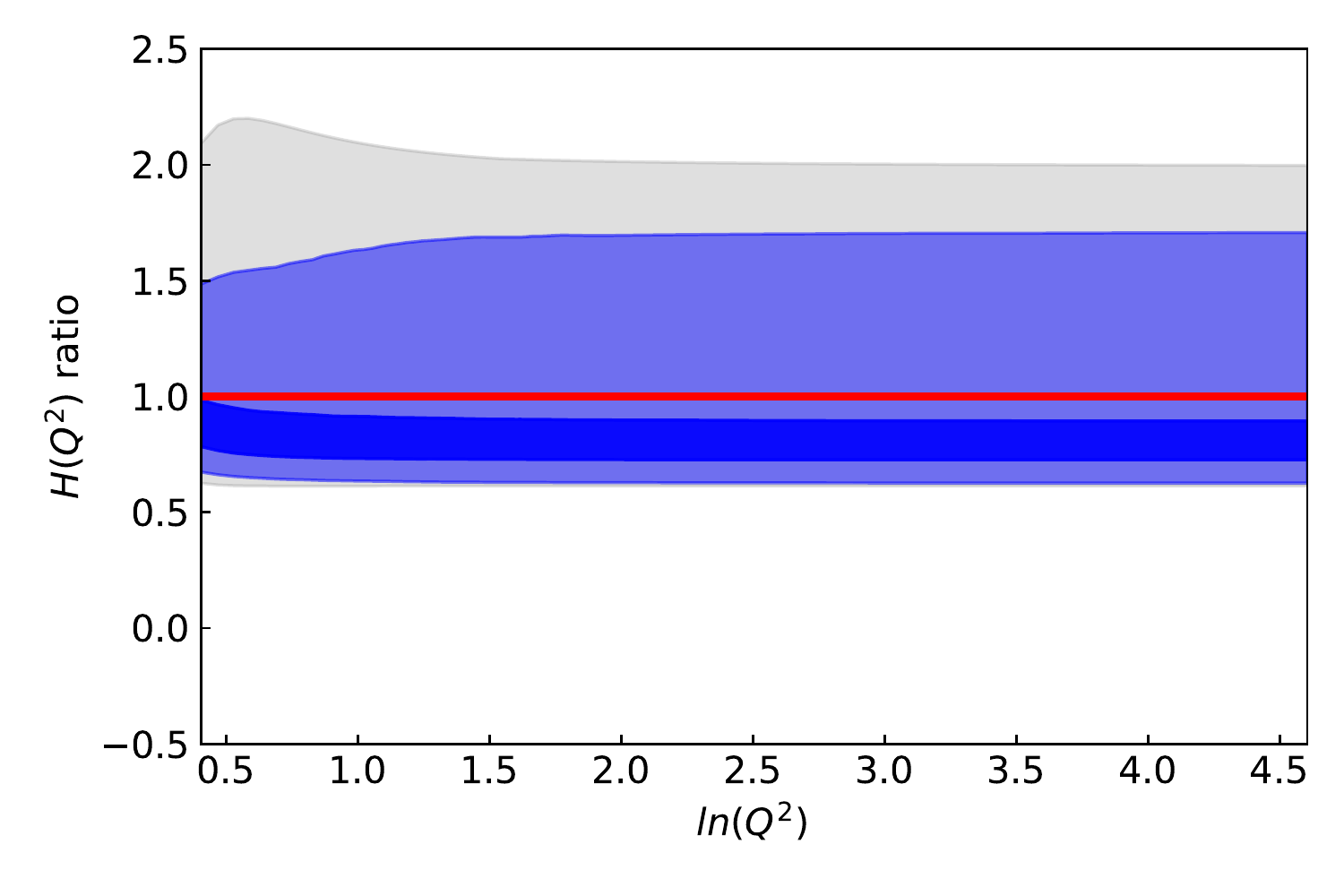}
	\includegraphics[width=0.48\textwidth]{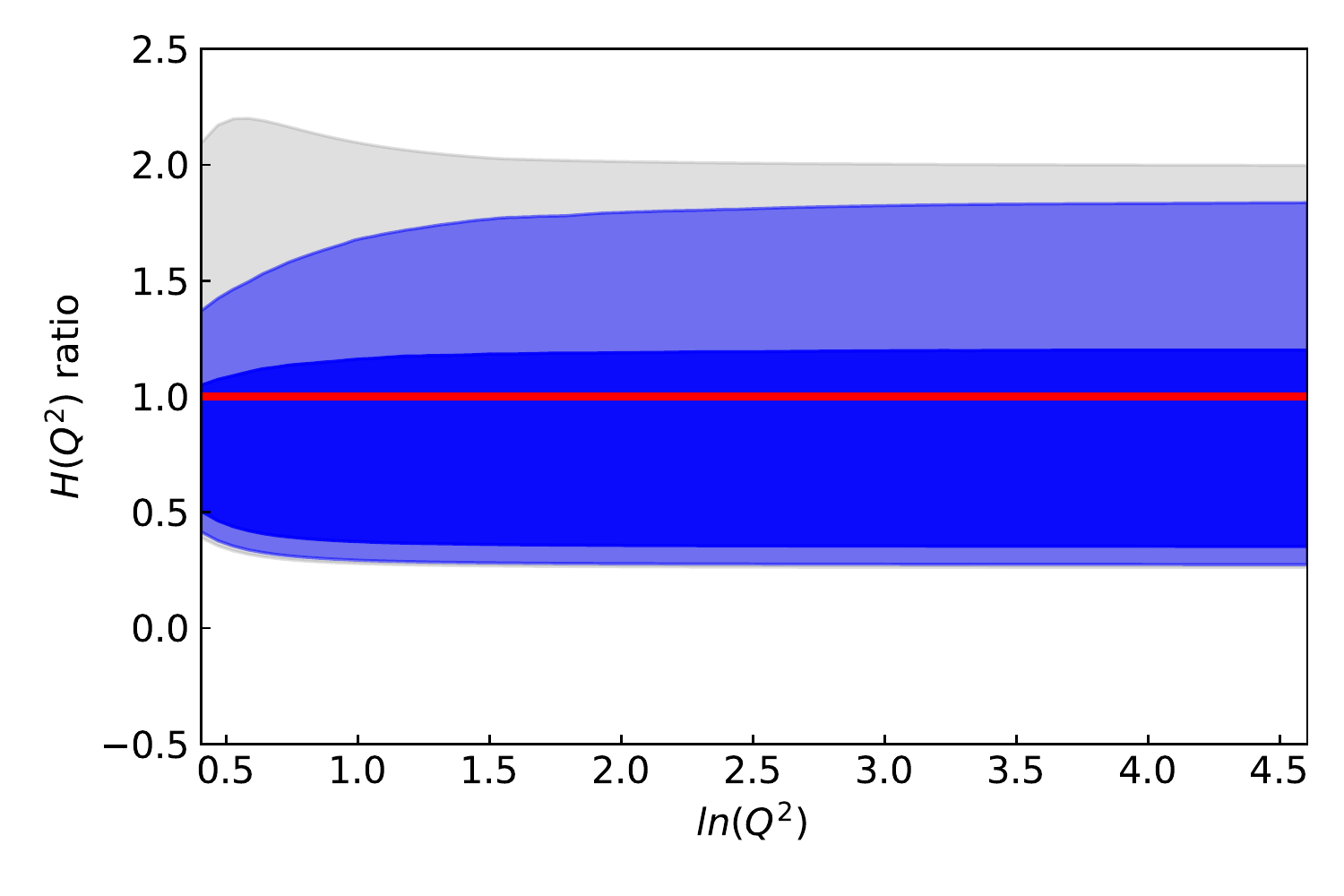}
	\caption[Closure tests of $c_1$ and $c_2$ plotted as the ratio of $H(Q^2)$ using 8 sets of mock data]{\label{fig:closure_matern_23} Closure tests of $c_1$ and $c_2$ plotted as the ratio of $H(Q^2)$ using 8 sets of mock data. The emulator is using the Mat\'{e}rn kernel with $\nu=5/2$ and $N_{PC}=6$.}
\end{figure}

The result of closure test using 8 random design points is shown in Fig.~\ref{fig:closure_matern_14} and Fig.~\ref{fig:closure_matern_23}. No qualitative difference is found compared to Fig.~\ref{fig:closure_exp_fluc_14} and Fig.~\ref{fig:closure_exp_fluc_23} which uses the RBF+white noise kernel. This is why $\langle\Delta\rangle$ or other metrics is needed for a quantitative comparison between different kernels.

\label{sec:appendix}

\nocite{*}
\bibliographystyle{ieeetr}
\bibliography{lit}

\end{document}